\documentclass[final,hyperfinal]{uoythesis}
\usepackage[scaled]{helvet}

\setsecheadstyle{\large\bfseries}
\setsubsecheadstyle{\normalsize\bfseries}
\setsubsubsecheadstyle{\normalsize\itshape}
\setparaheadstyle{\itshape}

\usepackage{uoythesisoptions}

\renewcommand{\chaptermark}[1]{\markboth{\thechapter.\ #1}{}}
\renewcommand{\sectionmark}[1]{\markright{\thesection.\ #1}}
\usepackage[utf8]{inputenc}  
\usepackage[T1]{fontenc}
\usepackage{lmodern}         

\usepackage{subfiles}
\graphicspath{
  {Figures/} 
}

\makeatletter
\newcommand*{\addFileDependency}[1]{
  \typeout{(#1)}
  \@addtofilelist{#1}
  \IfFileExists{#1}{}{\typeout{No file #1.}}
}
\makeatother

\usepackage{verbatim} 
\usepackage[math]{blindtext} 

\usepackage{xcolor} 
\usepackage[obeyDraft, textsize=small, colorinlistoftodos]{todonotes}
\usepackage{color}
\hypersetup{%
  colorlinks=true, 
 	allcolors={blue},    
    breaklinks=true
}

\usepackage{xspace} 
\usepackage{amsmath, amsthm}
\usepackage{amssymb, amsfonts}
\usepackage{mathrsfs} 
\usepackage{url}
\usepackage{leftindex}

\usetikzlibrary{arrows.meta, positioning}

\colorlet{mygray}{black!70}

\usepackage[eprint,          
            backend=biber,
            doi=true,
            date=year,
            eprint=true,
            sorting=none,
            url=false,
            citestyle=numeric-comp,
            maxnames=10,     
            giveninits=true, 
            safeinputenc     
           ]{biblatex}

\renewbibmacro{in:}{}

\usepackage{braket}

\usepackage{amsfonts}
\usepackage{amsmath}
\usepackage{amsthm}
\usepackage{amssymb}
\usepackage{siunitx}
\usepackage{graphicx}
\usepackage{lscape}
\usepackage[font=sf]{caption} 
\usepackage{dpfloat} 
 \usepackage{longtable}
\usepackage{fancyvrb}

\usepackage{doi}

\usepackage{url}

\usepackage{slashed}
\usepackage{faktor}
\usepackage{stackrel}
\usepackage{enumitem}
\usepackage{bbold}
\usepackage{csquotes}
\usepackage{dsfont}
\usepackage{listings}
\usepackage{microtype} 
\usepackage{blindtext}
\usepackage{multicol}

\usepackage{amsthm}
\usepackage{amsmath,amsfonts,amssymb,amsthm,epstopdf,titling,url,array,bm,mathtools}
\usepackage{xfrac}
\usepackage{braket}
\usepackage{physics}
\usepackage{thmtools}
\usepackage{thm-restate}

\usepackage{multirow}
\usepackage{graphicx}
\usepackage[table,xcdraw]{xcolor}
\usepackage{float}  
\usepackage{booktabs}
\usepackage{tikz-cd}
\usepackage{tikz}
\usetikzlibrary{decorations.markings,positioning}
\usetikzlibrary{shadows.blur}
\usepackage[export]{adjustbox}
\usepackage{standalone}
\usepackage[font=small]{caption}
\usepackage{subcaption}
\usepackage{tabularray}
\usepackage{enumitem}

\definecolor{lilac}{rgb}{0.2, 0.4, 0.2}
\definecolor{floralwhite}{rgb}{1.0, 0.98, 0.94}
\usepackage{footmisc}

\usepackage{todonotes}

\usepackage{soul}

\usepackage{titlesec}

\titleclass{\subsubsubsection}{straight}[\subsubsection]
\newcounter{subsubsubsection}[subsubsection]
\renewcommand\thesubsubsubsection{\thesubsubsection.\arabic{subsubsubsection}}
\titleformat{\subsubsubsection}
{\normalfont\normalsize\bfseries}{\thesubsubsubsection}{1em}{}
\titlespacing*{\subsubsubsection}{0pt}{3.25ex plus 1ex minus .2ex}{1.5ex plus .2ex}

\usepackage{setspace}

\usepackage{xcolor}
\usepackage[capitalize]{cleveref}

\newcommand{\ie}{i.e.\ }
\newcommand{\eg}{e.g.\ }

\newcommand{\textok}{\textsf{ok}}
\newcommand{\textfail}{\textsf{fail}}
\newcommand{\textyes}{\textsf{yes}}
\newcommand{\textno}{\textsf{no}}

\makeatletter
\@for\@tempa:=corollary,lemma,prop,definition,example,notation\do{%
  \expandafter\let\csname c@\@tempa\endcsname\relax}
\makeatother
 
\theoremstyle{plain}
\newtheorem{theorem}{Theorem}[chapter]

\theoremstyle{definition}

\theoremstyle{remark}
\newtheorem{remark}{Remark}[chapter]

\theoremstyle{definition}
\newtheorem{assumption}{Assumption}
 
\newtheorem{assumptionalt}{Assumption}[assumption]
\crefname{assumptionalt}{assumption}{assumptions}
\Crefname{assumptionalt}{Assumption}{Assumptions}
\crefrangeformat{assumptionalt}{assumptions~#3#1#4--#5#2#6}
\Crefrangeformat{assumptionalt}{Assumptions~#3#1#4--#5#2#6}

\newenvironment{assumptionp}[1]{
  \renewcommand\theassumptionalt{#1}
  \assumptionalt
}{\endassumptionalt}

\newcommand{\AOEtext}{AOE}
\newcommand{\AOEfulltext}{Absoluteness of Observed Events}

\newcommand{\BornCAOEText}{Born Compatibility}
\newcommand{\BornCPersKText}{Born Compatibility}
\newcommand{\UniversalityText}{Universality of Quantum Theory}
\newcommand{\PersKText}{Personal Knowledge}
\newcommand{\ClassicalTText}{Classical Agreement}

\newcommand{\possBornText}{possibilistic Born rule}

\newcommand{\AOE}{\hyperlink{AOE}{\color{darkgray}{AOE}}}
\newcommand{\AOEfull}{\hyperlink{AOE}{\color{darkgray}{Absoluteness of Observed}}\allowbreak\hyperlink{AOE}{\color{darkgray}{ Events}}}
\newcommand{\possBorn}{\hyperlink{poss_Born}{\color{darkgray}{possibilistic}}\allowbreak\hyperlink{poss_Born}{\color{darkgray}{ Born rule}}}
\newcommand{\BornCAOE}{\hyperlink{Born_Compat_AOE}{\color{darkgray}{Born}}\allowbreak\hyperlink{Born_Compat_AOE}{\color{darkgray}{ Compatibility}}}
\newcommand{\BornCPersK}{\hyperlink{Born_Compat_PersK}{\color{darkgray}{Born}}\allowbreak\hyperlink{Born_Compat_PersK}{\color{darkgray}{ Compatibility}}}
\newcommand{\universality}{\hyperlink{Universality}{\color{darkgray}{Universality of Quantum}}\allowbreak\hyperlink{Universality}{\color{darkgray}{ Theory}}}
\newcommand{\PersK}{\hyperlink{Pers_K}{\color{darkgray}{Personal}}\allowbreak\hyperlink{Pers_K}{\color{darkgray}{ Knowledge}}}
\newcommand{\ClassicalT}{\hyperlink{Classical_T}{\color{darkgray}{Classical}}\allowbreak\hyperlink{Classical_T}{\color{darkgray}{ Agreement}}}
\newcommand{\ROE}{\hyperlink{ROE}{\color{darkgray}{Relativity of Observed}}\allowbreak\hyperlink{ROE}{\color{darkgray}{ Events}}}
\newcommand{\BornP}{\hyperlink{BornP}{\color{darkgray}{Born}}\allowbreak\hyperlink{BornP}{\color{darkgray}{ Practicality}}}

\DeclareMathOperator{\arccosh}{arccosh}

\newcommand{\nocontentsline}[3]{}
\newcommand{\tocless}[2]{\bgroup\let\addcontentsline=\nocontentsline#1{#2}\egroup}

\usepackage{longtable, booktabs}

\usetikzlibrary{quantikz2}

\setsecheadstyle{\large\bfseries}
\setsubsecheadstyle{\normalsize\bfseries}
\setsubsubsecheadstyle{\normalsize\itshape}
\setparaheadstyle{\itshape}

\title{\textbf{
Foundational puzzles on quantum universality:}}
\subtitle{From Wigner's friend to black holes} 
\author{Laurens Walleghem}
\department{Mathematics} 
\qualification{PhD\todo[caption={Make it final}, color=red!60]{Change \texttt{draft} to \texttt{final} in the options passed to the class when everything will be ready}}
\submitdate{March 2026}  
\usepackage{etoolbox}
\usepackage{graphicx}

\newcommand{\YileYing}{Yìlè Y{\=\i}ng}

\begin{document}

\frontmatter

\makethesistitle

\chapter{Abstract}

Despite its remarkable empirical success and predictive power, quantum theory is riddled with mysteries.
One such conceptual puzzle is the measurement problem: why do quantum systems in superposition collapse to a single-valued outcome upon being measured?
The scenario of Wigner's friend illustrates this tension, with recent extensions deriving strong no-go results for the nature of reality. 
Another fundamental puzzle arises at the interface of quantum theory and general relativity---the black hole information paradox.
This thesis further explores these two foundational paradoxes.

Part I of this thesis focuses on extensions of the Wigner's friend scenario. 
We introduce several novel extended Wigner's friend arguments, and establish connections with notions of nonclassicality, i.e. the study of quantum phenomena that evade classical explanation. 
Key results are a refined version of the Frauchiger--Renner paradox, new extended Wigner's friend scenarios based on contextuality, and an argument against the absoluteness of free choices based on the Pusey--Barrett--Rudolph theorem.
The latter result complements the Local Friendliness no-go theorem, which challenged the absoluteness of experimentally obtained measurement outcomes.
In Part II of this thesis, we turn to black hole puzzles and the interplay of quantum theory and gravity.
We present a paradox that unifies extended Wigner's friend arguments with black hole puzzles, strengthening the cloning and firewall paradoxes. 
After discussing subtleties in these paradoxes, we next present an analysis combining elements of quantum theory with gravity, analyzing the detector response near a superposed black hole using elements of quantum reference frames.

\vspace{0.5cm}



\clearpage
\tableofcontents


\clearpage
\listoftables
\listoffigures


\chapter*{Acknowledgements}
\markboth{Acknowledgements}{Acknowledgements}
\addcontentsline{toc}{chapter}{Acknowledgements}

This work would not have been possible without the help of many.  
I would like to thank my supervisors Dr. Matt Pusey and Dr. Stefan Weigert for their guidance, especially for our interesting conversations on the Frauchiger--Renner paradox and our refinement thereof, and for teaching me how to present scientific results, and Prof. Chris Fewster and Roger Colbeck for their TAP support.
A special thanks goes to Dr. Rui Soares Barbosa; much of the work in the first part of my PhD arose from initial conversations with Rui. 
I would also like to thank my examiners Prof. Chris Fewster and Prof. Časlav Brukner for their useful comments, which significantly improved this thesis.
I thank my collaborators Rafael Wagner, \YileYing, David Schmid, Lorenzo Catani, Rutvij Bhavsar, Shashaank Khanna and Carlo Cepollaro, all of whom I learned a great deal from.
I would like to thank Raman Choudhary for the many interesting conversations on notions of nonclassicality, of crucial importance for my understanding thereof. 
I thank Dr. Benito Juárez--Aubry and Prof. Bernard Kay for many interesting discussions on black hole physics, and Prof. Chris Fewster for his always-helpful replies to my questions on (A)QFT. 
I thank Harkan Kirk-Karakaya for teaching me how to run programs on the UoY server.
I would like to thank Prof. Nicolò Defenu for furthering my understanding of quantum field theory and his astonishing intuitive understanding of QFT (condensed matter) systems. I have truly learned a lot from Nicolò.
I would like to thank Claudio Dappiaggi for helpful conversations on black hole physics and its viewpoint as a scattering problem.
I thank Dr. Ricardo Faleiro and Prof. Časlav Brukner for interesting discussions on  the firewall paradox, and Prof. David Mattingly for interesting discussions and exchanges on gravitational shielding and gauge invariance regarding quantum paradoxes.
I thank Dr. Ambroise Grau for his template for this thesis.

There is one person I have not mentioned above in my list of collaborators, and that is David Maibach, my partner in black hole crimes. 
Throughout my PhD, we have been having continuous conversations on the physics of black holes, which have advanced my understanding significantly.

Finally, I would like to express my sincere gratitude to my circle of family and friends, and all the new friends I made on the way.
My parents, and sister, for their continued support, and without whom my whole education would not have been possible.
Of the long list of my friends, I would like to thank Toon, Raman, Manisha, Quinten, Arne, Arno, Tanguy, Siebe, David, Jasper, Thomas, Yannick, Laurens, Michael, Lidewij, Marie, Joppe, João, Carolina, Anita, Davide, Angelos, Jaymin, Luisa, Osvaldo, Diogo, Eduardo, Toni, Leonardo, Lorenzo, Marco, Rafael, Timo, Nelson, Dieter, Kobe, Arne, Vincent, Jef, Bram, Will, Ari, Santiago, Alessandro, Baptiste, Tara, Justin, Zara, Gabriel, Zane, Máté, Rutvij, Peter, Shashaank, Kuntal, Filippo, Harkan, Vincenzo, Lewis, Ricky, Vasilis, Alistair, all the Bens and Sams in the office, Philip, Simen, Ross, Jamie, Sol, Dorsa, Ambroise, Ding, Alice, Simon, Samuel, Cody, Levent, Marie-Lou, Thomas, Victor, Tina, Lodewijk, Robin, Wilfried, Walter, Olivier, Marthe, Jan, Benjamin, Seba, David, Olivier L., Andreas V., and Khadija, for always supporting me, as well as those who I have not listed here.
They have supported me almost as much as Zane has continuously supported United, even in the most desperate of times.

I acknowledge support from the United Kingdom Engineering and Physical Sciences Research Council (EPSRC) through the DTP Studentship EP/W524657/1, and thank INL and its QLOC group in Braga, Portugal for the kind hospitality.

\begin{flushright}
Laurens Walleghem,\\
August 2026.
\end{flushright}

\chapter*{Author's declaration}
\markboth{Author's declaration}{Author's declaration}
\addcontentsline{toc}{chapter}{Author's declaration}


I declare that this thesis is a presentation of original work and I am the sole author. This work has not previously been presented for a degree or other qualification at this University or elsewhere. All sources are acknowledged as references.

\vspace{0.5cm}

The results in Part I of this thesis are based on the following works, in the order of appearance: 

\textbf{\Cref{chap:partI_refinedFR}}:

\begin{quotation}
\emergencystretch=1.5em

\noindent L. Walleghem, R. S. Barbosa, M. Pusey and S. Weigert. A refined Frauchiger--Renner paradox based on strong contextuality, \href{https://doi.org/10.22331/q-2026-05-26-2116}{Quantum \textbf{10}, 2116 (2026)}.
\end{quotation}

\textbf{\Cref{chap:part1_EWF_noncontextuality}}:

\begin{quotation}
\noindent L. Walleghem, R. Wagner,  Y. Yīng and D. Schmid. Extended Wigner's friend paradoxes do not require nonlocal correlations, \href{https://doi.org/10.1103/n4hv-rlgj}{Phys. Rev. A \textbf{112}, 022212 (2025)}.
\end{quotation}

\begin{quotation}
\noindent L. Walleghem, R. Wagner,  Y. Yīng and D. Schmid. Connecting extended Wigner's friend arguments and noncontextuality, \href{https://doi.org/10.22331/q-2025-07-31-1819}{Quantum \textbf{9}, 1819 (2025)}.
\end{quotation}

\begin{quotation}
\noindent L. Walleghem and L. Catani. An extended Wigner's friend no-go theorem inspired by generalized contextuality, \href{https://doi.org/10.48550/arXiv.2502.02461}{arXiv:2502.02461 (2025)}.
\end{quotation}

\textbf{\Cref{chap:part1_EWF_PBR}}:

\begin{quotation}
\noindent L. Walleghem. Questioning the absoluteness of free choices when internalized in Wigner's friend scenarios,    \href{https://doi.org/10.1103/hf5j-t83v}{Phys. Rev. A \textbf{114}, 032431 (2026)}.
\end{quotation}

\textbf{\Cref{chap:part1_sleepingbeauty}}:

\begin{quotation}
\noindent L. Walleghem. Stunned by Sleeping Beauty: How Prince Probability updates his forecast upon their fateful encounter,  \href{https://doi.org/10.48550/arXiv.2408.06797}{arXiv:2408.06797
(2024)}.
\end{quotation}

\vspace{0.5cm}

The results of Part II are based on the following works:

\textbf{\Cref{chap:part2_WFBH}}:

\begin{quotation}
\noindent L. Walleghem. Wigner's friend's black hole adventure: an argument for complementarity?, \href{https://doi.org/10.48550/arXiv.2507.05369}{arXiv:2507.05369 (2025)}.
\end{quotation}

\textbf{\Cref{chap:part2_detector}}:

\begin{quotation}
\noindent L. Walleghem and C. Cepollaro. Particle detector near a position-superposed black hole. \href{https://doi.org/10.48550/arXiv.2604.11897}{arXiv:2604.11897 (2026)}.
\end{quotation}

\vspace{0.5cm}

The following publication was left out of this thesis:

\begin{quotation}
\noindent L. Walleghem, S. Khanna and R. Bhavsar. Comment on a No-Go Theorem for $\psi$-Ontic Models, \href{https://doi.org/10.1007/s10701-025-00836-3}{Found Phys \textbf{55}, 23 (2025)}.
\end{quotation}

\vspace{0.5cm}

For proofreading parts of this work the online tools Grammarly and ChatGPT were used. I also acknowledge the use of ChatGPT (developed by OpenAI) and Claude (developed by Anthropic) as search engines, and, in the case of Claude, for improving code regarding numerical results.

\newpage

\chapter{Abbreviations}


\begin{longtable}{@{} p{3cm} p{10cm} @{}}
  \toprule
  \textbf{Abbreviation} & \textbf{Meaning} \\
  \midrule
  \endhead
  KS    & Kochen--Specker \\
  KSNC  & Kochen--Specker noncontextuality \\ 
  EWF   & Extended Wigner's friend \\
  FR    & Frauchiger--Renner \\
  LF    & Local Friendliness \\
  AOE   & Absoluteness of Observed Events \\
  LA    & Local Agency \\
  ROE   & Relativity of Observed Events \\
  NF    & Noncontextual Friendliness \\
  PBR   & Pusey--Barrett--Rudolph \\
  QFT   & Quantum field theory \\
  BH    & Black hole \\
  QRF   & Quantum reference frame(s) \\
  \bottomrule
\end{longtable}

\mainmatter

\ifSubfilesClassLoaded{ 
	\setcounter{chapter}{0} 
	\maketitle \tableofcontents \listoffigures \listoftodos
	}{}

\chapter[Introduction]{Introduction}\label{chap:intro}

\vspace{0.5cm}


   \hspace*{\fill} ``\textit{What is proved by impossibility proofs is lack of imagination.}''
   \newline
\hspace*{\fill} J.S. Bell, 1982~\cite{Bell:1982xg}.

\vspace{1.5cm}

\noindent

Two of the most prominent and persistent foundational puzzles in physics 
concern 
the measurement problem and the black hole information puzzle. 
These topics probe the limits of the two pillars of modern physics, quantum theory and general relativity. 
To date, there appear to be no fundamental constraints on the applicability of quantum theory, motivating the hypothesis of its universal validity.
The measurement problem pushes the universality of quantum theory into the macroscopic world, or at least into regimes where a macroscopic notion of an observer exists, while black hole paradoxes do so into the regime of astrophysical gravitational phenomena.
Recent results have sharpened the tension in the measurement problem by combining it with other distinctive features of quantum mechanics. 
In this thesis, we further develop and analyze these arguments in \Cref{part:universality_in_quantum_foundations}.
In \Cref{part:universality_black_holes}, we combine arguments of \Cref{part:universality_in_quantum_foundations} with black hole physics, leading to new black hole paradoxes, highlight subtleties in black hole paradoxes more generally, and move towards combining quantum aspects with gravity. 

In the remainder of this introduction, we provide some background for Parts I and II and present our main findings.

\section{Part I: Universality in quantum foundations}
\subsection{The measurement problem, a universality hypothesis and Wigner's friend}
The measurement problem~\cite{Maudlin1995,Pitowsky2007,BubPitowsky2010,brukner2017quantum,schlosshauer2005decoherence,legget20005measurement,Hance_2022}, noted since the earliest days of quantum theory and one of the key outstanding challenges in quantum foundations~\cite{Briggs:2013qxp}, arises from two types of time evolution that can be prescribed to a quantum system: (i) a deterministic, unitary evolution of a closed quantum system, and (ii) an indeterministic wavefunction collapse after a system undergoes a measurement. 
But why do quantum systems in a superposition collapse to a single-valued outcome, or more precisely, to the corresponding eigenstate, upon being measured?
What qualifies as an observer or a measurement, and which observers should apply evolution (ii) to a measured system, given that the system is ultimately part of a closed system governed by evolution (i)?

Besides Schrödinger's cat, another well-known illustration of this problem is the scenario of Wigner's friend~\cite{wigner1995remarks}.
In this thought experiment, involving Wigner and his friend, the friend measures a quantum system inside a sealed lab and claims to obtain a single-valued outcome. 
Wigner, remaining outside the sealed lab, considers his friend ultimately to be a quantum system as well, and describes the measurement as a unitary interaction of quantum systems. 
Consequently, Wigner describes his friend as being in a superposition of seeing different outcomes, at odds with the friend's claim.
Although the descriptions of Wigner and his friend differ, there is no immediate contradiction concerning the actual value obtained by the friend.
This is where recent extensions of the Wigner's friend scenario enter the discussion~\cite{brukner2017quantum,frauchiger2018quantum,bong2020strong,cavalcanti2021implications,haddara2022possibilistic,guerin2020no,utreras2023allowing,haddara2024local,szangolies2023quantum,walleghem2024strong,walleghem2023extended,walleghem2024connecting,walleghem2025extendedwignersfriendnogo,vilasini2019multi,vilasini2022general,ormrod2023theories,leegwater2022greenberger,schmid2023review,brukner2018no,ormrod2022no,nurgalieva2018inadequacy,montanhano2023contextuality,zukowski2021physics,walleghem2023extended,montanhano2023contextuality,guerin2020no,yile2023}, which derive contradictions concerning the actual outcome value obtained by the friend from seemingly innocuous assumptions.
The technical novelty of these extensions is that they combine Wigner's friend with Bell nonlocality~\cite{brunner2014bell,bell1964einstein,Bell_1976}, a form of `nonclassicality' of quantum theory, famously referred to as `spooky action at a distance' by Einstein~\cite{EinsteinBorn1971}.
In doing so, they eliminate the counterfactuality required to derive a contradiction in usual Bell nonlocality scenarios.

The first so-called \textit{extended Wigner's friend} argument, proposed by Brukner~\cite{brukner2017quantum}, laid the basis for the Frauchiger--Renner paradox~\cite{frauchiger2018quantum} and the Local Friendliness no-go theorem~\cite{bong2020strong,cavalcanti2021implications}, two of the most celebrated extended Wigner's friend paradoxes, and the main topics of the first part of this thesis.
The Frauchiger--Renner (FR) paradox involves several superobservers such as Wigner, who model their friends performing measurements inside sealed labs unitarily. 
The superobservers and friends reason about each other's outcomes and derive a contradiction. 

The Local Friendliness (LF) no-go theorem can be seen as a device-independent version of the Frauchiger--Renner paradox. Now, instead of reasoning about outcomes obtained by other agents\footnote{We will use the terms `agent' and `observer' interchangeably.}, each superobserver can freely choose whether to ask their friend for their outcome or to perform other operations on the entire friend's lab.  
Assuming a notion of locality\footnote{This `locality assumption' can also be seen as a causality requirement.}, i.e., that free choices can only affect outcome variables in their future, a contradiction is derived with the assumption that the measurement outcomes observed by the agents are absolute and single-valued, also referred to as Absoluteness of Observed Events\footnote{Occasionally, to improve readability, we will refer to the Absoluteness of Observed Events simply as Absoluteness.}. 

The Frauchiger--Renner paradox and Local Friendliness no-go theorem can be argued to probe slightly different questions. 
The Local Friendliness no-go theorem shows that a weak notion of locality and Absoluteness of Observed Events are incompatible, assuming only the experimental validity of quantum theory in addition to the existence of superobservers.
Arguments such as the Frauchiger--Renner paradox instead present us with the question of how agents can use quantum theory consistently, and, as we will argue, are particularly illuminating for going beyond the Absoluteness of Observed Events.

\subsection{Part I: Contributions}

Our main contributions in Part I of this thesis consist of a refined version of the Frauchiger--Renner paradox, connecting extended Wigner's friend scenarios to notions of nonclassicality, and an extended Wigner's friend argument based on the Pusey--Barrett--Rudolph theorem~\cite{pusey2012reality} that can be interpreted as an argument against the absoluteness of free choices. 
We now outline these results in more detail.

\vspace{0.5cm}

\textit{Refining the Frauchiger--Renner paradox.} In \Cref{chap:partI_refinedFR}, we present a refined version of the Frauchiger--Renner paradox~\cite{frauchiger2018quantum}.
In this paradox, several observers perform measurements in sealed labs, and are subsequently acted upon by their respective superobservers. 
The superobservers, having obtained their outcome, each make a prediction for the outcomes of other observers. After this, the superobservers meet up and share their predictions, and find that their predictions are contradictory. In contrast to the original Frauchiger--Renner paradox, a contradiction is obtained in each run of the protocol, and the required consistency condition for combining the reasoning of different agents is weakened.
Notably, FR's consistency condition can be argued to be already violated in a classical paradox, the Sleeping Beauty problem, which we discuss in \Cref{chap:part1_sleepingbeauty}, as opposed to our weaker consistency condition.
This weaker consistency requirement proves particularly interesting if one seeks to refute the Absoluteness of Observed Events; our refined FR paradox then suggests a principle, which we name the Relativity of Observed Events, to replace Absoluteness. 
Paraphrased as an extension of Peres' dictum~\cite{peres1978unperformed}, the principle of Relativity of Observed Events reads `\textit{Unperformed experiments have no results, and unknown results have no values}.'

\vspace{0.5cm}

\textit{Connecting Wigner's friend with noncontextuality.} In \Cref{chap:part1_EWF_noncontextuality}, we connect Wigner's friend arguments to noncontextuality~\cite{kochen1990onthe,budroni2022kochenspecker,abramsky2011sheaf,spekkens2005contextuality,schmid2020structure}.
Noncontextuality captures nonclassical features of quantum theory, i.e. how certain quantum phenomena admit no classical explanation. 
Two leading notions of noncontextuality are Kochen--Specker (KS) noncontextuality~\cite{kochen1990onthe,budroni2022kochenspecker,abramsky2011sheaf} and generalized noncontextuality~\cite{spekkens2005contextuality,schmid2020structure}.
KS noncontextuality concerns the impossibility to assign outcome values to a set of measurements, independently of the \textit{context} in which they take place. 
In other words, suppose we have measurements $A,B,C$, with experimental data from (i) context $\{A,B\}$ in which $A$ is measured jointly with $B$, and (ii) context $\{A,C\}$ in which  $A$ is measured jointly with $C$; KS noncontextuality then requires that the outcome value assigned to $A$ must be the same in each context.
There exist correlations in quantum theory that fail this condition, and we say that quantum theory is \textit{KS contextual}.
Bell nonlocality can be seen as a special case of KS contextuality, with the joint measurements arising from an $n$-party structure.
Generalized noncontextuality is defined within the framework of ontological models. 
Ontological models assign a space of ontic states to each system, with experimental preparations preparing a system in a distribution over ontic states, and measurements extracting information about that ontic state. 
Generalized noncontextuality then requires preparation procedures indistinguishable in any experiment, even when lacking full knowledge, to be represented identically in the ontological model, and similarly for transformations and measurement procedures.

To date, most extended Wigner's friend arguments are built on correlations that violate Bell nonlocality. 
In contrast, in this chapter we provide several new classes of extended Wigner's friend paradoxes employing noncontextual correlations, both Kochen--Specker~\cite{kochen1990onthe,budroni2022kochenspecker,abramsky2011sheaf} and generalized noncontextuality~\cite{spekkens2005contextuality,schmid2020structure}.
Furthermore, we connect conditions from the Local Friendliness assumptions on experimental correlations to conditions from Kochen--Specker noncontextuality.

\vspace{0.5cm}

\textit{An argument against the absoluteness of free choices.} In \Cref{chap:part1_EWF_PBR}, we provide an extended Wigner's friend argument based on the Pusey--Barrett--Rudolph (PBR) theorem~\cite{pusey2012reality}. 
The PBR theorem was originally presented as an argument against an epistemic interpretation of the quantum state, modulo some assumptions. 
We combine Wigner's friend and the scenario used by Pusey--Barrett--Rudolph to present a no-go theorem similar to the Local Friendliness no-go theorem, but as an argument against the absoluteness of free choices. 
Specifically, we use the same locality assumption of the Local Friendliness assumptions, but the relevant events in the assumption of Absoluteness of Observed Events, i.e. those that are unavailable after the protocol, concern free choices rather than measurement outcomes.

\vspace{0.5cm}

\textit{On Sleeping Beauty.} Finally, in \Cref{chap:part1_sleepingbeauty}, we briefly discuss the Sleeping Beauty problem.
Taking an operational viewpoint, we show how the tension in this problem arises from a difference in prior knowledge.

\section{Part II: Universality and black holes} 
\subsection{The black hole information problem and related paradoxes}
Arguably, black holes rank among the most fascinating phenomena in physics, where intriguing puzzles arise at the intersection of quantum theory and gravity.
In his seminal work, applying principles of quantum theory to black holes, Hawking showed that black holes radiate~\cite{hawking1975particle}. Moreover, he suggested this radiation to be purely thermal~\cite{hawking1976breakdown}, which would lead to a loss of information as the black hole evaporates. 
Even if one had started from an initial pure state of collapsing matter, the radiation left after evaporation is purely thermal, an apparent nonunitary evolution.
But where did this information go?

\vspace{0.5cm}

Other famous black hole puzzles involve objects and observers falling into the black hole, such as the black hole cloning argument~\cite{susskind1994gedanken,hayden2007black,harlow2016jerusalem} and the firewall paradox~\cite{braunstein2007quantum,braunstein2013better,almheiri2013black,harlow2016jerusalem,aaronson_complexity_2016}.
The black hole cloning paradox relies on the \textit{no-cloning} feature of quantum theory~\cite{wootters1982single,DIEKS1982271,barnum1996noncommuting,nielsen2002quantum}, which states that, in quantum theory, one cannot clone unknown quantum states.
That is, there cannot exist a quantum device which, given an arbitrary quantum state $\ket{\psi}$, outputs two copies of this state, i.e., $\ket{\psi}\otimes \ket{\psi}$.
In the black hole cloning paradox, one considers an observer Alice holding a quantum system $S$ in an arbitrary quantum state $\ket{\psi}$, falling into the black hole. 
An outside observer Bob gathers all Hawking radiation coming from the black hole until late times.
Under the standard assumptions of unitarity and that the information about the system $S$ can be decoded from the Hawking radiation, Bob decodes the quantum state $\ket{\psi}$ from the Hawking radiation. 
But this leads to a violation of the no-cloning theorem, with the quantum state $\ket{\psi}$ being present both inside and outside the black hole. 

The firewall paradox is derived from two main ingredients: (i) the monogamy of quantum entanglement~\cite{coffman2000distributed,Osborne:2006pmn,Koashi:2003pgf,Friis:2018nyl},
i.e. a system $A$ cannot be highly entangled with both a system $B$ and another disjoint system $C$, and (ii), the Page curve~\cite{page1993information,page2013time}.
The Page curve~\cite{page1993information} gives the entanglement entropy over time of a `typical' quantum subsystem $R$ that grows in a global system consisting of two subsystems $R$ and $B$ (the radiation and black hole). 
Assuming a global pure state at all times, the entropy increases as $R$ (the radiation) grows, reaches a maximum and then decreases again as purifying parts of $B$ become part of $R$. 
We note that this analysis stems from a toy quantum model; in actual quantum gravity, identifying the radiation subsystem is in fact a subtle and important point~\cite{laddha_2021,geng2022inconsistency,Uhlemann2022information,Uhlemann_2021,geng2021information}. 
The firewall argument then derives a contradiction as follows.
Assuming the Page curve for the radiation~\cite{page1993information}, late-time radiation modes just outside the black hole should be highly entangled with the early radiation.
However, according to quantum field theory on curved spacetime, which posits high entanglement across the horizon, these modes should also be highly entangled with modes inside the black hole, violating monogamy of entanglement.

\vspace{0.5cm}

While a full resolution of the black hole information puzzle remains elusive, much work has been done since, hinting that information is preserved.
After some early arguments against Hawking's claim of information loss~\cite{Banks:1983by,page1980is,tHooft:1990fkf,susskind1993stretched,kiem1995black}, 
a concrete argument in favor of unitarity came from AdS/CFT~\cite{maldacena1999large,harlow2016jerusalem}.
More recently, the proposed existence of `entanglement islands' inside the black hole yields a unitary Page curve for the radiation~\cite{penington2020entanglement,almheiri2020page,almheiri2019entropy,wang2024entanglement,Hartman_2020,wang2021page,balasubramanian2024microscopic,Hashimoto_2020,Balasubramanian_2021,almheiri2020entanglementislandsinhigher,kibe2022holographic,chen2020quantumextremalislandseasyI,chen2020quantumextremalislandseasyII,bousso_islands_2023,almheiri2023islandsoutsidehorizon,Hartman_2020islandsincosmology,Krishnan_2021,Van_Raamsdonk_2021,geng2021information,geng2021informationparadoxandits,geng2026seeingpagecurvesislands}, but whose physical interpretation remains heavily debated~\cite{geng2022inconsistency,Geng_2020,,geng2025revisitingrecentprogresskarchrandall,antonini2025apologiaislands,geng2026seeingpagecurvesislands,raju2022failure,geng2025makingcasemassiveislands,geng2026mechanisminformationencodingislands,Uhlemann2022information,Uhlemann_2021,geng2021information}.
Another interesting development is the identification of many extra degrees of freedom of asymptotically flat gravitating spacetimes, so-called \textit{soft hair}~\cite{hawking2016soft,strominger2018lecturesinfraredstructuregravity,carney_infrared_2017,carney_dressed_2018,strominger2017blackholeinformationrevisited,flanagan_order-unity_2021}, which could provide the missing piece Hawking was looking for when formulating his information-loss claim.
Namely, soft hair endows a black hole spacetime with many more degrees of freedom for the radiation to entangle with, and may thus be a crucial factor towards purification in a unitary evolution. 
Soft hair can be seen as arising from gauge transformations that act nontrivially on the boundary of spacetime~\cite{strominger2018lecturesinfraredstructuregravity}. 
Issues of gauge invariance, i.e. the nontrivial role played by apparent `redundant' degrees of freedom, comprising diffeomorphism and coordinate invariance, may also be crucial in resolving the cloning and firewall paradoxes, as we briefly comment upon below and in the chapters referred to there.

\vspace{0.5cm}

\subsection{Part II: Contributions}
By combining the extended Wigner's friend arguments of Part I with black hole physics, we sharpen the cloning and firewall paradoxes, using similar assumptions.
We criticize these assumptions by highlighting several important subtleties, which calls for more careful verification thereof using known techniques without committing to a specific theory of quantum gravity. 
We next present a calculation that uses some of these proposed methods in analyzing the detector response near a superposed black hole.

\vspace{0.5cm}

\textit{Wigner's friend's black hole adventure.} In \Cref{chap:part2_WFBH}, we combine extended Wigner's friend arguments with black hole physics into a unified argument. 
This is an extension of the work by Hausmann and Renner~\cite{hausmann2025firewallparadoxwignersfriend}, who drew analogies between Wigner's friend paradoxes and black hole puzzles. 
The cloning and firewall paradoxes raise sharp questions when combining the perspectives of different observers.
However, they leave room for a more general (hypothetical) theory that could describe the different viewpoints of the observers, which could accommodate (unobservable) cloning or non-monogamy of entanglement, for example.
The arguments of this chapter close this loophole, showing that no such theory exists if no observer can experimentally falsify the predictions of quantum theory.

In other words, we exhibit an inconsistency in the operational predictions prescribed by quantum theory under standard assumptions of black hole puzzles, such as the cloning and firewall paradoxes.
After presenting our no-go results, we criticize these assumptions by highlighting important subtleties.
Key elements in these subtleties include (i) quantum backreaction of the black hole leading to a superposed black hole, (ii) gauge invariance, i.e. the required invariance of true observables under gauge transformations, and (iii) observer-dependencies in describing the physics at hand. 
By emphasizing these issues, we invite a more careful verification of these assumptions towards resolving these puzzles.
Among the methods that can be employed for this purpose are the use of quantum reference frames and operational investigations of a black hole and its radiation. 
The next chapter illustrates a combination of these approaches in a simple, limited setting, without committing to a specific full theory of quantum gravity. 

\vspace{0.5cm}

\textit{Detector near a superposed black hole.} 
In \Cref{chap:part2_detector}, we analyze the response of a particle detector near a position-superposed BTZ black hole~\cite{banados_geometry_1993,Carlip_1995} using aspects of quantum reference frames. 
Conservation laws~\cite{compere_poincare_2020,flanagan_conserved_2017} suggest that an old black hole is highly superposed, for example with a highly superposed mass~\cite{flanagan_infrared_2021}, also advocated for in Refs.~\cite{page1980is,Bao_2018,calmet2022brief,akil2025quantumsuperpositionblackhole,pasterski_hps_2021,Hutchinson_2016}, and perhaps even more so when taking into account the vacuum degeneracy due to soft hair~\cite{strominger_gravitational_2016,strominger2018lecturesinfraredstructuregravity}.
This motivates the study of a superposed spacetime, and the particular choice of a position-superposed black hole as background (quantum) spacetime was motivated by the study of a detector's response near a mass-superposed black hole in literature by Foo et al.~\cite{foo2022quantum}. 
Our first result in this chapter is the derivation of a form of the Unruh--DeWitt interaction in a superposed spacetime using quantum reference frames.
Secondly, Foo et al.~\cite{foo2022quantum} found peaks in the detector response, argued to be reminiscent of black hole mass quantization.
By showing the absence of such peaks in a position-superposed black hole, this further strengthens the claim of such peaks being signatures of black hole mass quantization. 
Additionally, we derive analytical expressions for the spectrum probed by the detector, and show that, in the case of the mass-superposed black hole~\cite{foo2022quantum}, the peaks in the detector response arise from singularities in the spectrum.

\vspace{0.5cm}

\textit{Conclusions and outlook.} Finally, in \Cref{chap:conclusions&future-work}, we close with some concluding remarks and future directions for both parts of this thesis.

\section{Structure of thesis}
The remainder of this thesis is organized as follows. Part I starts with \Cref{chap:Q_prelims}, outlining preliminaries concerning Bell nonlocality, contextuality and extended Wigner's friend paradoxes, after which the main results of part I are presented in \Cref{chap:partI_refinedFR,chap:part1_EWF_noncontextuality,chap:part1_EWF_PBR,chap:part1_sleepingbeauty}.
Part II opens with some background on black hole paradoxes in \Cref{chap:BH_prelims}, after which we present the main results of part II in \Cref{chap:part2_WFBH,chap:part2_detector}.
Finally, we conclude in \Cref{chap:conclusions&future-work}.

This thesis is organized so that each chapter can be read individually, and we avoid extensive cross-referencing between chapters.  
Therefore, relevant assumptions are repeated in each chapter, occasionally with slightly different formulations to suit the specific result and purpose.

\end{document}

\part{Universality in quantum foundations}
\label{part:universality_in_quantum_foundations}

The first part of this thesis concerns foundational puzzles in quantum mechanics, more specifically, extensions of the Wigner's friend scenario, closely related to the measurement problem.
We will analyze, refine and present several novel extended Wigner's friend arguments, and examine a related classical paradox. 
After the preliminaries in \Cref{chap:Q_prelims}, we present the main results of this part.  

In \Cref{chap:partI_refinedFR} we refine the Frauchiger--Renner paradox, constructing a scenario that relies on the strongly nonlocal GHZ--Mermin scenario.
Whereas most existing extended Wigner's friend arguments rely on nonlocal correlations,
we present several extended Wigner's friend arguments based on noncontextuality in \Cref{chap:part1_EWF_noncontextuality}. 
Additionally, we investigate the link between Local Friendliness and noncontextuality inequalities.
Next, in \Cref{chap:part1_EWF_PBR} we combine Wigner's friend with the Pusey--Barrett--Rudolph theorem, which was originally constructed to question the validity of epistemic interpretations of the quantum state. 
The resulting paradox questions the absoluteness of free choices, in the same way that extended Wigner's friend arguments such as Local Friendliness question the absoluteness of outcomes of performed measurements under a locality assumption. 
Finally, in \Cref{chap:part1_sleepingbeauty} we present a discussion of the Sleeping Beauty problem, a classical paradox in probability theory and philosophy that bears similarities to Wigner's friend arguments, especially to the Frauchiger--Renner paradox. 
A key element of this analysis of the Sleeping Beauty problem is a novel operational viewpoint on the questions asked.

The results of Chapters \ref{chap:partI_refinedFR}, \ref{chap:part1_EWF_noncontextuality} and \ref{chap:part1_sleepingbeauty} are based on Refs.~\cite{walleghem2024strong},~\cite{walleghem2023extended,walleghem2024connecting,walleghem2025extendedwignersfriendnogo} and ~\cite{walleghem2024stunnedsleepingbeautyprince}, respectively. The results of \Cref{chap:part1_EWF_PBR} are based on Ref.~\cite{walleghem2026arefree}.

\ifSubfilesClassLoaded{ 
	\setcounter{chapter}{1} 
	\maketitle \tableofcontents \listoftodos
	}{}

\chapter[Quantum preliminaries]{Quantum preliminaries}\label{chap:Q_prelims}

In this section we outline the preliminaries to our work of Part I, which comprises extensions of Wigner's thought experiment.
We start with outlining the conventions we use regarding the formalism of quantum theory, after which we move to notions of nonclassicality, which help us to understand how quantum theory differs from classical theory, and how this quantum advantage could be harnessed. 
Next, we move to the measurement problem and Wigner's original thought experiment. 
We end by discussing some recent extended Wigner's friend arguments in literature.

\Cref{sec:Wignerfriend,sec:FR_overview}, \Cref{sec:LF} and \Cref{sec:gen_noncontext} contain parts copied from Ref.~\cite[Section 2]{walleghem2024strong}, Ref.~\cite[Section 2.2]{walleghem2024connecting} and Ref.~\cite[Section 3]{walleghem2025extendedwignersfriendnogo}, respectively.

\section{A short recap of quantum theory}\label{sec:QT}

\subsection{Notation in linear algebra}
Quantum theory associates systems with states, which can be composed, and upon which transformations can be performed and information can be extracted by means of measurements. 
For our introduction to quantum theory in this chapter, we restrict ourselves first to finite-dimensional systems, unless stated otherwise.

We denote the inner product of the Hilbert space by $\braket{\cdot}{\cdot}$, 
and the set of homomorphisms between two vector spaces $A,B$, i.e. the linear maps between $A$ and $B$, by $\text{Hom}(A,B)$.
A homomorphism from a vector space to itself is called an endomorphism. 
An endomorphism on a Hilbert space $\mathcal{H}$ is also called an operator on $\mathcal{H}$.
In the bra-ket notation, a vector $\alpha \in \mathcal{H}$ is associated with a \textit{ket} $\ket{\alpha}$, which can also be seen as an element in $\text{Hom}(\mathbb{C},\mathcal{H})$ as $\ket{\alpha}: \gamma \mapsto \alpha \gamma$. 
The adjoint $\ket{\alpha}^*$ of this mapping is called a \textit{bra}, and is an element of the dual space $\mathcal{H}^* = \text{Hom}(\mathcal{H},\mathbb{C})$ acting as $\bra{\alpha}: \beta \mapsto \bra{\alpha}\ket{\beta}$. It follows that $\bra{\alpha}\, \ket{\beta}=\braket{\alpha}{\beta}$. 
The composition $\ket{\beta}\bra{\alpha}$ is an operator on $\mathcal{H}$, and any operator can be written as a linear combination of such elements.
The tensor product is denoted by $\otimes$, the trace by $\operatorname{Tr}$ and the partial trace over a subsystem $B$ by $\operatorname{Tr}_B$.

\subsection{Quantum theory postulates} \label{sec:postulates}

The postulates of finite-dimensional\footnote{For infinite-dimensional quantum theory and in the case of infinitely many degrees of freedom (in local quantum field theories, for instance), we refer the reader to Refs.~\cite{moretti2017spectral,hall2013quantum,landsman2017foundations} and Refs.~\cite{haag2012local,wald1994quantum}, respectively; see also the paragraph \textit{Towards quantum field theory and beyond} in \Cref{sec:remarks-postulates} below. For a measurement scheme with state update rules in quantum field theory, see also Ref.~\cite{fewster2024measurementquantumfieldtheory}.} quantum theory can be summarized as follows. We refer the reader to Refs.~\cite{nielsen2002quantum,renner2020quantum,Wilde_2017} for more details.

\begin{quote}

\noindent \textit{a. State space.} The state of a quantum system (created through some preparation procedure) is associated with a \textit{density operator} $\rho$ on a complex Hilbert space, that is, a normalized positive semi-definite operator on $\mathcal{H}$, i.e. $\rho \geq 0$ and $\operatorname{Tr} \rho = 1$. The set of all density operators associated with a system is also referred to as the state space denoted by $S(\mathcal{H})$.
A density operator is called \textit{pure} if it can be written as $\rho = \ket{\psi}\bra{\psi}$. 
A system prepared probabilistically, i.e. with probability $p_i$ in state $\rho_i$, corresponds to the density operator $\sum_i p_i \rho_i$.

\noindent \textit{b. Composing quantum systems.} The composition of systems is facilitated by the tensor product $\otimes$. In a composite quantum system $A \otimes B$ with Hilbert space $\mathcal{H}_A \otimes \mathcal{H}_B$, the system $A$ with Hilbert space $\mathcal{H}_A$ is called a subsystem of $A \otimes B$. Given a state $\rho_{AB}$ for $A \otimes B$, the reduced state on system $A$ is given by the partial trace over $B$, $\rho_A = \operatorname{Tr}_B[\rho_{AB}]$. 
Given the composite system $A \otimes B$, not all states are \textit{separable}, i.e. there exist \textit{entangled} states $\rho_{AB}$ which cannot be written as $\sum_i p_i \rho_A^{(i)} \otimes \rho_B^{(i)}$.

\noindent \textit{c. Time evolution.} The time evolution of a closed quantum (sub)system corresponds to a unitary $U$ on the state space of that system, i.e. $\rho \mapsto U \rho U^\dagger$.

\noindent \noindent \textit{d. Measurements.} A projective measurement on a quantum (sub)system $A$ is associated with an \textit{observable} $O_A$, i.e. a Hermitian operator, decomposed into projectors as $O_A = \sum_{x \in X} x P_x$ with outcomes $x \in X$; $P_x$ is the projector onto the eigenspace with eigenvalue $x$. 
The probability of observing outcome $x$ when measuring $O_A$ on a density matrix $\rho_{AB}$ is determined by the Born rule: \begin{equation}
    p(x|O_A,\rho_{AB}) = \operatorname{Tr} [ P_x \otimes \mathbb{1}_B \,\rho_{AB} ].
\end{equation} 
The post-measurement state conditioned on this outcome is then given by \begin{equation}
    \rho_{AB|x} = \frac{1}{p(x|O_A,\rho_{AB})} (P_x \otimes \mathbb{1}_B) \, \rho_{AB} \, (P_x \otimes \mathbb{1}_B).
\end{equation} 
This state update is also referred to as the \textit{wavefunction collapse}.
More general measurements are introduced below.

\end{quote}

\subsection{Remarks} \label{sec:remarks-postulates}
We now outline some remarks about the above postulates. For the first five remarks concerning measurements and time evolution we refer to Refs.~\cite{renner2020quantum,Wilde_2017,nielsen2002quantum} for more details.

\vspace{0.5cm}

\textit{Open vs. closed systems.} The above postulates are presented in their traditional (restricted) form. 
In modern quantum information theory, states are represented by density operators and all physical processes (including measurements) are described by trace-preserving completely positive maps (TPCPMs). 
Closed systems evolve unitarily, while open systems evolve according to TPCPMs.
Further, the above measurement postulate concerns only sharp (or ideal or projective) measurements, while the most general form of a measurement is determined by a quantum instrument.
We briefly review these generalizations in the following four remarks.

\textit{Generalizing time evolution.}
The time evolution and measurement postulates above refer to isolated systems, i.e. when no other involved systems were discarded (traced over). More generally, the evolution on a subsystem is given by a trace-preserving completely positive map $\mathcal{E}$, where completely positive means that for any Hilbert space $\mathcal{H}_R$ the map $\mathcal{E} \otimes \mathcal{I}_R$ is positive. 
For any TPCPM on $\mathcal{H}_A$ (acting on density operators on $\mathcal{H}_A$), there always exists a Hilbert space $\mathcal{H}_E$ and a unitary $U$ on $\mathcal{H}_A \otimes \mathcal{H}_E$ such that $\mathcal{E}$ is the restriction of $U$ to subsystem $A$ (Stinespring dilation theorem): $E(\rho)=\operatorname{Tr}_E\left[U\left(\rho \otimes \sigma_E\right) U^{\dagger}\right]$. 

\textit{Operator-sum representation.} We can write any TPCPM $\mathcal{E} \in \text{Hom}(\mathcal{H}_A,\mathcal{H}_B)$ in the operator-sum representation as \begin{equation}
    \mathcal{E}: S_A \mapsto \sum_x M_x S_A M_x^\dagger
\end{equation} with $M_x \in \text{Hom}(\mathcal{H}_A,\mathcal{H}_B)$ and $\sum_x M_x^\dagger M_x = \mathbb{1}_A$; $\{M_x\}$ are called the Kraus operators. This association of $\{M_x\}_x$ to $\mathcal{E}$ is not unique.
Any mapping of this form is also a TPCPM.

\textit{Measurements as unitaries.} Any projective measurement $O_A = \sum_x x P_x$ on $\rho_{AB}$ can be seen as a TPCPM, and as a result, also as a unitary.
One constructs a Hilbert space $\mathcal{H}_X$ with orthonormal basis $\{\ket{e_x}\}_x$ corresponding to the outcomes $x \in X$. The post-measurement state is then given by \begin{equation}
    \rho'_{X AB } = \sum_x p(x) \ket{e_x}\bra{e_x} \otimes \rho_{AB|x} = \sum_x M_x \rho M_x^\dagger,
\end{equation} a TPCPM, with $p(x)=p(x|O_A,\rho_{AB})$ and $M_x = \ket{e_x} \otimes P_x \otimes \mathbb{1}_B$ satisfying $\sum_x M_x^\dagger M_x = \mathbb{1}_{AB}$.

\textit{Generalizing measurements.} Regarding measurements, we can allow more generally a measurement on $\rho \in S(\mathcal{H}_A)$ to be associated with a set of operators $\{ M_x \}_x$ on $\mathcal{H}_A$ satisfying $\sum_x  M_x^\dagger M_x = \mathbb{1}$.
Namely, letting $\mathcal{H}_X$ be a Hilbert space of dimension equal to the size of the outcome set with orthonormal basis $\{\ket{e_x}\}_x$, we can apply the isometry $V = \sum_x \ket{e_x} \otimes M_x$ acting on a state $\rho_A \in S(\mathcal{H}_A)$, and next applying the projective measurement $\{\ket{e_x}\bra{e_x} \otimes \mathbb{1}_A\}_x$ on $V \rho V^\dagger$. 
The resulting probabilities are given by $p(x)=\operatorname{Tr}[ M_x \rho_A M_x^\dagger ]$, and the post-measurement state by \begin{equation}
    \rho'_{A |x} = \frac{1}{p(x)} M_x \rho M_x^\dagger.
\end{equation}
Note that the probabilities are in fact determined by $N_x = M_x^\dagger M_x$ satisfying $\sum_x N_x = \mathbb{1}$ by \begin{equation}
    p(x) = \operatorname{Tr}[N_x \rho].
\end{equation}
Such positive Hermitian operators $\{N_x\}_x$ satisfying $\sum_x N_x = \mathbb{1}$ are called a positive operator-valued measure (POVM), and thus fully characterize the measurement probabilities. 
With any such POVM $N_x$ one can associate a CPM describing the measurement, for example by $M_x = \sqrt{N_x}$. This association is not unique, and as a result a POVM alone cannot be used to infer the post-measurement state; the latter depends on the Kraus operators or, equivalently, the chosen quantum instrument.

A quantum instrument, the most general description of a measurement that determines the post-measurement state, is a set $\{ \mathcal{E}_x \}$ of completely positive trace-nonincreasing maps such that $\mathcal{E} = \sum_x \mathcal{E}_x$ is trace-preserving. The outcome probabilities are given by $p(x)=\operatorname{Tr} [ \mathcal{E}_x \rho_A  ]$, and the post-measurement state by \begin{equation}
    \rho_{A|x} = \frac{1}{p(x)} \mathcal{E}_x (\rho_A).
\end{equation} Each CPM $\mathcal{E}_x$ is associated with a set of Kraus operators by $\mathcal{E}_x(\rho_A) = \sum_k M_{x,k} \rho_A M_{x,k}^\dagger$ and POVM elements $N_x=\sum_k M_{x, k}^{\dagger} M_{x, k}$. 
The sum over the index $k$ in the Kraus representation of each element of an instrument can intuitively be understood as summing over unobserved degrees of freedom (for instance arising from tracing out an environment), or as coarse-graining over a more fine-grained set of measurement outcomes~\cite{Wilde_2017}.
It follows from the above discussion that a quantum instrument can be implemented by a projective measurement on an extended system, and may thus be modeled unitarily as in the paragraph \textit{Measurements as unitaries} above.

\vspace{0.5cm}

\textit{Pure states.} Above, we have expressed a pure state as a density operator $\rho$ of the form $\ket{\psi}\bra{\psi}$; an equivalent condition is that $\operatorname{Tr} [\rho^2 ]= 1$. 
Often, a pure state is expressed simply as the normalized vector $\ket{\psi}$ in the Hilbert space. In this case, there is a phase ambiguity, in that $e^{i \theta} \ket{\psi}$ and $\ket{\psi}$ correspond to the same physical state, when considered isolated from everything else. 
This `redundant' phase degree of freedom leads to representations of (classical) symmetry groups on the Hilbert space being projective representations in quantum theory, and may lead to a type of `quantum anomaly', such as the anomalous central charge in a conformal quantum field theory for example~\cite{Weinberg_1995}. Any mixed state can be purified, i.e. for any state $\rho_A$ there exists a system $B$ and a pure state $\rho_{AB}$ on $\mathcal{H}_A \otimes \mathcal{H}_B$ such that $\operatorname{Tr}_B \rho_{AB} = \rho_A$.

\vspace{0.5cm}

\textit{The qubit.} A qubit is the quantum version of a classical bit, i.e. a system with a two-dimensional Hilbert space. A classical bit can be in two states, $\ket{0}$ and $\ket{1}$. These states also form a basis for the two-dimensional Hilbert space of a qubit, but now the qubit can be in a superposition of $\ket{0}$ and $\ket{1}$, i.e. a linear combination of these basis states, \begin{equation}
    \alpha \ket{0}+\beta \ket{1},
\end{equation} with $\alpha,\beta$ complex numbers, normalized such that $|\alpha|^2+|\beta|^2 = 1$.

\vspace{0.5cm}

\textit{Relating to the Hamiltonian.} In continuous-time systems, the time evolution postulate is sometimes phrased in terms of the Hamiltonian, a Hermitian operator whose eigenvalues are the energy levels of a system, reading as \begin{equation}
    i \hbar \partial_t \ket{\psi} = H \ket{\psi}.
\end{equation} 
In fact, unitary time evolution can be seen as arising from the interaction of a system with a clock, as in the Wheeler--DeWitt equation~\cite{wheeler1968super,dewitt1967quantum,page1983evolution,wootters1984time,giovannetti2015quantum,smith2019quantizingtime,boette2018history,gambini2007fundamental,Baumann2021generalized,marletto2017evolution,nikolova2018relational,diaz2019history,castro2020quantum,smith2020quantum,gambini2006relational,hohn2020switch,Castro_Ruiz_2017}, with the Hamiltonian generating the unitary representation~\cite{gogioso2019diagrammatic}.
In fact, doing the same for spatial coordinates leads to a quantum version of Einstein's rods and clocks~\cite{Blommaert_2019,geng2025quantumrodsclockgravitational,Giacomini_2021,wang2024relational,Castro_Ruiz_2017}, closely related to the paradigm of quantum reference frames that we will encounter in \Cref{part:universality_black_holes} of this thesis.


\vspace{0.5cm}

\textit{Alternative postulates.}  Above, we have adopted a standard way of expressing the postulates of quantum mechanics. 
However, there are alternative ways of axiomatizing or deriving quantum theory. 
In fact, an active area of research in quantum foundations concerns reconstructions of quantum theory and derivations from some core physical, informational or structural principles, aiming to answer the questions such as `What is quantum theory?' and `Why quantum theory?'. 
Developments in this field (including efforts reducing the number of postulates or towards a full derivation) can be categorized into operational derivations~\cite{hardy2001quantumtheoryreasonableaxioms,hardy2011reformulatingreconstructingquantumtheory,hardy2013reconstructingquantumtheory,Chiribella_2010,Chiribella_2011,Masanes_2011,M_ller_2015, barrett2007information,dakic2009quantumtheorybeyondentanglement,muller2021probabilistic,masanes2019measurement,Goyal_2010}, which posit physically motivated axioms on the structure of states, transformations and measurements; derivations that aim to derive quantum theory from information-theoretic principles on experimental correlations~\cite{popescu1994quantum,Brassard_2006,pawlowski2009information,Fritz_2013,Oppenheim_2010,SORKIN_1994,Brukner_2009,Clifton_2003,cabello2019quantum,Cabello_2019,chiribella2014measurementsharpnesscutsnonlocality}; agent-centric derivations~\cite{H_hn_2017,Hoehn:2016otu,H_hn_2017questions,M_ller_2020,Fuchs_2013}; and diagrammatic reconstructions~\cite{Selby_2021} inspired by categorical quantum mechanics~\cite{Coecke_2009,abramsky2004categorical,abramsky2009categorical}. 

\vspace{0.5cm}

\textit{Towards quantum field theory and beyond.}
Moving from finite-dimensional quantum theory to local quantum field theory (QFT), i.e. an infinite-dimensional quantum theory with infinitely many degrees of freedom $\phi(x,t)$ (formally) labeled by spacetime points, comes with many difficulties. 
The notions of subsystems, factorizations and tracing over subsystems must be treated with care, and instead of functions and operators, one works with fields as operator-valued distributions instead.
As a result, divergences in the calculation of certain quantities arise, and are typically handled by regularization and renormalization procedures, or by the introduction of novel concepts.
In the case of special-relativistic QFT, or a QFT with some other symmetry group, one must incorporate Lorentz invariance as well, and notions such as particles become observer-dependent.
In fact, Lorentz and Poincaré invariance, in the form of relativistic wave equations~\cite{fock1926uber,klein1927,Gordon:1926emj,dirac1928proceedings} or Poincaré representations~\cite{wigner1939onunitary}, and locality
can be argued to provide the conceptual leap from quantum mechanics to quantum field theory~\cite{Weinberg_1995}.
Conceptually, in light of the use of methods such as regularization and renormalization, one can view our use of (current) quantum field theories as effective theories~\cite{Weinberg_1995,Wilson:1971bg,POLCHINSKI1984269,Dupuis_2021}.

In curved spacetime, one has the additional difficulty of gauge symmetries associated with the curved spacetime background, namely diffeomorphism invariance and local Lorentz invariance in the presence of spin structures, the multitude of time parameters and the potential absence of a preferred one to define positive frequency modes and build Fock spaces,
and (quantum) backreaction of the background structure including the metric--- perhaps unavoidable crucial features rather than problems. 
Moreover, one must rule out some states on a physical basis, such as the Hadamard condition~\cite{KAY199149,wald1994quantum,Radzikowski:1996pa,Brunetti_2000,Fewster_2013}. 
Such Hadamard states have a short-distance singularity structure matching that of states in Minkowski spacetime with
finite energy density, in line with Einstein's equivalence principle, and is required for example to avoid infinite fluctuations of the quantum field~\cite{Fewster_2013}.

These many problems lead to a multitude of approaches to quantum field theory (on curved spacetime), such as canonical or Fock space quantization~\cite{birrell1984quantum,Weinberg_1995,wald1994quantum}, algebraic quantum field theory~\cite{haag2012local,Brunetti_2003,fewster2015algebraicquantumfieldtheory,fewster2024measurementquantumfieldtheory}, perturbative algebraic quantum field theory~\cite{Brunetti_2000,Rejzner2016,Hollands_2001}, path integral methods~\cite{FeynmanHibbs1965,Gibbons:1976ue,kleinert1995path}, deformation quantization~\cite{Kontsevich_2003,BAYEN197861}, with several possibilities of dealing with gauge symmetries~\cite{dirac2013lectures,faddeev1967feynman,henneaux1992quantization,BATALIN198127,Brunetti:2013maa,thiemann2001introductionmoderncanonicalquantum,Ashtekar_2021}.
Another recent approach to quantum field theory, still under active development, employs non-standard analysis formulated in the framework of categorical quantum mechanics~\cite{gogioso2018quantum,shaikh2022categoricalsemanticsfeynmandiagrams}, allowing for QFT postulates closely resembling those of standard quantum theory.

The approach we will use in \Cref{part:universality_black_holes} is the canonical quantization approach to quantum field theory. We will not present an outline of this approach, but refer the reader to the literature~\cite{birrell1984quantum,Weinberg_1995,wald1994quantum}.


\section{Notions of nonclassicality}\label{sec:nonclassicality}
Having outlined the foundational axioms of quantum theory, we move to the study of nonclassicality in quantum theory. 
The motivation for studying nonclassicality is twofold.
A first motivation is practical, in that it facilitates the harnessing of quantum advantage~\cite{howard2014contextuality,raussendorf2013contextuality,Lostaglio_2020,gupta2023quantum,shahandeh2021quantumcomputationaladvantageimplies,deutsch2020harnessing}, i.e., how the use of quantum theory's resources enables improved performance in certain tasks.
Secondly, studying quantum theory's nonclassicality deepens our understanding of quantum theory, how it defies our classical intuition, complementary to the study of reconstructions and derivations of quantum theory outlined in \Cref{sec:remarks-postulates}.

Why would nature, or at least a big part of it, be described by quantum theory? 
And how does quantum theory differ from classical theory? 
Is it possible that underlying quantum theory there is still some classical reality, potentially inaccessible to us, that explains these mysterious phenomena in nature? 
As Feynman famously expressed: `I think I can safely say that nobody understands quantum mechanics'~\cite{feynman1965}, referring to its counterintuitive properties. 

\textit{Notions of noncontextuality.} One such example defying our classical intuition is the non-existence of a common cause that classically and fully explains the correlations between distant quantum systems, without hidden causal connections, formalized as Bell nonlocality~\cite{bell1964einstein,brunner2014bell,Wood_2015,Pearl_2021,cavalcanti2014modifications}. 
Bell nonlocality can be seen as a special instance of Kochen--Specker (KS) contextuality~\cite{budroni2022kochenspecker}, which captures the nonexistence of hidden classical variables that predetermine the outcomes of experiments. 
Alternatively, also KS noncontextuality can be seen as capturing the idea of a classical causal model explaining empirical data, under the additional assumption that an operational symmetry is preserved in the classical model~\cite{cavalcanti2018classical,Pearl_2021}. 
Another leading notion of nonclassicality is generalized contextuality, which applies to preparations, transformations and measurements. 
KS noncontextuality can be seen as an instance of generalized noncontextuality under the additional assumption of outcome determinism for the measurements~\cite{spekkens2014status}. 

\textit{No fine-tuning.} Central to notions of nonclassicality are the existence of some `classical model' representing the experiment to explain the correlations, together with notions of no fine-tuning~\cite{catani2023mathematical,Wood_2015,cavalcanti2018classical,Pearl_2021}, with causal no fine-tuning for Bell locality and Kochen--Specker noncontextuality~\cite{Wood_2015,cavalcanti2018classical,Pearl_2021} and operational no fine-tuning for generalized noncontextuality~\cite{catani2023mathematical}.
No fine-tuning of a classical model expresses the idea that the model variables are nonconspiratorial, i.e. that the underlying classical model follows the causal structure of the experiment, or reflects properties of the operational theory at hand. 
In defining generalized noncontextuality, if two preparations, measurements or transformations are empirically indistinguishable, operational no fine-tuning requires that they are represented the same in the ontological models framework of classical realism, also known as Leibniz's principle of the identity of empirically indiscernibles~\cite{spekkens2019ontologicalidentityempiricalindiscernibles}.
Operational fine-tuning~\cite{catani2023mathematical} also captures the time-symmetry fine-tuning that a classical realism explanation of quantum theory requires~\cite{leifer_is_2017}.
Causal no fine-tuning or faithfulness of a causal model means that there are no additional conditional independence relations that are not implied by the causal diagram; if this were not the case, slightly varying parameters of the causal model generically leads to a violation of these additional conditional independences.
Bell nonlocality and KS contextuality demonstrate that quantum theory does not admit of a causal no fine-tuned classical model~\cite{Wood_2015,cavalcanti2018classical,Pearl_2021}, and generalized noncontextuality shows that quantum theory does not admit of an operational no-fine tuned classical (realist) model. 
Taken together, quantum theory does not admit of a classical realist explanation free of fine-tunings in either sense.

\vspace{0.5cm}

\textit{Beyond realism?} 
What these no-go theorems expressing the nonclassicality of quantum theory tell us is that, when adhering to the principle of no fine-tuning, classical, realism models cannot explain the experimental data of nature. 
Moreover, within the ontological models framework of classical realism, one can also ask whether the quantum wavefunction should be interpreted ontically or epistemically; does it capture the underlying reality or merely a lack of knowledge of a system? The PBR theorem~\cite{pusey2012reality,leifer2014quantum} demonstrates that the wavefunction cannot be epistemic, modulo some (relatively weak) assumptions. 
However, these classical models are \textit{realist} representations of the world, assuming the existence of `states of reality', independently of any observer, independently of how or by whom they are accessed, using a wording similar to Ref.~\cite{brukner2017quantum}. 
A possible resolution for a deeper understanding of quantum theory thus lies in allowing for relational descriptions and observer-dependencies.
This resolution can be argued to be in line with a resolution to Wigner's friend paradoxes that we will explore in \Cref{chap:partI_refinedFR}, the Relativity of Observed Events, expressing that certain facts may be relative. 

\subsection{Kochen--Specker contextuality and Bell nonlocality}
\label{sec:recap_KSNC}

This section follows Section 2 of Ref.~\cite{walleghem2024connecting}, nearly verbatim. 
We present a minimal amount of mathematics required; 
for a comprehensive introduction, we refer the reader to Refs.~\cite{amaral2018graph,budroni2022kochenspecker,abramsky2011sheaf}.

\vspace{0.5cm}

\textit{Kochen--Specker noncontextuality.} 
Let $\mathcal{M}$ be a finite set of measurements, of which some subsets are compatible, that is, jointly measurable~\cite{kunjwal2014quantum,kunjwal2014notejointmeasurabilitypovms}. 
These subsets are called \emph{contexts}, and a given measurement may appear in more than one context. A context is said to be maximal in a scenario if it is not a subset of any other context. For example, if we have 3 measurements labeled by $M_1,M_2,M_3$ in the same context (denoted as a set $\{M_1,M_2,M_3\}$), we can jointly measure them, obtaining a joint outcome $(m_1,m_2,m_3)$. A \emph{measurement scenario} 
$(\mathcal{M},\mathcal{C},O)$ is defined by a finite set of measurements $\mathcal{M}$, a finite set of maximal contexts $\mathcal{C}$, and a set of possible outcomes $O$ for each measurement.  

One can represent such a measurement scenario by a compatibility hypergraph---a hypergraph that associates a measurement to each node of the graph and a hyperedge to each context~\cite{amaral2018graph}. In specific measurement scenarios, it is sufficient to consider only the 2-skeleton of the compatibility hypergraph, known as the \emph{compatibility graph}~\cite[Def. 2.2, pg. 15]{amaral2018graph} (see also Ref.~\cite{xu2019necessary}), in which case contexts are denoted by cliques in the graph, see Figure~\ref{fig:cycles_prelims} for an example of this. 

\begin{figure}
    \centering
    \includegraphics[width=0.7\textwidth]{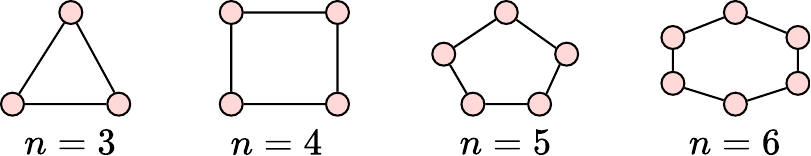}
    \caption[$n$-cycle compatibility hypergraphs]{\textbf{$n$-cycle compatibility hypergraphs.} Cycle graphs $C_n$ for $n=3,4,5,6$. Each graph defines a compatibility structure that can be associated with measurement scenarios; each node is associated with a measurement and the measurements connected by an edge are compatible and form a context. Graph $C_4$ depicts the compatibility structure of the bipartite Bell scenarios leading to the Clauser--Horne--Shimony--Holt (CHSH) inequality~\cite{clauser1969proposed}. Graph $C_5$ depicts the compatibility structure leading to the Klyachko--Can--Binicioğlu--Shumovsky (KCBS) inequality~\cite{klyachko2008simple}. }
    \label{fig:cycles_prelims}
\end{figure}

A behavior in such a measurement scenario is a collection of \emph{empirical correlations}, one for each context $C$, over the outcomes for the measurements in that context. 
An empirical correlation is said to satisfy \emph{no-disturbance} if, for any measurement appearing in more than one context, the marginal probabilities over its outcomes are independent of context.
For example, if $M_1$ is in the context with $C_1\coloneqq\{M_1,M_2\}$ and the context $C_2\coloneqq\{M_1,M_3\}$, namely, $M_1$ is compatible with $M_2$ and $M_3$, then the empirical correlation $p_{C_1}(m_1,m_2)$ of the first context and the empirical correlation $p_{C_2}(m_1,m_3)$ of the second context agree on their overlap\footnote{We say that distributions \emph{agree on their overlap} if they have the same marginals on the intersection of their domains.}, namely the probability for $m_1$:
\begin{align}
\sum_{m_2}p_{C_1}(m_1,m_2)=\sum_{m_3} p_{C_2}(m_1,m_3).
\end{align}

A behavior is said to be Kochen--Specker noncontextual (KSNC) if there is a single `global' probability distribution $p_{NC}(\cdot)$ over all measurements in $\mathcal{M}$ whose marginals reproduce the empirical correlations for each context. Note that any KSNC behavior automatically satisfies no-disturbance, so it only makes sense to consider KSNC for no-disturbing behaviors.
The existence of a global distribution reproducing all empirical correlations is equivalent to the existence of a \textit{deterministic hidden variable model}, or equivalently, a \textit{factorizable} hidden variable model.
Such deterministic hidden variable model is also referred to as an outcome-deterministic ontological model, as we define more carefully in \Cref{sec:gen_noncontext}. This means that there exists a hidden variable, or ontic state $\lambda$, potentially sampled over in each round of the experiment, that deterministically assigns an outcome to each possible measurement.
If for some particular compatibility hypergraph, there is a behavior (or correlations) with no such joint probability distribution, then one calls such a behavior KS contextual.

The set of all empirical correlations satisfying these assumptions in a given scenario is known as the Kochen--Specker noncontextuality polytope and is a function of the compatibility hypergraph for the scenario~\cite{abramsky2016possibilities,choudhary2024lifting,abramsky2011sheaf,barbosa2023closing,abramsky2012cohomology,barbosa2022continuous,araujo2013all,wagner2024inequalities}. 
The set of behaviors satisfying KS noncontextuality is characterized by \textit{KS noncontextuality inequalities}~\cite{Cabello_2008,klyachko2008simple,budroni2022kochenspecker,choudhary2024lifting,choudhary2025classical}.

\vspace{0.5cm}

\textit{Logical and strong contextuality.} 
Contextual correlations are called \emph{logically contextual} (or possibilistically contextual) if the proof that no joint probability distribution exists can be given while only keeping track of whether or not a given combination of measurement outcomes is possible or not, i.e., by looking at the support of the probability distributions in the argument~\cite{abramsky2016possibilities,abramsky2011sheaf}.
The contextuality is said to be \emph{strong} if moreover no single outcome assignment within a context can be extended to a global assignment compatible with the given behavior.

\vspace{0.5cm}

\textit{Example: the KCBS scenario}~\cite{klyachko2008simple}. Consider five binary-outcome measurements $\{M_i\}_{i\in\{1,2,...,5\}}$, where the pairs $\{(1,2)$, $(2,3)$, $(3,4)$, $(4,5)$, $(5, 1 )\}$ are jointly measurable, with compatibility graph $C_5$ as in \cref{fig:cycles_prelims}. 
Imagine that the system is prepared such that the observations for these five joint measurements satisfy
\begin{subequations} \label{eq:5_cycle_12+23_34+45_prelims}
\begin{align}
 p(1,1|1,2)=0, \label{eq:5_cycle_12_prelims} \\
 p(0,0|2,3)=0, \label{eq:5_cycle_23_prelims} \\
    p(1,1|3,4) =0, \label{eq:5_cycle_34_prelims} \\
    p(0,0|4,5)=0, \label{eq:5_cycle_45_prelims}
    \end{align}
\end{subequations}
and
\begin{align}   \label{eq:01_box_not0_prelims}
            p(0,1|5,1) \neq 0.
    \end{align}

Denoting the deterministic assignment of the outcome of measurement $M_i$ as $m_i$, Eq.~\eqref{eq:5_cycle_12+23_34+45_prelims} implies that 
\begin{equation}
\label{eq:mcontr_prelims}
   m_1=1 \Rightarrow m_2=0 \Rightarrow m_3=1 \Rightarrow m_4=0 \Rightarrow m_5=1.
\end{equation}
Therefore, in every run of the experiment where the outcome of $M_1$ is $1$, the outcome of $M_5$ must be $0$. However, \cref{eq:01_box_not0_prelims} tells us that in some runs of the protocol, the outcome of $M_5$ is $0$ and the outcome of $M_1$ is $1$. Thus, there is no Kochen--Specker noncontextual assignment for these runs. A quantum realization of this contextual behavior is given in \Cref{sec:5-cycle_paper1}.

\vspace{0.5cm}

\textit{Bell nonlocality.} A special case of a measurement scenario is a Bell scenario, in which the measurement contexts arise from a party structure where each party can choose from a certain set of measurements to perform on the system they own. 
An example is the two-party Bell scenario. 
Consider Alice and Bob, choosing measurement settings $x$ and $y$, obtaining outcomes $a$ and $b$, respectively. The behavior is then given by the probabilities $p(a,b|x,y)$.
The condition of no-disturbance is also referred to as \emph{no-signaling} in the case of a Bell scenario (with terminology derived from no-superluminal signaling), in this example given by $p(a|x,y)=p(a|x)$ and $p(b|x,y)=p(b|y)$.
The requirement of the existence of a global distribution for this behavior, or equivalently~\cite{fine1982hidden,abramsky2011sheaf}, a factorizable hidden variable model, corresponds to the requirement of Bell's local causality: \begin{equation}
    p(a,b|x,y) = \sum_\lambda p(\lambda) p(a|x,\lambda) p(b|y,\lambda)
\end{equation} with $\lambda \in \Lambda$ the hidden variables, or ontic states, of the system.
In this case, the Kochen--Specker noncontextuality polytope thus coincides with the Bell polytope~\cite{abramsky2016possibilities,choudhary2024lifting,abramsky2011sheaf,barbosa2023closing,abramsky2012cohomology,barbosa2022continuous,araujo2013all,wagner2024inequalities}---that is, the polytope of correlations consistent with Bell's local causality.

\vspace{0.5cm}

\textit{Compatibility, commutativity and no-disturbance for (un)sharp measurements.}
A finite set of measurements is said to be jointly measurable, or compatible, if they can all be obtained from coarse-grainings of a single measurement~\cite{Heinosaari_2008,kunjwal2014notejointmeasurabilitypovms,kunjwal2014quantum,kunjwal2019beyond}.
In quantum theory, sharp or ideal measurements correspond to projective measurements. 
More generally, a theory-independent definition of a sharp (or ideal) measurement is as follows~\cite{chiribella2014measurementsharpnesscutsnonlocality,xu2019necessary,Chiribella_2016}: a measurement is sharp (or ideal) if (i) it yields the same outcome when performed consecutive times, (ii) it only disturbs measurements that are incompatible with it, and (iii) all its coarse grainings have realizations satisfying (i) and (ii). 

In quantum theory, the observables of two compatible projective measurements commute.
The state update rule with projectors implies that two compatible projective measurements, that is, whose observables commute, are non-disturbing; applying the sequence of measurements $ABA$ for projective measurements $A,B$ implies that the second $A$-measurement yields the same outcome as the first. 
In fact, for projective measurements, the notion of commutativity, compatibility and no-disturbance are equivalent~\cite{Heinosaari_2008,budroni2022kochenspecker}.
This is not true anymore for generalized, non-projective measurements, also referred to as unsharp or nonideal measurements; for non-projective measurements such as POVMs, the notions of commutativity, joint measurability and no-disturbance are no longer equivalent~\cite{budroni2022kochenspecker}.
For example, for POVMs, pairwise compatibility of a set of measurements does not imply global compatibility. 
Pairwise commutativity of a set of POVMs does imply global compatibility, but the converse is not true~\cite{kunjwal2014notejointmeasurabilitypovms}.

\vspace{0.5cm}

\textit{Outcome-determinism and unsharp measurements.}\footnote{I thank Raman Choudhary for many interesting discussions on questions related to sharpness, outcome determinism and noncontextuality.} 
As we have seen, requiring KSNC can also be seen as requiring the existence of an outcome-deterministic ontological model.
It has been argued that KSNC, containing this assumption of outcome-determinism, cannot be applied to both sharp and unsharp measurements alike~\cite{spekkens2014status}.
One can motivate the assumption of outcome determinism for sharp quantum\footnote{Generalizing this proof of outcome-determinism from PVMs to theory-independent sharp measurements requires properties of perfect predictability of sharp measurements with eigenstates and operational indistinguishability for sharp measurements when mixing over all their respective eigenstates~\cite{spekkens2014status}.} measurements (and also drop the assumption of outcome determinism for unsharp  measurements) by appealing to the notion of generalized noncontextuality presented in \Cref{sec:gen_noncontext}~\cite{spekkens2005contextuality,liang2010specker,spekkens2014status}. 

However, regarding the question of when outcome-determinism is justified, we have seen that Bell nonlocality is an instance of KSNC, and in the case of Bell nonlocality, outcome-determinism is justified as an instance of Bell's local causality assumption, or as a causal no fine-tuning in a classical causal model~\cite{Wood_2015,cavalcanti2018classical,Pearl_2021}. Moreover, KSNC can be derived similarly under a causal no fine-tuning argument together with a symmetry assumption~\cite{cavalcanti2018classical,Pearl_2021}.
So, where could this story go wrong for unsharp measurements?

One potential viewpoint on this matter is the following, with a crucial role for no-disturbance. 
It could very well be that the description of POVMs is not enough to apply the causal no fine-tuning argument to argue for Bell locality and KS noncontextuality as plausible notions of classicality, as it does not capture the complete description of the experimental set-up. 
Namely, there exist many projective measurements on a larger system that give the same POVM. 
In particular, if one has two no-disturbing POVMs, it could be that in the larger system (including an ancilla and environment, for example) that comprises the experimental set-up described with PVMs, these additional systems introduce correlations leading to the PVMs violating the no-disturbance condition. 
If, on the other hand, in the full experimental set-up, the PVMs of which the POVMs are derived still satisfy the no-disturbance condition, then the causal no fine-tuning conditions employing the no-disturbance condition can be argued to be justified.
The crucial role for no-disturbance means that this argument is still compatible with the criticism of applying outcome determinism to just any POVM~\cite{spekkens2014status}.

This is in line with how literature deals with the case of unsharp measurements and experimental imperfections. 
One approach, on a case-by-case basis, is to pose assumptions on the noise in the hidden-variable model, while typically performing additional measurements to aid this characterization and potentially modifying the noncontextuality inequalities~\cite{budroni2022kochenspecker}.
Another is by devising noise-robust inequalities for KS scenarios within the framework of generalized noncontextuality~\cite{Kunjwal_2015,kunjwal2019beyond,kunjwal2019minimal}.



\subsection{Generalized noncontextuality} \label{sec:gen_noncontext}
Another notion of noncontextuality, referred to as generalized noncontextuality, was introduced by Spekkens in Ref.~\cite{spekkens2005contextuality}, and further developed in Refs.~\cite{spekkens2014status,schmid2018all,schmid2020structure,schmid2021characterization,mazurek2016experimental,kunjwal2019beyond,chaturvedi2021characterising,schmid2024noncontextualityinequalitiespreparetransformmeasurescenarios}. 
An operational theory consists of preparations, measurements and transformations. 
An ontological model for an operational theory is a classical realist description of a physical system as mediating between the preparation procedure and the measurement. 
Inspired by Leibniz principle of the ontological identity of empirical indiscernibles~\cite{spekkens2019ontologicalidentityempiricalindiscernibles}, the idea of generalized noncontextuality is that all operations, transformations and measurements that are statistically indistinguishable should have the same ontological representation.

\paragraph{An operational theory.} An \textit{operational theory} is a quadruple $T = (\mathcal{P},\mathcal{T},\mathcal{M},p)$, where $\mathcal{P},\mathcal{T},\mathcal{M}$ are sets of preparation, transformation and measurement procedures. Furthermore, for each preparation $P$, transformation $T$ and measurement $M$, there is a probability distribution $p(k|M,T,P)$ over the outcomes of the measurement $M$ on the $T$-transformed preparation $P$. 
The underlying physical intuition is that a system is prepared in the lab using procedure $P$, transformed under $T$ and measured in measurement $M$, giving outcome $k$ with probability $p(k|M,P)$. 
Notably, also unsharp measurements, i.e. POVMs in quantum theory, are also allowed. We also refer to the combination $(k,M)$ of a measurement $M$ and outcome $k$ as the effect $k|M$ or $[k|M]$. 

Probabilistic mixtures of preparations are considered again a preparation, denoted by $\sum_i p_i P_i$ for performing preparation $P_i$ with probability $p_i$.
Statistically indistinguishable preparations are called operationally equivalent, and similarly for measurements and transformations, as defined below in \textit{Operational equivalences and generalized noncontextuality.}

\vspace{0.5cm}

\textit{Ontological models.} The ontological models framework~\cite{harrigan2010einstein} formally provides a realist explanation of the statistics of a physical theory by postulating the existence of physical systems with properties (ontic states) represented by points $\lambda$ in a measurable set $\Lambda$ (the ontic state space). The statistics predicted by the theory are reproduced, in the ontological model, via the law of classical total probability: \begin{equation} p(k|M,P)= \sum_\lambda p(k|M,\lambda)p(\lambda|P), \end{equation} where the preparation $P$ is represented by a probability distribution $p(\lambda|P)$, and the measurement $M$ with outcome $k$ is represented by a probability distribution $p(k|M,\lambda)$, also referred to as the response or indicator function.
A transformation is represented by a transition matrix $p(\lambda|\lambda',T)$, giving the probability density for a transition from the ontic state $\lambda'$ to the ontic state $\lambda$.

The distributions $p(\lambda|P), p(k|M,\lambda), p(\lambda|\lambda',T)$ associated with a preparation procedure, measurement and transformations are often denoted by $\mu(\lambda|P), \xi(k|M,\lambda), \Gamma_T(\lambda|\lambda')$.

\vspace{0.5cm}

\textit{Operational equivalences and introducing generalized noncontextuality.}
Generalized noncontextuality requires that operational equivalences predicted by a physical theory correspond to identities in the ontological model of the theory. For instance, if two preparations $P$ and $P'$ are operationally equivalent, denoted $P \sim P'$, if they cannot be distinguished by any measurement, i.e. if for all possible measurements $M$ with outcomes $k$, $p(k|M,P) = p(k|M,P')$. 
An ontological model is then said to satisfy \emph{preparation noncontextuality} if it represents them with identical probability distributions, i.e. $p(\lambda|P) = p(\lambda|P')$ for all $P\sim P'$. 

A similar definition applies to measurements and transformations.
Two measurement effects $[k|M],[k|M']$ are operationally equivalent if they give the same results for any preparation, i.e. $[k|M] \sim [k|M']$ if their outcomes can be associated such that for all $P$ one has $p(k|M,P)=p(k|M',P)$. 
An ontological model is then said to satisfy \emph{measurement noncontextuality} if it equips them with equal response functions, i.e. $p(k|M,\lambda) = p(k|M',\lambda)$ for all $[k|M]\sim [k|M']$. 
Finally, two transformations $T,T'$ are operationally equivalent, denoted by $T \sim T'$ if $p(k|P,T,M)=p(k|P,T',M)$ for all preparations and measurements $P,M$. 
An ontological model is then said to satisfy \emph{measurement noncontextuality} if it equips them with equal transition matrix, i.e. $\mu(\lambda|\lambda', T) = \mu(\lambda|\lambda',T)$ for all $T \sim T'$.

In the paradigm of generalized noncontextuality, it is moreover assumed that probabilistic mixtures of preparations are represented as convex mixtures in the ontological model as well, i.e., probabilistically preparing $P_i$ with probability $p_i$, leading to measurement statistics $\sum_i p_i p(k|M,P_i)$, is represented as $\sum_i p_i \mu(\lambda|P_i)$, and similarly for measurements. I name this a \textit{convexity-preserving} assumption, and it is typically taken part of the definition of generalized noncontextuality definition itself.

\vspace{0.5cm}

\textit{Definition of generalized noncontextuality.} Summarizing, generalized noncontextuality requires that for an operational equivalence of preparations \begin{equation}
    \sum_j \alpha_{P_j} P_j \sim \sum_{j^{\prime}} \beta_{P_{j^{\prime}}} P_{j^{\prime}}
\end{equation} it holds that \begin{equation}
    \quad \sum_j \alpha_{P_j} \mu_{P_j}(\lambda)=\sum_{j^{\prime}} \beta_{P_{j^{\prime}}} \mu_{P_{j^{\prime}}}(\lambda) \text{ for all } \lambda,
\end{equation} with $\alpha_{P_j},\beta_{P_j}$ probability distributions, and for an operational equivalence of measurement effects \begin{equation}
    \sum_{i, m} \alpha_{k \mid M_i} \left[k \mid M_i\right] \sim \sum_{i^{\prime}, k^{\prime}} \beta_{k^{\prime} \mid M_{i^{\prime}}} \left[k^{\prime} \mid M_{i^{\prime}}\right]
\end{equation} it holds that
\begin{equation}
    \sum_{i, m} \alpha_{k \mid M_i} \xi_{k \mid M_i}(\lambda)=\sum_{i^{\prime}, k^{\prime}} \beta_{k^{\prime} \mid M_{i^{\prime}}} \xi_{k^{\prime} \mid M_{i^{\prime}}}(\lambda) \text{ for all } \lambda 
\end{equation}
with $p_{k i}=\alpha_{k \mid M_i}, q_{ki}=\beta_{k' \mid M_{i'}}$ probability distributions over $(k,i)$.

In a finite scenario, the set of correlations in an operational theory satisfying linear constraints derived from generalized noncontextuality forms a convex polytope and characterized by \textit{(generalized) noncontextuality inequalities}.~\cite{schmid2018all,schmid2024noncontextualityinequalitiespreparetransformmeasurescenarios}; techniques for addressing nonlinear constraints are yet to be developed~\cite{schmid2024noncontextualityinequalitiespreparetransformmeasurescenarios}.

\vspace{0.5cm}

\textit{A quantum example violating generalized noncontextuality.} The simplest proof that quantum theory manifests \textit{contextuality}, \textit{i.e.}, it does not admit of a generalized noncontextual ontological model, can be constructed in the \textit{simplest nontrivial scenario}~\cite{pusey2018robust,pusey2019,Khoshbin_2024}. This is a prepare-and-measure scenario consisting of two tomographically complete measurements, \textit{e.g.}, the Pauli $X$ and $Z$ measurements on a qubit, 
 and four preparations, \textit{e.g.}, $P_{\theta}= \cos(\theta/2)\ket{0}+\sin(\theta/2)\ket{1}$, for $\theta = \pi/4,3\pi/4,5\pi/4,7\pi/4$. These are represented on the Bloch disk in ~\cref{fig:Bloch_prelims}. 
An equal mixture of the preparations $\pi/4,5\pi/4$ yields the maximally mixed state, and similarly for the two preparations $3\pi/4,7\pi/4$.
Thus, they satisfy the following operational equivalence, \begin{align}
    &\frac{1}{2}p(k|M,P_\frac{\pi}{4}) +  \frac{1}{2}p(k|M,P_\frac{5\pi}{4}) =  \frac{1}{2}p(k|M,P_\frac{3\pi}{4}) +   \frac{1}{2}p(k|M,P_\frac{7\pi}{4}) \quad \forall [k|M].
\end{align} 
However, there does not exist an ontological model where these also provide an ontological identity.

\begin{figure}[h]
         \centering \includegraphics[width=0.29\textwidth]{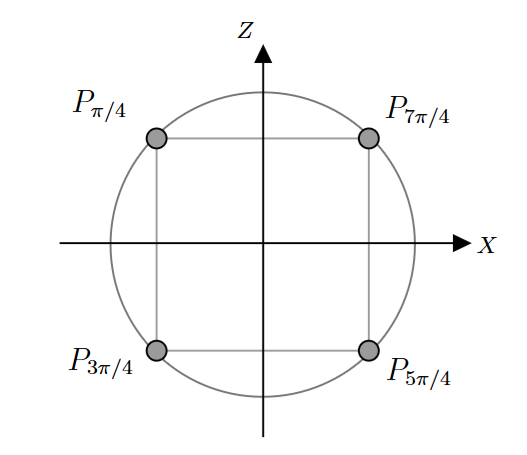}
         \caption[Simplest nontrivial scenario for generalized contextuality]{The simplest nontrivial scenario~\cite{pusey2018robust},
         involving four preparations (gray dots) and two measurements (the two axes) that provide a maximal violation of the noncontextuality inequalities.}
         \label{fig:Bloch_prelims}
\end{figure}

\paragraph{Outcome-deterministic ontological models} 
An ontological model for an operational theory $T$ is called \textit{outcome-deterministic} for measurement $M$ and ontic state $\lambda$ we have $\xi(k|M,\lambda) \in \{0,1\}$, and outcome-deterministic for a measurement $M$ if it is so for all states $\lambda$.

\section{The measurement problem, Wigner's friend and extensions}\label{sec:WF}
In its standard formulation, according to the textbook postulates of \Cref{sec:postulates}, quantum theory prescribes two distinct ways of time evolution: (i) the unitary, deterministic evolution of a closed system according to Schrödinger's equation, and (ii) the probabilistic, indeterministic evolution of a system after a measurement upon it, also referred to as the wavefunction collapse. 
But what causes the wavefunction to collapse? What precisely qualifies as a measurement such that evolution (ii) applies?

This puzzle, also referred to as the measurement problem\footnote{Often, one distinguishes several measurement problems; the one described here is also referred to as the `big' measurement problem~\cite{Pitowsky2007,BubPitowsky2010,brukner2017quantum}. Other related problems concern the question of why we do not observe macroscopic superpositions, in whose answer decoherence plays a prominent role~\cite{schlosshauer2005decoherence,brukner2017quantum,Pitowsky2007,BubPitowsky2010}.}~\cite{schlosshauer2005decoherence,legget20005measurement,Hance_2022,brukner2017quantum,BubPitowsky2010,Pitowsky2007}
has been a veil of mystery right since the birth of quantum mechanics. 
A famous illustration of this problem was formulated in 1935 in the form of Schrödinger's cat, where the cat's survival depends on a quantum event, and, according to the unitary Schrödinger evolution, the cat, in a sealed box, evolves into a superposition of dead and alive. 
However, when we open the box, the cat is found either dead or alive, and the wavefunction has collapsed. 
But what causes this collapse, and when? Is it the role of the observer, for the cat's wavefunction to collapse only when observed by us, or the chamber, or something else? 

Wigner's thought experiment~\cite{wigner1995remarks}, dating back to 1967, is another scenario showcasing the tension of the measurement problem, where an observer referred to as Wigner's friend resides in a sealed lab and measures a quantum system. 
According to Wigner, a \textit{superobserver}, the friend is ultimately a quantum system too and evolves into a superposition of seeing different outcomes, at odds with the friend's description, and it is only when Wigner opens the friend's lab to ask for their outcome that their descriptions match.

One can argue that this contradiction is only problematic if one believes the quantum state is a real physical state, with only one possible description, irrespective of anything else in the universe~\cite{Pitowsky2007}. 
However, the problem then shifts to the question of when, where and how a measurement happens, and which systems would count as measurement devices or observers.
Recent results have strengthened the tension in the measurement problem, obtaining a contradiction at the level of outcomes of performed measurements and observers' predictions.
More technically, these results consider extensions of Wigner's thought experiment combined with Bell nonlocality. 
Namely, in a Bell nonlocality scenario not all measurements are performed at once, and several rounds of the experiment are required in order to obtain all relevant joint correlations. 
In 2015, Brukner did away with this counterfactuality by proposing a thought experiment involving multiple Wigners and their friends, who share a Bell nonlocality state~\cite{brukner2017quantum}, laying the foundations for the Frauchiger--Renner paradox~\cite{frauchiger2018quantum} and Local Friendliness no-go theorem~\cite{bong2020strong}. 

In the rest of this subsection we outline Wigner's friend scenario, after which we present the Frauchiger--Renner paradox and Local Friendliness no-go theorem in \Cref{sec:FR_overview,sec:LF}, laying the basis for our work in the next chapters of Part I.

\subsection{Wigner's friend scenario}\label{sec:Wignerfriend}
The Wigner's friend scenario~\cite{wigner1995remarks} involves Wigner and his friend, Bob, who resides inside a sealed lab and measures a quantum system $S_{\!B}$.
Wigner treats Bob's lab itself as a quantum system, which Wigner, as a superobserver of Bob, may measure and control from the outside.
We denote by $L_{\!B}$ the system consisting of Bob's lab, including Bob himself but excluding $S_{\!B}$.
The setup is represented schematically in \cref{fig:Wignerfriend}.

\begin{figure}[h]
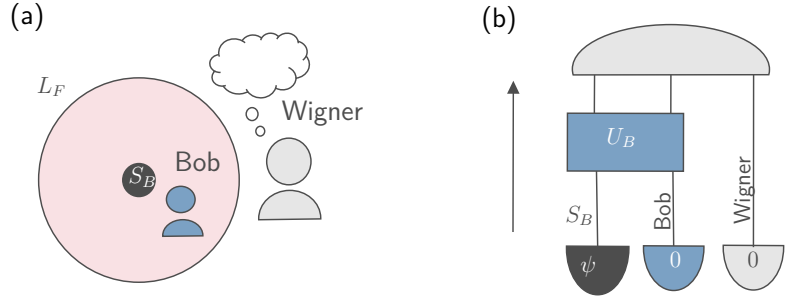
  \centering   \includestandalone[width=0.7\textwidth]{Figures/FINAL_WF} \caption[Wigner's friend scenario]{(a): Schematic sketch of Wigner's friend scenario. Bob performs a measurement on a system $S_{\!B}$ in a sealed lab $L_{\!B}$. Wigner is a superobserver of his friend Bob, modelling Bob's measurement unitarily as $U_B$. The friend sees a definite single-valued outcome, but Wigner describes his friend as being in a superposition of having obtained different outcomes. (b): Circuit representation of a Wigner's friend scenario.} \label{fig:Wignerfriend}
\end{figure}

The quantum system $S_{\!B}$ is initialised in the state $\ket{\psi}_{S_{\!B}} = \alpha \ket{0}_{S_{\!B}} + \beta \ket{1}_{S_{\!B}}$
with $\alpha, \beta \neq 0$.
The friend measures the system $S_B$ in the computational basis $\{ \ket{0}_{S_{\!B}}, \ket{1}_{S_{\!B}} \}$, labelling the outcomes  $0, 1$.
As a \emph{superobserver} for his friend, Wigner models Bob's measurement of the system $S_{\!B}$ as a unitary evolution of the joint quantum system $S_{\!B} L_{\!B}$\footnote{We will often omit tensor product when it is clearly implied, i.e. here $S_{\!B} L_{\!B}$ is an abbreviation of $S_{\!B} \otimes L_{\!B}$.}:
\begin{equation} \label{eq:unitary_friend}
    \ket{\psi}_{S_{\!B}} \ket{r}_{L_{\!B}} \;\;\overset{U_B}{\longmapsto}\;\; 
    \alpha \ket{00}_{S_{\!B} L_{\!B}} + \beta \ket{11}_{S_{\!B} L_{\!B}},
\end{equation}
where $\ket{0}_{L_{\!B}}$ and $\ket{1}_{L_{\!B}}$ correspond to the two possible outcomes as recorded in Bob's memory and elsewhere in his lab, and $\ket{r}_{L_{\!B}}$ denotes the initial (`ready') state of his lab.
From Wigner's perspective, the system $S_{\!B} L_{\!B}$ is in a superposition of
`$S_{\!B}$ is in state $\ket{0}$ and Bob has recorded the measured outcome $0$' and
`$S_{\!B}$ is in state $\ket{1}$ and Bob has recorded the measured outcome $1$'.
Note that the unitary evolution $U_B$ is described (on the relevant degrees of freedom) as a CNOT gate.
The friend sees a definite single-valued outcome, while Wigner would describe his friend as being in a superposition of states corresponding to different outcomes.

Both claims can in principle be tested, as in Deutsch's version of the Wigner's friend argument~\cite{deutsch1985quantum}; see also Ref.~\cite[Section 4]{nurgalieva2020testing}.
Indeed, the fact that the friend has obtained a definite, single-valued outcome can in principle be known to Wigner, namely by the friend passing a note to Wigner stating `I see a definite single-valued outcome' (but crucially without giving away any information about what the outcome actually is).
Wigner's description of the friend as being in a superposition state can also in principle be tested. 
Namely, as a superobserver, Wigner can perform state tomography on the entire friend's lab. One way for Wigner to perform such state tomography is the following~\cite{nurgalieva2020testing}.
As a superobserver, Wigner can \textit{undo} the friend's measurement, by applying the inverse unitary $U_B^\dagger$ to the friend's lab and the system, thereby resetting them back to their initial state.
Thus, by applying $U_B^\dagger$ and verifying whether the obtained initial state of $S_B L_B$ is indeed $|\psi \rangle_{S_B}|0 \rangle_{L_B}$ Wigner can verify his superposition assignment. 
Note that Wigner's \textit{undoing} of the friend's measurement is only effective, i.e. only resets the lab back to its initial state, if no information about the friend's measurement outcome escapes the lab.

An actual human playing the role of the friend may be practically infeasible, but an artificially intelligent program on a future quantum computer may be a reasonable physical realization of such an observer~\cite{wiseman2023thoughtful}.

This thought experiment shows the incompatibility between the existence of superobservers and an absolute, observer-independent notion of collapse~\cite{bong2020strong,nurgalieva2018inadequacy,deutsch1985quantum}.
However, even if one takes the wave function collapse not to be absolute, the possibility remains that the underlying fact of the friend's observed outcome be absolute~\cite{bong2020strong}.
To see this, suppose Wigner and his friend run the protocol many times. In some rounds Wigner verifies his superposition assignment of \cref{eq:unitary_friend} for the state of his friend Bob's lab, while in other rounds Wigner directly asks his friend Bob for the outcome. 
In that case, the two may disagree on (the timing of) the occurrence of collapse, but they agree on the underlying fact, namely which outcome the friend observed. 
While not asking his friend, Wigner may describe his friend as being in a superposition, but this does not rule out the possibility of there being an absolute fact regarding which outcome the friend obtained, even if unknown to or hidden from Wigner.

This is where extended Wigner's friend scenarios, involving more than one superobserver, come into play.
Since Brukner's scenario~\cite{brukner2017quantum}, many different extended Wigner's friend scenarios have been proposed~\cite{frauchiger2018quantum,bong2020strong,cavalcanti2021implications,haddara2022possibilistic,leegwater2022greenberger,schmid2023review,brukner2018no,ormrod2022no,ormrod2023theories,nurgalieva2018inadequacy,vilasini2019multi,montanhano2023contextuality,schmid2023review,leegwater2022greenberger,zukowski2021physics,walleghem2023extended,montanhano2023contextuality,gao2019quantum,guerin2020no,utreras2023allowing,szangolies2023quantum}. 
Among them, the Frauchiger--Renner paradox~\cite{frauchiger2018quantum} and Local Friendliness no-go theorem~\cite{bong2020strong,cavalcanti2021implications} have gained much attention.

\vspace{0.5cm}

\textbf{Terminology: superobservers and classical agents.}
For reasons of clarity, we briefly define what we mean by an agent, a measurement and introduce terminology regarding (super)observers and (classical) agents.

\paragraph*{Definition of an agent and measurement.} \label{def:agent}
An agent is any system that can obtain information through measurements, and store information. In quantum information theory using the Hilbert space formalism, the ability to store information means that an agent Bob has a personal preferred orthogonal basis in which he can store information. Storing information in these orthogonal `knowledge' states makes sure his information states are perfectly distinguishable for him. The basis must be a preferred one because not every quantum basis can be seen as encoding information for a single observer~\cite{brukner2021qubits,allam2023observer}. By a measurement an interaction of the agent Bob with a system $S$ is meant, that increases the information in his memory (i.e. in his knowledge states, orthogonal states in some (personal) preferred basis). The agent being able to obtain information thus means that the agent must be able to perform some measurements, and store the outcome in his knowledge states\footnote{Storing an outcome in a preferred basis might lead to production of entropy as in practice it corresponds to a state preparation, see for example~\cite{di2021arrow}, and thus might require suitable entropic conditions and an (agential) thermodynamic arrow of time. A preferred basis might also arise from interactions.}.

\paragraph*{Terminology regarding superobservers and classical agents.}
We will use the term observer and agent interchangeably, and by a superobserver, such as Wigner, we mean an observer who has extreme quantum control over another observer (their `friend'). As a result, a superobserver models the evolution of their friend within their sealed lab quantumly, i.e. as a unitary, including measurements performed by this friend. 
Within a given measurement protocol, by a classical agent we mean an agent who need not be modeled quantumly by anyone else in the protocol.

\section{The Frauchiger--Renner paradox}
\label{sec:FR_overview}

In this subsection we review the Frauchiger--Renner (FR) paradox.
We observe that the FR paradox
(i) can be strengthened by a slight modification in the reasoning stage, and
(ii) is based on the nonlocal Hardy model.
These observations suggest a stronger FR-like paradox that we describe in \Cref{chap:partI_refinedFR}, where we also carefully discuss the underlying assumptions.

\subsection{Entanglement version of the FR paradox}
We sketch the entanglement version of the FR paradox as outlined in Ref.~\cite{vilasini2022general}, credited to Lluis Masanes.
For the original prepare-and-measure version of the FR paradox we refer to Refs.~\cite{frauchiger2018quantum,vilasini2022general}.

The scenario is depicted in \cref{fig:FR_circuit}(a).
It involves two superobservers, Ursula  and Wigner, each with their own `friend', Alice and Bob respectively.
Each of the friends reside in a sealed lab with access to quantum systems $S_{\!A}$ and $S_{\!B}$ respectively, which are initialised in a shared entangled state.
As before, $L_{\!A}$ denotes Alice's lab including Alice herself but excluding the system $S_{\!A}$, and similarly $L_{\!B}$ for Bob.
All the agents -- Alice, Bob, Ursula, and Wigner -- perform measurements, obtaining outcomes $a,b,u,w \in \{0,1\}$ respectively.

\paragraph*{Measurement protocol.} 
The measurement protocol, sketched in circuit form in \cref{fig:FR_circuit}(b), proceeds as follows;
\begin{enumerate}[label=\arabic*.,leftmargin=*]
    \item \textbf{Initialisation.} The system $S_{\!A} S_{\!B}$, shared between Alice and Bob, starts in the specific entangled state $|\psi_0\rangle_{S_{\!A} S_{\!B}} = \frac{1}{\sqrt{3}} \left(\ket{00}+\ket{10}+\ket{11}\right)_{S_{\!A} S_{\!B}}$.
     The initial states of Alice's and Bob's labs are labelled by $\ket{r}_{L_{\!A}}$ and $\ket{r}_{L_{\!B}}$, respectively.
     The initial state of $S_{\!A} L_{\!A} S_{\!B} L_{\!B}$ is given by\footnote{\label{fn:tensorproduct}For notational ease, we will often omit the tensor product symbol: $\ket{00}_{AB}=\ket{0}_A \ket{0}_B = \ket{0} \otimes \ket{0}$.}
    \begin{align}
    \begin{split}
    \ket{\psi^{t=1}}_{S_{\!A} L_{\!A} S_{\!B} L_{\!B}}
    \;=\;&
    \frac{1}{\sqrt{3}}\left(\ket{0 r 0 r}+\ket{1r0r}+\ket{1r1r}\right)_{S_{\!A} L_{\!A} S_{\!B} L_{\!B}}.
    \end{split}
    \end{align}

\item \textbf{Alice and Bob measure.} Alice measures her system $S_{\!A}$ in the $\{ \ket{0}, \ket{1} \}$ basis, 
    and Bob similarly measures his system $S_{\!B}$.
    The unitary description of these measurements, as in \cref{eq:unitary_friend}, leads to the overall state
    \begin{align}\label{eq:FR_psi_t2}
    \begin{split}
    \ket{\psi^{t=2}}_{S_{\!A} L_{\!A} S_{\!B} L_{\!B}}
    \;=\;& \left(U_{A} \otimes U_{B} \right)\ket{\psi^{t=1}}_{S_{\!A} L_{\!A} S_{\!B} L_{\!B}}
     \\ \;=\;& \frac{1}{\sqrt{3}}(\ket{0000}+\ket{1100}+\ket{1111})_{S_{\!A} L_{\!A} S_{\!B} L_{\!B}}.
    \end{split}
    \end{align}
    where $U_{A}$ and $U_{B}$ are the unitaries describing Alice's and Bob's measurements, respectively.
    \item \textbf{Ursula and Wigner supermeasure.} Finally, the superobservers Ursula and Wigner respectively measure the systems $S_{\!A} L_{\!A}$ and $S_{\!B} L_{\!B}$ in a 
    basis with vectors\footnote{Note that a basis requires two more orthonormal vectors, but their corresponding outcomes have probability zero, so we do not need to specify the remaining vectors. \label{fn:fullbasis}}
    \begin{equation} \ket{\textok}=\frac{1}{\sqrt{2}}(\ket{00}-\ket{11}), \qquad \ket{\textfail}=\frac{1}{\sqrt{2}}(\ket{00} + \ket{11}) .\end{equation} Note that these measurements can be seen as the composition of (i) undoing the friend's measurements by applying the inverse unitaries $U_A^\dagger, U_B^\dagger$, followed by (ii) an $X$-measurement on $S_A,S_B$, respectively.
\end{enumerate}
The protocol is post-selected on both Ursula and Wigner obtaining the outcome $\textok$, denoted $u = \textok$ and $w = \textok$.
Observe that this happens with nonzero probability:
\begin{align}
&\abs*{\bra{\textok}_{S_{\!A} L_{\!A}}\otimes\bra{\textok}_{S_{\!B} L_{\!B}}\ket{\psi^{t=2}}_{{S_{\!A} L_{\!A} S_{\!B} L_{\!B}}}}^2
\\
\;=\; & \abs*{\frac{1}{2\sqrt{3}} (\bra{0000}-\bra{0011}-\bra{1100}+\bra{1111} ) (\ket{0000}+\ket{1100}+\ket{1111})}^2
\\ \;=\; & \abs*{\frac{1}{2\sqrt{3}}}^2 = \frac{1}{12}  \;>\; 0. 
\end{align}

\paragraph*{Reasoning steps.}
After a successful post-selected run of the protocol, Ursula reasons (i) about Bob's outcome $b$, (ii) about Bob's conclusion for Alice's outcome $a$, and (iii) about Bob's reasoning about Alice's reasoning about Wigner's outcome $w$.
Assuming that each agent uses the Born rule to make predictions, Ursula, using $u = \textok$, reasons that Bob must have obtained $b=1$, and thus that he would have predicted that $a=1$, so that Alice would have concluded that $w = \textfail$.
In summary,
$$ u = \textok \;\Rightarrow\; b = 1 \;\Rightarrow\; a = 1 \;\Rightarrow\; w = \textfail,$$
contradicting the initial post-selection on $u = \textok, w = \textok$.
We give additional details in the next section where we show how the argument may in fact be strengthened.

\begin{figure}
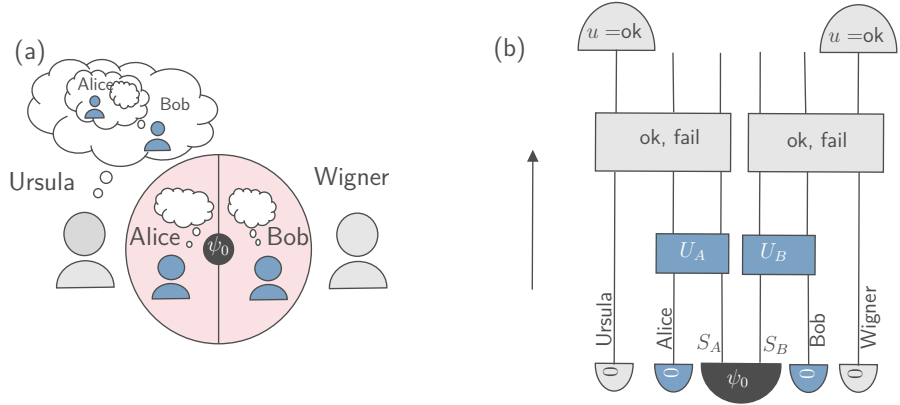
 \centering \includestandalone[width=0.8\textwidth]{Figures/FINAL_FR2}   \caption[Sketch of the FR paradox]{(a): Sketch of the FR paradox, where Ursula reasons about Bob who reasons about Alice who reasons about Wigner. (b): Circuit of the entanglement version of the FR paradox, as outlined in Ref.~\cite{vilasini2022general}, with the initial state of $S_{\!A}\otimes S_{\!B}$ given by $|\psi_0\rangle_{S_{\!A} S_{\!B}} = \frac{1}{\sqrt{3}} \left(\ket{00}+\ket{10}+\ket{11}\right)_{S_{\!A} S_{\!B}}$. 
} \label{fig:FR_circuit} 
\end{figure}

\subsection{A slightly stronger modified FR paradox}
\label{sec:stricter_FR}
We provide a stronger variant of the FR paradox in which only \emph{classical}\footnote{For a definition of a \emph{classical} agent, see \Cref{sec:Wignerfriend}.}
 agents reason.
This technique will be used in the GHZ--FR paradox of \Cref{chap:partI_refinedFR} as well.

The measurement protocol is identical to that of the FR paradox.
But an additional agent, Zeno,\footnote{The additional agent is needed here to reason about the joint outcomes of the friends Alice and Bob, but will not be necessary in the case of the GHZ--FR paradox of \Cref{sec:FR_refining}.} who does not perform any measurement, makes a prediction for Alice's and Bob's outcomes.
Ursula and Wigner, after obtaining their outcomes, reason about the outcomes obtained by Alice and Bob, respectively.
Ursula, and Wigner hand their predictions to Zeno, who derives a contradiction.\footnote{We note that Ref.~\cite{healey2018quantum} gives a presentation of the reasoning steps of the FR paradox that comes close to the one here.}

Post-selecting on $u=\textok,w=\textok$, which has a non-zero probability of occurring, we have the following predictions from applications of the Born rule: 
\begin{itemize}
    \item Ursula predicts that $b=1$ by excluding $b=0$ since 
\[\bra{\textok}_{S_{\!A} L_{\!A}} \bra{00}_{S_{\!B} L_{\!B}} \ket{\psi^{t=2}}_{S_{\!A} L_{\!A} S_{\!B} L_{\!B}}=0\,.\]

    \item Wigner predicts that $a=0$ by excluding $a=1$ since 
    \[\bra{11}_{S_{\!A} L_{\!A}}\bra{\textok}_{S_{\!B} L_{\!B}} \ket{ \psi^{t=2}}_{S_{\!A} L_{\!A} S_{\!B} L_{\!B}}=0\,\]
    \item Zeno predicts that $a=0, b=1$ can never occur since 
    \[\bra{00}_{S_{\!A} L_{\!A}}\bra{11}_{S_{\!B} L_{\!B}} \ket{\psi^{t=2}}_{S_{\!A} L_{\!A} S_{\!B} L_{\!B}}=0\,. \]
\end{itemize}
Ursula and Wigner classically communicate their predictions to Zeno.
Upon combining the predictions made by Ursula and Wigner with his own, Zeno finds a contradiction.
Thus, all reasoning and predicting is done by classical agents. 
Furthermore, each application of the Born rule refers to measurement outcomes that are obtained concurrently: its predictions could in principle be tested by interrupting the protocol before one or both supermeasurements.

\subsection{Logical non-locality powers the FR paradox}\label{sec:FR_non-locality}
In \Cref{sec:stricter_FR} we have seen how actually a contradiction can be obtained by having only \emph{classical} agents reason in the FR protocol.
The second crucial observation about the FR paradox is to recognise that it is underpinned
by the Hardy model~\cite{hardy1993nonlocality}, a proof of ``nonlocality without inequalities''. 
This approach has a logical flavour in that nonlocality is witnessed in terms of \emph{possibilities} alone rather than requiring full knowledge of the \emph{probabilities}.
In fact, Frauchiger and Renner~\cite{frauchiger2018quantum} explicitly stated that their protocol is motivated by Hardy's paradox, a connection that has also been emphasized in subsequent works such as Refs~\cite{frauchiger2018quantum,aaronson2018s,drezet2018wigner,vilasini2019multi,montanhano2023contextuality,vilasini2022general,fortin2019wigner}.

The Hardy model is a specific quantum realisation of the simple $(2,2,2)$ Bell scenario. This nonlocality scenario involves two parties, each able to choose between two measurement settings, with two outcomes $0$ and $1$.
The Hardy model, on which the FR paradox is based, arises from a specific choice of initial shared state and local measurements in the $(2,2,2)$ Bell scenario.
The two parties share a system in the entangled state  $\sfrac{1}{\sqrt{3}} (\ket{00}+\ket{10}+\ket{11}$,
which each can choose to measure locally in either the $\{\ket{0},\ket{1}\}$ or the $\{\ket{+},\ket{-}\}$ basis, known respectively as the $Z$- and the $X$-basis.\footnotemark

\cref{fig:FR_Rui} depicts a mapping between the $(2,2,2)$ Bell scenario and the extended Wigner's Friend scenario of the FR paradox.

\begin{figure}
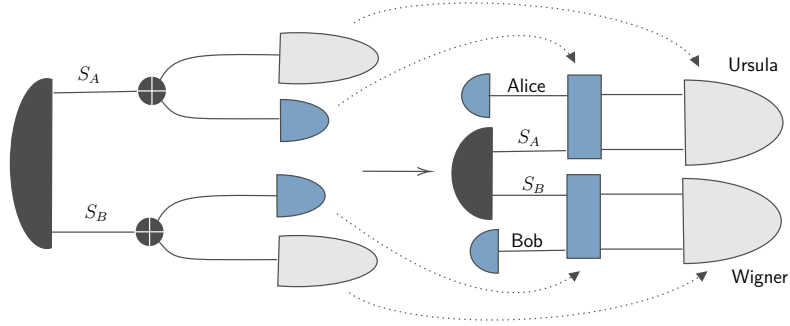
 \centering 
\includestandalone[width=0.7\textwidth]{Figures/tikz_FR_to_Bell}  \caption[Mapping a Bell nonlocality scenario to an FR scenario]{Mapping a Bell non-locality scenario (left) to an FR-like extended Wigner's friend scenario (right). In the Bell scenario each party can choose (pictured by a crossed circle) one out of two (incompatible) measurements, for example in the $Z$ basis $\{\ket{0},\ket{1}\}$ or in the $X$ basis $\{\ket{+},\ket{-}\}$.
In the corresponding extended Wigner's friend scenario, for each Bell party, one of the measurements, for example the computational basis $\ket{0},\ket{1}$ measurement, is performed by an observer and modelled unitarily as $U$, whereas the other measurement is performed by the corresponding superobserver in an entangled basis $U |\pm \rangle_{S_{\!F}} |0\rangle_{L_{\!F}}$.} \label{fig:FR_Rui} 
\end{figure}

The key idea is better explained by focusing on a \emph{single} party.
That is an experimenter who, in each run of the experiment, chooses between two (incompatible) dichotomic measurements to perform on their part of a shared system.
Such a situation can be translated into a Wigner's friend scenario:
the friend first performs one of the measurements, and then the superobserver Wigner undoes this measurement and performs the other.
The choice of measurement setting is thus translated to a choice between adopting the point of view of Wigner or that of the friend.
Applying the same idea to each party, the translation extends to the $(n,2,2)$ Bell scenario for any number $n$ of parties.

As discussed in \Cref{sec:nonclassicality}, Bell nonlocality scenarios are a particular case of contextuality scenarios, where maximal measurement contexts correspond to a choice of one measurement for each party.
In the $(2,2,2)$ Bell scenario there are four dichotomic measurements, two for each party. 
We choose to label Alice's measurements as $A$ and $U$ and Bob's as $B$ and $W$, foreshadowing the link to the FR scenario. The maximal measurement contexts are then
\[\{A,B\}, \quad \{U,B\}, \quad \{A,W\}, \quad \{U,W\} .\]
Using the Born rule we calculate the (two-party) joint outcome probabilities, yielding a probability distribution over joint outcomes for each context. 
In fact, in this case, the \emph{possibilistic} version of the Born rule suffices, which only distinguishes between \emph{possible} (probability greater than zero) and \emph{impossible} (probability zero) events.
\Cref{table:FR_table} indicates such possibilistic empirical data for the specific choice of state and measurements in the Hardy model:
each row is labelled by a measurement context and indicates the support of the probability distribution on the joint outcomes for that context; see e.g.~\cite{abramsky2011sheaf,abramsky2015contextuality}.

\footnotetext{We regard $\ket{0}$, $\ket{+}$ as corresponding to the outcome label $0$ and $\ket{1}$, $\ket{-}$ to the label $1$. Note that these are, respectively, the $+1$ eigenvectors and the $-1$ eigenvectors of the $Z$ and $X$ matrices.
Often, when regarding the $Z$ or $X$ matrices as measurement operators, one considers outcomes valued $\pm1$. Here, we use the bijection $+1 \mapsto 0$, $-1 \mapsto 1$.
Particularly relevant for the GHZ--Mermin argument, this bijection actually yields a group isomorphism between $\{+1,-1\}$ under multiplication and $\{0,1\}$ under addition modulo $2$, an instance of the isomorphism between the multiplicative group of complex $n$-th roots of unity  and the additive group of integers modulo $n$.\label{fn:01pm1}}

\newcommand{\tikzxmark}{%
\tikz[scale=0.18] {
    \draw[line width=0.7,line cap=round] (0,0) to [bend left=6] (1,1);
    \draw[line width=0.7,line cap=round] (0.2,0.95) to [bend right=3] (0.8,0.05);
}}
\newcommand{\tikzcmark}{%
\tikz[scale=0.18] {
    \draw[line width=0.7,line cap=round] (0.25,0) to [bend left=10] (1,1);
    \draw[line width=0.8,line cap=round] (0,0.35) to [bend right=1] (0.23,0);
}}

\begin{table}
\centering 
{%
\begin{tabular}{l|cccc|}
\cline{2-5}

                                & \footnotesize{$0,0$}            & \footnotesize{$0,1$}             & \footnotesize{$1,0$}             & \footnotesize{$1,1$}            \\ \hline
\multicolumn{1}{|l|}{$A,B$} & $1$ & 0 & $1$            & $1$  \\ 
\multicolumn{1}{|l|}{$A,W$} & $1$ & $1$             & 0  & $1$  \\ 
\multicolumn{1}{|l|}{$U,B$} & 0 & $1$ & $1$  & $1$               \\ 
\multicolumn{1}{|l|}{$U,W$} & $1$ & $1$ & $1$ & $1$ \\ \hline
\end{tabular}%
}%
\caption[(Im)possibility table for the Hardy model]{Support of joint outcome probability distributions for each context in the Hardy model or the FR paradox.
The rows are labelled by the four measurement contexts $\{A,B\},\{A,W\},\{U,B\},\{U,W\}$ and columns by the joint outcomes $(0,0),(0,1),(1,0),(1,1)$,
the combination of which gives an event: a specific joint outcome for a specific context of measurements.
The entries in each cell of this possibility table have two possible values: $1$ denotes an event being \textit{possible} while $0$ denotes an event being \textit{impossible}, \ie having zero probability.
Note that in the FR case the outcomes $u$ and $w$ are labelled by $\textok,\textfail$ instead of $0,1$.
This  table is logically contextual:
there is no consistent global assignment of values to $A,B,U,W$ that extends the possible event $\{U \mapsto 0, W \mapsto 0\}$ (in the bottom left);
starting from this partial assignment, row 3 forces $B \mapsto 1$, which used in row 1 gives $A \mapsto 1$, which used in turn in row 2 yields
$W \mapsto 1$, contradicting the initial assignment.} 
~

\label{table:FR_table}
\end{table}

Contextuality, of which non-locality is a special instance, occurs when there is no global probability distribution (on assignments of outcomes to all the measurements in the scenario) that marginalises to the empirical probability distributions within each measurement context.
The Hardy model~\cite{hardy1993nonlocality} witnesses \emph{logical} non-locality such that non-locality can be ascertained from the possibilistic information collected in \Cref{table:FR_table} alone.
The argument corresponds precisely to the logical reasoning chain in the FR paradox.
Specifically, the joint outcome assignment $\{u \mapsto 0, w \mapsto 0\}$, which occurs with positive probability,
cannot be extended to a global assignment for all $a,b,u,w$ which does not restrict to an impossible event in some other context (see caption of \Cref{table:FR_table} for details).
We note that the troublesome global assignments in the contextuality setting correspond to statements about (absolute) outcomes in FR-like scenarios.

The mapping from a Bell scenario to an extended Wigner's friend scenario as in \cref{fig:FR_Rui}
provides a recipe for producing extended Wigner's friend paradoxes based on nonlocal models.
We will exploit this recipe in the next section to produce an FR-like paradox based on the GHZ--Mermin model~\cite{greenberger1989going,mermin1990simple}.

\section{The Local Friendliness no-go theorem}\label{sec:LF}

The Local Friendliness scenario is similar to the Frauchiger--Renner protocol, with the difference that the two superobservers in the Local Friendliness scenario are each given a choice $x,y$ whether to ask their friend for their measurement outcome or to perform their own measurement which involves undoing the measurements of the friends. 
In this way, the measurement statistics $\wp(a,b|x,y)$ are obtained by Alice and Bob.\footnote{Similarly to Refs.~\cite{bong2020strong,walleghem2024connecting,walleghem2025extendedwignersfriendnogo}, when working in the context of the Local Friendliness no-go theorem or related results, we introduce a notational difference between empirical probability distributions that can be obtained by a single agent from the collected data, denoted by $\wp$, and theoretical probability distributions required to exist given certain assumptions, but potentially not observable by any agent, denoted by $p$.}
Instead of a paradox from agents reasoning about each other as in the FR paradox, a contradiction is now obtained with the Local Friendliness assumptions of Absoluteness of Observed Events and Local Agency.

An unfortunate convention in literature is that regarding the names of the involved agents different conventions are used in the literature on the Local Friendliness no-go theorem compared to the Frauchiger--Renner paradox. 
Specifically, the superobservers of the FR paradox are now called Alice and Bob in the LF scenario, as opposed to Ursula and Wigner, and the friends are called Charlie and Debbie. 
Whenever we will be concerned with the Local Friendliness results, we will thus use these conventions to be in line with literature.

We now outline the LF scenario and no-go theorem in some more detail, following Ref.~\cite{walleghem2025extendedwignersfriendnogo} closely.

\subsection{Local Friendliness scenarios} \label{sec:recap_LF_scenario}
The experimental setup, illustrated in \cref{fig:LF_cartoon}, is described as follows. We will present a theory-independent way of specifying the scenario, which we refer to as the $(N_A,N_B)$-setting LF scenario, while also outlining how the scenario for a quantum violation goes (taking $N_A=2=N_B$ for this).

Charlie and Debbie, spacelike separated, share a bipartite system in the state $\rho_{RS}$. 
They perform measurements denoted with $C,D$ on their respective subsystems, $R,S$, obtaining outcomes $c,d\in\{0,1\}$. 
In quantum theory, the measurements are modelled by unitaries $U_C,U_D$, respectively.
After Charlie has performed his measurement, Alice makes a choice $x \in \{1,\ldots,N_A\}$ with $N_A \geq 2$.  
If $x=1$ she asks Charlie for his outcome and sets her outcome equal to his, revealing his outcome to the world.
If $x\neq 1$ she applies another operation and measurement to Charlie's lab, choosing out of $N_A - 1$ measurements labelled by $x \in \{2,\ldots,N_A\}$. 
In particular, for the quantum scenario that will provide a violation of the LF assumptions, we can take $N_A = 2$ and for $x=2$ Alice acts as a superobserver and undoes Charlie's measurement by applying $U_C^\dagger$, thereby erasing the outcome record of $c$; subsequently she performs a measurement denoted with $A$ on $R$, obtaining outcome $a\in\{0,1\}$.
Similarly, Bob chooses $y\in\{1,\ldots, N_B\}$. For $y=1$, he takes Debbie's outcome to be his outcome. For $y\neq 1$ he applies another operation and measurement to Debbie's lab, choosing out of $N_B-1$ measurements. 
For the quantum scenario violating LF, we can take $N_B=2$ and for $y=2$ Bob undoes Debbie's measurement and subsequently performs a measurement denoted with $B$ on $S$, obtaining outcome $b\in\{0,1\}$.

\begin{figure}[h]
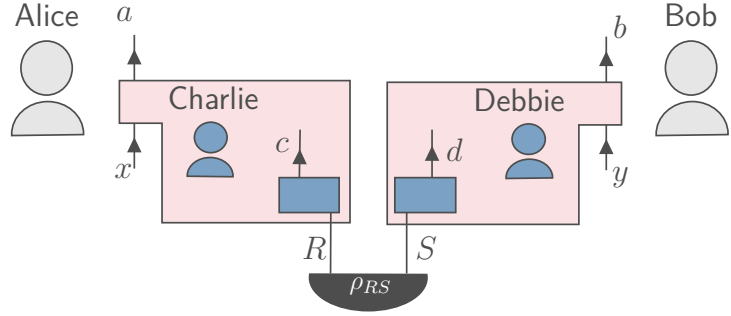

         \centering \includestandalone[width=0.65\textwidth]{LFsetup}
         \caption[Local Friendliness setup]{Local Friendliness setup~\cite{schmid2023review}. 
         }
         \label{fig:LF_cartoon}
\end{figure}

\subsection{No-go theorem}

Given the setup, we state the required assumptions of Absoluteness of Observed Events and Local Agency, whose conjunction is also referred to as `Local Friendliness' (LF).\footnote{Regarding terminology, the no-go theorem against the Local Friendliness assumptions as in \cref{th:LF} is typically referred to as the Local Friendliness no-go theorem, and scenarios used to prove it as Local Friendliness scenarios.}

\begin{assumption}[Absoluteness of Observed Events (AOE)] \label{assumption:AOE_prelims}
An observed event is an absolute single event, not relative to anything or anyone. 
\end{assumption}

This assumption can be motivated as an instance of realism and, in the LF scenario, it implies that there exist distributions $p(a,b,c,d|x,y)$ from which one can obtain the empirical probabilities, $\wp(a,b|x=2,y=2),\wp(a,d|x=2,y=1),\wp(c,b|x=1,y=2),\wp(c,d|x=1,y=1)$ via marginalisation.\footnote{To argue that Alice's result of asking Charlie for his outcome ís Charlie's outcome, an additional ``tracking assumption''~\cite{schmid2023review} is introduced, i.e. $p(a \mid c, x=1)=\delta_{a, c}$, in that `Alice's outcome tracks Charlie's exactly in the relevant runs of the protocol', and similarly for Bob and Debbie. For a discussion on why this is considered a minimal assumption, see~\cite{schmid2023review}.}

\begin{assumption}[Local Agency] \label{ass:LA_prelims}
The only relevant events correlated with an intervention are in its future light cone.
\end{assumption}

In the LF scenario, Local Agency implies, for example, that $p(c,b|x,y=2) = p(c,b|y=2)$, as none of $c,b$ are obtained in the future of the choice $x$.

We can now state the LF no-go theorem~\cite{bong2020strong,haddara2022possibilistic,cavalcanti2021implications}.
\begin{theorem}[LF no-go theorem] \label{th:LF}
If a superobserver can perform arbitrary quantum operations on an observer and its environment, then no physical theory can satisfy Local Friendliness.
\end{theorem} 
\begin{proof} Let us consider the LF scenario of figure \ref{fig:LF_cartoon} with the following specifications. Charlie and Debbie share the bipartite entangled state $(|01\rangle-|10\rangle)/\sqrt{2}$, and perform an $(X+Z)/\sqrt{2}$ and $Z$ measurement, respectively. When $x=2$, Alice undoes Charlie's measurement and performs an $(Z-X)/\sqrt{2}$ measurement; when $y=2$, Bob undoes Debbie's measurement and performs an $X$ measurement (any set of four measurements and an entangled state that produce correlations violating a Bell inequality would also work~\cite{walleghem2024connecting,haddara2024local}).  
There are four empirical distributions: 
\begin{equation} \label{eq:empirical_prob_LF_prelims}
    \begin{split}
        &\wp(c,d \mid x=1, y=1), \\
&\wp(c,b\mid x=1, y=2), \\
&\wp(a,d \mid x=2, y=1), \\
&\wp(a,b\mid x=2, y=2). 
    \end{split}
\end{equation}
By AOE and Local Agency, we can rewrite these
empirical correlations as follows, 
     \begin{align}
      \label{eq:empirical_to_global_corr_1}
        &\wp(c,d \mid x=1, y=1) = p(c,d|x=2,y=2), \\  \label{eq:empirical_to_global_corr_2}
&\wp(c,b\mid x=1, y=2)  = p(c,b|x=2,y=2), \\  \label{eq:empirical_to_global_corr_3}
&\wp(a,d \mid x=2, y=1)  = p(a,d|x=2,y=2), \\  \label{eq:empirical_to_global_corr_4}
&\wp(a,b\mid x=2, y=2) = p(a,b|x=2,y=2). 
    \end{align}
AOE is used to consider joint probability distributions when $c$ or $d$ are involved and possibly erased. To obtain Eq.~\eqref{eq:empirical_to_global_corr_2} we first appeal to an instance of Local Agency that enforces no-superdeterminism, $p(c|xy) = p(c)$, yielding $p(c|x=1,y=2)=p(c|x=2,y=2)$; we can then apply the rule of conditional probability and consider another instance of Local Agency expressing that neither $b$ nor $c$ occur in the future light cone of $x$: $p(b|c,x,y){=}p(b|c,y)$, 
thus obtaining \[\begin{split}&\wp(c,b|x{=}1,y{=}2) = p(c|x{=}1,y{=}2)p(b|c,x{=}1,y{=}2) \\&=p(c|x{=}2,y{=}2)p(b|c,x{=}2,y{=}2)=p(c,b|x{=}2,y{=}2).\end{split}\]

Similar instances of Local Agency lead to Eq.~\eqref{eq:empirical_to_global_corr_3}.  
Eq.~\eqref{eq:empirical_to_global_corr_1} derives from an instance of Local Agency enforcing no-superdeterminism, $p(c,d|x,y)=p(c,d)$.

By assuming AOE, we can assert that all the correlations in Eqs.~\eqref{eq:empirical_to_global_corr_1}, \eqref{eq:empirical_to_global_corr_2} ,\eqref{eq:empirical_to_global_corr_3} and \eqref{eq:empirical_to_global_corr_4} can be obtained
from a single joint probability distribution $p(a, b, c, d|x =
2, y = 2)$ via marginalization. 
However, given the choices of preparations and measurements made, by Fine's theorem~\cite{fine1982hidden,abramsky2011sheaf}, such distribution cannot exist.
Indeed, these give exactly the empirical correlations that lead to a maximal violation of the Bell inequalities. 
\end{proof}

In terms of rigor, the LF no-go theorem has the same status as Bell’s theorem, since it is precise, theory-independent, and -- at least in principle -- empirically testable~\cite{wiseman2023thoughtful}.

\ifSubfilesClassLoaded{ 
  \printbibliography[heading=bibintoc, title=References]
  }{}
\end{document}

\ifSubfilesClassLoaded{ 
	\setcounter{chapter}{1} 
	\maketitle \tableofcontents \listoftodos
	}{}

\chapter[A refined Frauchiger--Renner paradox based on strong contextuality]{A refined Frauchiger--Renner paradox based on strong contextuality}
\chaptermark{A refined FR paradox based on strong contextuality}
\label{chap:partI_refinedFR}

in this chapter we present a refined version of the Frauchiger--Renner paradox, where the consistency assumption is weakened significantly. 
Namely, a contradiction is derived when agents who meet and communicate declare their Born rule predictions to each other. 
In contrast, in the FR paradox, an agent needs to trust the reasoning and predictions of an agent in a  sealed lab with whom they cannot meet and communicate about their predictions. 
This is particularly interesting as already some classical paradoxes such as the Sleeping Beauty problem violate FR's consistency condition, but do not violate our weakened consistency requirement. 
We discuss the Sleeping Beauty problem and a particularly simple resolution as an intermezzo at the end of this Part, in \Cref{chap:part1_sleepingbeauty}. 
Our result implies that the route taken to tackle the FR paradox by rejecting its consistency requirement needs a more radical resolution, as it needs to reject even our weakened consistency condition. 
We will shortly discuss potential ways out of the (refined) FR paradox, and argue how, when taking the route of rejecting the Absoluteness of Observed Events, our paradox suggests a relational principle which we name the `Relativity of Observed Events'. 

A key objective of this chapter is to rigorously define the underlying principles and assumptions. To aid clarity, assumptions and their references throughout this chapter are highlighted using color and hyperlinks.
This chapter follows Ref.~\cite{walleghem2024strong}, nearly verbatim.

\vspace{0.5cm}

\section{Refining the Frauchiger--Renner paradox}\label{sec:FR_refining}

In this section we present the protocol of our refined version of the Frauchiger--Renner (FR) paradox, based on the strongly nonlocal GHZ--Mermin model. 
In the previous chapter we have seen how the Bell nonlocality scenario with two parties with each two measurements underlies the FR paradox (and the Local Friendliness no-go theorem). 
We present an FR-like paradox but based on a model with three parties that has a stronger form of nonlocality. 
As a consequence, a contradiction is derived in each round of the protocol, doing away with the post-selection step of the FR paradox. 
Moreover, as already hinted at in \Cref{sec:stricter_FR}, the reasoning steps are formulated differently from the original FR scenario, leading to a weaker consistency assumption to derive a stronger no-go theorem.

\subsection{Strong contextuality: the GHZ--Mermin argument} \label{sec:GHZ_recap}
\paragraph*{Strong contextuality.}
As recapped in \Cref{sec:recap_KSNC}, the sheaf-theoretic approach to non-locality and Kochen--Specker (KS) contextuality~\cite{abramsky2011sheaf,abramsky2015contextuality,barbosa2015contextuality} introduced a hierarchy of contextuality of increasing strength: probabilistic, logical (or possibilistic), and strong contextuality.
In quantum theory, each of these levels of the hierarchy is exemplified by a well-known model:
\[
    \text{Bell--CHSH (probabilistic)} \;\prec\; \text{Hardy (logical)} \;\prec\; \text{GHZ--Mermin (strong)} .
\]

The contextual model underlying the FR paradox is the logically contextual Hardy model: contextuality is witnessed at the level of the \emph{support} of probability distributions. Namely, there exists a \emph{possible} assignment of outcomes to a context which cannot be extended to a global assignment (to all measurements) consistent with the model, \ie whose restriction to each context is deemed possible.

Nevertheless, in the Hardy model, there exists a local assignment that can be extended globally; \eg the global assignment $\{a \mapsto 1, b \mapsto 1, u \mapsto 1, w \mapsto 1\}$ is consistent with the observed possibilities in \Cref{table:FR_table}.
\emph{Strong contextuality} holds when not a single such global assignment exists.\footnotemark

\footnotetext{This property coincides with \emph{maximal contextuality} in the sense that the fraction of the model that can be explained by noncontextual hidden variables is zero, or that the algebraic (no-signalling) maximum of a noncontextuality inequality violation is attained~\cite{abramsky2017contextual}.}

\paragraph*{The GHZ--Mermin model.}
A paradigmatic example of quantum-realizable strong contextuality is given by the GHZ--Mermin model~\cite{greenberger1989going,mermin1990quantum}, a specific realisation of the Bell scenario with three parties, each having two measurements with each two measurement outcomes. 
Following the presentation in Ref.~\cite{abramsky2015contextuality}, we consider three parties, $A$, $B$, and $C$, that share an entangled three-qubit system prepared in the GHZ state,
\[
\ket{ \psi^{\textsf{GHZ}} }_{ABC}  = \frac{1}{\sqrt{2}} \left(\ket{000}+\ket{111}\right)_{ABC}.
\] 
Each party performs either an $X$ or a $Y$ measurement on their qubit,
\ie a measurement in either the $\{\ket{+} , \ket{-}\}$ or the $\{\ket{+i}, \ket{-i}\}$ basis,
where $\ket{\pm } = (\ket{0} \pm \ket{1})/\sqrt{2}$ and $\ket{\pm i} = (\ket{0} \pm i\ket{1})/\sqrt{2}$. This leads to measurements $X_A,Y_A,X_B,Y_B,X_C,Y_C$ with outcomes $x_A, y_A, x_B, y_B, x_C, y_C \in \{0,1\}$.
The outcomes corresponding to $\ket{+}$ and $\ket{+i}$ are labeled $0$, those corresponding to  $\ket{-}$ and $\ket{-i}$ are labeled $1$.

A measurement context in this scenario consists of a choice of measurement for each party. We abbreviate the context $\{X_A, X_B, X_C\}$ as $XXX$ and analogously for the others.
Using the Born rule to distinguish possible from impossible events for measurement contexts $XXX$, $XYY$, $YXY$ and $YYX$, one obtains Table \ref{table:GHZ}.
Each row is labeled by a measurement context and indicates the support of a probability distribution on the respective joint outcomes; see e.g.~\cite{abramsky2011sheaf,abramsky2015contextuality}.
The outcomes satisfy the parity equations
\begin{equation} \label{eq:GHZ_Mermin}
x_A \oplus x_B \oplus x_C=0, \quad
x_A \oplus y_B \oplus y_C=1, \quad
y_A \oplus x_B \oplus y_C=1, \quad
y_A \oplus y_B \oplus x_C=1, 
\end{equation}
where $\oplus$ denotes addition modulo $2$.\footnote{This argument is also often presented with multiplication of $\pm1$-labeled outcomes; see \cref{fn:01pm1}.}
Assigning definite outcome values to all six measurements simultaneously, summing these four equations yields $0=1$, a contradiction. 
Therefore, no global assignment of outcomes to all six measurements is compatible with the possible observations for all measurement contexts -- the GHZ--Mermin model is strongly contextual, and as it is a nonlocality scenario, strongly nonlocal.

\begin{table}
\centering
{%
\begin{tabular}{l|cccccccc|}
\cline{2-9}
                                & \footnotesize{$0,0,0$}            & \footnotesize{$0,0,1$}  & \footnotesize{$0,1,0$} & \footnotesize{$0,1,1$} & \footnotesize{$1,0,0$} & \footnotesize{$1,0,1$} & \footnotesize{$1,1,0$} & \footnotesize{$1,1,1$}            \\ \hline
\multicolumn{1}{|l|}{$XXX$} & $1$ & $0$ & $0$ & $1$ & $0$ & $1$ & $1$ & $0$  \\ 
\multicolumn{1}{|l|}{$XYY$} & $0$ & $1$ & $1$ & $0$ & $1$ & $0$ & $0$ & $1$  \\ 
\multicolumn{1}{|l|}{$YXY$} & $0$ & $1$ & $1$ & $0$ & $1$ & $0$ & $0$ & $1$  \\ 
\multicolumn{1}{|l|}{$YYX$} & $0$ & $1$ & $1$ & $0$ & $1$ & $0$ & $0$ & $1$ \\ \hline
\end{tabular}%
}%
\caption[(Im)possibility table for the GHZ--Mermin scenario]{Table  with outcome (im)possibilities for measurements by three parties who share a GHZ state. The rows correspond to the support of probability distributions over the outcomes (columns 1--8) that arise when the measurements specified in the first column are performed ($XXX,XYY,YXY,YYX$). 
}
\label{table:GHZ}
\end{table}

\subsection{Description of the GHZ--FR paradox} \label{sec:GHZ_protocol}

Mapping Bell nonlocality scenarios to extended Wigner's friend scenarios, as described in \Cref{sec:FR_non-locality}, we build an FR-like paradox based on the GHZ--Mermin model.
The strong contextuality of this nonlocal model does away with the need for post-selection in the GHZ--FR paradox.

The scenario is depicted in \cref{fig:GHZ_final}(a). It involves three observer-superobserver pairs -- Alice and Ursula, Bob and Valentina, and Charlie and Wigner -- corresponding to the three parties in the GHZ--Mermin model.
The friends each reside in a sealed lab with access to quantum systems $S_{\!A}$, $S_{\!B}$, and $S_{\!C}$ respectively, which are initialized in a shared entangled state.
As before, $L_{\!A}$ denotes Alice's lab including Alice herself but excluding the system $S_{\!A}$, and similarly $L_{\!B}$ and $L_{\!C}$ for Bob and Charlie.

Alice, Bob, Charlie and their respective superobservers Ursula, Valentina, and Wigner perform measurements with outcomes $a,b,c,u,v,w \in \{0,1\}$. 
A circuit representation of the measurement protocol is shown in \cref{fig:GHZ_final}(b).
First, the observers $A, B, C$ perform $Y$-measurements on their part of a GHZ state. 
Then, Ursula performs an $X$-like measurement on Alice and her qubit,
and writes down her predictions for the outcomes of $B, C$.
One intuition for why it might be natural for Ursula to formulate predictions about $b,c$ is in case she does not know about Valentina's and Wigner's actions (or existence even); in that case she believes to still be able to gather Bob's outcome record, for example by asking him for his outcome.
Valentina and Wigner act analogously.
Once the protocol has been carried out, Ursula, Valentina and Wigner meet up and share their predictions, deriving a contradiction from them, as we will explain below.

\begin{figure}
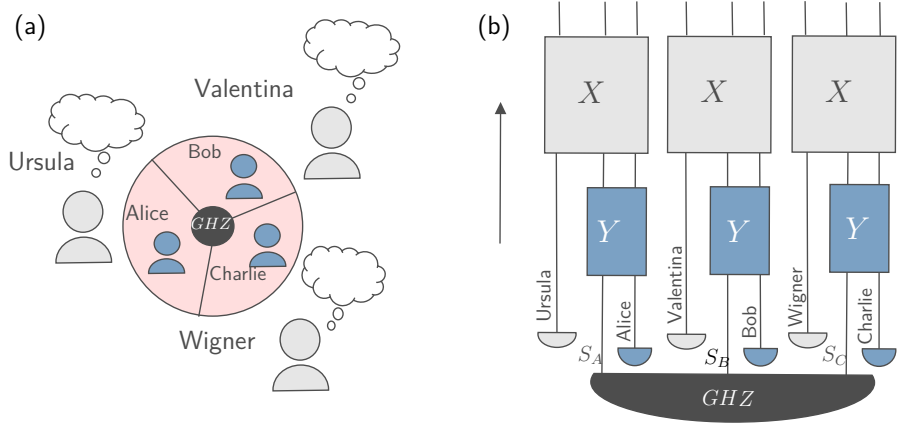
 \centering 
\includestandalone[width=0.8\textwidth]{Figures/FINAL_GHZ_WF}  \caption[The GHZ--FR protocol]{(a): Pictorial representation of the scenario GHZ--FR protocol, with superobservers $U,V,W$ and respective observers $A,B,C$, each having access to part of a GHZ state. (b): The GHZ--FR protocol in circuit form. Alice, Bob and Charlie each have access to one part of a three-qubit GHZ state and measure their qubit in the $Y$-basis. Ursula, Valentina and Wigner (red), superobservers for Alice, Bob and Charlie, respectively, each perform an $X$-like measurement on Alice, Bob and Charlie and their qubits.} \label{fig:GHZ_final}
\end{figure}

We first describe the measurement protocol in more detail, after which we present the reasoning steps through which they derive a contradiction.

\paragraph*{Measurement protocol.}
The measurement protocol, sketched in circuit form in \cref{fig:GHZ_final}, proceeds as follows:
\begin{enumerate}[label=\arabic*.,leftmargin=*]
\item \textbf{Initialisation.} The system $S_{\!A} S_{\!B}  S_{\!C}$, shared between Alice, Bob, and Charlie, starts in the GHZ state $|\psi^{\textsf{GHZ}}\rangle_{S_{\!A} S_{\!B}  S_{\!C}}=\frac{1}{\sqrt{2}} \left(\ket{000}+\ket{111}\right)_{S_{\!A} S_{\!B} S_{\!C}}$. The initial states of the three labs are labeled by $\ket{r}_{L_{\!A}}$, $\ket{r}_{L_{\!B}}$, $\ket{r}_{L_{\!C}}$ (with $r$ from `ready').
The initial state of $S_{\!A} L_{\!A} S_{\!B} L_{\!B} S_{\!C} L_{\!C}$ is thus given by
    \begin{align*}
    \begin{split}
    \ket{\psi^{t=1}}_{S_{\!A} L_{\!A} S_{\!B} L_{\!B} S_{\!C} L_{\!C}} = 
    \frac{1}{\sqrt{2}}\left(\ket{0r0r0r}+\ket{1r1r1r}\right)_{S_{\!A} L_{\!A} S_{\!B} L_{\!B} S_{\!C} L_{\!C}}.
    \end{split}
    \end{align*}

\item \textbf{Alice, Bob, and Charlie measure.} Alice, Bob and Charlie perform a measurement in the $Y$-basis $\ket{\pm i}$ on their respective qubit $S_{\!A},S_{\!B},S_{\!C}$, with outcomes $0,1$. Superobserver Ursula describes Alice's measurement by the unitary evolution $U_A$ that,  similar to \cref{eq:unitary_friend} must satisfy: \begin{equation} \label{eq:GHZ_best_unitary}
    \begin{split}
        \ket{+i }_{S_{\!A}} \ket{r}_{L_{\!A}} \overset{U_A}{\longmapsto} \ket{+i}_{S_{\!A}} \ket{0}_{L_{\!A}}, \qquad
        \ket{-i }_{S_{\!A}} \ket{r}_{L_{\!A}} \overset{U_A}{\longmapsto} \ket{-i }_{S_{\!A}} \ket{1}_{L_{\!A}}
    \end{split}
\end{equation} and similarly for Bob and Charlie where the states $\ket{0}_{L_{\!A}},\ket{1}_{L_{\!B}}$ denote the states of Alice's lab in which she obtains outcomes 0 and 1, respectively. This leads to the overall state
\begin{equation}
    \ket{\psi^{t=2}}_{S_{\!A} L_{\!A} S_{\!B} L_{\!B} S_{\!C} L_{\!C}}
    \;=\; \left(U_{A} \otimes U_{B} \otimes U_{C} \right)\ket{\psi^{t=1}}_{S_{\!A} L_{\!A} S_{\!B} L_{\!B}S_{\!C} L_{\!C}} .
\end{equation}

\item \textbf{Ursula, Valentina, and Wigner supermeasure.}
Finally, the superobservers Ursula, Valentina, and Wigner respectively measure the systems 
$S_{\!A} L_{\!A}$, $S_{\!B} L_{\!B}$, and $S_{\!C} L_{\!C}$ in a basis with vectors\footnote{The two other basis vectors need not be specified since $\ket{+i, 1}_{S_{\!A} L_{\!A}}$  and $\ket{-i, 0}_{S_{\!A} L_{\!A}}$ do not occur.}
    \begin{equation}
     \ket{\textyes}_{S_{\!A} L_{\!A}} =  U_A \ket{+}_{S_{\!A}} \ket{0}_{L_{\!A}}, \qquad \ket{\textno}_{S_{\!A} L_{\!A}} = U_A \ket{-}_{S_{\!A}} \ket{0}_{L_{\!A}} .   
    \end{equation}
 We label the outcomes of $\ket{\textyes}, \ket{\textno}$ by $0,1$.

\end{enumerate}

The probabilities for Ursula's measurement in the $| \textyes \rangle, | \textno\rangle$-basis are equal to those which one would obtain by undoing Alice's measurement on $S_{\!A}$, followed by an $X$-measurement on the qubit $S_{\!A}$, and similarly for Valentina's and Wigner's measurements. The probabilities for $| \textyes \rangle, | \textno \rangle$ measurements by superobservers thus correspond to probabilities for $X$-measurements on the original GHZ state.
For example, the unitary description of Alice's measurement followed by Ursula's measurement effect yields the probabilities
\begin{equation}
     \abs*{\bra{\pm}_{S_{\!A}} \bra{0}_{L_{\!A}} U_A^\dagger U_A \ket{\ldots}_{S_{\!A}}\ket{0}_{L_{\!A}}}^2 = \abs*{\bra{\pm}_{S_{\!A}}  \bra{0}_{L_{\!A}} \ket{\ldots}_{S_{\!A}}\ket{0}_{L_{\!A}} }^2 = \abs*{\bra{\pm}_{S_{\!A}} \ket{\ldots}_{S_{\!A}}}^2 .
\end{equation}

Possible lightcone structures of this scenario are sketched in \cref{fig:FR_GHZ_lightcone}.
Our results are independent of the exact causal structure, but we consider the natural causal structure arising from relativistic spacetime.
More generally, we could embed the GHZ--FR scenario into a process-theoretic framework like Categorical Quantum Mechanics~\cite{abramsky2009categorical,coecke_kissinger_2017} or the process operator formalism~\cite{oreshkov2012quantum}.

\begin{figure}[H]
\centering 
\includegraphics[width=0.6\textwidth]{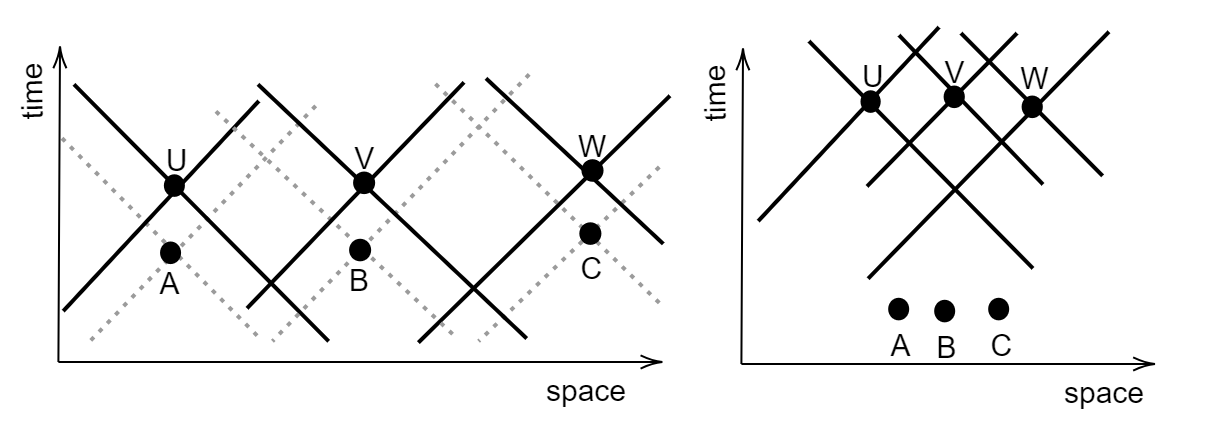}
\caption[Possible lightcone structures for the GHZ--FR protocol]{Possible lightcone structures of the agents in the GHZ--FR scenario, with Ursula, Valentina, and Wigner being superobservers for, respectively, Alice, Bob, and Charlie, who share a GHZ state. The GHZ--FR paradox occurs in both lightcone structures. A mixture of these lightcone structures is also possible. } \label{fig:FR_GHZ_lightcone} 
\end{figure}

\paragraph*{Reasoning steps.}
Using the possibilistic Born rule, Ursula reasons about her outcome together with Bob's and Charlie's outcomes, and concludes: "\textit{Having obtained $u$, I know that $b\oplus c=1 \oplus u$}", since jointly considering the outcomes of Ursula, Bob and Charlie corresponds to the context $XYY$ in the GHZ--Mermin model.
For example, if Ursula obtained $u=0$ then she finds that $b=c=0$ is excluded as: \begin{equation} \begin{split}
    p(u=0,b=0,c=0) &= \operatorname{Tr}\ket{\Psi} \bra{ \Psi } = 0,  
\end{split}
\end{equation} where \begin{equation}
    \ket{\Psi } = \bra{+}_{S_{\!A}} \bra{0}_{L_{\!A}} U_A^\dagger U_A \bra{ +i }_{S_{\!B}} \bra{  +i }_{S_{\!C}}  \ket{\psi^{GHZ}}_{S_{\!A} S_{\!B} S_{\!C}} \ket{000}_{L_{\!A}L_{\!B}L_{\!C}}.  
\end{equation}
Valentina and Wigner make similar reasoning statements, leading to \begin{equation} \label{eq:abc_predictions_GHZ_FR1}
        \begin{split}
            b \oplus c = 1 \oplus u, \quad
            a \oplus c = 1 \oplus v, \quad
            a \oplus b = 1 \oplus w.
        \end{split}
\end{equation} with $\oplus$ meaning summation modulo 2.
Applying the possibilistic Born rule to the  superobservers' measurements with classically recorded outcomes, one obtains \begin{equation} \label{eq:uvw_GHZ}
    u \oplus v \oplus w = 0.
\end{equation} 
For instance, we have \begin{equation} \begin{split}
    p(u=0,v=0,w=1) &= \operatorname{Tr}\ket{\Psi_{u=0,v=0,w=1}} \bra{ \Psi_{u=0,v=0,w=1} } = 0,  
\end{split}
\end{equation} where \begin{equation}
\begin{split}
    \ket{\Psi_{u=0,v=0,w=1} } = \bra{+}_{S_{\!A}}  \bra{0}_{L_{\!A}}  \bra{+}_{S_{\!B}}  \bra{0}_{L_{\!B}} \bra{-}_{S_{\!C}}  \bra{0}_{L_{\!C}} & U_A^\dagger U_A U_B^\dagger U_B U_C^\dagger U_C \\ & \ket{\psi^{GHZ}}_{S_{\!A} S_{\!B} S_{\!C}} \ket{000}_{L_{\!A}L_{\!B}L_{\!C}}.   
\end{split}
\end{equation} Namely, this is the state obtained when Alice, Bob and Charlie perform their $Y$-measurements on their systems $S_A, S_B, S_C$, modeled unitarily by Ursula, Valentina and Wigner, who next undo these measurements and perform $X$-measurements with outcomes $0,0,1$, respectively. The proofs for other combinations of values for $u,v,w$ with $u \oplus v \oplus w$ odd are entirely analogous.
As Ursula, Valentina and Wigner come together to share their predictions, they find \cref{eq:abc_predictions_GHZ_FR1,eq:uvw_GHZ}. 
These equations correspond to \cref{eq:GHZ_Mermin} from the GHZ--Mermin scenario, with $x_A,x_B,x_C,y_A,y_B,y_C$ substituted for $u,v,w,a,b,c$. Namely, summing all of the equations of \cref{eq:abc_predictions_GHZ_FR1,eq:uvw_GHZ} modulo 2 one finds $0=1$. Thus they find a contradiction.\footnote{Note that post-selected on $u,v,w$, the GHZ--FR argument gives a Specker's triangle~\cite{liang2010specker}. For instance, 
    for $u=v=w=0$ we find  \begin{equation}
        b \oplus c = 1, \quad a \oplus c = 1, \quad a \oplus b = 1.
    \end{equation}
}

\section{Two GHZ--FR no-go theorems} \label{sec:GHZ_FR_nogo}
Having described the GHZ--FR paradox, we spell out the required assumptions in detail that lead to a contradiction.
We construct two no-go theorems of increasing strength, which we name the GHZ--FR \textit{truth} and \textit{agreement} no-go theorems. 
The truth no-go theorem is easier to grasp as its assumptions are more natural.
Moreover, it connects to a wider body of literature as it involves the assumption of Absoluteness of Observed Events, stating that performed measurements have absolute, single-valued outcomes, which is often invoked in extended Wigner's friend paradoxes~\cite{bong2020strong,cavalcanti2021implications,haddara2022possibilistic,leegwater2022greenberger,schmid2023review,brukner2018no,ormrod2022no,ormrod2023theories,nurgalieva2018inadequacy,vilasini2019multi,montanhano2023contextuality,schmid2023review,leegwater2022greenberger,zukowski2021physics,walleghem2023extended,walleghem2024connecting,montanhano2023contextuality,gao2019quantum,guerin2020no}.
The agreement no-go theorem replaces Absoluteness of Observed Events by a weaker notion of agreement among classical agents. 
The fact that the paradox remains points towards a new principle to resolve extended Wigner's friend paradoxes. 
For each assumption, we state the general principle and specify its application to the GHZ--FR paradox.

\subsection{GHZ--FR truth no-go theorem} \label{sec:GHZ_nogo}

\tocless\subsubsubsection{Assumptions} \label{sec:subsubsec_truth_nogo}
The name `the GHZ--FR \textit{truth} no-go theorem' refers to the assumption of an underlying `absolute truth' about the observations made in each \emph{single run} of the protocol, formalised as {\AOEfull} (\cref{assumption:AOE}) below
\cite{haddara2022possibilistic,cavalcanti2021implications,bong2020strong}.

We begin by presenting a list of assumptions that capture how we usually think about outcomes of performed measurements in experiments, namely the conjunction of {\possBorn}, {\AOEfull}, and {\BornCAOE}.
Once we have established this common ground, we add the assumption that superobservers exist and that they can be described as in \Cref{sec:Wignerfriend}, \cref{eq:unitary_friend}, formalised as the assumption of {\universality}.

In quantum theory, the Born rule is used to obtain probability distributions for measurement outcomes.
In situations that involve multiple observers and superobservers,
one may expect the Born rule not to be applicable to all combinations of (actually performed) measurements.
For example, it would be too much to expect it to apply to the measurements of Wigner and his friend, as Wigner may even undo the friend's measurement.
In the present context however, only a minimal use of the Born rule is needed.
We only require the Born rule to hold in situations in which agents believe they could still test their predictions.
For example, in the GHZ--FR protocol Ursula makes predictions for Bob and Charlie, believing that
she could check their outcomes, \ie that she could still bring all relevant outcomes together.
Indeed, for all she knows, Valentina and Wigner might not have performed their measurements yet; in fact, Ursula need not know even about their existence and thinks she can still obtain the outcomes of Bob and Charlie, making it natural for her to formulate a prediction about $b,c$.
By contrast, one \textit{cannot} apply the Born rule to the joint outcomes of measurements performed by a superobserver-observer pair, for example to Valentina's and Bob's measurements, in a single use of the rule as nobody could ever reasonably believe to obtain such outcomes jointly; see also \cref{fig:example_restricted_Born rule}.
The Born rule may be valid more broadly, but we only require it be applicable in these instances.

Moreover, we only need \emph{possibilistic} use of the Born rule.


\begin{figure}
    \centering
    \tikzset{every picture/.style={line width=0.75pt}} 

\begin{tikzpicture}[x=0.75pt,y=0.75pt,yscale=-1,xscale=1]

\draw  [color={rgb, 255:red, 77; green, 77; blue, 77 }  ,draw opacity=1 ][line width=0.75]  (178.58,108.52) .. controls (179.9,101.67) and (175.56,94.89) .. (167.41,91.05) .. controls (159.26,87.21) and (148.72,87) .. (140.26,90.49) .. controls (137.27,86.51) and (131.79,83.76) .. (125.48,83.07) .. controls (119.17,82.39) and (112.77,83.85) .. (108.22,87.01) .. controls (105.67,83.4) and (100.67,80.97) .. (94.98,80.59) .. controls (89.29,80.21) and (83.72,81.93) .. (80.26,85.14) .. controls (75.65,81.31) and (68.32,79.7) .. (61.43,81) .. controls (54.55,82.3) and (49.35,86.29) .. (48.09,91.23) .. controls (42.44,92.32) and (37.74,95.08) .. (35.19,98.81) .. controls (32.65,102.54) and (32.51,106.86) .. (34.82,110.67) .. controls (29.25,115.78) and (27.96,122.59) .. (31.4,128.56) .. controls (34.84,134.52) and (42.51,138.75) .. (51.55,139.67) .. controls (51.62,145.27) and (55.96,150.41) .. (62.92,153.1) .. controls (69.88,155.8) and (78.36,155.63) .. (85.1,152.67) .. controls (87.97,159.37) and (96.04,164.31) .. (105.83,165.34) .. controls (115.62,166.37) and (125.38,163.31) .. (130.88,157.49) .. controls (137.62,160.36) and (145.71,161.19) .. (153.33,159.78) .. controls (160.95,158.38) and (167.44,154.86) .. (171.36,150.03) .. controls (178.25,150.59) and (184.91,148.07) .. (188.04,143.71) .. controls (191.16,139.35) and (190.09,134.08) .. (185.35,130.51) .. controls (191.5,127.95) and (194.63,122.89) .. (193.12,117.94) .. controls (191.62,113) and (185.8,109.31) .. (178.72,108.79) ; \draw  [color={rgb, 255:red, 77; green, 77; blue, 77 }  ,draw opacity=1 ][line width=0.75]  (185.35,130.51) .. controls (182.45,131.72) and (179.1,132.26) .. (175.74,132.08)(171.35,150.03) .. controls (169.91,149.91) and (168.5,149.65) .. (167.15,149.27)(130.88,157.49) .. controls (131.9,156.41) and (132.74,155.26) .. (133.41,154.06)(85.1,152.67) .. controls (84.58,151.44) and (84.24,150.18) .. (84.09,148.91)(51.55,139.67) .. controls (51.49,133.7) and (56.28,128.24) .. (63.88,125.63)(34.82,110.67) .. controls (36.05,112.7) and (37.93,114.5) .. (40.3,115.93)(48.09,91.23) .. controls (47.88,92.05) and (47.78,92.88) .. (47.8,93.72)(80.26,85.14) .. controls (81.41,86.1) and (82.35,87.16) .. (83.07,88.31)(108.22,87.01) .. controls (108.84,87.88) and (109.29,88.8) .. (109.59,89.75)(140.26,90.49) .. controls (138.48,91.23) and (136.82,92.12) .. (135.34,93.15)(178.58,108.52) .. controls (178.39,109.47) and (178.11,110.4) .. (177.72,111.31) ;
\draw  [color={rgb, 255:red, 77; green, 77; blue, 77 }  ,draw opacity=1 ][fill={rgb, 255:red, 230; green, 230; blue, 230 }  ,fill opacity=1 ][line width=0.75]  (111.04,213.23) .. controls (111.04,204.94) and (117.55,198.22) .. (125.57,198.22) .. controls (133.59,198.22) and (140.09,204.94) .. (140.09,213.23) .. controls (140.09,221.53) and (133.59,228.25) .. (125.57,228.25) .. controls (117.55,228.25) and (111.04,221.53) .. (111.04,213.23) -- cycle ;
\draw  [color={rgb, 255:red, 77; green, 77; blue, 77 }  ,draw opacity=1 ][fill={rgb, 255:red, 230; green, 230; blue, 230 }  ,fill opacity=1 ][line width=0.75]  (105.84,250.76) .. controls (105.83,250.56) and (105.82,250.37) .. (105.82,250.17) .. controls (105.68,240.22) and (114.67,232.02) .. (125.89,231.85) .. controls (137.12,231.68) and (146.34,239.61) .. (146.48,249.56) .. controls (146.48,249.78) and (146.48,249.99) .. (146.48,250.21) -- cycle ;
\draw  [color={rgb, 255:red, 77; green, 77; blue, 77 }  ,draw opacity=1 ][fill={rgb, 255:red, 249; green, 225; blue, 228 }  ,fill opacity=1 ][line width=0.75]  (154.84,223.81) .. controls (154.84,187.46) and (184.27,158) .. (220.58,158) .. controls (256.89,158) and (286.32,187.46) .. (286.32,223.81) .. controls (286.32,260.15) and (256.89,289.61) .. (220.58,289.61) .. controls (184.27,289.61) and (154.84,260.15) .. (154.84,223.81) -- cycle ;
\draw  [color={rgb, 255:red, 77; green, 77; blue, 77 }  ,draw opacity=1 ][line width=0.75]  (136.8,173.83) .. controls (136.8,171.53) and (138.66,169.66) .. (140.96,169.66) .. controls (143.25,169.66) and (145.11,171.53) .. (145.11,173.83) .. controls (145.11,176.12) and (143.25,177.99) .. (140.96,177.99) .. controls (138.66,177.99) and (136.8,176.12) .. (136.8,173.83) -- cycle ;
\draw  [color={rgb, 255:red, 77; green, 77; blue, 77 }  ,draw opacity=1 ][line width=0.75]  (132.09,184.82) .. controls (132.09,183.1) and (133.48,181.7) .. (135.2,181.7) .. controls (136.93,181.7) and (138.32,183.1) .. (138.32,184.82) .. controls (138.32,186.55) and (136.93,187.94) .. (135.2,187.94) .. controls (133.48,187.94) and (132.09,186.55) .. (132.09,184.82) -- cycle ;
\draw  [color={rgb, 255:red, 77; green, 77; blue, 77 }  ,draw opacity=1 ][fill={rgb, 255:red, 230; green, 230; blue, 230 }  ,fill opacity=1 ][line width=0.75]  (292.83,151.68) .. controls (292.83,143.39) and (299.33,136.67) .. (307.35,136.67) .. controls (315.38,136.67) and (321.88,143.39) .. (321.88,151.68) .. controls (321.88,159.97) and (315.38,166.69) .. (307.35,166.69) .. controls (299.33,166.69) and (292.83,159.97) .. (292.83,151.68) -- cycle ;
\draw  [color={rgb, 255:red, 77; green, 77; blue, 77 }  ,draw opacity=1 ][fill={rgb, 255:red, 230; green, 230; blue, 230 }  ,fill opacity=1 ][line width=0.75]  (287.62,189.2) .. controls (287.61,189.01) and (287.61,188.81) .. (287.61,188.62) .. controls (287.46,178.67) and (296.45,170.46) .. (307.68,170.3) .. controls (318.91,170.13) and (328.12,178.06) .. (328.27,188.01) .. controls (328.27,188.23) and (328.27,188.44) .. (328.26,188.66) -- cycle ;
\draw  [color={rgb, 255:red, 77; green, 77; blue, 77 }  ,draw opacity=1 ][fill={rgb, 255:red, 123; green, 161; blue, 196 }  ,fill opacity=1 ][line width=0.75]  (233.83,179.9) .. controls (233.83,175.22) and (238.03,171.43) .. (243.22,171.43) .. controls (248.41,171.43) and (252.61,175.22) .. (252.61,179.9) .. controls (252.61,184.57) and (248.41,188.36) .. (243.22,188.36) .. controls (238.03,188.36) and (233.83,184.57) .. (233.83,179.9) -- cycle ;
\draw  [color={rgb, 255:red, 77; green, 77; blue, 77 }  ,draw opacity=1 ][fill={rgb, 255:red, 123; green, 161; blue, 196 }  ,fill opacity=1 ][line width=0.75]  (230.46,203.64) .. controls (230.45,203.51) and (230.45,203.39) .. (230.45,203.26) .. controls (230.35,197.65) and (236.17,193.02) .. (243.43,192.93) .. controls (250.69,192.83) and (256.65,197.3) .. (256.74,202.91) .. controls (256.75,203.05) and (256.74,203.19) .. (256.74,203.33) -- cycle ;
\draw  [color={rgb, 255:red, 77; green, 77; blue, 77 }  ,draw opacity=1 ][fill={rgb, 255:red, 77; green, 77; blue, 77 }  ,fill opacity=1 ][line width=0.75]  (205.12,223.81) .. controls (205.12,215.94) and (212.04,209.57) .. (220.58,209.57) .. controls (229.12,209.57) and (236.04,215.94) .. (236.04,223.81) .. controls (236.04,231.67) and (229.12,238.05) .. (220.58,238.05) .. controls (212.04,238.05) and (205.12,231.67) .. (205.12,223.81) -- cycle ;
\draw [color={rgb, 255:red, 77; green, 77; blue, 77 }  ,draw opacity=1 ][line width=0.75]    (175.83,174.93) -- (220.58,223.81) ;
\draw [color={rgb, 255:red, 77; green, 77; blue, 77 }  ,draw opacity=1 ][line width=0.75]    (220.58,234.42) -- (210.28,290) ;
\draw [color={rgb, 255:red, 77; green, 77; blue, 77 }  ,draw opacity=1 ][line width=0.75]    (220.58,223.81) -- (281.1,198.81) ;

\draw [color={rgb, 255:red, 77; green, 77; blue, 77 }  ,draw opacity=1 ][line width=0.75]    (180,58.74) -- (167.16,78.04) ;
\draw [shift={(165.5,80.54)}, rotate = 303.63] [fill={rgb, 255:red, 77; green, 77; blue, 77 }  ,fill opacity=1 ][line width=0.08]  [draw opacity=0] (8.93,-4.29) -- (0,0) -- (8.93,4.29) -- cycle    ;
\draw [color={rgb, 255:red, 190; green, 68; blue, 53 }  ,draw opacity=1 ][fill={rgb, 255:red, 1; green, 0; blue, 0 }  ,fill opacity=1 ][line width=1.5]    (56,66.74) -- (156,166.74) ;
\draw [color={rgb, 255:red, 190; green, 68; blue, 53 }  ,draw opacity=1 ][fill={rgb, 255:red, 1; green, 0; blue, 0 }  ,fill opacity=1 ][line width=1.5]    (143.22,71.43) -- (61.5,159.54) ;

\draw (57,167.92) node [anchor=north west][inner sep=0.75pt]  [font=\large,color={rgb, 255:red, 77; green, 77; blue, 77 }  ,opacity=1 ] [align=left] {{\large Ursula}};
\draw (205.5,114.75) node [anchor=north west][inner sep=0.75pt]  [font=\large,color={rgb, 255:red, 77; green, 77; blue, 77 }  ,opacity=1 ] [align=left] {{\large Valentina}};
\draw (193.8,163.31) node [anchor=north west][inner sep=0.75pt]  [font=\large,color={rgb, 255:red, 77; green, 77; blue, 77 }  ,opacity=1 ] [align=left] {Bob};
\draw (42,98.74) node [anchor=north west][inner sep=0.75pt]  [color={rgb, 255:red, 77; green, 77; blue, 77 }  ,opacity=1 ] [align=left] {Using the Born rule \\I think about what \\$\displaystyle \ \ \ \ \ \ b$ and $\displaystyle v$ are};
\draw (78,37.74) node [anchor=north west][inner sep=0.75pt]  [color={rgb, 255:red, 77; green, 77; blue, 77 }  ,opacity=1 ] [align=left] {\textbf{Not allowed per Born Compatibility}};

\end{tikzpicture}
    \caption[An example of a reasoning situation following the GHZ--FR protocol that is not allowed by \possBornText]{An example of a reasoning situation following the GHZ--FR protocol as in \cref{fig:GHZ_final} that is not allowed using the {\protect\possBorn}: Valentina is a superobserver for Bob, performing a supermeasurement on Bob's lab. Ursula cannot use the Born rule to make a single prediction involving both Bob's and Valentina's outcomes $b$ and $v$, as she can never store both simultaneously in a memory.}
    \label{fig:example_restricted_Born rule}
\end{figure}
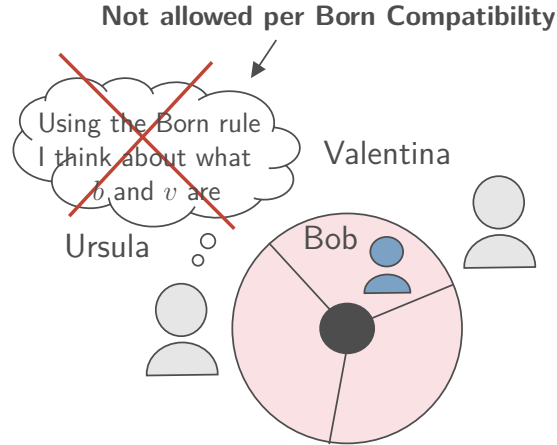

The simplest requirement regarding the possibilistic Born rule is to require its experimental validity, i.e. that a classical agent can never experimentally disprove the possibilistic Born rule by gathering outcomes of measurements, possibly performed by other agents. We will refer to this principle as the \hypertarget{poss_Born}{\possBorn}.
We assume that the agent using the Born rule has, of course, a correct description of the measurements and states involved.

The principle {\possBorn} allows an agent to make predictions about joint measurement outcomes which are indeed borne out if they gather the outcomes of all involved measurements.
The stronger requirement that such predictions made by an agent be \emph{valid} even if the agent does not actually bring those measurement outcomes together can be phrased as follows:
\begin{quote}
    Every performed measurement has an absolute, single-valued outcome about which the possibilistic Born rule as used by agents provides valid predictions.
\end{quote}
We can split this principle up into two philosophical principles:
the existence of an absolute truth, formalised in {\AOEfull},
and {\possBorn} as used by agents vis-à-vis this absolute truth,
regardless of whether its predictions are verified by gathering all involved outcomes, formalised in {\BornCAOE}.\footnotemark

\footnotetext{In epistemic logic,  {\AOEfull} could be formalised as the Truth axiom~\cite{vilasini2019multi,nurgalieva2018inadequacy,montanhano2023contextuality}; see  \Cref{remark:epistemic} ahead for the formulation of key assumptions in our no-go results in the language of epistemic modal logic.}

\begin{assumptionp}{A1}[\AOEfulltext \text{ }(AOE)]   
\label{assumption:AOE}
\hypertarget{AOE}{}
Every performed measurement has an absolute, single-valued outcome. 

Concretely, in each run of the protocol, there exists an assignment $f: \mathcal{M} \rightarrow O$,
where $\mathcal{M}$ is the (finite) set of all the measurements performed by agents in the protocol,
mapping each performed measurement to its observed outcome.\footnotemark
\end{assumptionp}

\footnotetext{For simplicity, we take all measurements to be valued on the same outcome set $O$. If each measurement $m$ has its own set $O_M$ of (a priori) possible outcomes, then the type of $f$ would more properly be that of a dependent function, or a tuple, $f \in \prod_{m \in \mathcal{M}}O_m$, but we might as well take $O = \cup_{m \in \mathcal{M}} O_m$.}

We now require the absolute outcomes of {\AOE} to be compatible with an agent's use of the Born rule, making the agent's statements obtained from their application of the Born rule true regardless of whether the agent actually brings the involved outcomes together, but only when the agent believes to be able to gather the relevant outcomes in a joint memory.

\begin{assumptionp}{A2}[\BornCAOEText]
\hypertarget{Born_Compat_AOE}{} 
\label{assumption:borncompatibility}
    Outcomes assigned per {\AOE} to a set of measurements that an agent $A$ believes she can gather jointly in a memory must be compatible with $A$'s use of the possibilistic Born rule.
    \newline 
    Concretely, the assignment $f: \mathcal{M} \rightarrow O$ per {\AOE} is compatible with all allowed uses of the {\possBorn} by agents. 
\end{assumptionp}

Using {\AOEfull} and {\BornCAOE} an agent may translate their predictions from {\possBorn} to conditions on $f: \mathcal{M} \rightarrow O$.
For example, if the Born rule deems impossible (\ie assigns probability zero to) the joint outcome $(a_1,\ldots,a_n)$ of the measurements $M_1,\ldots,M_n$, then in each run of the protocol the `observed outcome' assignment $f$ cannot satisfy $f(M_i)=a_i$ for all $i=1,\ldots,n$.

Without any further restriction, the {\AOE} assumption is mathematically trivial, \ie a value assignment $f:\mathcal{M}\rightarrow O$ can always be found in each round of a measurement protocol such as the GHZ--FR.
However, {\BornCAOE} introduces restrictions on such value assignments, forcing them to conform with testable predictions, thereby rendering the underlying absoluteness requirement non-trivial.

\begin{remark}[{\AOEtext} and noncontextuality]
    The assumption {\AOEfull} closely resembles that of noncontextuality in the sheaf-theoretic framework~\cite{abramsky2011sheaf}.
    In FR-like paradoxes the different contexts required for a contradiction are operationally brought together through the statements made by different agents.
    Physically, the difference with respect to Bell nonlocality scenarios lies in the fact that in FR-like paradoxes all the measurements that are assigned outcomes are actually performed in each single round of the protocol.
    The assumptions {\AOEfull} and {\BornCAOE} do not yield contradictions in quantum experiments without superobservers,\footnotemark 
     such as standard Bell or contextuality scenarios.
    In such setups, after each run of the experiment there exists a consistent assignment of outcome values to all the measurements that \textit{have been performed} in that run. 
    Therefore, {\AOEfull} does not involve any counterfactual statements about the outcomes of unperformed measurements, as in Bell locality or noncontextuality assumptions, since all involved measurements are actually performed by some agent.
    The only `counterfactual' aspect here is captured by the fact that the {\possBorn} is required to be valid regardless of whether the outcomes of the performed measurements are actually brought together, as per {\BornCAOE}. 
\end{remark}

\footnotetext{Even with a single observer-superobserver pair as in Wigner's original thought experiment, these two assumptions are not contradictory, as after each single run of the experiment there always exists a consistent history of outcome assignments.}
    
The two \cref{assumption:AOE,assumption:borncompatibility} capture how we normally reason in (quantum) experiments.
Our final assumption introduces superobservers.

\begin{assumptionp}{A3}[\UniversalityText]
\hypertarget{Universality}{} 
\label{assumption:universality}
An observer can perform a measurement on a quantum system. Furthermore, a sealed lab in which an observer performs a measurement on some system can be described as a closed quantum system that evolves unitarily, upon which a superobserver can apply quantum operations, including measurements.\\
Concretely, this assumption ensures that the GHZ--FR protocol including superobservers (pictured in circuit form in \cref{fig:GHZ_final}, with the parties situated as in \cref{fig:FR_GHZ_lightcone}) can be performed. 
More specifically: first, there exists a GHZ state, upon which Alice, Bob, and Charlie can perform their $Y$-measurements; these measurements and everything else happening in each of the observers' sealed labs can be modeled as a known unitary by a superobserver; secondly, superobservers can perform supermeasurements on observers, that is, Ursula, Valentina, and Wigner can perform the $X$-like supermeasurements described in \Cref{sec:GHZ_protocol}. 
\end{assumptionp}

We are now in a position to state the GHZ--FR truth no-go theorem.

\begin{theorem}[GHZ--FR truth no-go theorem] \label{th:GHZ_FR_no_go}
   The GHZ--FR paradox shows that the  {\universality}, {\AOEfull} and {\BornCAOE} are incompatible.
\end{theorem}
\begin{proof} 
    The assumption {\universality} ensures that the GHZ--FR measurement protocol can be performed. 
    For the reasoning stage of the paradox, the following assumptions are required. 
    In each run of the protocol, by {\AOEfull}, there exists a function $f \colon \mathcal{M}\to O$ assigning an outcome to each performed measurement.
    Writing $A, B, C, U, V, W$ for the measurements performed in the GHZ--FR paradox, such assignment $f$ consists of outcome values $f(A),f(B),f(C),f(U),f(V),f(W)$, which were denoted by $a,b,c,u,v,w$, respectively, in \Cref{sec:GHZ_protocol}.

    Using {\BornCAOE} (which can be applied since Ursula has no knowledge about Valentina and Wigner and thus believes she is able to gather $u,b,c$ jointly), Ursula reasons that this assignment must satisfy \begin{equation} f(B) \oplus f(C) = 1 \oplus f(U), \end{equation} and similarly Valentina and Wigner respectively reason that
    \begin{align} f(A)  \oplus f(C) &= 1\oplus f(V), \\ f(A) \oplus f(B) &= 1 \oplus f(W). \end{align}
    Finally, applying {\possBorn} (implied by {\BornCAOE}) to the classically recorded outcomes of $U,V,W$, which are available after the protocol, we know that these must satisfy \begin{equation}
        f(U) \oplus f(V) \oplus f(W)=0.
    \end{equation}
    But no assignment $f$ can simultaneously satisfy the four required equations, leading to a contradiction.
\end{proof}

\tocless\subsubsubsection{Ways to resolve the truth no-go theorem} 
\label{sec:ways_to_resolve_truth}

The assumptions of the GHZ--FR truth no-go theorem (\Cref{th:GHZ_FR_no_go}) are shown in \cref{fig:GHZ_FR_assumptions}.
The theorem forces us to reject at least one of these three assumptions. We discuss each of the three possibilities in turn.

\begin{figure}[H]
\centering
\tikzset{every picture/.style={line width=0.75pt}} 

\begin{tikzpicture}[x=0.75pt,y=0.75pt,yscale=-1,xscale=1]

\draw    (641.5,186) -- (651.16,174.56) -- (665.56,157.49) ;
\draw [shift={(667.5,155.2)}, rotate = 130.17] [fill={rgb, 255:red, 0; green, 0; blue, 0 }  ][line width=0.08]  [draw opacity=0] (8.93,-4.29) -- (0,0) -- (8.93,4.29) -- cycle    ;
\draw    (243.94,212.79) -- (186,159.22) ;
\draw [shift={(183.8,157.19)}, rotate = 42.75] [fill={rgb, 255:red, 0; green, 0; blue, 0 }  ][line width=0.08]  [draw opacity=0] (8.93,-4.29) -- (0,0) -- (8.93,4.29) -- cycle    ;
\draw    (164.26,130.33) -- (164.26,73.6) ;
\draw [shift={(164.26,70.6)}, rotate = 90] [fill={rgb, 255:red, 0; green, 0; blue, 0 }  ][line width=0.08]  [draw opacity=0] (8.93,-4.29) -- (0,0) -- (8.93,4.29) -- cycle    ;
\draw    (243.94,212.79) -- (243.44,74.64) ;
\draw [shift={(243.43,71.64)}, rotate = 89.79] [fill={rgb, 255:red, 0; green, 0; blue, 0 }  ][line width=0.08]  [draw opacity=0] (8.93,-4.29) -- (0,0) -- (8.93,4.29) -- cycle    ;
\draw    (19.19,130.38) -- (19.42,73.6) ;
\draw [shift={(19.43,70.6)}, rotate = 90.23] [fill={rgb, 255:red, 0; green, 0; blue, 0 }  ][line width=0.08]  [draw opacity=0] (8.93,-4.29) -- (0,0) -- (8.93,4.29) -- cycle    ;
\draw  [fill={rgb, 255:red, 0; green, 0; blue, 0 }  ,fill opacity=1 ] (-7.5,30) -- (271.5,30) -- (271.5,70) -- (-7.5,70) -- cycle ;
\draw  [fill={rgb, 255:red, 46; green, 90; blue, 140 }  ,fill opacity=1 ] (-42.5,103.57) -- (80.87,103.57) -- (80.87,157.19) -- (-42.5,157.19) -- cycle ;
\draw  [fill={rgb, 255:red, 46; green, 90; blue, 140 }  ,fill opacity=1 ] (101.55,103.52) -- (226.98,103.52) -- (226.98,157.14) -- (101.55,157.14) -- cycle ;
\draw  [fill={rgb, 255:red, 46; green, 90; blue, 140 }  ,fill opacity=1 ] (176.85,185.98) -- (310.5,185.98) -- (310.5,239.6) -- (176.85,239.6) -- cycle ;
\draw    (668.5,135.2) -- (668.5,72.2) ;
\draw [shift={(668.5,69.2)}, rotate = 90] [fill={rgb, 255:red, 0; green, 0; blue, 0 }  ][line width=0.08]  [draw opacity=0] (8.93,-4.29) -- (0,0) -- (8.93,4.29) -- cycle    ;
\draw    (626.5,206) -- (580.05,158.64) ;
\draw [shift={(577.95,156.5)}, rotate = 45.56] [fill={rgb, 255:red, 0; green, 0; blue, 0 }  ][line width=0.08]  [draw opacity=0] (8.93,-4.29) -- (0,0) -- (8.93,4.29) -- cycle    ;
\draw    (561.25,127.55) -- (561.25,71.61) ;
\draw [shift={(561.25,68.61)}, rotate = 90] [fill={rgb, 255:red, 0; green, 0; blue, 0 }  ][line width=0.08]  [draw opacity=0] (8.93,-4.29) -- (0,0) -- (8.93,4.29) -- cycle    ;
\draw    (637.62,210.3) -- (637.11,71.66) ;
\draw [shift={(637.1,68.66)}, rotate = 89.79] [fill={rgb, 255:red, 0; green, 0; blue, 0 }  ][line width=0.08]  [draw opacity=0] (8.93,-4.29) -- (0,0) -- (8.93,4.29) -- cycle    ;
\draw    (412.72,127.55) -- (412.96,71.4) ;
\draw [shift={(412.97,68.4)}, rotate = 90.24] [fill={rgb, 255:red, 0; green, 0; blue, 0 }  ][line width=0.08]  [draw opacity=0] (8.93,-4.29) -- (0,0) -- (8.93,4.29) -- cycle    ;
\draw  [fill={rgb, 255:red, 0; green, 0; blue, 0 }  ,fill opacity=1 ] (384.5,28) -- (708.5,28) -- (708.5,68) -- (384.5,68) -- cycle ;
\draw  [fill={rgb, 255:red, 46; green, 90; blue, 140 }  ,fill opacity=1 ] (350.5,100.64) -- (474.95,100.64) -- (474.95,154.45) -- (350.5,154.45) -- cycle ;
\draw  [fill={rgb, 255:red, 46; green, 90; blue, 140 }  ,fill opacity=1 ] (497.99,100.64) -- (624.51,100.64) -- (624.51,154.45) -- (497.99,154.45) -- cycle ;
\draw  [fill={rgb, 255:red, 189; green, 16; blue, 224 }  ,fill opacity=1 ] (571.04,185.12) -- (648.5,185.12) -- (648.5,239.2) -- (571.04,239.2) -- cycle ;
\draw  [fill={rgb, 255:red, 189; green, 16; blue, 224 }  ,fill opacity=1 ] (645.96,101.34) -- (724.5,101.34) -- (724.5,155.43) -- (645.96,155.43) -- cycle ;

\draw (38.37,39.63) node [anchor=north west][inner sep=0.75pt]  [font=\normalsize,color={rgb, 255:red, 255; green, 255; blue, 255 }  ,opacity=1 ] [align=left] {GHZ--FR truth no-go theorem};
\draw (190.92,197.94) node [anchor=north west][inner sep=0.75pt]  [font=\normalsize,color={rgb, 255:red, 255; green, 255; blue, 255 }  ,opacity=1 ] [align=left] { Absoluteness of \\Observed \ Events};
\draw (100.14,119.58) node [anchor=north west][inner sep=0.75pt]  [font=\normalsize,color={rgb, 255:red, 255; green, 255; blue, 255 }  ,opacity=1 ] [align=left] { Born Compatibility};
\draw (-33.08,112.52) node [anchor=north west][inner sep=0.75pt]  [font=\normalsize,color={rgb, 255:red, 255; green, 255; blue, 255 }  ,opacity=1 ] [align=left] { \ Universality of \\Quantum Theory};
\draw (-49,-5) node [anchor=north west][inner sep=0.75pt]   [align=left] {(a)};
\draw (380,-2) node [anchor=north west][inner sep=0.75pt]   [align=left] {(b)};
\draw (439.37,38.63) node [anchor=north west][inner sep=0.75pt]  [font=\normalsize,color={rgb, 255:red, 255; green, 255; blue, 255 }  ,opacity=1 ] [align=left] {GHZ--FR agreement no-go theorem};
\draw (497.13,119.58) node [anchor=north west][inner sep=0.75pt]  [font=\normalsize,color={rgb, 255:red, 255; green, 255; blue, 255 }  ,opacity=1 ] [align=left] { Born Compatibility};
\draw (359.53,110.75) node [anchor=north west][inner sep=0.75pt]  [font=\normalsize,color={rgb, 255:red, 255; green, 255; blue, 255 }  ,opacity=1 ] [align=left] { \ Universality of \\Quantum Theory};
\draw (575.19,193.45) node [anchor=north west][inner sep=0.75pt]  [font=\normalsize,color={rgb, 255:red, 255; green, 255; blue, 255 }  ,opacity=1 ] [align=left] { Personal \\Knowledge};
\draw (650.17,109.45) node [anchor=north west][inner sep=0.75pt]  [font=\normalsize,color={rgb, 255:red, 255; green, 255; blue, 255 }  ,opacity=1 ] [align=left] { Classical \\Agreement};

\end{tikzpicture}  \caption[Assumptions of the GHZ--FR no-go theorems]{Assumptions leading to (a) the GHZ--FR truth no-go theorem of \Cref{sec:GHZ_nogo} and (b) the GHZ--FR agreement no-go theorem of \Cref{sec:GHZ_nogo_trust}, where AOE is replaced by Personal Knowledge and Classical Agreement. An arrow from one assumption box to another denotes that the former is used to formulate the latter.} \label{fig:GHZ_FR_assumptions} 
\end{figure}
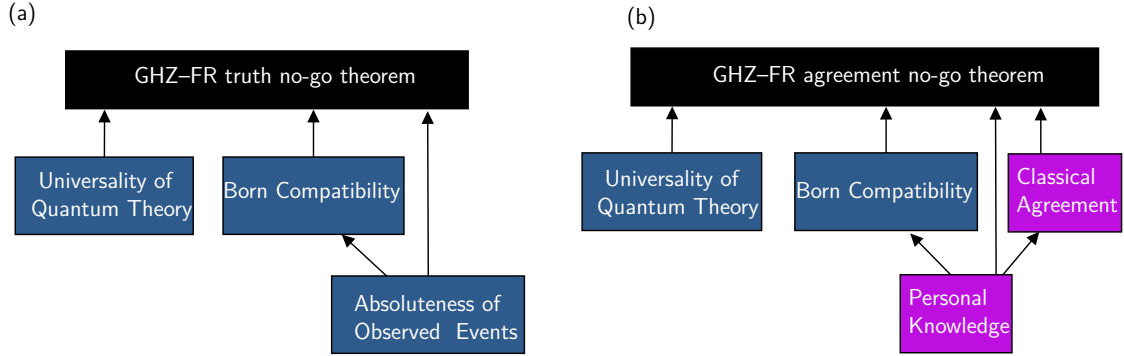

\paragraph*{a. Reject {\AOEfull}}
Refuting {\AOEfull} means one cannot simply talk about single-valued outcomes obtained in measurements performed by other agents as being absolute, i.e. a function $f:\mathcal{M} \rightarrow O$ assigning an outcome to each performed measurement cannot exist. Still, we would like agents to be able to formulate predictions correctly. 
Therefore, we may ask `When can an agent assign a single-valued outcome to a performed measurement?'
Naturally, the experimenter who performs the measurement can assign a single-valued outcome, namely the one they observe. 
This extends also to any agent who learns about the outcome, i.e. any agent obtaining a copy of the outcome record and storing it in their memory\footnote{For a more elaborate definition of \emph{learning about an outcome}, see \Cref{sec:possible_resolution} below.}. 
Can other agents assign an outcome -- that is not a priori absolute -- too?
We probe this possibility in the GHZ--FR agreement no-go theorem in the next section, where we weaken {\AOEfull} to allow for each agent to assign a personal single-valued outcome to performed measurements, which may a priori differ among different agents. The stronger no-go result obtained indicates that this approach may not lead to a resolution,
and that perhaps only the experimenter and agents learning about the outcome can assign an outcome value to a performed experiment.

\paragraph*{b. Reject {\BornCAOE}} As {\BornCAOE} assumes {\AOE}, refuting {\AOE} implies refuting {\BornCAOE}.
However, one may also consider accepting {\AOE} but refuting {\BornCAOE}.
In that case, there is still an absolute single-valued outcome assigned to each performed measurement,
but the use of the Born rule per {\possBorn} does not provide valid predictions about such assignment. 
In other words, more cautious use of the Born rule is required. Therefore, one needs an additional prescription for how all agents can use the Born rule. This is the route typically taken by (contextual) hidden variable theories. In this case, the observers' memories may also be described by hidden variables, such as in Bohmian theories~\cite{sudbery2017single,lazarovici2019quantum,sudbery2019hidden}, making correct use of the Born rule more subtle.
We will discuss this option further in  \Cref{sec:GHZ_nogo_trust}.

When refuting {\BornCAOE}, one can instead go further as well and refute the experimental validity of the Born rule ({\possBorn}) as well, i.e. modify the Born rule for all applications rather than only the non-experimental ones.

\paragraph*{c. Reject {\universality}} 
To reject the existence of superobservers, however, comes with its own complications.
If we adopt the viewpoint that only classical observers can perform measurements, where do we draw the line?
Would a hypothetical quantum computer showing signs of artificial intelligence not qualify as an observer? 
Or would the notion of a classical observer arise as an emergent concept?
Alternatively, one may reject the notion of measurement altogether.

In the foundations of quantum theory, the above questions deal with where to place the Heisenberg cut -- or whether it even exists.
In fact, Wigner's original thought experiment probed precisely this question.
However, with the FR and GHZ--FR paradoxes, the problem gains new importance, due to the stronger implications of accepting superobservers.
Namely, seemingly innocent use of quantum theory leads to contradictory statements regarding not only the question of \textit{when} a measurement does happen and classical outcomes are produced, but also regarding  \textit{what} actual outcome values were observed. 
In short, inconsistent histories may be obtained by different observers.

\medskip

In other words, if one accepts the experimental validity of the (possibilistic) Born rule and the existence of superobservers {\ref{assumption:universality}}, the contradiction in the GHZ--FR paradox arises through the assumed existence of a function $f\colon\mathcal{M} \rightarrow O$ assigning absolute, single-valued outcomes to all performed measurements which are compatible with the agents' allowed uses of the Born rule. 
In that case,  one must refute either the existence of such an absolute assignment $f$, thus rejecting {\AOE} {\ref{assumption:AOE}}, or the compatibility of $f$ with the agents' use of the Born rule, rejecting {\BornCAOE} {\ref{assumption:borncompatibility}} and require more cautious use of the Born rule.

We refer to \Cref{sec:comparison_FR} for a discussion of the FR assumptions, responses in literature and their relevance for the GHZ--FR no-go theorems.

\subsection{GHZ--FR agreement no-go theorem} \label{sec:GHZ_nogo_trust}
\tocless\subsubsubsection{Assumptions} \label{sec:subssubsec_trust_nogo}
We now replace
{\AOEfull} by a weaker condition, split up into two assumptions.
We introduce {\PersK}, which a priori allows different agents to assign different single-valued outcomes to the same measurement,
but subject it to a consistency condition called {\ClassicalT}, which stipulates that a statement known to be true by a classical agent can be taken to be true by any other classical agent to whom it is communicated.
As in the truth no-go theorem,
we require the outcome assignments per {\PersK} to be compatible with the Born rule. 
These assumptions allow superobservers Ursula, Valentina and Wigner to formulate their predictions about outcomes and share them with each other, obtaining a contradiction.


We stress that only classical agents need to reason and communicate;
there is no need for classical agents to trust the reasoning of quantum agents as in Ref.~\cite{frauchiger2018quantum,nurgalieva2018inadequacy,montanhano2023contextuality}; see also \Cref{remark:epistemic} below.

\begin{assumptionp}{A1$^\prime$}[\PersKText]
\hypertarget{Pers_K}{} 
\label{assumption:personalknowledge}
Every performed measurement can be assigned a single-valued outcome by an agent who believes they are able to gather that outcome.\footnote{For example, in the GHZ--FR paradox Ursula applies {\PersK} to assign values to Bob's and Charlie's measurements, whose outcomes she believes she could still obtain  as she does not need to know about Valentina's or Wigner's existence or operations.}
\newline
Concretely, in each run of the protocol, for each agent $A$, there exists an assignment $f_A: \mathcal{M}_A \rightarrow O$,
where $\mathcal{M}_A$ is the set of performed measurements whose outcomes could be gathered from $A$'s perspective,
mapping each performed measurement to an outcome value assigned by $A$.
\end{assumptionp}
The agents (superobservers) assigning outcomes in the GHZ--FR paradox only do so \emph{after} having performed their own measurement, i.e. based on what they believe is available to them then.

We recall that {\BornCAOE} builds on {\AOE}, but we can also state an analogous Born rule compatibility requirement for this weakened version of {\AOE}.

\begin{assumptionp}{A2$^\prime$(i)}[\BornCPersKText]
\hypertarget{Born_Compat_PersK}{} 
\label{assumption:borncompatibility_truth}
Outcomes assigned \textit{by an agent} $A$ per {\PersK} to a set of measurements must be compatible with $A$'s use of the possibilistic Born rule.
\newline
Concretely, for every agent $A$, the assignment $f_A \colon \mathcal{M}_A \rightarrow O$ per {\PersK} is compatible with all allowed uses of the {\possBorn} by $A$.
\end{assumptionp}

In {\PersK}, each function $f_A$ assigns outcomes to measurements, but, crucially, these outcomes need not be absolute but can be relative to agent $A$. 
Using {\BornCPersK}, agent $A$ may translate her predictions from {\possBorn} to conditions on $f_A:\mathcal{M}_A \rightarrow O$. For example, if the Born rule deems impossible the joint outcome $(a_1,\ldots,a_n)$ of the measurements $M_1,\ldots, M_n$, then the assignment $f_A$ cannot satisfy $f_A(M_i) = a_i$ for all $i=1,\ldots,n$. Agent $A$ may conclude:  \begin{quote}
`I (agent $A$) know that there exists a measurement $M_i \in \{M_1,\ldots,M_n\}$, possibly performed by other agents, that does not have outcome $a_i$ (for me),
regardless of whether I ask the involved agent(s) for it.
\end{quote}

Next, we state a minimal practical requirement on the outcome assignments across different classical agents.
This assumption was not needed for the GHZ--FR truth no-go theorem, as it is implied by {\AOE} and {\BornCAOE}.

\begin{assumptionp}{A2$^\prime$(ii)}[\ClassicalTText]
\hypertarget{Classical_T}{} 
\label{assumption:classicaltrust}
The outcome assignments (per \PersK) of classical agents who communicate classically must agree on their overlap.
\newline
Concretely, if classical agents $A$ and $B$ classically communicate their outcome assignments, then these assignments $f_A$ and $f_B$ must agree on their overlap $\mathcal{M}_A \cap \mathcal{M}_B$.
\end{assumptionp}

In fact, the way we use this assumption is that, for example, Ursula and Valentina, must agree on their outcome assignment for the measurement $C$: $f_U(C)=f_V(C)$. In fact, in a version of the protocol where Wigner performs his measurement later, both Ursula and Valentina can ask Charlie for his outcome value. They will obtain the same value, motivating this assumption. In our particular protocol, {\PersK} and {\ClassicalT} will lead to agents $U,V,W$ having to agree on their assignments so that again a single assignment $f$ to all performed measurements can be built that must satisfy the GHZ--Mermin equations.

\begin{theorem}[GHZ--FR agreement no-go theorem] \label{th:GHZ_FR_no_go_trust}
   The GHZ--FR paradox shows that the assumptions  {\universality}, {\PersK}, {\ClassicalT}, and {\BornCPersK}  are incompatible.
\end{theorem}
\begin{proof}
    The first part of the proof is analogous to the proof of \Cref{th:GHZ_FR_no_go}, replacing the absolute outcome assignment $f$ by the outcome assignments $f_U,f_V,f_W$ of Ursula, Valentina and Wigner per {\PersK}. For example, as Ursula has no knowledge about Valentina and Wigner and thus believes she is able to gather $u,b,c$ jointly, she can assign them outcome values $f_U(B),f_U(C)$ (and $f_U(U)$ of course), compatible with her Born rule use per {\PersK} and {\BornCPersK}. Then {\ClassicalT} requires these outcome assignments to agree on their overlaps as Ursula, Valentina and Wigner are treated classically in the protocol and can physically meet and communicate as much as they like, so that a single global outcome assignment can be built that satisfies the GHZ--Mermin equations. Details can be found in Appendix B of Ref.~\cite{walleghem2024strong}.
\end{proof}

\begin{remark}[Heisenberg cuts]
    In literature, sometimes $A$ applying the Born rule to set of jointly performed measurements is also referred to as $A$ choosing a Heisenberg cut, having the performed measurements outside the cut. In that terminology, our assumptions of {\PersK} and {\BornCPersK} can also be phrased as: `An agent $A$ can choose a Heisenberg cut, placing a set of measurements she thinks she can still gather jointly all outside the cut.' The paradox then arises if it is understood that different choices of Heisenberg cuts, when chosen by classical agents, must not lead to contradictions about the actual outcome values, which is another way of phrasing {\ClassicalT}. 
    Interpretations of quantum mechanics which resolve the measurement problem by determining or eliminating the Heisenberg cut will resolve EWF arguments by providing an unambiguous description for the entire protocol.
   We discuss Heisenberg cuts in more detail in \Cref{sec:previous_work}.
\end{remark}

\begin{remark}[The GHZ--FR assumptions in  modal logic] \label{remark:epistemic}
    We briefly outline a discussion of the GHZ--FR paradox in terms of epistemic logic, and refer to Appendix C of Ref.~\cite{walleghem2024strong}.

    Syntactically, in epistemic logic~\cite{fagin2004reasoning,blackburn2001modal}, besides the usual logical connectives, formulas can be built using a unary modal operator $K_A$ for each agent $A$, with a formula $K_A \phi$ interpreted as `agent $A$ believes (or knows) that $\phi$ is True'. 
    Different sets of axioms are typically considered in epistemic logic systems, varying with the intended interpretation.
    
    In Ref.~\cite{nurgalieva2018inadequacy}, the \textit{Distribution Axiom} of modal logic (a.k.a.\ axiom \textbf{K}) is listed as assumption (D) of the FR paradox
    (see \Cref{sec:comparison_FR}), and
    used to justify the use of \textit{modus ponens} by quantum agents in the FR paradox. 
    In the GHZ--FR paradox, however, as in the version of the FR paradox from \Cref{sec:stricter_FR}, only classical agents need to reason, 
    dispensing with the reasoning by quantum agents considered in the earlier treatments of the FR paradox~\cite{nurgalieva2018inadequacy,vilasini2019multi,montanhano2023contextuality,frauchiger2018quantum}.
    The fact that we need only consider reasoning by classical agents in the protocol is the reason why we do not state such reasoning as a separate assumption in our no-go theorems.
    In modal logical terms, this is reflected in the fact that we do not need the epistemic modalities $K_A$ for quantum agents:
    the whole argument can be phrased in a logical system with modalities only for each \textit{classical} agent in the protocol. 
    It could perhaps be argued that this somewhat weakens the appeal of a formalisation of the paradoxes using modal logic.

    Still, the core assumptions of our GHZ--FR no-go theorems can be phrased in these terms.
    The existence of an absolute truth, per \AOEfull, in the GHZ--FR truth no-go theorem corresponds to the Truth axiom: if an agent \textit{knows} something, then it is true: $K_A \phi \rightarrow \phi$. 
    The agreement no-go theorem requires a form of agreement only among classical agents.
    It can be seen as a considerable weakening of the epistemic trust structures identified for the FR paradox in~\cite{nurgalieva2018inadequacy,vilasini2019multi,montanhano2023contextuality}, in that we do not need classical agents to believe the reasoning of quantum agents.
\end{remark}

\tocless\subsubsubsection{Ways to resolve the agreement no-go theorem} \label{sec:ways_to_resolve_trust}
The truth and agreement no-go theorems differ only in that the {\AOE} assumption in the truth no-go theorem is replaced by {\PersK} and {\ClassicalT} in the agreement no-go theorem (cf. \cref{fig:GHZ_FR_assumptions}).
If one opts to resolve the truth no-go theorem by rejecting {\AOE}, then one must reject one of the assumptions {\PersK} and {\ClassicalT} to resolve the agreement no-go theorem.
We argue that, in that case, the most practical resolution is to further weaken {\PersK}.
Other resolutions are of course possible.

We now discuss the repercussions of separately rejecting each of the assumptions of the agreement no-go theorem. The consequences of rejecting {\universality} are the same as in the truth no-go theorem, as discussed in \Cref{sec:ways_to_resolve_truth}.

\paragraph*{a. Reject {\PersK}}\label{para:reject_Pers_Knowledge} 
Rejecting {\PersK} implies that one also rejects {\AOEfull} of the truth no-go theorem.
If the outcome of a performed measurement is not absolute, one must ask: `Who can assign a single-valued outcome to a performed measurement?'\footnote{As the outcome of a performed measurement is considered to be classical information, a physical entity assigning an outcome to a measurement
needs a preferred basis in its quantum storage system, as argued in Refs.~\cite{brukner2021qubits,allam2023observer}, which constitutes our definition of an agent.}
Certainly an experimenter can assign a single value to the outcome of a measurement performed by themself and test the Born rule in this way -- but who else?

Suppose that an agent $A$'s prediction about outcomes of performed measurements is valid `no matter whether $A$ actually brings the involved outcomes together'. Then, one is assuming the existence of absolute outcomes or outcomes relative to the agent $A$ as in {\PersK}.
Therefore, rejecting {\AOE} and {\PersK} suggests that only the experimenter themself can assign a single-valued outcome to a performed measurement, and by extension any other agent learning about the outcome.
This proposal suggests a new principle: 
 {\ROE}\footnote{
 The seed of this principle originated in a discussion of L.W. with E. Cavalcanti about the implications of the Local Friendliness no-go theorem~\cite{bong2020strong,cavalcanti2021implications}, but it is not necessarily what E. G. Cavalcanti believes to be a resolution.}:
\begin{quote}
    \hypertarget{ROE}{\ROE}
    Every performed measurement can be assigned a single-valued outcome by an agent who learns about that outcome.
    \end{quote}

Consequently, an agent $A$ can reason about the outcome $b$ of a measurement performed by another agent $B$ only to make statements about the value of this outcome $b$ \textit{conditioned on} agent $A$ asking $B$ for it, formalised in the following principle:
\begin{quote}
\hypertarget{BornP}{\BornP} An agent $A$ can translate predictions based on the Born rule only into statements about outcomes that $A$ would obtain when $A$ learns about these outcomes.

\end{quote}
To \emph{learn about an outcome}, an agent performs an operation such that they, directly or indirectly, get to know the outcome result of the measurement through interactions with the experimenter (or other agents who know the result), measured system or environment, namely by (i) directly asking the experimenter, (ii) measuring the measured system in the same measurement basis or (iii) indirectly obtaining the measurement outcome through traces in the environment (which can also include being told about it by another agent).

\paragraph*{b. Reject {\BornCPersK}} \label{par:reject_Born_compatibility} If one rejects Born Compatibility, one can still keep {\AOE} (and {\PersK}).
However, in that case, as the GHZ--FR paradox shows, one needs a rule for when to expect Born rule predictions to be valid, as agents need to be able to phrase their Born rule predictions correctly.
When holding on to the experimental validity of the Born rule per {\possBorn}, as a principle for when such predictions are valid, one can use {\BornP} as well.
If an agent has more knowledge of the underlying theory or ontology their world is subject to, they can perhaps make more precise statements. 
For example, one can imagine a theory or ontology where AOE is satisfied (as in Bohmian mechanics), and where an agent can say their prediction is valid as long as they can gather all involved outcomes.
However, one can imagine a theory where additionally actually asking for the outcomes changes the actual outcomes through hidden variables. 
Therefore, without further knowledge, in theories rejecting {\BornCPersK}, agents can use {\BornP} for correct predictions.

\paragraph*{c. Reject {\ClassicalT}} \label{par:reject_classical_trust} Rejecting {\ClassicalT} would mean that classically communicating agents (communicating in whichever way they prefer) disagree on the truth value of a statement about what happened in a measurement.
In principle, every performed measurement whose outcome is not stored in some classical agent's memory can be part of a paradox à la GHZ--FR. Thus, disagreement among classically communicating agents about what happened in the measurement $M$ topples a basic assumption of science: we cannot trust other scientists' reports of what happened in an experiment.
To see this difficulty in more detail, we consider the following situation. Alice performs a measurement, and Bob makes a statement about Alice's outcome, claiming for example that Alice obtained $a=0$. Bob could have arrived at this statement by reasoning or by explicitly asking Alice. 
Holding on to {\PersK} and {\BornCPersK}, there is no need for Bob to specify how he obtained the statement, which is true \textit{for him}. 
Therefore, when Bob tells us that Alice obtained $a=0$, we simply cannot know whether his statement (or any statement derived from it) is true \textit{for us}. 
A potential way out is that Bob conveys more details about how he obtained his conclusion, and then we decide whether or not to believe him, as argued in QBism~\cite{debrota2020respecting}, for example.
In our everyday world, such disagreement between agents does not arise, because of the absence of superobservers.


\subsection{A possible resolution} \label{sec:possible_resolution}

In \Cref{sec:ways_to_resolve_trust} we proposed {\BornP} to resolve the GHZ--FR paradox, and by extension all extended Wigner's friend paradoxes, if one accepts the existence of superobservers. 
The principle says that observers should state their Born rule predictions \emph{conditioned} on learning about the involved outcomes.
In theories that reject {\AOE}, this principle can be implemented as {\ROE}, as explained in \Cref{para:reject_Pers_Knowledge}.
This resolution can be seen as an agent-based extension of Peres's dictum that `Unperformed experiments have no results'~\cite{peres1978unperformed},
\begin{quote}
    Agent-based Peres's dictum:
    `Unperformed experiments have no results, and unknown results have no values.'
\end{quote}
We can see this as an extension of Peres's dictum, by considering learning about the outcome result (of a measurement performed by another agent) as an experiment too. 
Namely, you cannot \emph{a priori} assign outcome values to measurements or experiments, possibly performed by other agents, you can only do so upon actually learning the outcome. 
In a sense, this principle is a natural classical vs.\ quantum separation.
If a superobserver Wigner can model his friend's measurement unitarily, \ie as a quantum system, then Wigner cannot assign classical information to his friend's measurement outcome, as this information is still quantum for Wigner. 
Instead, Wigner should only assign a single value to his friend's outcome when he learns about that outcome.
Here is an example of {\BornP} in use for a simple Wigner's friend scenario:
\begin{quote}
    Wigner finds that $f=1$ is impossible (has probability zero) with $f$ denoting his friend's outcome and may conclude:
    ``Using the Born rule I (Wigner) know that my friend would never reply that she obtained $f=1$ \emph{if} I were to ask her for her outcome, (as long as my friend's memory record and mine are not wiped or updated).''
\end{quote}

\paragraph*{Definition of `learning about an outcome'.}
    By `learning about an outcome of a measurement', or `asking an experimenter for the outcome of their performed measurement', we mean the following. Consider an agent $A$ performing a measurement on a qubit system $S$ in the basis $\ket{0}, \ket{1}$. Let $E$ be the environment, namely the whole closed system excluding $A$ and $S$. We can model this measurement as a unitary: \begin{equation}
        \big( \alpha \ket{0} + \beta \ket{1} \big)_S \otimes |\text{ready} \rangle_A \otimes | \text{ready} \rangle_E \rightarrow \alpha \ket{0}_S \otimes \ket{0}_A \otimes \ket{0}_E + \beta \ket{1}_S \otimes \ket{1}_A \otimes \ket{1}_E
    \end{equation} where we assume that the agent and environment are initialized in the $|\text{ready}\rangle_A, |\text{ready} \rangle_E$. Here $\ket{0}_E, \ket{1}_E$ denote the traces of the measurement outcome $0$ or $1$ in the environment. Learning about the outcome of this measurement then includes: \begin{itemize}
        \item Asking the experimenter $A$ for the outcome, corresponding to measuring $A$ in the basis $\ket{0}_A,\ket{1}_A$;
        \item Measuring the system $S$ in the measurement basis $\ket{0}_S,\ket{1}_S$;
        \item Obtaining the measurement outcome through traces in the environment, corresponding to measuring the environment $E$ in the basis $\ket{0}_E,\ket{1}_E$;
        \item classical communication about outcomes through another agent who learns about the outcome in one of the three ways stated above.
    \end{itemize}

\paragraph{Resolving the paradox.} Let us briefly explain how assuming {\BornP} instead of {\AOE} (or instead of {\PersK}) and {\BornCAOE} resolves the GHZ--FR paradox. 
The FR paradox is resolved analogously.

We assume that the GHZ--FR measurement protocol has been performed, considering a run with the (classically available) outcomes $u=0,v=0,w=0$, say.
Ursula, making her predictions, finds using {\BornP}:
\begin{quote}
    Ursula: `if I (Ursula) were to ask Bob and Charlie for their outcomes $b,c$, then I would find that $b \oplus c = 1$, as long as my memory and theirs are not wiped or updated.' 
\end{quote} 
Bob and Charlie argue similarly to conclude that $a \oplus c = 1$ and $a \oplus b = 1$ when they ask for $(a,c)$ and $(a,b)$, respectively. 
However, none of these superobservers have actually asked the other observers Alice, Bob and Charlie for their outcomes, so that no paradox arises.
All precise statements, as well as those for the FR paradox, can be found in Appendix D of Ref.~\cite{walleghem2024strong}.

\section{Comparison with the FR paradox} \label{sec:comparison_FR}
We compare the GHZ--FR paradox to the FR paradox in \Cref{sec:comparison_FR_GHZ_FR}. 
Then, in \Cref{sec:previous_work}, we place our work within the broader literature on the subject: we briefly review previous responses to the FR paradox, as well as some work closely related to the GHZ--FR paradox, where similar measurement protocols have been proposed but with different analyses.

\subsection{Why the GHZ--FR paradox is stronger than the FR paradox} \label{sec:comparison_FR_GHZ_FR} 
To establish the GHZ--FR paradox, in stark contrast to the FR paradox~\cite{frauchiger2018quantum,vilasini2022general}, (i) only \emph{classical} agents reason, and (ii) \emph{no post-selection} on outcomes is required.
The Born rule is only used to determine that certain (joint) measurement outcomes are \emph{impossible}.
It need not be used to assert that certain outcomes are \emph{possible}, that is, that they occur with (some) nonzero probability.
While the former are definitive statements about \emph{every} run of the experiment -- they state that a particular combination of outcomes is \emph{never} witnessed --, the latter inform about events that occur only in \emph{some} runs, which would need to be post-selected for the derived contradiction to apply.
This difference is at the root of distinctive feature (ii) above.

To compare the assumptions of the FR no-go theorem and those of the GHZ--FR agreement no-go theorem, the strongest of the two no-go theorems from \Cref{sec:GHZ_FR_nogo}, we begin by listing the assumptions of both theorems (cf. \cref{fig:assumptions_FR_GHZ_FR}).
For the FR theorem, we rely on the presentation in Ref.~\cite{vilasini2022general}, which
considers the following list of assumptions: (C) Consistency, (D) Distribution axiom of modal logic, (Q) Quantum Born rule statements, (U) Unitarity, and (S) Single-valuedness of outcomes.

Distinctive feature (i) -- that only classical agents reason -- is reflected in the weakening of assumptions (C), (D), and (Q) in the GHZ--FR paradox.
Distinctive feature (ii) -- that no post-selection is required -- is not reflected in the assumptions. 
It corresponds to the fact that the contradiction at the heart of the GHZ--FR paradox occurs in \emph{every} single round of the protocol,
whereas the FR paradox only obtains a contradiction in \emph{some} (postselected) rounds in which a specific value of the final (super)measurements is observed.

\paragraph*{The FR assumptions.} We first summarize the assumptions of the FR no-go theorem in our own wording. All uses of the word `statement' are shorthand for `statement about outcome values observed by agents'.\footnote{In the FR, one refers to outcome values observed by a single agent in these statements. In the GHZ--FR multiple agents are involved in such statements, for example Ursula reasons about $b,c$ observed by Bob and Charlie, respectively; Ursula treats these joint outcomes as an outcome of a joint measurement by Bob and Charlie. However, in the GHZ--FR no-go theorem we could also allow $A,B,C$ to share their outcome values with each other so that both $b,c$ are observed by Bob and Charlie for example, and have a superobserver later undo this sharing before other measurements are performed. We could even have all these measurements $A,B,C$ be performed by the same agent, with $U,V,W$ acting upon different parts of that agent's memory, so that when Ursula talks about outcomes $b,c$ this refers to a joint measurement performed by that single agent.}
\begin{itemize}
\item[(Q)] An agent $A$ can translate the use of the possibilistic quantum Born rule applied to a set of performed (not necessarily all by $A$) measurements, to make statements about what outcomes are obtained or not.
\item[(C)] If an agent $B$ knows that statement $s_1$ is true from reasoning with a theory that an agent $A$ accepts\footnote{By a \emph{theory that an agent $A$ accepts}, we mean reasoning according to these five rules.} and $A$ knows about this reasoning of $B$ about $s_1$, then $s_1$ is true for agent $A$ as well. 
\item[(D)] If an agent $A$ knows a statement $s_1$ and also knows that $s_1$ implies statement $s_2$ then agent $A$ can conclude `I (A) know $s_2$'. 
\footnote{Assumption (D) refers to the Distribution axiom of Modal Logic and was first identified in Ref.~\cite{nurgalieva2018inadequacy}, along with assumption (U).}
\item[(S)] Outcomes of performed measurements from an agent's viewpoint who uses (Q) are single-valued.
\item[(U)] An agent $A$ can model another agent $B$ measuring a system as a unitary process.
\end{itemize}
Let us give an example of (Q) in use. If agent $A$ applies the Born rule to a set of measurements that can be simulated in parallel and finds that a certain outcome $(a_1,\ldots,a_n)$ cannot occur, then $A$ can conclude that some $a_i$ is not obtained (in a measurement performed possibly by another agent).\footnote{We rephrased assumption (Q) here somewhat compared to FR's original phrasing, as unwanted interpretations of the latter may (wrongly) conclude collapse to be necessary for the FR paradox~\cite{aaronson2018s,schmid2023review,tausk2018brief,bub2021understanding,mucino2020wigner}.}

\paragraph*{Rephrasing the GHZ--FR assumptions} We now rephrase the GHZ--FR assumptions in order to simplify the comparison with the assumptions of the FR paradox, \emph{emphasizing} the differences.\footnote{In fact, this paraphrasing of our no-go theorem has slightly stronger assumptions than needed. We did so to tailor them more to the phrasing of the FR assumptions. To weaken them, we could allow the (Q) assumption to have statements about outcome values assigned by the agent making the prediction, subject to \ClassicalT.}
\begin{itemize}
    \item[(Q')] A \emph{classical}\footnote{We recall that by a \textit{classical} agent we mean an agent that need not be modeled quantumly by any other agent in the protocol.} agent $A$ can translate the use of the possibilistic quantum Born rule applied to a set of performed (not necessarily all by $A$) measurements, to make statements about what outcomes are obtained or not.
    \item[(C')] If a \emph{classical} agent $B$ knows that statement $s_1$ is true from reasoning with a theory that a \emph{classical} agent $A$ accepts, and \emph{$B$ communicates his reasoning to $A$ through classical communication} (so that $A$ knows about this reasoning of $B$ about $s_1$), then $s_1$ is true for agent $A$ as well. 
    \item[(D')] If a \emph{classical} agent $A$ knows a statement $s_1$ and also knows that $s_1$ implies statement $s_2$ then agent $A$ can conclude `I (A) know $s_2$'. 
\end{itemize}

\paragraph*{Comparing FR and GHZ--FR.} Comparing the FR and GHZ--FR assumptions, (U) and (S) are the same, but the GHZ--FR assumptions (C'), (D'), (Q') are weaker versions of the FR assumptions (C), (D), (Q), respectively (cf. \cref{fig:assumptions_FR_GHZ_FR}).
Namely, only \emph{classical agents} need to produce statements and reason, and accept statements classically communicated among them.\footnote{As described in \Cref{sec:stricter_FR}, a slight modification of the reasoning in the FR paradox can also produce a no-go theorem from these weaker assumptions. A similar modification of the reasoning in the FR paradox can be found in Ref.~\cite{healey2018quantum}.} This is an important weakening, as for example the Sleeping Beauty problem of Ref.~\cite{jones_thinking_2024} shows how FR's assumption (C) might be problematic already in a classical setup, undermining the motivation for this assumption (C) in addition to establishing that a violation of (C) is certainly not what is especially quantum about extended Wigner's friend paradoxes. Their argument does not apply to its weakened version (C') in the GHZ--FR paradox.

\paragraph*{Explaining the rephrasing of the GHZ--FR assumptions.}
We have argued that the GHZ--FR paradox requires assumptions (U), (S), (C'), (D'), and (Q'), modeled after those used in discussions of the FR paradox.
We now explain how they relate to the concepts used to describe the GHZ--FR agreement no-go theorem of \Cref{sec:GHZ_nogo_trust}; see also \cref{fig:assumptions_FR_GHZ_FR}.
Concerning the truth no-go theorem, we recall that the assumptions {\ClassicalT} and {\PersK} of the agreement no-go theorem are replaced by the \textit{a priori} stronger {\AOEfull} in the truth no-go theorem. 
\begin{itemize}[]
\item[(U)] is the equivalent of the GHZ--FR assumption {\universality}. 
\item[(S)] is captured in {\PersK}, and {\AOEfull}, allowing an agent to assign a single-valued outcome to performed measurements. 
\item[(C')] is a (slightly stronger) paraphrasing of {\ClassicalT}. 

\item[(D')] allows for classical agents to reason and is assumed implicitly in the GHZ--FR agreement no-go theorem. Classical agents are just like us, using elementary rules of logic such as the \textit{modus ponens} as expressed by assumption (D').

\item[(Q')] is captured in the conjunction of {\AOE} or {\PersK} and {\BornCAOE}. 
The assumption (Q') assumes the validity of the Born rule. But in fact, (Q') needs more, as it allows an agent $A$ to make statements about outcomes obtained in a measurement performed by another agent $B$, regardless of whether $A$ tests this Born rule prediction or not, captured by {\PersK} and {\BornCPersK}, respectively. 
\end{itemize}

\begin{figure}[h]
\tikzset{every picture/.style={line width=0.75pt}} 

\begin{tikzpicture}[x=0.75pt,y=0.75pt,yscale=-1,xscale=1]

\draw  [fill={rgb, 255:red, 46; green, 90; blue, 140 }  ,fill opacity=1 ] (383.5,257) -- (603.5,257) -- (603.5,285) -- (383.5,285) -- cycle ;
\draw  [fill={rgb, 255:red, 46; green, 90; blue, 140 }  ,fill opacity=1 ] (382.5,97) -- (601.5,97) -- (601.5,125) -- (382.5,125) -- cycle ;
\draw  [color={rgb, 255:red, 208; green, 2; blue, 27 }  ,draw opacity=1 ][fill={rgb, 255:red, 225; green, 121; blue, 134 }  ,fill opacity=1 ] (84.5,112.2) .. controls (84.5,103.36) and (140.91,96.2) .. (210.5,96.2) .. controls (280.09,96.2) and (336.5,103.36) .. (336.5,112.2) .. controls (336.5,121.04) and (280.09,128.2) .. (210.5,128.2) .. controls (140.91,128.2) and (84.5,121.04) .. (84.5,112.2) -- cycle ;
\draw  [color={rgb, 255:red, 208; green, 2; blue, 27 }  ,draw opacity=1 ][fill={rgb, 255:red, 225; green, 121; blue, 134 }  ,fill opacity=1 ] (84.5,164.2) .. controls (84.5,155.92) and (141.36,149.2) .. (211.5,149.2) .. controls (281.64,149.2) and (338.5,155.92) .. (338.5,164.2) .. controls (338.5,172.48) and (281.64,179.2) .. (211.5,179.2) .. controls (141.36,179.2) and (84.5,172.48) .. (84.5,164.2) -- cycle ;
\draw  [color={rgb, 255:red, 208; green, 2; blue, 27 }  ,draw opacity=1 ][fill={rgb, 255:red, 225; green, 121; blue, 134 }  ,fill opacity=1 ] (84.5,217) .. controls (84.5,205.95) and (96.7,197) .. (111.75,197) .. controls (126.8,197) and (139,205.95) .. (139,217) .. controls (139,228.05) and (126.8,237) .. (111.75,237) .. controls (96.7,237) and (84.5,228.05) .. (84.5,217) -- cycle ;
\draw  [color={rgb, 255:red, 208; green, 2; blue, 27 }  ,draw opacity=1 ][fill={rgb, 255:red, 225; green, 121; blue, 134 }  ,fill opacity=1 ] (84.5,270) .. controls (84.5,258.95) and (96.7,250) .. (111.75,250) .. controls (126.8,250) and (139,258.95) .. (139,270) .. controls (139,281.05) and (126.8,290) .. (111.75,290) .. controls (96.7,290) and (84.5,281.05) .. (84.5,270) -- cycle ;
\draw  [color={rgb, 255:red, 208; green, 2; blue, 27 }  ,draw opacity=1 ][fill={rgb, 255:red, 225; green, 121; blue, 134 }  ,fill opacity=1 ] (84.5,323) .. controls (84.5,311.95) and (96.7,303) .. (111.75,303) .. controls (126.8,303) and (139,311.95) .. (139,323) .. controls (139,334.05) and (126.8,343) .. (111.75,343) .. controls (96.7,343) and (84.5,334.05) .. (84.5,323) -- cycle ;
\draw  [color={rgb, 255:red, 208; green, 2; blue, 27 }  ,draw opacity=1 ][fill={rgb, 255:red, 225; green, 121; blue, 134 }  ,fill opacity=1 ] (282.5,217) .. controls (282.5,205.95) and (294.7,197) .. (309.75,197) .. controls (324.8,197) and (337,205.95) .. (337,217) .. controls (337,228.05) and (324.8,237) .. (309.75,237) .. controls (294.7,237) and (282.5,228.05) .. (282.5,217) -- cycle ;
\draw  [color={rgb, 255:red, 208; green, 2; blue, 27 }  ,draw opacity=1 ][fill={rgb, 255:red, 225; green, 121; blue, 134 }  ,fill opacity=1 ] (282.5,270) .. controls (282.5,258.95) and (294.7,250) .. (309.75,250) .. controls (324.8,250) and (337,258.95) .. (337,270) .. controls (337,281.05) and (324.8,290) .. (309.75,290) .. controls (294.7,290) and (282.5,281.05) .. (282.5,270) -- cycle ;
\draw  [color={rgb, 255:red, 208; green, 2; blue, 27 }  ,draw opacity=1 ][fill={rgb, 255:red, 225; green, 121; blue, 134 }  ,fill opacity=1 ] (282.5,323) .. controls (282.5,311.95) and (294.7,303) .. (309.75,303) .. controls (324.8,303) and (337,311.95) .. (337,323) .. controls (337,334.05) and (324.8,343) .. (309.75,343) .. controls (294.7,343) and (282.5,334.05) .. (282.5,323) -- cycle ;
\draw  [fill={rgb, 255:red, 46; green, 90; blue, 140 }  ,fill opacity=1 ] (383.5,307) -- (603.5,307) -- (603.5,335) -- (383.5,335) -- cycle ;
\draw  [fill={rgb, 255:red, 46; green, 90; blue, 140 }  ,fill opacity=1 ] (383.5,144) -- (604.5,144) -- (604.5,172) -- (383.5,172) -- cycle ;
\draw  [fill={rgb, 255:red, 46; green, 90; blue, 140 }  ,fill opacity=1 ] (383.5,213) -- (604.5,213) -- (604.5,241) -- (383.5,241) -- cycle ;
\draw [color={rgb, 255:red, 74; green, 144; blue, 226 }  ,draw opacity=1 ]   (337,112) -- (376,112) ;
\draw [shift={(379,112)}, rotate = 180] [fill={rgb, 255:red, 74; green, 144; blue, 226 }  ,fill opacity=1 ][line width=0.08]  [draw opacity=0] (8.93,-4.29) -- (0,0) -- (8.93,4.29) -- cycle    ;
\draw [color={rgb, 255:red, 74; green, 144; blue, 226 }  ,draw opacity=1 ]   (337,164) -- (373.05,157.53) ;
\draw [shift={(376,157)}, rotate = 169.82] [fill={rgb, 255:red, 74; green, 144; blue, 226 }  ,fill opacity=1 ][line width=0.08]  [draw opacity=0] (8.93,-4.29) -- (0,0) -- (8.93,4.29) -- cycle    ;
\draw [color={rgb, 255:red, 74; green, 144; blue, 226 }  ,draw opacity=1 ]   (337,217) -- (379.19,161.39) ;
\draw [shift={(381,159)}, rotate = 127.18] [fill={rgb, 255:red, 74; green, 144; blue, 226 }  ,fill opacity=1 ][line width=0.08]  [draw opacity=0] (8.93,-4.29) -- (0,0) -- (8.93,4.29) -- cycle    ;
\draw [color={rgb, 255:red, 74; green, 144; blue, 226 }  ,draw opacity=1 ]   (337,217) -- (376.05,224.44) ;
\draw [shift={(379,225)}, rotate = 190.78] [fill={rgb, 255:red, 74; green, 144; blue, 226 }  ,fill opacity=1 ][line width=0.08]  [draw opacity=0] (8.93,-4.29) -- (0,0) -- (8.93,4.29) -- cycle    ;
\draw [color={rgb, 255:red, 74; green, 144; blue, 226 }  ,draw opacity=1 ]   (337,270) -- (377,270) ;
\draw [shift={(380,270)}, rotate = 180] [fill={rgb, 255:red, 74; green, 144; blue, 226 }  ,fill opacity=1 ][line width=0.08]  [draw opacity=0] (8.93,-4.29) -- (0,0) -- (8.93,4.29) -- cycle    ;
\draw [color={rgb, 255:red, 74; green, 144; blue, 226 }  ,draw opacity=1 ]   (337,323) -- (377,323) ;
\draw [shift={(380,323)}, rotate = 180] [fill={rgb, 255:red, 74; green, 144; blue, 226 }  ,fill opacity=1 ][line width=0.08]  [draw opacity=0] (8.93,-4.29) -- (0,0) -- (8.93,4.29) -- cycle    ;

\draw (42,48) node [anchor=north west][inner sep=0.75pt]   [align=left] {\textbf{FR no-go theorem}};
\draw (304,48) node [anchor=north west][inner sep=0.75pt]   [align=left] {\textbf{GHZ--FR agreement (truth) no-go theorem}};
\draw (199,103) node [anchor=north west][inner sep=0.75pt]  [color={rgb, 255:red, 255; green, 255; blue, 255 }  ,opacity=1 ] [align=left] {(U)};
\draw (420,263) node [anchor=north west][inner sep=0.75pt]  [color={rgb, 255:red, 255; green, 255; blue, 255 }  ,opacity=1 ] [align=left] {Classical Agreement (AOE)};
\draw (386,104) node [anchor=north west][inner sep=0.75pt]  [color={rgb, 255:red, 255; green, 255; blue, 255 }  ,opacity=1 ] [align=left] {Universality of Quantum Theory};
\draw (222,230) node [anchor=north west][inner sep=0.75pt]  [font=\Large,rotate=-180] [align=left] {$\displaystyle < $};
\draw (222,336) node [anchor=north west][inner sep=0.75pt]  [font=\Large,rotate=-180] [align=left] {$\displaystyle < $};
\draw (223,283) node [anchor=north west][inner sep=0.75pt]  [font=\Large,rotate=-180] [align=left] {$\displaystyle < $};
\draw (198,155) node [anchor=north west][inner sep=0.75pt]  [color={rgb, 255:red, 255; green, 255; blue, 255 }  ,opacity=1 ] [align=left] {(S)};
\draw (100,208) node [anchor=north west][inner sep=0.75pt]  [color={rgb, 255:red, 255; green, 255; blue, 255 }  ,opacity=1 ] [align=left] {(Q)};
\draw (100,261) node [anchor=north west][inner sep=0.75pt]  [color={rgb, 255:red, 255; green, 255; blue, 255 }  ,opacity=1 ] [align=left] {(C)};
\draw (100,314) node [anchor=north west][inner sep=0.75pt]  [color={rgb, 255:red, 255; green, 255; blue, 255 }  ,opacity=1 ] [align=left] {(D)};
\draw (298,208) node [anchor=north west][inner sep=0.75pt]  [color={rgb, 255:red, 255; green, 255; blue, 255 }  ,opacity=1 ] [align=left] {(Q')};
\draw (298,261) node [anchor=north west][inner sep=0.75pt]  [color={rgb, 255:red, 255; green, 255; blue, 255 }  ,opacity=1 ] [align=left] {(C')};
\draw (298,314) node [anchor=north west][inner sep=0.75pt]  [color={rgb, 255:red, 255; green, 255; blue, 255 }  ,opacity=1 ] [align=left] {(D')};
\draw (402,312) node [anchor=north west][inner sep=0.75pt]  [color={rgb, 255:red, 255; green, 255; blue, 255 }  ,opacity=1 ] [align=left] {\textit{only classical agents reason}};
\draw (406,150) node [anchor=north west][inner sep=0.75pt]  [color={rgb, 255:red, 255; green, 255; blue, 255 }  ,opacity=1 ] [align=left] {Personal Knowledge (AOE)};
\draw (424,219) node [anchor=north west][inner sep=0.75pt]  [color={rgb, 255:red, 255; green, 255; blue, 255 }  ,opacity=1 ] [align=left] {Born Compatibility };

\end{tikzpicture}  \caption[Comparing the assumptions of the FR and GHZ--FR paradoxes]{Comparing the assumptions of the FR paradox to those of the GHZ--FR no-go theorems. An assumption of the FR no-go theorem being strictly stronger than a corresponding assumption of the GHZ--FR agreement no-go theorem is denoted by $>$. In red we compare the assumptions (U), (S), (Q), (C) and (D) of the FR no-go theorem to the rephrased GHZ--FR assumptions (U), (S), (Q'), (C') and (D'). How these rephrased assumptions correspond to the GHZ--FR assumptions of \Cref{sec:GHZ_FR_nogo}, which we have used throughout this work, is shown by the arrows between the assumptions in red circles and blue boxes. The assumptions of the agreement no-go theorem are listed, as the agreement no-go theorem is stronger than the truth one. In the truth no-go theorem, {\protect\PersK} and {\protect\ClassicalT} are replaced by the stronger {\protect\AOEfull}, as listed in brackets.} \label{fig:assumptions_FR_GHZ_FR} 
\end{figure}
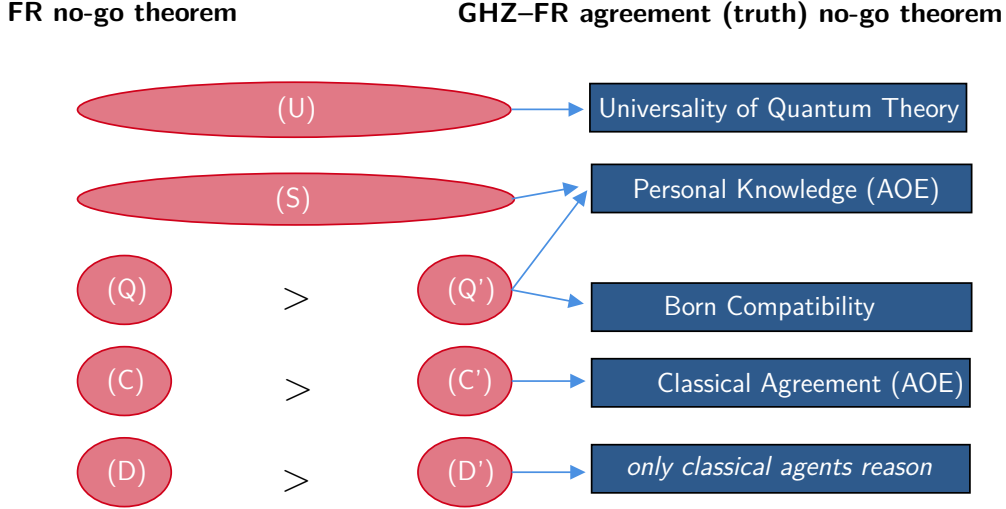

\section{Related work on the FR paradox} \label{sec:previous_work}
In this section, we will compare the GHZ--FR paradox to previous work on the FR paradox and protocols related to the GHZ--FR paradox.

\subsection{Previous responses to the FR paradox}
The FR no-go theorem~\cite{frauchiger2018quantum,vilasini2022general} forces us to refute one of its assumptions, and has been discussed extensively in the literature. 
We are not able to survey all relevant works; instead, we aim to provide a concise overview.
We classify previous responses to the FR paradox into three main categories.\footnote{Some works belong to more than one category, so we mention them more than once.} For more extended reviews we point the reader to Refs.~\cite{vilasini2022general,nurgalieva2020testing}.

\medskip

First, there is work that questions and re-analyses the assumptions used in the FR paradox~\cite{vilasini2022general,nurgalieva2018inadequacy,sudbery2019hidden,nurgalieva2020testing,montanhano2023contextuality,boge2019quantum,vilasini2019multi,fraser2020fitch,schmid2023review}.
Some implicit assumptions have been identified in Refs.~\cite{nurgalieva2018inadequacy,sudbery2019hidden}. 
A link between paradoxes in FR-like scenarios and multi-agent knowledge structures, epistemic logic~\cite{fagin2004reasoning,blackburn2001modal}, and contextuality has been investigated in Refs.~\cite{montanhano2023contextuality,boge2019quantum,nurgalieva2018inadequacy,vilasini2019multi,fraser2020fitch,corti2023logico,baltag2023logic}.

Other works claim that the FR paradox contains implicit assumptions that can be deemed invalid a priori, thereby questioning the relevance or novelty of the FR paradox and no-go theorem~\cite{aaronson2018s,healey2018quantum,araujo2018flaw,fortin2019wigner,drezet2018wigner}. 
Aaronson~\cite{aaronson2018s} argued in a blog post that the reasoning statements made by the friends are invalid at the end of the protocol as the brains of the friends have been `Hadamarded' in the process. 
It is not clear to us whether this argument claims that (i) only reasoning or (ii) also mere experiencing of the measurement result by the unitarily-modeled quantum observers is invalid. 
Case (i) has been criticized by del Rio and Renner~\cite{del2024reply}, and anyway does not apply to the GHZ--FR paradox, as only `classical' agents reason.
Case (ii) could correspond to rejecting {\universality}, but may also suggest a relativity of observed events, depending on its exact interpretation.
Healey~\cite{healey2018quantum} argues that an additional assumption of Intervention Sensitivity is required for reasoning about outcomes obtained by quantum agents,
and Drezet~\cite{drezet2018wigner} asserts that in a Bohrian interpretation reasoning by quantum agents may be problematic.
Vilasini and Woods~\cite{vilasini2022general} suggest that the arguments of Refs.~\cite{healey2018quantum,araujo2018flaw} also refer to the post-selection step, but this depends on the exact reading of these two references.
Criticisms of post-selection and reasoning by quantum agents do not apply to the GHZ--FR paradox as no post-selection or reasoning by quantum agents is needed.  
Ara\'ujo~\cite{araujo2018flaw} seems to argue that both collapse and no-collapse modeling of a measurement is needed for the FR argument. However, we note that the state-update rule is not needed~\cite{fortin2019wigner}, and only unitary modeling and the Born rule for predictions are used. Namely, each use of the Born rule in the FR involves only measurements that can be performed jointly.\footnote{We thank David Schmid and Yìlè Y\={\i}ng for discussions on this matter.} 
Fortin and Lombardi~\cite{fortin2019wigner} argue that the FR argument is either illegitimate or uninteresting and simply another proof of contextuality as agents make reasoning statements that correspond to different contexts when mapping the FR protocol to a Bell nonlocality scenario, but do not respond to how agents can then use quantum theory consistently in the presence of superobservers.

\medskip

Secondly, the FR paradox has been analysed from the viewpoint of specific interpretations of quantum theory~\cite{nurgalieva2020testing,lazarovici2019quantum,drezet2018wigner,sudbery2017single,sudbery2019hidden,di2021stable,cavalcanti2023consistency,gambini2019single,debrota2020respecting,losada2019frauchiger}, and arguments have been {put forward for which assumption of the FR paradox to refute.
The FR paradox has been studied in Bohmian and Bell--Bohmian theories in Refs.~\cite{lazarovici2019quantum,drezet2018wigner,sudbery2017single}, in a relational quantum theory setting in Ref.~\cite{di2021stable}, from a QBist perspective in Ref.~\cite{debrota2020respecting}, and in the consistent-histories interpretation in Ref.~\cite{losada2019frauchiger}.
An overview of the FR paradox in different interpretations of quantum mechanics can be found in Ref.~\cite{nurgalieva2020testing}. 
\begin{itemize}
\item The (Bell--)Bohmian analyses of the FR paradox~\cite{lazarovici2019quantum,drezet2018wigner,sudbery2017single,sudbery2019hidden} argue that the 
agents apply the Born rule (Q) wrongly, as they do not take into account the full state, \ie the pilot wavefunction.
But still, one would like the agents to be able to make correct predictions, even if they cannot take an outsider's view. 
We have provided a possible way to phrase their predictions correctly in \Cref{sec:ways_to_resolve_trust} under `Reject Born Compatibility'.

\item The relational and QBist perspectives propose to refute FR's consistency assumption (C),
allowing for different agents to assign different truth values to statements~\cite{nurgalieva2020testing,di2021stable,debrota2020respecting}. One may argue that also the consistent-histories approach to the FR~\cite{losada2019frauchiger} refutes (C)~\cite{nurgalieva2020testing}, as inferences are made between inconsistent sets of histories.
When refuting (C), one would still like to know when one agent can trust another agent's statement or reasoning. 
For example, in the GHZ--FR paradox, one can imagine that Ursula's, Valentina's, and Wigner's measurements are all performed by the same agent (sequentially in their proper time), who then makes statements similar to those of Ursula, Valentina, and Wigner: must this agent then not trust their own statement? 
In the QBist approach to FR by DeBrota, Fuchs, and Schack~\cite{debrota2020respecting}, it is for example argued that when to trust another agent's statement can be decided on a case-by-case basis.
Our GHZ--FR no-go theorem shows that even classical communication cannot be trusted when allowing different truth values for different agents, and we have argued that perhaps it is easier to instead restrict the validity of statements using the Born rule for predictions, while holding on to consistency between agents. 
The fact that no quantum agents need to reason in the GHZ--FR paradox is crucial for this argument, and poses sharper restrictions on building consistent reasoning principles.\footnote{Whether the relative-facts approach of Ref.~\cite{di2021stable}, which refutes (C) and argues to not believe the reasoning of the quantum agents, resolves the paradoxes presented in this chapter or not depends on its exact interpretation. For a more elaborate discussion see Appendix E of Ref.~\cite{walleghem2024strong}.} 

\item Other works~\cite{relano2018decoherence,relano2020decoherence,kastner2020unitary,zukowski2021physics,gambini2019single} argue for refuting the existence of superobservers because of decoherence or absolute collapse, rendering the FR protocol -- involving superobservers -- impossible, even in principle. 
However, decoherence may be part of an explanation of the quantum-to-classical transition, but it is not a full answer to the measurement problem~\cite{schlosshauer2019quantum,schlosshauer2005decoherence,schlosshauer2007quantum}. Absolute collapse requires an explanation of how or why the collapse happens, as in spontaneous collapse theories~\cite{ghirardi1985model,ghirardi1986unified} or gravity-induced collapse~\cite{penrose1996gravity,diosi1987universal}. 
Experiments may shed light on whether collapse theories, which have different dynamics than ordinary quantum theory, are likely to be correct~\cite{bassi2003dynamical,forstner2020nanomechanical,arnquist2022search,donadi2021novel,carlesso2022present}.
\end{itemize}

\medskip

Thirdly, there is some work towards responding to Renner's challenge of building a consistent set of reasoning principles for quantum theory
that evades contradictions like the FR paradox~\cite{renes2021consistency,narasimhachar2020agents,polychronakos2022quantum,vilasini2022general}. 
Here we note that Vilasini and Woods~\cite{vilasini2022general} propose that one be explicit about which Heisenberg cuts are used in each statement concerning outcomes of performed measurements. This leaves open the question of the physical meaning of Heisenberg cuts, and why statements should be formulated as dependent on them? Is it merely a choice of what to model inside and outside the quantum theory? If so, then still desirably no contradictions about the outcome values as in the (GHZ--)FR paradox should arise when making different choices, if it is merely a choice. However, a resolution à la {\ROE} proposes a reason of why such mentioning would be necessary: it refers to the statement being conditional on physically obtaining (a copy of) the outcome record(s) one is talking about; a prediction by agent $A$ is meant for when $A$ actually obtains those outcome records (and the protocol be performed truthfully per agent $A$'s knowledge).}
Moreover, we also note that the GHZ--FR no-go theorem invalidates the set of reasoning principles from Ref.~\cite{polychronakos2022quantum}. For a proof see Appendix E.2 of Ref.~\cite{walleghem2024strong}.

\medskip

Our work belongs in the first category, as it provides a stronger no-go theorem than the FR. It also touches on the third category by suggesting ways to resolve paradoxes involving superobservers in quantum theory.

\subsection{Related work on the GHZ--FR paradox}
A protocol similar to the one underlying the GHZ--FR paradox has been proposed by {\.Z}ukowski and Markiewicz~\cite{zukowski2021physics} and by Leegwater~\cite{leegwater2022greenberger}. While the measurements performed in the protocols (without reasoning steps) in Refs.~\cite{zukowski2021physics,leegwater2022greenberger} are the same,
the exact way in which a contradiction is obtained differs, as do the ensuing analyses and no-go theorems. 
In Ref.~\cite{zukowski2021physics}, an explicit motivation for how the different measurement contexts that lead to a contradiction are chosen and considered together is not provided. 
In Ref.~\cite{leegwater2022greenberger} this motivation comes from simultaneity in special relativity. 
However, in cases where the motivation from special relativity of Ref.~\cite{leegwater2022greenberger} cannot be applied, the GHZ--FR paradox still yields a contradiction; for more detail see Appendix F of Ref.~\cite{walleghem2024strong}.

Going beyond Ref.~\cite{leegwater2022greenberger}, we investigate the relationship of the GHZ--FR protocol to the FR paradox more thoroughly, and expose the hierarchy of their underlying nonlocality models.
Unlike Refs.~\cite{zukowski2021physics,leegwater2022greenberger} our agreement no-go theorem requires no absolute outcomes, due to the reasoning part of our protocol being different. 
This important weakening allowed us to propose the {\ROE} resolution as replacing {\AOEfull} in \Cref{sec:GHZ_nogo_trust,sec:possible_resolution}.
Unlike these works~\cite{zukowski2021physics,leegwater2022greenberger}, we move away from having a bird's-eye view on the protocol as an outsider, and instead consider the viewpoints from observers inside the protocol who try to make predictions and reason, as was the goal in the original FR paradox. To relate to the notion of \textit{Born Inaccessible correlations}\footnote{Born Inaccessible Correlations is defined in Ref.~\cite{schmid2023review} as: ``It requires that when two measurements are made in parallel, the two outcomes always arise with a joint frequency given by the Born rule, even if no single observer could access both outcomes, even in principle.''} of Ref.~\cite{schmid2023review}, we acknowledge that there will always be correlations inaccessible to any agent. However, we assign the label `inaccessible' as we can take the outsider's bird's-eye view on the protocol; it might be impossible for agents inside to have this knowledge in all situations. Were we to make a no-go theorem using the notion of Born Inaccessible correlations, we would find a no-go theorem similar to the GHZ--FR truth no-go theorem, replacing Born Compatibility with Born Inaccessible correlations.
For a further discussion of the works~\cite{leegwater2022greenberger,zukowski2021physics} see Appendix F in Ref.~\cite{walleghem2024strong}. 

A similar relational resolution as {\ROE} in the context of Relational Quantum Mechanics~\cite{rovelli1996relational,rovelli2018space}
was discussed by Cavalcanti, Di Biagio and Rovelli~\cite{cavalcanti2023consistency} for a scenario involving the same measurement protocol as the GHZ--FR paradox, based on Refs.~\cite{zukowski2021physics,lawrence2023relative}. 
Our resolution per {\ROE} accords with their proposed resolution, except that our proposal does not invoke the existence of cross-perspective links~\cite{adlam2022information}.
Furthermore, as noted by Cavalcanti and Wiseman in Ref.~\cite{cavalcanti2021implications}, to resolve the Local Friendliness no-go theorem, by giving up on {\AOE} for a notion of relative events one can maintain compatibility with Leibniz's principle.

\subsection{Relation to the Local Friendliness no-go theorem} 
The Local Friendliness no-go theorem~\cite{bong2020strong,haddara2022possibilistic,cavalcanti2021implications,schmid2023review} is based on a protocol similar to the FR's but is phrased in a device-independent fashion. 
It uses the Born rule only for outcomes that are obtained by us (\ie that are available after the protocol),
but in addition to {\AOE} it also requires a notion of free choices satisfying `Local Agency', a locality constraint not invoking hidden variables as in Bell's local causality.
If in the GHZ--FR paradox of \Cref{sec:FR_refining}
we give the superobservers a choice between performing their measurement or asking their respective observers for their outcomes, we obtain a Local Friendliness scenario~\cite{bong2020strong,haddara2022possibilistic}. 
In \Cref{chap:part1_EWF_noncontextuality} we clarify how also contextuality scenarios may be used to produce a Local Friendliness no-go theorem.

\section{Conclusion} \label{sec:conclusion_chap3}
We conclude by summing up the main results regarding the GHZ--FR paradox and outline some future research directions.

\subsection{Summary}
Motivated by the question of what the FR paradox~\cite{frauchiger2018quantum} implies for quantum foundations, we presented a similar but strictly stronger paradox, the GHZ--FR paradox, based on the strongly contextual GHZ--Mermin model~\cite{greenberger1989going,mermin1990simple}.
Outlining carefully the required assumptions, we obtained two no-go theorems of increasing strength, which we named the GHZ--FR truth and agreement no-go theorems.
The easiest to grasp is the GHZ--FR truth no-go theorem, whose assumptions are summarized in \Cref{table:GHZ_FR_truth_assumptions}. 
The agreement no-go theorem weakens {\AOEfull} replacing it by {\PersK} and {\ClassicalT}; see \Cref{table:AOE_trust_truth}: instead of \emph{absolute} outcome assignments to performed measurements, it posits outcome assignments that may a priori differ per agent, subject to the condition that classically communicating agents must agree on their outcome assignments.

\begin{table}[h]
\resizebox{\textwidth}{!}{%
\begin{tabular}{|
>{\columncolor[HTML]{F0EEEC}}l 
>{\columncolor[HTML]{F0EEEC}}l |}
\hline
\multicolumn{2}{|l|}{\cellcolor[HTML]{656565}{\color[HTML]{FFFFFF} }}                                                       \\
\multicolumn{2}{|l|}{\multirow{-2}{*}{\cellcolor[HTML]{656565}{\color[HTML]{FFFFFF} \textbf{GHZ--FR truth no-go theorem}}}} \\ \hline
                                           &                                                                                                 \\
Absoluteness of Observed Events &
  \begin{tabular}[c]{@{}l@{}}Every performed measurement has an absolute, single-valued outcome. \end{tabular}\\ (AOE) &
          \\
Born Compatibility &
  \begin{tabular}[c]{@{}l@{}} Outcomes assigned (per \text{AOE}) to a set of measurements that an agent $A$ \\ believes she can gather jointly in a memory  must be compatible with $A$'s use \\ of the possibilistic Born rule.
 \end{tabular} \\  &
                           \\
Universality of Quantum Theory             & Superobservers exist (in principle) and quantum theory is correct.             \\
                                                       
  \hline
\end{tabular}%
}
\caption[Assumptions of the GHZ--FR truth no-go theorem]{Assumptions of the GHZ--FR truth no-go theorem; more precisely stated assumptions can be found in \Cref{sec:GHZ_nogo}.}
 \label{table:GHZ_FR_truth_assumptions}
\end{table}

\begin{table}[h]
\resizebox{\textwidth}{!}{%
\begin{tabular}{|
>{\columncolor[HTML]{F0EEEC}}l 
>{\columncolor[HTML]{F0EEEC}}l 
>{\columncolor[HTML]{F0EEEC}}l |}
\hline
\cellcolor[HTML]{656565}{\color[HTML]{FFFFFF} }                                                       & \multicolumn{2}{l|}{\cellcolor[HTML]{656565}{\color[HTML]{FFFFFF} }}                                                                                                                                             \\
\multirow{-2}{*}{\cellcolor[HTML]{656565}{\color[HTML]{FFFFFF} \textbf{GHZ--FR truth no-go theorem}}} & \multicolumn{2}{l|}{\multirow{-2}{*}{\cellcolor[HTML]{656565}{\color[HTML]{FFFFFF} \textbf{GHZ--FR agreement no-go theorem}}}}                                                                                       \\ \hline
\multicolumn{1}{|l|}{\cellcolor[HTML]{F0EEEC}}                                                        & Personal Knowledge & \begin{tabular}[c]{@{}l@{}}Every performed measurement can be assigned a \\ single-valued outcome by an agent who believes \\ they are able to gather that outcome.\end{tabular} \\
\multicolumn{1}{|l|}{\multirow{-2}{*}{\cellcolor[HTML]{F0EEEC}Absoluteness of Observed Events}}       &                    &                                                                                                                                                                                             \\
\multicolumn{1}{|l|}{\cellcolor[HTML]{F0EEEC}}                                                        & Classical Agreement    & \begin{tabular}[c]{@{}l@{}}The outcome assignments of classical agents 
 who \\ communicate classically must agree on their overlap. \end{tabular}                                                      \\ \hline
\end{tabular} } 
\caption[Assumptions of the GHZ--FR agreement no-go theorem]{The GHZ--FR agreement no-go theorem is based on the same assumptions as the GHZ--FR truth no-go theorem, but with Personal Knowledge and Classical Agreement replacing Absoluteness of Observed Events.}
\label{table:AOE_trust_truth}
\end{table}

Pursuing the route of rejecting {\AOE} or {\BornCAOE}, as discussed in \Cref{sec:GHZ_nogo_trust}, the agreement no-go theorem suggests a principle that we call {\BornP}, defined in \Cref{sec:possible_resolution}, meaning that an agent $A$ can use the Born rule only to formulate predictions that condition on $A$ actually learning about all involved outcomes.
This principle can be seen as an agent-based extension of Peres's dictum~\cite{peres1978unperformed} `Unperformed experiments have no results':
\begin{quote}
   `Unperformed experiments have no results, and unknown results have no values'.
\end{quote}

One may view this as a natural quantum-vs-classical statement.
Therefore, as a possible resolution to extended Wigner's friend paradoxes we propose {\BornP} as a principle governing agents' use of quantum theory in a world where {\universality} holds.
In a relational framework, rejecting {\AOE}, {\BornP} can be implemented by having single-valued outcomes only defined relative to the experimenter and other agents who learn about the outcome, which we named {\ROE} in \Cref{sec:GHZ_nogo_trust}. 
In a hidden-variable framework, rejecting {\BornCAOE}, one may instead have hidden variables determining absolute, single-valued outcomes of all performed measurements but agents must follow {\BornP} for correct predictions of the Born rule.
Of course, one may also question the existence of superobservers, or simply allow for classically communicating agents to have inconsistent truth assignments. The latter, however, seems impractical, as we argued in \Cref{sec:ways_to_resolve_trust,sec:previous_work}.

\subsection{Outlook}
To conclude, our work opens up various directions for future work.
In upcoming work, we will discuss the implications of the GHZ--FR paradox for the measurement problem, for different interpretations of quantum theory, and how it suggests a resolution to Renner's challenge of providing a consistent set of reasoning rules in quantum settings.
Another intriguing question is what implications the GHZ--FR paradox and other extended Wigner's friend paradoxes have for the discussion on the epistemic or ontic nature of quantum states~\cite{harrigan2010einstein,pusey2012reality,hardy2013quantum,leifer2014quantum} or ontologies of quantum mechanics.
A deeper investigation of the relationship between {\ROE} or {\BornP} and other proposed resolutions of extended Wigner's friend paradoxes,
including bubbles and quantum influences~\cite{ormrod2024quantum,cavalcanti2021view}, the relational approach of `fact-nets'~\cite{martin2023fact}, and quantum reference frames~\cite{de_la_Hamette_2020,giacomini2019quantum} is left for future work.

Another interesting avenue to explore is to combine
Wigner's friend with other nonclassical quantum features such as KS noncontextual models that are not nonlocality models, steering~\cite{wiseman2007steering,schrodinger1935discussion}, incompatibility~\cite{guhne2023colloquium}, generalized contextuality~\cite{spekkens2005contextuality,schmid2020structure}, coherence~\cite{streltsov2017colloquium,galvao2020quantum}, etc.
We do so for Kochen--Specker, generalized noncontextuality and the Pusey--Barrett--Rudolph argument~\cite{pusey2012reality} in the next chapters.

\ifSubfilesClassLoaded{ 
  \printbibliography[heading=bibintoc, title=References]
  }{}
\end{document}

\ifSubfilesClassLoaded{ 
	\setcounter{chapter}{1} 
	\maketitle \tableofcontents \listoftodos
	}{}

\chapter[Extended Wigner's friend arguments and noncontextuality]{Extended Wigner's friend arguments and noncontextuality}\label{chap:part1_EWF_noncontextuality}

As mentioned in \Cref{sec:WF}, previous extended Wigner's friend (EWF) arguments are primarily based on Bell nonlocality. 
In the next four sections, we present links between extended Wigner's friend arguments and other notions of nonclassicality, such as Kochen--Specker (KS) and generalized noncontextuality, and introduce a class of new EWF arguments. 

Specifically, in \Cref{sec:EWF_on_KS_not_nonlocal}, we present an extended Wigner's friend argument using a single system, based on KS noncontextuality, by suitably generalizing an assumption used in previous EWF arguments. 
Next, in \Cref{sec:LFasKSNC}, we focus on the Local Friendliness (LF) no-go theorem. We show how constraints on scenarios, arising from the Local Friendliness assumptions, are mathematically related to Kochen--Specker noncontextuality requirements on the relevant measurements. 
This allows to find many more inequalities that the LF polytope (i.e. the operational correlations fulfilling the Local Friendliness assumptions) satisfies.
Moreover, in \Cref{sec:KSNC_to_LF} we construct scenarios violating Local Friendliness based solely on Kochen--Specker contextuality. 

These findings raise the question of whether a yet more general notion of nonclassicality, generalized contextuality, can be used to derive EWF arguments. 
In \Cref{sec:EWF_based_on_generalised_contextuality}, we answer this question affirmatively, under the condition that one generalizes the Local Friendliness assumptions to a more spacetime-independent form. 
Namely, the LF assumptions always refer to a background spacetime where one assumes the causality of Special or General Relativity to hold. However, one might be interested in an EWF no-go theorem that does not make use of such background, for example when working with background-independent quantum gravity theories, or to generalize to cases where the precise locations are unknown. 
By doing so, we uncover that behind the intuition of Local Friendliness and EWF arguments more generally is a principle of no fine-tuning, a well-known concept in quantum foundations, under the assumption that the Absoluteness of Observed Events holds.

More specifically, the Local Friendliness assumptions consist of two main parts: Absoluteness of Observed Events and Local Agency. Local Agency captures a notion of locality, that each choice can only be correlated with events in its future. 
Local Agency can be motivated from the postulate that no physical system can (locally) travel faster than the speed of light in special and general relativity.
However, it can also be motivated from the empirical finding of no-signaling. 
No-signaling here means that no physical message (a choice $x$) can be sent faster than the speed of light; all events $a,b,...$ outside of $x$'s future lightcone must be uncorrelated with $x$.
Local Agency can be motivated by requiring that no-signaling still holds for those joint correlations existing by Absoluteness of Observed Events, even when it cannot be verified by a single agent.
Our generalization, named Operational Agency, captures the idea that operational equivalences, i.e. empirically observed independencies of choices, must still hold for the joint correlations that exist by Absoluteness of Observed Events even when the relevant events are not brought together (i.e. not observed by a single agent).
In other words, it expresses a notion of no-conspiracy, or no fine-tuning: observed independencies must remain true even when unobserved (but whose correlations exist when assuming Absoluteness of Observed Events).
Quantum theory violates Local Friendliness and Operational Friendliness, assuming its universality in that superobservers exist.

Beyond simply providing additional proofs of Local Friendliness, these works contribute by identifying the broader nonclassical structure underlying such (LF) results and distinguishing essential assumptions from features specific to the original LF construction. Moreover, they situate the LF assumptions within established notions of noncontextual classical explanation, clarify their dependence on background spacetime structure and extend EWF arguments to scenarios outside the standard multipartite Bell framework. 

\vspace{0.5cm}

Rather than providing a single discussion at the end, we will provide brief discussions in the different sections of this chapter on the relevant notions introduced, when necessary.

\vspace{0.5cm}

As we are concerned with the Local Friendliness no-go theorem in this chapter and the next, we follow the conventions in literature and introduce a notational difference between empirical probability distributions that can be obtained by a single agent from the collected data, denoted by $\wp$, and theoretical probability distributions required to exist given certain assumptions, but potentially not observable by any agent, denoted by $p$.

\Cref{sec:EWF_on_KS_not_nonlocal} is based on Ref.~\cite{walleghem2023extended}, \Cref{sec:KSNC_to_LF,sec:LFasKSNC} on Ref.~\cite{walleghem2024connecting} and \Cref{sec:EWF_based_on_generalised_contextuality} on Ref.~\cite{walleghem2025extendedwignersfriendnogo}.

\section[An extended Wigner's friend paradox based on KS contextuality]{An extended Wigner's friend paradox based on KS contextuality} \label{sec:EWF_on_KS_not_nonlocal}
Thus far, we have seen how EWF arguments are based on Bell nonlocality models. As KS contextuality can be seen as a generalisation of Bell nonlocality, one can ask whether one can construct an EWF argument that involves measurements only on a single system based on KS contextuality. 
In this subsection we answer this question affirmatively, and produce an EWF argument based on the 5-cycle KCBS contextual model~\cite{klyachko2008simple}.

\subsection{Assumptions}

A number of assumptions we consider here are common to most EWF arguments, the first two being the Universality of Quantum Theory and Absoluteness of Observed Events~\cite{bong2020strong,haddara2022possibilistic,cavalcanti2021implications}, introduced in \Cref{chap:Q_prelims,chap:partI_refinedFR}, which we repeat here. 

\begin{assumption}[Universality of Quantum Theory] \label{assumption:Universality_chap4}
A sealed lab in which an observer performs a measurement on some system can be described as a closed quantum system that evolves unitarily, upon which a superobserver can apply arbitrary quantum operations, including measurements. 
\end{assumption}

\begin{assumption}[Absoluteness of Observed Events (AOE)] \label{assumption:AOE_chap4}
An observed event is an absolute single event, not relative to anything or anyone. 
\end{assumption}

Another standard assumption for EWF no-go theorems is that the outcome for any single measurement (or outcomes for measurements done in parallel, if those outcomes are jointly observable) occurs at frequencies obeying the Born rule, in accordance with operational quantum theory. Here we only need the \emph{possibilistic Born rule}, demanding that such outcomes never occur if the Born rule assigns probability 0 to them, and that otherwise, they sometimes occur. 
Often this assumption is not stated explicitly, as it simply concerns the experimental validity of quantum theory.

We will make use of all the assumptions introduced above. Our no-go theorem also makes use of one novel assumption called \emph{Commutation Irrelevance}, which will be introduced in due time.\newline

\subsection{The $5$-cycle noncontextuality scenario} \label{sec:5-cycle_paper1}
We briefly recap the 5-cycle noncontextuality no-go theorem from Ref.~\cite{cabello2013simple,klyachko2008simple,santos2021conditions}, as presented in \Cref{sec:recap_KSNC}, and provide a quantum realization upon which our EWF argument will be based. 
Recall that a Kochen--Specker noncontextual assignment~\cite{budroni2022kochenspecker} for a set of measurements is a deterministic assignment of outcomes to those measurements, such that the assignment for each measurement is independent of which other compatible measurements are performed jointly with it.

Consider five binary-outcome measurements $\{M_i\}_{i\in\{1,2,...,5\}}$, where the pairs $\{(1,2)$, $(2,3)$, $(3,4)$, $(4,5)$, $(5, 1 )\}$ are jointly measurable, with compatibility graph of \cref{fig:NC_5cycle} and outcome probabilities satisfying
\begin{subequations} \label{eq:5_cycle_12+23_34+45}
\begin{align}
 \wp(1,1|1,2)=0, \label{eq:5_cycle_12} \\
 \wp(0,0|2,3)=0, \label{eq:5_cycle_23} \\
    \wp(1,1|3,4) =0, \label{eq:5_cycle_34} \\
    \wp(0,0|4,5)=0, \label{eq:5_cycle_45}
    \end{align}
\end{subequations}
and
\begin{align}   \label{eq:01_box_not0}
            \wp(0,1|5,1) \neq 0.
    \end{align}
A deterministic KS assignment of the outcome of measurement $M_i$ as $m_i$, Eq.~\eqref{eq:5_cycle_12+23_34+45} satisfies 
\begin{equation}
\label{eq:mcontr}
   m_1=1 \Rightarrow m_2=0 \Rightarrow m_3=1 \Rightarrow m_4=0 \Rightarrow m_5=1.
\end{equation}
However, \cref{eq:01_box_not0} tells us that in some runs of the protocol, the outcome of $M_5$ is $0$ and the outcome of $M_1$ is $1$. Thus, there is no Kochen--Specker noncontextual assignment for these runs. 

\begin{flushleft}  
\begin{figure}[]
\begin{center}
\tikzset{every picture/.style={line width=0.75pt}} 

\begin{tikzpicture}[x=0.75pt,y=0.75pt,yscale=-1,xscale=1]

\draw   (179.28,158.07) -- (163.08,208.74) -- (109.88,208.98) -- (93.21,158.46) -- (136.11,127) -- cycle ;
\draw  [fill={rgb, 255:red, 251; green, 214; blue, 219 }  ,fill opacity=1 ] (129.86,127.1) .. controls (129.81,123.64) and (132.56,120.8) .. (136.01,120.75) .. controls (139.47,120.7) and (142.3,123.46) .. (142.36,126.91) .. controls (142.41,130.36) and (139.65,133.2) .. (136.2,133.25) .. controls (132.75,133.3) and (129.91,130.55) .. (129.86,127.1) -- cycle ;
\draw  [fill={rgb, 255:red, 251; green, 214; blue, 219 }  ,fill opacity=1 ] (86.97,158.56) .. controls (86.91,155.11) and (89.67,152.27) .. (93.12,152.21) .. controls (96.57,152.16) and (99.41,154.92) .. (99.46,158.37) .. controls (99.52,161.82) and (96.76,164.66) .. (93.31,164.71) .. controls (89.86,164.77) and (87.02,162.01) .. (86.97,158.56) -- cycle ;
\draw  [fill={rgb, 255:red, 251; green, 214; blue, 219 }  ,fill opacity=1 ] (103.63,209.07) .. controls (103.58,205.62) and (106.34,202.78) .. (109.79,202.73) .. controls (113.24,202.68) and (116.08,205.43) .. (116.13,208.89) .. controls (116.18,212.34) and (113.43,215.18) .. (109.98,215.23) .. controls (106.52,215.28) and (103.68,212.53) .. (103.63,209.07) -- cycle ;
\draw  [fill={rgb, 255:red, 251; green, 214; blue, 219 }  ,fill opacity=1 ] (156.83,208.83) .. controls (156.77,205.38) and (159.53,202.54) .. (162.98,202.49) .. controls (166.43,202.44) and (169.27,205.19) .. (169.33,208.64) .. controls (169.38,212.1) and (166.62,214.94) .. (163.17,214.99) .. controls (159.72,215.04) and (156.88,212.28) .. (156.83,208.83) -- cycle ;
\draw  [fill={rgb, 255:red, 251; green, 214; blue, 219 }  ,fill opacity=1 ] (173.03,158.17) .. controls (172.98,154.72) and (175.74,151.88) .. (179.19,151.82) .. controls (182.64,151.77) and (185.48,154.53) .. (185.53,157.98) .. controls (185.59,161.43) and (182.83,164.27) .. (179.38,164.32) .. controls (175.93,164.37) and (173.09,161.62) .. (173.03,158.17) -- cycle ;

\end{tikzpicture}  
\caption[The compatibility graph of the 5-cycle scenario]{The compatibility graph of the 5-cycle scenario. The pairs $\{(1,2),(2,3),(3,4),$ $(4,5),(5, 1 )\}$ are jointly measurable.
} \label{fig:NC_5cycle} 
\end{center}
\end{figure}
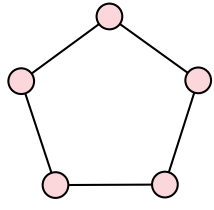
\end{flushleft}

The set of correlations in Eq.~\eqref{eq:5_cycle_12+23_34+45} and Eq.~\eqref{eq:01_box_not0} has a quantum realization as follows.
Consider a qutrit prepared in the state \begin{equation}\label{eq:eta}
    |\eta \rangle := \sqrt{\frac{1}{3}} (|0\rangle+\vert 1\rangle+\vert 2\rangle)\equiv \sqrt{\frac{1}{3}} (1,1,1)^T,
\end{equation} where $T$ denotes transposition, and define measurements 
\begin{equation}
\label{eq:M}
M_i:=\{\vert v_i\rangle \langle v_i \vert ,\mathbb{1}-\vert v_i\rangle \langle v_i \vert \},
\end{equation}
with the states $|v_i \rangle$ defined as \begin{equation}\label{eq: KCBS measurements}
    \begin{split}
        &|v_1 \rangle = \sqrt{\frac{1}{3}} (1,-1,1)^T, \\
        &|v_2 \rangle = \sqrt{\frac{1}{2}} (1,1,0)^T, \\
         &|v_3 \rangle = (0,0,1)^T, \\
        &|v_4 \rangle = (1,0,0)^T, \\
        &|v_5 \rangle = \sqrt{\frac{1}{2}} (0,1,1)^T.
    \end{split}
\end{equation} 
We take outcome $0$ of a given measurement to correspond to $\mathbb{1}-\vert v_i\rangle \langle v_i \vert $ and outcome $1$ to correspond to $\vert v_i \rangle \langle v_i \vert $. In this case, Eqs.~\eqref{eq:5_cycle_12}-\eqref{eq:5_cycle_45} are satisfied  (regardless of the state that the measurements act on) 
while $\wp(0,1|5,1) = \big|  \langle \eta | (\mathbb{1}-|v_5\rangle \langle v_5 |) |v_1 \rangle \big|^2 =  1/9\neq 0$, leading to Eq.~\eqref{eq:01_box_not0}. Hence, quantum theory is not consistent with noncontextuality.

\subsection{An EWF scenario with measurements on a single system} \label{sec:EWF_single-system_scenario}

We now leverage these 5-cycle correlations to construct an EWF no-go theorem. The argument involves five friends (denoted $F_1$, ..., $F_5$) who perform the five measurements $M_i$ defined in \cref{eq:M} sequentially on a single system $S$ prepared in the state $\ket{\eta}$ of \cref{eq:eta}, and a superobserver Wigner who undoes the first three of these measurements at particular times during the protocol.

By the assumption of Universality of Quantum Theory, each observer can be modeled as a quantum system---and indeed, as a qubit, since only two orthogonal (coarse-grained) states of the observer are relevant to the argument.
If each measurement $M_i$ is performed in a sufficiently isolated environment, Universality of Quantum Theory further implies that it can be described by a unitary. In particular, if the measurements are performed in a manner that has minimal disturbance on system $S$, then each measurement $M_i$ can be modeled by a CNOT gate $U_{M_i}$ between the system $S$ and the agent $A_i$, namely,{ 
\begin{align}
\label{eq_uni}
   U_{M_i}&:={\mathrm CNOT}_{SF_i} \\ \nonumber
   &=(\mathbb{1}-\ket{v_i}\bra{v_i})_{\mathrm S} \otimes\mathbb{1}_{\mathrm F_i}+   \ket{v_i}\bra{v_i}_{\mathrm S} \otimes ( \ket{1}\bra{0} + \ket{0}\bra{1})_{\mathrm F_i}.
\end{align}
}
where $\ket{0}_{F_i}$ is the coarse-grained state of $F_i$ having observed outcome 0 and $\ket{1}_{F_i}$ is for having observed outcome 1. We follow a standard abuse of notation by letting $\ket{0}_{F_i}$ also denote the coarse-grained `ready' state of the observer---that is, the initial state of each observer (prior to the measurement).

The superobserver Wigner is assumed to have perfect quantum control over the joint system of $F_1$, $F_2$, $F_3$, and $S$. In particular, we assume he has the (extreme) technological capabilities to implement the inverse of the unitary operations for the first three friends' measurements.

\begin{figure}[htb!]
\begin{center}
\includestandalone[width=0.45\columnwidth]{Figures/5cycleEWF1}
    \caption[Schematic depiction of an EWF scenario on a single system using 5-cycle contextuality]{\textbf{Schematic depiction of the scenario of \Cref{sec:EWF_single-system_scenario}}. Five friends $F_1, \dots, F_5$ perform the five measurements from a 5-cycle noncontextuality no-go theorem on a system $S$, while a superobserver Wigner undoes the first three of these measurements at particular times in between.  Each measurement $M_i$ is modeled unitarily as $U_{M_i}$, following the assumption of Universality of Quantum Theory. The unitaries from $U_{M_2}$ to $U^{\dagger}_{M_3}$ are collectively denoted as $U$. In each measurement $M_i$, agent $F_i$ obtains an absolute outcome $f_i$, following the assumption of Absoluteness of Observed Events. Here the three small horizontal lines \rotatebox{90}{\begin{quantikz} \ground{} \end{quantikz}} on a vertical line (a wire) means the system corresponding to the wire can be discarded, i.e. no more operations must be performed on it.}    \label{scenario}
    \end{center}
\end{figure}
The exact sequence of operations is shown in the unitary circuit representation of the protocol in Fig.~\ref{scenario}.
Specifically, $M_1$ is the first measurement on $S$, followed by $M_2$, after which the superobserver undoes the first measurement by applying the unitary $U_{M_{1}}^{\dagger}$. Then $M_3$ is performed, followed by the undoing $U_{M_{2}}^{\dagger}$, followed by $M_4$, followed by the undoing $U_{M_{3}}^{\dagger}$, followed by $M_5$.

From Absoluteness of Observed Events, each agent $A_i$ observes an absolute outcome, denoted $f_i$, during their measurement $M_i$. We label the $\ket{v_i}\bra{v_i}$ outcome by $f_i=1$ and the other by $f_i=0$. 

\subsection{No-go theorem}

Measurements $M_1$ and $M_2$ are done immediately in sequence without any processes occurring in between, so the joint outcome $(f_1,f_2)$ can be directly observed (up until the time when $U_{M_2}^{\dagger}$ is performed).\footnote{ In fact, even though the records of the outcomes $f_1$ and $f_2$ may be erased later, it is nevertheless possible to obtain empirical evidence that the joint outcome $(f_1,f_2)$ agrees with the Born rule predictions. For example, we could add an extra instruction to friend $F_2$ in the experimental protocol: if she observes $f_1 = f_2 = 1$, in contradiction with \cref{eq_a12}, she should open the door of her lab and request that the experiment be halted immediately. Furthermore, in the case where $(f_1,f_2)$ is as expected, $F_2$ could write a note stating ``agrees with the Born rule'' and pass it to Wigner through a tiny slit in her lab door that is small enough and opened for a short enough time so that no other interaction between the inside and outside of the lab can occur. This note can be retained by Wigner even if the records of $(f_1,f_2)$ are later erased, and the rest of the experimental protocol can proceed unaffected. The inclusion of this note is analogous to the one in Deutsch's version of Wigner's friend~\cite[Sec. 8]{deutsch1985quantum}, the one in Ref.~\cite[Sec.~IV.B]{copenhagen}, and the one in Ref.~\cite[Sec. 4.2.2]{yile2023}.} 
 The Born rule predicts that 
\begin{equation}
\label{eq_a12}
    \wp(f_1=1,f_2=1)=0,
\end{equation}
just as in \cref{eq:5_cycle_12}, the correlation for the $(1,2)$ pair of measurements in the 5-cycle noncontextuality scenario. Moreover, this Born rule prediction---and those that follow---can be made without appealing to the projection postulate, as we explain later. This correlation $f_1,f_2$ is available to observers up until $U_{M_1}^\dagger$ is applied (which happens after measurement $M_2$).

Measurements $M_2$ and $M_3$ are not done immediately in sequence; rather, the unitary $U_{M_1}^{\dagger}$ is applied in between them. Nevertheless, the joint outcome $(f_2,f_3)$ is still directly observable (up until $U_{M_2}^{\dagger}$ is performed). The Born rule predicts that
\begin{equation}
\label{eq_a23}
    \wp(f_2=0,f_3=0)=0,
\end{equation}
just as in \cref{eq:5_cycle_23}, the correlation for the $(2,3)$ pair of measurements in the 5-cycle noncontextuality scenario.
This is the case because the unitary $U_{M_1}^{\dagger}$ commutes with the $M_3$ measurement (as shown explicitly in Appendix E of Ref.~\cite{walleghem2023extended}), and so it does not affect the Born rule prediction for $\wp(f_2,f_3)$. 

Similarly, the Born rule predicts that 
\begin{align}
    \wp(f_3=1,f_4=1)=0, \label{eq_a34} \\
    \wp(f_4=0,f_5=0)=0, \label{eq_a45}
\end{align}
where both refer to joint outcomes that can be directly observed, and the only process happening between each pair of measurements is a unitary that commutes with those measurements (and so cannot affect their statistics). \cref{eq_a34,eq_a45} are analogous to \cref{eq:5_cycle_34,eq:5_cycle_45}, respectively, in the 5-cycle noncontextuality scenario.

However, the same argument cannot be applied to the joint outcome $(f_1,f_5)$, since it necessarily cannot be observed by any observer, even in principle. This is because the outcome $f_1$ is erased by the unitary $U_{M_1}^{\dagger}$ (meaning all outcome records of $f_1$, direct and indirect, are physically erased) prior to the outcome $f_5$ being generated. However, if the process $U:= U_{M_3}^{\dagger} U_{M_4} U_{M_2}^{\dagger} U_{M_3} U_{M_1}^{\dagger} U_{M_2}$ would be applied after $M_5$ instead of before $M_5$, the correlation $(f_1,f_5)$ would be available to a single agent. Consequently, even though the process  $U$, which is done between $M_1$ and $M_5$, commutes with $M_5$, the assertion that it does not affect the correlation between $M_1$ and $M_5$ is {\em not} guaranteed by the correctness of operational quantum theory (in particular, it does not follow from our assumption of Possibilistic Born rule). Different interpretations of quantum theory may make different predictions for unobservable correlations, while still being consistent with operational quantum theory, as was highlighted already in, e.g., Ref.~\cite{schmid2023review}.

Therefore, to complete the argument, we require an assumption that the commutation relation $U_{M_5}UU_{M_1}=UU_{M_5}U_{M_1}$ implies $p_{U_{M_5}UU_{M_1}}(f_1,f_5)=p_{UU_{M_5}U_{M_1}}(f_1,f_5)$, where the former denotes the joint distribution when $U$ is performed between $M_1$ and $M_5$, and the latter denotes the joint distribution when $U$ is performed after $M_5$. Let us call this assumption {\em Commutation Irrelevance}.\footnote{ We note that someone with operationalist leanings might object to this assumption on the grounds that it constrains correlations that are unobservable in principle. However, we note that it does not suffice to retreat to the position that `unperformed experiments have no results'~\cite{peres1978unperformed}, since the relevant correlations in an EWF argument refer to outcomes of {\em actually performed measurements}. This point was already made in e.g., Refs.~\cite{bong2020strong,cavalcanti2021implications,schmid2023review}.}
\begin{assumption}[Commutation Irrelevance]
    Any unitary process performed between two measurements does not affect the correlations between their outcomes, provided that the unitary commutes with at least one of the two measurements---{\em even} if the outcomes of the two measurements are not jointly observable to any observer.
\end{assumption}

In the latter case, where $U$ is carried out {\em after} $M_5$, the correlation $\wp(f_1,f_5)$ is observable, and so is constrained by operational quantum theory to be
$p_{UU_{M_5}U_{M_1}}(f_1=1,f_5=0) \neq 0$. Consequently,
assuming Commutation Irrelevance, the unobservable correlations for the actual experiment (where $U$ is done between $M_1$ and $M_5$) are  constrained in the same way:
\begin{equation}
    p(f_1=1,f_5=0)\neq 0. \label{eq_a15}
\end{equation}
This is analogous to \cref{eq:01_box_not0} in the 5-cycle noncontextuality scenario.

\cref{eq_a12,eq_a23,eq_a34,eq_a45} imply that the actually observed outcomes must satisfy
\begin{equation}
    f_1=1\Rightarrow f_2=0 \Rightarrow f_3=1 \Rightarrow f_4=0 \Rightarrow f_5=1.
\end{equation} 
However, \cref{eq_a15} tells us that the event $(f_1=1,f_5=0)$ occurs in some runs of the protocol. 

Thus, we have a contradiction and consequently a no-go theorem against the conjunction of our assumptions: Universality of Quantum Theory, Absoluteness of Observed Events, possibilistic Born Rule and Commutation Irrelevance.

In Appendix D of Ref.~\cite{walleghem2023extended}, this scenario is extended to an infinite family of EWF paradoxes based on $n$-cycle proofs of the failure of Kochen--Specker noncontextuality for $n\geq 4$. There we also introduce a unified description of possibilistic $n$-cycle contextuality models for both even and odd $n$.

\subsection{Discussion and outlook}

\textit{How well motivated is Commutation Irrelevance?}

No-go results are only as interesting as their underlying assumptions are compelling.
Hence, it is necessary to question whether the assumption of Commutation Irrelevance is well-motivated. 
This is the key assumption beyond the standard assumptions common to EWF arguments, and plays a role analogous to the auxiliary assumptions of Timing Irrelevance or Local Agency in the Frauchiger--Renner and Local Friendliness arguments, respectively. 

In fact, the assumption of Commutation Irrelevance is a minor extension of the assumption of Timing Irrelevance~\cite{schmid2023review} required in the Frauchiger--Renner (and the Pusey--Masanes~\cite{PuseyYoutube,LeiferYoutube}) arguments. Timing Irrelevance is the assumption that the correlations obtained in two quantum circuits whose only difference is the timing of the measurements must be identical (even if the difference in timings matters for the question of whether or not the correlations in question are in principle observable). Timing Irrelevance is a special case of Commutation Irrelevance, since the identity channel (representing the action of waiting) commutes with any operation.

Both Timing Irrelevance and Commutation Irrelevance can be motivated by the assumption that quantum theory is {\em complete} in the sense that there is no deeper theory or deeper set of facts (like hidden variables) that determine what measurement outcomes occur. This assumption is endorsed by researchers sympathetic to the Copenhagen school of thought. On the other hand, as argued in Ref.~\cite{schmid2023review}, for interpretations that violate Completeness, or researchers who are less certain that quantum theory is the deepest possible description of nature, Timing Irrelevance and Commutation Irrelevance may not be compelling assumptions.

One explicit example of an interpretation that violates Commutation Irrelevance is Bohmian mechanics. This follows from the fact that Bohmian mechanics violates even the weaker assumption of Timing Irrelevance, as discussed in Ref.~\cite{schmid2023review}. One can also see Bohmian mechanics's violation of Commutation Irrelevance in the scenario discussed here by noticing that Bohmian mechanics reproduces quantum predictions for {\em observable} correlations. 
For our scenario (where $U$ is done between $M_1$ and $M_5$), Bohmian mechanics reproduces the observable predictions that $f_1=1\Rightarrow f_2=0 \Rightarrow f_3=1 \Rightarrow f_4=0 \Rightarrow f_5=1$.
Hence, $p_{U_{M_5}UU_{M_1}}(f_1=1,f_5=0)=0$ holds. However, for the case where $U$ is done after $M_5$, so that the outcomes of $M_1$ and $M_5$ can be jointly observed, Bohmian mechanics reproduces the quantum prediction for this correlation, namely, that $p_{UU_{M_5}U_{M_1}}(f_1=1,f_5=0)\neq 0$. So in Bohmian mechanics, $p_{U_{M_5}UU_{M_1}}(f_1=1,f_5=0)\neq p_{UU_{M_5}U_{M_1}}(f_1=1,f_5=0)$.

\vspace{0.5cm}

\textit{Related works}
There are a few related works that construct EWF arguments on a single system~\cite{guerin2020no,nurgalieva2023multi,szangolies2023quantum}. Refs.~\cite{nurgalieva2023multi,szangolies2023quantum} implement constructions similar to ours based on contextual correlations, but do not identify or motivate the critical assumption that (we will argue) is required.  In addition, Ref.~\cite{nurgalieva2023multi} applies the projection postulate to processes that are {\em also} treated unitarily in the argument, which (as mentioned just above) undermines the apparent paradox.  

Ref.~\cite{guerin2020no} also builds an EWF argument on a single system---one that is not based on contextual correlations, but rather on an assumption of a kind of linearity~\cite{guerin2020no}. This assumption can be criticized on interpretation-independent grounds~\cite{schmid2023review}.
Although the novel assumption required in our construction can also be criticized and would be rejected in some interpretations (as we discuss later), we argue that it may be motivated within others (and in particular, the same ones under which Frauchiger--Renner's assumptions are motivated). Still, understanding the problems with the linearity assumption in Ref.~\cite{guerin2020no} gives interesting insights---e.g., into the nature of quotiented representations of quantum processes~\cite{schmid2023review}, and so is worth study. Similarly, we hope that our no-go theorem will help provide insights into the metaphysical assumptions going into the Frauchiger--Renner argument and related arguments (with consequent implications for interpretations of quantum theory).


\section{Deriving Local Friendliness inequalities from KSNC constraints}
\label{sec:LFasKSNC}
\sectionmark{Deriving LF inequalities from KSNC constraints}

In this section, we show how techniques from the study of Kochen--Specker noncontextuality can be used to derive many Local Friendliness inequalities in a wide range of Local Friendliness scenarios, including multipartite systems and sequential measurements.
We do so by considering Kochen--Specker inequalities relative to {\em subgraphs} of the compatibility graph naturally associated to the Local Friendliness scenario in question.

\subsection{Technical consequences of the Local Friendliness assumptions for LF scenarios}
\label{sec:LFnogo}

We first introduce some terminology regarding the $N_A,N_B$-setting LF scenario of \Cref{sec:LF}. Recall that in this scenario, we have two friends, Charlie and Debbie, performing a measurement, which we denote by $A_1$ and $B_1$, after which Alice and Bob each choose $x \in \{1,2,\ldots, N_A\}, y \in \{1,2,\ldots, N_B\}$.
For $x=1$, Alice opens Charlie's lab and reveals his outcome $a_1$ to the world. 
For $x \in \{2,\ldots,N_A\}$, Alice performs some operation on Charlie's lab (for example undoing Charlie's measurement in the quantum scenario), after which Alice instructs Charlie to perform another measurement on his system; which measurement she instructs him to perform is determined by $x \in \{2,\ldots, N_A\}$. 
Charlie's new measurement is denoted $A_x$, and his respective outcome is denoted $a_x$. For instance, if $x=2$, the new measurement performed by Charlie is $A_2$ with outcome $a_2$. After Charlie has performed his new measurement, Alice reveals his outcome $a_x$ to the world. 
Bob and Debbie act analogously.

At the end of the experiment, we collect statistics for $a_x$, $b_y$, $x$, and $y$, which constitute the empirical probabilities $\wp(a_x,b_y|x,y)$. 
These probabilities constitute the empirical correlations (or behaviors) in this LF scenario. For simplicity, we also denote these empirical probabilities as $\wp(a_x,b_y)$ without the explicit conditionals since the conditionals are actually redundant --- the empirical correlation between $a_x$ and $b_y$ is only defined in the runs with the corresponding choices $x$ and $y$.

We now look in some more detail at the technical consequences of the Local Friendliness assumptions for the $(N_A,N_B)$-setting LF scenario. 

In each individual run of the LF scenario, either two, three, or four measurements are performed, depending on Alice's and Bob's choices. For example, when $x=1,y=1$, only two measurements are performed while when  $x=2,y=2$, four measurements are performed in a single run. Thus, depending on the choices, in each run, either two, three, or four outcomes are observed by the agents. AOE implies that for a given value of $x$ and a given value of $y$, there exists a joint distribution over the outcomes observed in the corresponding runs. For example, for $x=i,y=1$, measurements $A_1,A_i,B_1$ are performed and their respective outcomes exist in a single run. By AOE, for these runs, one can assign a well-defined joint distribution over their outcomes $a_1,a_i,b_1$, which we denote as $p(a_1,a_i,b_1|x{=}i,y{=}1)$.\footnote{Here we follow the common abuse of notation in the literature that the same letter such as $a_1$ is used to denote both the random variable representing the outcome and the value of the random variable. If one wishes to write the distribution in a more explicit way, one could use $\mathcal{A}_1, \mathcal{A}_i,\mathcal{B}_1$ for the random variables denoting the respective outcomes and use the lowercase letters for their values; then, the joint distribution can be written as $p(\mathcal{A}_1=a_1,\mathcal{A}_i=a_i,\mathcal{B}_1=b_1|x{=}i,y{=}1)$. } By the experimental setup, the empirical correlation $\wp(a_i,b_1)$ is its marginal, i.e., $p(a_1,a_i,b_1|x{=}i,y{=}1) = \wp(a_i,b_1)$. 
For all given pairs of values $x,y$ of choices, we have the following respective joint distributions:

\begin{subequations}
\label{eq:AOEdis}
\begin{align}
    p(a_1,b_1| x=1,y=1),\\
    p(a_1,a_i,b_1|x=i,y=1),\\
    p(a_1,b_1,b_j|x=1,y=j), \\
    p(a_1,a_i,b_1,b_j|x=i,y=j), \label{eq:AOEij}
\end{align}
\end{subequations}
for $i\in\{2,3,\dots,N_A\}$ and $j\in\{2,3,\dots,N_B\}$.

While these joint probability distributions are not empirically accessible\footnote{For instance, the joint probability distribution $p(a_1,b_1,b_2|x=1,y=2)$ is not empirically accessible as for the setting $y=2$ Alice reverses Debbie's measurement $B_1$, thereby erasing outcome $b_1$ and destroying every record of it. Consequently, for $y=2$ no observer can jointly access records of both outcomes $b_1$ and $b_2$.}, according to the experimental setup they must satisfy
\begin{subequations}
\label{eq_emp}
\begin{align}
    p(a_1,b_1|x=1,y=1)=\wp(a_1,b_1), \label{eq:AOEem} \\
  \sum_{a_1}p(a_1,a_i,b_1|x=i,y=1)=\wp(a_i,b_1),  \\
  \sum_{b_1}p(a_1,b_1,b_j|x=1,y=j)=\wp(a_1,b_j), \label{eq:empa1bj}\\
 \sum_{a_1,b_1}p(a_1,a_i,b_1,b_j|x=i,y=j)=\wp(a_i,b_j). \label{eq:marg11}
\end{align}
\end{subequations}

We will now apply the other Local Friendliness assumption, Local Agency to this setting.

\begin{assumption}[Local Agency] \label{ass:LA_chap4}
The only relevant events correlated with an intervention are in its future light cone.
\end{assumption}

Since in LF scenarios, we assume that none of $a_i$, $a_1$ or $b_1$ is in the future light cone of Bob's measurement setting $y$, and none of $b_j$, $a_1$ or $b_1$ is in the future light cone of Alice's measurement setting $x$, Local Agency implies that 
\begin{subequations}
    \label{eq:LA}
\begin{align}
p(a_1,b_1|x,y)&=p(a_1,b_1) \\
p(a_1,a_i,b_1|x=i,y) &= p(a_1,a_i,b_1|x=i), \label{eq:LAaib1}\\
p(a_1,b_1,b_j|x,y=j)&=p(a_1,b_1,b_j|y=j).\label{eq:LAa1bj}
\end{align}
\end{subequations}

The existence of the distributions in \cref{eq:AOEdis} and the conditions in \cref{eq_emp,eq:LA} constitute the \emph{LF constraints} on the
empirical correlations in the LF scenario, leading to the so-called \emph{LF inequalities}.
The set of empirical correlations consistent with all LF inequalities is termed the \emph{LF polytope}. 
The Local Friendliness no-go theorem~\cite{bong2020strong,wiseman2023thoughtful,cavalcanti2021implications}, as we have seen in \Cref{sec:LF}, then shows that some LF inequalities can be violated in a proposed quantum realization of the LF scenario.
In the sections below, we will explore the details of LF inequalities for various LF scenarios, including generalizations to the ones defined in \Cref{sec:recap_LF_scenario}.

\subsection{The compatibility graph for a $N_A,N_B$-setting LF scenario}
\label{sec:compatibility_graph}

We start our results by applying the notion of a compatibility graph introduced in \Cref{sec:recap_KSNC} to the $N_A,N_B$-setting LF scenario defined in \Cref{sec:recap_LF_scenario}. In a $N_A,N_B$-setting LF scenario, all measurements done on Alice's side, i.e., $A_1,\dots,A_{N_{A}}$, are compatible with all measurements done on Bob's side, i.e., $B_1,\dots,B_{N_{B}}$. However, $A_1,\dots,A_{N_{A}}$ measurements are not necessarily compatible with each other, and similarly, $B_1,\dots,B_{N_{B}}$ measurements are not necessarily compatible with each other. Thus, the corresponding compatibility graph for the $N_A,N_B$-setting LF scenario is as shown in \cref{fig:NaBa} where each of $A_1,\dots,A_{N_{A}}$ is connected with each of $B_1,\dots,B_{N_{B}}$, but no connections among  $A_1,\dots,A_{N_{A}}$ or among $B_1,\dots,B_{N_{B}}$.

\begin{figure}[h!]
\begin{center}
\includegraphics[width=0.22\textwidth]{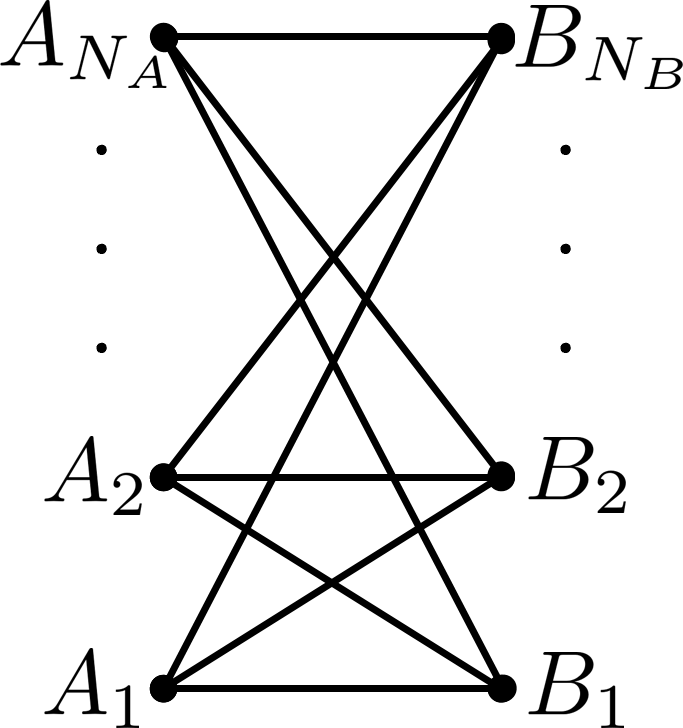}  
\caption[Compatibility graph for a $N_A,N_B$-setting LF scenario]{Compatibility graph for a $N_A,N_B$-setting LF scenario. Each node is associated with a measurement. The measurements connected by an edge are compatible and form a context. Every node on one side is connected by an edge with every node on the other side, forming a complete bipartite graph.} 
\label{fig:NaBa} 
\end{center}
\end{figure}

Such a compatibility graph is equivalent to a compatibility graph of a bipartite Bell scenario where in each run, Alice implements one of the measurements $A_1,\dots,A_{N_{A}}$ while Bob implements one of the measurements $B_1,\dots,B_{N_{B}}$.  We refer to such a bipartite Bell scenario as \emph{the corresponding Bell scenario for the $N_A,N_B$-setting LF scenario}. 

Note that the constraints coming from Local Friendliness, presented in \Cref{sec:LFnogo}, imply the no-disturbance constraints (cf.~\Cref{sec:recap_KSNC}) on the empirical  correlations in a $N_A,N_B$-setting LF scenario. These no-disturbance constraints are
\begin{subequations}
\label{eq:NS}
\begin{align}
    \sum_{b_l}\wp(a_k,b_l) = \sum_{b_{l'}}\wp(a_k,b_{l'}) \\
    \sum_{a_k}\wp(a_k,b_l) = \sum_{a_{k'}}\wp(a_{k'},b_{l}),
\end{align}
\end{subequations}
for all $k,k'\in\{1,2,\dots,N_A\}$ and $l,l'\in\{1,2,\dots,N_B\}$.
They are obtained by taking the appropriate marginals of distributions in \cref{eq:LA} and then connecting them with empirical correlations via \cref{eq_emp}.
For instance, to prove that $\sum_{b_1} \wp(a_1,b_1)=\sum_{b_2} \wp(a_1,b_2)$, \begin{equation}
\begin{split}
        \sum_{b_1} \wp(a_1,b_1) &= \sum_{b_1} p(a_1,b_1|x=1,y=1)  = \sum_{b_1} p(a_1,b_1|x=1,y=2) \\ &=  \sum_{b_1,b_2} p(a_1,b_1,b_2|x=1,y=2) \\ &= \sum_{b_2} p(a_1,b_2|x=1,y=2) = \sum_{b_2} \wp(a_1,b_2)
\end{split}
\end{equation} where the first and last equalities hold by \cref{eq_emp} and the second and second-to-last by \cref{eq:LA}, and the third and fourth equalities by marginalization of probabilities.
Furthermore, all (expected) empirical correlations in a quantum experimental realization of an LF scenario satisfy these no-disturbance relations (which can also be seen as no-signaling here).

As the LF constraints imply no-disturbance, it makes sense to further ask if LF constraints can be understood as KSNC constraints in some sense, which is what we do next in this section. In particular, we will show that the LF constraints on the empirical correlations in the LF scenario sometimes (and we will see when) coincide with the KSNC constraints on the corresponding compatibility graph for some $N_A,N_B$ and always imply KSNC constraints on subgraphs of that graph. In the cases where they coincide (cf. \Cref{sec:22,sec:32}), the KSNC constraints on the corresponding compatibility graph provide all LF inequalities, and the LF polytope coincides with the Bell polytope of the corresponding Bell scenario. In other cases (cf. \Cref{sec:33}), the LF constraints on the empirical correlations in the LF scenario are \emph{strictly weaker} than the KSNC constraints on the corresponding compatibility graph. In such scenarios, some (not all) LF inequalities can be derived via demanding KSNC constraints on \emph{subgraphs} of the corresponding graph and the LF polytope strictly contains the Bell polytope. Finally, we generalize the LF scenario recapped in \Cref{sec:LFnogo} to multipartite sequential LF scenarios and we characterize a wide class of such scenarios where all the LF inequalities are KSNC constraints on the corresponding graph, meaning that the LF polytope coincides with the Bell polytope for the corresponding multipartite Bell scenario. 

Summarizing, we generalize Local Friendliness scenarios, point out connections between Local Friendliness and KS noncontextuality and thereby derive a wide class of Local Friendliness inequalities in these scenarios.

Before deriving these results, we will first provide an alternative expression in \Cref{sec:equi} of the LF constraints defined in \Cref{sec:LFnogo}, which will aid in our investigation of the relationship between LF constraints and KSNC constraints.

\subsection{An alternative expression of the LF constraints}
\label{sec:equi}

We saw in \Cref{sec:LFnogo} that the constraints implied by LF in the $N_A,N_B$-setting LF scenario are the existence of probability distributions as in \cref{eq:AOEdis} and satisfying \cref{eq_emp,eq:LA}.
Next, we introduce an alternative expression of these constraints; one which will later be useful for making the connection to KSNC. Two of its immediate consequences are shown in \Cref{sec:1stLFasKSNC} and \Cref{sec:Bell_inside_LF}.

This alternative expression is the following. There must exist a joint distribution $p_{ij}(a_1, b_1, a_i, b_j)$ for any $i \in \{2,\ldots,N_A\},$ and $j \in \{2,\ldots,N_B\}$ with the following two properties: 1) each $p_{ij}(a_1,a_i,b_1, b_j)$ has as its marginals the four pairwise empirical correlations $\wp(a_i,b_j)$, $\wp(a_1,b_j)$, $\wp(a_i,b_1)$ and $\wp(a_1,b_1)$; 2) they agree on their overlaps, i.e., 
\begin{subequations}
\label{eq:agree}
\begin{align}
\sum_{a_i,b_j}p_{ij}(a_1,a_i,b_1,b_j)&=\sum_{a_{i'},b_{j'}}p_{i'j'}(a_1,a_{i'},b_1,b_{j'}), \label{eq:agreeab}\\
\sum_{b_j}p_{ij}(a_1,a_i,b_1,b_j)&= \sum_{b_{j'}}p_{ij'}(a_1,a_i,b_1,b_{j'}), \label{eq:agreeb}\\
\sum_{a_i}p_{ij}(a_1,a_i,b_1,b_j)&=\sum_{a_{i'}}p_{i'j}(a_1,a_{i'},b_1,b_j). \label{eq:agreea}
\end{align}
\end{subequations}

We will now prove that this is equivalent to the LF constraints defined in \Cref{sec:LFnogo}. 

First, we will show that the LF constraints imply the existence of the distributions $\{p_{ij}(a_1, b_1, a_i, b_j)\}_{i \in \{2,\ldots,N_A\},j \in \{2,\ldots,N_B\}}$ with the two properties mentioned above.

The LF constraints include the existence of $p(a_1,a_i,b_1,b_j|x=i,y=j)$ for any $i\in\{2,3,\dots,N_A\}$ and $j\in\{2,3,\dots,N_B\}$ as per \cref{eq:AOEij}, a consequence of the assumption of AOE in the $N_A,N_B$-setting LF scenario. Now, let 
\begin{align}
\label{eq:defPij}
    p_{ij}(a_1,a_i,b_1,b_j) = p(a_1,a_i,b_1,b_j|x=i,y=j).
\end{align}
Then, \cref{eq_emp,eq:LA} together imply that
\begin{subequations}
    \label{eq:LAij}
\begin{align}
     \sum_{a_1,b_1} p_{ij}(a_1,a_i,b_1,b_j) 
    =\wp(a_i,b_j),  \label{eq:LA0}\\
     \sum_{a_i,b_1} p_{ij}(a_1,a_i,b_1,b_j) 
    =\wp(a_1,b_j),  \label{eq:LA1}\\
     \sum_{a_1,b_j} p_{ij}(a_1,a_i,b_1,b_j) 
    =\wp(a_i,b_1), \label{eq:LA2} \\
     \sum_{a_i,b_j} p_{ij}(a_1,a_i,b_1,b_j) 
    =\wp(a_1,b_1). \label{eq:LA3}
\end{align}
\end{subequations}
Here, \cref{eq:LA0} is equivalent to \cref{eq:marg11} using \cref{eq:defPij}, and \cref{eq:LA1} comes from the fact that
\begin{align}
    &\sum_{a_i,b_1} p_{ij}(a_1,a_i,b_1,b_j) \nonumber \\
    = &\sum_{a_i,b_1} p(a_1,a_i,b_1,b_j|x=i,y=j) \nonumber \\
    =& p(a_1,b_j|x=i,y=j) \nonumber \\
    = & p(a_1,b_j|x=1,y=j) \nonumber \\
    = & \wp(a_1,b_j), 
\end{align}
where we used \cref{eq:LAa1bj} for the third equality and \cref{eq:empa1bj} for the fourth equality; the proof for \cref{eq:LA2} or \cref{eq:LA3} is similar.

Eqs.~\eqref{eq:LAij} mean that the joint distribution $p_{ij}(a_1,a_i,b_1,b_j)$ has the four pairwise empirical correlations $\wp(a_i,b_j)$, $\wp(a_1,b_j)$, $\wp(a_i,b_1)$ and $\wp(a_1,b_1)$ as its marginals, proving the first property.

Furthermore, the LF constraints due to the assumption of Local Agency in Eqs.~\eqref{eq:LA} directly imply Eqs.~\eqref{eq:agree} by \cref{eq:defPij}, meaning that the distributions in $$\{p_{ij}(a_1, b_1, a_i, b_j)\}_{i \in \{2,\ldots,N_A\},j \in \{2,\ldots,N_B\}}$$ agree on their overlaps, proving the second property.

\vspace{0.5cm}

Now, we will prove the converse direction: that the existence of $$\{p_{ij}(a_1, b_1, a_i, b_j)\}_{i \in \{2,\ldots,N_A\},j \in \{2,\ldots,N_B\}}$$ with the above mentioned two properties implies all LF constraints.

For any $i \in \{2,\ldots,N_A\}$ and $j \in \{2,\ldots,N_B\}$, define 
\begin{subequations}
    \label{eq:AOEconverse}
\begin{align}
    &p(a_1,b_1|x=1,y=1)\coloneqq \sum_{a_2,b_2} p_{22}(a_1, a_2,b_1,b_2) \\
    &p(a_1, a_i,b_1|x=i,y=1)\coloneqq \sum_{b_2} p_{i2}(a_1, a_i,b_1,b_2) \\
    &p(a_1, b_1,b_j|x=1,y=j) \coloneqq  \sum_{a_2} p_{2j}(a_1, a_2,b_1,b_j)\\
    &p(a_1, a_i,b_1,b_j|x=i,y=j) \coloneqq p_{ij}(a_1, a_i,b_1,b_j)
\end{align}
\end{subequations}

These are the distributions required to exist by AOE in Eqs.~\eqref{eq:AOEdis}. They reproduce the empirical correlations as in Eqs.~\eqref{eq_emp} due to the first property of $$\{p_{ij}(a_1, b_1, a_i, b_j)\}_{i \in \{2,\ldots,N_A\},j \in \{2,\ldots,N_B\}}$$ as in Eqs.~\eqref{eq:LAij}. 

Furthermore, the LF conditions in \cref{eq:LA} are also satisfied due to the second property of $\{p_{ij}(a_1, b_1, a_i, b_j)\}_{i \in \{2,\ldots,N_A\},j \in \{2,\ldots,N_B\}}$ that these distributions agree on their overlaps. For example, to prove \cref{eq:LAaib1}, it suffices to note that for any $i \in \{2,\ldots,N_A\}$, $j \in \{2,\ldots,N_B\}$ and $j'\in\{1,2,\dots,N_B\}$, we have
\begin{align}
    &p(a_1,a_i,b_1 |x=i,y=j) \nonumber\\
    = &\sum_{b_j} p_{ij}(a_1,a_i,b_1,b_j) \nonumber\\
    = &\sum_{b_{j'}} p_{ij'}(a_1,a_i,b_1,b_j') \nonumber\\
    = &p(a_1,a_i,b_1|x=i,y=j'), 
\end{align}
where the second equality is due to the fact that $p_{ij}(a_1,a_i,b_1,b_j)$ and $p_{ij'}(a_1,a_i,b_1,b_j')$ agree on their overlap.

By expressing the LF constraints in terms of the existence of distributions in $\{p_{ij}(a_1, b_1, a_i, b_j)\}_{i \in \{2,\ldots,N_A\},j \in \{2,\ldots,N_B\}}$ and the two properties they have, we can easily obtain two insights into the features of the LF inequalities, as we now show.

\tocless\subsubsection{A class of tight LF inequalities that are KSNC constraints on subgraphs} \label{sec:1stLFasKSNC}

The first insight that immediately follows from the alternative expression just derived is the existence of a class of LF inequalities that are KSNC constraints on subgraphs of the compatibility graph for a $N_A,N_B$-setting LF scenario, i.e.,  \cref{fig:NaBa}. Specifically, all the LF inequalities in this class are KSNC constraints on graphs of the form shown in \cref{fig:a1aib1bj}.

\begin{figure}[h!]
\begin{center}
\includegraphics[width=0.2\textwidth]{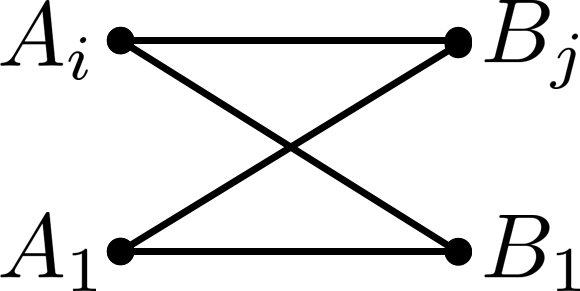}  
\caption[The compatibility graph for $A_1$, $A_i$, $B_1$ and $B_j$]{The compatibility graph for $A_1$, $A_i$, $B_1$ and $B_j$ with $i\in \{2,3,\dots,N_A\}$ and $j\in\{2,3,\dots,N_B\}$. It is a subgraph of \cref{fig:NaBa}.} 
\label{fig:a1aib1bj} 
\end{center}
\end{figure}

To see this, first recall from \Cref{sec:recap_KSNC} that demanding KSNC constraints on a compatibility graph is equivalent to demanding the existence of a joint probability distribution over the outcomes of all measurements associated with vertices of the graph, such that the distribution has the empirical correlation associated with each edge as its marginals.
For the compatibility graph in \cref{fig:a1aib1bj}, from \cref{eq:LAij}, we see that $p_{ij}(a_1,a_i,b_1,b_j)$ is a joint distribution over the outcomes of the four measurements in the graph and has as its marginals the four empirical correlations for the four respective edges, namely, $\wp(a_i,b_j)$, $\wp(a_1,b_j)$, $\wp(a_i,b_1)$ and $\wp(a_1,b_1)$. 

Thus, the LF constraints require the KSNC constraints on graphs in the form of the one in \cref{fig:a1aib1bj}, which are subgraphs of \cref{fig:NaBa}, the full compatibility graph for a $N_A,N_B$-setting LF scenario. Furthermore, since \cref{fig:NaBa} corresponds to a bipartite Bell scenario where both Alice and Bob have binary measurement settings, the KSNC constraints on 
\cref{fig:NaBa} are CHSH inequalities. Therefore, for each $i\in \{2,3,\dots,N_A\}$ and $j\in\{2,3,\dots,N_B\}$, we can derive LF inequalities that are CHSH inequalities on $\wp(a_i,b_j)$, $\wp(a_1,b_j)$, $\wp(a_i,b_1)$ and $\wp(a_1,b_1)$. 

Furthermore, these LF inequalities are in fact \emph{tight}, meaning that they can be saturated by correlations that satisfy the LF assumptions.  This is because of the second immediate insight we gain from the alternative expression of the LF constraints, which, as we will explain next, is that the LF inequalities are not stricter than the Bell inequalities for the corresponding Bell scenario, or in other words, the LF polytope always contains the corresponding Bell polytope. This fact has been previously shown in Refs.~\cite{bong2020strong,cavalcanti2021implications,yile2023,haddara2024local} (and indirectly also in Ref.~\cite{woodhead2014}.)

\tocless\subsubsection{The LF polytope contains the Bell polytope} \label{sec:Bell_inside_LF}

Here we give an alternative proof of the fact that the LF polytope contains the Bell polytope in the corresponding Bell scenario to the ones in previous literature such as Refs.~\cite{bong2020strong,yile2023}.

The Bell polytope in the Bell scenario corresponding to our LF scenario is precisely characterized by the KSNC constraints on the compatibility graph of the LF scenario as in \Cref{sec:compatibility_graph}. That is, it comes from the requirement that there must exist a global distribution $p(a_1,b_1,a_2,b_2,\ldots,a_{N_A},b_{N_B})$ over the outcomes of all measurements that reproduces all empirical correlations, namely, all $\wp(a_x,b_y)$ with $ x\in\{1,2,\dots,N_A\}$ and $y\in\{1,2,\dots,N_A\}$.

When such a $p(a_1,b_1,a_2,b_2,\ldots,a_{N_A},b_{N_B})$ exists, it further implies the existence of all distributions in $\{p_{ij}(a_1, b_1, a_i, b_j)\}_{i \in \{2,\ldots,N_A\},j \in \{2,\ldots,N_B\}}$ with the two properties described at the beginning of \Cref{sec:equi}. Specifically, we can define $p_{ij}(a_1, b_1, a_i, b_j)$ simply by marginalising $p(a_1,b_1,a_2,b_2,\ldots,a_{N_A},b_{N_B})$. These distributions evidently reproduce the four empirical correlations $\wp(a_i,b_j)$, $\wp(a_1,b_j)$, $\wp(a_i,b_1)$ and $\wp(a_1,b_1)$; they also agree on their overlap by construction (as they all stem from the same global distribution). 

Thus, the KSNC constraints (which are also Bell inequalities) for the corresponding Bell scenario imply the respective LF constraints, which also means that the LF polytope always contains the corresponding Bell polytope.

\vspace{0.5cm}

\subsection{The 2,2-setting LF scenario}
\label{sec:22}
In this scenario, Alice and Bob have binary measurement settings. Thus, in total, we have four measurements $A_1$, $A_2$, $B_1$ and $B_2$ with measurement outcomes $a_1$, $a_2$, $b_1$ and $b_2$. The compatibility graph for the 2,2-setting LF scenario is shown in \cref{fig:22}.

\begin{figure}[h!]
\begin{center}
\includegraphics[width=0.2\textwidth]{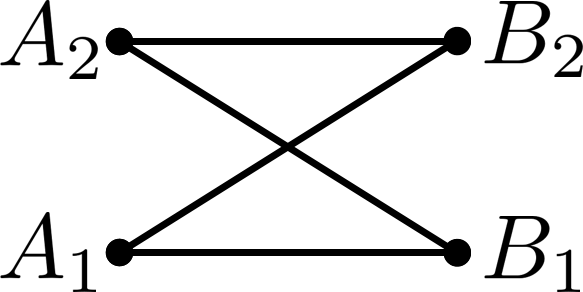}  
\caption[The compatibility graph for the 2,2-setting LF scenario]{The compatibility graph for the 2,2-setting LF scenario, which is a special case of the graph in \cref{fig:NaBa}.} 
\label{fig:22} 
\end{center}
\end{figure}

From the alternative expression described in \Cref{sec:equi} of the LF constraints, we know that there exists a joint distribution $p_{22}(a_1,a_2,b_1,b_2)$ such that it has as its marginals $\wp(a_2 ,b_2)$, $\wp(a_1,b_2)$, $\wp(a_2,b_1)$ and $\wp(a_1,b_1)$, which are all the empirical correlations in the 2,2-setting LF scenario.
Thus, as discussed in \Cref{sec:1stLFasKSNC}, the four pairwise correlations $\wp(a_x, b_y)$ for $x=1,2$ and $y=1,2$ must obey KSNC constraints on the compatibility graph in \cref{fig:22}, which give rise to the CHSH inequalities on these four correlations. Furthermore, since the LF polytope must contain the corresponding Bell polytope, as was proven in \Cref{sec:Bell_inside_LF}, in this 2,2-setting scenario, the LF polytope coincides with the Bell polytope whose nontrivial facets are characterized by these CHSH inequalities.

\vspace{0.5cm}

\subsection{The 3,2-setting LF scenario}
\label{sec:32}

Now consider the case where Alice has 3 settings and Bob has 2 settings. There are now six empirical pairwise correlations, namely $\wp(a_x, b_y)$ for $x=1,2,3$ and $y=1,2$.

Assuming Local Friendliness, following \Cref{sec:equi}, we have that
\begin{itemize}
    \item $p_{22}(a_1,a_2,b_1,b_2)$ has as its marginals $\wp(a_2, b_2)$, $\wp(a_1,b_2)$, $\wp(a_2,b_1)$ and $\wp(a_1,b_1)$. Thus, the four pairwise correlations $\wp(a_x, b_y)$ for $x=1,2$ and $y=1,2$ must obey KSNC constraints on the compatibility graph in \cref{fig:a12b120}. 
    \item  $p_{32}(a_1,a_3,b_1,b_2)$ has as its marginals $\wp(a_3,b_2)$, $\wp(a_1,b_2)$, $\wp(a_3,b_1)$ and $\wp(a_1,b_1)$. Thus, the four pairwise correlations $\wp(a_x, b_y)$ for $x=1,3$ and $y=1,2$ must obey KSNC constraints on the compatibility graph in \cref{fig:a13b120}. 
\end{itemize}

\begin{figure}[h!]
    \centering
        \begin{subfigure}[b]{0.16\textwidth}
         \centering
         \includegraphics[width=\textwidth]{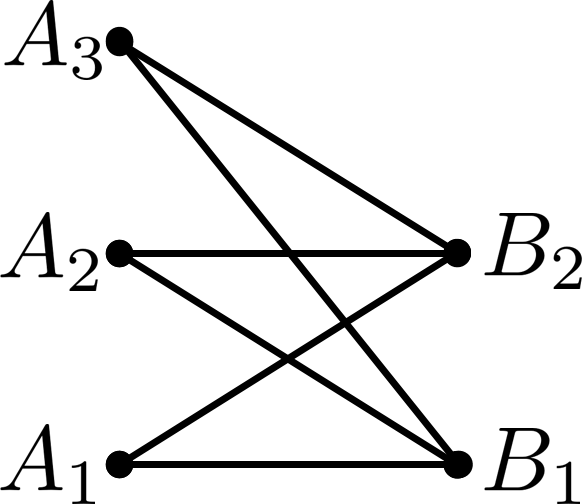}
         \caption{}
         \label{fig:a123b12m}
     \end{subfigure}
    \hspace{0.4mm}
    \begin{subfigure}[b]{0.16\textwidth}
         \centering
         \includegraphics[width=\textwidth]{Figures/a12b12.png}
         \caption{}
    \label{fig:a12b120}
     \end{subfigure}
    \hspace{0.4mm}
    \begin{subfigure}[b]{0.16\textwidth}
         \centering
         \includegraphics[width=\textwidth]{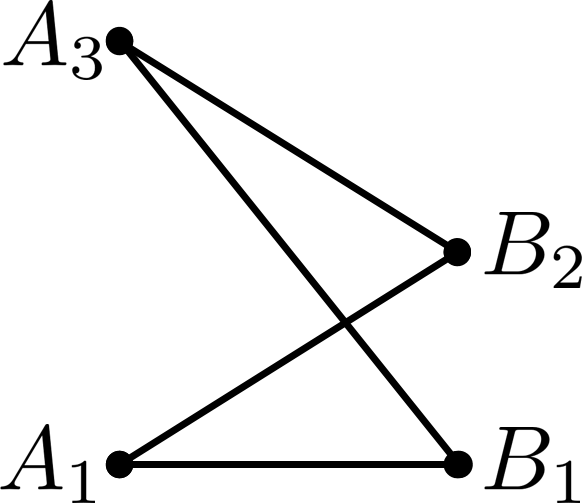}
         \caption{}
         \label{fig:a13b120}
     \end{subfigure}
\caption[The compatibility graph of the 3,2-setting LF scenario]{(a): The compatibility graph of the 3,2-setting LF scenario. (b) and (c): Subgraphs of the graph in (a) that are special cases of the graph in \cref{fig:a1aib1bj}. In \Cref{sec:glue} we explain how LF assumptions allow us to `glue' the distributions for (b) and (c) together to form a joint distribution for (a), using the fact that LF requires the distributions for (b) and (c) to agree on their overlap, i.e., have the same marginals on the intersection of their domain.}
\end{figure}

So far, we have \emph{not} shown that the four pairwise correlations $\wp(a_x,b_y)$ for $x=2,3$ and $y=1,2$ must satisfy KSNC constraints on a corresponding compatibility graph, and in fact one cannot do so by an argument analogous to the above. 

Nevertheless, we will now show that the Local Friendliness assumptions do require KSNC constraints for these correlations as well, and in fact, that the Local Friendliness assumptions imply KSNC constraints on the compatibility graph for the whole 3,2-LF scenario as shown in \cref{fig:a123b12m}, meaning that all six pairwise correlations $\wp(a_x,b_y)$ for $x=1,2,3$ and $y=1,2$ are marginals of a single joint probability distribution.

\tocless\subsubsection{The `gluing' trick}
\label{sec:glue}

The aforementioned joint distribution of which all six pairwise correlations $\wp(a_x,b_y)$ for $x=1,2,3$ and $y=1,2$ are marginals can be constructed using $p_{22}(a_1,a_2,b_1,b_2)$ and $p_{32}(a_1,a_3,b_1,b_2)$ via 
\begin{align}
\label{eq:2232}
    & p_{22,32} 
    \coloneqq  \frac{p_{22}(a_1,a_2,b_1,b_2)p_{32}(a_1,a_3,b_1,b_2)}{\sum_{a_3} p_{32}(a_1,a_3,b_1,b_2)}.
\end{align}
We call the trick used in \cref{eq:2232} for constructing $p_{22,32}$ the \emph{gluing} operation\footnote{A similar technique was used in, e.g., Refs.~\cite{fine1982joint,haddara2024local}.}, which constructs a larger joint probability distribution (i.e., $p_{22,32}$)  from smaller distributions (i.e., $p_{22}(a_1,a_2,b_1,b_2)$ and $p_{32}(a_1,a_3,b_1,b_2)$) in such a way that (as we will prove soon) the larger one has as its marginals all empirical pairwise correlations that are marginals of the smaller ones. We call it `gluing' since in terms of the compatibility graphs upon which the KSNC constraints are demanded by the respective joint distributions, such a procedure `glues' together two smaller graphs (i.e., \cref{fig:a12b120} and \cref{fig:a13b120}) to a bigger one (i.e., \cref{fig:a123b12m}); in terms of the probability distributions, it `glues' together two marginal distributions to create a bigger joint distribution.

 The proof for $p_{22,32}$ to recover all six pairwise correlations uses the fact that 
$p_{22}(a_1,a_2,b_1,b_2)$ and $p_{32}(a_1,a_3,b_1,b_2)$ agree on their overlap:
\begin{align}
\label{eq:bcd2232}
\sum_{a_2} p_{22}(a_1,a_2,b_1,b_2)
    =  \sum_{a_3} p_{32}(a_1,a_3,b_1,b_2), 
\end{align}
which is a special case of \cref{eq:agreea}.  See Appendix B of Ref.~\cite{walleghem2024connecting} for the explicit proof.

As such, all six pairwise empirical correlations in the 3,2-setting LF scenario can be recovered as marginals of a single joint distribution $p_{22,32}$. Thus, these six pairwise correlations must satisfy the KSNC constraints on the compatibility graph in \cref{fig:a123b12m}. As this compatibility graph also represents a Bell scenario, it follows that these constraints give rise to Bell inequalities. Furthermore, these Bell inequalities are tight LF inequalities since the LF polytope must contain the Bell polytope as proven in \Cref{sec:Bell_inside_LF}. Since these six pairwise correlations are all the observable correlations in the 3,2-setting LF scenario, then, the Local Friendliness polytope coincides with the Bell polytope in this case.

A trivial corollary of the above is that the four pairwise correlations $\wp(a_x,b_y)$ for $x=2,3$ and $y=1,2$ can be recovered as marginals of a single joint distribution, namely, $p_{22,32}$, and so must satisfy CHSH inequalities, and equivalently, the constraints implied by KSNC for any four projective measurements satisfying the compatibility relations in \cref{fig:a1aib1bj} (with the appropriate relabeling). 

Following an analogous proof to the one we just did here for the 3,2-setting LF scenario, one can further show that a $(N_A,2)$-setting or a $(2,N_B)$-setting LF scenario for any integer $N_A,N_B\geq 2$ has its LF constraints being equivalent to the KSNC constraints on the compatibility graph for that LF scenario, which give rise to the Bell inequalities for the corresponding Bell scenario. 

\tocless\subsubsection{The minimal LF scenario}
\label{sec:minimal}
An analogous argument can be constructed to explain why the LF polytope for the so-called `minimal LF scenario' studied in Ref.~\cite{wiseman2023thoughtful,yile2023} also coincides with the corresponding Bell polytope. 

In the minimal LF scenario, there is no Debbie. Alice has a binary measurement setting. She and Charlie proceed as usual. Bob, however, now chooses between two possible measurements $B_y$ on the other half of the bipartite system depending on his binary setting choice $y$. The empirical correlations in this scenario are the four pairwise correlations: $\wp(a_2, b_2)$, $\wp(a_1,b_2)$, $\wp(a_2,b_1)$ and $\wp(a_1,b_1)$.

Here, AOE demands the existence of $p(a_1,b_1|x=1,y=1)$, $p(a_1a_2b_1|x=2,y=1)$ and $p(a_1a_2b_2|x=2,y=2)$ for the corresponding runs of the experiment.  As shown in Appendix B of Ref.~\cite{walleghem2024connecting}, by applying the gluing trick to $p(a_1a_2b_1|x=2,y=1)$ and $p(a_1a_2b_2|x=2,y=2)$, we obtain a joint probability distribution that recovers all four pairwise correlations, and thus explains why the LF polytope coincides with the corresponding Bell polytope in the minimal scenario.

Furthermore, one can generalize the minimal LF scenario and show that if Bob's measurement choice is not binary but instead with higher cardinality for any integer $N_B\geq 2$, it is still true that all LF constraints in this scenario are equivalent to the KSNC constraints (or equivalently the Bell inequalities) on a corresponding Bell scenario. 

\vspace{0.5cm}

\subsection{The 3,3-setting LF scenario}
\label{sec:33}
Now consider the case where both Alice and Bob have 3 settings. In this case, the Local Friendliness polytope does {\em not} coincide with the Bell polytope, as was shown in Ref.~\cite{bong2020strong}, which used a software to derive the LF polytope for binary outcomes explicitly. We will give a more algebraic derivation and explanation of this fact (and without the restriction to binary outcomes).

Assuming Local Friendliness, following \Cref{sec:equi}, we have that $p_{22}(a_1,a_2,b_1,b_2)$ has as its marginals $\wp(a_2,b_2)$, $\wp(a_1,b_2)$, $\wp(a_2,b_1)$ and $\wp(a_1,b_1)$. Thus, the four pairwise correlations $\wp(a_x,b_y)$ for $x=1,2$ and $y=1,2$ must obey the KSNC constraints on the compatibility relations in \cref{fig:a12b12}. Since this compatibility graph in \cref{fig:a12b12} corresponds to a Bell scenario, these KSNC constraints give rise to Bell inequalities. Similar reasoning holds for $p_{32}(a_1,a_3,b_1,b_2)$, $p_{23}(a_1,a_2,b_1,b_3)$ and $p_{33}(a_1,a_3,b_1,b_3)$, depicted in \cref{fig:a13b12,fig:a12b13,fig:a13b13}.

\begin{figure}[!h]
\captionsetup[subfigure]{aboveskip=-2pt,belowskip=-1pt}
    \centering
    \begin{subfigure}[b]{0.16\textwidth}
         \centering
         \includegraphics[width=\textwidth]{Figures/a12b12.png}
         \caption{}
    \label{fig:a12b12}
     \end{subfigure}
    \hspace{0.4mm}
    \begin{subfigure}[b]{0.16\textwidth}
         \centering
         \includegraphics[width=\textwidth]{Figures/a13b12.png}
         \caption{}
         \label{fig:a13b12}
     \end{subfigure}
     \hspace{0.4mm}
    \begin{subfigure}[b]{0.16\textwidth}
         \centering
         \includegraphics[width=\textwidth]{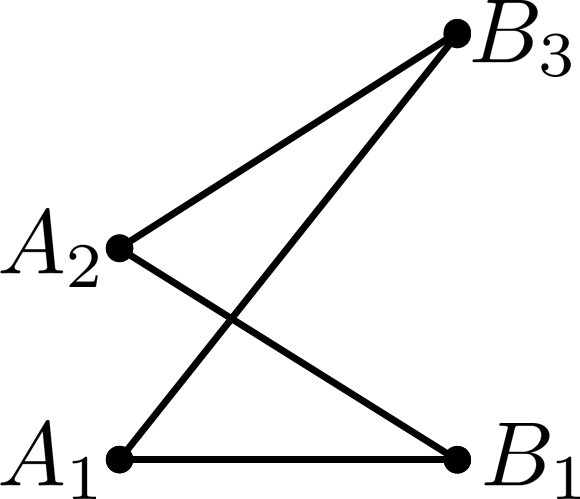}
         \caption{}
         \label{fig:a12b13}
     \end{subfigure}
    \hspace{0.4mm}
    \begin{subfigure}[b]{0.16\textwidth}
         \centering
         \includegraphics[width=\textwidth]{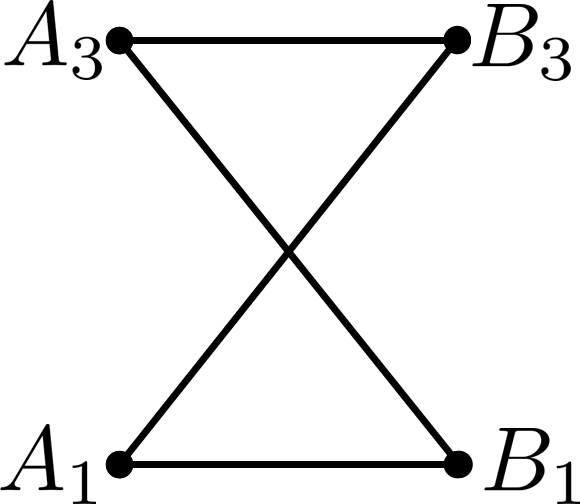}
         \caption{}
         \label{fig:a13b13}
     \end{subfigure}
    \caption[Special cases of the compatibility graph in \cref{fig:NaBa}]{Special cases of the compatibility graph in \cref{fig:NaBa} that are subgraphs of the compatibility graph of the 3,3-setting LF scenario. }
    \label{fig:4pair}
\end{figure}

Furthermore, as proven in \Cref{sec:equi}, these four distributions, $p_{22},p_{23},p_{32},p_{33}$ (where we omit their arguments for compactness), agree on their overlaps. 
Then, by gluing various three out of the four distributions together, following an argument like that for \cref{eq:2232}, we obtain the following. Gluing together $p_{22},p_{32},p_{23}$ we find that $p_{22,32,23} \! \coloneqq \! 
    \frac{p_{22}p_{32}p_{23}}{\sum_{a_3} p_{32} \sum_{b_3} p_{23}}$ has as its marginals all pairwise correlations $\wp(a_x,b_y)$ for $x=1,2,3$ and $y=1,2,3$ \emph{except} for $\wp(a_3,b_3)$, and thus, we also denote $p_{22,32,23}$ as $p_{\lnot a_3b_3}$. As such, these pairwise correlations must obey the KSNC constraints on the compatibility graph in \cref{fig:noa3b3}. The same reasoning holds for $p_{\lnot a_2b_2}, p_{\lnot a_2b_3}, p_{\lnot a_3b_2}$ with compatibility graphs in \cref{fig:noa2b2,fig:noa2b3,fig:noa3b2}.

\begin{figure}[!h]
\captionsetup[subfigure]{aboveskip=-2pt,belowskip=-1pt}
    \centering
    \begin{subfigure}[b]{0.16\textwidth}
         \centering
         \includegraphics[width=\textwidth]{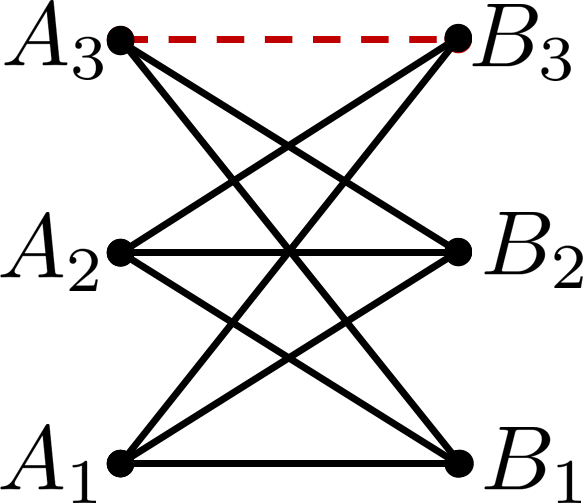}
         \caption{}
    \label{fig:noa3b3}
     \end{subfigure}
    \hspace{0.4mm}
    \begin{subfigure}[b]{0.16\textwidth}
         \centering
         \includegraphics[width=\textwidth]{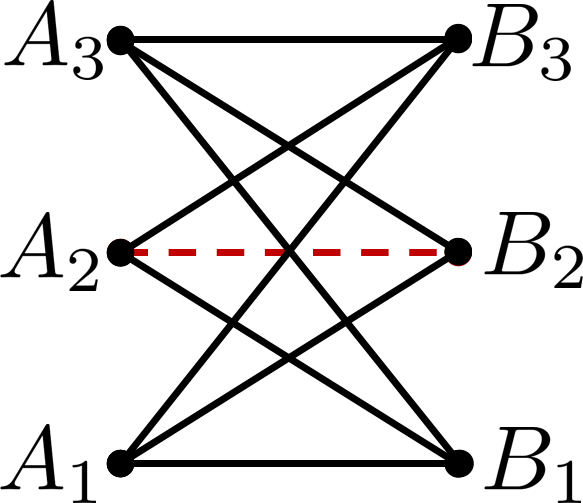}
         \caption{}
         \label{fig:noa2b2}
     \end{subfigure}
     \hspace{0.4mm}
    \begin{subfigure}[b]{0.16\textwidth}
         \centering
         \includegraphics[width=\textwidth]{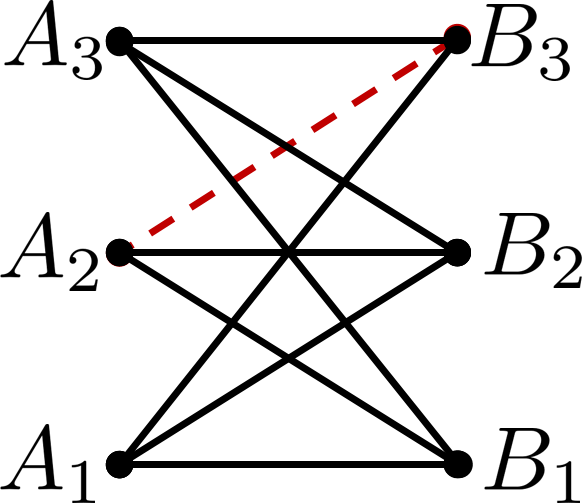}
         \caption{}
         \label{fig:noa2b3}
     \end{subfigure}
    \hspace{0.4mm}
    \begin{subfigure}[b]{0.16\textwidth}
         \centering
         \includegraphics[width=\textwidth]{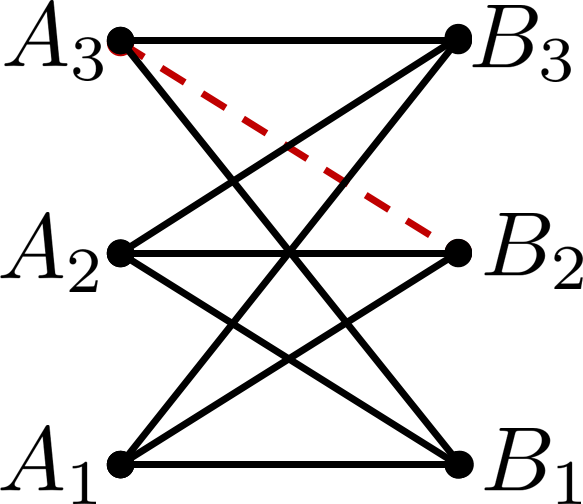}
         \caption{}
         \label{fig:noa3b2}
     \end{subfigure}
    \caption[Subgraphs of the compatibility graph of the 3,3-setting LF scenario]{Subgraphs of the compatibility graph of the 3,3-setting LF scenario. Each of them has 8 edges, i.e., 8 pairwise compatibility relations. The red dashed line indicates the missing compatibility relation for deriving Bell inequalities on all six measurements in each graph.}
    \label{fig:8pair}
\end{figure}

Unlike all of the previous cases, these compatibility graphs in \cref{fig:8pair} are {\em not} the complete compatibility graphs of any Bell scenario, because of the missing edges indicated by the red dashed lines in \cref{fig:8pair}; rather, they are subgraphs. 
Consequently,  the inequalities derived from KSNC constraints in this case will not generally be tight Bell inequalities. 
Nevertheless, some of these KSNC inequalities coincide with tight Bell inequalities because the compatibility graph in \cref{fig:8pair} includes subgraphs that are the complete compatibility graphs for some other Bell scenarios. 
For example, \cref{fig:noa3b3} includes the graphs in \cref{fig:a12b12,fig:a13b12,fig:a12b13} and \cref{fig:a123b12m} as subgraphs, and they are complete compatibility graphs for some Bell scenarios; in addition, it includes a compatibility subgraph for $\{A_2,A_3,B_1,B_3\}$, one for $\{A_1,A_3,B_2,B_3\}$, one for $\{A_2,A_3,B_1,B_2\}$, one for $\{A_1,A_2,B_2,B_3\}$ and one for $\{A_2,A_3,B_1,B_2,B_3\}$, of which each represents a complete compatibility graph for a Bell scenario. 
Thus, the KSNC constraints on the compatibility graphs in  \cref{fig:8pair} include Bell inequalities for these compatibility subgraphs.

The compatibility graphs in \cref{fig:8pair} are also \emph{not} the compatibility graph of the 3,3-setting LF scenario because of the missing edges depicted as red dashed lines. In fact, the Local Friendliness assumptions do not imply the existence of a joint probability distribution that can have as its marginals all the 9 pairwise empirical correlations $\wp(a_x,b_y)$ for $x=1,2,3$ and $y=1,2,3$ in the 3,3-setting LF scenario, and hence, the LF polytope here is strictly bigger than the Bell polytope, as pointed out in Ref.~\cite{bong2020strong}.  In Appendix B of Ref.~\cite{walleghem2024connecting}, we provide further intuitions on why this is the case by considering how the gluing trick introduced in \Cref{sec:glue} works.

Since any $(N_A,N_B)$-setting LF scenario with $N_A,N_B\geq 3$ has the 3,3-setting LF scenario as a subscenario, we can further see that a $(N_A,N_B)$-setting LF scenario for any integer $N_A,N_B\geq 3$ will have its LF constraints being strictly weaker than the KSNC constraints on the compatibility graph for that LF scenario, meaning that the LF polytope is strictly bigger than the corresponding Bell polytope. 

Note that the KSNC constraints on the four graphs in \cref{fig:8pair} do not have all constraints given by Local Agency. In particular, Local Agency further demands certain constraints on the relationship between  $p_{\lnot a_3b_3}$, $p_{\lnot a_2b_2}$, $p_{\lnot a_2b_3}$ and $p_{\lnot a_3b_2}$, due to the fact that any two of the building blocks of the gluing trick (namely, $p_{22}$, $p_{23}$, $p_{32}$ and $p_{33}$) must agree on their overlap.
Thus, the inequalities we have derived here are not necessarily tight.
This can also be seen by comparing to the results in Ref.~\cite{bong2020strong} (where tight inequalities were derived numerically for a specific scenario); see Appendix C of Ref.~\cite{walleghem2024connecting} for more details.

Although the inequalities we have just derived are not necessarily tight (such as some of the ones in \Cref{sec:33}), and so do not fully characterize the set of correlations satisfying LF assumptions, they are nonetheless tight enough to admit violations within unitary quantum theory. Every LF inequality (whether it is tight or not) is also a valid Bell inequality, as a consequence of the fact that the LF polytope always contains the Bell polytope for the same setting and outcome variable cardinalities. 
If such a Bell inequality can be violated by quantum theory in a Bell scenario, then it, as a LF inequality, can also be violated by quantum theory.
This is because one can translate any quantum proposal for a $N_A,N_B$-setting Bell scenario violating a Bell inequality to a quantum proposal for the $N_A,N_B$-setting Local Friendliness scenario that violates the corresponding LF inequality. See Appendix D of Ref.~\cite{walleghem2024connecting} for more details.  

\vspace{0.5cm}

\subsection{Generalizations to multipartite sequential LF scenarios}
\label{sec:generalisation_LF_Bell}

So far, we have considered bipartite LF scenarios and characterized the relationship between the LF polytope in these scenarios and their corresponding KSNC or Bell polytopes. This analysis can be extended to more general LF scenarios involving multiple superobserver–observer pairs performing measurements sequentially (with arbitrarily long sequences). Furthermore, we can characterize the conditions under which all LF constraints in such a scenario can also be interpreted as KSNC constraints, and in particular, as Bell inequalities. We refer interested readers to Appendix E of Ref.~\cite{walleghem2024connecting} for more details.

\section[Constructing new LF scenarios from KSNC arguments]{Constructing new LF scenarios from KS noncontextuality arguments}
\sectionmark{New LF scenarios from KS noncontextuality arguments}
\label{sec:KSNC_to_LF}

In \Cref{sec:LFasKSNC}, we focused on showing how, given a specific LF scenario, one can derive many LF inequalities using known tools and results from the literature on KSNC, and on clarifying the relation between Bell local causality and LF. 
In this section, we show how given a possibilistic proof of the failure of KSNC one can construct an extended Wigner's friend argument---a no-go theorem against the Local Friendliness assumptions of AOE and Local Agency. 

Compared to the measurements of the underlying KSNC arguments, some additional measurements are introduced. 
Without these additional measurements, no-go theorems against AOE can be formulated, but only together with the assumption of Commutation Irrelevance of \Cref{sec:EWF_on_KS_not_nonlocal} replacing Local Agency.
Details can be found in Appendix K of Ref.~\cite{walleghem2024connecting}.

\vspace{0.5cm}

\subsection{The 5-cycle example} \label{sec:5-cycle}
In \Cref{sec:EWF_on_KS_not_nonlocal}, it was shown that the 5-cycle noncontextuality scenario~\cite{cabello2013simple,klyachko2008simple,santos2021conditions} can be used to construct an extended Wigner's friend argument with the assumption of Commutation Irrelevance, which is not one of the LF assumptions. 
Here, we show that the protocol of \Cref{sec:EWF_on_KS_not_nonlocal} can in fact be adapted to construct a novel scenario where one can prove the LF no-go theorem~\cite{bong2020strong} (cf. \Cref{th:LF}), that is, a no-go theorem based on the LF assumptions without using Commutation Irrelevance.

\tocless\subsubsection{The LF 5-cycle protocol}
\label{sec:5cycleprotocol}

\begin{figure}[htb!]
\centering
\tikzset{every picture/.style={line width=0.75pt}} 

\begin{tikzpicture}[x=0.75pt,y=0.75pt,yscale=-1,xscale=1]

\draw  [fill={rgb, 255:red, 255; green, 241; blue, 241 }  ,fill opacity=1 ] (150,35.53) .. controls (150,31.93) and (152.93,29) .. (156.53,29) -- (224.97,29) .. controls (228.57,29) and (231.5,31.93) .. (231.5,35.53) -- (231.5,202.27) .. controls (231.5,205.87) and (228.57,208.8) .. (224.97,208.8) -- (156.53,208.8) .. controls (152.93,208.8) and (150,205.87) .. (150,202.27) -- cycle ;
\draw  [fill={rgb, 255:red, 255; green, 241; blue, 241 }  ,fill opacity=1 ] (234.5,266.28) .. controls (234.5,265.57) and (235.07,265) .. (235.78,265) -- (360.22,265) .. controls (360.93,265) and (361.5,265.57) .. (361.5,266.28) -- (361.5,279.72) .. controls (361.5,280.43) and (360.93,281) .. (360.22,281) -- (235.78,281) .. controls (235.07,281) and (234.5,280.43) .. (234.5,279.72) -- cycle ;
\draw  [fill={rgb, 255:red, 255; green, 241; blue, 241 }  ,fill opacity=1 ] (375,129.32) .. controls (375,128.48) and (375.68,127.8) .. (376.52,127.8) -- (394.98,127.8) .. controls (395.82,127.8) and (396.5,128.48) .. (396.5,129.32) -- (396.5,145.28) .. controls (396.5,146.12) and (395.82,146.8) .. (394.98,146.8) -- (376.52,146.8) .. controls (375.68,146.8) and (375,146.12) .. (375,145.28) -- cycle ;
\draw    (216.11,233.47) -- (245.5,250) ;
\draw [shift={(213.5,232)}, rotate = 29.36] [fill={rgb, 255:red, 0; green, 0; blue, 0 }  ][line width=0.08]  [draw opacity=0] (10.72,-5.15) -- (0,0) -- (10.72,5.15) -- (7.12,0) -- cycle    ;
\draw    (201.5,209) -- (201.5,226) ;
\draw    (201.5,226) -- (217.5,235) ;
\draw    (369.76,228.34) -- (346.5,250) ;
\draw [shift={(371.95,226.29)}, rotate = 137.04] [fill={rgb, 255:red, 0; green, 0; blue, 0 }  ][line width=0.08]  [draw opacity=0] (10.72,-5.15) -- (0,0) -- (10.72,5.15) -- (7.12,0) -- cycle    ;
\draw    (382.5,147) -- (382.5,217.39) ;
\draw    (382.5,217.39) -- (369.77,229.24) ;
\draw    (245.5,162.4) -- (245.5,177.39) ;
\draw    (220.5,162.4) -- (245.5,162.4) ;
\draw [shift={(217.5,162.4)}, rotate = 360] [fill={rgb, 255:red, 0; green, 0; blue, 0 }  ][line width=0.08]  [draw opacity=0] (8.93,-4.29) -- (0,0) -- (8.93,4.29) -- cycle    ;
\draw    (245.5,118.4) -- (245.5,133.39) ;
\draw    (220.5,118.4) -- (245.5,118.4) ;
\draw [shift={(217.5,118.4)}, rotate = 360] [fill={rgb, 255:red, 0; green, 0; blue, 0 }  ][line width=0.08]  [draw opacity=0] (8.93,-4.29) -- (0,0) -- (8.93,4.29) -- cycle    ;
\draw    (130.5,184.4) -- (130.5,199.39) ;
\draw [shift={(130.5,181.4)}, rotate = 90] [fill={rgb, 255:red, 0; green, 0; blue, 0 }  ][line width=0.08]  [draw opacity=0] (8.93,-4.29) -- (0,0) -- (8.93,4.29) -- cycle    ;
\draw    (130.5,199.39) -- (158.5,199.39) ;
\draw    (130.5,116.4) -- (130.5,131.39) ;
\draw [shift={(130.5,113.4)}, rotate = 90] [fill={rgb, 255:red, 0; green, 0; blue, 0 }  ][line width=0.08]  [draw opacity=0] (8.93,-4.29) -- (0,0) -- (8.93,4.29) -- cycle    ;
\draw    (130.5,131.39) -- (158.5,131.39) ;
\draw    (131.5,71.4) -- (131.5,86.39) ;
\draw [shift={(131.5,68.4)}, rotate = 90] [fill={rgb, 255:red, 0; green, 0; blue, 0 }  ][line width=0.08]  [draw opacity=0] (8.93,-4.29) -- (0,0) -- (8.93,4.29) -- cycle    ;
\draw    (131.5,86.39) -- (159.5,86.39) ;
\draw    (389.5,149.4) -- (389.5,164.39) ;
\draw [shift={(389.5,146.4)}, rotate = 90] [fill={rgb, 255:red, 0; green, 0; blue, 0 }  ][line width=0.08]  [draw opacity=0] (8.93,-4.29) -- (0,0) -- (8.93,4.29) -- cycle    ;
\draw    (386.5,112.4) -- (386.5,127.39) ;
\draw [shift={(386.5,109.4)}, rotate = 90] [fill={rgb, 255:red, 0; green, 0; blue, 0 }  ][line width=0.08]  [draw opacity=0] (8.93,-4.29) -- (0,0) -- (8.93,4.29) -- cycle    ;
\draw    (245.5,250) -- (245.5,265.6) ;
\draw    (346.5,250) -- (346.5,265.6) ;
\draw    (246.5,72.4) -- (246.5,87.39) ;
\draw    (221.5,72.4) -- (246.5,72.4) ;
\draw [shift={(218.5,72.4)}, rotate = 360] [fill={rgb, 255:red, 0; green, 0; blue, 0 }  ][line width=0.08]  [draw opacity=0] (8.93,-4.29) -- (0,0) -- (8.93,4.29) -- cycle    ;
\draw    (130.5,158.4) -- (130.5,173.39) ;
\draw [shift={(130.5,155.4)}, rotate = 90] [fill={rgb, 255:red, 0; green, 0; blue, 0 }  ][line width=0.08]  [draw opacity=0] (8.93,-4.29) -- (0,0) -- (8.93,4.29) -- cycle    ;
\draw    (130.5,173.39) -- (158.5,173.39) ;
\draw    (134.5,27.4) -- (134.5,42.39) ;
\draw [shift={(134.5,24.4)}, rotate = 90] [fill={rgb, 255:red, 0; green, 0; blue, 0 }  ][line width=0.08]  [draw opacity=0] (8.93,-4.29) -- (0,0) -- (8.93,4.29) -- cycle    ;
\draw    (134.5,42.39) -- (162.5,42.39) ;

\draw (122,261) node [anchor=north west][inner sep=0.75pt]   [align=left] {Alice $\displaystyle \&$ Friends$ $};
\draw (369,262) node [anchor=north west][inner sep=0.75pt]   [align=left] {Bob};
\draw (240,177.8) node [anchor=north west][inner sep=0.75pt]   [align=left] {$\displaystyle x_{1}$};
\draw (240,133.8) node [anchor=north west][inner sep=0.75pt]   [align=left] {$\displaystyle x_{2}$};
\draw (104,178.8) node [anchor=north west][inner sep=0.75pt]   [align=left] {$\displaystyle f_{1}$};
\draw (103,107.8) node [anchor=north west][inner sep=0.75pt]   [align=left] {$\displaystyle f_{3}$};
\draw (105,64.8) node [anchor=north west][inner sep=0.75pt]   [align=left] {$\displaystyle f_{4}$};
\draw (160,186.8) node [anchor=north west][inner sep=0.75pt]   [align=left] {$\displaystyle U_{1}$};
\draw (194,136.8) node [anchor=north west][inner sep=0.75pt]   [align=left] {$\displaystyle U_{1}^{\dagger }$};
\draw (194,92.8) node [anchor=north west][inner sep=0.75pt]   [align=left] {$\displaystyle U_{2}^{\dagger }$};
\draw (161,75.8) node [anchor=north west][inner sep=0.75pt]   [align=left] {$\displaystyle U_{4}$};
\draw (387,164.8) node [anchor=north west][inner sep=0.75pt]   [align=left] {$\displaystyle y$};
\draw (377,89.8) node [anchor=north west][inner sep=0.75pt]   [align=left] {$\displaystyle b$};
\draw (329,245.8) node [anchor=north west][inner sep=0.75pt]   [align=left] {$\displaystyle T$};
\draw (248,247.8) node [anchor=north west][inner sep=0.75pt]   [align=left] {$\displaystyle S$};
\draw (241,87.8) node [anchor=north west][inner sep=0.75pt]   [align=left] {$\displaystyle x_{3}$};
\draw (158,163.8) node [anchor=north west][inner sep=0.75pt]   [align=left] {$\displaystyle U_{2}$};
\draw (105,149.8) node [anchor=north west][inner sep=0.75pt]   [align=left] {$\displaystyle f_{2}$};
\draw (159,120.8) node [anchor=north west][inner sep=0.75pt]   [align=left] {$\displaystyle U_{3}$};
\draw (194,49.8) node [anchor=north west][inner sep=0.75pt]   [align=left] {$\displaystyle U_{3}^{\dagger }$};
\draw (164,31.8) node [anchor=north west][inner sep=0.75pt]   [align=left] {$\displaystyle U_{5}$};
\draw (108,20.8) node [anchor=north west][inner sep=0.75pt]   [align=left] {$\displaystyle f_{5}$};

\end{tikzpicture}
\caption[The schematic representation of the LF 5-cycle scenario]{The schematic representation of the LF 5-cycle scenario, see also \Cref{fig_cartoon5cycle}.}
\label{fig_setusp5cycle}
\end{figure}
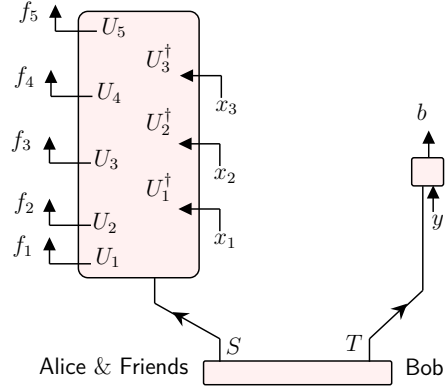

\begin{figure}[htb!]
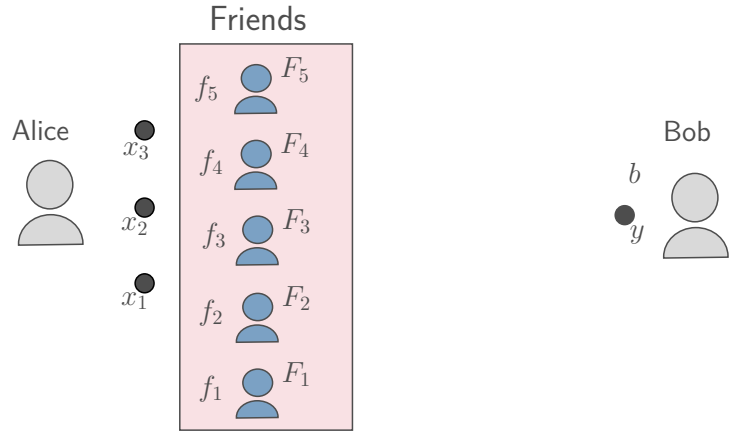

\centering
\includestandalone[width=0.65\textwidth]{Figures/Chap4_LF_5cycle}
\caption[Cartoon representation of the LF 5-cycle scenario]{Cartoon representation of the 5-cycle LF setup, with Alice a superobserver for $F_1,F_2,F_3,F_4
,F_5$, who perform measurements. Alice makes choices $x_1,x_2,x_3$ whether to reveal measurement outcomes and Bob, spacelike separated, makes measurement choice $y$ and obtains outcome $b$.} 
\label{fig_cartoon5cycle}
\end{figure}

We leverage the 5-cycle KSNC argument of \Cref{sec:5-cycle_paper1} to construct a novel scenario in which one can prove the LF no-go theorem.  Our presentation of this LF 5-cycle protocol highlights the differences from the protocol in \Cref{sec:EWF_on_KS_not_nonlocal}.

This new LF scenario involves two groups of agents. The first group consists of Alice (denoted $A$) and five friends (denoted $F_1$, ..., $F_5$). The second group consists of a single agent Bob.  See \Cref{fig_cartoon5cycle} for a cartoon representation of the set-up.  Compared to the protocol of \Cref{sec:EWF_on_KS_not_nonlocal}, we have one more agent here, namely, Bob; furthermore, the superobserver Alice here has measurement choices. These additions are crucial for deriving a no-go theorem based on LF assumptions without using the assumption of Commutation Irrelevance of \Cref{sec:EWF_on_KS_not_nonlocal}.

In the following, we will describe the quantum protocol that will be used to prove the no-go theorem. 
However, the scenario could also be described in a theory-independent way as discussed in Appendix A of Ref.~\cite{walleghem2024connecting}.  

At the beginning of the experiment, $F_1$ and Bob share a bipartite system $ST$ where $S$ is a qutrit and $T$ is a qubit. The joint system $ST$ is prepared in the following entangled state:\footnote{\label{ft:mp}One way to construct this entangled state \eqref{eq_5cycleini} is to let $F_1$ perform measurement $A_1$ on $\ket{\eta}_S$ and prepare $\ket{0}_T$ if her measurement outcome corresponds to $\Pi_{v_1^{\perp}}$ or $\ket{1}_T$ if her measurement outcome corresponds to $\ket{v_1}$, and then $F_1$ sends system $T$ to Bob. In fact, instead of starting with $F_1$ and $B$ sharing the entangled state \eqref{eq_5cycleini}, the protocol could start with $F_1$ having system $S$ prepared in $\ket{\eta}_S$, followed by the above measure-prepare procedure. All subsequent operations by Alice, the friends $F_i$ and Bob remain the same. In this alternative protocol, $F_1$'s operation is in the past-light-cone of everyone else's operations, while Bob's operations are still space-like separated from all operations on Alice's side except for $F_1$'s. This is similar to how the measure-prepare version of the Frauchiger--Renner argument~\cite{frauchiger2018quantum} as it was originally presented, is equivalent to the simpler entanglement-based scenario (see, e.g., Refs.~\cite{vilasini2022general,schmid2023review}).
}
\begin{equation} \label{eq_5cycleini}
    \begin{split}
        \ket{\Psi}_{ ST}
    = & \Pi_{v_1^\perp} \ket{\eta}_S \otimes \ket{0}_T + \braket{v_1}{\eta} \ket{v_1}_S   \otimes \ket{1}_T 
    \end{split}
\end{equation}
where $\Pi_{v_1^{\perp}}:=\mathbf{1}-\ketbra{v_1}$. System $S$ is sent to Alice's side while system $T$ is sent to Bob.  

On Alice's side, the protocol is similar to the one of \Cref{sec:EWF_on_KS_not_nonlocal}. 
Alice has five friends, $F_1$, ..., $F_5$, who perform the five measurements $A_i$ defined in \cref{eq:M} sequentially on the system $S$, with the superobserver's reversal of their measurements in between, i.e. applying the inverse unitary, as shown in Fig.~\ref{fig_setusp5cycle}. From AOE, each friend $F_i$ observes an absolute outcome, denoted $f_i$, during their measurement $A_i$. We label the $\ketbra{v_i}$ outcome by $f_i=1$ and the other outcome by $f_i=0$. In addition, the superobserver Alice has a binary choice $x_i$ right before she implements the inverse unitary $U^{\dagger}_i$ for $i=1,2,3$. If $x_i=1$, Alice opens the door to $F_i$'s and $F_{i+1}$'s labs and reveals their respective measurement outcome $f_i$ and $f_{i+1}$ to the world; if $x_i=2$, she simply continues the protocol.\footnote{Note that in the $x_i=2$ case, all records of the outcome $f_i$ are erased by the inverse unitaries applied by Alice; in contrast, in the $x_i=1$ case, the act of revealing the outcome $f_i$ to the world implies making and distributing many classical copies of it. In that case, the inverse unitaries performed by Alice do not have the effect of reverting the state of $S$ and the Friends to their initial states. They moreover will strongly impact outcomes observed by friends later in the protocol. However, all this is irrelevant, because in those cases the later outcomes of the friends are not used in the proof of the LF no-go theorem. Indeed, one could avoid these subtleties simply by stipulating that when $x_i=1$, Alice not only reveals the two immediately prior measurement outcomes to the world but also terminates the protocol.} When $x_1=x_2=x_3=2$, the $f_4$ and $f_5$ outcomes are revealed to the world at the end of the experiment.  

On Bob's side, Bob has a binary choice $y$. If $y=1$, he makes a measurement $B_{0/1}$ on system $T$, which is a computational basis measurement;
if $y=2$, he makes a measurement $B_{\pm}$ on system $T$, which is a $\pm$ basis measurement, i.e, $\{ \ketbra{+}, \ketbra{-}  \}$.
We denote Bob's absolute outcome by $b$.

We assume that all operations on Alice's side are done at space-like separation from Bob's measurement setting and outcome.

\begin{figure}[]
\begin{center}
\tikzset{every picture/.style={line width=0.75pt}} 

\begin{tikzpicture}[x=0.75pt,y=0.75pt,yscale=-1,xscale=1]

\draw  [fill={rgb, 255:red, 0; green, 0; blue, 0 }  ,fill opacity=1 ] (251.79,221) .. controls (251.79,218.24) and (253.81,216) .. (256.31,216) .. controls (258.8,216) and (260.82,218.24) .. (260.82,221) .. controls (260.82,223.76) and (258.8,226) .. (256.31,226) .. controls (253.81,226) and (251.79,223.76) .. (251.79,221) -- cycle ;
\draw  [fill={rgb, 255:red, 0; green, 0; blue, 0 }  ,fill opacity=1 ] (281.6,179) .. controls (281.6,176.24) and (283.62,174) .. (286.11,174) .. controls (288.61,174) and (290.63,176.24) .. (290.63,179) .. controls (290.63,181.76) and (288.61,184) .. (286.11,184) .. controls (283.62,184) and (281.6,181.76) .. (281.6,179) -- cycle ;
\draw  [fill={rgb, 255:red, 0; green, 0; blue, 0 }  ,fill opacity=1 ] (272.56,123) .. controls (272.56,120.24) and (274.59,118) .. (277.08,118) .. controls (279.57,118) and (281.6,120.24) .. (281.6,123) .. controls (281.6,125.76) and (279.57,128) .. (277.08,128) .. controls (274.59,128) and (272.56,125.76) .. (272.56,123) -- cycle ;
\draw  [fill={rgb, 255:red, 0; green, 0; blue, 0 }  ,fill opacity=1 ] (239.14,84) .. controls (239.14,81.24) and (241.17,79) .. (243.66,79) .. controls (246.15,79) and (248.18,81.24) .. (248.18,84) .. controls (248.18,86.76) and (246.15,89) .. (243.66,89) .. controls (241.17,89) and (239.14,86.76) .. (239.14,84) -- cycle ;
\draw  [fill={rgb, 255:red, 0; green, 0; blue, 0 }  ,fill opacity=1 ] (196.69,64) .. controls (196.69,61.24) and (198.71,59) .. (201.21,59) .. controls (203.7,59) and (205.72,61.24) .. (205.72,64) .. controls (205.72,66.76) and (203.7,69) .. (201.21,69) .. controls (198.71,69) and (196.69,66.76) .. (196.69,64) -- cycle ;
\draw    (286.11,179) -- (256.31,221) ;
\draw    (277.08,123) -- (286.11,179) ;
\draw    (243.66,84) -- (277.08,123) ;
\draw    (201.21,64) -- (242.76,86) ;
\draw  [fill={rgb, 255:red, 0; green, 0; blue, 0 }  ,fill opacity=1 ] (72.95,104) .. controls (72.95,101.24) and (74.97,99) .. (77.46,99) .. controls (79.96,99) and (81.98,101.24) .. (81.98,104) .. controls (81.98,106.76) and (79.96,109) .. (77.46,109) .. controls (74.97,109) and (72.95,106.76) .. (72.95,104) -- cycle ;
\draw  [fill={rgb, 255:red, 0; green, 0; blue, 0 }  ,fill opacity=1 ] (70.19,203) .. controls (70.19,200.24) and (72.22,198) .. (74.71,198) .. controls (77.2,198) and (79.23,200.24) .. (79.23,203) .. controls (79.23,205.76) and (77.2,208) .. (74.71,208) .. controls (72.22,208) and (70.19,205.76) .. (70.19,203) -- cycle ;
\draw    (201.21,64) -- (256.31,221) ;
\draw    (201.21,64) -- (77.46,104) ;
\draw    (256.31,221) -- (77.46,104) ;
\draw    (201.21,64) -- (74.71,203) ;
\draw    (256.31,221) -- (74.71,203) ;
\draw    (286.11,179) -- (77.46,104) ;
\draw    (277.08,123) -- (77.46,104) ;
\draw    (243.66,84) -- (77.46,104) ;
\draw    (286.11,179) -- (74.71,203) ;
\draw    (277.08,123) -- (74.71,203) ;
\draw    (243.66,84) -- (74.71,203) ;
\draw  [fill={rgb, 255:red, 0; green, 0; blue, 0 }  ,fill opacity=1 ] (592.6,221.72) .. controls (592.6,218.95) and (594.62,216.72) .. (597.11,216.72) .. controls (599.61,216.72) and (601.63,218.95) .. (601.63,221.72) .. controls (601.63,224.48) and (599.61,226.72) .. (597.11,226.72) .. controls (594.62,226.72) and (592.6,224.48) .. (592.6,221.72) -- cycle ;
\draw  [fill={rgb, 255:red, 0; green, 0; blue, 0 }  ,fill opacity=1 ] (622.4,179.72) .. controls (622.4,176.95) and (624.42,174.72) .. (626.92,174.72) .. controls (629.41,174.72) and (631.43,176.95) .. (631.43,179.72) .. controls (631.43,182.48) and (629.41,184.72) .. (626.92,184.72) .. controls (624.42,184.72) and (622.4,182.48) .. (622.4,179.72) -- cycle ;
\draw  [fill={rgb, 255:red, 0; green, 0; blue, 0 }  ,fill opacity=1 ] (613.37,123.72) .. controls (613.37,120.95) and (615.39,118.72) .. (617.89,118.72) .. controls (620.38,118.72) and (622.4,120.95) .. (622.4,123.72) .. controls (622.4,126.48) and (620.38,128.72) .. (617.89,128.72) .. controls (615.39,128.72) and (613.37,126.48) .. (613.37,123.72) -- cycle ;
\draw  [fill={rgb, 255:red, 0; green, 0; blue, 0 }  ,fill opacity=1 ] (579.95,84.72) .. controls (579.95,81.95) and (581.97,79.72) .. (584.47,79.72) .. controls (586.96,79.72) and (588.98,81.95) .. (588.98,84.72) .. controls (588.98,87.48) and (586.96,89.72) .. (584.47,89.72) .. controls (581.97,89.72) and (579.95,87.48) .. (579.95,84.72) -- cycle ;
\draw  [fill={rgb, 255:red, 0; green, 0; blue, 0 }  ,fill opacity=1 ] (537.5,64.72) .. controls (537.5,61.95) and (539.52,59.72) .. (542.01,59.72) .. controls (544.51,59.72) and (546.53,61.95) .. (546.53,64.72) .. controls (546.53,67.48) and (544.51,69.72) .. (542.01,69.72) .. controls (539.52,69.72) and (537.5,67.48) .. (537.5,64.72) -- cycle ;
\draw    (626.92,179.72) -- (597.11,221.72) ;
\draw    (617.89,123.72) -- (626.92,179.72) ;
\draw    (584.47,84.72) -- (617.89,123.72) ;
\draw    (542.01,64.72) -- (583.56,86.72) ;
\draw  [fill={rgb, 255:red, 0; green, 0; blue, 0 }  ,fill opacity=1 ] (413.75,104.72) .. controls (413.75,101.95) and (415.78,99.72) .. (418.27,99.72) .. controls (420.77,99.72) and (422.79,101.95) .. (422.79,104.72) .. controls (422.79,107.48) and (420.77,109.72) .. (418.27,109.72) .. controls (415.78,109.72) and (413.75,107.48) .. (413.75,104.72) -- cycle ;
\draw  [fill={rgb, 255:red, 0; green, 0; blue, 0 }  ,fill opacity=1 ] (411,203.72) .. controls (411,200.95) and (413.02,198.72) .. (415.52,198.72) .. controls (418.01,198.72) and (420.03,200.95) .. (420.03,203.72) .. controls (420.03,206.48) and (418.01,208.72) .. (415.52,208.72) .. controls (413.02,208.72) and (411,206.48) .. (411,203.72) -- cycle ;
\draw    (542.01,64.72) -- (418.27,104.72) ;
\draw    (597.11,221.72) -- (418.27,104.72) ;
\draw    (542.01,64.72) -- (415.52,203.72) ;
\draw    (597.11,221.72) -- (415.52,203.72) ;
\draw    (626.92,179.72) -- (418.27,104.72) ;
\draw    (617.89,123.72) -- (418.27,104.72) ;
\draw    (584.47,84.72) -- (418.27,104.72) ;
\draw    (626.92,179.72) -- (415.52,203.72) ;
\draw    (617.89,123.72) -- (415.52,203.72) ;
\draw    (584.47,84.72) -- (415.52,203.72) ;

\draw (257.05,219) node [anchor=north west][inner sep=0.75pt]  [font=\Large] [align=left] {$\displaystyle A_{1}$};
\draw (290.47,170) node [anchor=north west][inner sep=0.75pt]  [font=\Large] [align=left] {$\displaystyle A_{2}$};
\draw (284.53,102) node [anchor=north west][inner sep=0.75pt]  [font=\Large] [align=left] {$\displaystyle A_{3}$};
\draw (252.53,65) node [anchor=north west][inner sep=0.75pt]  [font=\Large] [align=left] {$\displaystyle A_{4}$};
\draw (200.76,33) node [anchor=north west][inner sep=0.75pt]  [font=\Large] [align=left] {$\displaystyle A_{5}$};
\draw (41.75,92) node [anchor=north west][inner sep=0.75pt]  [font=\Large] [align=left] {$\displaystyle B_{\pm }$};
\draw (33,192) node [anchor=north west][inner sep=0.75pt]  [font=\Large] [align=left] {$\displaystyle B_{0/1}$};
\draw (597.85,219.72) node [anchor=north west][inner sep=0.75pt]  [font=\Large] [align=left] {$\displaystyle A_{1}$};
\draw (631.27,170.72) node [anchor=north west][inner sep=0.75pt]  [font=\Large] [align=left] {$\displaystyle A_{2}$};
\draw (627.34,105.72) node [anchor=north west][inner sep=0.75pt]  [font=\Large] [align=left] {$\displaystyle A_{3}$};
\draw (591.34,63.72) node [anchor=north west][inner sep=0.75pt]  [font=\Large] [align=left] {$\displaystyle A_{4}$};
\draw (537.56,35.72) node [anchor=north west][inner sep=0.75pt]  [font=\Large] [align=left] {$\displaystyle A_{5}$};
\draw (382.56,94.72) node [anchor=north west][inner sep=0.75pt]  [font=\Large] [align=left] {$\displaystyle B_{\pm }$};
\draw (373.81,194.72) node [anchor=north west][inner sep=0.75pt]  [font=\Large] [align=left] {$\displaystyle B_{0/1}$};

\end{tikzpicture}  \caption[A compatibility graph for the measurements in the LF 5-cycle protocol]{
Left: A compatibility graph for the measurements in the LF 5-cycle protocol of \Cref{sec:5cycleprotocol}. Right: The 
 constraints from LF assumptions do not impose KSNC relative to the compatibility graph on the left, but they do effectively impose the same constraints as KSNC relative to the subgraph obtained when the compatibility relation between $A_1$ and $A_5$ is absent, as shown on the right. (They may impose further constraints in addition to these.)} \label{fig:LF_5-cycle_graphs} 
\end{center}     
\end{figure}
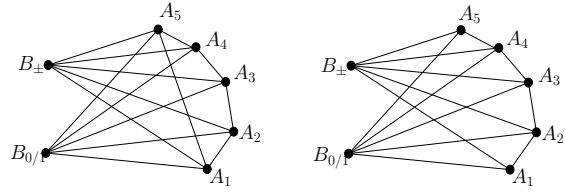

One can naturally associate a compatibility graph with this 5-cycle LF scenario: namely the one shown in the left subfigure of \cref{fig:LF_5-cycle_graphs}. The Local Friendliness assumptions on this LF 5-cycle scenario \emph{do not} imply KSNC on this compatibility graph, nor on the compatibility graph for the 5-cycle scenario (shown in \cref{fig:NC_5cycle}), since the joint distribution over $A_1$ and $A_5$ measurement outcomes is not observable.  Rather, the LF assumptions imply KSNC constraints on the graph in the right figure of \cref{fig:LF_5-cycle_graphs} (where the edge between $A_1$ and $A_5$ is absent). Nevertheless, as we show now, the possibilistic proof of the failure of KSNC in the 5-cycle scenario can still be mathematically turned into a proof of the Local Friendliness no-go theorem. 

\tocless\subsubsection{The no-go theorem}
\label{sec:5cycle_proof}

\begin{figure}[htb!]
\begin{center}
\includegraphics[width=0.4\textwidth]{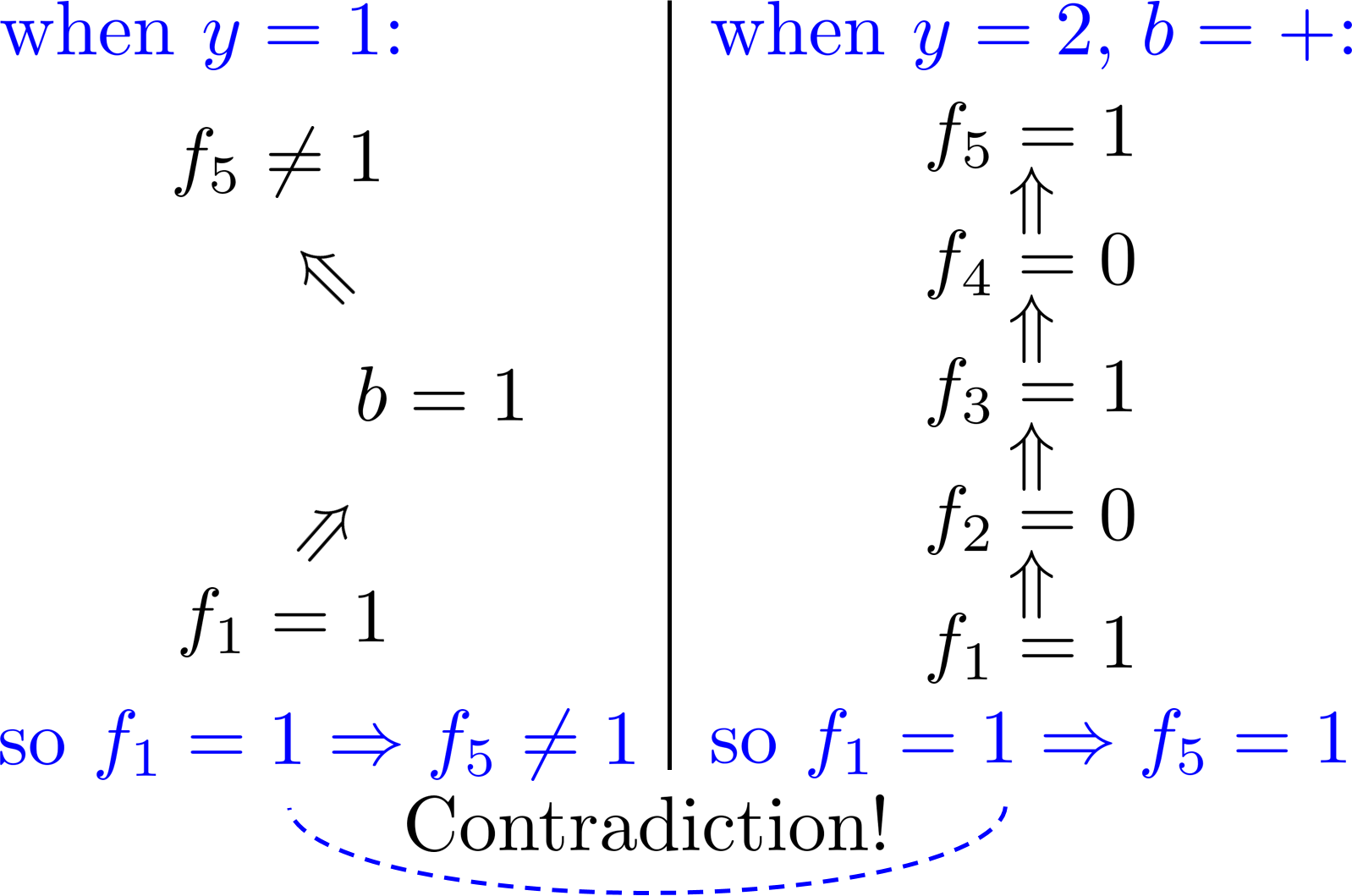}
\caption[The schematic proof using the LF 5-cycle protocol for the LF no-go theorem]{The schematic proof using the LF 5-cycle protocol for the LF no-go theorem.  The key steps in the proof are as follows. When Bob measures his qubit in the computational basis (i.e., when $y=1$, as in the left side of the figure),  his outcome $b$ must be the same as the $f_1$ outcome, which in turn constrains the correlation over $f_1$ and $f_5$. When Bob measures his qubit in the $\pm$ basis (i.e., when $y=2$) and obtains outcome $b=+$ (as in the right side of the figure), the effective state on Alice and her friends' side becomes $\ket{\eta}_{S}$, i.e., the state needed in the 5-cycle proof of KSNC theorem, and one obtains a contradictory constraint on the correlation between $f_1$ and $f_5$ under the Local Friendliness assumptions as Local Agency requires the correlation $f_1,f_5$ to be independent of $y$. 
}
\label{fig_summary5cycle}
\end{center}
\end{figure}

A schematic of the proof is given in Figure~\ref{fig_summary5cycle}. The proof begins by noting that AOE demands the existence of the conditional probability distribution
\begin{align}
\label{eq:aoe5cycle}
p(f_1,f_2,f_3,f_4,f_5,b|x_1,x_2,x_3,y),
\end{align}
exists (although of course, it is not accessible to any agent). Next, one considers the predictions made by quantum theory for various \emph{empirical} correlations in the LF 5-cycle protocol, and then applies Local Agency to them to obtain that
\begin{subequations} 
\begin{align}
& p(f_1=1,f_2=1|  x_{1-3}=2, y=2,b=+) = 0, \label{eq:ma12}\\
&p(f_2=0,f_3=0| x_{1-3}=2, y=2,b=+) = 0, \label{eq:ma23}\\
&P (f_3=1,f_4=1|  x_{1-3}=2, y=2,b=+) = 0, \label{eq:ma34} \\
&P (f_4=0,f_5=0| x_{1-3}=2, y=2,b=+)=0, \label{eq:ma45}
\end{align}
\end{subequations} 
and 
\begin{align}
\label{eq:ma15}
p(f_1=1,f_5=1|x_{1-3}=2,y=2,b=+)=0.
\end{align} (A derivation of these equations is given below.)
These five equations constitute constraints on five marginals of the joint distribution 
\begin{align}
    p(f_1,f_2,f_3,f_4,f_5|x_{1-3}=2,y=2,b=+),
\end{align}
which is well-defined due to the existence of \cref{eq:aoe5cycle} and the Born rule prediction that 
\begin{align}\label{eq:b+}
    \wp (b=+|x_{1-3}=2,y=2)>0
\end{align}
(a prediction that is empirically accessible).

However, these five constraints, i.e., \cref{eq:ma12,eq:ma23,eq:ma34,eq:ma45,eq:ma15},  cannot be simultaneously satisfied. This fact is the basis of the KSNC proof in the 5-cycle scenario; one can see it most easily by noting that these equations imply (respectively) that when $x_{1-3}=2, y=2,b=+$, we have 
\begin{align}
\label{eq:imp5}
    f_1=1\Rightarrow f_2=0 \Rightarrow f_3=1 \Rightarrow f_4=0 \Rightarrow f_5=1.
\end{align} 
and
\begin{align}
    f_1=1 \Rightarrow f_5\neq 1,
\end{align}
which are in contradiction. Therefore there can be no joint distribution of the form in Eq.~\eqref{eq:ma15} consistent with the empirical predictions given by the Born rule and Local Agency. Hence the no-go theorem.

It remains only to establish Eqs.~\eqref{eq:ma12}-\eqref{eq:ma15}. 

We begin with \cref{eq:ma12,eq:ma23,eq:ma34,eq:ma45}.
Consider the following empirical predictions given by the Born rule for the cases where Bob measures in $\pm$ basis and obtains outcome $+$:
\begin{subequations} 
\begin{align}
&\wp (f_1=1,f_2=1| x_{1}=1,x_2,x_3, y=2,b=+) = 0, \label{eq:em12}\\
&\wp(f_2=0,f_3=0| x_{1}=2,x_2=1,x_3, y=2,b=+) = 0, \label{eq:em23}\\
&\wp (f_3=1,f_4=1| x_{1-2}=2,x_3=1, y=2,b=+) = 0, \label{eq:em34} \\
&\wp(f_4=0,f_5=0| x_{1-3}=2, y=2,b=+)=0. \label{eq:em45}
\end{align}
\end{subequations} 
These are analogous to \cref{eq:5_cycle_12,eq:5_cycle_23,eq:5_cycle_34,eq:5_cycle_45}; indeed, a quick way to see that these are the Born rule predictions in the present scenario is to realize that when $y=2$ and $b=+$, the state $S$ on Alice's side is effectively $\ket{\eta}$ defined in \cref{eq:eta}.

For \cref{eq:em12}, since both $f_1$ and $f_2$ are in the past light cone of all Alice's choices,  they are not in the future light cone of $x_1,x_2,x_3$ and according to Local Agency, we have
\begin{align}
   & \wp (f_1,f_2| x_{1}=1,x_2,x_3, y=2,b=+) \nonumber \\
   = & p(f_1,f_2| x_{1-3}=2,y=2,b=+) 
\end{align}

Similarly, for \cref{eq:em23,eq:em34}, Local Agency demands that 
\begin{align}
    & \wp (f_2,f_3| x_{1}=2,x_2=1,x_3, y=2,b=+) \nonumber \\
   = & p(f_2,f_3| x_{1-3}=2,y=2,b=+),\\
    & \wp (f_3,f_4| x_{1-2}=2,x_3=1, y=2,b=+) \nonumber \\
   = & p(f_3,f_4| x_{1-3}=2,y=2,b=+)
\end{align}

Finally, \cref{eq:em45} is identical to Eq.~\eqref{eq:ma45}.

Therefore, together with \cref{eq:em12,eq:em23,eq:em34,eq:em45}, one obtains \cref{eq:ma12,eq:ma23,eq:ma34,eq:ma45}.

Now, we prove Eq.~\eqref{eq:ma15}.
To do so, we start with the following empirical predictions by the Born rule
\begin{align}
   & \wp(f_1=1,b=0|x_{1}=1,x_2, x_3,y=1)=0, \label{eq:ema1b}\\
   &\wp(f_5=1,b=1|x_{1-3}=2, y=1)=0. \label{eq:ema5b}
\end{align}
\cref{eq:ema1b} can be seen by noticing the perfect correlation between the outcome of $A_1$ on system $S$ and the outcome of computational basis measurement on system $T$ given by the initial state $\ket{\Psi_{ST}}$. \cref{eq:ema5b} can be seen by noticing that when the outcome on $T$ corresponds to $\ket{1}_T$, the state of $S$ is effectively $\ket{v_1}_S$, which is orthogonal to $\ket{v_5}_S$.

To relate these two empirical correlations, we use Local Agency again: Since neither $f_1$ nor $b$ is in the future light cone of $x_1,x_2,x_3$, Local Agency demands that \cref{eq:ema1b} satisfies
\begin{align}
    &\wp(f_1=1,b=0|x_{1}=1,x_2, x_3,y=1) \nonumber \\
    =& p(f_1=1,b=0|x_{1-3}=2,y=1)=0. \label{eq:a1b}
\end{align}

Now notice that both $f_1$ and $f_5$ are space-like separated from $y$; thus, according to Local Agency,
\begin{align}
    &p(f_1=1,f_5=1|x_{1-3}=2,y=1) \nonumber\\
    = &p(f_1=1,f_5=1|x_{1-3}=2,y=2)=0.
\end{align}
This implies that 
\begin{align}
    p(f_1=1,f_5=1,b=+|x_{1-3}=2,y=2)=0.
\end{align}
Using \cref{eq:b+}, we arrive at
\begin{align}
p(f_1=1,f_5=1|x_{1-3}=2,y=2,b=+)=0,
\end{align}
as claimed.

\vspace{0.5cm}

\subsection{The Peres--Mermin example} \label{sec:Peres-Mermin}

\tocless\subsubsection{The Peres--Mermin noncontextuality argument}
\label{sec:recap_PM}

We briefly recap the Peres--Mermin magic square~\cite{mermin1990simple,mermin1993hidden,peres1997quantum} proof of (state-independent) contextuality.
Consider the following square of quantum observables for a 4-dimensional qudit: \begin{equation} \label{eq:Peres_Mermin_array}
\left[\begin{array}{lll}
A & a & \alpha \\
B & b & \beta \\
C & c & \gamma
\end{array}\right]=\left[\begin{array}{ccc}
\sigma_z \otimes \mathbf{1} & \mathbf{1} \otimes \sigma_z & \sigma_z \otimes \sigma_z \\
1 \otimes \sigma_x & \sigma_x \otimes \mathbf{1} & \sigma_x \otimes \sigma_x \\
\sigma_z \otimes \sigma_x & \sigma_x \otimes \sigma_z & \sigma_y \otimes \sigma_y
\end{array}\right] .
\end{equation} 
Crucially, the triples of observables in a given column or row are all compatible with each other, and thus can be jointly measured. Thus the scenario consists of six possible joint measurements, called measurement contexts---one for each row and one for each column---and each observable appears in exactly two contexts (namely in one row and in one column). Moreover, the product of all three observables in any given context is  $\pm\mathbf{1}$:
\begin{equation} \label{eq:column_row_PM}
    \begin{split}
        A a \alpha = \mathbf{1}, \quad  B b \beta  = \mathbf{1}, \quad  C c \gamma = \mathbf{1}, \\
         ABC  = \mathbf{1}, \quad  abc  = \mathbf{1}, \quad \alpha \beta \gamma  = -\mathbf{1}.
    \end{split}
\end{equation} 

By KSNC, the values assigned to mutually commuting observables must obey equalities satisfied by the observables themselves, and so the product of the values assigned to the three observables in each row or column must be $\pm1$. More specifically, denoting the assignment to observable $O$ by $v_O\in \{-1,1\}$, one must have 
\begin{equation} 
    \begin{split}
         v_{A} v_{a} v_{\alpha}  = 1, \quad  v_{B} v_{b} v_{\beta}  = 1, \quad v_{C} v_{c} v_{\gamma} = 1, \\
          v_{A}v_{B}v_{C}  = 1, \quad  v_{a}v_{b}v_{c}  = 1, \quad v_{\alpha} v_{\beta} v_{\gamma} = -1.
    \end{split}
\end{equation} 
But these six equations cannot be satisfied by any nine $\pm1$-valued assignments, since the first three equalities imply that
\begin{equation}
v_{A}v_{a}v_{\alpha}v_{B}v_{b}v_{\beta}v_{C}v_{c}v_{\gamma} = 1,
\end{equation} 
while the last three imply that 
\begin{equation}
v_{A}v_{a}v_{\alpha}v_{B}v_{b}v_{\beta}v_{C}v_{c}v_{\gamma}=-1.
\end{equation}

The above arguments also establish a fact that we will utilize later, which is that the following 6 probability distributions 
\begin{subequations}
\label{eq:PMdistr}
\begin{align}
    p(v_Av_av_{\alpha}=-1)=0,\\ 
    p(v_Bv_bv_{\beta}=-1)=0, \\
    p(v_Cv_cv_{\gamma}=-1)=0, \\
    p(v_Av_Bv_C=-1)=0,\\
    p(v_av_bv_{c}=-1)=0, \\
    p(v_{\alpha}v_{\beta}v_{\gamma}=1)=0
\end{align}
\end{subequations}
cannot simultaneously be marginals of a joint probability distribution
\begin{align}
    p(v_A,v_a,v_{\alpha},v_B,v_b,v_{\beta},v_C,v_c,v_{\gamma}).
\end{align} 
(Indeed, every KSNC argument can be cast as a marginal problem of this sort~\cite{abramsky2011sheaf}.)

\tocless\subsubsection{The LF Peres--Mermin protocol} \label{sec:PeresMerminprotocol}

We now leverage the Peres--Mermin KSNC argument to construct a novel scenario in which one can prove the LF no-go theorem.\footnote{Another extended Wigner's friend argument based on the Peres--Mermin KSNC argument was introduced in Ref.~\cite{szangolies2023quantum}; see also Ref.~\cite{walleghem2023extended}, where we argue that an additional assumption that we call Commutation Irrelevance is required there.    
Furthermore, unlike the LF Peres–Mermin protocol we develop here, the protocol in 
Ref.~\cite{szangolies2023quantum} cannot be used to prove a LF no-go theorem. The crucial difference is that our protocol includes an additional group of agents, namely, the Bobs, and allows the superobserver Alice to have measurement choices.} 
This LF Peres--Mermin scenario involves 8 agents. The first two agents are Alice (denoted $A$) and her Friend (denoted $F$), of which Alice is the superobserver of the Friend. The remaining six agents are Bobs, denoted by $B_A$, $B_a$, $B_{\alpha}$, $B_{B}$, $B_{b}$, $B_{\beta}$ respectively. See \Cref{fig_cartoon_LF_PM} for a cartoon representation of the set-up. 
In the following, we will describe the quantum protocol that will be used to prove the no-go theorem and use the assumption of Universality of Quantum Theory and the Born rule for empirical correlations; however, the scenario could also be described in a theory-independent way if one wished to derive robust inequalities for empirical data rather than simply a no-go theorem.

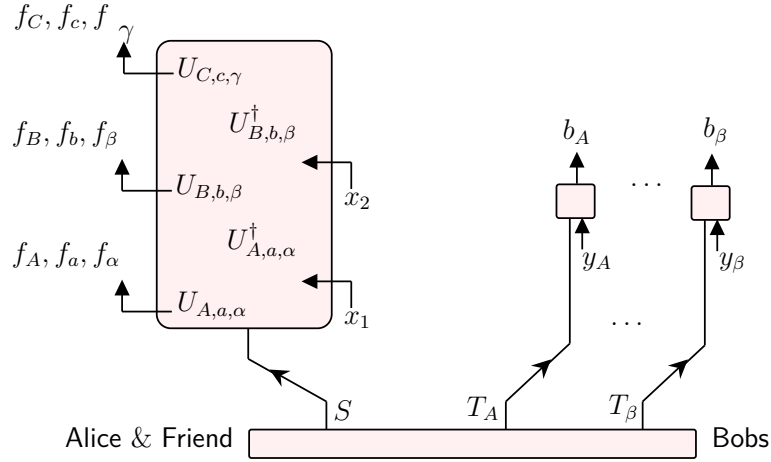
\begin{figure}[]
\centering
\tikzset{every picture/.style={line width=0.75pt}} 

\begin{tikzpicture}[x=0.75pt,y=0.75pt,yscale=-1,xscale=1]

\draw  [fill={rgb, 255:red, 255; green, 241; blue, 241 }  ,fill opacity=1 ] (130,46.9) .. controls (130,42.54) and (133.54,39) .. (137.9,39) -- (220.6,39) .. controls (224.96,39) and (228.5,42.54) .. (228.5,46.9) -- (228.5,192.9) .. controls (228.5,197.26) and (224.96,200.8) .. (220.6,200.8) -- (137.9,200.8) .. controls (133.54,200.8) and (130,197.26) .. (130,192.9) -- cycle ;
\draw  [fill={rgb, 255:red, 255; green, 241; blue, 241 }  ,fill opacity=1 ] (182,259.08) .. controls (182,258.37) and (182.57,257.8) .. (183.28,257.8) -- (432.22,257.8) .. controls (432.93,257.8) and (433.5,258.37) .. (433.5,259.08) -- (433.5,272.52) .. controls (433.5,273.23) and (432.93,273.8) .. (432.22,273.8) -- (183.28,273.8) .. controls (182.57,273.8) and (182,273.23) .. (182,272.52) -- cycle ;
\draw  [fill={rgb, 255:red, 255; green, 241; blue, 241 }  ,fill opacity=1 ] (355,121.32) .. controls (355,120.48) and (355.68,119.8) .. (356.52,119.8) -- (374.98,119.8) .. controls (375.82,119.8) and (376.5,120.48) .. (376.5,121.32) -- (376.5,137.28) .. controls (376.5,138.12) and (375.82,138.8) .. (374.98,138.8) -- (356.52,138.8) .. controls (355.68,138.8) and (355,138.12) .. (355,137.28) -- cycle ;
\draw  [fill={rgb, 255:red, 255; green, 241; blue, 241 }  ,fill opacity=1 ] (431,122.32) .. controls (431,121.48) and (431.68,120.8) .. (432.52,120.8) -- (450.98,120.8) .. controls (451.82,120.8) and (452.5,121.48) .. (452.5,122.32) -- (452.5,138.28) .. controls (452.5,139.12) and (451.82,139.8) .. (450.98,139.8) -- (432.52,139.8) .. controls (431.68,139.8) and (431,139.12) .. (431,138.28) -- cycle ;
\draw    (196.11,225.47) -- (225.5,242) ;
\draw [shift={(193.5,224)}, rotate = 29.36] [fill={rgb, 255:red, 0; green, 0; blue, 0 }  ][line width=0.08]  [draw opacity=0] (10.72,-5.15) -- (0,0) -- (10.72,5.15) -- (7.12,0) -- cycle    ;
\draw    (181.5,201) -- (181.5,218) ;
\draw    (181.5,218) -- (197.5,227) ;
\draw    (349.76,220.34) -- (326.5,242) ;
\draw [shift={(351.95,218.29)}, rotate = 137.04] [fill={rgb, 255:red, 0; green, 0; blue, 0 }  ][line width=0.08]  [draw opacity=0] (10.72,-5.15) -- (0,0) -- (10.72,5.15) -- (7.12,0) -- cycle    ;
\draw    (362.5,139) -- (362.5,209.39) ;
\draw    (362.5,209.39) -- (349.77,221.24) ;
\draw    (426.76,220.34) -- (410.94,235.07) -- (403.5,242) ;
\draw [shift={(428.95,218.29)}, rotate = 137.04] [fill={rgb, 255:red, 0; green, 0; blue, 0 }  ][line width=0.08]  [draw opacity=0] (10.72,-5.15) -- (0,0) -- (10.72,5.15) -- (7.12,0) -- cycle    ;
\draw    (438.5,140) -- (438.5,210.39) ;
\draw    (438.5,210.39) -- (425.77,222.24) ;
\draw    (239.5,174.4) -- (239.5,189.39) ;
\draw    (214.5,174.4) -- (239.5,174.4) ;
\draw [shift={(211.5,174.4)}, rotate = 360] [fill={rgb, 255:red, 0; green, 0; blue, 0 }  ][line width=0.08]  [draw opacity=0] (8.93,-4.29) -- (0,0) -- (8.93,4.29) -- cycle    ;
\draw    (239.5,107.4) -- (239.5,122.39) ;
\draw    (214.5,107.4) -- (239.5,107.4) ;
\draw [shift={(211.5,107.4)}, rotate = 360] [fill={rgb, 255:red, 0; green, 0; blue, 0 }  ][line width=0.08]  [draw opacity=0] (8.93,-4.29) -- (0,0) -- (8.93,4.29) -- cycle    ;
\draw    (110.5,176.4) -- (110.5,191.39) ;
\draw [shift={(110.5,173.4)}, rotate = 90] [fill={rgb, 255:red, 0; green, 0; blue, 0 }  ][line width=0.08]  [draw opacity=0] (8.93,-4.29) -- (0,0) -- (8.93,4.29) -- cycle    ;
\draw    (110.5,191.39) -- (138.5,191.39) ;
\draw    (110.5,108.4) -- (110.5,123.39) ;
\draw [shift={(110.5,105.4)}, rotate = 90] [fill={rgb, 255:red, 0; green, 0; blue, 0 }  ][line width=0.08]  [draw opacity=0] (8.93,-4.29) -- (0,0) -- (8.93,4.29) -- cycle    ;
\draw    (110.5,123.39) -- (138.5,123.39) ;
\draw    (111.5,42.4) -- (111.5,57.39) ;
\draw [shift={(111.5,39.4)}, rotate = 90] [fill={rgb, 255:red, 0; green, 0; blue, 0 }  ][line width=0.08]  [draw opacity=0] (8.93,-4.29) -- (0,0) -- (8.93,4.29) -- cycle    ;
\draw    (111.5,57.39) -- (139.5,57.39) ;
\draw    (369.5,141.4) -- (369.5,156.39) ;
\draw [shift={(369.5,138.4)}, rotate = 90] [fill={rgb, 255:red, 0; green, 0; blue, 0 }  ][line width=0.08]  [draw opacity=0] (8.93,-4.29) -- (0,0) -- (8.93,4.29) -- cycle    ;
\draw    (445.5,142.4) -- (445.5,157.39) ;
\draw [shift={(445.5,139.4)}, rotate = 90] [fill={rgb, 255:red, 0; green, 0; blue, 0 }  ][line width=0.08]  [draw opacity=0] (8.93,-4.29) -- (0,0) -- (8.93,4.29) -- cycle    ;
\draw    (366.5,104.4) -- (366.5,119.39) ;
\draw [shift={(366.5,101.4)}, rotate = 90] [fill={rgb, 255:red, 0; green, 0; blue, 0 }  ][line width=0.08]  [draw opacity=0] (8.93,-4.29) -- (0,0) -- (8.93,4.29) -- cycle    ;
\draw    (442.5,105.4) -- (442.5,120.39) ;
\draw [shift={(442.5,102.4)}, rotate = 90] [fill={rgb, 255:red, 0; green, 0; blue, 0 }  ][line width=0.08]  [draw opacity=0] (8.93,-4.29) -- (0,0) -- (8.93,4.29) -- cycle    ;
\draw    (225.5,242) -- (225.5,257.6) ;
\draw    (326.5,242) -- (326.5,257.6) ;
\draw    (403.5,242) -- (403.5,257.6) ;

\draw (77,255) node [anchor=north west][inner sep=0.75pt]   [align=left] {Alice $\displaystyle \&$ Friend$ $};
\draw (441,255) node [anchor=north west][inner sep=0.75pt]   [align=left] {Bobs};
\draw (234,189.8) node [anchor=north west][inner sep=0.75pt]   [align=left] {$\displaystyle x_{1}$};
\draw (234,122.8) node [anchor=north west][inner sep=0.75pt]   [align=left] {$\displaystyle x_{2}$};
\draw (47,150.8) node [anchor=north west][inner sep=0.75pt]   [align=left] {$\displaystyle f_{A} ,f_{a} ,f_{\alpha }$};
\draw (47,82.8) node [anchor=north west][inner sep=0.75pt]   [align=left] {$\displaystyle f_{B} ,f_{b} ,f_{\beta }$};
\draw (48,16.8) node [anchor=north west][inner sep=0.75pt]   [align=left] {$\displaystyle f_{C} ,f_{c} ,f_{ \begin{array}{{>{\displaystyle}l}}
\gamma \\
\end{array}}$};
\draw (140,178.8) node [anchor=north west][inner sep=0.75pt]   [align=left] {$\displaystyle U_{A,a,\alpha }$};
\draw (168,139.8) node [anchor=north west][inner sep=0.75pt]   [align=left] {$\displaystyle U_{A,a,\alpha }^{\dagger }$};
\draw (139,112.8) node [anchor=north west][inner sep=0.75pt]   [align=left] {$\displaystyle U_{B,b,\beta }$};
\draw (169,73.8) node [anchor=north west][inner sep=0.75pt]   [align=left] {$\displaystyle U_{B,b,\beta }^{\dagger }$};
\draw (140,46.8) node [anchor=north west][inner sep=0.75pt]   [align=left] {$\displaystyle U_{C,c,\gamma }$};
\draw (368,156.8) node [anchor=north west][inner sep=0.75pt]   [align=left] {$\displaystyle y_{A}$};
\draw (445,156.8) node [anchor=north west][inner sep=0.75pt]   [align=left] {$\displaystyle y_{\beta }$};
\draw (436,80.8) node [anchor=north west][inner sep=0.75pt]   [align=left] {$\displaystyle b_{\beta }$};
\draw (357,81.8) node [anchor=north west][inner sep=0.75pt]   [align=left] {$\displaystyle b_{A}$};
\draw (303,237.8) node [anchor=north west][inner sep=0.75pt]   [align=left] {$\displaystyle T_{A}$};
\draw (383,236.8) node [anchor=north west][inner sep=0.75pt]   [align=left] {$\displaystyle T_{\beta }$};
\draw (228,239.8) node [anchor=north west][inner sep=0.75pt]   [align=left] {$\displaystyle S$};
\draw (395,117.8) node [anchor=north west][inner sep=0.75pt]   [align=left] {$\displaystyle \dotsc $};
\draw (384,195.8) node [anchor=north west][inner sep=0.75pt]   [align=left] {$\displaystyle \dotsc $};

\end{tikzpicture}
\caption[Schematic diagram of the LF Peres--Mermin scenario]{Schematic diagram of the LF Peres--Mermin scenario, see also \Cref{fig_cartoon_LF_PM}. All operations performed by Alice and Friend are space-like separated from all operations performed by the six Bobs; all operations performed by the six Bobs are further mutually space-like separated.
}
\label{fig_setuspPM}
\end{figure}

At the beginning of the experiment, $F$ and the six Bobs share a 7-partite system $ST$ where $S$ is a qudit and $T$ consists of 6 subsystems (denoted $T_j$ for $j\in \mathbf{j}\coloneqq\{A,a,\alpha, B,b,\beta\}$),
each of which is a qubit. The joint system $ST$ is prepared in the following entangled state:\footnote{This initial entangled state, i.e., \cref{eq:initialgiant}, can be constructed by
\begin{align}
\label{eb_construini0}
   \ket{\psi_0}_{ST} = \left( \prod_{j=1}^6 {\mathrm CNOT}_{ ST_j} \right) \ket{\psi} _{S} \bigotimes_{j=1}^{6} \ket{0}_{T_j}
\end{align}
where 
\begin{align}
\label{eb_creatST0}
   & {\mathrm CNOT}_{ ST_j} \coloneqq (\Pi_{j^{\perp}})_S\otimes\mathbf{1}_{T_j}+ (\Pi_j)_{ S}\otimes ( \ketbra{1}{0} + \ketbra{0}{1})_{T_j}. 
\end{align}
Similar to \cref{ft:mp}, this entangled state can be constructed in a measure-prepare fashion around which the thought experiment could alternatively be constructed (although some extra complexity is introduced by the fact that since, as illustrated in \cref{fig_setuspPM}, the $B,b,\beta$ measurements are not done at the initial step of the protocol, and thus, more care needs to be taken regarding the space-time arrangement of the procedures).
} 
\begin{align}
\label{eq:initialgiant}
   &\ket{\psi_0}_{ST} \coloneqq  \sum_{k_j= 0, 1, \forall j \in \mathbf{j}}
   \left( \prod_{j\in\mathbf{j}}\Pi_{{j}^{k_j}} 
\ket{\psi} \right)_{S} \bigotimes_{j\in\mathbf{j}} \ket{k_j}_{T_j},
\end{align}
where $\Pi_{j^{0}}=\Pi_{j^{\perp}}$ is the projector into the subspace of the qudit $S$ that yields outcome $-1$ for the observable $j\in \mathbf{j}$, while $\Pi_{j^{1}}=\Pi_{j}=\mathbf{1}-\Pi_{j^{\perp}}$, and $\ket{\psi}$ can be \emph{any} 4-qubit quantum state. 

System $S$ is sent to the Friend while each $T_j$ is sent to the corresponding $B_j$. On Alice's side, the Friend will measure the 9 observables $\{i\}_{i\in\mathbf{i}\coloneqq\{A,a,\alpha, B,b,\beta,C,c,\gamma\}}$ on system $S$. Specifically, she will first measure $\{A,a,\alpha\}$, then $\{B,b,\beta\}$, and finally $\{C,c,\gamma\}$. Whether she implements \emph{each} triple of observables using a joint measurement (since they commute) or in a minimally disturbing sequential manner is not important for our proof of the LF no-go theorem. In either case, from AOE, the Friend will obtain an outcome, denoted $f_i$, for each observable $i\in\mathbf{i}$. (In the case where each triple of observables is measured using a single joint measurement, the $f_i$s are the corresponding coarse-grained outcomes of the joint measurement.)  We label the $-1$ outcome for the observable $i$ by $f_i=-1$, and the other outcome by $f_i=1$. 

Under the assumption of Universality of Quantum Theory, we use $U_{A,a,\alpha}$, $U_{B,b,\beta}$, and $U_{C,c,\gamma}$ to denote the respective unitaries modeling the measurements for each triple of observables. The superobserver Alice is assumed to have perfect quantum control over the joint system of $F$ and $S$. She will implement the inverse unitaries $U_{A,a,\alpha}^{\dagger}$ and $U_{B,b,\beta}^{\dagger}$ at specific points in time during the protocol.

The exact sequence of these quantum operations is shown in the schematic representation of the protocol in Fig.~\ref{fig_setuspPM}.
Specifically, first, $F$ measures $A$, $a$, and $\alpha$ on $S$, after which the superobserver Alice undoes these measurements by applying $U_{A,a,\alpha}^{\dagger}$. Then $F$ measures $B$, $b$ and $\beta$ on $S$, followed by the undoing $U_{B,b,\beta}^{\dagger}$. Finally, $F$ measures $C$, $c$, and $\gamma$ on $S$.

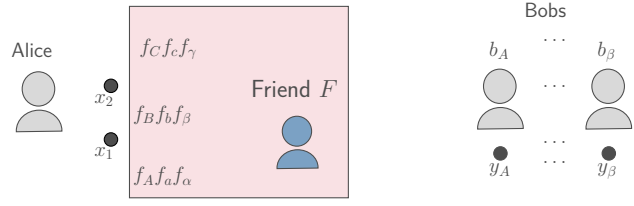
\begin{figure}[]
\centering
\tikzset{every picture/.style={line width=0.75pt}} 

\begin{tikzpicture}[x=0.75pt,y=0.75pt,yscale=-1,xscale=1]

\draw  [color={rgb, 255:red, 77; green, 77; blue, 77 }  ,draw opacity=1 ][fill={rgb, 255:red, 249; green, 225; blue, 228 }  ,fill opacity=1 ][line width=0.75]  (154.5,108.2) -- (358.5,108.2) -- (358.5,286.6) -- (154.5,286.6) -- cycle ;
\draw  [color={rgb, 255:red, 77; green, 77; blue, 77 }  ,draw opacity=1 ][fill={rgb, 255:red, 217; green, 217; blue, 217 }  ,fill opacity=1 ][line width=0.75]  (54.5,185.14) .. controls (54.5,176.32) and (61.29,169.17) .. (69.67,169.17) .. controls (78.05,169.17) and (84.85,176.32) .. (84.85,185.14) .. controls (84.85,193.96) and (78.05,201.11) .. (69.67,201.11) .. controls (61.29,201.11) and (54.5,193.96) .. (54.5,185.14) -- cycle ;
\draw  [color={rgb, 255:red, 77; green, 77; blue, 77 }  ,draw opacity=1 ][fill={rgb, 255:red, 217; green, 217; blue, 217 }  ,fill opacity=1 ][line width=0.75]  (49.05,225.05) .. controls (49.04,224.84) and (49.04,224.64) .. (49.03,224.44) .. controls (48.89,213.85) and (58.28,205.13) .. (70.01,204.95) .. controls (81.75,204.77) and (91.38,213.2) .. (91.53,223.79) .. controls (91.53,224.01) and (91.53,224.24) .. (91.52,224.46) -- cycle ;
\draw  [color={rgb, 255:red, 77; green, 77; blue, 77 }  ,draw opacity=1 ][fill={rgb, 255:red, 217; green, 217; blue, 217 }  ,fill opacity=1 ][line width=0.75]  (484.46,182.67) .. controls (484.46,173.85) and (491.25,166.7) .. (499.63,166.7) .. controls (508.02,166.7) and (514.81,173.85) .. (514.81,182.67) .. controls (514.81,191.49) and (508.02,198.64) .. (499.63,198.64) .. controls (491.25,198.64) and (484.46,191.49) .. (484.46,182.67) -- cycle ;
\draw  [color={rgb, 255:red, 77; green, 77; blue, 77 }  ,draw opacity=1 ][fill={rgb, 255:red, 217; green, 217; blue, 217 }  ,fill opacity=1 ][line width=0.75]  (479.01,222.57) .. controls (479,222.37) and (479,222.17) .. (479,221.96) .. controls (478.85,211.38) and (488.24,202.65) .. (499.97,202.47) .. controls (511.71,202.29) and (521.34,210.73) .. (521.49,221.31) .. controls (521.49,221.54) and (521.49,221.76) .. (521.48,221.99) -- cycle ;
\draw  [fill={rgb, 255:red, 77; green, 77; blue, 77 }  ,fill opacity=1 ] (131,232.25) .. controls (131,228.8) and (133.8,226) .. (137.25,226) .. controls (140.7,226) and (143.5,228.8) .. (143.5,232.25) .. controls (143.5,235.7) and (140.7,238.5) .. (137.25,238.5) .. controls (133.8,238.5) and (131,235.7) .. (131,232.25) -- cycle ;
\draw  [fill={rgb, 255:red, 77; green, 77; blue, 77 }  ,fill opacity=1 ] (131,182.25) .. controls (131,178.8) and (133.8,176) .. (137.25,176) .. controls (140.7,176) and (143.5,178.8) .. (143.5,182.25) .. controls (143.5,185.7) and (140.7,188.5) .. (137.25,188.5) .. controls (133.8,188.5) and (131,185.7) .. (131,182.25) -- cycle ;
\draw  [color={rgb, 255:red, 77; green, 77; blue, 77 }  ,draw opacity=1 ][fill={rgb, 255:red, 77; green, 77; blue, 77 }  ,fill opacity=1 ] (494,245.25) .. controls (494,241.8) and (496.8,239) .. (500.25,239) .. controls (503.7,239) and (506.5,241.8) .. (506.5,245.25) .. controls (506.5,248.7) and (503.7,251.5) .. (500.25,251.5) .. controls (496.8,251.5) and (494,248.7) .. (494,245.25) -- cycle ;
\draw  [color={rgb, 255:red, 77; green, 77; blue, 77 }  ,draw opacity=1 ][fill={rgb, 255:red, 217; green, 217; blue, 217 }  ,fill opacity=1 ][line width=0.75]  (585.46,182.67) .. controls (585.46,173.85) and (592.25,166.7) .. (600.63,166.7) .. controls (609.02,166.7) and (615.81,173.85) .. (615.81,182.67) .. controls (615.81,191.49) and (609.02,198.64) .. (600.63,198.64) .. controls (592.25,198.64) and (585.46,191.49) .. (585.46,182.67) -- cycle ;
\draw  [color={rgb, 255:red, 77; green, 77; blue, 77 }  ,draw opacity=1 ][fill={rgb, 255:red, 217; green, 217; blue, 217 }  ,fill opacity=1 ][line width=0.75]  (580.01,222.57) .. controls (580,222.37) and (580,222.17) .. (580,221.96) .. controls (579.85,211.38) and (589.24,202.65) .. (600.97,202.47) .. controls (612.71,202.29) and (622.34,210.73) .. (622.49,221.31) .. controls (622.49,221.54) and (622.49,221.76) .. (622.48,221.99) -- cycle ;
\draw  [color={rgb, 255:red, 77; green, 77; blue, 77 }  ,draw opacity=1 ][fill={rgb, 255:red, 77; green, 77; blue, 77 }  ,fill opacity=1 ] (595,245.25) .. controls (595,241.8) and (597.8,239) .. (601.25,239) .. controls (604.7,239) and (607.5,241.8) .. (607.5,245.25) .. controls (607.5,248.7) and (604.7,251.5) .. (601.25,251.5) .. controls (597.8,251.5) and (595,248.7) .. (595,245.25) -- cycle ;
\draw  [color={rgb, 255:red, 77; green, 77; blue, 77 }  ,draw opacity=1 ][fill={rgb, 255:red, 123; green, 161; blue, 196 }  ,fill opacity=1 ][line width=0.75]  (296.92,224.55) .. controls (296.92,216.84) and (303.04,210.6) .. (310.59,210.6) .. controls (318.13,210.6) and (324.25,216.84) .. (324.25,224.55) .. controls (324.25,232.25) and (318.13,238.5) .. (310.59,238.5) .. controls (303.04,238.5) and (296.92,232.25) .. (296.92,224.55) -- cycle ;
\draw  [color={rgb, 255:red, 77; green, 77; blue, 77 }  ,draw opacity=1 ][fill={rgb, 255:red, 123; green, 161; blue, 196 }  ,fill opacity=1 ][line width=0.75]  (292.03,259.41) .. controls (292.02,259.23) and (292.01,259.05) .. (292.01,258.86) .. controls (291.88,249.62) and (300.33,242) .. (310.89,241.84) .. controls (321.46,241.69) and (330.13,249.05) .. (330.26,258.3) .. controls (330.26,258.5) and (330.26,258.7) .. (330.25,258.9) -- cycle ;

\draw (43,140.77) node [anchor=north west][inner sep=0.75pt]  [font=\large,color={rgb, 255:red, 77; green, 77; blue, 77 }  ,opacity=1 ] [align=left] {{\large Alice}};
\draw (521.23,103.22) node [anchor=north west][inner sep=0.75pt]  [font=\large,color={rgb, 255:red, 77; green, 77; blue, 77 }  ,opacity=1 ] [align=left] {{\large Bobs}};
\draw (156.33,199.51) node [anchor=north west][inner sep=0.75pt]  [font=\large,color={rgb, 255:red, 77; green, 77; blue, 77 }  ,opacity=1 ] [align=left] {$\displaystyle f_{B} f_{b} f_{\beta }$};
\draw (162.33,136.51) node [anchor=north west][inner sep=0.75pt]  [font=\large,color={rgb, 255:red, 77; green, 77; blue, 77 }  ,opacity=1 ] [align=left] {$\displaystyle f_{C} f_{c} f_{\gamma }$};
\draw (120.33,188.51) node [anchor=north west][inner sep=0.75pt]  [font=\large,color={rgb, 255:red, 77; green, 77; blue, 77 }  ,opacity=1 ] [align=left] {$\displaystyle x_{2}$};
\draw (120.33,237.51) node [anchor=north west][inner sep=0.75pt]  [font=\large,color={rgb, 255:red, 77; green, 77; blue, 77 }  ,opacity=1 ] [align=left] {$\displaystyle x_{1}$};
\draw (488.25,253.25) node [anchor=north west][inner sep=0.75pt]  [font=\large,color={rgb, 255:red, 77; green, 77; blue, 77 }  ,opacity=1 ] [align=left] {$\displaystyle y_{A}$};
\draw (488.25,139.25) node [anchor=north west][inner sep=0.75pt]  [font=\large,color={rgb, 255:red, 77; green, 77; blue, 77 }  ,opacity=1 ] [align=left] {$\displaystyle b_{A}$};
\draw (266.33,174.51) node [anchor=north west][inner sep=0.75pt]  [font=\Large,color={rgb, 255:red, 77; green, 77; blue, 77 }  ,opacity=1 ] [align=left] {Friend $\displaystyle F$};
\draw (156.33,256.51) node [anchor=north west][inner sep=0.75pt]  [font=\large,color={rgb, 255:red, 77; green, 77; blue, 77 }  ,opacity=1 ] [align=left] {$\displaystyle f_{A} f_{a} f_{\alpha }$};
\draw (589.25,253.25) node [anchor=north west][inner sep=0.75pt]  [font=\large,color={rgb, 255:red, 77; green, 77; blue, 77 }  ,opacity=1 ] [align=left] {$\displaystyle y_{\beta }$};
\draw (589.25,139.25) node [anchor=north west][inner sep=0.75pt]  [font=\large,color={rgb, 255:red, 77; green, 77; blue, 77 }  ,opacity=1 ] [align=left] {$\displaystyle b_{\beta }$};
\draw (538.25,180.25) node [anchor=north west][inner sep=0.75pt]  [font=\large,color={rgb, 255:red, 77; green, 77; blue, 77 }  ,opacity=1 ] [align=left] {$\displaystyle \dotsc $};
\draw (538.25,232.25) node [anchor=north west][inner sep=0.75pt]  [font=\large,color={rgb, 255:red, 77; green, 77; blue, 77 }  ,opacity=1 ] [align=left] {$\displaystyle \dotsc $};
\draw (538.25,251.25) node [anchor=north west][inner sep=0.75pt]  [font=\large,color={rgb, 255:red, 77; green, 77; blue, 77 }  ,opacity=1 ] [align=left] {$\displaystyle \dotsc $};
\draw (538.25,137.25) node [anchor=north west][inner sep=0.75pt]  [font=\large,color={rgb, 255:red, 77; green, 77; blue, 77 }  ,opacity=1 ] [align=left] {$\displaystyle \dotsc $};

\end{tikzpicture}
\caption[Cartoon representation of the LF Peres--Mermin scenario]{Cartoon representation of the LF Peres--Mermin scenario, see also \Cref{fig_setuspPM}. Alice is a superobserver for her Friend, making choices $x_1,x_2$ whether to reveal the measurement outcomes obtained by the Friend, $f_A,f_a,f_\alpha$ and $f_B,f_b,f_\beta$, respectively. The spacelike separated Bobs make measurement choices $y_A,\ldots,y_\beta$ and obtain outcomes $b_A,\ldots,b_\beta$.}
\label{fig_cartoon_LF_PM}
\end{figure}

In addition, the superobserver Alice has two binary choices $x_1$ and $x_2$. The $x_1$ choice is made right before she implements the inverse unitary $U_{A,a,{\alpha}}^{\dagger}$, and the $x_2$ choice is made right before she implements the inverse unitary $U_{B,b,{\beta}}^{\dagger}$. If $x_1=1$, Alice reveals measurement outcomes $f_A$, $f_a$ and $f_{\alpha}$ to the world; if $x_1=2$, she simply continues the protocol.  Similarly, when $x_2=1$, Alice reveals measurement outcomes $f_B$, $f_b$ and $f_{\beta}$ to the world; if $x_2=2$, she simply continues the protocol. 
If $x_1=x_2=2$, the $f_C$, $f_c$ and $f_{\gamma}$ outcomes are revealed to the world at the end of the experiment.\footnote{Similar to the LF 5-cycle case, any measurement outcome obtained after $x_n=2$ is not used for the proof of the LF no-go theorem, so equivalently, whenever $x_n=2$, Alice could also terminate the protocol on her side.} 

For each Bob $B_j$ with any $j\in\mathbf{j}$, he has a binary measurement choice $y_j$. If $y_j=1$, he measures system $T_j$ in the computational basis; if $y_j=2$, he measures system $T$ in the $\pm$ basis, i.e.
\begin{align}
    \{ \ketbra{+}, \ketbra{-}  \}.
\end{align}
We denote $B_j$'s respective absolute outcome by $b_j$. 

We assume that all operations on Alice's side (including Alice and her Friend) are done at space-like separation from all operations by the Bobs; furthermore, all of the operations by the different Bobs are mutually space-like separated from each other.

Similar to the LF 5-cycle scenario, one can show that the Local Friendliness constraints on the LF Peres--Mermin scenario do not imply KSNC constraints on the compatibility graph associated with either the LF Peres--Mermin scenario or the KSNC Peres--Mermin scenario. Nevertheless, now we will show in \Cref{sec:proof_LFPM} that the possibilistic proof of the failure of KSNC in the Peres--Mermin scenario can be mathematically turned
into a proof of Local Friendliness no-go theorem.

\vspace{0.5cm}

\tocless\subsubsection{The no-go theorem}
\label{sec:proof_LFPM}

In this protocol, AOE demands the existence of the following conditional probability distribution:
\begin{align}
\label{eq:PMaoe}
    P&(f_{\mathbf{i}}, b_{\mathbf{j}}|x_1,x_2, y_{\mathbf{j}}) \\ & \coloneqq 
    p(f_A, f_a, f_{\alpha}, f_B, f_b, f_{\beta}, f_C, f_c, f_{\gamma}, b_A, b_a, b_{\alpha}, b_B, b_b, b_{\beta} 
    |x_1,x_2, y_A, y_a, y_{\alpha}, y_B, y_b, y_{\beta}) \nonumber
\end{align}
where $f_{\mathbf{i}}$,  $b_{\mathbf{j}}$, and $y_{\mathbf{j}}$ are the respective short-hand notations for \emph{all} elements in $\{f_i\}_{i\in\mathbf{i}}$, $\{b_j\}_{j\in\mathbf{j}}$, and that in $\{y_i\}_{j\in\mathbf{j}}$. 

Below, we will consider the predictions made by quantum theory for various \emph{empirical} correlations in the Peres--Mermin LF protocol, and we will then apply Local Agency to them to show that
\begin{subequations}
    \begin{align}
        p(f_A  f_{a} f_{\alpha} = -1  &|x_1=x_2=2, y_{\mathbf{j}}=2, b_{\mathbf{j}}=+)=0, \label{eq:Aaalpha}\\
        p(f_B  f_b f_{\beta} = -1  &|x_1=x_2=2, y_{\mathbf{j}}=2, b_{\mathbf{j}}=+)=0, \label{eq:Bbbeta} \\
        p(f_C  f_c f_{\gamma} = -1  &|x_1=x_2=2, y_{\mathbf{j}}=2, b_{\mathbf{j}}=+)=0, \label{eq:Ccgamma}\\
        p(f_A  f_B f_C = -1  &|x_1=x_2=2, y_{\mathbf{j}}=2, b_{\mathbf{j}}=+)=0, \label{eq:ABC} \\
        p(f_a  f_b f_c = -1 &|x_1=x_2=2, y_{\mathbf{j}}=2, b_{\mathbf{j}}=+)=0, \label{eq:abc}\\
        p(f_{\alpha}  f_{\beta} f_{\gamma} = 1  &|x_1=x_2=2, y_{\mathbf{j}}=2, b_{\mathbf{j}}=+)=0, \label{eq:alphabetagamma}
    \end{align}
\end{subequations}
where $y_{\mathbf{j}}=2$ is the short-hand notation for $y_j=2, \forall j \in \mathbf{j}$, and similarly for $b_{\mathbf{j}}=+$. 

These six equations constitute constraints on six marginals of the joint distribution
\begin{align}
\label{eq:x1y1qplus}
    p(&f_{\mathbf{i}}|x_1=x_2=2, y_{\mathbf{j}}=2, b_{\mathbf{j}}=+),
\end{align}
which is well-defined due to the existence of \cref{eq:PMaoe} together with the empirical predictions given by the Born rule that
\begin{align}
\label{eq:qplus}
    \wp(b_{\mathbf{j}}=+|x_1=x_2=2, y_{\mathbf{j}}=2 )>0.
\end{align}

However, from the Peres--Mermin KSNC proof in \Cref{sec:recap_PM}, we know that these six constraints, i.e., \cref{eq:Aaalpha,eq:Bbbeta,eq:Ccgamma,eq:ABC,eq:abc,eq:alphabetagamma}, cannot be simultaneously satisfied (as they are analogous to Eqs~\eqref{eq:PMdistr}).
 Therefore, there can be no joint distribution of the form in \cref{eq:x1y1qplus} that is consistent with the empirical predictions given by the Born rule and Local Agency. Hence the no-go theorem.

It remains only to establish \cref{eq:Aaalpha,eq:Bbbeta,eq:Ccgamma,eq:ABC,eq:abc,eq:alphabetagamma}. The steps for proving \cref{eq:Aaalpha,eq:Bbbeta,eq:Ccgamma} are similar to those for proving \cref{eq:ma12,eq:ma23,eq:ma34,eq:ma45} in \Cref{sec:5cycle_proof}, and the steps for proving \cref{eq:ABC,eq:ABC,eq:alphabetagamma} are similar to those for proving \cref{eq:ma15} in \Cref{sec:5cycle_proof}. A proof can be found in Appendix I of Ref.~\cite{walleghem2024connecting}.

\vspace{0.5cm}

\subsection{Translating any possibilistic KSNC argument into a LF argument } \label{sec:Commutation_Friendlinessmain}

We have now seen two examples of how possibilistic proofs of the failure of KSNC can be translated into LF arguments. More generally, one has the following:

\begin{restatable}{theorem}{LFpossKSNC}
    \label{th:LFpossKSNC}
Every quantum possibilistic proof of the failure of Kochen--Specker noncontextuality can be mathematically translated into a proof of the failure of Local Friendliness in a corresponding extended Wigner's friend scenario.
\end{restatable} 

We prove this in Appendix J of Ref.~\cite{walleghem2024connecting}, using a construction analogous to the above constructions (where all Bobs only need to perform either the computational basis measurement or the complementary $\pm$ measurement). 

Note that in these LF constructions, one introduces extra measurements beyond those in the contextuality scenarios\footnote{It is commonplace for translations between different scenarios to involve slightly different states and measurements; see for example Ref.~\cite{Kunjwal_2015,PhysRevLett.129.240401,cabello2021converting} for other examples.};
in particular, the measurements made by Bobs on qubits $T_j$. 
However, one can avoid the addition of these additional measurements (and any choices of Alice or of the Bobs) if one appeals to the assumption of \textit{Commutation Irrelevance} of \Cref{sec:EWF_on_KS_not_nonlocal} rather than Local Agency. 
For details, see Appendix K of Ref.~\cite{walleghem2024connecting}. 

One could alternatively provide a different sort of translation from KSNC arguments to LF arguments by first translating the KS construction into a Bell scenario using known procedures~\cite{cabello2021converting,cabello2010proposal,aolita2012fully,cabello2011proposal}, 
and then using the mapping from Bell scenarios to LF scenarios given by reversing the translation given in Section~\ref{sec:generalisation_LF_Bell}. (Note that the translation from a KSNC argument to a Bell argument also involves adding additional measurements in the case where the initial KSNC argument is state-dependent~\cite{cabello2021converting}.)

\section{An extended Wigner's friend paradox based on generalized contextuality} \label{sec:EWF_based_on_generalised_contextuality}
\sectionmark{EWF paradox based on generalized contextuality}

The Local Friendliness (LF) no-go theorem~\cite{bong2020strong,cavalcanti2021implications,haddara2022possibilistic} shows a contradiction between the predictions of quantum theory and the assumptions of Absoluteness of Observed Events -- ascribing single and objective values to observed measurement outcomes  -- and Local Agency -- a strengthened version of no-signaling that also applies to situations involving outcome correlations that cannot be verified by a single agent.
As discussed in \Cref{chap:Q_prelims}, the LF no-go theorem can be conceptualized as combining a Wigner's friend experiment with a Bell scenario and, assuming that the LF scenario can be realized, results in an even stronger theorem than Bell's theorem, given that the LF assumptions are weaker than the ones of Bell's theorem~\cite{wiseman2023thoughtful}. One way to ensure that the LF scenario can be realized is to assume the Universality of Quantum Theory, in that superobservers exist.
The sense in which the LF no-go theorem is said to be stronger than Bell's theorem is exemplified in Ref.~\cite{haddara2024local}, where the authors show that, in standard LF scenarios, the polytope of correlations consistent with Local Friendliness (the conjunction of Absoluteness of Observed  Events and Local Agency) is strictly larger than the polytope of locally causal correlations.

In this section, we combine Wigner's friend experiments with scenarios manifesting generalized contextuality~\cite{spekkens2005contextuality}, one of the leading notions of nonclassicality. 
Specifically, we consider the simplest scenario showing generalized contextuality~\cite{pusey2018robust}, which is a prepare-and-measure scenario on a single system involving four preparations and two measurements, and we include quantum observers.
This theorem demonstrates the inconsistency of quantum theory with the joint assumptions of Absoluteness of Observed Events and \textit{Noncontextual Agency}, the latter being a weaker version of noncontextuality, in the same way that Local Agency is a weaker version of local causality. Our result generalizes the Local Friendliness no-go theorem and, granted that the scenario can be realized, is stronger than no-go theorems based on generalized noncontextuality.
In designing the experiment we leverage a known correspondence between proofs of nonlocality in Bell scenarios and proofs of contextuality in prepare-and-measure scenarios~\cite{liang2010specker,coecke_kissinger_2017,Henaut2018,CataniFaleiro2023, Wright_2023}. In the same way in which, under the assumption that the scenario can be realized, the LF no-go theorem  is stronger than Bell's theorem, we obtain a stronger no-go theorem than the one associated with generalized contextuality.

This section contains several copied parts of Ref.~\cite{walleghem2025extendedwignersfriendnogo}.

\subsection{Bell nonlocality and generalized contextuality}  
In a Bell scenario, two parties, Alice and Bob, share a bipartite system and each perform a measurement on their respective parts.
Alice's and Bob's measurement choices are denoted by $x$ and $y$, respectively, with outcomes $a$ and $b$.
Repeating the protocol many times allows for the collection of measurement statistics $\wp(a,b|x,y)$.\footnote{We here introduce, similarly as in the previous section, a notational difference between empirical probability distributions that can be obtained by a single agent from the collected data, denoted with $\wp$, and theoretical probability distributions required to exist given certain assumptions, but potentially not observable by any agent, denoted with $p$.} These correlations are consistent with the existence of an ontological model satisfying \textit{local causality}~\cite{bell1964einstein} if they can be expressed as follows:
\begin{equation}
\label{eq:locality}
    \wp(a,b|x,y) = \sum_\lambda p(\lambda) p(a|x,\lambda)p(b|y,\lambda),
\end{equation}
where $\lambda$ represents the ontic state in the ontological model. 
An equivalent way to require that Eq.~\eqref{eq:locality} is satisfied is to demand the existence of a joint distribution 
$p(a_1,a_2,b_1,b_2)$, where $a_x=a|x, b_y = b|y$, 
for all possible measurement outcomes and settings, from which the observed probabilities $\wp(a,b|x,y)$ can be obtained by marginalizing over the unperformed measurements~\cite{fine1982joint,abramsky2011sheaf}, for example, $\wp(a,b|x=2,y=2) = \sum_{a_1,b_1}p(a_1,a_2,b_1,b_2)$. 
As we have encountered several times in the previous sections and chapters, Bell's theorem famously proves that quantum theory manifests \textit{nonlocality}, \textit{i.e.}, it does not admit of a locally causal ontological model~\cite{bell1964einstein}.

A Bell scenario can also be conceptualized as a prepare-and-measure scenario, where the preparation stage on Bob's side is determined by steering from Alice's setting and outcome $(x,a)$, and the measurement stage consists of Bob's measurement $y$ and outcome $b$~\cite{coecke_kissinger_2017,Henaut2018,CataniFaleiro2023,Wright_2023} (see \cref{fig:Bell_to_PM}). In this setting, the nonlocality of quantum theory turns into the inconsistency of quantum theory with a generalized noncontextual ontological model~\cite{spekkens2005contextuality}.

\begin{figure}
         \centering \includegraphics[width=0.55\textwidth,height=0.13\textheight]{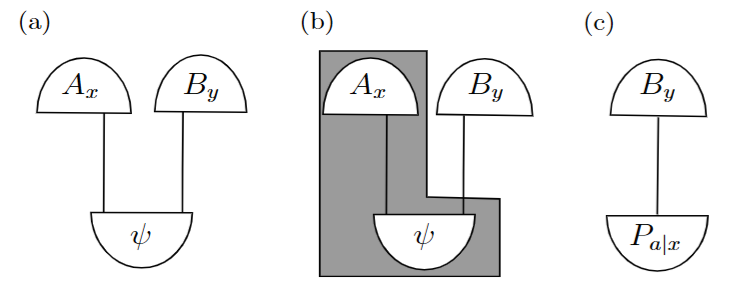}
         \caption[From Bell nonlocality to prepare-and-measure scenarios]{(a) Bell scenario; (b) Bell scenario as a prepare-and-measure scenario; (c) prepare-and-measure scenario. Subscripts denote the preparation and measurement variables, and $\psi$ the bipartite state.} 
         \label{fig:Bell_to_PM}
\end{figure}


As introduced in \Cref{sec:nonclassicality}, generalized noncontextuality requires that operational equivalences predicted by a physical theory are preserved in the ontological model of the theory. 
In the context of local causality, no-signaling is the operational equivalence that is preserved in the ontological model. Specifically, with respect to $a$ in a Bell scenario,  no-signaling implies $\wp(a|x,y)=\wp(a|x)$ and its preservation in the ontological model reads as $p(a|\lambda,x,y) = p(a|\lambda,x).$ 
This assumption, known as parameter independence, combined with outcome independence -- \textit{e.g.}, with respect to $a$,   $p(a|\lambda,b,x,y)=p(a|\lambda,x,y)$ -- defines local causality\footnote{{Note that the assumption of measurement independence, $p(\lambda|x,y)=p(\lambda)$, which is also required to prove Bell’s theorem, is already built into the definition of ontological models~\cite{catani2023mathematical}.}}.

The simplest proof that quantum theory manifests \textit{contextuality}, \textit{i.e.}, does not admit of a generalized noncontextual ontological model, can be constructed in the \textit{simplest nontrivial scenario}~\cite{pusey2018robust,pusey2019,Khoshbin_2024}. This is a prepare-and-measure scenario consisting of two tomographically complete measurements, \textit{e.g.}, the Pauli $X$ and $Z$ measurements on a qubit, 
 and four preparations, \textit{e.g.}, $P_{\theta}= \cos(\theta/2)\ket{0}+\sin(\theta/2)\ket{1}$, for $\theta = \pi/4,3\pi/4,5\pi/4,7\pi/4$. These are represented on the Bloch disk in ~\cref{fig:Bloch}. 
The preparations satisfy the following operational equivalence, \begin{align}
    &\frac{1}{2}\wp(k|M,P_\frac{\pi}{4}) +  \frac{1}{2}\wp(k|M,P_\frac{5\pi}{4}) \nonumber
    =  \frac{1}{2}\wp(k|M,P_\frac{3\pi}{4}) +   \frac{1}{2}\wp(k|M,P_\frac{7\pi}{4}).
\end{align} 
However, there does not exist an ontological model where these also provide an ontological equivalence.

\begin{figure}[h]
         \centering \includegraphics[width=0.29\textwidth]{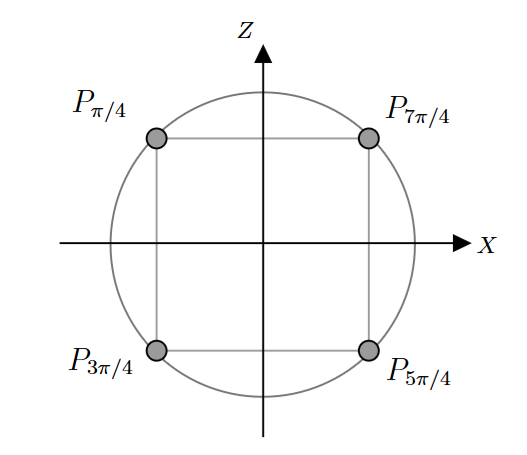}
         \caption[Simplest nontrivial scenario for generalized contextuality]{The simplest nontrivial scenario~\cite{pusey2018robust},
         involving four preparations (gray dots) and two measurements (the two axes) that provide a maximal violation of the noncontextuality inequalities.}
         \label{fig:Bloch}
\end{figure}

A strategy obtaining a maximum quantum violation in Bell scenario can be mapped to the simplest scenario with the specific choices of preparations and measurements above. In the former case, one shows nonlocality; in the latter, preparation contextuality.

\subsection{Rephrasing the Local Agency assumption}

Recall the Local Friendliness scenario of \Cref{sec:LF}, illustrated in \cref{fig:LF_cartoon_Chap4}, involving Charlie and Debbie and their superobservers Alice and Bob who make choices $x,y$.

\begin{figure}[h]
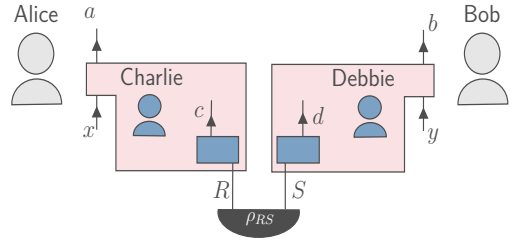

         \centering \includestandalone[width=0.45\textwidth,height=0.142\textheight]{LFsetup}
         \caption[Local Friendliness set-up]{Local Friendliness setup~\cite{schmid2023review}. 
         }
         \label{fig:LF_cartoon_Chap4}
\end{figure}

Given the setup, let us rephrase the assumption of Local Agency.

\begin{assumption}[Local Agency] \label{ass:LA}
No-signaling outside the future light cone, which would be verified by a hypothetical agent with access to all the relevant variables, still holds even if it cannot be verified by a single agent.
\end{assumption}

\noindent The above formulation of Local Agency is equivalent to the standard one~\cite{cavalcanti2021implications}: \begin{quote}
``The only relevant events correlated with an intervention are in its future light cone.''
\end{quote}

\noindent Indeed, the choice settings (``the interventions'') can only be correlated  with the relevant events that lie within their future light cone, \textit{even if all the events cannot be observed by a single agent}. 
In this sense, Local Agency can be read as a stronger version of standard no-signaling.
In the LF scenario, Local Agency implies, for example, that $p(c,b|x,y=2) = p(c,b|y=2)$, as none of $c,b$ are obtained in the future of the choice $x$. 
Notice how this is different from requiring the disjoint conditions to hold: $\wp(b|x,y=2)=\wp(b|y=2)$ -- an instance of no-signaling -- and $\wp(c|x,y)=\wp(c)$ -- an instance of no-signaling to the past, or no-superdeterminism.
Local Agency is stronger than standard no-signaling and no-superdeterminism insofar as it requires the joint independence, which, in addition to the no-signaling and no-superdeterminism above, also requires $p(b|c,x,y=2)=p(b|c,y=2)$.

Local Agency can be motivated by a commitment to relativistic principles and the belief that these should hold even if they cannot be directly verified by a single agent. It can also be justified as an instance of no operational fine-tuning ~\cite{catani2023mathematical} in a universe where AOE holds: if events are absolute and a theory prescribes no-signaling, the latter should also hold even if it cannot be verified by a single agent. 
Put differently, no-signaling should hold even if its verification were possible only for a hypothetical agent {who could access all relevant variables by cheating the protocol.} Let us call this agent Eve and imagine that she has access to outcomes $b,c$ 
and choices $x$ for $y=2$. 
Beyond the standard no-signaling instances $\wp(b|x,y=2) = \wp(b|y=2)$ and $\wp(c|x,y)=\wp(c)$, {Eve can now verify an additional no-signaling instance}: $p(b,c|x=2,y=2) = p(b,c|x=1,y=2)$, \textit{i.e.}, if Eve were to access only $b,c,y=2$, she would not get any information on $x$. 
Local Agency demands that such no-signaling conditions verified by Eve 
must remain valid even in her absence, {and even if the relevant events are jointly inaccessible in the sense that they are not jointly observed by any agent, yet still exist by virtue of AOE.}

It is worth noting that while Local Agency is stronger than no-signaling, it is weaker than local causality, as shown in Ref.~\cite{cavalcanti2021implications}. 
This is further evident in scenarios where the polytope of correlations implied by Local Friendliness (the conjunction of AOE and Local Agency) is strictly larger than the polytope of locally causal correlations, as demonstrated in Ref.~\cite{haddara2024local}.

\subsection{Noncontextual Friendliness no-go theorem}  
The experimental setup, illustrated in \cref{fig:OF_tryout}, is described as follows. 
Alice prepares a qubit $S$ in the state $P_a$, for values of $a=0,1$ each happening with probability $1/2$.
She sends the prepared system to her friend Charlie, who performs a binary measurement $C$ on $S$, where the two possible outcomes are labeled by $0$ and $1$. The measurement is modeled unitarily as $U_C$. 
After Charlie has performed his measurement, depending on the value of the variable $x\in\{1,2\}$, Alice makes one of two choices.
If $x=1$ she takes Charlie's outcome to be her outcome. If $x=2$ she acts as a superobserver and undoes Charlie's measurement by applying $U_C^\dagger$, thereby erasing the outcome record of $c$. Notice that, in this latter case, unlike the LF setup, she does not perform a measurement $A$ on $S$, after erasing the outcome of Charlie. 
Subsequently, the system is passed to Debbie who performs a measurement $D$ on the system. This measurement is modeled unitarily as $U_D$.
After Debbie has performed her measurement, depending on the value of the variable $y\in\{1,2\}$, Bob makes one of two choices.
If $y=1$ he takes Debbie's outcome to be his outcome. If $y=2$ he acts as a superobserver and undoes Debbie's measurement by applying $U_D^\dagger$, thereby erasing the outcome record of $d$, and subsequently performs a measurement denoted with $B$ on $S$ yielding outcome $b$.

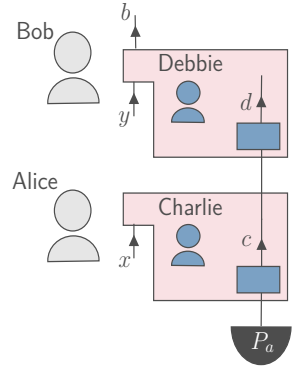
\begin{figure}[h!]
         \centering \tikzset{every picture/.style={line width=0.75pt}} 

\begin{tikzpicture}[x=0.75pt,y=0.75pt,yscale=-1,xscale=1]

\draw  [color={rgb, 255:red, 77; green, 77; blue, 77 }  ,draw opacity=1 ][fill={rgb, 255:red, 77; green, 77; blue, 77 }  ,fill opacity=1 ][line width=0.75]  (282.32,260.61) .. controls (282.56,276.9) and (293.49,290.02) .. (307,290.05) .. controls (320.51,290.08) and (331.55,277.01) .. (331.89,260.72) -- cycle ;
\draw  [color={rgb, 255:red, 77; green, 77; blue, 77 }  ,draw opacity=1 ][fill={rgb, 255:red, 230; green, 230; blue, 230 }  ,fill opacity=1 ][line width=0.75]  (135.5,169.67) .. controls (135.5,160.85) and (142.29,153.7) .. (150.67,153.7) .. controls (159.05,153.7) and (165.85,160.85) .. (165.85,169.67) .. controls (165.85,178.49) and (159.05,185.64) .. (150.67,185.64) .. controls (142.29,185.64) and (135.5,178.49) .. (135.5,169.67) -- cycle ;
\draw  [color={rgb, 255:red, 77; green, 77; blue, 77 }  ,draw opacity=1 ][fill={rgb, 255:red, 230; green, 230; blue, 230 }  ,fill opacity=1 ][line width=0.75]  (130.05,209.58) .. controls (130.04,209.37) and (130.04,209.17) .. (130.03,208.97) .. controls (129.89,198.38) and (139.28,189.66) .. (151.01,189.48) .. controls (162.75,189.3) and (172.38,197.73) .. (172.53,208.32) .. controls (172.53,208.54) and (172.53,208.77) .. (172.52,208.99) -- cycle ;
\draw  [color={rgb, 255:red, 77; green, 77; blue, 77 }  ,draw opacity=1 ][fill={rgb, 255:red, 230; green, 230; blue, 230 }  ,fill opacity=1 ][line width=0.75]  (135.46,46.2) .. controls (135.46,37.38) and (142.25,30.23) .. (150.63,30.23) .. controls (159.02,30.23) and (165.81,37.38) .. (165.81,46.2) .. controls (165.81,55.02) and (159.02,62.17) .. (150.63,62.17) .. controls (142.25,62.17) and (135.46,55.02) .. (135.46,46.2) -- cycle ;
\draw  [color={rgb, 255:red, 77; green, 77; blue, 77 }  ,draw opacity=1 ][fill={rgb, 255:red, 230; green, 230; blue, 230 }  ,fill opacity=1 ][line width=0.75]  (130.01,86.1) .. controls (130,85.9) and (130,85.7) .. (130,85.49) .. controls (129.85,74.91) and (139.24,66.18) .. (150.97,66) .. controls (162.71,65.82) and (172.34,74.26) .. (172.49,84.84) .. controls (172.49,85.07) and (172.49,85.29) .. (172.48,85.52) -- cycle ;
\draw  [color={rgb, 255:red, 77; green, 77; blue, 77 }  ,draw opacity=1 ][fill={rgb, 255:red, 249; green, 225; blue, 228 }  ,fill opacity=1 ] (192.68,156) -- (192.68,181.2) -- (217.88,181.2) -- (217.88,240) -- (329,240) -- (329,156) -- cycle ;
\draw  [color={rgb, 255:red, 77; green, 77; blue, 77 }  ,draw opacity=1 ][fill={rgb, 255:red, 123; green, 161; blue, 196 }  ,fill opacity=1 ] (256.34,189.61) .. controls (256.34,184.63) and (252.02,180.6) .. (246.68,180.6) .. controls (241.35,180.6) and (237.02,184.63) .. (237.02,189.61) .. controls (237.02,194.58) and (241.35,198.61) .. (246.68,198.61) .. controls (252.02,198.61) and (256.34,194.58) .. (256.34,189.61) -- cycle ;
\draw  [color={rgb, 255:red, 77; green, 77; blue, 77 }  ,draw opacity=1 ][fill={rgb, 255:red, 123; green, 161; blue, 196 }  ,fill opacity=1 ] (259.82,212.84) .. controls (259.83,212.71) and (259.83,212.58) .. (259.84,212.45) .. controls (259.93,206.49) and (253.95,201.57) .. (246.49,201.46) .. controls (239.02,201.36) and (232.89,206.12) .. (232.79,212.09) .. controls (232.79,212.23) and (232.79,212.37) .. (232.79,212.52) -- cycle ;
\draw [color={rgb, 255:red, 77; green, 77; blue, 77 }  ,draw opacity=1 ]   (307.5,217) -- (307.5,176) ;
\draw [shift={(307.5,191.5)}, rotate = 90] [fill={rgb, 255:red, 77; green, 77; blue, 77 }  ,fill opacity=1 ][line width=0.08]  [draw opacity=0] (8.93,-4.29) -- (0,0) -- (8.93,4.29) -- cycle    ;
\draw [color={rgb, 255:red, 77; green, 77; blue, 77 }  ,draw opacity=1 ]   (307.5,217) -- (307.11,260) ;
\draw  [color={rgb, 255:red, 77; green, 77; blue, 77 }  ,draw opacity=1 ][fill={rgb, 255:red, 123; green, 161; blue, 196 }  ,fill opacity=1 ] (322.57,213.36) -- (287.14,213.36) -- (287.14,234) -- (322.57,234) -- cycle ;
\draw  [color={rgb, 255:red, 77; green, 77; blue, 77 }  ,draw opacity=1 ][fill={rgb, 255:red, 249; green, 225; blue, 228 }  ,fill opacity=1 ] (192.68,44) -- (192.68,69.2) -- (217.88,69.2) -- (217.88,128) -- (329,128) -- (329,44) -- cycle ;
\draw  [color={rgb, 255:red, 77; green, 77; blue, 77 }  ,draw opacity=1 ][fill={rgb, 255:red, 123; green, 161; blue, 196 }  ,fill opacity=1 ] (256.34,77.61) .. controls (256.34,72.63) and (252.02,68.6) .. (246.68,68.6) .. controls (241.35,68.6) and (237.02,72.63) .. (237.02,77.61) .. controls (237.02,82.58) and (241.35,86.61) .. (246.68,86.61) .. controls (252.02,86.61) and (256.34,82.58) .. (256.34,77.61) -- cycle ;
\draw  [color={rgb, 255:red, 77; green, 77; blue, 77 }  ,draw opacity=1 ][fill={rgb, 255:red, 123; green, 161; blue, 196 }  ,fill opacity=1 ] (259.82,100.84) .. controls (259.83,100.71) and (259.83,100.58) .. (259.84,100.45) .. controls (259.93,94.49) and (253.95,89.57) .. (246.49,89.46) .. controls (239.02,89.36) and (232.89,94.12) .. (232.79,100.09) .. controls (232.79,100.23) and (232.79,100.37) .. (232.79,100.52) -- cycle ;
\draw [color={rgb, 255:red, 77; green, 77; blue, 77 }  ,draw opacity=1 ]   (201.5,96) -- (201.5,69) ;
\draw [shift={(201.5,77.5)}, rotate = 90] [fill={rgb, 255:red, 77; green, 77; blue, 77 }  ,fill opacity=1 ][line width=0.08]  [draw opacity=0] (8.93,-4.29) -- (0,0) -- (8.93,4.29) -- cycle    ;
\draw [color={rgb, 255:red, 77; green, 77; blue, 77 }  ,draw opacity=1 ]   (203.01,43) -- (202.52,17) ;
\draw [shift={(202.67,25)}, rotate = 88.92] [fill={rgb, 255:red, 77; green, 77; blue, 77 }  ,fill opacity=1 ][line width=0.08]  [draw opacity=0] (8.93,-4.29) -- (0,0) -- (8.93,4.29) -- cycle    ;
\draw [color={rgb, 255:red, 77; green, 77; blue, 77 }  ,draw opacity=1 ]   (306.98,106) -- (307.5,64) ;
\draw [shift={(307.3,80)}, rotate = 90.71] [fill={rgb, 255:red, 77; green, 77; blue, 77 }  ,fill opacity=1 ][line width=0.08]  [draw opacity=0] (8.93,-4.29) -- (0,0) -- (8.93,4.29) -- cycle    ;
\draw [color={rgb, 255:red, 77; green, 77; blue, 77 }  ,draw opacity=1 ]   (201.5,207) -- (201.5,180) ;
\draw [shift={(201.5,188.5)}, rotate = 90] [fill={rgb, 255:red, 77; green, 77; blue, 77 }  ,fill opacity=1 ][line width=0.08]  [draw opacity=0] (8.93,-4.29) -- (0,0) -- (8.93,4.29) -- cycle    ;
\draw [color={rgb, 255:red, 77; green, 77; blue, 77 }  ,draw opacity=1 ]   (307.5,177) -- (307.5,120) ;
\draw  [color={rgb, 255:red, 77; green, 77; blue, 77 }  ,draw opacity=1 ][fill={rgb, 255:red, 123; green, 161; blue, 196 }  ,fill opacity=1 ] (322.57,101.36) -- (287.14,101.36) -- (287.14,122) -- (322.57,122) -- cycle ;

\draw (99,138.3) node [anchor=north west][inner sep=0.75pt]  [font=\large,color={rgb, 255:red, 77; green, 77; blue, 77 }  ,opacity=1 ] [align=left] {Alice};
\draw (102.23,20.75) node [anchor=north west][inner sep=0.75pt]  [font=\large,color={rgb, 255:red, 77; green, 77; blue, 77 }  ,opacity=1 ] [align=left] {Bob};
\draw (297.11,263) node [anchor=north west][inner sep=0.75pt]  [font=\large,color={rgb, 255:red, 255; green, 255; blue, 255 }  ,opacity=1 ] [align=left] {$\displaystyle P_{a}$};
\draw (289.33,187.44) node [anchor=north west][inner sep=0.75pt]  [font=\large,color={rgb, 255:red, 77; green, 77; blue, 77 }  ,opacity=1 ] [align=left] {$\displaystyle c$};
\draw (219.85,158.07) node [anchor=north west][inner sep=0.75pt]  [font=\large,color={rgb, 255:red, 77; green, 77; blue, 77 }  ,opacity=1 ] [align=left] {Charlie};
\draw (187.25,204) node [anchor=north west][inner sep=0.75pt]  [font=\large,color={rgb, 255:red, 77; green, 77; blue, 77 }  ,opacity=1 ] [align=left] {$\displaystyle x$};
\draw (288.33,76.44) node [anchor=north west][inner sep=0.75pt]  [font=\large,color={rgb, 255:red, 77; green, 77; blue, 77 }  ,opacity=1 ] [align=left] {$\displaystyle d$};
\draw (219.85,46.07) node [anchor=north west][inner sep=0.75pt]  [font=\large,color={rgb, 255:red, 77; green, 77; blue, 77 }  ,opacity=1 ] [align=left] {Debbie};
\draw (187.25,91) node [anchor=north west][inner sep=0.75pt]  [font=\large,color={rgb, 255:red, 77; green, 77; blue, 77 }  ,opacity=1 ] [align=left] {$\displaystyle y$};
\draw (189.25,6) node [anchor=north west][inner sep=0.75pt]  [font=\large,color={rgb, 255:red, 77; green, 77; blue, 77 }  ,opacity=1 ] [align=left] {$\displaystyle b$};

\end{tikzpicture}
         \caption[Noncontextual Friendliness setup]{Noncontextual Friendliness setup.}          
         \label{fig:OF_tryout}
\end{figure}

Given this setup, let us state our assumption of Noncontextual Agency, replacing Local Agency. Accordingly, we name the conjunction of AOE and Noncontextual Agency \textit{Noncontextual Friendliness} (NF).

\begin{assumption}
[Noncontextual Agency] 
\label{Ass:OA}
Any operational equivalence, which would be verified by a hypothetical agent with access to all the relevant variables, still holds even if it cannot be verified by a single agent.
\end{assumption}


Notice the similarity between Noncontextual Agency and Local Agency of Assumption \ref{ass:LA}: the former can be viewed as a special case of the latter, where the operational equivalence considered is specifically no-signaling. Just as Local Agency is a stronger version of no-signaling, Noncontextual Agency is a stronger version of standard operational equivalence.\footnote{{As is the case for Noncontextuality, Noncontextual Agency relies on operational equivalences that are defined once the theory -- and hence the relevant systems and the set of in principle distinguishing measurements -- has been specified; it is therefore subject to the ``system loophole'' formulated in~\cite{Gitton_2024}.}} In the context of the setup considered here, standard operational equivalence pertains to setting choices. More precisely, they are about the preparations associated with choices $x=1$ and $x=2$: $\wp(d|x,y=1)=\wp(d|y=1)$ and $\wp(b|x,y=2)=\wp(b|y=2)$.
Noncontextual Agency additionally implies that these equalities also hold when conditioned on $c$, \textit{i.e.}, $p(d|c,x,y=1)=p(d|c,y=1)$ and $p(b|c,x,y=2)=p(b|c,y=2)$, despite the fact that no single observer can verify them, since when $x=2$, the outcome $c$ is erased. Thus, similarly to Local Agency, Noncontextual Agency lifts the disjoint conditional independences of $d|y$ and $c|y$ on $x$ to a joint conditional independence of $c,d|y$ on $x$ (and similarly for $b,c|x,y$).

{Regarding the name “Noncontextual Agency”, the term noncontextual is used similarly to generalized noncontextuality: just as generalized noncontextuality requires that operational equivalences are preserved as ontological identities within an ontological model explaining the physical theory, here noncontextual indicates that operational equivalences 
are preserved at the level of the Absoluteness of Observed Events -- and thus remain valid even when they are operationally inaccessible, in the sense that no single agent can verify them. In other words, it represents a form of noncontextuality formulated not within the framework of ontological models~\cite{harrigan2010einstein}, but under the sole assumption that the underlying reality satisfies AOE. This terminology parallels that of Local Agency, where local is inspired by the notion of local causality in Bell’s theorem (specifically, its parameter independence aspect, sometimes referred to as locality~\cite{wiseman2017causarum}), but where no-signaling is required to hold only at the AOE level -- \textit{i.e.}, in a world where observed events are absolute -- rather than in an ontological model.}

Noncontextual Agency can be seen as an instance of no operational fine-tuning~\cite{catani2023mathematical} in a universe where AOE holds. Indeed, as with Local Agency, one can motivate it by arguing that if a theory prescribes an operational equivalence, such an equivalence should also hold even if it cannot be verified by a single agent. Put differently, an operational equivalence should hold even if its verification were possible only for a hypothetical agent {who could access all relevant variables by cheating the protocol}.  

Let us call this agent Eve, and imagine that she has access to outcomes $c,d$ and choices $x,y$. 
Since $c$ lies in the past of $x,y$, Eve can verify, in accordance with the quantum predictions, that $\wp(c|x,y)=\wp(c)${; in fact, this independence can also be verified without referring to Eve, as shown in Section 4.2.2 of Ref.~\cite{yile2023}.  If, moreover, Eve cheats the protocol by storing Charlie’s outcome \(c\) in a device before any potential undoing action is performed on Charlie’s lab, she can then verify that \(p(d|c,x,y) = p(d|c,y)\).}
Indeed, if Eve has access to $c$ and $d$, the probabilities $p(d|c,x,y)$ correspond to preparing system $S$ (the system Charlie measured) in the post-measurement state $P_c$ associated with outcome $c$, followed by Alice's transformation, which depends on $x$, and Debbie's measurement. 
For $x=1$, Alice takes Charlie's outcome to be hers thus not affecting the system $S$, and for $x=2$ she applies $U_C^\dagger$, erasing $c$. 
However, as Eve has already copied 
$c$ into her memory, which remains unaffected by Alice’s operations, the undoing $U_C^\dagger$ does not alter $P_c$. Thus, in both cases $x=1$ and $x=2$, Eve observes that $P_c$ is unaltered until Debbie applies her measurement, confirming that $p(d|c,x,y) = p(d|c,y)$. A similar argument establishes that $p(b|c,x,y)=p(b|c,y)$.
Noncontextual Agency demands that the operational equivalences verified by Eve 
must remain valid even in her absence, {and even if the relevant events are jointly inaccessible in the sense that they are not jointly observed by any agent, yet still exist by virtue of AOE.}


It is worth noting that while Noncontextual Agency is stronger than operational equivalence, it is weaker than {generalized noncontextuality:} as shown in Appendix B of Ref.~\cite{walleghem2025extendedwignersfriendnogo}, one can construct a scenario showing that the polytope of NF correlations is strictly larger than the polytope of noncontextual correlations, mirroring the proof strategy of Ref.~\cite{haddara2024local}.

We can now state the NF no-go theorem.
\begin{theorem}[Noncontextual Friendliness no-go theorem] \label{th:NCF}
If a superobserver can perform arbitrary quantum operations on an observer and its environment, then no physical theory can satisfy Noncontextual Friendliness. 
\end{theorem}

\proof
Let us consider the NF scenario of  \cref{fig:OF_tryout} with the following specifications (the preparations and measurements employed will be the same as in the simplest nontrivial scenario). Alice chooses preparations  $P_{0}= P_{\pi/4}$ and $P_1=P_{5\pi/4}$, for $a=0,1$, respectively.
Charlie measures in the basis given by $P_{3\pi/4}$ and $P_{7\pi/4}$.
Debbie's measurement $D$ is the Pauli $Z$ measurement. 
When $y=2$, Bob undoes Debbie's measurement and performs a Pauli $X$ measurement on $S$.
At the end of the experiment the following empirical correlations of observed outcomes are obtained, \begin{equation} \label{eq:empirical_prob_LF}
    \begin{split}
        &\wp(c,d \mid x=1, y=1), \\
&\wp(c,b\mid x=1, y=2), \\
&\wp(a,d \mid x=2, y=1), \\
&\wp(a,b\mid x=2, y=2). 
    \end{split}
\end{equation}

By AOE and Noncontextual Agency, we can rewrite these empirical correlations as follows, 

\begin{align} \label{eq:empirical_PandM_EWF_1}
        &\wp(c,d \mid x=1, y=1) = p(c,d|x=2,y=2), \\ \label{eq:empirical_PandM_EWF_2}
&\wp(c,b\mid x=1, y=2)  = p(c,b|x=2,y=2), \\ \label{eq:empirical_PandM_EWF_3}
&\wp(a,d \mid x=2, y=1)  = p(a,d|x=2,y=2), \\ \label{eq:empirical_PandM_EWF_4}
&\wp(a,b\mid x=2, y=2) = p(a,b|x=2,y=2). 
\end{align}

AOE is used to consider joint probability distributions when $c$ or $d$ are involved and possibly erased.
To obtain Eq.~\eqref{eq:empirical_PandM_EWF_1}, we first appeal to an instance of Noncontextual Agency that enforces no-superdeterminism, $p(c,d|x,y){=}p(c,d|x)$, yielding $p(c,d|x{=}1,y{=}1){=}p(c,d|x{=}1,y{=}2)$; we can then apply the rule of conditional probability and consider two other instances of Noncontextual Agency (the first again enforcing no-superdeterminism): $p(c|x,y){=}p(c)$ and $p(d|c,x,y){=}p(d|c,y)$, thus obtaining \[\begin{split}&\wp(c,d|x{=}1,y{=}1) = p(c|x{=}1,y{=}1)p(d|c,x{=}1,y{=}1) \\&=p(c|x{=}2,y{=}2)p(d|c,x{=}2,y{=}2)=p(c,d|x{=}2,y{=}2).\end{split}\]

Similar instances of Noncontextual Agency, but with  variables $c,b$ -- namely $p(c|x,y)=p(c)$ and $p(b|c,x,y)=p(b|c,y)$ --  lead to Eq.~\eqref{eq:empirical_PandM_EWF_2},  
\[\begin{split}&\wp(c,b|x{=}1,y{=}2) = p(c|x{=}1,y{=}2)p(b|c,x{=}1,y{=}2) \\ &=p(c|x{=}2,y{=}2)p(b|c,x{=}2,y{=}2)=p(c,b|x{=}2,y{=}2).
\end{split}\]

Eq.~\eqref{eq:empirical_PandM_EWF_3} derives from an instance of Noncontextual Agency enforcing no-super\-determinism, $p(a,d|x,y)=p(a,d|x)$.

By assuming AOE, we can assert that all the correlations in Eqs.~\eqref{eq:empirical_PandM_EWF_1},\eqref{eq:empirical_PandM_EWF_2},\eqref{eq:empirical_PandM_EWF_3} and \eqref{eq:empirical_PandM_EWF_4} can be obtained from a single joint probability distribution $p(a,b,c,d|x=2,y=2)$ via marginalization. 
However, recall that the preparations and measurements used to obtain these correlations correspond to those in the simplest scenario that achieve the maximum violation of noncontextuality inequalities~\cite{pusey2018robust}. Through the mapping to the Bell scenario, they are also equivalent to those achieving the maximum violation of Bell inequalities. By Fine's theorem~\cite{fine1982hidden,abramsky2011sheaf}, such a joint probability distribution cannot exist.
Let us stress that the NF assumptions are weaker than those used in the noncontextuality no-go theorem proven in the simplest scenario~\cite{pusey2018robust}. Only the empirical correlations in the scenarios are the same. 
Notice how the present proof mirrors the proof of the LF no-go theorem, with Noncontextual Agency replacing Local Agency.

\subsection{Discussion} 
In this section, we have shown that the Local Friendliness no-go theorem can be generalized to the Noncontextual Friendliness no-go theorem. 
Just as no-signaling is a specific instance of an operational equivalence, Local Agency is a specific instance of Noncontextual Agency.
Furthermore, just as the LF no-go theorem is stronger than Bell’s theorem, granted that the LF scenario can be realized, the NF no-go theorem is stronger than theorems based on generalized noncontextuality, particularly the one leveraging the simplest scenario,  granted that the NF scenario can be realized.
Local Agency strengthens no-signaling by requiring it to hold even when it cannot be verified by a single agent. Similarly, Noncontextual Agency extends operational equivalence to cases where no single agent can verify the equivalence.

Key to establishing these connections is the rephrasing of Local Agency, highlighting how it can be understood as a stronger version of no-signaling. Specifically, Local Agency can be interpreted as no-signaling verified by a hypothetical agent who is capable of breaking the protocol, accessing all relevant variables, and confirming the condition. Recognizing that no-signaling is itself an instance of operational equivalence naturally leads to the mirrored definition of Noncontextual Agency.
The assumption of Noncontextual Agency encompasses the newly introduced assumption of Commutation Irrelevance of \Cref{sec:EWF_on_KS_not_nonlocal}.

It is worth discussing here the conceptual significance of the result of this section, which primarily arises from the conceptual status of Noncontextual Agency. 
In our formulation, the assumption of Noncontextual Agency constitutes an instance of the methodological principle of no fine-tuning: in a world where AOE holds, it would be conspiratorial for operational equivalences to apply only when verified by a single agent, but to fail otherwise. It is indeed natural (i.e., non-conspiratorial) to assume that, in a world satisfying AOE, the theory continues to predict equivalences to hold for correlations involving jointly inaccessible outcomes. 
Moreover, as we emphasize through the introduction of the hypothetical agent, if one were hypothetically able to access all relevant variables simultaneously, one would indeed verify the operational equivalence (albeit in a different protocol that breaks the one under consideration).

Notice that the above considerations also apply to the case of Local Agency, which is an instance of Noncontextual Agency. 
Consequently, Local Agency shifts from being justified by a metaphysical assumption about the nature of the underlying causal structure—namely, that it must be relativistic—to being an instance of the methodological principle of no fine-tuning. As such, it carries a lighter ontological weight.
In turn, the lesson drawn from the LF and NF no-go theorems becomes clear: if one believes that the LF and NF scenarios can indeed be realized -- for instance, by endorsing the universality of quantum theory\footnote{In support of the universality of quantum theory, we note that there is currently no experimental evidence against it, nor any indication of regimes in which it might fail. Moreover, even if quantum theory were not universal, one would still be required to specify precisely the domains of its applicability and to provide a clear account of how the transition to alternative physical descriptions occurs in other regimes.} -- and accepts the no fine-tuning (or no-conspiracy) principle -- which, following many others~\cite{spekkens2019ontologicalidentityempiricalindiscernibles,catani2023mathematical,schmid2021unscrambling}, we regard as methodologically indispensable -- then one is compelled to conclude that observed events are relative; that is, one must abandon AOE, the sole remaining metaphysical assumption in the theorem.

Interpreting Local Agency as an instance of Noncontextual Agency also allows us to dig deeper into the justification of Local Agency itself. Even the justification based on the belief in an underlying relativistic causal structure---one that is assumed to hold also for inaccessible outcomes---can only be motivated by an appeal to the no fine-tuning principle in a world where AOE holds. Otherwise, there would be no reason to endorse it. Why should one believe that no-signaling also holds for jointly inaccessible outcomes, and that the relativistic causal structure extends to them? Because, in a world where AOE holds, it would otherwise be conspiratorial not to.
 
As far as we are aware, the only work that explicitly discusses no fine-tuning in the context of an extended Wigner’s friend scenario is Ref.~\cite{yile2023}, although the authors do so within the framework of causal models. They show that, even when one allows for a nonclassical causal model and assumes AOE, it remains impossible to account for the statistics of the LF scenario without invoking causal fine-tuning~\cite{Wood_2015,cavalcanti2018classical} -- a result which bears a close relation to ours. We leave a detailed investigation of this connection for future work.

Another interesting avenue would be to explore proofs of the NF no-go theorem in scenarios beyond the one considered here, which is based on the simplest scenario. 
More broadly, it would be valuable to generalize the NF no-go theorem by considering examples of fine-tunings that go beyond those linked to contextuality~\cite{catani2023mathematical}, such as violations of bounded ontological distinctness~\cite{Chaturvedi_2020,Khoshbin_2024}. Additionally, it would be worthwhile to investigate NF scenarios that are realizable within post-quantum theories -- for example those proposed in~\cite{vilasini2019multi,ormrod2023theories,nurgalieva2025theoryadmitswignersfriend}.

\ifSubfilesClassLoaded{ 
  \printbibliography[heading=bibintoc, title=References]
  }{}
\end{document}

\ifSubfilesClassLoaded{ 
	\setcounter{chapter}{1} 
	\maketitle \tableofcontents \listoftodos
	}{}

\chapter[Are free choices absolute, when Pusey--Barrett--Rudolph are friends of Wigner?]{Are free choices absolute, when Pusey--Barrett--Rudolph are friends of Wigner?}\label{chap:part1_EWF_PBR}
\chaptermark{Are free choices absolute, when PBR are friends of Wigner?}

The Local Friendliness no-go theorem, which we have encountered several times in the previous chapters, can be read as an argument against the absolute outcomes of performed measurements, under a locality and universality assumption.
Proofs of this no-go theorem relied on correlations violating Bell locality, and, as we have shown in \Cref{chap:part1_EWF_noncontextuality}, proofs can be found based on noncontextuality as well.
In this chapter, we provide an entirely different extended Wigner's friend scenario based on the Pusey--Barrett--Rudolph scenario, which tackled questions on the reality of the quantum state. 
This scenario can be interpreted as an argument against the absoluteness of free choices, under the same locality assumption as in
the Local Friendliness no-go theorem.

\section{The Pusey--Barrett--Rudolph argument} \label{sec:PBR}

The Pusey--Barrett--Rudolph theorem~\cite{pusey2012reality} is a result against an epistemic reading of the quantum state in the \textit{ontological models framework}.
As reviewed in \Cref{sec:gen_noncontext}, in the framework of ontological models, a physical system is associated with one or more ontic states $\lambda$ about which a measurement can reveal certain information. 
It may occur that full knowledge of a preparation $P$ of the system does not specify a unique ontic state $\lambda$, but only a distribution $\mu_P(\lambda)$ over such ontic states, referred to as the ontic distribution.
A measurement $M$ is represented by a probability distribution $p(\xi|\lambda,M)$ over its outcomes $\xi$ and the empirical probability of obtaining $\xi$ upon a measurement $M$ given preparation $P$ is given by $p(\xi|M,P) = \sum_\lambda p(\xi|M,\lambda)\mu_P(\lambda)$.

The question of whether the quantum state is real or not can be interpreted as follows within this framework.
The quantum state is said to be real, or \textit{ontic}, if each pair of distinct, pure quantum states $\ket{\psi_0},\ket{\psi_1}$ have non-overlapping distributions, $\mu_{\psi_0}(\lambda) \mu_{\psi_1}(\lambda) = 0$.
The quantum state is said to be \textit{epistemic} if it is not ontic, i.e. if there exists a pair of operationally distinguishable quantum states  $\ket{\psi_0},\ket{\psi_1}$ that have overlapping ontic distributions, $\mu_{\psi_0}(\lambda) \mu_{\psi_1}(\lambda) > 0$.
A quantum state being epistemic captures the idea that it represents mere knowledge; if the ontic distributions $\mu_{\psi_0}(\lambda), \mu_{\psi_1}(\lambda)$ corresponding to preparations of two different pure quantum states $\ket{\psi_0},\ket{\psi_1}$ have overlapping support, then these quantum states cannot refer to a physical property of the physical system.


The Pusey--Barrett--Rudolph argument against epistemic states in this framework goes as follows, following Ref.~\cite{pusey2012reality} closely.
Consider two copies of a preparation device that operate independently and can each prepare a quantum system in the state $\ket{0}$ or $\ket{+}$.  
Here $\ket{0},\ket{1}$ is a basis of a qubit, $\ket{+}=(\ket{0}+\ket{1})/\sqrt{2}$ and for later convenience we also define $\ket{-}=(\ket{0}-\ket{1})/\sqrt{2}$.
The possible preparations arising from the two devices jointly are thus $|0\rangle \otimes|0\rangle,|0\rangle \otimes|+\rangle,|+\rangle \otimes|0\rangle \text { and }|+\rangle \otimes|+\rangle$.

Beyond the ontological models framework assumption $(i)$, it is assumed that $(ii)$ a form of \textit{preparation independence} holds~\cite{leifer2014quantum,mansfield_reality_2016}, i.e., that the joint ontic distribution of multipartite preparations leading to product quantum states factorizes, $\mu_{P_A P_B}(\lambda_A,\lambda_B) = \mu_{P_A}(\lambda_A)\mu_{P_B}(\lambda_B)$. Note that this property is only required for product states, and need not be assumed for entangled states. 

Next, an entangled measurement is performed upon the two systems, which projects onto the four orthogonal states 
\begin{equation} \label{eq:definition_xi_PBR}
\begin{aligned}
& \left|\xi_1\right\rangle=\frac{1}{\sqrt{2}}(|0\rangle \otimes|1\rangle+|1\rangle \otimes|0\rangle), \\
& \left|\xi_2\right\rangle=\frac{1}{\sqrt{2}}(|0\rangle \otimes|-\rangle+|1\rangle \otimes|+\rangle), \\
& \left|\xi_3\right\rangle=\frac{1}{\sqrt{2}}(|+\rangle \otimes|1\rangle+|-\rangle \otimes|0\rangle), \\
& \left|\xi_4\right\rangle=\frac{1}{\sqrt{2}}(|+\rangle \otimes|-\rangle+|-\rangle \otimes|+\rangle).
\end{aligned}
\end{equation}
Now if the ontic distributions of the quantum states $\ket{0},\ket{+}$ overlap, then there is a nonzero probability $q^2 > 0$ that an ontic state $\lambda$ is prepared that is compatible with all possible preparations listed above.
Each of these states $\ket{\xi_i}$ is orthogonal to a different one of the given preparations, implying that the corresponding measurement outcome $\xi_i$ has probability zero of occurring when a prepared state is orthogonal to $\ket{\xi_i}$.
Namely, $\ket{\xi_1}$ is orthogonal to $\ket{0}\otimes \ket{0}$,  $\ket{\xi_2}$ is orthogonal to $\ket{0}\otimes \ket{+}$,  $\ket{\xi_3}$ to $\ket{+}\otimes \ket{0}$ and  $\ket{\xi_4}$ to $\ket{+}\otimes \ket{+}$.
This fact immediately gives the contradiction; when given such $\lambda$ compatible with all possible preparations, the measurement device is unsure which preparation had been performed, and thus runs the risk of giving a measurement outcome for which the actual preparation had probability zero of occurring. 

This shows that the quantum states $\ket{0},\ket{+}$ must have zero overlap in their ontic distributions. A generalization of this argument shows that in fact any pair of distinguishable quantum states must have zero overlap~\cite{pusey2012reality}, proving that the quantum state must be ontic, modulo the assumptions $(i)-(ii)$ presented above.
Arguments that loosen the preparation independence assumption $(ii)$ can be found in Refs.~\cite{spekkens2012not,mansfield_reality_2016,leifer2014quantum}.

\section{Are free choices absolute?}  \label{sec:PBR_WF_scenario}
The scenario that we will consider, sketched in \cref{fig:PBR_LCF_scenario}, combines Wigner's friend with the Pusey--Barrett--Rudolph argument to derive a contradiction with the Local Friendliness assumptions, now applied to free choices.

Two friends, Alice and Bob, residing in sealed labs and modeled as quantum systems by their superobserver Wigner, each make a `free choice' $a,b \in \{ 0, 1 \}$. 
For now, we assume that Alice and Bob choose $0$ and $1$ with equal probabilities $1/2$.\footnote{In fact, more precisely, we assume for now that the amplitudes of their choices are given by $1/\sqrt{2}$.}
They next encode their choices $a,b$ in the computational bases of qubits $T_A,T_B$ that they send to Charlie and Debbie, respectively.\footnote{The systems $T_A,T_B$ are the only systems that can escape Alice's and Bob's otherwise sealed labs.} 

We can model the state at this point by \begin{equation}
    \label{eq:init_state_protocol_1} 
    \ket{\Phi} = \ket{\phi_0}_{T_A L_A} \otimes \ket{\phi_0}_{T_B L_B},
\end{equation} with, for $I=A,B$,\footnote{Here and later, for ease of notation, we omit the tensor product, i.e. we write $\ket{00}$ or $\ket{0}\ket{0}$ for $\ket{0}\otimes \ket{0}$.} \begin{equation}
    \label{eq:init_state_protocol_1_each_part}
        \ket{\phi_0}_{T_I L_I} = \frac{1}{\sqrt{2}} \bigg( \ket{00}_{T_I L_I} + \ket{11}_{T_I L_I} \bigg).
\end{equation}
Here $\ket{0}_{L_A}$ and $\ket{1}_{L_A}$ denote the states of Alice's lab when she chose $a=0$ and $a=1$, respectively, and similarly for Bob.

After this, Wigner chooses $z=1,2$, where for $z=1$ he reveals Alice's and Bob's choices (for example by measuring their labs in the computational basis) and for $z=2$ he performs the entangled PBR measurement on Alice's and Bob's labs $L_A,L_B$, namely measuring $L_A \otimes L_B$ in the basis $\ket{\xi_1}$, $\ket{\xi_2}$, $\ket{\xi_3}$, $\ket{\xi_4}$. 
Wigner's outcomes are denoted by $w \in \{ \xi_1, \xi_2, \xi_3, \xi_4 \}$.

Having received the qubits $T_A, T_B$ from Alice and Bob, Charlie and Debbie choose measurement settings $x$ and $y$, respectively. 
For $x=1$, Charlie decodes Alice's choice by measuring $T_A$ in the computational basis, obtaining outcome $c \in \{0,1\}$. 
For $x=2$, Charlie measures $T_A$ in the $X$-basis, obtaining outcome $c \in \{+,-\}$, and similarly for Debbie choosing $y=1,2$ to measure $T_B$ in the $Z$-- or $X$--basis.
Charlie and Debbie do so spacelike separated from each other and from Wigner.

\begin{figure}[h]
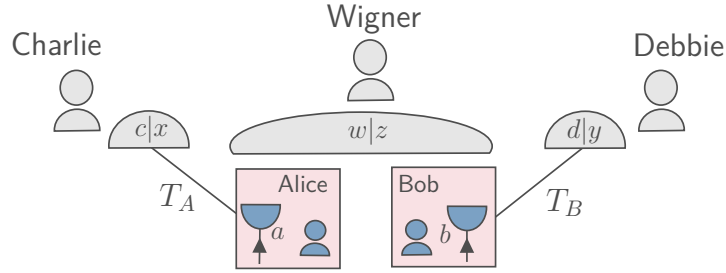

         \centering
\includestandalone[width=0.65\textwidth]{Figures/PBR_LCF_new}
         \caption[Combining Wigner's friend with the Pusey--Barrett--Rudolph argument]{Protocol that combines Wigner's friend and the PBR argument to violate Local Friendliness. Two friends Alice and Bob each make a choice $a,b \in \{ 0, 1\}$. They encode their choices in the computational basis of qubits that they send to Charlie and Debbie, who choose $x,y$ whether to decode their outcomes or measure their received qubit in the $X$-basis. Wigner is a superobserver for Alice and Bob, and chooses $z=1,2$. For $z=1$ he asks Alice and Bob for their choices while for $z=2$ he measures Alice's and Bob's labs jointly in an entangled PBR measurement. The only events unavailable after the protocol are the `free' choices $a,b$ of Alice and Bob.}
         \label{fig:PBR_LCF_scenario}
\end{figure}

\vspace{0.5cm}

We next derive a contradiction of this scenario with the Local Friendliness assumptions. 
Regarding Absoluteness of Observed Events, the only two events unavailable after the protocol are Alice's and Bob's free choices.
All other measurement outcomes and choice settings are available after the protocol. 
As a result, we only need the assumption that the freely chosen settings $a,b$ of Alice and Bob, even if unavailable after the protocol, are absolute and single-valued, alongside the outcomes and choices available after the protocol.

\begin{assumption}[Absoluteness of Observed Events (AOE)] 
\label{assumption:AOE_PBR}
An observed event is an absolute single event, not relative to anything or anyone. 
\end{assumption}


\begin{assumption}[Local Agency] \label{ass:LA_PBR}
    The only relevant events correlated with an intervention are in its future light cone.
\end{assumption}

\vspace{0.5cm}

A violation of Local Friendliness is then obtained as follows. 
If $a=0,b=0$ and $x=1,y=1$, then the choices of Alice and Bob are revealed by Charlie and Debbie\footnote{Here we use a weak `tracking assumption', see Ref.~\cite{schmid2023review}.} and the probabilities of Wigner for $z=2$ are those of performing the PBR measurement on the state $\ket{0}\otimes\ket{0}$, finding that\footnote{In this section, we denote empirical probabilities by $\wp$, to distinguish them from probabilities $p$ which are required to exist by AOE.} \begin{equation}
    \label{eq:PBR_LCF_1}  \wp(a{=}0,b{=}0,w{=}\xi_1 | x{=} 1, y{=} 1,z{=}2) = 0.
\end{equation}
If $x=1,y=2$ and $a=0,d=+$, then Charlie reveals Alice's choice by measuring the qubit $T_A$ she received from Alice in the computational basis, which turns out to be $a=0$, while Bob's choice is not revealed by Debbie who measured $T_B$ in the $X$-basis obtaining $d=+$.
Therefore, if $x=1,y=2,d=+$ and $a=0$, then the probabilities of Wigner for $z=2$ are those of performing the PBR measurement on $\ket{0}\otimes\ket{+}$, finding that 
\begin{equation}
        \label{eq:PBR_LCF_2}  \wp(a{=}0,w{=}\xi_2 | x{=} 1, y{=} 2,d{=}+,z{=}2) = 0.
\end{equation}
Similarly, if $b=0$ and $x=2,y=1,z=2,c=+$, the probabilities of Wigner are those of performing the PBR measurement on $\ket{+}\otimes\ket{0}$, finding that 
\begin{equation}
        \label{eq:PBR_LCF_3}  \wp(b{=}0,w{=}\xi_3 | x{=} 1, y{=} 2,c{=}+,z{=}2) = 0.
\end{equation}
Finally, if $x=2,y=2,z=2,c=+,d=+$, then the probabilities of Wigner are those of performing the PBR measurement on the $\ket{+}\otimes\ket{+}$ state, finding 
\begin{equation}
        \label{eq:PBR_LCF_4}  \wp(w{=}\xi_4 | x{=}2, y{=} 2,c{=}+,d{=}+,z{=}2) = 0.
\end{equation}

Next, we will use the Local Friendliness assumptions. In this case, AOE and Local Agency imply that $p(a,b,w|x,y,z)=p(a,b,w|z)$, i.e. the joint correlations of $a,b,w$, which always exist by AOE, should be independent of $x$ and $y$. Moreover, they also imply that $p(a,b,w,c|x,y,z)=p(a,b,w,c|x,z)$ and $p(a,b,w,d|x,y,z)=p(a,b,w,d|y,z)$.

Applying AOE and Local Agency to \cref{eq:PBR_LCF_1}, we obtain that 
\begin{equation}
    \label{eq:PBR_LCF_1_1} p(a{=}0,b{=}0,w{=}\xi_1 | x{=} 2, y{=} 2,z{=}2) = 0,
\end{equation} and, using $p(c{=}+,d{=}+|x{=}2,y{=}2,z{=}2) > 0$, that 
\begin{equation}
    \label{eq:PBR_LCF_1_final}  p(a{=}0,b{=}0,w{=}\xi_1 | x{=} 2, y{=} 2, c{=} +, d{=}+,z{=}2) = 0.
\end{equation}
Similarly, applying AOE and Local Agency to \cref{eq:PBR_LCF_2}, we obtain that
\begin{equation}
        \label{eq:PBR_LCF_2_2}  p(a{=}0,w{=}\xi_2 | x{=} 2, y{=} 2,d{=}+,z{=}2) = 0,
\end{equation} and, using $p(c{=}+,d{=}+|x{=}2,y{=}2,z{=}2) > 0$, that
\begin{equation}
        \label{eq:PBR_LCF_2_final}  p(a{=}0,b{=}0,w{=}\xi_2 | x{=} 2, y{=} 2,c{=}+,d{=}+,z{=}2) = 0.
\end{equation}
The same reasoning for \cref{eq:PBR_LCF_3} leads to
\begin{equation}
        \label{eq:PBR_LCF_3_final}  p(a{=}0,b{=}0,w{=}\xi_3 | x{=} 2, y{=} 2,c{=}+,d{=}+,z{=}2) = 0.
\end{equation}

Summarizing, we have the following four conditions (from \cref{eq:PBR_LCF_1_final,eq:PBR_LCF_2_final,eq:PBR_LCF_3_final,eq:PBR_LCF_4}) on $p(a,b,w|E)$ with $E=(x{=}2,y{=}2,c{=}+,d{=}+,z{=}2)$:
\begin{equation} \label{eq:PBR_LF_together}
    \begin{split}
         p(a=0,b=0,w=\xi_1 | E) = 0, \\
         p(a=0,b=0,w=\xi_2 |E) = 0, \\
          p(a=0,b=0,w=\xi_3 | E) = 0, \\
          p(a=0,b=0,w=\xi_4 | E) = 0.
    \end{split}
\end{equation}
We thus find that \begin{equation} p(a{=}0,b{=}0,c{=}+,d{=}+|x{=}2,y{=}2,z{=}2)=0. \end{equation}
However, this cannot be true, as by Local Agency we have that $p(a,b,c,d|x,y,z)=p(a,b,c,d|x,y)$ and \begin{equation}
    \wp(a{=}0,b{=}0,c{=}+,d{=}+|x{=}2,y{=}2,z{=}1) > 0.
\end{equation}
We thus obtain a contradiction, finding a violation of the conjunction of Local Agency and the assumption that the events of Alice's and Bob's free choices, being the only unavailable values at the end of the protocol, are absolute and single-valued.

\section[Generalizing to arbitrary choice amplitudes]{Generalizing to arbitrary choice amplitudes}\label{sec:generalizing_choice_amplitudes_PBR_WF}
In the protocol outlined above, we have assumed that Alice chooses $a=0,1$ with equal probabilities, or, more correctly, with equal amplitudes $1/\sqrt{2}$, leading to the state in \cref{eq:init_state_protocol_1_each_part} after her encoding of her choice in $T_A$, and similarly for Bob. 
But what if Alice's choice probabilities (and amplitudes) were different from $1/2$ (and $1/\sqrt{2}$)? 
In general, she could choose $a=0,1$ with arbitrary probabilities and amplitudes, leading to replacing \cref{eq:init_state_protocol_1_each_part} by \begin{equation} \label{eq:psi_choices_remark}
    \ket{\psi}_{T_A L_A} = \alpha \ket{00}_{T_A L_A}+\beta \ket{11}_{T_A L_A}.
\end{equation}
Then, to obtain a contradiction, we proceed as follows. 
We let Alice and Bob make their choices several times, while Charlie and Debbie perform state tomography on Alice's and Bob's labs when they have made their choices. This allows Charlie and Debbie to know $\alpha$ and $\beta$. Then, we run the protocol, and there are two possibilities. 

The first possibility is that for $x=1$, Charlie still decodes Alice's choice by measuring $T_A$ in the computational basis, whereas for $x=2$ he now measures $T_A$ in a basis that contains the state \begin{equation}
    \ket{\Gamma} = \gamma \ket{0} + \delta \ket{1},
\end{equation} with 
\begin{equation} \gamma^* \alpha = \delta^* \beta.
\end{equation}
In this way, when Charlie obtains $c=\Gamma$, the post-selected state for $L_A$ is $\ket{+}$, and similarly for Debbie choosing $y$, leading again to the relevant states $\ket{00},\ket{+}\ket{0},\ket{0}\ket{+},\ket{+}\ket{+}$ of the PBR argument. Wigner acts as before, and a contradiction is obtained in the same way as in \Cref{sec:PBR_WF_scenario}.

The second possibility is that Charlie and Debbie still measure in the $Z$- or $X$-basis, but that Wigner now performs a PBR measurement that discriminates between $\ket{0}$ and $\alpha \ket{0} + \beta \ket{1}$. For this, we actually need several additional agents to Alice and Bob and Charlie and Debbie. Namely, we need $n$ agents making their choice and encoding their choice to send to some faraway observers who measure in the $Z$- or $X$-basis, with $n$ determined by the general joint PBR measurement as in the original PBR argument~\cite{pusey2012reality}.

\section{On the modeling of Alice's and Bob's choices}  \label{sec:modelling_choices_Alice_Bob_WF_PBR}
\sectionmark{On the modeling of Alice's and Bob's choices}  

In the above argument, regarding state tomography by Charlie and Debbie, we have assumed that each time that Alice and Bob make a choice which they encode in the computational bases of the qubits, the same coefficients $\alpha$ and $\beta$ appear in \cref{eq:psi_choices_remark}. 
The motivation for this is that, assuming we initialized the state of Alice's lab each time in exactly the same state, and take the same exterior conditions, the unitary evolution of the lab is the same and always leads to \cref{eq:psi_choices_remark}. One could argue that there is some tension with the intuitive concept of a free choice in this case, indeed. Perhaps this suggests that one aspect of the free choice, the choice of the frequencies, only truly appears in open systems.
Now, if we fed in different initial conditions for Alice's lab, a slightly different environment, then we would need Charlie and Debbie to additionally keep track of this environment control in order to obtain the coefficients $\alpha$ and $\beta$ that depend on this environment. Then, in the actual protocol, when they choose their basis $\ket{\Gamma},\ket{\lnot \Gamma}$ they would need to additionally control on the environment states in order for the protocol to succeed.

\section{Use of correlations} \label{sec:use_of_correlations_PBR_WF}
Note that our scenario can also be seen as Wigner swapping the entanglement between Alice and Charlie and Bob and Debbie to entanglement between Charlie and Debbie. The correlations used violate Bell nonlocality, as we show now.

The level of the contradiction in the Wigner's friend PBR argument is at the possibilistic level, so to check whether nonlocality fuels the contradiction, we need to check for nonlocality at the possibilistic level.
We can see the obtained measurement outcomes in the Wigner's friend PBR argument as a tripartite Bell scenario, involving Charlie, Debbie and Wigner, choosing settings $x,y,z$ and obtaining outcomes $c,d,w$. The initial state is given as in \cref{eq:init_state_protocol_1}. 
Recall that Charlie chooses $x=1,2$ to measure $T_A$ in the $Z$- or $X$-basis, and similarly for Debbie measuring $T_B$. 
Wigner makes a choice $z=1,2$. For $z=2$ he performs the joint PBR measurement on $L_A \otimes L_B$ with outcome $w \in  \{ \xi_i \}_{i=1,\ldots,4}$. For $z=1$ he reveals Alice's and Bob's choices, meaning he measures $L_A \otimes L_B$ in the computational basis. Upon Wigner asking Alice and Bob for their outcome, let us label the corresponding outcome for $\ket{ij}_{L_A L_B}$ by $w= (i,j) | z= 1$ for $i,j \in \{0,1\}$.

\vspace{0.5cm}

The empirical probabilities that are used in the scenario are \begin{equation} \label{eq:proof_nonlocality1}
    \begin{split}
        &\wp(c{=}0,d{=}0,w{=}\xi_1|x{=}1,y{=}1,z{=}2)=0, \\
        &\wp(c{=}0,d{=}+,w{=}\xi_2|x{=}1,y{=}2,z{=}2)=0, \\
        &\wp(c{=}+,d{=}0,w{=}\xi_3|x{=}2,y{=}1,z{=}2)=0, \\
        &\wp(c{=}+,d{=}+,w{=}\xi_4|x{=}2,y{=}2,z{=}2)=0, \\
    \end{split}
\end{equation}
and \begin{equation} \label{eq:proof_nonlocality2}
    \wp(c{=}+,d{=}+,w{=} (0,0) |x{=}2,y{=}2,z{=}1) > 0.
\end{equation}
Moreover, to argue that $c|x=1$ is actually Alice's choice $a$ as revealed by Wigner when $z=1$, and similarly for Debbie and Bob, we have that \begin{equation} \label{eq:w_delta_ab}
    \wp(c,d,w|x{=}1,y{=}1,z{=}1) = \delta_{w,(c,d)} \wp(w|z{=}1).
\end{equation}
Now it is easy to see that these correlations are logically contextual (or, equivalently in this case, logically nonlocal, cf. \Cref{sec:recap_KSNC}). 
To simplify stating the argument, we label the outcome $c$ when performing measurement $x=1,2$ (where for $x=1$ Charlie measures Alice's qubit in the computational basis, decoding her choice $a$) by $c_1,c_2$, i.e., $c_x= (c|x)$ and similarly $d_y=(d|y)$ and $w_z = (w|z)$.
Then, we find from \cref{eq:proof_nonlocality2} that the assignment $(c_2{=}+,d_2{=}+,w_1{=}(0,0))$ is possible in the context $C=(x{=}2,y{=}2,z{=}1)$. We now prove that there exists no global assignment extending this possible assignment $(c_2{=}+,d_2{=}+,w_1{=}(0,0))$, meaning the empirical correlations are logically contextual. 
By \cref{eq:w_delta_ab}, $w_1=(0,0)$ implies that $c_1=0=d_1$.
However, this assignment cannot be extended to a global assignment; one cannot assign a value to $w_2$ consistent with \cref{eq:proof_nonlocality1}.
Namely, as $c_1=0=d_1$, by the first equation in \cref{eq:proof_nonlocality1}, we cannot have $w_2 = \xi_1$. 
As $c_1=0,d_2=+$, we cannot have $w_2=\xi_2$ by the second equation in \cref{eq:proof_nonlocality1}. 
Finally, as $c_2=+,d_1=0$, by the third equation in \cref{eq:proof_nonlocality1} we cannot have $w_2=\xi_3$, and the fourth equation and $c_2=+=d_2$ imply that $w_2\neq\xi_4$.

\section{Discussion} \label{sec:discussion_PBR_WF}
We have provided a scenario that violates the Local Friendliness assumptions based on a scenario combining Wigner's friend with the PBR theorem. 
In fact, the only relevant variables that are unavailable after the protocol, are Alice's and Bob's free choices, and thus, under the same locality assumption of the Local Friendliness no-go theorem, this provides an argument against the absoluteness of free choices.
In order to do so, we had to model the free choices being made by Alice and Bob in a particular way. 
We have tried not to subscribe to any interpretation of free choices a priori, i.e. whether free choice is an emergent concept or not, and our argument is not an argument against the notion of free choice, only against the absoluteness of it, modulo quantum universality and a locality assumption.
It might feel unnatural to model free choices in the way we did, and perhaps, in any way other than from the perspective of the entity who made that choice.
In case one objects to such modeling, at least, our argument invites a careful operational rethinking of free choices and their role in physics.

\ifSubfilesClassLoaded{ 
  \printbibliography[heading=bibintoc, title=References]
  }{}
\end{document}

\ifSubfilesClassLoaded{ 
	\setcounter{chapter}{1} 
	\maketitle \tableofcontents \listoftodos
	}{}

\chapter[Intermezzo: the Sleeping Beauty problem]{Intermezzo: the Sleeping Beauty problem}\label{chap:part1_sleepingbeauty}

In \Cref{chap:partI_refinedFR} we briefly encountered the Sleeping Beauty problem. 
In the Sleeping Beauty problem, Sleeping Beauty is put asleep, and a coin is tossed. 
For Tails, Sleeping Beauty is woken up on Monday, put asleep again and woken up again on Tuesday (with no recollection of her awakening on Monday). 
For Heads, Sleeping Beauty is woken up on Monday only. 
Each time Sleeping Beauty is woken up, she is asked what her belief is that the outcome was Heads.     
What should Sleeping Beauty reply?
In literature arguments have been made for both 1/3 and 1/2 as the correct answer.
In this chapter, I argue that 1/3 is the correct answer, and not 1/2, as Beauty being awake is nontrivial information that updates her beliefs about Heads to 1/3. 
This claim is strengthened by considering an additional observer, Prince Probability, who may or may not meet Sleeping Beauty. 
If he meets an awake Sleeping Beauty, he lowers his credence in Heads to 1/3.

Moreover, as argued in \Cref{chap:partI_refinedFR}, the Sleeping Beauty problem serves as a classical example that violates FR's consistency condition, and thereby raises the question of what is genuinely  \textit{quantum} or \textit{nonclassical} about the FR paradox~\cite{jones_thinking_2024}. 
Indeed, as we will see, Beauty and the Prince have different (probability) assignments.
However, our refined version of the FR paradox, presented in \Cref{chap:partI_refinedFR}, requires a weaker consistency condition, namely that observers need only agree when they meet and communicate about their thoughts. This condition is not violated in the Sleeping Beauty problem: when Beauty and the Prince meet, they will agree on their assignments.

In the remainder of this chapter, we discuss an operational argument that explains how the probabilities 1/2 and 1/3 arise, identify when each applies, and briefly relate to literature.
This chapter is based on Ref.~\cite{walleghem2024stunnedsleepingbeautyprince}.

\section{Introduction}
\paragraph{The Sleeping Beauty problem} I follow Ref.~\cite[Appendix C]{jones_thinking_2024} for the description of the Sleeping Beauty\footnote{We will sometimes abbreviate Sleeping Beauty as SB or Beauty.} problem\footnote{The SB problem was first written down by Elga in \cite{elga2020self}, who attributed it to Robert Stalnaker and Arnold Zuboff. Similar problems were posed in \cite{aumann1997forgetful,piccione1997interpretation}.}.
\begin{quote}
    \textit{The Sleeping Beauty problem: }
    Sleeping Beauty participates in an experiment lasting 48 hours in which she is put to sleep on Sunday, and a fair coin is tossed. If the outcome is Heads, she will be woken up on Monday only and put back to sleep. If the outcome is Tails, she will be woken up on Monday and Tuesday. More specifically, for outcome Tails Sleeping Beauty is woken up on Monday and put back to sleep with her memory of her awakening erased, to be woken up on Tuesday again. 
    Sleeping Beauty cannot distinguish between any of these awakenings. 
    When woken up, Sleeping Beauty is asked for her credence that the outcome of the coin toss was Heads.  
\end{quote}

As a proposal to make the question somewhat clearer, we may ask Sleeping Beauty, when she is woken up, the following question: `What do you now think the outcome was, Heads or Tails? And what is your credence now for Heads?'

\vspace{0.5cm}

Two possibilities for Sleeping Beauty to answer have been debated heavily in literature. The Halfers argue that 1/2 is the correct answer, the main argument being the absence of any new information for Beauty when she wakes up so that she should stick to her credence 1/2 she had on Sunday, whereas the Thirders argue that 1/3 is the correct answer.
Many works in favour of Elga's argument for 1/3~\cite{elga2020self}, and against Lewis's 1/2 arguments~\cite{lewis2001sleeping} have appeared \cite{elga2020self,elga2004defeating,hitchcock2004beauty,draper2008diachronic,arntzenius2003some,dorr2002sleeping,van2013sleeping,horgan2004sleeping,weintraub2004sleeping,monton2002sleeping,titelbaum2012embarrassment,titelbaum2013quitting,neal2006puzzles,stalnaker2010our,groisman2008end,bradley2006betting,rosenthal}.
The arguments for 1/2 mostly claim that no extra information is present for Beauty when she is woken up.\footnote{The fact that this is false was already noted (among others) in \cite{weintraub2004sleeping,horgan2004sleeping,hitchcock2004beauty}. In this chapter we give further arguments for this claim, by considering a concrete operationalization of the problem and considering the viewpoint of an outsider who may or may not meet Sleeping Beauty.} 
Here we argue that this claim is incorrect, in \Cref{sec:SB_Thirder}, using simple Bayesian probability theory and a slight extension of the protocol, with an outsider Prince Probability who may or may not meet Sleeping Beauty. We also consider the credence of a Sleeping Beauty who knows that she is dreaming. We believe that the strength of our argument is its simplicity. 
Of particular relevance is Rosenthal's argument~\cite{rosenthal}, which gives an elementary conditional-probability argument for 1/3 by introducing an auxiliary dime in the Sleeping Beauty experiment. Our approach is related, but operationalizes the relevant evidence as Beauty’s being awake now and makes the corresponding update explicit from the perspective of an external observer.
We consider variations of the SB problem in \Cref{sec:SB_variations}, and relate to Elga's argument and literature in \Cref{sec:literature}.

\section{A Thirder's Tale of Beauty and the Prince} \label{sec:SB_Thirder}

In this section we will argue how Sleeping Beauty knowing that she is awake is useful, nontrivial information for Sleeping Beauty, leading to the Thirder position. First, we consider Beauty's point of view, but perhaps most convincing is the story including an outsider Prince Probability, who may or may not meet Beauty and updates his credence in Heads accordingly, which we discuss subsequently. 
Finally, we argue what credence Beauty would assign if she knew she were dreaming (and thus asleep), the same credence that Prince would assign upon encountering Sleeping Beauty whilst she is asleep.

\vspace{0.5cm}

\paragraph{Beauty knowing that she is awake is nontrivial information.} In the Sleeping Beauty protocol, Beauty is asked to give her credence for Heads when being woken up. 
Importantly, here the fact that Beauty is awake is nontrivial: she could have been in a state of sleep too.
Suppose for example that each time Beauty is woken up she is awake for z hours (with $z < 24$). The event of her being awake under Heads then has probability \begin{equation}
    p(\text{SB awake}|H) =  z/(2\cdot 24) = \frac{1}{2}x,
\end{equation} with $x= z/24$, as she is awake for only $z$ hours in the two days (48 hours) of the experiment.\footnote{Here and further on, the events $H$ and $T$ stand for the outcome of the coin toss being Heads and Tails, respectively.}
On the other hand, under Tails we have \begin{equation}
\begin{split}
            p(\text{SB awake}|T) &= \frac{2z}{48} = x.
\end{split}
\end{equation} 
The probability of Beauty being awake is thus \begin{equation}
    p(\text{SB awake}) = p(H) p(\text{SB awake}|H) + p(T) p(\text{SB awake}|T) = \frac{1}{2} \cdot \frac{1}{2}x+\frac{1}{2} \cdot x = \frac{3}{4}x.
\end{equation} 
Therefore, the credence Beauty gives to Heads while being awake is \begin{equation} \label{eq:Heads_prob_main}
    p(H|\text{SB awake}) = \frac{p(H) p(\text{SB awake}|H)}{p(\text{SB awake})} = \frac{1/2 \cdot 1/2 x}{3/4 x } = 1/3.
\end{equation} 
Analogously, we find that $p(T|\text{SB awake})=2/3$. 
Thus indeed, while being awake Sleeping Beauty gives credence 1/3 to Heads, using the fact that she is awake, because she is more likely to be awake when the outcome is Tails. 

This argument can be easily extended to an experiment lasting $N$ days, where Beauty is woken up on $N$ consecutive days for Tails, whereas only once for Heads. In that case, when woken up, Beauty's belief in Heads is $1/(N+1)$. Furthermore, if the coin toss is also unfair, with $P(H)=h$, it is easy to verify that our method that led to \cref{eq:Heads_prob_main} gives that $P(H|\text{SB awake}) = \frac{1}{N\cdot \frac{1-h}{h} + 1}$.

\vspace{0.5cm}

\paragraph{Perspective of outsider Prince Probability.} 
Consider an outsider, Prince Probability\footnote{We will also refer to Prince Probability simply as Prince or PP.}, and for concreteness let Sleeping Beauty always reside in a sealed lab. When not partaking in the protocol, Prince Probability assigns credence 1/2 to Tails. Prince Probability has no idea of which day it is, and at any instance Prince Probability can choose to visit Beauty's lab. If he does so, he sees Beauty either asleep or awake. If Prince Probability sees Beauty being awake, he will lower his credence to 1/3 for Heads because under Tails she is awake for more time. The calculation is analogous to the one right above in \cref{eq:Heads_prob_main}. 
Upon meeting the Prince, Beauty does not seem to get any relevant information for Heads versus Tails, suggesting that she should indeed always assign credence 1/3 to Heads when she is awake. We have sketched this protocol in \cref{fig:PrinceProbability_SB}. 

This is in line with arguments in literature by Neal, Stalnaker and Horgan ~\cite{stalnaker2010our,neal2006puzzles,Horgan2017Exchange}. They considered an additional agent also part of the experiment (called Prince, Sleeping Ugly and Sleeping Betty, respectively), who was woken up both on Monday and Tuesday. 
These arguments are similar to ours, with the difference that our Prince Probability is truly an outsider to the experiment, not put to sleep.

\begin{figure}[t]
         \centering
\includegraphics[width=0.7\columnwidth]{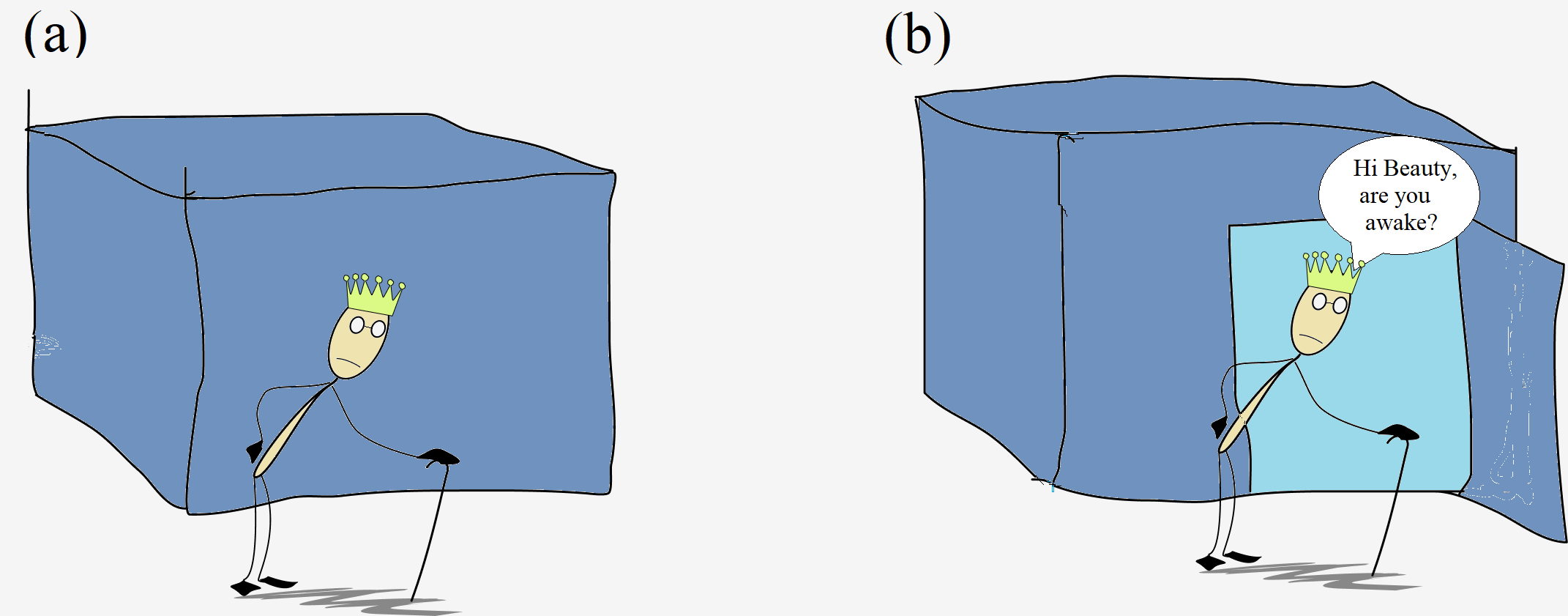}
         \caption[Prince Probability and Sleeping Beauty]{A schematic overview of the extension of the Sleeping Beauty problem that includes an outsider's perspective (Prince Probability's) as described in \Cref{sec:SB_Thirder}. (a): Sleeping Beauty resides in a sealed lab and an outsider Prince Probability assigns a probability 1/2 to Heads. (b): Prince Probability opens Sleeping Beauty's lab and can now see whether she is awake or asleep. If she is awake, Prince Probability lowers his credence for Heads to 1/3. If she is asleep, he increases his credence for Heads.
          }
         \label{fig:PrinceProbability_SB}
\end{figure}

\vspace{0.5cm}

\paragraph{What if Sleeping Beauty knew she was dreaming?} 
Consider the Sleeping Beauty problem, but now Beauty knows she is dreaming (and thus asleep, assuming she always dreams while asleep). Using similar Bayesian arguments as above, she will now increase her credence in Heads. 
Suppose Sleeping Beauty is awake for $x$ hours each time she is woken up (so $x < 24$ and we assume $x$ to be known to her). We have that $P(\text{SB asleep}|H) = (48-x)/48 = 1-y$ and $P(\text{SB asleep}|T) = (48-2x)/48=1-2y$ where we have defined $y=x/48$ (with now $y < 1/2$). 
Therefore, \begin{equation}
\begin{split}
    P(\text{SB asleep}) &= P(H)  P(\text{SB asleep}|H) + P(T)  P(\text{SB asleep}|T) \\ &= 1/2 \cdot (1-y) + 1/2 \cdot (1-2y) = 1/2 \cdot (2-3y),
\end{split}
\end{equation} and thus \begin{equation}
    P(H|\text{SB asleep}) = \frac{P(H) P(\text{SB asleep}|H)}{P(\text{SB asleep})} = \frac{1/2 \cdot (1-y)}{1/2 \cdot (2-3y)} = \frac{1-y}{2-3y}.
\end{equation} 
Thus, knowing that she is currently dreaming, Sleeping Beauty assigns credence $(1-y)/(2-3y)$ to Heads in this scenario.
This is also the credence an outsider Prince Probability would give to Heads when he sees that SB is asleep.
Importantly, through $y$ this result does depend on the time span Sleeping Beauty is awake for each time when woken up.
For example, if each time she wakes up, SB is awake for 23 (1) hours, she gives credence 0.926... (0.505...) to Heads while knowing that she is dreaming.
This difference and dependence on $y$ make sense intuitively: for $x=23$ Beauty is asleep for 25 hours if Heads and only 2 hours if Tails, whereas for $x=1$ Beauty is asleep for 47 hours if Heads and for 46 hours if Tails, a much smaller difference.

\section{A variation of the Sleeping Beauty problem}
\label{sec:SB_variations}
We now discuss a slightly altered version of the copy-SB version of Ref.~\cite{jones_thinking_2024}, similar to the Sailor's Child problem~\cite{neal2006puzzles} and Rosenthal’s sleeping-twins variant~\cite{rosenthal}, which has a resolution similar to the one we proposed above in \Cref{sec:SB_Thirder} for the original SB problem.

\vspace{0.5cm}

\paragraph{The copy-SB protocol} 
There are two identical sealed labs, indistinguishable from the outside and inside. 
Sleeping Beauty resides in one of those labs.
A fair coin is tossed, with the outcome not revealed to Sleeping Beauty. 
If the outcome is Tails, a copy of Sleeping Beauty is made and put in the second lab, so that in that case both labs contain a Sleeping Beauty. Again Sleeping Beauty knows the protocol.
Each version of Sleeping Beauty is asked whether she thinks the outcome of the coin toss was Heads or Tails, and what her credence for Heads is.
We will consider again the perspective of an outsider Prince Probability. 
All Prince Probability knows is the protocol: he knows that there are two labs, but he does not know in which lab the original Sleeping Beauty resided. He knows there was a coin toss that may have led to the copying of Sleeping Beauty so that both labs would contain a Sleeping Beauty, but he does not know the outcome of the coin toss, i.e. he does not know whether both labs contain a Sleeping Beauty or only one and which lab contains a Sleeping Beauty.

\vspace{0.5cm}

\paragraph{Resolution of the copy-SB problem} 
We will argue that if SB is woken up and is asked which outcome the coin toss had, she would assign 1/3 probability for Heads.
An outsider, Prince Probability, who does not know the outcome of the coin toss, gives credence 1/2 to Heads prior to any action he undertakes. 
But when he meets a SB, he changes his credence to 1/3 for Heads. 
Therefore, when Prince Probability and Beauty meet, they agree on their probability assignments for Heads, namely 1/3. Before they have met, however, Prince Probability and Beauty can each correctly assign different credences to Heads, namely 1/2 by Prince Probability and 1/3 by Beauty.

\vspace{0.5cm}

\paragraph{Prince Probability's point of view}
The probability that opening a lab Prince Probability finds Beauty is \begin{equation}
\begin{split}
    p(\text{PP finds SB}) &= p(T) p(\text{PP finds SB}|T)+ p(H) p(\text{PP finds SB}|H) \\ &= \frac{1}{2} \cdot 1 + \frac{1}{2} \frac{1}{2} = \frac{3}{4}    
\end{split}
\end{equation} and similarly, the probability of Prince Probability finding a Beauty upon opening the door of a lab under outcome Heads (which we refer to as finding a Heads SB) is $p(\text{PP finds SB} \land H) = \frac{1}{4}$.
 Therefore, if Prince Probability finds a Sleeping Beauty, the probability that the outcome of the coin toss was Heads is equal to $p(\text{PP finds SB}\land H|\text{PP finds SB}) = \frac{1}{3}$.

Thus, before Prince Probability enters a lab, he argues that the probability for Heads is 1/2. 
But, when he enters a lab, he changes his credence in Heads, depending on whether or not there is a Beauty. 
Namely, if he finds a Beauty, his credence in Heads becomes 1/3. 
This is not surprising: Prince Probability has gotten new information, namely that he sees a Beauty. 
For outcome Tails he is more likely to find a Beauty. Thus, conditioned on Prince Probability finding a Beauty, he gives credence 1/3 to Heads. 
Still, Prince Probability assigns probability 1/2 to Heads before he undertakes any action.

\paragraph{What if Prince Probability finds no Sleeping Beauty?}
If there are only two labs and if Prince Probability opens a lab and finds no Beauty, he will know the outcome was Heads for sure. 
If there were additional empty labs (so under Tails only still two labs contain a Beauty), then Prince Probability will have credence higher than 1/2 but less than 1 that the outcome was Heads, as he is more likely to find no Beauty if the outcome were Heads. His credence here thus depends on the number of labs there are, similar to the discussion of Prince Probability finding Sleeping Beauty asleep in \Cref{sec:SB_Thirder}.

\paragraph{Sleeping Beauty's point of view} 
From the above discussion, we found that when Prince Probability and a Beauty meet, Prince Probability assigns probability 1/3 to Heads. 
But from Beauty's point of view, Prince Probability visiting her does not give her useful information: Prince Probability simply chooses to enter her lab or the other one, with equal probability. 
This seems to suggest that Beauty indeed always assigns probability 1/3.
Namely, Beauty names the two labs Lab 1 and Lab 2, and without loss of generality assumes she is in Lab 2. So the fact that she exists is the event `There is a SB in Lab 2' or shortly `SB in Lab 2'. 
However, while outside the labs, Prince Probability can name the two labs 1 and 2 to his liking, but he cannot make a statement such as `There is a SB in Lab 2', he can only say `Lab 1 or Lab 2 contain a SB'.
Thus, Sleeping Beauty's credence (of 1/3) in Heads is conditional on the event `SB in Lab 2', a statement that Prince Probability having credence 1/2 before entering the labs cannot make. This is an almost trivial observation for Sleeping Beauty, but one that matters. 
Namely, if the outcome of the coin toss is Tails, the probability of `SB in Lab 2' is given by \begin{equation}
    p(\text{SB in Lab 2}|T) = 1, 
\end{equation} whereas for Heads there is a Beauty either in Lab 1 or 2, but not both, \begin{equation}
    p(\text{SB in Lab 2}|H) = \frac{1}{2}.
\end{equation} We also have that \begin{equation}
    p(\text{SB in Lab 2}) = p(T) p(\text{SB in Lab 2}|T) + p(H) p(\text{SB in Lab 2}|H) = \frac{1}{2} \cdot 1 + \frac{1}{2} \cdot \frac{1}{2} = \frac{3}{4}.
\end{equation}
Therefore, the credence for Beauty for Tails is \begin{equation}
    p(T|\text{SB in Lab 2}) = \frac{p(T) p(\text{SB in Lab 2})}{p(\text{SB in Lab 2})} = \frac{1/2 \cdot 1}{3/4} = 2/3.
\end{equation} The credence Sleeping Beauty gives to Heads is thus 1/3, where she uses the fact that she exists (i.e. Lab 2 contains her).

\section{Relating to literature on Sleeping Beauty} \label{sec:literature}
Elga's argument using the Principle of Indifference for self-locating belief leads to the same conclusion as ours in \Cref{sec:SB_Thirder}, namely that Sleeping Beauty should assign credence 1/3 to Heads when woken up \cite{elga2020self}. Furthermore, also in variations of the protocol, such as an unfair coin toss and many awakenings, Elga's resolution method gives the same answer as ours, and accords with the updating of the outsider Prince Probability's point of view when meeting a Sleeping Beauty who is awake. 
We can see this fact as an argument for the correctness of Elga's reasoning.

After repeating Elga's argument and showing how it agrees with our arguments, we briefly discuss what arguments on the Sleeping Beauty problem have been given in literature, and counter some of the prominent Halfer arguments.

\subsection{Elga's argument for 1/3 and the Principle of Indifference} 
Elga \cite{elga2020self} uses the Principle of Indifference for self-locating belief (see also Ref.~\cite{elga2004defeating}) to argue that the correct answer for Sleeping Beauty to give when woken up is 1/3.
More precisely, in Ref.~\cite{elga2004defeating} Elga states the Principle of Indifference for self-locating belief as follows (quoting from Ref.~\cite{elga2004defeating}): 
\begin{quote}
    \textsc{Principle of Indifference:} Similar centered worlds deserve equal credence.
\end{quote}

If Sleeping Beauty is woken up, she knows she is in one of three events: 
\begin{itemize}
    \item[$H_1$:] Heads and Monday,
    \item[$T_1$:] Tails and Monday,
    \item[$T_2$:] Tails and Tuesday.
\end{itemize}
Elga's resolution then proceeds by arguing that $P(T_1) = P(T_2)$ and $P(T_1)=P(H_1)$. Namely, given Tails, $T_1$ and $T_2$ take place in similar centered worlds, so by the Principle of Indifference for self-locating belief $P(T_1|T_1 \text{ or } T_2)=P(T_2|T_1 \text{ or } T_2)$, so that $P(T_1)=P(T_2)$. 
Alternatively to the Sleeping Beauty problem where the coin is flipped \emph{first} on Sunday, we could also imagine the protocol where Sleeping Beauty is woken up on Monday and \emph{then} a coin is flipped determining whether Sleeping Beauty is also woken up on Tuesday (Tails) or not (Heads).
Elga argues that in both cases, if Sleeping Beauty were told that it is Monday, she must assign the same credence to Heads, namely 1/2, i.e. $P(H_1|H_1 \text{ or } T_1) = 1/2$, leading to $P(H_1) = P(T_1)$. 
Using $P(T_1)=P(T_2)$ and $P(H_1)=P(T_1)$ together with the normalisation $P(T_1) + P(T_2) + P(H_1)=1$, we obtain that indeed $P(H)=P(H_1)=1/3$. Elga thus argues that when woken up Sleeping Beauty should give credence 1/3 to Heads, in agreement with our resolution.

\vspace{0.5cm}

Importantly, the Principle of Indifference does \emph{not} allow you to argue that, as $T_1,T_2,H_1$ are all subjectively indistinguishable, they should all get equal credence 1/3. For example, in the case of an unfair coin toss, Elga's argument will correctly divert from credence 1/3 for Heads. For example, if the unfair coin toss gives Heads with probability $h=1/4$, Elga's argument goes as follows. Still, $T_1$ and $T_2$ take place in similar centered worlds, so $P(T_1) = P(T_2)$. 
Suppose it is Monday, so we condition on the event $H_1 \text{ or } T_1$ (SB waking up on Monday). Then, by the same argument as above, we could also take the coin to be tossed after Sleeping Beauty waking up on Monday and being put back to sleep, for which then Heads has probability $h=1/4$. 
So $P(H_1|H_1 \text{ or } T_1) = h$, giving $P(H_1) = h P(H_1) + h P(T_1)$, and thus $P(H_1) = h/(1-h) P(T_1)$. Combining this with $P(T_1)=P(T_2)$ and the normalisation $P(H_1)+P(T_1)+P(T_2)=1$, we obtain,  for example for $h=1/4$, that $P(H_1)=1/7$. It is easy to verify that also in the general case of $N$ awakenings and an unfair coin toss with $P(H)=h$, Elga's resolution accords with ours and Rosenthal's~\cite{rosenthal}, namely $P(H|\text{SB awake}) = \frac{1}{N\cdot \frac{1-h}{h} + 1}.$

\vspace{0.5cm}

As noted by Elga himself using an entertaining story involving Dr Evil \cite{elga2004defeating}, a consequence of the Principle of Indifference that seems somewhat unsettling is the fact that it favors situations where there are many duplicates of ourselves, see also Ref.~\cite{titelbaum2013ten}. 
Similar difficulties arise when applying the Principle of Indifference to the many-worlds interpretation of quantum mechanics \cite{titelbaum2013ten} and Boltzmann Brain universes\footnote{Another uneasiness about Boltzmann Brains is the following. Boltzmann Brains arise in our models of physics and cosmology, but Boltzmann Brains have untrustworthy memories, i.e. their memories of `observed' data are not in line with the universe they live in \cite{carroll2020boltzmann}. Thus if we were Boltzmann Brains, we should not trust our memories of `observed' data, undermining the use of our physical theories and thus the argument for the existence of Boltzmann Brains.} in cosmology \cite{jones_thinking_2024,titelbaum2013ten}. We will not explore these interesting questions further here.

\subsection{The Thirders vs. the Halfers}
Lewis argues in \cite{lewis2001sleeping} against Elga's resolution of 1/3, and favors 1/2, based on the apparent (but false) idea that no new relevant information is gained by Sleeping Beauty waking up. 
In \Cref{sec:SB_Thirder} we have countered this thought by operationalising the SB scenario somewhat more. In that case it is clear that by knowing that she is awake, Beauty does obtain relevant new information, i.e. Beauty being woken up is a nontrivial fact during the two days of the experiment. 
The fact that this piece of information is nontrivial, namely `I (SB) am awake \emph{now}', has also been noted in (among others) Refs.~\cite{weintraub2004sleeping,horgan2004sleeping,hitchcock2004beauty}, as an argument against 1/2. 
Monton argued for a change in information too, namely that when Beauty wakes up compared to before she was put to sleep on Sunday Beauty has a loss of her temporal location \cite{monton2002sleeping}. 
Moreover, using a variation of the protocol, Dorr \cite{dorr2002sleeping} and Arntzenius \cite{arntzenius2003some} argued how indeed at Beauty's awakening there is nontrivial new information for her.
Rosenthal~\cite{rosenthal} uses conditional probabilities while introducing an additional dime to argue for 1/3; his arguments and results are in line with ours, and his derivation may be viewed as complementary to ours.
Our more operational setting and introduction of the outsider's view Prince Probability who may or may not meet Beauty thus provide strong support for these arguments.

\vspace{0.5cm}

White provided a generalized Sleeping Beauty problem in Ref.~\cite{white2006generalized} that, in his view, seems to argue for 1/2. However, for his argument in favor of 1/2 using this generalized Sleeping Beauty problem, White uses a probability conditioned on the event `SB is being woken up at least once', which is different from the self-locating event where Sleeping Beauty notices that she has been woken up, also noted in Refs.~\cite{van2013sleeping,horgan2007synchronic}.
Namely, before Prince Probability enters any lab (and thus assigns credence 1/2 to Heads), he obviously knows that Beauty is woken up at least once. It is only when he actually enters a lab and meets Beauty while she is awake that he updates his credence in Heads to 1/3. 
The conditioned event $W=\text{ `SB awake at least once'}$ is different from `SB awake \emph{now}', corresponding to the statements `SB in Lab 1 or 2' versus `SB in Lab 2' of \Cref{sec:SB_variations}.


\vspace{0.5cm}

The Bayesian rule of updating (conditionalization) for self-locating beliefs has also been questioned in favor of 1/2 answers for the SB problem in Ref.~\cite{meacham2008sleeping}. However, having given arguments in favor of 1/3, especially our argument with the outsider Prince Probability who when meeting an awake SB changes his credence to 1/3 for Heads, questions such proposed change for the Bayesian update rule for self-locating beliefs as in Ref.~\cite{meacham2008sleeping}.

\vspace{0.5cm}

For more literature on the Sleeping Beauty problem, we refer the reader to Ref.~\cite{van2013sleeping,titelbaum2013ten}.

\section{Conclusion} \label{sec:conclusion_chap6}
In this chapter we have given arguments employing simple Bayesian probability that suggest 1/3 to be the correct answer for Sleeping Beauty to reply when she is woken up and asked for her credence in Heads. 
Sleeping Beauty does have extra useful information, which is why 1/2 is not her correct answer. 
An outsider, Prince Probability, assigns credence 1/2 to Heads, but when Prince Probability meets Sleeping Beauty being awake, he lowers his credence in Heads to 1/3.
Furthermore, if Prince Probability meets Sleeping Beauty while asleep, he instead increases his credence in Heads.

\ifSubfilesClassLoaded{ 
  \printbibliography[heading=bibintoc, title=References]
  }{}
\end{document}

\part{Universality and black holes}
\label{part:universality_black_holes}

The central assumption of \Cref{part:universality_in_quantum_foundations} was that quantum theory applies equally to microscopic laboratory systems, measurement devices and macroscopic observers.
Under this hypothesis, we explored several foundational results that challenge our naive understanding of reality. 
We now turn to another question of universality. 
How can quantum theory, assuming its universal applicability extends to gravity,
resolve the puzzles that black holes pose?
While we will not fully answer this question, we extend these paradoxes further, point out subtleties concerning some standard assumptions, and present a calculation in which aspects of quantum theory are combined with general relativity in the form of a superposed black hole.  

After briefly outlining black hole paradoxes and proposed resolutions in literature in \Cref{chap:BH_prelims}, the remainder of this part is organized as follows.
Bridging Part I and Part II of this thesis, we present several scenarios in \Cref{chap:part2_WFBH} that merge extended Wigner's friend arguments with black hole physics.
These scenarios yield paradoxes that strengthen the cloning and firewall paradoxes, and rely on standard assumptions of black hole puzzles.
We highlight several important subtleties in these assumptions, inviting a more careful verification thereof towards resolving these puzzles.
Among the methods that can be employed for this purpose are the use of quantum reference frames and operational investigations of a black hole and its radiation.
In \Cref{chap:part2_detector}, we illustrate the combination of these approaches by analyzing the response of a particle detector near a position-superposed black hole using aspects of quantum reference frames. 
This analysis was motivated by the study of a detector's response near a mass-superposed black hole in literature~\cite{foo2022quantum}.
\Cref{chap:part2_WFBH,chap:part2_detector} are based on Refs.~\cite{walleghem2025wignersfriendsblackhole} and \cite{walleghem2026particle}, respectively.

We end with a short discussion and conclusion of both parts in \Cref{chap:conclusions&future-work}.

\ifSubfilesClassLoaded{ 
	\setcounter{chapter}{1} 
	\maketitle \tableofcontents \listoftodos
	}{}

\chapter[A brief introduction to black hole puzzles]{A brief introduction to black hole puzzles}\label{chap:BH_prelims}

Arguably, few phenomena in the universe are as captivating as black holes, where two of the most foundational theories of modern science -- general relativity and quantum theory -- intersect in profound ways. 
Applying principles of quantum theory to black holes, Hawking proved in his seminal paper~\cite{hawking1975particle} that black holes radiate, and by energy conservation thereby evaporate over time. 
However, Hawking’s calculations suggested this radiation to be approximately thermal, which seemed to indicate a loss of information: an initial pure state of matter collapsing to a black hole would evolve nonunitarily to a mixed state of radiation after evaporation~\cite{hawking1976breakdown}.
This enigma is also referred to as the black hole information  puzzle~\cite{harlow2016jerusalem,Polchinski_2016}.  

Page gave a general argument of how a black hole system might evolve, based on the assumption that the black hole and radiation combined are always in a random pure state \cite{page1993average,page1993information}. 
The resulting curve for the (renormalised~\cite{holzhey_geometric_1994,harlow2016jerusalem}) entanglement entropy\footnote{We generally distinguish two types of entropy: the entanglement entropy, also referred to as the von Neumann or fine-grained entropy, describing the entanglement of one system with another, and the thermodynamic entropy, a coarse-graining thereof~\cite{almheiri2021entropy}.} is estimated to decrease somewhat past halfway of the evaporation~\cite{page1993information,page2013time}, where it diverges from the entanglement entropy per Hawking's calculation.

From Page's unitary entropy curve, late-time radiation should be highly entangled with the early radiation. Moreover, the equivalence principle and quantum field theory (QFT) in curved spacetime suggest high entanglement across the horizon~\cite{eisert2010colloquium}. Therefore, late-time radiation just outside the black hole should be highly entangled with both the early radiation and modes inside the black hole, apparently violating monogamy of entanglement, also known as the firewall paradox~\cite{braunstein2007quantum,braunstein2013better,almheiri2013black,mathur2009information}.

Another black hole puzzle is a no-cloning argument  analyzed by Susskind and Thorlacius~\cite{susskind1994gedanken} and later by Hayden and Preskill~\cite{hayden2007black}, suggesting a form of black hole complementarity~\cite{Stephens_1994,susskind1993stretched,susskind1994gedanken}. 
The firewall argument is sometimes presented as an argument against complementarity~\cite{harlow2016jerusalem}, but additionally assumes the approximate validity of semiclassical gravity to argue for high entanglement across the horizon.

In the rest of this chapter, we briefly outline arguments in literature arguing for unitarity, and comment upon an aspect overlooked by Hawking when originally formulating his information loss paradox, a black hole's soft hair.

\section{Literature on the black hole information puzzle} 
\label{sec:BH_puzzles}

The resolution of the black hole information puzzle remains heavily debated and inconclusive, but much work has been done since, hinting that information is preserved. 
Expectations that black hole evaporation is a unitary process come from AdS/CFT~\cite{maldacena1999large,harlow2016jerusalem}, and more recently from the proposed existence of `entanglement islands' inside the black hole giving a Page curve for the radiation~\cite{penington2020entanglement,almheiri2020page,almheiri2019entropy,wang2024entanglement,Hartman_2020,wang2021page,balasubramanian2024microscopic,Hashimoto_2020,Balasubramanian_2021,almheiri2020entanglementislandsinhigher,kibe2022holographic,chen2020quantumextremalislandseasyI,chen2020quantumextremalislandseasyII,bousso_islands_2023,almheiri2023islandsoutsidehorizon,Hartman_2020islandsincosmology,Krishnan_2021,Van_Raamsdonk_2021,geng2021information,geng2021informationparadoxandits,geng2026seeingpagecurvesislands}.\footnote{Concerns have been expressed that such Page curve from islands only arises in massive gravity~\cite{geng2022inconsistency,Geng_2020,raju2022failure,geng2025revisitingrecentprogresskarchrandall,geng2021information,geng2025makingcasemassiveislands}; in turn, a reply can be found in Ref.~\cite{antonini2025apologiaislands}, and a reply to this reply in Ref.~\cite{geng2026seeingpagecurvesislands}. See also footnote \ref{fn:replica_subtility} in \Cref{chap:part2_WFBH}.} 
Replica trick calculations of entanglement entropy in 1+1d Euclidean gravity~\cite{mertens2023solvable} suggest that additional saddle points —-- so-called replica wormholes~\cite{almheiri2020replica,penington2022replica,Goto_2021}---appear when entanglement islands form.
Such replica wormholes lack a clear interpretation, but have been proposed to relate to black hole hair~\cite{calmet2024blackholeinformationreplica,An2023replica}.
Unitarity has also been argued for from two-dimensional investigations~\cite{ashtekar_information_2008,ashtekar_surprises_2011,Ashtekar_2011,barenboim2024no}.

Another proposal for information recovery consists of stable or long-lived remnants, objects that look small from the outside but store an enormous amount of information in their interior~\cite{Chen_2015,ong2025caseblackholeremnants}. An example of such proposal is black-to-white hole tunneling~\cite{DAmbrosio:2020mut,Christodoulou:2016vny,rovelli2014planck,Bianchi:2018mml,BenAchour:2020gon,Rovelli:2024sjl,Haggard:2014rza,Han:2023wxg}. Arguments in favor of remnants argue that there is a notion of volume that is very large for old black holes~\cite{Christodoulou:2014yia,Christodoulou:2016tuu}. 
Remnants are, however, often argued to encounter problems such as infinite pair production~\cite{Giddings_1995,susskind1995troubleremnants,Giddings:1993km}\footnote{For counterarguments see~\cite{ong2025caseblackholeremnants,Chen_2015,Bianchi:2018mml}.}. A related idea consists of baby universes that form inside black holes~\cite{Parentani_1994,Easson_2001}. If a (Planck-scale) remnant held the information, there would be no need for the Hawking radiation to purify (until the remnant potentially decays and spits out the information it holds in low-energetic radiation), and thus also no Page curve for the (pre-remnant) Hawking radiation, no cloning and firewall paradoxes (and thus also no black hole Wigner's friend arguments of this work). We will not discuss this option further here.

Proposals that change the structure or existence of the horizon include, in addition to firewalls, fuzzballs~\cite{Mathur_2005}, nonviolent nonlocality~\cite{giddings2013nonviolent}, gravastars as an alternative to black holes~\cite{mazur2004gravitational} and that evaporation may be halted earlier than expected leading to `dark stars', which, however, by itself may not yield a full resolution to the information puzzle~\cite{KAWAI_2013,Baccetti_2017,Baccetti_2018,Carballo_Rubio_2018,cardoso2019testing,Chen_2018}.

One recent idea regarding a physical mechanism for unitary black hole evaporation includes soft hair, which we discuss more in \Cref{sec:soft_hair}, low-energy gravitons dressing black hole spacetimes that could purify the Hawking radiation~\cite{hawking2016soft,strominger2018lecturesinfraredstructuregravity,carney_infrared_2017,carney_dressed_2018,strominger2017blackholeinformationrevisited,flanagan_order-unity_2021,calmet2022quantuma}. 
Soft hair and soft theorems~\cite{weinberg_infrared_1965} are deeply tied to (gravitational) memory~\cite{Zeldovich:1974gvh,christodoulou1991nonlinear,thorne1992gravitational,Bieri_2024,maibach2026balancefluxlawsgeneral} and asymptotic symmetries~\cite{bondi1962gravitational,sachs1962gravitational,compere2019advancedlecturesgeneralrelativity,flanagan_conserved_2017,compere_poincare_2020,compere_asymptotic_2023,satishchandran2019asymptotic,prabhu_infrared_2024} through the infrared triangle~\cite{strominger2018lecturesinfraredstructuregravity}, and indeed argued to be of crucial importance for infrared problems in massless scattering~\cite{prabhu_infrared_2022,Sommerfeld:1931qaf,fierz1936kunstliche,bloch1937note,Mott_1931,satishchandran2019asymptotic,prabhu_infrared_2024}.
Related proposals are quantum hair from higher-order curvature corrections~\cite{calmet2022quantumb,calmet2023quantum,calmet2024blackholeinformationreplica,calmet2022quantuma} and black-hole gravitational condensates~\cite{dvali2011blackholesquantumnportrait},
and a black hole wavefunction, which can be seen as a form of hair, that spreads or branches out~\cite{Eyheralde_2017,Eyheralde_2020,brustein2013restoring,alberte2015density,flanagan_order-unity_2021,calmet2024blackholeinformationreplica,calmet2022quantuma}.

\section{A black hole's soft hair} \label{sec:soft_hair}

Near infinity, far away from the black hole, what a particle detector such as the Unruh--DeWitt one probes is the radiation density matrix $\rho_{w,w'}$ defined by~\cite{Eyheralde_2017,Eyheralde_2022,alberte2015density,agullo2007short} \begin{equation} \label{eq:radiation-density_matrix}
\rho_{w,w'}=\int_0^{\infty} \mathrm{d} \omega'' \beta_{w w''} \beta_{w' w''}^*,
\end{equation} with $\beta$ the Bogoliubov coefficients relating past and future vacua and $w$ shorthand for the frequency and angular labels $(\omega, l, m)$.
This density matrix was suggested to be approximately thermal by Hawking~\cite{hawking1975particle,Eyheralde_2022}, being only non-zero for $\omega=\omega',l=l',m=m'$. Therefore, 
it cannot be pure, even if the black hole were formed from a pure state. 
As this calculation does not take into account any backreaction,
this leaves a mechanism for early radiation to get entangled with the black hole's properties, and thereby also with the late radiation.
When taking into account backreaction, conservation laws indeed suggest that the radiation's frequency and angular labels become entangled with the black hole mass and momentum, and thereby potentially influence later emissions. 
Yet, the no-hair theorems, or at least a naive reading thereof, tell us that the only relevant properties of a black hole are its mass, position, angular and linear momentum, and electromagnetic charge.
Therefore, the radiation’s infinite set of angular momentum $l$-labels only has the 11 black hole no-hair degrees of freedom (in 4d) to correlate with, too few to purify its state. 
This (apparent) problem of nonunitary evolution is also referred to as the black hole information puzzle.

\vspace{0.5cm}

However, in the last 15 years it has become clear that there are potentially many more degrees of freedom of the black hole spacetime that the radiation can get entangled with, the black hole's soft hair~\cite{hawking2016soft,donnay_supertranslations_2016,carney_infrared_2017}. 
In fact, already since the pioneering work of Bondi, van der Burg, Metzner and Sachs~\cite{bondi1962gravitational,sachs1962gravitational}, it has been known that the asymptotic symmetry group of asymptotically flat spacetimes (in 4d) is not just the Poincaré group of translations, rotations and boosts, but an infinite-dimensional group, enlarged with supertranslations, angle-dependent translations (and moreover potentially  
extended to encompass superrotations and superboosts~\cite{barnich_symmetries_2010,campiglia_asymptotic_2014,compere_poincare_2020,strominger2018lecturesinfraredstructuregravity}).
These additional symmetries endow the spacetime with many more nontrivial charges, and stem from diffeomorphisms that can locally be gauged away 
but act nontrivially at infinity.
Such asymptotic symmetries are deeply tied to the memory effect~\cite{Zeldovich:1974gvh,christodoulou1991nonlinear,Bieri_2024,heisenberg2025unifyingordinarynullmemory}, the lasting displacement of the relative positions of detectors after the passing of a gravitational wave, and soft theorems~\cite{weinberg_infrared_1965} through the infrared triangle~\cite{strominger2018lecturesinfraredstructuregravity,strominger_gravitational_2016,choi2025classicalsuperrotationinfraredtriangle}. 
Related is a long-standing problem in massless quantum scattering, where the $S$-matrix was long known to suffer from infrared divergences~\cite{prabhu_infrared_2022,Sommerfeld:1931qaf,fierz1936kunstliche,bloch1937note,Mott_1931}. 
Historically, this has been dealt with by calculating inclusive cross sections, focusing on observables agnostic to the infrared sector~\cite{YENNIE1961379,bloch1937note,weinberg_infrared_1965}, but a complete, satisfactory treatment remains problematic.
Incorporating memory effects appears crucial for developing such a framework~\cite{prabhu_infrared_2024}.
The field of quantum reference frames~\cite{Vanrietvelde_2020,vanrietvelde_switching_2023,Giacomini_2021,giacomini2019quantum,hoehn_trinity_2021,castroruiz2025interpretingquantumreferenceframe,H_hn_2022,Araujo_Regado_2025,kabel2023quantum,kabel2024identification,Cepollaro_2024} may have an enlightening role to play in this story~\cite{Araujo_Regado_2025,kabel2023quantum}, 
with Goldstone modes labeling the degenerate vacua 
suggested to arise from the relation between internal and external reference frames~\cite{Araujo_Regado_2025}. 
Therefore, a black hole's soft hair was an aspect overlooked by Hawking when phrasing his argument for the information puzzle, and thus might play an important role in its resolution, or at least in a complete quantum treatment of a black hole's evolution.

\section{What about paradoxes for infalling observers?} \label{sec:infalling_paradoxes}

Despite these arguments for unitarity, a clear, robust physical understanding of black hole paradoxes for infalling observers such as the firewall and cloning paradoxes remains elusive. 
One proposal is black hole complementarity~\cite{Stephens_1994,susskind1993stretched,susskind1994gedanken,Lowe_1995,kiem1995black} where the inside of a black hole and its radiation outside are not independent, and can be motivated from nontrivial relations between infalling matter and outgoing modes by backreaction~\cite{kiem1995black}.
Complementarity is only a principle, for which the firewall paradox was originally presented as a challenge~\cite{almheiri2013black,susskind2013blackholecomplementarityharlowhayden}, and can be formalized in different ways for an actual theory of the interior~\cite{harlow2016jerusalem}. 
Modeling black hole evaporation through quantum circuits suggests the importance of computational complexity and quantum error correction~\cite{harlow_quantum_2013,aaronson_complexity_2016,susskind2016computational,kim_ghost_2020,yoshida_firewalls_2019,akers2022blackholeinteriornonisometric}, where it has been argued that computational complexity protects the semiclassical causal structure of spacetime~\cite{kim_ghost_2020}, i.e. computational complexity preventing someone from creating a `firewall' of horizon excitations by manipulating the Hawking radiation.

However, even with complexity~\cite{aaronson_complexity_2016,harlow_quantum_2013,harlow2016jerusalem,Oppenheim_2014,akers2022blackholeinteriornonisometric,kim_ghost_2020,susskind2016computational,akers2022blackholeinteriornonisometric} and waiting time~\cite{hayden2007black} arguments to avoid operational contradictions for the firewall and cloning paradoxes, quantum theory cannot give a single description of two quantum clones or a violation of monogamy of entanglement.
Specific proposals for the interior include 
state-dependence~\cite{papadodimas2014state,papadodimas2014black,Harlow_2014},
also argued for in a milder form in the paradigm of entanglement wedge reconstructions and islands~\cite{penington2020entanglement,Cotler_2019entanglement}, final state projection~\cite{Lloyd_2014,Bousso_2014}, strong complementarity~\cite{harlow2016jerusalem} and proposals that modify aspects around the horizon, either violently such as firewalls, or (potentially) more subtly through nonviolent nonlocality~\cite{Giddings_2013,giddings2013nonviolent} and string-theoretic fuzzballs~\cite{Mathur_2005}, see also Refs.~\cite{harlow2016jerusalem,cardoso2019testing} for more discussion on other proposed compact quantum objects.
Other proposals view the interior or spacetime as emergent concepts~\cite{concepcion2024complementaritydynamicalblackhole,dvali2011blackholesquantumnportrait}, with Ref.~\cite{concepcion2024complementaritydynamicalblackhole} shown to exhibit a form of complementarity and state-dependence.

\vspace{0.5cm}

We will revisit these paradoxes for infalling observers in \Cref{chap:part2_WFBH}, more carefully analyzing their assumptions, and combining them with extended Wigner's friend paradoxes.
As we will discuss in \Cref{sec:implications_WFBH} of \Cref{chap:part2_WFBH}, important subtleties for paradoxes featuring infalling observers may be superposed quantum geometries and issues of locality and gauge-invariance.

\ifSubfilesClassLoaded{ 
  \printbibliography[heading=bibintoc, title=References]
  }{}
\end{document}

\ifSubfilesClassLoaded{ 
	\setcounter{chapter}{1} 
	\maketitle \tableofcontents \listoftodos
	}{}

\chapter[Wigner's friend's black hole adventure]{Wigner's friend's black hole adventure}\label{chap:part2_WFBH}

At the heart of both black hole physics and Wigner’s friend scenarios lies the question of unitarity. 
In Wigner’s friend setups, sealed-lab measurements are modeled unitarily, probing the measurement problem.
In black hole physics, the unitarity problem concerns information preservation in evaporation. 
Inspired by recent work of Hausmann and Renner~\cite{hausmann2025firewallparadoxwignersfriend}, who draw analogies between Wigner's friend and black hole paradoxes, we combine extensions of the Wigner's friend scenario with black hole physics to derive no-go theorems that strengthen the black hole no-cloning and firewall paradoxes.
More specifically, the cloning paradox leaves room for a theory that could describe the entire black hole physics of the interior and exterior while satisfying the assumptions of these paradoxes but without operational contradictions concerning quantum theory's predictions. 
We close this loophole, showing that no such theory exists modulo the standard assumptions of the cloning paradox.
Similarly, we lift the firewall argument to a paradox showing that no such theory exists when assuming persistent outcome records in the black hole interior. 
We also give an argument for why the firewall paradox assumptions give rise to a violation of no-signalling.  
We will employ standard assumptions of the cloning and firewall paradoxes for our protocols regarding unitarity and the decoding of Hawking radiation, but provide a more critical assessment in \Cref{sec:implications_WFBH} by briefly outlining several important subtleties that question the validity of these assumptions.

The assumption that superobservers as in the Wigner's friend scenario exist can be argued to be not much more exotic than the capability of an observer to reconstruct the initial state of a complicated quantum system from its Hawking radiation.
In our protocols and no-go theorems, we assume that the superobserver can perform arbitrary quantum operations on an observer and its environment, using the wording of Ref.~\cite{bong2020strong}.
This chapter focuses on 4-dimensional asymptotically flat Schwarzschild black holes, but our arguments can be extended to other types of black holes.

The remainder of this chapter is organized as follows. 
We present our two main protocols in \Cref{sec:combining_FR_and_BH,sec:EWF_firewall}, combining extended Wigner's friend paradoxes with the black hole cloning and firewall paradoxes. 
Next, we relate to the work of Hausmann and Renner~\cite{hausmann2025firewallparadoxwignersfriend} in \Cref{sec:HR}, and conclude by highlighting some subtleties underlying black hole puzzles in \Cref{sec:implications_WFBH}.

\begin{figure*}[t]
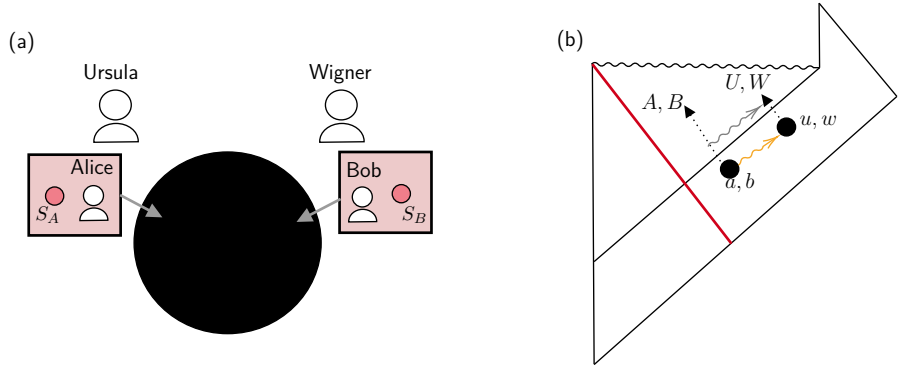

         \centering
\includestandalone[width=0.8\textwidth]{Figures/setup_and_penrose}
         \caption[Combining extended Wigner's friend with black hole cloning]{Initial set-up of our first protocol (a) and pictured in a Penrose diagram (b) for a black hole formed from collapse. Alice and Bob measure $S_A,S_B$ and fall into the black hole. Ursula and Wigner catch their Hawking radiation (yellow wiggly arrow) upon which they perform an operation, after which they fall into the black hole to retrieve the outcomes of Alice and Bob (green wiggly arrow). As pointed out in \Cref{sec:critical_note_assumptions} and discussed further in \Cref{sec:implications_WFBH}, one may question, among others, whether the location of the retrieval and decoding of Hawking radiation can be well into the bulk, whether Ursula and Wigner can still fall into the black hole after this decoding, and the correct causal structure. We also provide a theorem (\Cref{th:nogo_EWFcloning_notmeet}) where Ursula and Wigner remain outside the black hole.}
         \label{fig:radiation_mirror_timelike_path}
\end{figure*}

\section[Combining Wigner's friend and black hole cloning]{Combining extended Wigner's friend paradoxes and black hole cloning}
\label{sec:combining_FR_and_BH}

In this section, we construct a protocol that combines the Frauchiger--Renner and black hole cloning paradoxes and derive two no-go theorems. 
We briefly compare to the cloning paradox. 
Implications 
are discussed in \Cref{sec:implications_WFBH}.
Before we start, we comment upon assumptions from standard black hole puzzles that we adopt, and point out critical subtleties therein that we will discuss later in \Cref{sec:implications_WFBH}.

\subsection{Critical note about black hole assumptions} \label{sec:critical_note_assumptions}

In the remainder of this section and in \Cref{sec:EWF_firewall}, we provisionally adopt the standard assumptions used in black-hole cloning and firewall arguments. 
These include the existence of suitable black hole radiation subsystems, the possibility of decoding them in the required spacetime region, an approximate inside/outside factorization, persistent interior records, and the identification of late radiation with near-horizon modes. These assumptions are not derived here and are known to be conceptually delicate in quantum gravity. The no-go results below should therefore be read as conditional results about the strengthened consequences of these assumptions. In \Cref{sec:implications_WFBH} we return to these assumptions and highlight key issues that cast doubt upon the validity of these assumptions. 

A finite-dimensional quantum-information model of the operationally relevant subsystems is assumed, with qubits and finite memories representing effective or code-subspace degrees of freedom. 
The no-go results are therefore conditional on the possibility of identifying, decoding, and coherently manipulating such effective subsystems, including subsystems encoded in the Hawking radiation.
From the viewpoint of an observer outside the black hole, the black hole dynamics is modeled, for the purposes of the protocol, as implementing a unitary scrambling/encoding map $U_R$ on the black hole, infalling systems and relevant environmental degrees of freedom, à la Hayden–Preskill~\cite{hayden2007black}. 
It is further assumed that an appropriate recovery operation on the accessible Hawking radiation (possibly together with previously collected radiation or auxiliary systems) reconstructs the relevant code-subspace state of the laboratory, after which the inverse of the laboratory measurement can be applied as in the Wigner's friend scenario. 
Where (and when) and how such information recovery would occur is not prescribed, and these questions constitute an important point of scrutiny, as noted above and further discussed in \Cref{sec:implications_WFBH}. 

It is further assumed that Alice and Bob hold systems that are sufficiently isolated within their lab and can be effectively modeled as qubits, upon which they will perform measurements. 
(This last assumption concerns only the relevant effective degrees of freedom, and does not require their gravitational fields to be perfectly shielded from their surroundings, for instance.)

\subsection{Protocol and no-go theorem} \label{sec:FR_BH_protocol}

Alice and Bob, residing in sealed labs, each measure their part of a Hardy state\footnote{We will sometimes omit the tensor product symbol for ease of notation, i.e. write $\ket{00}$ or $\ket{0}\ket{0}$ for $\ket{0} \otimes \ket{0}$.}~\cite{hardy1993nonlocality} \begin{equation} \label{eq:hardy} \ket{\psi_0}_{S_A S_B} = \frac{1}{\sqrt{3}}(\ket{00}+\ket{10}+\ket{11})_{S_A S_B} \end{equation} in the computational basis with outcomes $a,b \in \set{0,1}$, modeled unitarily as $U_A,U_B$, and next fall into an old black hole, i.e. a black hole past the Page time. Ursula retrieves and decodes the Hawking radiation in which the state of all of Alice's sealed lab is encoded (assuming an argument à la Hayden--Preskill~\cite{hayden2007black}, where for an outsider the black hole dynamics is assumed to be unitary and rapidly mixing), ``undoes'' Alice's measurement by applying the inverse unitary $U_A^\dagger$ and measures $S_A$ in the $\ket{\pm}=\sfrac{1}{\sqrt{2}} (\ket{0}\pm\ket{1})$ basis with outcome $u \in \{\pm\}$.\footnote{One could also allow Alice (and Bob) to perform their measurement inside the black hole, and in that case question whether Ursula would still have to apply $U_A^\dagger$. I thank Eric Cavalcanti for this comment. This protocol may be more subtle, as it also involves the question of how Alice's (and Bob's) measurement inside is modeled from the outside.} Wigner acts analogously on Bob's Hawking radiation, obtaining $w \in \{\pm\}$. 
Alice and Bob, once inside the black hole, try to send their outcomes $a,b$ towards Ursula and Wigner who fall into the black hole after having performed their measurements.\footnote{Note that for Ursula and Wigner to obtain a message from Alice and Bob, Ursula and Wigner must decode the Hawking radiation at locations where they can still fall into the (evaporating) black hole, and thus cannot do this at late times near null or timelike future infinity. One can also derive a contradiction when Ursula and Wigner remain outside the black hole, which is the content of \Cref{th:nogo_EWFcloning_notmeet}. We comment more upon this in \Cref{sec:implications_WFBH}.}

\vspace{0.5cm}

Upon obtaining $u=-,w=-$, which has a nonzero probability of occurring, Ursula and Wigner reason as follows:
\begin{itemize}
    \item $u=-$ implies $b = 1$ as $\langle - |_{S_A} \langle 0|_{S_B}  \ket{\psi_0}_{S_A S_B} = 0$,
    \item $w=-$ implies $a = 0$ as $\langle 1|_{S_A} \bra{-}_{S_B} \ket{\psi_0}_{S_A S_B} = 0$,
    \item but $a=0,b=1$ cannot occur as $\langle 01|_{S_A S_B} \ket{\psi_0}_{S_A S_B}=0$,
\end{itemize} and thus a contradiction is obtained. This contradiction becomes operational if Alice and Bob, once inside the black hole, succeed in sending their outcomes and predictions to Ursula and Wigner. We could also have formulated the protocol by having Alice and Bob reason about the outcomes of Ursula and Wigner.

\vspace{0.5cm}

To explain in some more detail how the probabilities above are obtained, let us focus on the first, Ursula's claim about Bob's outcome $b$. 
 We start with the initial Hardy state $\ket{\psi_0}$ of \cref{eq:hardy} for the systems $S_A \otimes S_B$, and Alice's and Bob's memories initialized in some states, the $\ket{0}$ states for example, \begin{equation} \label{eq:Psi_0} \ket{\Psi_0} := \ket{\psi_0}_{S_A S_B} \ket{0}_A \ket{0}_B. \end{equation} 
 Alice and Bob perform their measurements, which are modeled unitarily as $U_A$ and $U_B$ acting on \cref{eq:Psi_0}, in this case CNOT gates with controls on $S_A$ and $S_B$, as explained in \Cref{sec:WF}, obtaining $U_A U_B \ket{\Psi_0}$.

We next make explicit the idealized recovery assumption used in the protocol. Let $L=S_AAS_BB$ denote the relevant joint laboratory degrees of freedom of Alice and Bob, let $B_0$ denote the black hole before the relevant scrambling and evaporation, let $E$ denote previously emitted radiation, let $R$ denote the subsequently emitted radiation accessible to Ursula and Wigner, and let $B'$ denote the remaining black hole. 
 From the outsider's point of view (Ursula's and Wigner's point of view), Alice's and Bob's labs get scrambled by the black hole into Hawking radiation by evaporation, implemented by a unitary as 
\begin{equation}
V_{\mathrm{BH}}:
\mathcal H_L\otimes\mathcal H_{B_0}
\longrightarrow
\mathcal H_R\otimes\mathcal H_{B'} .
\end{equation}
We assume that, on the relevant laboratory code subspace $L$, there exists a coherent decoding isometry
\begin{equation}
W_{\mathrm{dec}}:
\mathcal H_R\otimes\mathcal H_E
\longrightarrow
\mathcal H_{\widetilde L}\otimes\mathcal H_J
\end{equation}
that decodes the relevant information from the radiation $RE$, i.e. 
\begin{equation}
\label{eq:coherent_recovery}
\begin{split}
(W_{\mathrm{dec}}\otimes\mathbb 1_{B'})
(V_{\mathrm{BH}}\otimes\mathbb 1_E)
\ket{\phi}_L\ket{\Omega}_{B_0 E}
 =
\ket{\phi}_{\widetilde L}\ket{\chi}_{JB'}
\end{split}
\end{equation}
for every state $\ket{\phi}_L$ in that code subspace $L$. Here, $\widetilde L$ denotes the reconstructed laboratory, $J$ denotes unused output degrees of freedom of the decoder, and the state $\ket{\chi}_{JB'}$ is independent of $\ket{\phi}_L$. We henceforth identify the reconstructed laboratory degrees of freedom $\widetilde L$ with the corresponding original labels $L$.
 
 We consider now the case where Bob obtains $b=0$, giving the state
 \begin{equation} \label{eq:Psi_0_step2}
 \begin{split}
    & \ket{0}_{B} \bra{0}_{B} U_A U_B \ket{\Psi_0} \\ &= \ket{00}_{S_B B} \bra{0}_{S_B} U_A \ket{\psi_0}_{S_A S_B} \ket{0}_A 
 \end{split}
 \end{equation} 
 up to normalization, where we used that \begin{equation} \ket{0}_B \bra{0}_B U_B \ket{0}_B = \ket{00}_{S_B B} \bra{0}_{S_B}.\end{equation} 
 Next, Alice's and Bob's labs get scrambled by $V_{BH}$, and Ursula and Wigner apply the decoding map $W_\mathrm{dec}$ to the radiation (which includes both the early and late radiation $E$ and $R$), and thus obtain \cref{eq:Psi_0_step2} after decoding by \cref{eq:coherent_recovery}.
 Ursula next undoes Alice's measurement by applying $U_A^\dagger$, performs her measurement, where we consider the case where her outcome is $u=-$, giving  \begin{equation} \begin{split}
 \label{eq:Psi_final_prob0} & \ket{00}_{S_B B} \bra{-}_{S_A} U_A^\dagger \bra{0}_{S_B} U_A  \ket{\psi_0}_{S_A S_B} \ket{0}_A \\ &= \ket{00}_{S_B B} \bra{-}_{S_A} \bra{0}_{S_B} \ket{\psi_0}_{S_A S_B} \ket{0}_A  \\ &= 0,    
 \end{split} \end{equation}  using that $U_A^\dagger U_A = \mathbb{1}$ and $\bra{-}_{S_A} \bra{0}_{S_B}  \ket{\psi_0}_{S_A S_B} =0$, resulting in a probability 0, $p(u=-,b=0)=0$. 
 The other probabilities used in Ursula's and Wigner's reasoning are obtained analogously.

\vspace{0.5cm}

We have thus proved the following no-go theorem.
\begin{theorem} \label{th:theorem_FR_BH}
    If a superobserver can decode the Hawking radiation and reconstruct the quantum state of an infalling observer, and, when falling in, still receive a short message from that observer, sent inside the black hole and assumed to be unaltered by actions outside, an operational contradiction is found with quantum theory's predictions.
\end{theorem}

Compared to usual EWF arguments, here all of the outcomes $a,b,u,w$ seem available on a single spatial slice, as can be seen on \cref{fig:radiation_mirror_timelike_path}(b), and \Cref{th:theorem_FR_BH} does not require the assumption of absolute outcomes of performed measurements: 
it involves only empirical correlations accessible to a single observer.

\vspace{0.5cm}

If, as suggested in literature on the black hole cloning paradox (see \Cref{sec:BHcloning}), an observer catching the Hawking radiation from an infalling observer $A$ cannot obtain any message inside from $A$ due to time constraints, and so Ursula and Wigner are unable to obtain the outcomes of Alice and Bob inside, then we can still use our protocol to derive another no-go theorem.

\begin{theorem} \label{th:nogo_EWFcloning_notmeet}
    If a superobserver can decode the Hawking radiation and reconstruct the quantum state of an infalling observer, a contradiction is found if (Ci) the outcome records of an observer inside the black hole remain coherent and untouched by operations outside, (Cii) observers can make valid Born rule predictions about measurement outcomes (without actually obtaining those outcomes, but where those outcome records remain untouched) and (Ciii) observers who can communicate and meet have to agree on their predictions.
\end{theorem}

\subsection{Do the outcome records of observers inside get altered by operations outside?}

In the protocol above, the correlation $p(u,w)$ can be obtained experimentally by Ursula and Wigner, who remain outside the black hole, and the correlation $p(a,b)$ by Alice and Bob who fall into the black hole together.
However, if Alice and Bob fail to send a message with their outcomes inside the black hole to Ursula and Wigner, the correlations $p(u,b),p(a,w)$ are not verified experimentally as no single agent obtains both $u,b$, for example, which led to \Cref{th:nogo_EWFcloning_notmeet}.
So, in this case, Ursula and Wigner could just as well remain outside the black hole. But why does Ursula argue then that the joint correlations $p(u,b)$ of her prediction are those she would obtain if Bob had remained outside the black hole to compare her outcome to $b$? (Similarly for Wigner.) 

Similarly to the Local Friendliness no-go theorem~\cite{bong2020strong}, one can try to verify this prediction in certain rounds of the protocol\footnote{In these rounds of the protocol, in addition to the systems $S_A,S_B$, now also Alice and Bob and their lab environments are initialized in the same known initial states.} (when Ursula and Wigner do not fall into the black hole), by allowing Alice to make a choice $x=0,1$ whether to leave her outcome outside the black hole or not (similarly for Bob with a choice $y$). 
In rounds where $x=0$ she leaves a copy of her outcome $a_O$ outside the black hole and the empirical correlations $p(a_O,w)$ are obtained outside the black hole. 
If Wigner can also access Alice's outcome $a$ inside, he finds that $p(a_O,w,a|x=0,y=1) = \delta_{a_O,a} p(a_O,w|x=0,y=1)$,  confirming consistency between inside and outside outcomes. 
Other choices of $x,y$ test other statements.
Implementing this idea in our first protocol we arrive at a no-go theorem where an assumption of no-signalling is used to argue that all these statements remain true regardless of the choices $x,y$. 

An alternative way of motivating the validity of the quantum correlations $p(a,w)$, $p(u,b)$ is to note that if Bob fell into the black hole much earlier than Alice, Wigner could decode Bob's Hawking radiation and perform his measurement with outcome $w$, while Alice remains outside still, so that they can verify the correlation $p(a,w)$. An analogous argument holds for Alice falling in earlier and the verification of $p(u,b)$.

\vspace{0.5cm}

The arguments above suggest that the correlations $p(a,w),p(u,b),p(a,b),p(u,w)$ are indeed those predicted by quantum theory as explained in \cref{eq:Psi_final_prob0}, if one can assign outcome values to the measurements of Alice and Bob who fell into the black hole. 
A possibility is that Ursula's and Wigner's actions erase Alice's and Bob's physical outcome records as in Wigner's friend scenarios.
However, semiclassical causality arguments would argue that Ursula's and Wigner's actions outside the black hole cannot alter the experiences of Alice and Bob inside.
Nevertheless, refuting this assumption, referred to as $(Ci)$ in \Cref{th:theorem_FR_BH,th:nogo_EWFcloning_notmeet} and also present in the cloning paradox, 
still seems one of the most reasonable ways out if one believes in unitarity, compared to refuting assumptions $(Cii)$ or $(Ciii)$ of \Cref{th:nogo_EWFcloning_notmeet}. 
As we will discuss in \Cref{sec:subtleties}, subtleties regarding black hole puzzles for infalling observers can be argued to make this option more natural.

\subsection{Comparison with the no-cloning argument} \label{sec:BHcloning}

The cloning paradox analyzed by Susskind and Thorlacius~\cite{susskind1994gedanken} and Hayden and Preskill~\cite{hayden2007black} is similar in spirit to our protocol, but does not directly provide a logical contradiction on the level of operational probability predictions.
Our result rules out any logically consistent theory that can describe the viewpoints of the different observers if they can use quantum theory for their own predictions, assuming memories inside the black hole remain unaltered by actions on the outside radiation.

In the cloning paradox, Alice falls into the black hole while carrying a quantum system $A$. 
Bob catches the Hawking radiation required to reconstruct $A$'s state, and dives into the black hole.
Alice, once inside the black hole, sends her system $A$ (or quantum information about it) to Bob, who would observe a violation of quantum theory's no-cloning.
Hayden and Preskill~\cite{hayden2007black} suggested that Bob, who collects the Hawking radiation outside to reconstruct a system $A$'s state, can just never retrieve the system $A$ inside. 
Therefore, an operational violation of no-cloning seems protected.
Namely, for Alice to send her qubit $A$ to Bob, according to Hayden and Preskill, she must operate within a Planck time scale on her proper time to send information about her qubit to Bob for it to reach him before he hits the singularity\footnote{This is assuming the causal structure of the Schwarzschild black hole, and need not be the case for the classical causal structure of the rotating and charged black holes. However, the actual (quantum) interior of the latter is likely quite different from the classical one~\cite{zilberman2022quantum,juarez2022quantumkicks,Ju_rez_Aubry_2015,juarez-aubry_quantum_2023,Lanir_2019,hollands2020quantum,Hollands_2020,Papadodimas_2020,cardoso2018quasi,shrivastava2020quantumaspectschargedblack,Dias_2019,Dias_2019strong,Emparan_2020,mcmaken2023,zilberman2020quantumfluxes}.}. But this means that Alice must send her information in super-Planckian frequencies, in a strong gravity regime, likely much different from low-energy gravity. 
The same argument against an actual operational contradiction holds for the protocol of \Cref{sec:FR_BH_protocol}.
Nevertheless, the existence of these two copies in the cloning paradox remains problematic if one believes both views should be described by one single quantum theory.
Similarly, in the protocol above, even if Ursula and Wigner can never get any messages from Alice and Bob inside the black hole, a puzzle remains, as stated in \cref{th:nogo_EWFcloning_notmeet}.
In analogy with \Cref{th:nogo_EWFcloning_notmeet}, we could phrase the no-cloning theorem as: 
\begin{quote}
    If an observer can reconstruct a system $S$'s quantum state from the Hawking radiation of an infalling quantum system, a violation of no-cloning is found if $(Ci)$ the state of the quantum system $S$ that fell into the black hole remains coherent and unaltered by operations outside, and $(iv)$ quantum theory can describe the physics of the whole spacetime, including both the inside and outside of the black hole.
\end{quote}
One could propose to refute $(iv)$ here, and instead postulate a theory that describes the evolution of the whole spacetime. Our protocol and no-go theorem above show that this is impossible if no single observer can falsify quantum theory's operational predictions. 
We discuss these paradoxes further in \Cref{sec:implications_WFBH}.

\subsection{Combining other extended Wigner's friend paradoxes with the cloning paradox}
From the scenario outlined in \Cref{sec:FR_BH_protocol}, it is clear that we can combine other extended Wigner's friend arguments with the black hole cloning argument to produce no-go theorems. 
For example, we could come up with a scenario that combines the Local Friendliness no-go theorem with the black hole cloning paradox by having Ursula and Wigner choose $x,y$ whether to reveal the outcome of Alice and Bob when they have decoded their Hawking radiation.
Similarly, we could combine the Pusey--Barrett--Rudolph Wigner's friend argument of \Cref{chap:part1_EWF_PBR} with the black hole cloning paradox by having Wigner perform the PBR measurement on the decoded Hawking radiation from two friends Alice and Bob, who made some choices earlier while falling into the black hole.

\section[Combining Wigner's friend and the firewall argument]{Combining extended Wigner's friend paradoxes and the firewall argument}\label{sec:EWF_firewall}

In this section we first recap the firewall paradox, after which we construct a protocol and no-go theorem that combines the firewall paradox and extended Wigner's friend arguments. Implications are discussed in \Cref{sec:implications_WFBH}.

\subsection{Preliminaries: the firewall argument} \label{sec:firewall_recap}
\textbf{The firewall argument.} We follow the presentation of Ref.~\cite{hausmann2025firewallparadoxwignersfriend}. An infalling referee Bob holds a (distilled) qubit $Q_B$ right outside an old black hole (i.e. past the Page time), being part of the late-time Hawking radiation. Two observers, Alice and Charlie, starting outside and inside the black hole, respectively, collaborate to provide two predictions $(P,\overline{P})$ to the referee Bob, for when he measures his qubit $Q_B$ in the $Z$- and $X$-bases. Alice and Charlie win if the referee's test succeeds, i.e. if the referee Bob is given the correct predictions; this is called the \textit{complementarity game}~\cite{hausmann2025firewallparadoxwignersfriend}. Charlie, inside the black hole, can distill a qubit $Q_C$ (approximately) maximally entangled with $Q_B$ by the assumption of high entanglement across the horizon~\cite{eisert2010colloquium,reeh1961bemerkungen} (assuming semiclassical gravity), and Alice, holding the early-time Hawking radiation outside the black hole, can distill a qubit $Q_A$ (approximately) maximally entangled with $Q_B$ by Page's argument~\cite{page1993information,page2013time}. Alice and Charlie measure their qubits $Q_A,Q_C$ in the $X$- and $Z$-bases, respectively, obtaining outcomes $a,c$. 
As Charlie's qubit is maximally entangled with $Q_B$, he predicts the referee's outcome $b$ to be the same as his outcome $c$ when he measures his qubit in the $Z$-basis, and similarly for Alice predicting outcome $a$ for the referee when he measures in the $X$-basis. They send their predictions $(a,c)$ to the referee who falls into the black hole (to retrieve Charlie's message and outcome). The predictions $P=a,\overline{P}=c$ yield a winning probability of 1, violating the quantum bound\footnote{The `quantum bound' means the maximal probability of winning the complementarity game achievable by any strategy permitted by quantum theory.} of $\frac{1}{2}+\frac{1}{\sqrt{8}} < 1$~\cite{hausmann2025firewallparadoxwignersfriend,Renes_2017} for winning the complementarity game. This violation is based on a violation of monogamy of entanglement.

We can also construct an extension of the firewall argument by giving Charlie and the referee Bob measurement choices $x,y$, resulting in a violation of no-signalling based on the violation of monogamy of entanglement (see Appendix C of Ref.~\cite{walleghem2025wignersfriendsblackhole}).

If some observer can obtain all outcomes of Alice, Bob and Charlie, an operational violation of quantum theory's formalism is found in the firewall argument. Such operational violation is more plausible than in the black hole cloning paradox, as no observer needs to wait for the Hawking radiation of an infalling observer to have come out. 
In fact, in typical treatments of the firewall argument, Alice, Charlie and the referee Bob are played by a single observer; this is possible if Alice's distilling operation can be performed sufficiently rapidly so she can still obtain Charlie's mode inside the black hole.
A potential problem in both of these paradoxes lies in the complexity of Alice's operation of distilling the qubit $Q_A$ from the early radiation, which could require a time exponentially long in the black hole entropy~\cite{harlow_quantum_2013,harlow2016jerusalem,aaronson_complexity_2016} compared to its polynomial evaporation time. 
However, with exponentially long preprocessing time such long decoding time could be avoided, as suggested in a toy black hole representation in Ref.~\cite{Oppenheim_2014}.

\vspace{0.5cm}

\textbf{The firewall paradox as a no-go theorem.} The firewall argument can be stated as a no-go theorem in the following way.

\begin{theorem}[Firewall paradox] \label{th:firewall} The firewall argument shows that the following four assumptions are incompatible:

    (Fi) the early radiation is maximally entangled with the late radiation---motivated by unitarity and the Page curve---and (Fii) an observer can distill a qubit $Q_A$ of the early radiation maximally entangled with a qubit mode $Q_B$ of the late radiation that is just outside the horizon; 
    
    (Fiii) the late radiation modes just outside the horizon are highly entangled with modes inside the black hole such that inside one can distill a qubit $Q_C$ maximally entangled with $Q_B$---motivated by the equivalence principle and high entanglement across the horizon as in QFT on curved spacetime; 

    (iv) quantum theory can describe the whole evolution comprising both the inside and outside of the black hole (including all radiation). 
\end{theorem}
One way to evade this no-go theorem would be to alter assumption $(iv)$\footnote{In typical treatments of the firewall and cloning paradoxes, assumption $(iv)$ is not explicitly stated, but implicitly assumed.}, by considering a more general theory than quantum theory that describes the evolution of both the inside and outside of the black hole. 
This route will not work if one accepts the existence of superobservers along with the persistence of outcome records of the black hole interior as in $(Ci)$: any theory that would do so would find an experimental violation of quantum theory's predictions. 
We prove this claim in the next section by presenting a protocol that combines Wigner's friend with the firewall paradox. 
Alternatively, one can introduce measurement choices $x,y$ for Charlie and the referee Bob in the firewall paradox, and use the monogamy of entanglement violation to derive a contradiction of $(Fi)-(Fiii)$ with no-signalling (as defined in the semiclassical causal structure) as we prove in Appendix \ref{sec:strengthened_firewall}. 

\subsection{Protocol and no-go theorem}

Consider the following protocol, laid out stepwise and sketched in \cref{fig:firewall_EWF}, which can be seen as combining Wigner's friend arguments and the firewall paradox.

\vspace{0.5cm}

\textbf{Initial set-up.} The argument involves four observers Alice, Bob\footnote{Compared to the firewall paradox in the complementarity game presentation above in \Cref{sec:firewall_recap}, Bob is not a referee in this protocol, just an agent performing a measurement.}, Ursula and Wigner, who all start outside the black hole.
Similarly as in the firewall paradox, Bob holds a qubit mode of the late radiation close to an old black hole, which he gathered in free fall. Alice holds the early radiation, and decodes a qubit $S_A$ maximally entangled with Bob's qubit $S_B$, the same starting set-up as in the firewall paradox, such that the systems they hold are in the Bell state $(\ket{00}+\ket{11})_{S_A S_B}/\sqrt{2}$.
We also assume that Alice and Bob are each located in a sealed lab.
The measurements that will be performed on the systems $S_A, S_B$ have correlations that violate the CHSH inequality~\cite{clauser1969proposed}, such that there exists no global distribution reproducing all of these correlations~\cite{fine1982hidden,fine1982joint,abramsky2011sheaf,brunner2014bell}.

\textbf{Step 1.} Hovering outside the black hole (having stopped his free fall), Bob measures his system $S_B$ in the $\ket{\pm}$-basis, with outcome labelled with $b$, while Alice measures in the basis $\ket{a_0},\ket{a_1}$ with outcome denoted $a=0,1$ and \begin{equation}
\left|a_0\right\rangle=\cos \theta|0\rangle+\sin \theta|1\rangle, \quad\left|a_1\right\rangle=-\sin \theta|0\rangle+\cos \theta|1\rangle,
\end{equation} with $\theta = \pi/8$.
Next, Alice and Bob come together and compare their outcomes, verifying the correlation $p(a,b)$ (verifying the consistency of the joint outcomes $a,b$ with quantum theory's prediction for $p(a,b)$).
Ursula, having perfect control over everything that happens in Bob's lab, subsequently undoes this verification\footnote{As Ursula has perfect quantum control over Bob's sealed lab, she can apply the inverse of the unitary verification operation by Alice and Bob.}.

\textbf{Step 2.} Alice falls into the black hole and measures a qubit $S_W$ that is maximally entangled with Bob's qubit $S_B$, i.e. whose joint state is given by $(\ket{00}+\ket{11})_{S_W S_B}/\sqrt{2}$; this is the use of maximal entanglement across the horizon as in the firewall paradox.\footnote{Here we neglect the potential issue of long decoding times of the Hawking radiation as mentioned in \Cref{sec:firewall_recap}.} 
Alice measures the qubit $S_W$ in the computational $\ket{0},\ket{1}$ basis, labeling her outcome by $a_w$. 
She then verifies the correlation $p(a,a_w)$.

\textbf{Step 3.} In the meantime, outside the black hole, Ursula decodes Alice's Hawking radiation, and undoes her measurement by restoring Alice's lab back to its initial state. 
Next, she performs the measurement $\ket{u_0},\ket{u_1}$ with outcome denoted $u$ and \begin{equation}
\left|u_0\right\rangle=\cos \theta|0\rangle-\sin \theta|1\rangle, \quad\left|u_1\right\rangle=\sin \theta|0\rangle+\cos \theta|1\rangle,
\end{equation} with $\theta = \pi/8$.
Having obtained her outcome $u$, she asks Bob for his outcome and they verify the correlation $p(u,b)$.
After this, Wigner undoes this verification and lets Bob fall into the black hole. 

\textbf{Step 4.} Wigner decodes the Hawking radiation from Bob's lab outside the black hole, and undoes Bob's measurement to reset his lab back to its initial state.
After this, Wigner measures Bob's qubit $S_B$ in the computational $\ket{0},\ket{1}$ basis, obtaining an outcome labelled by $w$.
Ursula then asks for Wigner's outcome and they verify the correlation $p(u,w)$. 
Now, because $S_B$ and $S_W$ were maximally entangled as $(\ket{00}+\ket{11})_{S_W S_B}/\sqrt{2}$, and $w$ and $a_w$ were obtained by measuring $S_B$ and $S_W$ in the computational basis, quantum theory tells us that $a_w = w$. 
We use this fact to argue that the correlation $p(a,a_w)$ would be the same as the correlation $p(a,w)$, which arises from outcomes $a,w$ of measurements in the $\ket{a_0},\ket{a_1}$ and computational basis, respectively, on the maximally entangled state $(\ket{00}+\ket{11})_{S_A S_B}/\sqrt{2}$.

\textbf{Deriving a contradiction.} Thus, in the course of this scenario, all of the correlations $p(a,b),p(u,b),p(a,a_w),p(u,w)$ were verified, with moreover $a_w=w$. 
In addition, all of these outcomes $a,b,u,a_w,w$ still exist, some inside the black hole and some outside. 
Thus, they seemingly coexist on a spatial slice still, suggesting that all of these correlations should come from a joint probability distribution $p(a,b,u,a_w,w)$.
However, this cannot be true as the verified correlations are chosen to be of the sort that violate the CHSH inequality, such that no such global distribution $p(a,b,u,a_w,w)$ reproducing these correlations $p(a,b),p(u,b),p(a,a_w),p(u,w)$ with $a_w=w$ exists~\cite{fine1982hidden,fine1982joint,abramsky2011sheaf,brunner2014bell,clauser1969proposed}. We thus find a contradiction.

\vspace{0.5cm}

This argument leads to the following no-go theorem.

\begin{theorem}[EWF and firewalls] \label{th:EWF_firewalls}
If a superobserver can decode the Hawking radiation and reconstruct the quantum state of an infalling observer, then under the assumptions (Fi)-(Fiii) of the firewall paradox of \Cref{th:firewall} and (Ci) of \Cref{th:nogo_EWFcloning_notmeet}, a contradiction is found with quantum theory's predictions for joint correlations.
\end{theorem}

\begin{figure*}[t]
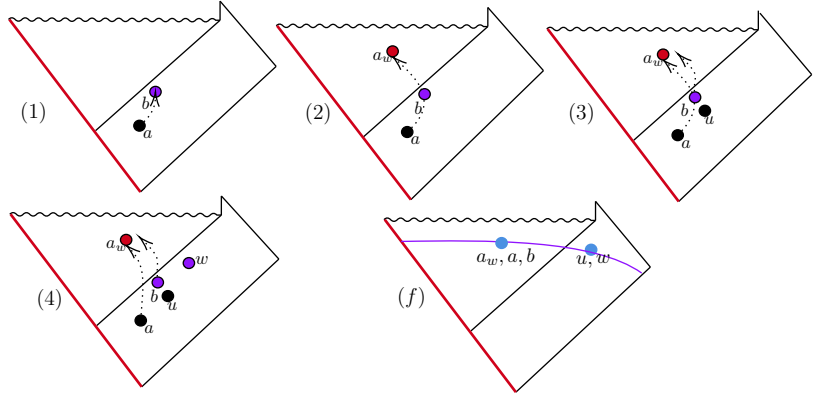

         \centering
\includestandalone[width=0.8\textwidth]{Figures/EWF_firewall_steps}
         \caption[Combining extended Wigner's friend with the firewall argument]{Stepwise explanation of the EWF-firewall protocol. (1) Alice and Bob hold a maximally entangled state, obtained from high entanglement between early and late radiation, perform their measurements with outcomes $a,b$ and verify the correlation $p(a,b)$; this verification action is then undone by Ursula. (2) Alice falls into the black hole and measures a qubit $S_W$ that is maximally entangled with Bob's qubit, obtained from high entanglement across the horizon, verifying the correlation $p(a,a_w)$, while Bob remains outside. (3) Ursula catches the Hawking radiation of Alice's lab, undoes her measurement and performs a measurement with outcome $u$ so that she can verify the correlation $p(u,b)$ with Bob. Next, Wigner undoes this verification, and Bob falls into the black hole. (4) Wigner catches the Hawking radiation from Bob's lab, undoes Bob's measurement and performs a measurement with outcome $w$, verifying the correlation $p(u,w)$ with Ursula. Summarizing, in steps (1), (2), (3) and (4), the correlations $p(a,b),p(a,a_w),p(u,b)$ and $p(u,w)$ are verified. The final picture $(f)$ shows how seemingly all outcomes $a_w,a,b,u,w$ coexist, contradicting the fact that the verified correlations violate the CHSH inequality and cannot arise from a single, global probability distribution. We comment on these Penrose diagrams and locations of information recovery in \Cref{sec:implications_WFBH}.}
         \label{fig:firewall_EWF}
\end{figure*}

\section{Relating to the work by Hausmann and Renner} \label{sec:HR}

Hausmann and Renner (HR)~\cite{hausmann2025firewallparadoxwignersfriend} describe two Wigner's friend protocols and two protocols involving a black hole, and draw analogies between Wigner's friend and black hole paradoxes.
We have extended their work by combining an extended Wigner's friend protocol~\cite{frauchiger2018quantum,bong2020strong,cavalcanti2021implications,haddara2022possibilistic,walleghem2024strong} with black hole physics into a unified argument.
HR's first black hole argument is similar to the cloning paradox, and can be seen as combining a single Wigner's friend scenario with the cloning paradox. 
Their second no-go theorem, which we have described in \Cref{sec:firewall_recap}, is a rephrasing of the firewall argument where the non-monogamy of entanglement is used to violate the quantum bound of the complementarity game.

\vspace{0.5cm}

\textbf{How Hausmann and Renner relate to Wigner's friend.} \Cref{th:theorem_FR_BH} assumes that infalling observers can still send a message to the observers catching their Hawking radiation who fall into the black hole later, an assumption that may be problematic as argued in the context of the black hole cloning paradox (see \Cref{sec:BHcloning}), while \Cref{th:nogo_EWFcloning_notmeet} derives a no-go theorem without such assumption, based on the same protocol.
In the firewall paradox, the possibility of one single observer bringing the problematic aspects inside and outside the black hole together is more plausible as no observer needs to wait outside the black hole to catch the Hawking radiation of an infalling observer before falling into the black hole.
Hausmann and Renner~\cite{hausmann2025firewallparadoxwignersfriend} present a no-go theorem based on the firewall paradox as paraphrased in \Cref{sec:firewall_recap}, but with assumptions phrased differently than the ones of \Cref{th:firewall}.
A key assumption in their no-go theorem is a consistency assumption ($\mathcal{C}$), which allows Alice, inside the black hole, to reason about Bob outside the black hole. 
We can paraphrase the consistency assumption ($\mathcal{C}$), also used in EWF arguments such as the FR paradox~\cite{frauchiger2018quantum}, as follows: \begin{quote}
    ($\mathcal{C}$): If an observer $B$, modeled classically by observer $A$ (i.e., above $A$'s Heisenberg cut), knows that a statement $s$ about some measurement outcome(s) is true from reasoning with a theory that $A$ accepts, and $A$ knows about this reasoning of $B$, then $s$ is true for $A$ as well.
\end{quote}
Notably, assumption ($\mathcal{C}$) is already violated in certain classical paradoxes~\cite{jones_thinking_2024,walleghem2024stunnedsleepingbeautyprince}, and is thus perhaps too strong. 
Therefore, one could apply the same criticism to Hausmann and Renner's theorems~\cite[Theorems 3 and 5]{hausmann2025firewallparadoxwignersfriend}.
However, the FR paradox can be slightly rephrased such that only a weaker consistency assumption is required~\cite{walleghem2024strong}, which we refer to as ($\mathcal{C}'$), where observers $A$ and $B$ only need to take each other's statements to be true if they can communicate and physically meet. 
We use a similar condition to ($\mathcal{C}'$) in \Cref{th:nogo_EWFcloning_notmeet}, the combination of assumptions $(Cii)$ and $(Ciii)$ there.
In the EWF-firewall argument of \Cref{th:EWF_firewalls}, we do not need a condition for the consistency of reasoning or predictions such as ($\mathcal{C}'$), as all involved correlations are actually observed, similarly as in typical formulations of the firewall paradox.

In fact, in the firewall paradox formulation of \Cref{sec:firewall_recap} (per Hausmann and Renner~\cite{hausmann2025firewallparadoxwignersfriend}), the assumption ($\mathcal{C}'$) could be used instead of ($\mathcal{C}$) if Bob were to fall into the black hole to meet Alice.
However, Hausmann and Renner refrain from letting Bob fall into the black hole, as they model his reasoning while he remains outside the black hole, used in the assumptions that as an outsider he can model the Hawking radiation process as a typical unitary for the early radiation and the modes at the stretched horizon (which will constitute the late radiation). 
Assuming no-retrocausality, one could argue that Bob only needs to model this unitary up to the point where he could verify his claims by asking the referee holding the modes at the stretched horizon, to use ($\mathcal{C}'$) instead of ($\mathcal{C}$).
We will not discuss this matter in more detail here.

\vspace{0.5cm}

The refined version of the FR paradox that uses ($\mathcal{C}'$) instead of ($\mathcal{C}$)~\cite{walleghem2024strong}, as well as other EWF arguments~\cite{bong2020strong,cavalcanti2021implications,frauchiger2018quantum,brukner2018no}, can be used to argue against absolute, single-valued outcomes of performed measurements, in favor of relationalism~\cite{rovelli1996relational,rovelli2018space,cavalcanti2023consistency,walleghem2024strong,debrota2020respecting,ormrod2024quantum}, assuming the existence of superobservers.
However, the no-go theorems in this chapter serve another purpose. Naive treatments of the interior and exterior of the black hole suggest that all relevant outcomes coexist on a single spatial slice, and thus arguing that these outcomes would not coexist as in EWF paradoxes such as the (refined) FR paradox seems somewhat odd here. 
Therefore, these no-go theorems rather suggest that these naive treatments of black hole physics must be false.

\section{Discussion} \label{sec:implications_WFBH}

We have presented two scenarios combining Wigner's friend and black hole physics into a unified argument, resulting in \Cref{th:theorem_FR_BH,th:nogo_EWFcloning_notmeet,th:EWF_firewalls} that extend the black hole cloning and firewall paradoxes. 
Specifically, the cloning paradox leaves room for a theory that describes the interior and exterior of the black hole to evade the no-cloning argument while not giving up on the standard black hole assumptions without violating quantum theory's predictions. 
With the scenarios in \Cref{sec:combining_FR_and_BH} we close this loophole, and show that no such theory exists if observers cannot witness an experimental violation of quantum theory's predictions.
Similarly, the firewall paradox leaves room for such a theory evading the firewall argument while not giving up on the standard black hole assumptions going into the firewall paradox without violating quantum theory's predictions. 
In \Cref{sec:EWF_firewall} we show that such theory must violate no-signalling, and gives a violation of quantum theory's predictions when assuming the persistence of outcome records in the black hole interior.
Consequently, the problem must lie in other assumptions of the firewall and cloning paradoxes to prevent such violations.

In \Cref{chap:BH_prelims} we have argued that there are three answers to the black hole information puzzle, (i) information loss\footnote{A central argument for information loss is that, semiclassically, the entanglement of the Hawking radiation with degrees of freedom behind the horizon persists after evaporation, so that a hypersurface containing the radiation in the post-evaporation region is not a Cauchy surface and the state restricted to it is mixed~\cite{Unruh_2017}. Whether such argument survives quantum gravity and superposed spacetimes remains unclear. For example, holography~\cite{Marolf_2009,raju2022lessons,laddha_2021} and the potential failure of the split property in quantum gravity~\cite{raju2022failure} would imply that the `interior' degrees of freedom are not necessarily independent of those in the (post-evaporation) region where the Hawking radiation resides.}~\cite{Unruh_2017}, (ii) remnant-like objects holding the  information, 
and (iii) aspects that were overlooked in Hawking's calculations, such as soft hair, superposed geometries, replica wormholes and islands, for example. 
It is within (iii) that paradoxes for infalling observers such as the cloning and firewall paradoxes take place.
In the remainder of this section, we will highlight subtleties in these paradoxes, and point toward further directions.
As commented upon in \Cref{chap:BH_prelims}, one possibility is to invoke modifications at the level of the horizon, or its status, such as firewalls, fuzzballs and gravastars. 
We will instead focus on subtleties that are present without a priori invoking such new structures, and briefly discuss their implications.

\subsection{Subtleties in black hole paradoxes} \label{sec:subtleties}

\textit{a. Conservation laws and superposed geometries.}\label{sec:conservation_laws_superp_geometries} First, by conservation laws~\cite{compere_poincare_2020,flanagan_conserved_2017}, one expects an old black hole to be highly superposed, for example with a highly superposed mass~\cite{flanagan_infrared_2021}, also advocated for in Refs.~\cite{page1980is,Bao_2018,calmet2022brief,akil2025quantumsuperpositionblackhole,pasterski_hps_2021}.
Moreover, a black hole's soft hair in 4d asymptotically flat spacetime implies that the classical vacuum is highly degenerate~\cite{strominger_gravitational_2016,strominger2018lecturesinfraredstructuregravity}.
Therefore, drawing Penrose diagrams\footnote{Additionally, it is unclear how to continue a Penrose diagram post-evaporation as the near-singularity region is one of strong quantum gravity, see for example \cref{fig:Penrose_post_evap} for two different proposals. 
This is why in later Penrose diagrams of \cref{fig:critique_cloning_firewall} we will add a question mark `$?$' for how exactly the post-evaporation spacetime should be pictured.} with single-spacetime slices into the bulk of an old black hole may be problematic, and observers falling into the black hole can get entangled with its properties through the gravitational interaction, or even measure them.\footnote{One possibility would be to construct an effective spacetime metric from the correlators in the superposed spacetimes~\cite{Kempf_2021,saravani2016spacetime,foo2022quantum}; this requires further exploration and may not necessarily yield the standard black hole metric. See also Ref.~\cite{akil2023semiclassical} for semiclassical effective metrics as averages over superposed metrics.}

\vspace{0.5cm}

\textit{b. Locality and gauge invariance.}\label{sec:locality_gauge_invariance} Secondly, the notions of locality and gauge invariance\footnote{I thank David Mattingly for many interesting discussions and exchanges on these issues.}  in (quantum) gravity are subtle, and may have substantive implications for black hole (infaller) paradoxes~\cite{JACOBSON_2013,Jacobson_2019,Papadodimas:2015xma,papadodimas2014state,JACOBSON_2013,geng2026mechanisminformationencodingislands,raju2019toy,Chakraborty_2022}.
To make a localized observable gauge-invariant, one can use relational observables or gravitational dressings~\cite{Rovelli:1990ph,Brown_1995,Giddings:2019hjc,Giddings:2022hba,Giddings:2025xym,Giddings:2025bkp,Donnelly:2016rvo,Donnelly:2018nbv,Giddings_2006,goeller2022diffeomorphisminvariantobservablesdynamicalframes,Jacobson_2019,Giddings_2019,Giddings:2024qcf,JACOBSON_2013,Papadodimas:2015xma,papadodimas2014state,goeller2022diffeomorphisminvariantobservablesdynamicalframes,deboer2022frontiersquantumgravityshared,geng2022inconsistency}, with nontrivial effects regarding spacetime locality. 
Such dressings have been argued to be of crucial importance for the mechanism behind islands~\cite{geng2026mechanisminformationencodingislands}.
A related problem is the impossibility to locally factorize the Hilbert spaces of the outside radiation and inside of the black hole in the bulk in quantum gravity~\cite{JACOBSON_2013,Jacobson_2019,raju2022failure}.

\vspace{0.5cm}

\textit{c. Where is a Page curve obtained? At future null infinity?}\label{sec:where_page_curve}
Related to the previous point, one could argue that unitarity from the radiation and the Page curve\footnote{\label{fn:replica_subtility}If gravity is holographic~\cite{hooft2009dimensionalreductionquantumgravity,Susskind_1995,witten1998antisitterspaceholography,polchinski1999smatricesadsspacetime,susskind1999holography,maldacena1999large,giddings2000flat,Aharony_2000,de_Boer_2003,Dappiaggi_2004,Arcioni_2004,Arcioni_2003,DAPPIAGGI_2006,papadodimas2014state,Marolf_2009,Marolf_2013,JACOBSON_2013,Giddings_2018gauge,Giddings_2020,Donnay_2022carrollian,donnay_celestial_2023,pasterski2023chaptercelestialholography,laddha_2021,raju2019toy,raju2022failure,Chowdhury_2022,Araujo_Regado_2023}, one might have to couple the boundary, such as the AdS one in recent replica-trick calculations~\cite{almheiri2019entropy,almheiri2020replica,almheiri2021entropy,penington2020entanglement,penington2022replica,geng2021information}, to a non-gravitational bath to store the radiation and obtain a Page curve, argued to lead to massive gravitons through a Higgs mechanism in Refs.~\cite{raju2022failure,geng2022inconsistency,geng2025revisitingrecentprogresskarchrandall,Geng_2020,geng2025makingcasemassiveislands,geng2026mechanisminformationencodingislands}. Another possibility is to couple the black hole to a gravitational bath, and define the black hole and bath system (the radiation) through a non-geometric split~\cite{Uhlemann2022information,Uhlemann_2021,geng2021information}.} are only obtained near or at future (null) infinity in asymptotically flat spacetimes, which is also where the radiation was collected in an asymptotically flat study of islands~\cite{Hartman_2020}, and that gauge-invariantly identifying the near horizon modes and their evolution to form the late radiation arriving at future (null) infinity has nontrivial implications.
Closer to an old black hole, one can still have the ability to measure or get entangled with the (superposed) gravitational field, isolating the radiation and decoding information from it would involve a nontrivial construction in the bulk, with the backreaction of that construction being potentially non-negligible~\cite{Hui_2014}, and one has the failure of factorization per \hyperref[sec:locality_gauge_invariance]{$b$} there;  
see also \cref{fig:critique_cloning_firewall}.

Therefore, regarding the cloning paradox, even if the Page curve suggests that one would only require a few qubits from an old black hole to reconstruct the quantum state of a system $S$ that just fell into the black hole~\cite{hayden2007black}, one might need to recover those qubits at or near future (null) infinity, with no `nice' slices~\cite{preskill1992black,Lowe_1995,giddings2006black,Puhm_2017,Giddings_2006,polchinski1995stringtheoryblackhole,harlow2016jerusalem} hugging future null infinity until late times in an evaporating black hole spacetime while running into the black hole interior containing the system $S$ that fell in, as in \cref{fig:critique_cloning_firewall}.
Additionally, potential issues of nonlocal and non-adiabatic effects on nice slices have been expressed before~\cite{Giddings_2006,Lowe_1995,Puhm_2017}.

If information recovery and a Page curve are only to be expected at or near future (null) infinity $I^+$, a possibility could be that operations on the radiation do change the memory of an infalling observer, but, with the same causality of Wigner's friend; only when that memory arrives at future (null) infinity encoded in the radiation.
Such causality picture, with experiences during infall taking place earlier than the collection of that information at infinity, was advocated for recently in Ref.~\cite{concepcion2024complementaritydynamicalblackhole}.
Naturally, questions of the full experience of an infalling observer are still limited by the black hole singularity, and dependent upon its potential resolution. 
Speculations aside, the issues highlighted in this section call for an operational view on causality in quantum gravity~\cite{Hardy_2007,oreshkov2012quantum,chiribella2013quantum,castroruiz2018dynamics,Parker_2022}.

\begin{figure}[h]
         \centering
\tikzset{every picture/.style={line width=0.75pt}} 

\begin{tikzpicture}[x=0.75pt,y=0.75pt,yscale=-1,xscale=1]

\draw [color={rgb, 255:red, 0; green, 0; blue, 0 }  ,draw opacity=1 ][line width=0.75]    (122.81,196.25) -- (198.52,135.6) ;
\draw    (121.92,134.48) .. controls (123.61,132.84) and (125.28,132.86) .. (126.92,134.55) .. controls (128.56,136.24) and (130.23,136.27) .. (131.92,134.63) .. controls (133.61,132.99) and (135.27,133.01) .. (136.92,134.7) .. controls (138.57,136.39) and (140.23,136.41) .. (141.92,134.77) .. controls (143.61,133.13) and (145.27,133.15) .. (146.92,134.84) .. controls (148.56,136.53) and (150.23,136.56) .. (151.92,134.92) .. controls (153.61,133.28) and (155.27,133.3) .. (156.92,134.99) .. controls (158.57,136.68) and (160.23,136.7) .. (161.92,135.06) .. controls (163.61,133.42) and (165.28,133.45) .. (166.92,135.14) .. controls (168.57,136.83) and (170.23,136.85) .. (171.92,135.21) .. controls (173.61,133.57) and (175.27,133.59) .. (176.92,135.28) .. controls (178.56,136.97) and (180.22,136.99) .. (181.91,135.35) .. controls (183.6,133.71) and (185.27,133.74) .. (186.91,135.43) .. controls (188.56,137.12) and (190.22,137.14) .. (191.91,135.5) .. controls (193.6,133.86) and (195.26,133.88) .. (196.91,135.57) -- (198.52,135.6) -- (198.52,135.6) ;
\draw [color={rgb, 255:red, 255; green, 255; blue, 255 }  ,draw opacity=1 ]   (186.97,133.81) .. controls (188.64,132.16) and (190.31,132.17) .. (191.97,133.84) .. controls (193.62,135.52) and (195.29,135.53) .. (196.97,133.88) .. controls (198.64,132.23) and (200.31,132.24) .. (201.97,133.91) .. controls (203.63,135.58) and (205.3,135.59) .. (206.97,133.94) .. controls (208.64,132.29) and (210.31,132.3) .. (211.97,133.97) .. controls (213.63,135.64) and (215.3,135.65) .. (216.97,134) .. controls (218.65,132.35) and (220.32,132.36) .. (221.97,134.04) .. controls (223.63,135.71) and (225.3,135.72) .. (226.97,134.07) .. controls (228.64,132.42) and (230.31,132.43) .. (231.97,134.1) .. controls (233.63,135.77) and (235.3,135.78) .. (236.97,134.13) .. controls (238.64,132.48) and (240.31,132.49) .. (241.96,134.16) .. controls (243.61,135.84) and (245.28,135.85) .. (246.96,134.2) -- (249.46,134.21) -- (249.46,134.21) ;
\draw [color={rgb, 255:red, 0; green, 0; blue, 0 }  ,draw opacity=1 ]   (122.51,242.49) -- (241.27,148.18) ;
\draw [color={rgb, 255:red, 208; green, 2; blue, 27 }  ,draw opacity=1 ][line width=1.5]    (122.23,134.48) -- (188.03,189.91) ;
\draw [color={rgb, 255:red, 0; green, 0; blue, 0 }  ,draw opacity=1 ]   (198.27,110.62) -- (241.27,148.18) ;
\draw [color={rgb, 255:red, 0; green, 0; blue, 0 }  ,draw opacity=1 ]   (122.51,242.49) -- (122.23,134.48) ;
\draw [color={rgb, 255:red, 0; green, 0; blue, 0 }  ,draw opacity=1 ]   (198.27,110.62) -- (198.52,135.6) ;
\draw [color={rgb, 255:red, 0; green, 0; blue, 0 }  ,draw opacity=1 ][line width=0.75]    (312.6,206.09) -- (367.44,160.39) ;
\draw [color={rgb, 255:red, 255; green, 255; blue, 255 }  ,draw opacity=1 ]   (375.01,144.15) .. controls (376.68,142.49) and (378.35,142.5) .. (380,144.18) .. controls (381.66,145.85) and (383.33,145.86) .. (385,144.21) .. controls (386.68,142.56) and (388.35,142.57) .. (390,144.25) .. controls (391.66,145.92) and (393.33,145.93) .. (395,144.28) .. controls (396.67,142.63) and (398.34,142.64) .. (400,144.31) .. controls (401.66,145.98) and (403.33,145.99) .. (405,144.34) .. controls (406.67,142.69) and (408.34,142.7) .. (410,144.37) .. controls (411.65,146.05) and (413.32,146.06) .. (415,144.41) .. controls (416.67,142.76) and (418.34,142.77) .. (420,144.44) .. controls (421.66,146.11) and (423.33,146.12) .. (425,144.47) .. controls (426.67,142.82) and (428.34,142.83) .. (430,144.5) .. controls (431.66,146.17) and (433.33,146.18) .. (435,144.53) -- (437.5,144.55) -- (437.5,144.55) ;
\draw [color={rgb, 255:red, 0; green, 0; blue, 0 }  ,draw opacity=1 ]   (312.9,253) -- (410.88,170.04) ;
\draw [color={rgb, 255:red, 208; green, 2; blue, 27 }  ,draw opacity=1 ][line width=1.5]    (311.19,160.13) -- (369.94,205.44) ;
\draw [color={rgb, 255:red, 0; green, 0; blue, 0 }  ,draw opacity=1 ]   (310.55,83.4) -- (410.88,170.04) ;
\draw [color={rgb, 255:red, 0; green, 0; blue, 0 }  ,draw opacity=1 ]   (312.9,253) -- (310.55,83.4) ;
\draw  [draw opacity=0][fill={rgb, 255:red, 208; green, 2; blue, 27 }  ,fill opacity=1 ] (309.44,159.71) .. controls (316.07,153.44) and (326.83,149.33) .. (339,149.27) .. controls (351.66,149.21) and (362.85,153.54) .. (369.49,160.18) -- cycle ;
\draw  [draw opacity=0][fill={rgb, 255:red, 208; green, 2; blue, 27 }  ,fill opacity=1 ] (367.98,159.78) .. controls (361.17,164) and (351.28,166.68) .. (340.26,166.72) .. controls (328.72,166.77) and (318.39,163.92) .. (311.51,159.41) -- cycle ;
\draw    (311.51,159.41) .. controls (313.19,157.75) and (314.86,157.76) .. (316.51,159.44) .. controls (318.17,161.11) and (319.84,161.12) .. (321.51,159.47) .. controls (323.19,157.82) and (324.86,157.83) .. (326.51,159.51) .. controls (328.17,161.18) and (329.84,161.19) .. (331.51,159.54) .. controls (333.18,157.89) and (334.85,157.9) .. (336.51,159.57) .. controls (338.17,161.24) and (339.84,161.25) .. (341.51,159.6) .. controls (343.19,157.95) and (344.86,157.96) .. (346.51,159.64) .. controls (348.17,161.31) and (349.84,161.32) .. (351.51,159.67) .. controls (353.18,158.02) and (354.85,158.03) .. (356.51,159.7) .. controls (358.16,161.38) and (359.83,161.39) .. (361.51,159.74) .. controls (363.18,158.09) and (364.85,158.1) .. (366.51,159.77) -- (367.98,159.78) -- (367.98,159.78) ;

\draw (99,102.83) node [anchor=north west][inner sep=0.75pt]   [align=left] {$\displaystyle ( a)$};
\draw (285.46,111.09) node [anchor=north west][inner sep=0.75pt]   [align=left] {$\displaystyle ( b)$};
\draw (324.64,151.63) node [anchor=north west][inner sep=0.75pt]  [font=\footnotesize,color={rgb, 255:red, 255; green, 255; blue, 255 }  ,opacity=1 ] [align=left] {$\displaystyle QG$};

\end{tikzpicture}
         \caption[Continuing the spacetime post-evaporation]{To discuss the evaporation of black holes and unitarity, the post-evaporation spacetime must be pictured as well. Two possibilities are (a): continuing past the singularity, and (b) the same continuation but for a nonsingular black hole; see also Refs.~\cite{Ashtekar_2005,hawking2016soft,Schindler_2020,Ashtekar_2020,Ashtekar_2025,hiscock1981models,hiscock1981parttwo,Hayward_2006,Ashtekar_2011,ashtekar_surprises_2011,Stephens_1994,barenboim2024no,Carballo-Rubio:2025fnc,almheiri2019entropy}. 
         }
         \label{fig:Penrose_post_evap}
\end{figure}
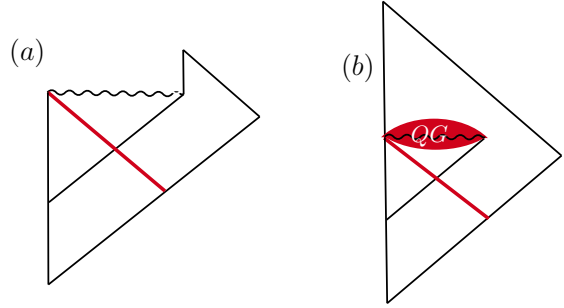

\vspace{0.5cm}

\textit{d. More subtleties and observer-dependencies in the firewall paradox.} \label{sec:more_subtleties_firewall} Fourth, as for the firewall paradox, the appearance of superposed geometries for old black holes was already mentioned above in \hyperref[sec:conservation_laws_superp_geometries]{$a$}, conflicting with the assumption $(Fiii)$ in \Cref{th:firewall} of high entanglement across the horizon derived from QFT on a single spacetime, as well as the non-factorization regarding locality and gauge invariance per \hyperref[sec:locality_gauge_invariance]{$b$} for specifying the early radiation and near horizon (interior and exterior) modes.

Moreover, also in QFT, the experience of an infalling observer, and the entanglement they see, may be very different from an observer remaining outside (for longer)~\cite{Barbado_2011,Barbado_2016}.
Consequently, the descriptions of the black hole's evolution and entanglement profile for different observers could be quite distinct~\cite{Alsing_2012,devuyst2025gravitationalentropyobserverdependent}, with entanglement known to be a frame-dependent notion~\cite{giacomini2019quantum} and potential subtleties regarding the equivalence principle for infalling observers~\cite{Barbado_2016,Barbado_2011,smerlak2013new,ng2022little,preciado2024more}.
As mentioned in the previous point \hyperref[sec:where_page_curve]{\textit{c}}, assumption $(Fii)$ of the firewall paradox as presented in \Cref{th:firewall} can also be questioned, with late-time radiation modes only required to purify the early radiation when arriving at infinity, and need not necessarily be trivially identified with near-horizon modes.

\vspace{0.5cm}

\textit{e. Infaller's backreaction and initial states.}\label{sec:infaller_backreaction} Fifth, the backreaction of an infalling object or observer is a nontrivial effect~\cite{Dray:1985yt,hooft1987strings,kiem1995black,Stephens_1994,yoshida_firewalls_2019,gaddam_soft_2024,penington2020entanglement,Ferrari:1988cc,Sfetsos_1995,shenker2014black,Betzios:2016yaq,Gaddam:2021zka,BenTov:2017kyf,polchinski2015chaosblackholesmatrix,pasterski_hps_2021}. 
Backreaction due to infalling objects and aspects of ultrarelativistic scattering have been investigated using shockwaves~\cite{HOOFT198761,hooft1987strings,Dray:1985yt,Ferrari:1988cc,Sfetsos_1995,BenTov:2017kyf,Verlinde_1992,polchinski2015chaosblackholesmatrix,shenker2014black}, but can also be probed using (black hole) eikonal approximations to scattering~\cite{gaddam_soft_2024,Gaddam:2020rxb,Gaddam:2021zka,divecchia2024gravitationaleikonalparticlestring,Amati_2008,Veneziano_2008,Veneziano_2008II,Marchesini_2008high,Kabat_1992,Di_Vecchia_2020,Ciafaloni_2008,Dvali_2015,Addazi_2017,ciafaloni2019infrared,banks1999modelhighenergyscattering}. 

Furthermore, 
maintaining the coherence of the quantum information encoded in an infalling system as observed by a co-infalling observer, a crucial assumption in the cloning paradox for example, requires careful shielding of that quantum information to prevent leakage and decoherence. 
This is complicated by the standard assumption that gravity does not allow for shielding~\cite{bertolami2006generaltheoryrelativitysurvive,savrov2012gravitational,Dukehart:2024bsn}, implying that such quantum information should be encoded in other degrees of freedom.
Recent arguments of decoherence of superpositions due to (soft) radiation leaking into horizons or null infinity demonstrate such complications~\cite{Danielson:2022tdw,Danielson:2024yru,Belenchia_2018,danielson2025horizonssoftquantuminformation}.

Relatedly, a full treatment of the unitarity problem of black hole evolution requires specification of the initial \textit{quantum} state for black hole formation, where literature separates into collapsing shells~\cite{petr_kay_kuchar1992quantum,hajicek1992quantum,hajivcek2001unitary,louko_hamiltonian_1998,corichi2002quantum,vaz2007quantum,Vaz_2022,Eyheralde_2017,baccetti2019black,gaztanaga2025gravitational,berezin1997quantum,Malafarina_2017,ziprick2016polymer,rovelli2014planck}, high-energy collisions of two particles~\cite{Penroses1974unpublished,Matschull_1999,Eardley_2002,rychkov2004black,hirotaka2005improved,giddings2004black,Hsu_2003,Veneziano_2008,Veneziano_2008II,Marchesini_2008high,Amati_2008,divecchia2024gravitationaleikonalparticlestring,Di_Vecchia_2020,Cardoso_2015exploring,Choptuik_2010,sperhake_cardoso_2009cross,death1992gravitational,ciafaloni2015emerging,ciafaloni2019infrared,Addazi_2019,Bern_2022,damour2018high,NEILL2013177,yoshiblackno2003,east2013ultra}, holographic investigations of collapsing shells~\cite{Anous_2016} and colliding particles~\cite{Haehl_2023}, and Euclidean path integral preparations of black holes in two-dimensional gravity~\cite{mertens2023solvable,kourkoulou2017purestatessykmodel}.

\begin{figure}[]
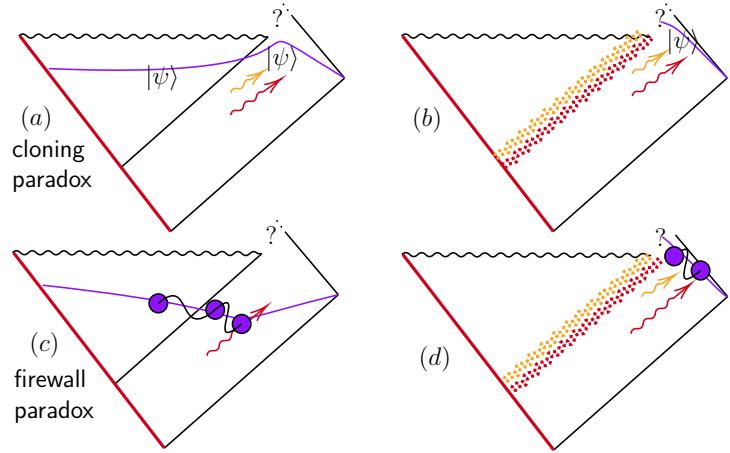

         \centering
\includestandalone[width=0.65\textwidth]{Figures/critique_cloning_firewall}
         \caption[A critique of the cloning and firewall paradoxes]{A critique of the cloning paradox (a), with the state $\ket{\psi}$ cloned on a spatial surface, and the firewall paradox (c), with three systems violating monogamy of entanglement, the inside mode highly entangled with the mode just outside the horizon and the early radiation also highly entangled with the same late mode just outside the horizon. Diagrams (b) and (d) give an alternative views for the cloning and firewall paradox, respectively. Namely, in order to recover the information, it is only at or near $I^+$ that one must be able to reconstruct the state $\ket{\psi}$ of an infalling system from the radiation and potentially remaining gravitational degrees of freedom. Similarly, it is only at or near $I^+$ that the early radiation must be highly entangled with the late radiation. Furthermore, the experiences of infalling observers might be quite different as they get entangled with the superposed, backreacted geometry (in which a notion of single spatial slices may be questionable as well).}
         \label{fig:critique_cloning_firewall}
\end{figure}

\subsection{Implications and future directions} \label{sec:future_directions}
The considerations above raise several doubts upon typical assumptions in black hole paradoxes for infalling observers, which also underlie our no-go theorems. 
Specifically, the locality and gauge-invariance issues of \hyperref[sec:locality_gauge_invariance]{$b$}, further complicate the superposed spacetime of an old black hole per \hyperref[sec:conservation_laws_superp_geometries]{$a$}, and the related problems of causality, and how and where information is recovered from the radiation per \hyperref[sec:where_page_curve]{$c$}, as well as the infaller's backreaction in \hyperref[sec:infaller_backreaction]{$e$}, question how independent the relevant interior and exterior operators in black hole paradoxes are. 
These elements thus question the validity of assumption $(Ci)$ in the cloning paradox and \Cref{th:theorem_FR_BH,th:nogo_EWFcloning_notmeet}, that quantum information `inside' remains unaltered by operations `outside' the black hole, with inside and outside referring to a semiclassical picture.

Furthermore, superposed geometries and soft hair, non-factorization, the location of information retrieval and observer-dependencies as in \hyperref[sec:conservation_laws_superp_geometries]{$a$}, \hyperref[sec:locality_gauge_invariance]{$b$}, \hyperref[sec:where_page_curve]{$c$} and \hyperref[sec:more_subtleties_firewall]{$d$} present challenges for assumptions $(Fii)$ and $(Fiii)$ of the firewall paradox and \Cref{th:EWF_firewalls}, namely the relevance of high entanglement across the horizon and the identification of late-time (purifying) radiation modes with near horizon modes.

Therefore, with the aim of progressing towards a better physical understanding, operational investigations of assumptions and explicit modeling of infalling quantum systems in the context of black hole puzzles are an interesting avenue, with potential for lab experiments in analogue gravity~\cite{barcelo2005analogue,unruh1995sonic,steinhauer2016observation,lahav2014realisation,drori2019observation,Steinhauer_2014,bose2025massive,Munoz2019observation,liberati2019information,chowdhury2022sachdev}. 
One such possibility would entail the use of quantum detectors, where theoretical studies with detectors in (nonclassical) spacetimes have given us a rich plethora of results already~\cite{Barbado_2011,Barbado_2016,ng2022little,henderson2018harvesting,Ng_2014,bhattacharya_probing_2024,Smith_2014,Barbado_2011,Hodgkinson_2012,hodgkinson2013particledetectorscurvedspacetime,paczos2023hawking,henderson2018harvesting,Henderson_2020,Henderson_2022,Ju_rez_Aubry_2014,preciado2024more,wang2024singular,spadafora2024deepknottedblackhole,Foo_2020,PhysRevResearch.3.043056,foo2021entanglement,foo2023quantum,goel2024accelerateddetectorsuperposedspacetime,foo2022quantum,suryaatmadja2023signaturesrotatingblackholes,niermann2024particle,paczos2023hawking,Stritzelberger_2021,Hotta_2020,Stritzelberger_2020,Cong_2020,Ng_2018,henderson2018harvesting,Gallock_Yoshimura_2021,Mendez_Avalos_2022,Tjoa_2020,robbins2020entanglementamplificationrotatingblack,chakraborty2024entanglementharvestingquantumsuperposed,vsoda2022acceleration,pan2024gravity,chen2024quantum,chen2023quantum,foo2022quantum,paczos2023hawking}.
Furthermore, as black hole puzzles concern different observers' perspectives, questions of gauge invariance, dressings, locality and soft and edge modes, with moreover radiation leaking into or across (reference frames at) null infinity, the paradigm of quantum reference frames~\cite{hoehn_trinity_2021,giacomini2019quantum,Giacomini_2021,vanrietvelde_switching_2023,rovelli_quantum_1990,goeller2022diffeomorphisminvariantobservablesdynamicalframes,kabel2023quantum,kabel2024identification,Fewster_2024,ahmad2024quantum,Carrozza_2024,kabel2023quantum,Araujo_Regado_2025} may provide a useful framework for further investigations of some of these subtleties.  
We move towards such investigations in the next chapter, combining elements of quantum reference frames with detectors in a nonclassical gravitational setting.

\vspace{0.5cm}

Finally, computational complexity and quantum error correction have been associated with black holes, holography and the decoding of Hawking radiation on numerous occasions~\cite{harlow_quantum_2013,aaronson_complexity_2016,kim_ghost_2020,Oppenheim_2014,susskind2013blackholecomplementarityharlowhayden,susskind2016computational,brown2019pythonslunchgeometricobstructions,harlow2016jerusalem,akers2022blackholeinteriornonisometric,Brown_2017,BAIGUERA20261,engelhardt2023algebraicereprcomplexitytransfer,Belin_2022,almheiri2018holographicquantumerrorcorrection,Verlinde_2013,Hayden_2019,Almheiri_2015,Pastawski_2015,Cotler_2017chaos}, but have not been probed in the context of Wigner's friend.  
Where are \textit{our} superobservers? Why can we trust our memories, for example for formulating physical laws? If Wigner can manipulate the brain of his friend, Wigner could implant memories and experiences that would lead the friend to different physical laws, or even nonsensical results.
We can imagine a similar situation when an observer, let's say a bee, falls into a fire. 
It burns to ashes, but we expect that when carefully keeping track of all relevant pieces of information, we can reconstruct the bee in the (classical) state right before it fell in. 
Now if we could perform such a complex operation, then perhaps we could also manipulate its brain into thinking that right before it fell into the fire the world was filled with flying crocodiles. 
Could it be that emergence and computational complexity are important concepts regarding consistency, Wigner's friend experiments and related (self-referential) puzzles such as Boltzmann brains~\cite{jones_thinking_2024,carroll2020boltzmann}?

The apparent absence of superobservers~\cite{rivlin2025emergenceclassicalitywignersfriend} is implied by the emergence of (kinematic and dynamical) classicality~\cite{zurek1991decoherence,zurek2003decoherence,schlosshauer2007quantum,schlosshauer2005decoherence,Zurek:2009ymk,Brand_o_2015,Allahverdyan_2013,Kofler_2007,brukner2017quantum,bibak2025classicallimitquantummechanics,omnes1994interpretation,gell-mann1993classical,Lombardo_1996,Halliwell_1998}, which have been linked to classical chaos and random matrices~\cite{zurek1994decoherencechaos,brukner2017quantum,Strasberg_2023}.
Moreover, topics such as computational complexity, chaos, quantum error correction and scrambling appear in measurement-induced phase transitions~\cite{Choi_2020,Gullans_2020,Suzuki_2025,Skinner_2019,Li_2019}. It would be interesting to further probe connections between these areas, and review the concept of an observer in this light.

\vspace{0.5cm}

\ifSubfilesClassLoaded{ 
  \printbibliography[heading=bibintoc, title=References]
  }{}
\end{document}

\ifSubfilesClassLoaded{ 
	\setcounter{chapter}{1} 
	\maketitle \tableofcontents \listoftodos
	}{}

\chapter[Particle detector near position-superposed black hole]{Particle detector near position-superposed black hole}\label{chap:part2_detector}

In the previous chapter, we have outlined several subtleties in standard assumptions of black hole paradoxes. 
Among the methods that can be employed to further probe these assumptions are quantum reference frames (QRFs) and detectors. 
In this chapter, we present an analysis combining these elements, by calculating the response of an Unruh--DeWitt detector in a 2+1d spacetime that contains a BTZ black hole in a superposition of locations as viewed from a reference frame adapted to the detector. 
We do so without relying on the details of a full theory of quantum gravity, but showing that this physical situation can be equivalently (and more conveniently) described in a different reference frame as a detector in a superposition of different locations in a fully classical gravitational field. 
In this frame, we can calculate the detector's evolution without involving any quantum property of gravity. Next, we perform a second QRF transformation to return to the original frame, where the black hole is in superposition, and examine the resulting dynamics. 
Our findings demonstrate that a joint measurement of the detector and the black hole's position can reveal effects of the superposition, manifesting as interference fringes. 
We define a measurement whose outcome probabilities contain a nonclassical contribution that would be absent for a black hole described by a classical mixture of positions. 

Our methods and setup are inspired by Ref.~\cite{foo2022quantum}, which studies a BTZ black hole in superposition of different masses. 
By contrast, we place the black hole in superposition of locations (as viewed from a reference frame adapted to the detector). 
Crucially, we find similar but different results: the resulting interference pattern is smooth - showing no sharp peaks---unlike the pronounced peaks presented in Ref.~\cite{foo2022quantum}.
This further supports the interpretation in Ref.~\cite{foo2022quantum} that such peaks are a consequence of black hole mass quantization~\cite{bekenstein2020quantum,hod1998bohr,Bekenstein_1995,bekenstein1998quantumblackholesatoms,deppe2024echoesbeyonddetectinggravitational}. 
We show analytically how this difference arises from singularities in the spectrum probed by the detector.    

\vspace{0.5cm}

The rest of this chapter is organized as follows. 
In \Cref{sec:bottom_up_QRF_detectors}, we further motivate our approach, combining bottom-up quantum gravity, quantum reference frames and detectors.
Next, in \Cref{sec:setup} we introduce the necessary concepts for our calculations: we provide a short pedagogical introduction to the BTZ black hole, Unruh--DeWitt detectors and quantum reference frames.
The reader who is already familiar with these concepts can directly read \Cref{sec:new_setup_and_ref_fields,sec:new_results_and_numerical_analysis}, where we describe the physical scenario we consider, and present our theoretical, numerical results and corresponding analytical expressions. 
Finally, we conclude in \Cref{sec:discussion}.
We work in natural units, i.e. $\hbar = k_B = c = 8G = 1$.

\section{Motivation: bottom-up quantum gravity, quantum reference frames and detectors}\label{sec:bottom_up_QRF_detectors}
\sectionmark{Motivation: bottom-up quantum gravity, QRFs and detectors}

Black holes serve as a prime example of a natural setting where quantum theory intersects with general relativity. Ideally, a full theory of quantum gravity is needed to completely describe the behaviour of black holes as specific cases within a broader framework. This approach, often referred to as \enquote{top-down}, has so far struggled to yield concrete predictions due to the challenges in making calculations with existing models.

Crucially, one of the most intriguing aspects of black holes—--their emission of Hawking radiation—was predicted using a different, \enquote{bottom-up} approach. Without relying on a full theory of quantum gravity, but instead using only established principles from general relativity and quantum theory, Hawking predicted that black holes emit (approximately) thermal radiation~\cite{hawking1975particle}. This discovery led to fundamental puzzles in theoretical physics, such as the black hole information puzzle~\cite{hawking1976breakdown,harlow2016jerusalem,Polchinski_2016}, and the cloning~\cite{susskind1993stretched,hayden2007black} and firewall paradoxes~\cite{page1993average,page1993information,braunstein2007quantum,braunstein2013better,almheiri2013black}, which we discussed in \Cref{chap:BH_prelims,chap:part2_WFBH}.

Other examples of \enquote{bottom-up} approaches include theoretical proposals to study quantum-gravitational effects that would be absent if gravity was classical, such as table-top quantum gravity experiments~\cite{bose2017spin,marletto2017gravitationally,bose2022mechanism,tino2021testing,marshman2020locality,moore2021searching,margalit2021realization,Bengyat_2024,tobar2024detecting}, or the study of quantum superpositions of semiclassical spacetimes~\cite{foo2023quantum,foo2022quantum,suryaatmadja2023signaturesrotatingblackholes,foo_schrodingers_2021,Foo_2022}. 
In the last framework, quantum reference frames (QRFs)~\cite{giacomini2019quantum,de_la_Hamette_2023,de_la_Hamette_2020,Vanrietvelde_2020,delahamette2021perspectiveneutralapproachquantumframe,cepollaro2024entanglement,Loveridge_2018,carette2024operationalquantumreferenceframe,Fewster_2024,kabel2024identification,Cepollaro_2024,zych2018relativityquantumsuperpositions,zych2018quantum,kabel2023quantum,goeller2022diffeomorphisminvariantobservablesdynamicalframes,Araujo_Regado_2025} have caught recent attention, as a tool to study quantum gravitational phenomena: when spacetime exhibits non-classical behaviour, classical coordinates may no longer provide an adequate description of the physical situation, necessitating the use of a quantum framework for reference frames. 
Moreover, QRFs have also been linked to asymptotic symmetries~\cite{bondi1962gravitational,sachs1962gravitational,compere2019advancedlecturesgeneralrelativity,compere_poincare_2020}, memory~\cite{Zeldovich:1974gvh,christodoulou1991nonlinear,thorne1992gravitational,Bieri_2024,maibach2026balancefluxlawsgeneral}, and soft and edge modes~\cite{Freidel_2020,Donnelly_2015,Donnelly_2016,Speranza_2018} in Refs.~\cite{kabel2023quantum,Araujo_Regado_2025,Carrozza_2022,gomes2025boundariesframesissuephysical}, which endow asymptotically flat spacetimes with soft hair through infrared physics~\cite{strominger2018lecturesinfraredstructuregravity}, argued to play an important role for the black hole information puzzle, for example~\cite{hawking2016soft,Donnay_2016,carney_infrared_2017,strominger2017blackholeinformationrevisited}.

An operational way to analyze the behaviour of black
holes and Hawking radiation is by probing the radiation with a particle detector, such as the Unruh--DeWitt detector~\cite{birrell1984quantum}. 
This approach has recently regained much attention: work has been done on the response of detectors
around black holes~\cite{Ng_2014,bhattacharya_probing_2024,Smith_2014,Barbado_2011,Hodgkinson_2012,hodgkinson2013particledetectorscurvedspacetime,paczos2023hawking,henderson2018harvesting,Henderson_2020,Henderson_2022,juarez2022quantumkicks,Ju_rez_Aubry_2014,Lifschytz_1994} and in AdS spacetime~\cite{jennings2010response}, infalling detectors crossing the black hole horizon~\cite{preciado2024more,ng2022little,wang2024singular,spadafora2024deepknottedblackhole,Barbado_2016}, detectors in a superposed trajectory~\cite{Foo_2020,PhysRevResearch.3.043056,foo2021entanglement}, detectors in superposed quotiented Minkowski~\cite{foo2023quantum} and Rindler spacetimes~\cite{goel2024accelerateddetectorsuperposedspacetime}, in superposed BTZ black holes~\cite{foo2022quantum,suryaatmadja2023signaturesrotatingblackholes}, superposed detectors in de Sitter and Schwarzschild spacetime~\cite{niermann2024particle,paczos2023hawking}, entanglement harvesting~\cite{Stritzelberger_2021,Hotta_2020,Stritzelberger_2020,Cong_2020,Ng_2018,henderson2018harvesting,Gallock_Yoshimura_2021,Mendez_Avalos_2022,Tjoa_2020,robbins2020entanglementamplificationrotatingblack,chakraborty2024entanglementharvestingquantumsuperposed,dubey2025harvesting,Membrere_2023,Barman_2023,wang2025harvestinginformationhorizon}, acceleration- and gravity-induced transparency effects~\cite{vsoda2022acceleration,pan2024gravity}, anti-Unruh~\cite{brenna2016anti,garay2016thermalization} and anti-Hawking~\cite{Henderson_2020,CAMPOS2021136198} phenomena, covariantly smeared detectors in spacetime~\cite{Mart_n_Mart_nez_2020}, etc. 
Furthermore, more complex models for detectors have been proposed in ~\cite{giacomini2022second,wood2022quantized,sudhir2021unruh,gale2023relativistic}, used for example for investigating the Unruh effect~\cite{barbado2020unruh,sudhir2021unruh}.

\section{Introducing BTZ black holes, Unruh--DeWitt detectors and quantum reference frames}\label{sec:setup}
\sectionmark{Introducing BTZ black holes, UDW detectors, and QRFs}

\subsection{BTZ Black hole} \label{sec:BTZ}
The BTZ black hole~\cite{Banados_1992,Carlip_1995,carlip2023quantumgravity21dimensions} is a 2+1-dimensional black hole solution to Einstein's equation with negative cosmological constant $\Lambda = -1/l^2$, asymptotically anti-de Sitter ($AdS_3$) with radius $l$. It is particularly interesting as its field correlation functions have analytically closed forms~\cite{Lifschytz_1994,Carlip_1995,Hodgkinson_2012,carlip2003quantum,foo2022quantum,Smith_2014,henderson2018harvesting}, making theoretical explorations more tractable.

We consider here a non-rotating and uncharged BTZ black hole, with metric
\begin{equation} \label{eq:BTZ_metric}
    \mathrm{d} s^2=-f(r) \, \mathrm{d} t^2+f(r)^{-1} \mathrm{~d} r^2+r^2 \mathrm{~d} \varphi^2,
\end{equation} with $f(r)=r^2/l^2-M$ and coordinate range $\sqrt{M}l < r < \infty,-\infty < t < \infty, \varphi \in [0,2\pi)$. 

The black hole has an event horizon at $r_h=\sqrt{M}l$, with maximally extended Penrose diagram as in \cref{fig:BTZ_Penrose}. Notably, being a 2+1-dimensional gravitational spacetime, the BTZ black hole has no Newtonian limit~\cite{DESER1984220,barrow1986three,cornish1991gravitation,Carlip_1995}, but other limits such as the Newton--Hooke one can be considered~\cite{Bacry:1968zf,Gibbons_2003,Bizo__2018,Pal_2025}.

\begin{figure}[h]  \centering   \tikzset{every picture/.style={line width=0.75pt}} 

\begin{tikzpicture}[x=0.75pt,y=0.75pt,yscale=-1,xscale=1]

\draw   (83.97,124) -- (188,124) -- (188,228.03) -- (83.97,228.03) -- cycle ;
\draw    (83.97,124) -- (188,228.03) ;
\draw    (83.97,228.03) -- (188,124) ;
\draw    (83.97,122.5) -- (188,122.5)(83.97,125.5) -- (188,125.5) ;
\draw    (83.97,226.53) -- (188,226.53)(83.97,229.53) -- (188,229.53) ;

\draw (139.08,152.19) node [anchor=north west][inner sep=0.75pt]  [font=\scriptsize,rotate=-315] [align=left] {$\displaystyle r=r_{h}$};
\draw (109.1,129.91) node [anchor=north west][inner sep=0.75pt]  [font=\scriptsize,rotate=-45] [align=left] {$\displaystyle r=r_{h}$};
\draw (126.22,106.54) node [anchor=north west][inner sep=0.75pt]  [font=\scriptsize] [align=left] {$\displaystyle r=0$};
\draw (124.22,232.54) node [anchor=north west][inner sep=0.75pt]  [font=\scriptsize] [align=left] {$\displaystyle r=0$};
\draw (190.22,174.54) node [anchor=north west][inner sep=0.75pt]  [font=\scriptsize] [align=left] {$\displaystyle r=\infty $};
\draw (49.22,174.54) node [anchor=north west][inner sep=0.75pt]  [font=\scriptsize] [align=left] {$\displaystyle r=\infty $};

\end{tikzpicture} \caption[Maximally extended Penrose diagram of the BTZ black hole]{The (maximally) extended Penrose diagram of the BTZ black hole~\cite{Carlip_2005}.} \label{fig:BTZ_Penrose}
\end{figure}
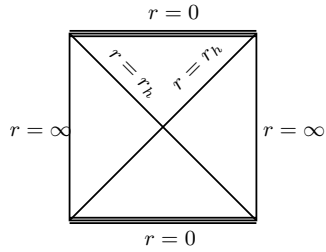

Spatial infinity is timelike, so that information can enter and leave at spatial infinity. Therefore, boundary conditions must be imposed for a field theory on the BTZ spacetime. 
In general, 2+1d gravity has no propagating degrees of freedom, but it is due to this boundary that nontrivial dynamics can occur, and so-called `boundary gravitons' can appear in a quantum theory of gravity~\cite{carlip2003quantum,carlip2023quantumgravity21dimensions,Carlip_2005}.
The BTZ black hole can be obtained from
the universal covering space of $AdS_3$~\cite{Kraus2008} by means of the quotient under $\Gamma_M:\varphi \rightarrow \varphi+2 \pi \sqrt{M}$, see \Cref{sec:app_BTZ_AdS}. This fact will allow for correlation functions of quantum fields in BTZ black hole spacetime to be calculated precisely using correlations in $AdS_3$. 
The temperature assigned\footnote{By a `temperature assigned to a black hole', we mean the mutually consistent `temperature' parameter, identified through the thermal spectrum of exterior quantum fields and, in equilibrium, as the KMS temperature of the exterior Hartle--Hawking state~\cite{Lifschytz_1994,KAY199149,hawking1975particle}, the periodicity of Euclidean time required for regularity at the horizon~\cite{Gibbons:1976ue,Banados_1992}, or from thermodynamic considerations~\cite{Wald:1999vt,Carlip_1995}.} to a BTZ black hole equals $T_0  = \sqrt{M}/(2\pi l) $~\cite{Carlip_1995}, and indeed a static detector around a BTZ black hole detects a red- or blueshifted local Tolman temperature~\cite{Santiago_2019,Lima_2019} $T_0/\sqrt{-g_{00}}$~\cite{Hodgkinson_2012,hodgkinson2013particledetectorscurvedspacetime}.

A test particle at fixed radius $R$ in space moving only in time has proper time that relates to coordinate time $t$ as $g_{00} (dt/d\tau)^2=-1$ so \begin{equation} \label{eq:tau_to_t}
    \tau = \sqrt{-g_{00}} t = \sqrt{f(R)} t =: \gamma(R) t = \sqrt{\frac{R^2}{l^2}-M} t,
\end{equation} with $\gamma(R)$ the redshift factor. Thus, when we consider two clocks at fixed, but different radii $R_1$ and $R_2$ from the black hole, there will be a relative time dilation between them. 

We introduce a particle field that interacts with the gravitational field of the black hole, and later we will measure the particle field with an Unruh--DeWitt detector. Specifically, we consider a massless scalar field $\phi(x)$ conformally coupled to the metric with equation of motion \begin{equation}
    \Box \phi -\frac{1}{8}R \phi = 0,
\end{equation} with vacuum analogous to the Hartle--Hawking state~\cite{Carlip_1995,Lifschytz_1994}.

The measurement probabilities of the Unruh--DeWitt detector, as explained below, depend on the two-point correlator of the scalar field in the BTZ black hole background. Such correlator can be calculated from the $AdS_3$ correlators (see~\cite{foo2022quantum,Hodgkinson_2012,Lifschytz_1994,hodgkinson2013particledetectorscurvedspacetime} for details):
\begin{equation} \label{eq:W_BTZ_from_image_sum}
    W_{\mathrm{BTZ}}\left(\mathrm{x}, \mathrm{x}^{\prime}\right)=\sum_n \Upsilon^n W_{AdS}\left(\mathrm{x}, \Gamma^n \mathrm{x}^{\prime}\right)
\end{equation} with $\Upsilon=+1,-1$ for an untwisted and twisted field~\cite{Lifschytz_1994}. We will consider the untwisted case $\Upsilon=1$ unless stated otherwise.
The correlation function in the scalar vacuum of $AdS_3$ is given by \begin{equation} \label{eq:BTZ_correlation}
    \begin{split}
        W_{\mathrm{AdS}}\left(x, x^{\prime}\right)=\frac{1}{4 \pi l \sqrt{2}}\left[\frac{1}{\sqrt{\rho\left(x, x^{\prime}\right)}}-\frac{\zeta}{\sqrt{\rho\left(x, x^{\prime}\right)+2}}\right],
    \end{split}
\end{equation} with $\rho(x,x')$ the squared geodesic distance in the 2+2 embedding space of $AdS_3$, and $\zeta \in\{-1,0,1\}$ specifies the boundary conditions at asymptotic infinity: Neumann, transparent or Dirichlet~\cite{carlip2003quantum,Lifschytz_1994,Shiraishi_1994}. We will take transparent boundary
conditions $\zeta = 0$, unless stated otherwise.
Neumann and Dirichlet boundary conditions are reflective boundary conditions, implemented through requirements on the derivative of the field and the field itself, respectively, when viewing the covering space of $AdS$ as half of an Einstein static universe~\cite{Lifschytz_1994,avis1978quantum}. 
In elementary wave-mechanical analogy of a free and a fixed endpoint, Neumann and Dirichlet boundary conditions lead to an in-phase and out-of-phase reflection of a wave. 
Transparent boundary conditions for a massless conformally coupled scalar field on (the covering space of) $AdS$, which is conformally isometric to half of an Einstein static universe, arise when taking the quantization of the Einstein static universe with no reflecting boundary conditions on $AdS$~\cite{avis1978quantum}. 
This leads to effective two-time Cauchy surfaces in the case of transparent boundary conditions for $AdS$ and a recirculation of energy and angular momentum across conformal infinity~\cite{avis1978quantum}. 
As the (eternal) BTZ black hole is a quotient of an open region of $AdS_3$, the implications of transparent boundary conditions for specifying initial data and conservation or flux-balance laws in the BTZ spacetime are not immediately clear and are beyond the purpose of this work; however, the BTZ two-point function for a massless conformally coupled scalar field can be calculated from the $AdS_3$ two-point function through image sums as explained above in \cref{eq:W_BTZ_from_image_sum}.
Other boundary conditions are possible as well; for instance, Refs.~\cite{Bussola_2017,CAMPOS2021136198} consider Robin-type boundary conditions for the BTZ spacetime.
We refer to Appendix \ref{sec:app_W_BTZ} for more details on the BTZ two-point correlation functions.

\subsection{Unruh--DeWitt detectors} \label{sec:UDW}
In flat spacetime, the usual definition of a particle relies on excitations of fields, decomposed into positive and negative frequency modes --- a decomposition that is possible when a preferred notion of time exists.
On a general curved spacetime, nevertheless, this straightforward definition is not possible, due to the lack of a preferred notion of time. A possible way of defining particles in a more operational way was introduced by Unruh~\cite{unruh1976notes}, suggesting that \enquote{a particle is what a particle detector detects}. 

Unruh modeled a particle detector as a scalar field in a cavity, and showed what we now call the Unruh effect~\cite{Crispino_2008}: an accelerated detector in Minkowski spacetime registers a thermal bath of particles. DeWitt~\cite{dewitt1979} proposed a simplified detector model, replacing the scalar field in a cavity by a simple two-level system, obtaining the Unruh--DeWitt detector~\cite{birrell1984quantum,martin2013wavepacket}.

Specifically, an Unruh--DeWitt (UDW) detector
consists of a detector modeled as a quantum system following a path
$x(\tau)$ in some curved spacetime that interacts linearly with a scalar field.\footnote{Extensions to detectors for fermionic or other boson fields exist~\cite{hummer2016renorm,ali2021unruhdewittdetectorresponsescomplex}, as well as explicit models for $n$-level detectors~\cite{Guedes_2024,lima2023unruh}, detectors with quantized center of mass~\cite{Stritzelberger_2020,sudhir2021unruh}, second-quantized detectors~\cite{giacomini2022second,gale2023relativistic} and detector models derived relativistic-covariantly using the influence functional \cite{torres24particle}. For simplicity, we will use a 2-level detector model.} The spacetime path\footnote{Usually the spacetime path is some classical trajectory on a background spacetime; later, we will allow the detector in a superposition of paths (meaning the path is an operator with a state highly peaked around two classical trajectories). We will denote the corresponding coordinate position operator that controls the paths by the hatted version $\hat{\mathbf{x}}$.} can be freely chosen, geodesic or non-geodesic driven by some external force. The detector has internal degrees of freedom, typically taken to be an $n$-level system with energy gaps, that \enquote{click} when a particle is measured, 
and external degrees of freedom, corresponding to the position of the detector.  It interacts with the particle field through a linear coupling, with the Schrödinger-picture interaction Hamiltonian in the detector's rest frame being 
\begin{equation} \label{eq:H_int}
 \hat H_\text{int}^\tau = \lambda \eta(\tau)  \hat \mu \otimes \phi(\mathbf{x}(\tau)),
\end{equation} with $x(\tau)$ the detector's prescribed spacetime trajectory and $\hat \mu$ the monopole moment which has off-diagonal elements in the detector's energy eigenbasis so that this interaction allows the detector to switch between its internal energy levels. Here $\hat H^\tau_\text{int}$ is written with respect to the proper time of the detector; choosing a different time $t$ to describe the evolution  yields the transformed Hamiltonian~\cite{Mart_n_Mart_nez_2020} \begin{equation} 
\hat H^t_\text{int} = \frac{d\tau}{dt} \hat H^\tau_\text{int}. \end{equation}
The coupling strength $\lambda$ will be assumed to be small, and $\eta(\tau)$ is a switching (-on and -off) function of the detector, which we will take to be a Gaussian $\eta(\tau) = e^{-\frac{\tau^2}{2\sigma^2}}$ with variance much smaller than the timescale $\Delta \tau$ of the experiment. We say the detector has detected a particle when the detector transitions to a higher internal energy level. An example of a detector could be an atom as a detector for particles of the electromagnetic field.

In the remainder of this chapter we will also use an idealized quantum-controlled extension of this Unruh--DeWitt detector model~\cite{Foo_2020,PhysRevResearch.3.043056,foo2021entanglement} in a BTZ spacetime, in which the spatial coordinate specifying the prescribed detector path can be in quantum superposition. We denote the corresponding (spatial) coordinate-position operator controlling the paths by $\hat{\mathbf{x}}_D$, with spectral measure $E_{\hat{\mathbf{x}}_D}(d \mathbf{x}_D) $ so that $\hat{\mathbf{x}}_D = \int \mathbf{x}_D  E_{\hat{\mathbf{x}}_D} (d \mathbf{x}_D)$, also denoted formally as $\ket{\mathbf{x}_D}\bra{\mathbf{x}_D} d\nu(\mathbf{x}_D)$. We take the different trajectories to be prescribed externally and neglect the dynamics of the control degree of freedom and apparatus maintaining them; see for instance Refs.~\cite{Stritzelberger_2020,wood2022quantized,gale2023relativistic,sudhir2021unruh} for treatments including dynamics of the quantized centre-of-mass.
We denote by $x_{\mathbf{x}_D}(\tau)$ the prescribed spacetime trajectory associated with the spatial-coordinate value $\mathbf{x}_D$, which we will take to be static trajectories $x_{\mathbf{x}_D}(\tau) = (t_{\mathbf{x}_D}(\tau),\mathbf{x}_D)$ held fixed spatially at $\mathbf{x}_D$; see also \cref{sec:new_reference_fields} for a treatment modeling the reference fields explicitly.
The quantum-controlled extension of the Unruh--DeWitt interaction becomes~\cite{Foo_2020,PhysRevResearch.3.043056,foo2021entanglement,Stritzelberger_2020,gale2023relativistic,wood2022quantized,sudhir2021unruh} (with pre-subscript $I$ for interaction picture)  \begin{equation} \label{eq:H_int_I_pic}
    { }_I \hat{H}_{\mathrm{int}}^\tau(\tau)=\int E_{\hat{\mathbf{x}}_D} (d\mathbf{x}_D) \otimes \lambda \eta(\tau) \hat{\mu}(\tau) \otimes \hat{\phi}\left(x_{\mathbf{x}_D}(\tau)\right)
\end{equation} with $\hat \mu(\tau)= e^{i \hat H_0 \tau}\, \hat \mu \, e^{-i\hat H_0\tau}$ with $\hat H_0$ the detector's monopole moment Hamiltonian so that $\hat H_0 \ket{E}_\mu = E \ket{E}_\mu$, and $\tau$ the detector's proper time.
We will denote states of the detector's position and internal degrees of freedom by $\ket{\mathbf{x}_D}_D,\ket{E}_\mu$, respectively.
For finitely many branches for $\mathbf{x}_D$ the integral reduces to a sum; for instance, for two sharply localized branches $\ket{\mathbf{x}_1},\ket{\mathbf{x}_2}$ this reduces to 
\begin{equation}
\sum_i \ket{\mathbf{x}_i} \bra{\mathbf{x}_i}_D \otimes {}_I \hat H_{\text{int}, i}^\tau (\tau),
\end{equation} the quantum-controlled form also used in Refs.~\cite{Foo_2020,foo2023superpositionsthermalisationstatesrelativistic}, with our control states having physical (coordinate) position labels,  and $ {}_I \hat H_{\text{int}, i}^\tau (\tau) = \lambda \eta(\tau)  \hat \mu \otimes \phi(x_i(\tau))$. Here $x_i(\tau)$ is the detector worldline of the detector held externally fixed at spatial $\mathbf{x}_i$, i.e.,  $x_i(\tau):=x_{\mathbf{x}_i}(\tau)$.
In our BTZ set-up of this chapter, we will take all paths to be static with equal fixed angular coordinates, and the radial part of $\mathbf{x}_D$ will be the relational radial coordinate of the detector in a reference frame adapted to the black hole, or alternatively, under a QRF transformation, a radial coordinate indicating the black hole position in a frame adapted to the detector; \cref{sec:new_reference_fields} gives this control variable a physical interpretation using quantum reference fields.

This detector can be used to measure properties of the scalar field to gain information on the semiclassical gravitational field.
To explain how a detector `clicks', let us consider one single classical trajectory $x(\tau)$ of the detector, with the detector initialized in its ground state and field $\phi$ in the Hartle--Hawking vacuum state $\ket{0}_\phi$. The detector \enquote{clicks} when a transition to another internal state $\ket{E}_\mu$ happens, due to the interaction with the particle field. This leads to a transition amplitude (see Appendix~\ref{sec:app_interaction} for details) 
\begin{equation} \label{eq:single_detector_amplitude}
   \begin{split}
       \bra{\vartheta,E} \ket{\psi(\tau_f)} &= i \lambda \langle E |\mu |E_0\rangle_\mu   \times \\ &\times \int_{\tau_i}^{\tau_f} d\tau \eta(\tau) e^{i(E-E_0)\tau} \langle \vartheta| \phi(x(\tau)) |0\rangle_{\phi},
   \end{split}
\end{equation} 
with $\ket{\vartheta}$ a final scalar field state that we will sum over, and later we will take $\tau_i = - \tau_f$.
The probability of the transition $E_0 \rightarrow E$ happening, independently of the final state of the field, is
\begin{align} \label{eq:transition_prob}
    P &= \int d\vartheta \left | \langle{\vartheta,E |\psi(\tau_f)}\rangle\right|^2 \nonumber \\
    &=  |\langle E | \mu |E_0\rangle|^2 \lambda^2  \mathcal{F}(E-E_0),
\end{align}
with the detector response function given by \begin{equation}
\begin{split}
    \mathcal{F}(\Omega) =& \int_{\tau_i}^{\tau_f} d\tau \int_{\tau_i}^{\tau_f}  d\tau' \eta(\tau) \eta(\tau') e^{-i \Omega(\tau-\tau')} \\& \langle 0 | \phi(x(\tau)) \phi(x(\tau')) |0\rangle \\ &= \int_{\tau_i}^{\tau_f} d\tau \int_{\tau_i}^{\tau_f} d\tau' \eta(\tau) \eta(\tau') e^{-i \Omega (\tau-\tau')} \\ & W_{BTZ}(x(\tau), x(\tau'))  . 
 \end{split}
\end{equation} 
Here $W_{BTZ}(x,x')$ is the scalar field two-point correlator in a BTZ black hole background.

In \Cref{sec:new_setup_and_ref_fields,sec:new_results_and_numerical_analysis} we will specialize this quantum-controlled detector construction to branches in which the detector is held at a fixed spatial coordinate $\mathbf{x}_i$, so that each branch corresponds to a prescribed static non-geodesic trajectory.

\subsection{Quantum Reference Frames} \label{sec:QRFs}
When describing physical systems, we typically do so relative to a chosen reference frame. A compelling question emerges when considering the possibility that the reference itself exhibits quantum properties. This idea has sparked significant research into quantum reference frames \cite{giacomini2019quantum,de_la_Hamette_2023,de_la_Hamette_2020,Vanrietvelde_2020,delahamette2021perspectiveneutralapproachquantumframe,Loveridge_2018,carette2024operationalquantumreferenceframe,Fewster_2024,cepollaro2024entanglement,kabel2024identification,Cepollaro_2024}, which investigate how the physical world is described when taking the reference frame to be a quantum system. Notably, it has been shown that fundamental quantum features, such as superposition and entanglement, depend on the chosen frame~\cite{giacomini2019quantum,cepollaro2024entanglement}: a system that appears to be in superposition with respect to one quantum reference frame may be localized in another. This idea has been especially valuable in the study of quantum-gravitational phenomena, as it allows for a description within a reference frame where gravity is entirely classical. As a result, it offers insights into such phenomena without requiring a complete theory of quantum gravity.

In order to introduce these ideas, let us first consider three classical physical systems, $B$, $D$, and $F$ as particles on a Cartesian spatial line, with the usual translation structure available. 
From the reference frame of $B$, we say that $D$ and $F$ are at position $\mathbf{x}_D$ and $\mathbf{x}_F$ respectively. On the other hand, when taking $D$ as a reference, one assigns positions $\mathbf{q}_B=-\mathbf{x}_D$ and $\mathbf{q}_F=\mathbf{x}_F-\mathbf{x}_D$ to $B$ and $F$ respectively. 
Can we extend this notion to quantum systems? Is it possible to define a transformation between \emph{quantum} reference frames?

If the three systems are now quantum, it could be that $D$ is in a superposition state of two locations $\mathbf{x}_1$ and $\mathbf{x}_2$. How can we transform to the reference frame of $D$ in this scenario? A natural idea is to extend the classical transformations by linearity, namely applying the classical transformation for each branch of the superposition.
The previous situation, from the reference frame of $B$, is described by 
\begin{equation}
    \ket{\psi}^{(B)}=\frac{1}{\sqrt{2}}\bigg( \ket{\mathbf{x}_1 }_{\!D} + \ket{\mathbf{x}_2}_{\!D} \bigg) \ket{\mathbf{x}_F}_{F}.
\end{equation} 
A QRF transformation, i.e. a coherent translation controlled on system $D$'s position, is implemented through the unitary transformation 
\begin{equation}\label{eq:QRF_transformation}
    \hat S^{(B)\to(D)} = \int d\mathbf{x} \,\ket{-\mathbf{x}}_{B}\bra{\mathbf{x}}_D \otimes T_F^\dagger(\mathbf{x}),
\end{equation} with $T_F$ the translation operator on the Hilbert space of the system $F$.
The state in the perspective of $D$ is then
\begin{align}
    \ket{\psi}^{(D)} &=  S^{(B)\to(D)} \ket{\psi}^{(B)} \nonumber \\
    &= \frac{\ket{-\mathbf{x}_1}_B\ket{\mathbf{x}_F-\mathbf{x}_1}_F + \ket{-\mathbf{x}_2}_B\ket{\mathbf{x}_F-\mathbf{x}_2}_F}{\sqrt{2}}.
\end{align} 
This describes the same physical setup but when choosing the origin to be system $D$'s location\footnote{Sometimes the origin state is also written explicitly, i.e. for example $\ket{\mathbf{0}}_D$ in this case.}, see also \cref{fig:QRF_BH}. 

\begin{figure}[h]  \centering   \includegraphics[width=0.6\textwidth]{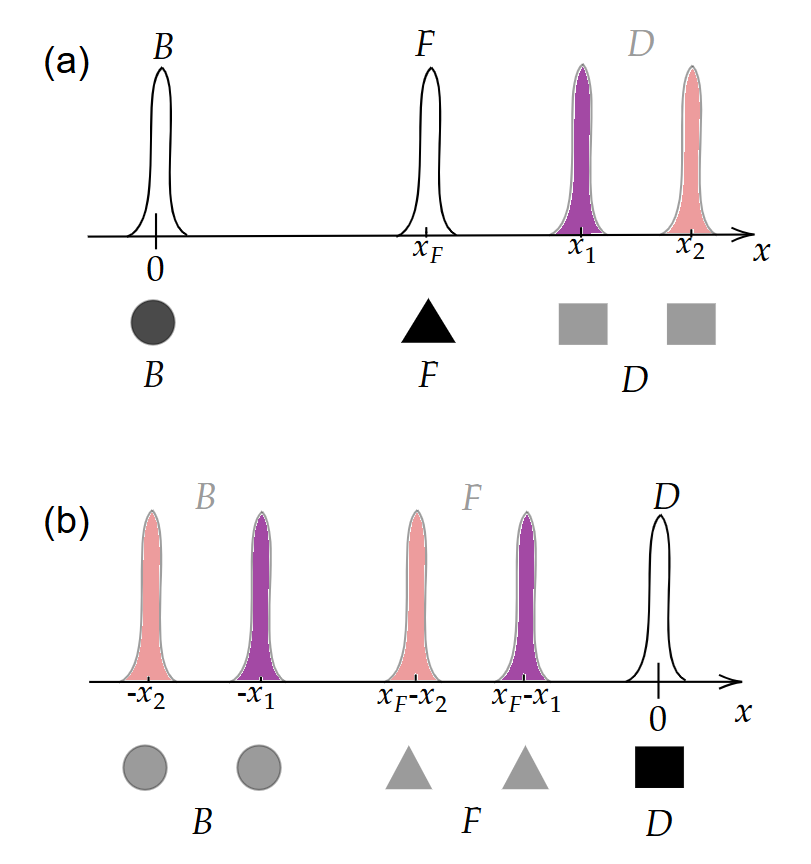} \caption[Action of a quantum reference frame transformation]{(a): Systems $B$ and $F$, near a position-superposed system $D$, which under a QRF transformation can be mapped to (b), where $B$ and $F$ are in indefinite, entangled locations but $D$ has a definite single-valued location. Adapted from Ref.~\cite{giacomini2019quantum}.} \label{fig:QRF_BH}
\end{figure}

Although the introductory example above uses ordinary spatial translations, the analogous QRF map $ \hat S^{(B)\to(D)} $ used later in this work in the BTZ spacetime is a branch-dependent change of coordinate reference-field coordinates, making use of diffeomorphism invariance rather than a translation symmetry (there is no radial translation symmetry) or isometry of the BTZ geometry.

Consider now the situation where not only we want to (spatially) `translate' the physical description, with a translation defined in the coordinate space using reference fields of \Cref{sec:new_reference_fields}, but also change the time coordinate we use in Schrödinger's equation. To do so, the (classical) reference frame change involves two steps: we first apply the translation, and then transform the Hamiltonian to account for the new time parameter. Under this change of classical reference frame, the Hamiltonian transforms as
\begin{equation}
    \hat H^{(D)} = \frac{dt}{d\tau} T \hat H^{(B)} T^\dagger, 
\end{equation}
where $T$ implements the classical spatial translation, $\tau$ is the new time parameter and $t$ is the old one. Applied to our set-up, if we choose $t$ to be the coordinate time, and $\tau$ to be the proper time in the detector frame, we obtain
\begin{align}
    \tau &= \gamma(\textbf{x}_D) \,t, \\
    \gamma(\textbf{x}_D)&=\sqrt{-g_{00}(\textbf{x}_D)} = \sqrt{\frac{\textbf{x}_D^2}{l^2}-M}.
\end{align}
We now generalize this transformation to superposition of translations, as we have explained before. The resulting Hamiltonian is (see for instance Ref.~\cite{giacomini2022second} for a similar treatment)
\begin{equation}
    \hat H^{(D)} =  \hat S^{(B) \to (D)} \gamma^{-1} (\hat{\textbf{x}}_D) \hat H^{(B)} \hat S^{\dagger (B) \to (D)}
\end{equation}
The reparametrization of time is now promoted to an operator: when $D$ is in superposition with respect to $B$, their proper times are in superposition with respect to each other.
In the rest of this chapter we apply these ideas to the quantum-controlled extension of the Unruh-DeWitt detector scenario described in the previous section, involving a detector $D$, a black hole $B$ and a scalar field $\phi$. The operator $\hat{\mathbf{x}}_D$ is the same coordinate-position/control operator controlling the paths as introduced in the previous section, and the interaction depends on the field through the local operator $\phi(\hat x_D)$, namely the field evaluated at the detector position. 
Since $\phi$ is a scalar, the change of frame introduces no additional internal or tensorial transformation on the field degrees of freedom. Its only effect is the corresponding change of the argument at which the field is evaluated, which is already accounted for by the transformation of the detector position. Therefore, the QRF transformation acts non-trivially only on the relative position degrees of freedom of $B$ and $D$, while acting as the identity on the field Hilbert space: 
\begin{equation}\label{eq:QRF_transformation_field}
    \hat S^{(B)\to(D)} = \int d\nu (\mathbf{x}) \,\ket{-\mathbf{x}}_{B}\bra{\mathbf{x}}_D \otimes \mathbb{1}_\phi.
\end{equation}
We will provide a realization of this transformation using reference fields in \Cref{sec:new_reference_fields}, which makes the relational meaning of the coordinate-position operator and the branchwise coordinate map explicit.


\section{Set-up, QRF transformation and reference fields} \label{sec:new_setup_and_ref_fields}

\subsection{Setup} \label{sec:setup_scenario}

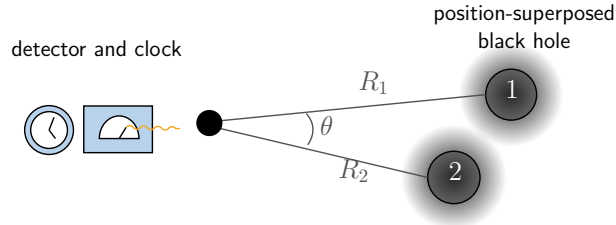
\begin{figure}[h]  \centering   \tikzset{every picture/.style={line width=0.75pt}} 

\begin{tikzpicture}[x=0.75pt,y=0.75pt,yscale=-1,xscale=1]

\draw  [color={rgb, 255:red, 0; green, 0; blue, 0 }  ,draw opacity=1 ][fill={rgb, 255:red, 184; green, 210; blue, 235 }  ,fill opacity=1 ] (85.73,147.02) .. controls (85.73,137.49) and (93.46,129.77) .. (102.98,129.77) .. controls (112.51,129.77) and (120.23,137.49) .. (120.23,147.02) .. controls (120.23,156.54) and (112.51,164.27) .. (102.98,164.27) .. controls (93.46,164.27) and (85.73,156.54) .. (85.73,147.02) -- cycle ;
\draw  [fill={rgb, 255:red, 184; green, 210; blue, 235 }  ,fill opacity=1 ] (127.5,130.03) -- (176,130.03) -- (176,163) -- (127.5,163) -- cycle ;
\draw  [color={rgb, 255:red, 0; green, 0; blue, 0 }  ,draw opacity=1 ][fill={rgb, 255:red, 255; green, 255; blue, 255 }  ,fill opacity=1 ] (90,147.02) .. controls (90,139.84) and (95.81,134.03) .. (102.98,134.03) .. controls (110.16,134.03) and (115.97,139.84) .. (115.97,147.02) .. controls (115.97,154.19) and (110.16,160) .. (102.98,160) .. controls (95.81,160) and (90,154.19) .. (90,147.02) -- cycle ;
\draw    (106.5,138.03) -- (102.98,147.02) ;
\draw    (107.5,153.03) -- (102.98,147.02) ;
\draw    (157.5,145.03) -- (152.5,153.03) ;
\draw  [fill={rgb, 255:red, 255; green, 255; blue, 255 }  ,fill opacity=1 ] (138.5,153.03) .. controls (138.5,153.03) and (138.5,153.03) .. (138.5,153.03) .. controls (138.5,144.2) and (144.77,137.05) .. (152.5,137.05) .. controls (160.23,137.05) and (166.5,144.2) .. (166.5,153.03) -- cycle ;
\draw    (157.5,145.03) -- (152.5,153.03) ;
\draw  [fill={rgb, 255:red, 74; green, 74; blue, 74 }  ,fill opacity=0.5 ][blur shadow={shadow xshift=0pt,shadow yshift=0pt, shadow blur radius=14.25pt, shadow blur steps=19 ,shadow opacity=100}] (411.5,122.02) .. controls (411.5,112.08) and (419.55,104.03) .. (429.48,104.03) .. controls (439.42,104.03) and (447.47,112.08) .. (447.47,122.02) .. controls (447.47,131.95) and (439.42,140) .. (429.48,140) .. controls (419.55,140) and (411.5,131.95) .. (411.5,122.02) -- cycle ;
\draw  [fill={rgb, 255:red, 74; green, 74; blue, 74 }  ,fill opacity=0.53 ][blur shadow={shadow xshift=0pt,shadow yshift=0pt, shadow blur radius=14.25pt, shadow blur steps=19 ,shadow opacity=100}] (370.5,180.52) .. controls (370.5,170.31) and (378.78,162.03) .. (388.98,162.03) .. controls (399.19,162.03) and (407.47,170.31) .. (407.47,180.52) .. controls (407.47,190.72) and (399.19,199) .. (388.98,199) .. controls (378.78,199) and (370.5,190.72) .. (370.5,180.52) -- cycle ;
\draw [color={rgb, 255:red, 74; green, 74; blue, 74 }  ,draw opacity=0.95 ]   (207.5,142.06) -- (411.5,122.02) ;
\draw [color={rgb, 255:red, 74; green, 74; blue, 74 }  ,draw opacity=0.95 ]   (207.5,142.06) -- (370.5,180.52) ;
\draw  [draw opacity=0] (284.63,135.36) .. controls (286.97,137.77) and (288.59,140.73) .. (289.17,144.08) .. controls (290,148.8) and (288.65,153.49) .. (285.74,157.43) -- (265.96,148.15) -- cycle ; \draw  [color={rgb, 255:red, 74; green, 74; blue, 74 }  ,draw opacity=0.95 ] (284.63,135.36) .. controls (286.97,137.77) and (288.59,140.73) .. (289.17,144.08) .. controls (290,148.8) and (288.65,153.49) .. (285.74,157.43) ;  
\draw  [fill={rgb, 255:red, 0; green, 0; blue, 0 }  ,fill opacity=1 ] (207.5,142.06) .. controls (207.5,137.23) and (211.41,133.32) .. (216.24,133.32) .. controls (221.07,133.32) and (224.98,137.23) .. (224.98,142.06) .. controls (224.98,146.89) and (221.07,150.8) .. (216.24,150.8) .. controls (211.41,150.8) and (207.5,146.89) .. (207.5,142.06) -- cycle ;
\draw [color={rgb, 255:red, 245; green, 166; blue, 35 }  ,draw opacity=1 ]   (157.5,145.03) .. controls (159.21,143.41) and (160.87,143.45) .. (162.5,145.16) .. controls (164.13,146.87) and (165.79,146.91) .. (167.5,145.29) .. controls (169.21,143.67) and (170.88,143.72) .. (172.49,145.43) .. controls (174.12,147.14) and (175.78,147.18) .. (177.49,145.56) .. controls (179.2,143.94) and (180.86,143.98) .. (182.49,145.69) .. controls (184.12,147.4) and (185.78,147.44) .. (187.49,145.82) .. controls (189.2,144.2) and (190.86,144.24) .. (192.49,145.95) -- (195.5,146.03) -- (195.5,146.03) ;
\draw [color={rgb, 255:red, 255; green, 255; blue, 255 }  ,draw opacity=1 ]   (342,214) -- (425.5,224.03) ;

\draw (75,82) node [anchor=north west][inner sep=0.75pt]   [align=left] {detector and clock};
\draw (373,57) node [anchor=north west][inner sep=0.75pt]   [align=left] {position-superposed \\ \ \ \ \ \ \ black hole};
\draw (293,136) node [anchor=north west][inner sep=0.75pt]  [font=\large,color={rgb, 255:red, 74; green, 74; blue, 74 }  ,opacity=0.95 ] [align=left] {$\displaystyle \theta $};
\draw (320,104) node [anchor=north west][inner sep=0.75pt]  [font=\large,color={rgb, 255:red, 74; green, 74; blue, 74 }  ,opacity=0.95 ] [align=left] {$\displaystyle R_{1}$};
\draw (306,166) node [anchor=north west][inner sep=0.75pt]  [font=\large,color={rgb, 255:red, 74; green, 74; blue, 74 }  ,opacity=0.95 ] [align=left] {$\displaystyle R_{2}$};
\draw (424,111) node [anchor=north west][inner sep=0.75pt]  [font=\large,color={rgb, 255:red, 255; green, 255; blue, 255 }  ,opacity=1 ] [align=left] {$\displaystyle 1$};
\draw (384,169) node [anchor=north west][inner sep=0.75pt]  [font=\large,color={rgb, 255:red, 255; green, 255; blue, 255 }  ,opacity=1 ] [align=left] {$\displaystyle 2$};

\end{tikzpicture} \caption[Set-up: detector near position-superposed black hole]{A detector with a clock is used to measure scalar particles created some fixed distance away from a position-superposed black hole. Relative to the detector, the black hole is in a superposition of radial distances $R_1$ and $R_2$.} \label{fig:setup2}
\end{figure}

\textit{Setup.} Consider the following scenario, illustrated in \cref{fig:setup2}. A particle detector is used to measure a massless conformally coupled scalar field near a 2+1d BTZ black hole that can be interpreted as being in a superposition of locations as viewed from a reference frame adapted to the detector. Such a black hole can, for example, result from the collapse of a shell in a superposition of two different positions in this frame. The detector comes with a clock and is held stationary rather than freely falling; this is the particular prescribed non-geodesic trajectory considered here, maintained by an external force whose dynamics we neglect. The detector is turned on for a finite proper time interval according to its clock. 

How does the detector respond to this situation? Can it witness quantum effects of the superposed spacetime?
\vspace{0.5cm}

\textit{Approximations.} 
In our calculations, we neglect the gravitational backreaction of the scalar field and detector, as well as reaction effects of the detector's clicking and we assume the two `translated' frame-dressed black hole metrics to be orthogonal\footnote{Or more precisely, we assume different eigenstates for the operator specifying the relational radial coordinate distance between the detector and black hole to be orthogonal, see also \Cref{sec:new_reference_fields}.}. We also assume the black hole metric to be static, i.e. we neglect potential evaporation. As introduced in \Cref{sec:UDW}, we work with a point-like detector, i.e. we assume the detector to be highly localized in each branch of the evolution. Moreover, we also neglect a potential conjugate variable to the black hole position~\cite{Eyheralde_2017,Eyheralde_2020,louko_hamiltonian_1998}.

\vspace{0.5cm}

\textit{Using QRFs.} In order to model the setup described above, we cannot employ the usual treatment of BTZ black holes and Unruh--DeWitt detectors: since the black hole is in superposition of locations in the chosen detector's reference frame, the evolution of the particle field will be in a quantum superposition of spacetimes, and the usual description of QFT on curved spacetime does not suffice to describe this situation.

Despite not subscribing to any specific theory of quantum gravity, we assume here that any such theory must have a semiclassical limit, where perfectly distinguishable gravitational states are orthogonal, and such that in each branch classical general relativity applies. We assign to each of these states a vector on a Hilbert space $\mathcal{H}_B = L^2(\mathbb{R})$, labeled by the position of the black hole with respect to the detector.

Under these assumptions, we can model the physical setup in the perspective of the detector, before it is switched on, as 
\begin{equation} \label{eq:initial_state_BH_superp}
    \ket{\psi(\tau_i)}^{(D)} = \frac{\ket{\mathbf{x}_1 }_{\!B} + \ket{\mathbf{x}_2}_{\!B} }{\sqrt{2}} \ket{0}_\phi \ket{E_0}_{\mu}.
\end{equation}
Here $\ket{\mathbf{x}_{1/2} }_{\!B}$ are the two orthogonal semiclassical states of the black hole, corresponding to two different coordinate positions relative to the detector (see also \Cref{sec:new_reference_fields}) with $\ket{0}_\phi$ the vacuum state of the scalar field in the gravitational BTZ background, 
evolved according to standard QFT on curved spacetime, and $\ket{E_0}_{\mu}$ represents the internal degrees of freedom of the detector before it is switched on.

In order to model the response of the detector, we need to define the Hamiltonian under which it evolves in this situation. A possible solution is to postulate it, as was done in Ref.~\cite{foo2022quantum}. We follow here an alternative approach, using QRF transformations to derive it.

As described in the previous section, the physical setup of the black hole in superposition of locations with respect to a detector can be described in the reference frame of the black hole as the detector being in superposition of locations and the black hole being localized (cf.~\cref{fig:QRF_BH}). In other words, by applying the QRF change defined in Eq.~\eqref{eq:QRF_transformation_field}, we obtain the state
\begin{align}
    \ket{\psi(t_i)}^{(B)} = \frac{\ket{-\mathbf{x}_1}_D + \ket{-\mathbf{x}_2}_D}{\sqrt{2}} \ket{0}_\phi \ket{E_0}_\mu.
\end{align}
where we choose the initial coordinate time to be $t_i=\tau_i=0$.
In this reference frame, there is no source of gravitational field in quantum superposition, and hence we can employ the usual tools of QFT in curved space time. Specifically, the UDW interaction for a localized black hole and quantum-controlled detector position introduced in \Cref{sec:UDW} has the following form with respect to the coordinate time~\cite{giacomini2022second}: 
\begin{equation} \label{eq:H_int_time_changes}
    H^{(B),t}_{\text{int}} = \gamma(\hat{\textbf{x}}_D) \lambda\, \eta(\gamma(\hat{\textbf{x}}_D)t)\,\mu \otimes \phi(\hat{\mathbf{x}}_D).
\end{equation} 
Hence we can perform the QRF transformation~\cref{eq:QRF_transformation_field} to obtain the Hamiltonian in the detector's frame with respect to its proper time as
\begin{equation} \label{eq:H_int_QRF_transformed}
    H_\text{int}^{(D),\tau}=\lambda \, \eta(\tau) \mu \, \otimes \phi(-\hat{ \textbf{x}}_B),
\end{equation} see also \Cref{app:H_int_superp_BH}.
This Hamiltonian is analogous to the one that was postulated in Ref.~\cite{foo2022quantum}, in the case of a mass-superposed black hole.

\subsection{A treatment of the position-superposition using reference fields} \label{sec:new_reference_fields}

This section presents a reference-field realization of the restricted quantum reference frame construction used above. It provides a relational and diffeomorphism-invariant interpretation of the coordinate-position operator $\hat{\mathbf{x}}$ appearing in the quantum-controlled UDW interaction and derives the branchwise coordinate relabelling underlying \cref{eq:QRF_transformation_field}. This section is not intended as a complete dynamical treatment of quantum reference fields, and operates within the approximations stated below.
Readers not interested in these details can skip this section. In subsequent later sections we comment on when and how this introduction of reference fields changes certain calculations or expressions.

In the previous section we described the set-up of a detector-black hole system where the relative coordinate distance between the detector and black hole is in a superposition. 
To give an operational meaning to such a position superposition, and to clarify the meaning of the radial specialization of the control operator $\hat{\mathbf{x}}$ as a relational radial coordinate distance between the black hole and detector, we introduce coordinate reference fields~\cite{kabel2024identification,chen2026quantumreferencefieldstransformations,hoehn2023matterrelativequantumhypersurfaces,thiemann2026quantumreferenceframesrelational}. 
This can be seen as a more precise kinematical treatment of quantum reference frames as outlined in the previous section, and clarifies the meaning of the coordinate operator $\mathbf{\hat{x}}$ in the Unruh--DeWitt interaction and its QRF transform. 
While we restrict to our specific set-up of two quantum reference fields restricted to specific sets of states, a quantum-controlled detector worldline, a fixed metric and conformally coupled scalar field and no backreaction, seeking a QRF transformation that works in this limited setting, a more complete quantum treatment of QRFs and their transformations is an active topic of research in literature; we comment on what a more complete dynamical treatment may involve, and refer the reader to Refs.~\cite{chen2026quantumreferencefieldstransformations,aguilargutierrez2026relationalpathintegraleffective,thiemann2026quantumreferenceframesrelational} for work in this direction. We comment upon this in the paragraph \textit{More complete treatment} below.

\vspace{0.5cm}

\textit{QRFs of reference fields.} In 2+1 dimensions, a classical frame of reference fields can be given by a locally invertible map $X^A: U \subset M \rightarrow X(U) \subset \mathbb{R}^3$ of three reference fields $X^0,X^1,X^2$ that we take to be three scalar fields. 
In a complete quantum treatment of reference fields, a quantum coordinate reference frame consists of quantum scalar fields,
restricted to a subspace formally spanned by field eigenstates with eigenfunctions that define locally valid coordinates. 
In our limited treatment, we will restrict the fields to states $\ket{\Psi_X}$ sharply peaked around a classical field configuration $X^A$ with negligible fluctuation $\delta \hat{X}^A = \hat{X}^A - X^A$. We will nevertheless allow for these quantum reference fields to be in states sharply peaked around a superposition of classical field configurations, i.e. $\ket{\Psi_{X,1}}+\ket{\Psi_{X,2}}$.
We will neglect the backreaction of the reference fields on all other systems including the black hole spacetime, and will assume factorization of the different systems and fields involved.\footnote{For a treatment in linearized gravity that does not neglect backreaction see Ref.~\cite{chen2026quantumreferencefieldstransformations}.}

A reference frame can be used to dress fields and observables into gauge-invariant operators (i.e. invariant under diffeomorphisms on $M$)~\cite{goeller2022diffeomorphisminvariantobservablesdynamicalframes,chen2026quantumreferencefieldstransformations,Loveridge:2016tnh,Dittrich_2007,Tambornino_2012}. 
For instance, given a spacetime scalar $F$, its dressing by $X$ has value $F_X(\xi^A)$ at the physical coordinate reading $X^A = \xi^A$ given by $F_X(\xi^A) = F(X^{-1}(\xi^A))$. 
Note that $F_X$ is indeed invariant under diffeomorphisms on $M$, as both $F$ and $X$ transform.
In the case of a quantum reference frame $\hat{X}^A$, such dressing can be defined formally through its eigenfunctions\footnote{One can define such eigenfunctions and eigenvectors using equal-time hypersurfaces in an auxiliary foliation~\cite{chen2026quantumreferencefieldstransformations}. In the case of quantum gravity with superposed metrics, for instance, such a foliation and timelike-spacelike split could be defined branch per branch.} $\ket{X}_X$, $F_{\hat{X}}(\xi) \ket{X}_X = F(\hat{X}^{-1}(\xi)) \ket{X}_X = F({X}^{-1}(\xi)) \ket{X}$. The inverse $\hat{X}^{-1}$ is thus defined branchwise, applying the inverse of the classical field configuration, $X^{-1}$, in each spectral branch.
A similar construction holds for spacetime tensors, with the $X$-dressing obtained by pulling back the tensor along $X^{-1}$; for instance, $(g_X)_\xi(u,v) = (X^{-1})^* g_(\xi)(u,v)$. 
The metric $g$ on $M$ that we will consider is classical (but its dressing might change (quantumly) under (quantum) dressings).

\textit{The detector worldline.} The detector's path $z: I \subset \mathbb{R} \rightarrow M: \tau \rightarrow z(\tau)$\footnote{Here $z(\tau)$ denotes the underlying spacetime worldline; its coordinate representation is the $x(\tau)$ used in \Cref{sec:UDW}.} will be presented in a specific reference frame by specifying $z^{(X)} := X \circ z: I \subset \mathbb{R} \rightarrow \mathbb{R}^3$.
We will allow $z^{(X)}$ to be a quantum operator $\hat{z}^{(X)}$ with, formally, eigenfunctions denoting detector worldlines $z^{(X)}$. In particular, we will take the radial coordinate position of the detector $z_r$ to be an operator, and consider detector paths that remain at fixed values of the spatial parts $X^1,X^2$ of the reference frame. 
This radial component $\hat{z}_r$ is the radial specialization of the coordinate-position/control operator $\hat{\mathbf{x}}_D$ introduced in \Cref{sec:UDW}.
We take the detector path to be externally prescribed, but potentially in a quantum superposed semiclassical state, neglecting quantum fluctuations around the semiclassical states, while ignoring all dynamics related to this external force or holding apparatus and path-control degree of freedom. Therefore, it can be seen as a detector model with quantum-controlled prescribed trajectories of Refs.~\cite{Foo_2020,PhysRevResearch.3.043056,foo2021entanglement}; see Refs.~\cite{gale2023relativistic,wood2022quantized,sudhir2021unruh} for treatments taking into account the dynamics of a detector with quantized center of mass.

\vspace{0.5cm}

We will consider reference frames $X_B$, adapted to the black hole, and $Y_{D,i}$, adapted to the detector in branch $i$, and construct a quantum reference frame transformation between them, which can be seen as the equivalent of \cref{eq:QRF_transformation_field}.

\vspace{0.5cm}

\textit{QRF adapted to the black hole.} The first reference frame $X_B^0,X_B^1,X_B^2$ that we consider is one adapted to the BTZ black hole exterior, and in the coordinate system $t,r,\varphi$ of \cref{eq:BTZ_metric}, we simply take these to be \begin{equation}
    X_B^0(x) = t(x), \quad X_B^1(x) = r(x), \quad X_B^2(x) = \varphi(x).
\end{equation} This can be seen as an eigenfunction in the formal eigenstate $\ket{X}$ of the operator $\hat{X}$.
In this treatment of reference fields, we do not require our reference fields to satisfy a prescribed equation of motion (and thus they can be considered `idealized' or `test' reference fields). If, instead, for instance, we wanted our reference fields to be massless scalar fields on the BTZ spacetime satisfying $\Box X_B^A = 0$, we would have to change $X_B^1$, for example taking $X_B^1(x) = C_1 \ln |1- \frac{r_h^2}{r(x)^2}| + C_2$, for some constants $C_1,C_2$.

Our choice of reference frame $X$ means the metric in $X$-coordinates (the $X$-dressed metric $g_X$) is given by \begin{equation}
     \mathrm{d} s^2=-f(r) \, \mathrm{d} t^2+f(r)^{-1} \mathrm{~d} r^2+r^2 \mathrm{~d} \varphi^2,
\end{equation} with $f(r)=r^2/l^2-M$.

\textit{Specifying the detector in the $X$-frame.} 
To specify the detector location, we start in the $X$-reference frame, and establish the detector location through the reference frame $X$, considering $z^{(X)}(\tau) = X \circ z (\tau)$.
All detector trajectories we consider take the detector to be held externally fixed at a certain radial and angular coordinate, rather than freely falling.
We will nevertheless allow our detector location to be in a superposition of radial locations $X^1 = r_{D,1}$ and $X^1 = r_{D,2}$. Therefore, the radial component of $z^{(X)}$ is promoted to an operator denoted $\hat{z}_{r}(\tau)$. We will write its value in branch $i$ as $r_{D,i}$.
Formally, $\hat{z}_r$ is an operator acting on some control system that controls the path of the detector (keeps it fixed at some quantum radial location), and so acts on the Hilbert space of a particle on the coordinate line $r$, i.e. $\mathcal{H} = L_2(\mathbb{R},dr)$ (or more precisely, taking an open part $I \subset \mathbb{R}$, $L_2(I,dr)$).
In this reduced representation its spectral measure will be denoted $E_{\hat{z}_r}(d r)$, also written formally as $E_{\hat{z}_r}(d r)=$ $|r\rangle\langle r| d r$.

The relational radial coordinate position between the detector (held fixed at certain radial and angular coordinate) and black hole can, for instance, be given meaning through the redshift factor on the detector worldline $z_i(\tau)$, \begin{equation}
    \frac{d\tau}{d X^0(z_i(\tau))} = \gamma(R_i),
\end{equation} which is a function of the radial coordinate on that worldline. 
Other ways to give it an operational meaning include using signals sent from a clock at the boundary.

\textit{Witnessing branch coherence.} To certify that the detector is in a superposition of worldlines, or under a QRF transformation, the black hole in a superposition of radial locations as seen from a frame adapted to the detector, one possible way to witness this coherence is to recombine the branches and perform a measurement that erases which-branch information, similarly to the $\ket{\pm}$ measurement of \Cref{sec:new_results}, or, more generally, perform a measurement sensitive to the off-diagonal coherence. 
In fact, the interference of \Cref{sec:new_results} can be viewed as constituting such superposition witness as it is absent in the case of the corresponding incoherent classical mixture.
Operationally, the measurement and initial preparation themselves may require additional control systems. 
If these require additional quantum reference systems but the corresponding operations employ idealized absolute quantities, the localizability of the relevant reference quantities (represented in general by POVMs) and the localization of the reference system's state determine how accurately the idealized measurement and state preparation can be implemented~\cite{Loveridge_2018,Loveridge:2016tnh,Busch:2016rui}. 
Additionally, repeated use of the same finite-resource reference system may degrade its state and consequently its suitability as a (good) reference~\cite{Bartlett_2006,Bartlett_2007}. 
We will not go deeper into these issues here.

\vspace{0.5cm}

\textit{QRF adapted to the detector.} The second reference frame $Y_D^0,Y_D^1,Y_D^2$ that we consider is a reference frame adapted to the detector, meaning it is well-localized with respect to the detector, taking in branch $i$ \begin{equation}
    Y_{D,i}^0(x) = t(x), \quad Y_{D,i}^1(x) = r(x)-r_{D,i}, \quad Y_{D,i}^2(x) = \varphi(x),
\end{equation} with $t,r,\phi$ the BTZ coordinates of \cref{eq:BTZ_metric}, with detector at $Y_{D,i}^1(x) = 0, \phi(x) = 0$. Note that the condition $Y_{D,i}^1(x) = 0$ means that $r(x) = r_{D,i}$ with $r_{D,i}$ the (static) radial location of the detector in the $t,r,\phi$-coordinates in branch $i$ (recall that the detector is held fixed at some fixed radial distance in each branch). 

Note that we could as well have adopted the choice to instead consider the first reference field $Y_{D,i}^0$ (timelike in the black hole exterior that we are probing) to be a reference field that reduces to the detector's proper time at the detector's location (and is appropriately extended away from it to the whole open region considered), i.e. $Y_{D,i}^0(x) = \tau(x)$ with $\tau(x)|_{\text{detector location}}=\tau$. Instead, we have opted to implement the additional redshift factor from relating the time coordinate and proper time of the Hamiltonians in each frame, as presented in \Cref{sec:QRFs}. Had we chosen this detector's proper time coordinate extension as the reference field $Y_{D,i}^0$, this branch-dependent redshift factor would result from the QRF transformation between these two frames of reference fields $X_B$ and $Y_{D,i}$ in each branch outlined below.

An alternative to two different reference fields in the two branches is to consider an operator $\hat{Y}$ restricted to a subspace generated by eigenvectors $\ket{Y}$ with eigenfunctions of the form $Y_{D,i}$ for some $r_{D,i}$, with $r_{D,i}$ the eigenvalues of $\hat{z}_r$, the radial component of the detector worldline in the $X$-frame $X \circ \hat{z}$.

Note that the two branches are not gauge-related because of the different relational distance $r_{D,i}=X^1_B( z_i)$ between the two branches. 

\vspace{0.5cm}

\textit{Transforming between QRFs of reference fields.} The classical reference frame transformation between these two frames in branch $i$ is determined by $X_B \circ Y_{D,i}^{-1}$, loosely stated, determined by a change of reference coordinate fields given by $r \rightarrow r+r_{D,i}$ in each branch. Now let us make this more precise.

Classically, a transformation from reference frame fields $X^A$ to $Y^A$ is determined by \begin{equation}
    Q_{X \rightarrow Y}: \mathbb{R}^3 \rightarrow \mathbb{R}^3: \xi^A \rightarrow X \circ Y^{-1} (\xi),
\end{equation} acting on dressed configuration space operators by its pullback.
For instance, the spacetime scalar $F$ with $F_X: \mathbb{R}^3 \rightarrow \mathbb{R}$ defined by $F_X(\xi) = F(X^{-1}(\xi))$ does transform correctly under this pullback: \begin{equation}
\begin{split}
        F_Y(\xi) &= F(Y^{-1}(\xi)) = F( X^{-1} \circ X \circ Y^{-1} (\xi)) \\ &= F\big( X^{-1} ( X \circ Y^{-1})(\xi) \big) = F_X (X \circ Y^{-1}(\xi)) \\ &= Q_{X \rightarrow Y}^* F_X (\xi) 
\end{split}
\end{equation} 
If, in our superposed BTZ case, we now apply this transformation coherently in each branch $i$, we get in branch $i$ the map \begin{equation}
    Q_{X_B \rightarrow {Y_D,i}} = X_B \circ Y_{D,i}^{-1}: \xi^A \rightarrow \xi^A + \delta^A_r \, r_{D,i}.
\end{equation}  We can thus make this work in each branch by considering the operator \begin{equation}
     \hat{Q}_{X \rightarrow \hat{Y}}: \xi^A \rightarrow \xi^A + \delta^A_r \,\hat{r}_D^{(X)},
\end{equation} with $\hat{r}_D^{(X)}$ the operator for the radial coordinate position between the detector and black hole.

In branch $i$ in the reference frame adapted to the detector, the $Y_{D,i}$-dressed metric $g_{Y_{D,i}}$ can be written as \begin{equation} \label{eq:BTZ_metric_2}
    \mathrm{d} s^2=-f_i(\overline{r}) \, \mathrm{d} t^2+f_i(\overline{r})^{-1} \mathrm{~d} \overline{r}^2+(\overline{r}+r_{D,i})^2 \mathrm{~d} \varphi^2,
\end{equation} with $f_i(\overline{r})=(\overline{r}+r_{D,i})^2/l^2-M$, with $t,\overline{r},\varphi$ now denoting $Y_{D,i}^0, Y_{D,i}^1, Y_{D,i}^2$. 
Our set-up can thus be interpreted as a black hole in a position-superposition as seen from the $Y$-frame.
The equations of motion are covariant under this change of frame.

Concluding, we thus see that this map $\hat{Q}_{X \rightarrow Y}$ and its pullback reproduce the branchwise coordinate relabelling underlying the QRF transformation of \cref{eq:QRF_transformation_field}: \begin{equation}
    \hat S^{(B)\to(D)} = \int d\nu(\mathbf{x}) \,\ket{-\mathbf{x}}_{B}\bra{\mathbf{x}}_D \otimes \mathbb{1}_\phi,
\end{equation} where the measure $d\nu(\mathbf{x})$ is a chosen measure for the reduced relational coordinate Hilbert space and could also be written equivalently using the spectral-measure notation of \Cref{sec:UDW}; it is not a spacetime volume measure.
The identity on $\phi$ arises as the QRF transformation acts as the identity on the abstract field-state factor (the undressed field $\phi$). Had we considered the dressed field $\phi_X$, it would change accordingly, as explained above.

The QRF map $\hat{Q}_{X \rightarrow \hat{Y}}$ we found is a reduction of the full quantum reference frame transformation, only intended to work correctly in the limited set-up we consider. 
It arises from the operator nature of $\hat{Y}$, and more specifically, $\hat{Y}^1$ through the definition of the detector worldline $\hat{z}^{X^1}$ in the $X$-reference frame. 
Note that this treatment clarifies the meaning of the coordinate operator $\hat{\mathbf{x}}$ used in \cref{eq:H_int_time_changes,eq:H_int_QRF_transformed} and the choice of assigning a Hilbert space $\mathcal{H}_B = L^2(\mathbb{R},dr)$ to the black hole-detector system that captures this relational degree of freedom in the radial coordinate; it is the relational parameter describing the transformation between the reference fields, which happens to be a translation in the radial $r$-coordinate, $r \rightarrow r+r_{D,i}$, which is why in the main text we refer to this as a `translation'. (It is thus not intended to refer to a translation symmetry as there is none in the BTZ spacetime.)

A more complete quantum reference frame transformation, not restricted to a specific set of reference field states, considers (pullbacks of) $\hat{Q}_{\hat{X} \rightarrow \hat{Y}} =  \hat{X} \circ \hat{Y}^{-1}$, which can be defined (formally) through its eigenfunctions.
Moreover, such QRF transformation may be defined through a perspective-neutral approach that takes into account the full dynamics and constraints~\cite{chen2026quantumreferencefieldstransformations,delahamette2021perspectiveneutralapproachquantumframe,Vanrietvelde_2020,aguilargutierrez2026relationalpathintegraleffective}, see also the paragraph \textit{More complete QRF treatment} below.
The frames we consider are idealized, sharply localized frames within the restricted configuration subspace considered here, resembling the ideal-frame limit of operational QRF constructions~\cite{Loveridge:2016tnh,Loveridge_2018,de_la_Hamette_2020,delahamette2021perspectiveneutralapproachquantumframe,carette2024operationalquantumreferenceframe}, although we do not construct a covariant PVM or invariant measure for the full diffeomorphism group. 

\vspace{0.5cm}

\textit{Connecting to literature.} We can connect this treatment also to that of Ref.~\cite{kabel2024identification}. Namely, being in the reference frame of $X_B$, we write the initial state as $$\ket{X_B}  (\ket{x_1}_D \ket{Y_{D,1}^A} + \ket{x_2}_D \ket{Y_{D,2}^A} )/\sqrt{2}.$$ Going to the reference frame of $Y_D$, we perform a branch-dependent diffeomorphism $d^{(i)}$ on the manifold $M$ (or the open region we care about), given by $r \rightarrow r+r_{D,i}$ in branch $i$, or more precisely, \begin{equation} 
\label{eq:define_diffeo_phi_i}
\begin{split}
    d^{(i)}:M \rightarrow M: p& = X_B^{-1}(t_p,r_p,\varphi_p) \\ &\mapsto d^{(i)}(p) = X_B^{-1}(t_p,r_p+r_{D,i},\varphi_p).
\end{split}
\end{equation} 
This sets indeed $\left(d^{(1)}\right)^*\left(Y_{D,1}\right)=\left(d^{(2)}\right)^*\left(Y_{D,2}\right)=\tilde{Y_D}_*$, so that we end up in a state where $\hat{Y}_D$ is single-valued (note that there are more diffeomorphisms that achieve this).
The branch-dependent diffeomorphisms are related to the above QRF transformation through $X_B \circ d^{(i)} \circ X_B^{-1} = X_B \circ Y_{D,i}^{-1}$.

\vspace{0.5cm}

\textit{More complete QRF treatment.} We have taken the reference fields to have zero backreaction, non-interacting with other fields and ignored their dynamics. In a more complete treatment, we should specify their dynamics and include their backreaction. 
Different formalisms then provide different counterparts of the reduced frame transformation we considered here.

In particular, the frame change map $\hat{X} \circ \hat{Y}^{-1}$ and the change of operator-dressing for the configuration variables may not determine the complete transformation; it may be necessary to further obtain its action on conjugate momenta, constraints, physical
Hamiltonians, or the physical inner product.
For instance, the perspective-neutral approach~\cite{Vanrietvelde_2020,hohn2020switch,Giacomini_2021,delahamette2021perspectiveneutralapproachquantumframe} considers the physical space encoding all different QRF perspectives, from which one reduces to a QRF $X$'s perspective through a reduction map $R_X$, with the QRF transformation given by $R_Y R_X^\dagger$, a unitary if the reduction maps are isometric isomorphisms onto the relevant reduced physical spaces.
This idea can be implemented canonically~\cite{chen2026quantumreferencefieldstransformations,Giacomini_2021,hohn2020switch,devuyst2025gravitationalentropyobserverdependent,Ali_Ahmad_2022,Vanrietvelde_2020}, through the path integral~\cite{aguilargutierrez2026relationalpathintegraleffective} and quantization of classical covariant phase-space methods~\cite{goeller2022diffeomorphisminvariantobservablesdynamicalframes,giesel2025linkingedgemodesgeometrical}. 
Other, related approaches are the operational or algebraic approach~\cite{carette2024operationalquantumreferenceframe,Loveridge_2018,Fewster_2024,fewster2025semilocalobservablesedgemodes} and the extra-particle approach~\cite{castroruiz2023relativesubsystemsquantumreference}, whose relation to the perspective-neutral approach remains an active topic of investigation~\cite{castroruiz2025interpretingquantumreferenceframe,Doat_2026}. 

Moreover, we have neglected the action of the reference frame transformation on boundary data; in the presence of (asymptotic) boundaries, edge modes might appear if a boundary is held fixed~\cite{Carrozza_2022,kabel2023quantum,giesel2025linkingedgemodesgeometrical}.

\section[Detector response near position-superposed black hole]{Main results: detector response near a position-superposed black 
hole} \label{sec:new_results_and_numerical_analysis}

\subsection{Results} \label{sec:new_results}

\textit{Measuring the final state.} We now have a full description of the physical situation: given the initial state and interaction Hamiltonian in \cref{eq:H_int_QRF_transformed,eq:initial_state_BH_superp}, we can calculate the final state of the system for a transition to energy level $E$ and final scalar field state $\vartheta$ (which we will sum over), to first order in the interaction coupling. The resulting state is simply a linear superposition, i.e. 
\begin{equation} \label{eq:transition_amplitude_superp}
    \langle{E,\vartheta}|{\psi(\tau_f)}^{(D)}\rangle = \frac{\ket{\mathbf{x}_1}_B \Theta_1 + \ket{\mathbf{x}_2}_B \Theta_2}{\sqrt{2}},
\end{equation}
where $\Theta_{1/2}$ are the single-branch transition amplitudes defined in \cref{eq:single_detector_amplitude} for detector position at the origin and black hole at relational position $\mathbf{x}_{1/2}$. 

In our limited treatment that neglects the gravitational dynamics of the black hole and the dynamics of the control variable or apparatus keeping the detector in place, the relative phase between the branches $\Delta \Phi$ is zero. In a treatment not neglecting these, an additional phase might arise.
The relative phase $\Delta \Phi$ in our case may have several potential origins or contributions in a more complete treatment. First, it may arise from the approximate evolution of the black hole as an approximate eigenstate of the gravitational Hamiltonian with its mass as eigenvalue with respect to coordinate time $t$ if we were to include the corresponding Hilbert space and state in our initial-state description. A relative phase difference when considering proper-time evolution might then appear as coordinate time relates differently to the detector's proper time in the two branches. 
The precise assumption in that case would thus be that for the purpose of the experiment the BTZ black hole spacetime remains approximately static and is an approximate eigenstate of the gravitational Hamiltonian, so that the phase coming from the black hole evolution is determined by its eigenvalue, presumably proportional to the black hole's ADM mass $M$ (in line with \cite{foo2022quantum}),
i.e. $H^t_{0} \ket{M}_{BH} = M \ket{M}_{BH}$. Such phase might thus also be particularly relevant in the case of a mass-superposed black hole where a phase difference between the two branches of the superposition can also develop as the mass differs between the two branches. As we consider our black hole to have the same mass in both branches of the superposition, we do not assign an explicit Hilbert space for the mass to the black hole, only for its position. In a more complete treatment, the spacetime location and energy-momentum of the black hole may not be independent~\cite{louko_hamiltonian_1998,Eyheralde_2017}. 
Approximating the black hole as developing only a phase independent of spatial location might also be an issue because of noncovariance, which would require a dynamical analysis taking into account the gravitational constraints and backreaction, potential non-factorization, etc. We neglect these complications here.
Secondly, we also neglect any additional relative phase associated with the external control degree of freedom and the apparatus responsible for keeping the detector on the prescribed static trajectory.
If one were to take into account these phases, we could also simply absorb them in the definition of the states.
Whether an additional phase would be visible in the experimental data depends on the precise implementation of the interferometric black hole position measurement which we describe below.

In order to witness interference effects, we must lose which-way information. For this reason, a simple measurement on the internal transition of the detector from $E_0$ to $E$ does not suffice, as it would produce a mixed signal $P_1+P_2$ with no interference effect. Instead, we measure a transition to an energy eigenstate $\ket{E}$, together with a measurement on the BH position that ignores which-way (or which-branch) information\footnote{If we model the physical reference fields of \cref{sec:new_reference_fields} as well, we would have to do a joint measurement that ignores which-branch information by including all systems $S$ that (might) contain such which-branch information $S_i$, including the physical reference fields, and, in fact, also the device holding the detector at fixed radial location (which is not a geodesic), meaning $\ket{\pm} =1/\sqrt{2} ( \ket{\mathbf{x}_1}_B \ket{S_1}_S \pm \ket{\mathbf{x}_2}_B \ket{S_2}_S)$. If we did not do so, systems holding which-branch information might decohere the superposition leading to suppression of the off-diagonal term $P_{12}$.} $\ket{\pm}_B = \frac{\ket{\mathbf{x}_1}_B \pm \ket{\mathbf{x}_2}_B}{\sqrt{2}}$.
This is the standard measurement of interferometry, where the two branches are recombined with a beam splitter, and subsequently the output ports of the interferometer are measured. This may be hard to realize for a cosmological black hole, but might be possible experimentally in the context of analogue gravity~\cite{barcelo2005analogue,braunstein2023analogue,unruh1981experimental,garay2000sonic,weinfurtner2011measurement}. The measurement probabilities are
\begin{equation}\label{eq:probabilities}
    P_\pm = \frac{1}{4} \big( P_1 + P_2 \pm 2 P_{12} \big),
\end{equation}
where $P_{1/2}$ are the probabilities defined in Eq.~\eqref{eq:transition_prob} for a detector in a single black hole spacetime (but with the black hole at a different location $\mathbf{x}_1,\mathbf{x}_2$ relative to the detector). $P_{12}$ is an interference term obtained as $P_{12} = \text{Re}[F_1^* F_2]$, with $F_i$ the single-branch amplitudes of Eq.~\eqref{eq:single_detector_amplitude}, now summed over the final field states, for the black hole at relational positions $\mathbf{x}_1,\mathbf{x}_2$, that is
\begin{align} \label{eq:interference_term}
    P_{12} = &|\langle E | \mu |E_0\rangle|^2 {\lambda^2} \nonumber \\ & \int_{\tau_i}^{\tau_f} d\tau d\tau'  \eta(\tau) \eta(\tau') e^{-i\Delta E(\tau-\tau')} W^{12}_{BTZ}(x(\tau),x(\tau'))
\end{align} with $\Delta E = E-E_0$ and $W^{(12)}_{BTZ}$ the two-point function \begin{equation}
    W^{(12)}_{BTZ}(x,x') = \langle 0_1 |_\phi \phi_1(x) \phi_2(x') |0_2\rangle_\phi,
\end{equation} 
and $\phi_i$ denotes the scalar field (and its vacuum $\ket{0_i}_\phi$) for a black hole at location $\mathbf{x}_i$ for $i=1,2$. 

We therefore find an interference term that is a genuine quantum effect extracted from the Unruh--DeWitt detector due to the black hole being in a quantum superposition. 

\subsection{Numerical analysis} \label{sec:results}

In \cref{fig:Fig1own} we plot the measurement probabilities given in Eq.~\eqref{eq:probabilities}, rescaled by a normalization factor $\mathcal{N}=\sigma \lambda^2/2 |\langle E |\mu |E_0\rangle|^2$ that depends on details of the detector's internal structure, as a function of the ratio $R_2/R_1$ of the radial coordinate distances between the detector and black hole in the branches of the superposition. 

\begin{figure}[h]  \centering   \includegraphics[width=0.45\textwidth]{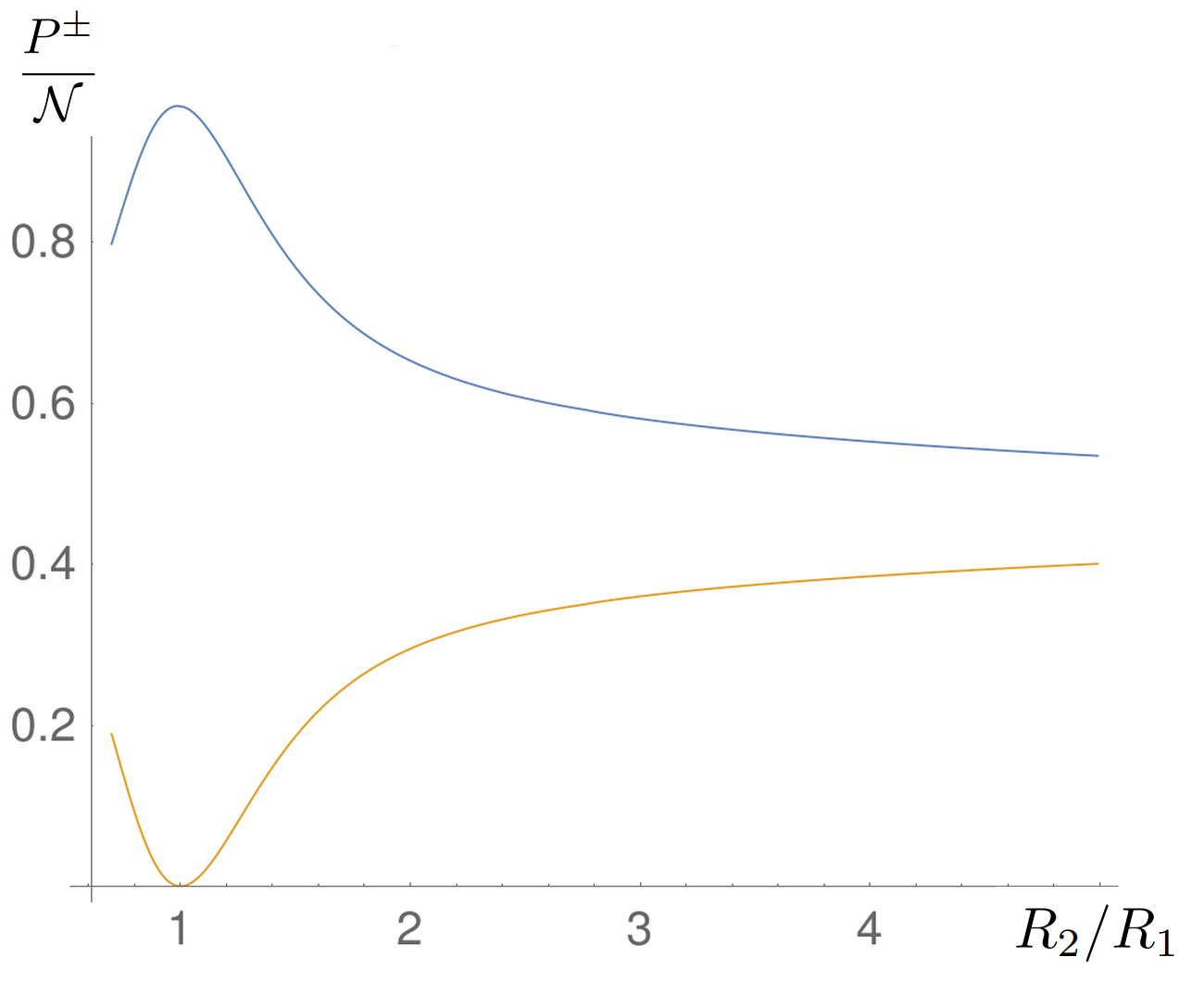} \caption[Interferometric and detector response measurement probabilities for position-superposed black hole]{Measurement probabilities $P_{E}^{\pm}/\mathcal{N}$ with $\mathcal{N}=\sigma \lambda^2/2 |\langle E |\mu |E_0\rangle|^2$ for a position-superposed black hole as a function of ${R_2/R_1}$. Here $P^+_E/\mathcal{N}$ is in blue and $P^-_E/\mathcal{N}$ in orange.
Parameters used: $l/\sigma = 5,\sigma=1$, $\tau_f = 10 \sigma = -\tau_i$ in the detector's own clock time, and $R_1/\sigma = 2 r_h,Ml^2 =4,E-E_0 = 0.00016$.} \label{fig:Fig1own}
\end{figure}
We see here the effect of the interference term, which is proportional to the difference between the two graphs. The interference is peaked around $R_1=R_2$, where the (coordinate) distance between the black hole and detector is the same in the two branches, and decreases with increasing ratio and as the distances between the detector and black hole increase in both branches, an effect in line with the decay of the conformally coupled scalar field two-point function with distance, i.e. due to $W^{12}_{BTZ}(R_1,R_2)$ in \cref{eq:interference_term}.
However, right at and very near to $R_1 = R_2$, this calculation cannot be trusted, as then the black hole is not in a superposition of locations (as seen from the detector). 
In this case, we have $\ket{+}_B = 2/\sqrt{2}\ket{x_1}, \ket{-} = 0$ meaning that the $\ket{\pm}$ measurement is ill-defined and the assumed initial state in \cref{eq:initial_state_BH_superp} with $\mathbf{x}_1 = \mathbf{x}_2$ is not normalized there, such that the calculation of the interference probabilities in \cref{eq:probabilities} cannot be trusted there. 

Furthermore, it is meaningful to compare our results with those of Ref.~\cite{foo2022quantum}, where a mass-superposed black hole was considered, instead of the position-superposed one we consider. In particular, we are comparing the measurement probabilities in Eq.~\eqref{eq:probabilities}, with the measurement probabilities of Eq.~(11) of Ref.~\cite{foo2022quantum}, that are analogous to the setup studied here, but when considering a black hole in superposition of masses. We plot their probabilities in \cref{fig:Fig1Foocompare}, reproducing Fig.~(1) of Ref.~\cite{foo2022quantum}. 
The authors argue that the sharp peaks that they observe are a consequence of having a superposition of different masses, and in particular interpret it in light of black hole mass quantization~\cite{bekenstein2020quantum,hod1998bohr,Bekenstein_1995,bekenstein1998quantumblackholesatoms,deppe2024echoesbeyonddetectinggravitational}. Our results further support this claim: when the black hole is in superposition of locations, but with single-valued mass, the sharp peaks are indeed absent. In the next subsection we will show in detail how this phenomenon arises from the spectrum probed by the detector.

\begin{figure}[h]  \centering   \includegraphics[width=0.45\textwidth]{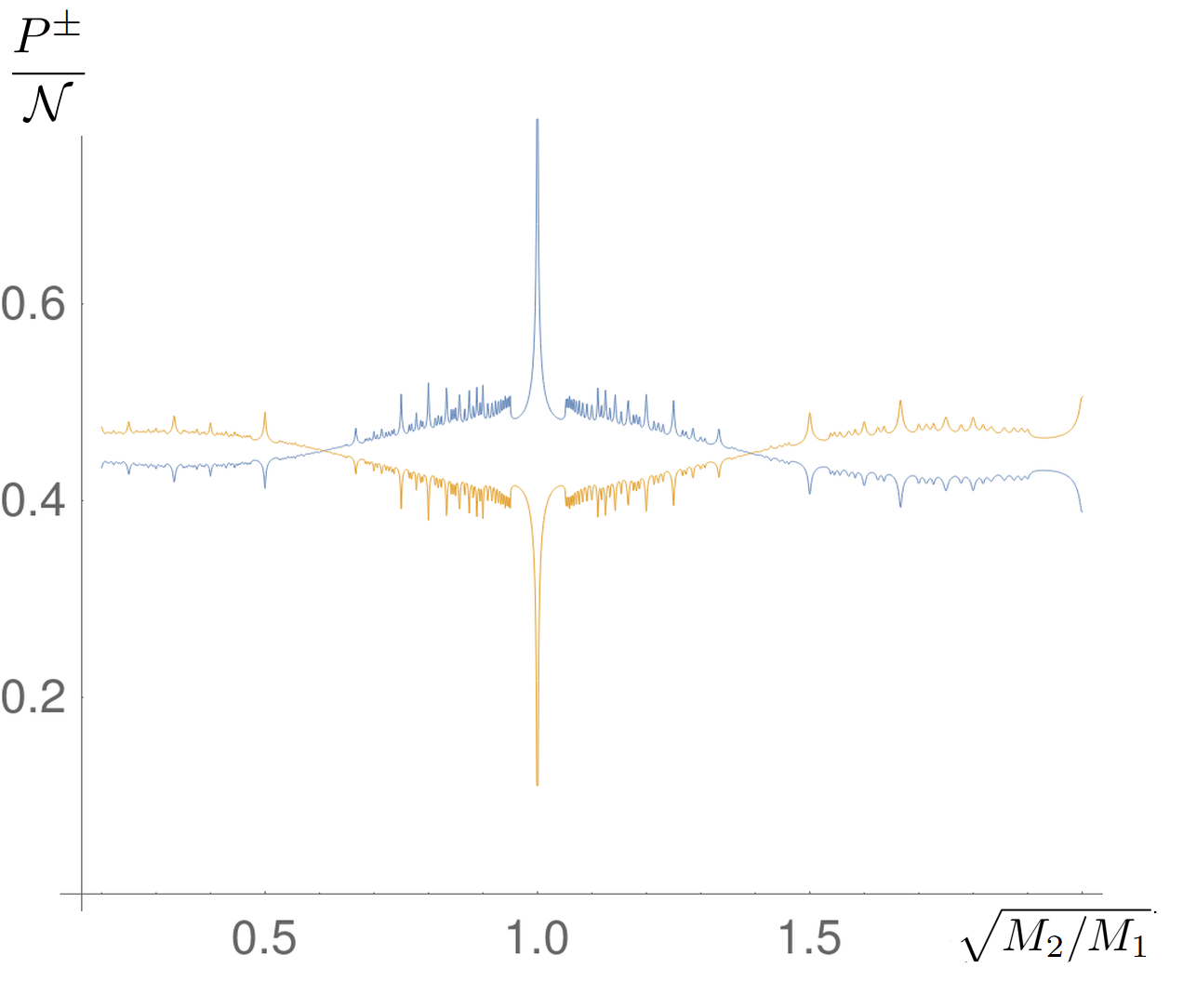} \caption[Interferometric and detector response measurement probabilities for mass-superposed black hole~\cite{foo2022quantum}]{Reproduction of Fig.~(1) of Ref.~\cite{foo2022quantum}: Measurement probabilities $P_{E}^{\pm}/\mathcal{N}$ with $\mathcal{N}=\sigma \lambda^2/2 |\langle E |\mu |E_0\rangle|^2$ for a mass-superposed black hole as a function of $\sqrt{M_2/M_1}$. Here $P^+_E/\mathcal{N}$ is in blue and $P^-_E/\mathcal{N}$ in orange. 
Parameters used: $l/\sigma = 5,\sigma=1$, $\tau_f = 5 \sigma = -\tau_i,R/\sigma = 25,M_1 l^2 =4,E-E_0 = 0.01 M_1 (= \Omega$ in Ref.~\cite{foo2022quantum}).} \label{fig:Fig1Foocompare}
\end{figure}

\subsection{Mass- vs. position-superposition} \label{sec:analytical}
In this subsection we derive an analytical expression for the spectrum probed by the detector in the case of a mass- and position-superposed black hole. We use it to identify the origin of the peaks that were observed numerically in the detector response for a mass-superposed black hole in~\cite{foo2022quantum} (see~\cref{fig:Fig1Foocompare}).  We find that the peaks can be traced to a coherent enhancement of the image sum at rational square-root mass ratios, and that additional singularities of the spectrum arise at certain rational mass ratios and corresponding position values in the mass-superposed case. Concretely, we will consider a generic mass- and position-superposed black hole, with position and mass $R_1,M_1$ and $R_2,M_2$ in the two branches of the superposition, respectively. 
Details can be found in Appendix \ref{sec:analytical_appendix}.

\vspace{0.5cm}

\textit{Single-spacetime detector response.} In a non-superposed spacetime, the detector response from \cref{eq:transition_prob} can be expressed as~\footnote{See Ref.~\cite{Fewster_2016} for a more precise treatment.}
\begin{equation} \label{eq:Fewster_eq_chi_hatW}
    \mathcal{F}(E)=\frac{1}{2 \pi} \int_{-\infty}^{\infty} d \omega \,|\widehat{\eta}({\scriptstyle\frac{\omega}{\gamma}})|^2 \,\widehat{W}_{\scriptstyle BTZ}(E \gamma +\omega),
\end{equation} with time integration limits pushed to $\pm \infty$ and the Fourier transform defined as
\begin{equation}
    \widehat{f}(\omega)=\int_{-\infty}^{\infty} d s f(s) \, \mathrm{e}^{-\mathrm{i} \omega s}.
\end{equation}
An analytical expression for $\widehat{W}_{BTZ}(K)$ can be found~\cite{Lifschytz_1994}, see Appendix \ref{sec:analytical_appendix} for details.

\textit{Interference term in detector response.} We can also express the interference term $\mathcal{F}_{12}(E)$, both for mass- and position-superposed black holes, in a similar form as \cref{eq:Fewster_eq_chi_hatW}, obtaining \begin{equation}
\begin{split}
    \mathcal{F}_{12}(E) = \frac{1}{2\pi} \int_{-\infty}^{+ \infty}  \mathrm{d}\omega  \hspace{0.05cm} \widehat{\eta}({\scriptstyle\frac{\omega}{\widetilde{\gamma_1}}})^* \hspace{0.05cm} \widehat{\eta}\left({\scriptstyle\frac{\omega}{\widetilde{\gamma_2}} + E [\frac{\widetilde{\gamma_1}}{\widetilde{\gamma_2}} -1] } \right) \\ \hspace{0.05cm} \widehat{W}^{(12)}_{BTZ}\left(\omega+\widetilde{\gamma_1} E\right). 
\end{split}
\end{equation}
Following Ref.~\cite{foo2022quantum}, we consider AdS time coordinates $\overline{t}$ related to the BTZ coordinates $t$ as $\overline{t}=t \sqrt{M}$ (see Supplemental Material of Ref.~\cite{foo2022quantum}, for instance).
Using the formulas in Appendix B of Ref.~\cite{Lifschytz_1994}, an analytical expression for $\widehat{W}^{(12)}_{BTZ}$ can be found, with the additional subtlety that for $\sqrt{M_1/M_2} \in \mathbb{Q} \backslash \{1\}$ and $R_1 /\sqrt{M_1} = R_2 /\sqrt{M_2}$ an additional pole arises, giving a singular term $\textsf{sing.}$:
\begin{equation} \label{eq:widehatW_12_BTZ_K}
    \begin{split}
        \widehat{W}^{(12)}_{BTZ}(K) = \textsf{sing.} + \frac{1}{2  \sqrt{\widetilde{\gamma_1} \widetilde{\gamma_2}}} \frac{1}{\sum_k \Upsilon^{2k}} \hspace{0.07cm} \frac{1}{e^{K/T_H^{AdS}}+1} \\ \sum_{m,n} \bigg[ P_{\frac{i K}{2 \pi T_H^{AdS}}-\frac{1}{2}}\big(\beta_{mn}^{(12)} \big) - \zeta  P_{\frac{i K}{2 \pi T_H^{AdS}}-\frac{1}{2}}\big(\beta_{mn}^{'(12)}\big) \bigg],
    \end{split}
\end{equation} 
with $T_H^{AdS}$ the Hawking temperature of the AdS–Rindler horizon with respect to AdS–Rindler time, $T_H^{AdS} = 1/(2 \pi l), \widetilde{\gamma_i}=\gamma(R_i,M_i)/\sqrt{M_i}$, and $\beta_{nm}^{(12)},\beta_{nm}^{'(12)}$ factors that depend on $M_1,M_2,R_1,R_2$. Here $P_\nu$ is a Legendre function of the first kind.
If $\sqrt{M_1/M_2} \notin \mathbb{Q} \backslash \{1\}$, then we have $\textsf{sing.} = 0$, else it gives the following contribution if also $R_1 = R_2 \sqrt{M_1/M_2}$:
\begin{equation} \label{eq:sing.}
\begin{split}
    \textsf{sing.} = & \frac{1}{4  \sqrt{\widetilde{\gamma_1}\widetilde{\gamma_2}} pq}  \\  &\text{ if } \sqrt{\frac{M_1}{M_2}} = \frac{p}{q} \in \mathbb{Q} \backslash \{1\} \text{ and } \frac{R_1}{R_2} =  \sqrt{\frac{M_1}{M_2}}  
\end{split}
\end{equation}
 with $\text{gcd}(p,q)=1$. Here we used a natural cut-off procedure for the infinite sums $\sum_{m,n}$ as suggested by the group structures that define the BTZ spacetime through the periodic identification of AdS--Rindler space, see also Appendix \ref{sec:analytical_appendix}.

For other values of $R_1,R_2$, one must resort to the `regular part' of $\widehat{W}^{(12)}_{BTZ}(K)$ of \Cref{eq:widehatW_12_BTZ_K}. For rational $\sqrt{M_1/M_2}=\frac{p}{q}$ and $R_2/R_1$ not this same rational number but $R_1,R_2$ fixed, $\beta_{mn}^{(12)}$ has the same value for terms in the sum over $m,n$ satisfying $m/n=\sqrt{M_1/M_2}$, and thus add up coherently. In addition, for $R_1,R_2$ both large compared to the horizon radius, $\beta_{mn}^{(12)} \rightarrow 1+$, the Legendre function $P_{-\frac{1}{2}-iK...}$ remains close to 1 until large $|K|$-values before decaying leading to sharper peaks.  
On the other hand, for irrational $\sqrt{M_1/M_2}$, for large $|K|$ oscillatory behavior (decaying with $|K|$) of the Legendre function dephases the sum leading to cancellations on average. For near-rational irrational $\sqrt{M_1/M_2}$ cut-off dependent peaks may occur at large $R_1,R_2$, but only limitedly as determined by the irrational offset, with more averaged cancellations for large $|K|$.
Additionally, large positive $K$-values are exponentially suppressed by the thermal factor, and thus main contributions to (peaked) behaviour come from large negative $K$-values. 

This leads to the peaks in the interfered response function in Ref.~\cite{foo2022quantum}, as pictured in \Cref{fig:Fig1Foocompare}.

\section{Discussion and outlook} \label{sec:discussion}

We have analyzed the transition probability of a static detector in a position-superposed BTZ black hole spacetime, where our measurement loses which-way information on the position of the black hole. To do so, we employed a quantum reference frame transformation, showing that the physical situation is equivalent to a quantum-controlled position-superposed Unruh--DeWitt detector evolving in a classical spacetime. As a result, we find a `nonclassical' term that would be absent for a black hole in a probabilistic mixture of locations.

The same setup but for a mass-superposed BTZ black hole was considered by Foo et al in Ref.~\cite{foo2022quantum}, where the interferometric measurement was performed upon superposed mass states for the black hole, and sharp peaks reminiscent of black hole quantization were found.
In our setup we find no such sharp peaks, providing further evidence that those peaks are a consequence of black hole mass quantization.
Moreover, we have provided analytical expression for the spectrum probed by an Unruh--DeWitt detector in the case of a mass- and position-superposed BTZ black hole. 
These expressions show that the sharp peaks originate from a coherent enhancement of the image sum at rational square-root mass ratios, and that at certain position ratios an additional pole singularity appears at rational square-root mass ratios.

\vspace{0.5cm}

Building on our results, and more broadly on the study of detectors in superposed spacetimes, several promising avenues for future research emerge. A natural direction for future work is to investigate the physical regimes in which our approximations break down. For instance, we have assumed that the two branches of the superposition are classically perfectly distinguishable, so that the corresponding black hole states can be treated as orthogonal. This assumption is expected to fail when the two superposed locations become arbitrarily close. A first step toward addressing this regime could be to apply linearized quantum gravity, along the lines of Refs.~\cite{chen2023quantum,chen2024quantum}. In addition, we might relax the assumption of gravity being sourced only by position. In the more general situation in which gravity is sourced by both position and momentum, a more refined treatment of semiclassical states is required, for example in terms of gravitational coherent states~\cite{carney_dressed_2018,ashtekar2015geometryphysicsnullinfinity,ashtekar_null_2018,Prabhu_2024,prabhu_infrared_2024,strominger2018lecturesinfraredstructuregravity,Thiemann_2001,klauder1985coherent,zhang1990coherent,Kiefer:2004xyv,hartle1983wave}.

Furthermore, another interesting possibility would be to investigate coherences in a detector with many energy levels near a superposed BTZ black hole, as was done for the Schwarzschild black hole in Ref.~\cite{paczos2023hawking}.
Moreover, we have investigated the first nontrivial perturbative order in the Unruh--DeWitt interaction, but a more complete understanding of this interaction requires the investigation of higher-order terms~\cite{bradler2018unitaryevolutionpairunruhdewitt,han2025reliablequantummasterequation} and nonperturbative treatments~\cite{Brown_2013,Polo_G_mez_2024}.

As the quantum BTZ black hole has an effective description as a two-dimensional conformal field theory (CFT) at infinity that can be coupled to classical matter~\cite{Carlip_2014,Carlip_2005,emparan1998quantization}, it would be interesting to investigate the results regarding transition amplitudes and probabilities of detectors in superposed BTZ black hole spacetimes in the related CFT.
One would expect similar effects as for the position- and mass-superposed BTZ black holes in other superposed spacetimes, such as superposed Schwarzschild black holes for example, and how the spectra of potential peaks in solvable 1+1d (black hole) spacetimes~\cite{mertens2023solvable} relate to CFTs on the boundary in AdS/CFT~\cite{maldacena1999large}. 
Our analytical results suggest that this could be achieved by probing the spectrum of interfered two-point functions between superposed spacetimes, and look for additional Hadamard-like singularities~\cite{KAY199149,wald1994quantum,Radzikowski:1996pa,Fewster_2013} in these interfered two-point functions that relate properties such as black hole mass with the quantized angular structure of the quantized field solutions.
In general, calculating Wightman functions in such superpositions might be challenging, but branch-dependent coordinate transformations, a form of a quantum reference frame transformation, or embeddings in higher-dimensional (Minkowski) spacetime may offer potential routes. 
An example of the latter is Schwarzschild in 6d Minkowski spacetime~\cite{fronsdal1959completion} in the GEMS (Global Embedding in Minkowski Spacetime) paradigm~\cite{russo2008accelerating,deser1997accelerated,deser1998equivalence,deser1999mapping,sheykin2019global,sheykin2021global}, even though it is not always clear when (in which cases) exactly this mapping works~\cite{Paston_2014}. 

\vspace{0.5cm}

Finally, this chapter combines aspects of quantum reference frames, quantum gravity and detector models, and thus works toward an operational understanding of quantum theories of gravity. 
Such operational approach allows to critically assess some common assumptions of black hole paradoxes~\cite{hawking1976breakdown,harlow2016jerusalem,Polchinski_2016,raju2022lessons}, especially when combining the viewpoints of different observers such as the cloning~\cite{susskind1994gedanken,hayden2007black} and firewall paradoxes~\cite{almheiri2013black,braunstein2007quantum,braunstein2013better,harlow2016jerusalem}, as mentioned in \Cref{sec:subtleties}.
Explicit calculations using quantum reference frames and detectors probing the radiation can further quantify these subtleties. 
For example, claims concerning the entanglement of subsystems can be probed operationally with entanglement harvesting~\cite{Stritzelberger_2021,Hotta_2020,Stritzelberger_2020,Cong_2020,Ng_2018,henderson2018harvesting,Gallock_Yoshimura_2021,Mendez_Avalos_2022,Tjoa_2020,robbins2020entanglementamplificationrotatingblack,chakraborty2024entanglementharvestingquantumsuperposed,dubey2025harvesting,Membrere_2023,Barman_2023,wang2025harvestinginformationhorizon}, including claims of early and late radiation entanglement, and which observers would detect high entanglement across the horizon and in the radiation from entangled black holes~\cite{Maldacena_2013,susskind2014ereprghzconsistencyquantum,Susskind_2016}.

\vspace{0.5cm}

\ifSubfilesClassLoaded{ 
  \printbibliography[heading=bibintoc, title=References]
  }{}
\end{document}

\ifSubfilesClassLoaded{ 
	\setcounter{chapter}{1} 
	\maketitle \tableofcontents \listoftodos
	}{}

\chapter[Conclusions and future work]{Conclusions and future work}\label{chap:conclusions&future-work}

\section{Conclusions}\label{sec:conclusions-conclusions}

In this thesis, we have discussed several paradoxes and setups that push quantum theory's applicability to its limits. For each part, we summarize our contributions and outline future directions.

\section{Part I: discussion and outlook}

In \Cref{part:universality_in_quantum_foundations}, we have analyzed paradoxes that sharpen the tension within the measurement problem. 
The central assumption was the Universality of Quantum Theory, under which measurements performed in sealed laboratories can be modeled unitarily.
In other words, we assumed quantum theory applies to all systems alike,  leading to the scenario of Wigner's friend and extensions thereof. 
Our main contributions of \Cref{part:universality_in_quantum_foundations} include a refined version of the Frauchiger--Renner paradox, connections between extended Wigner's friend scenarios and notions of nonclassicality, and an argument against the absoluteness of free choices. 
We briefly discuss the implications of these paradoxes by categorizing potential resolutions, after which we indicate directions for future work. 

\vspace{0.5cm}

\textit{Categorizing resolutions to extended Wigner's friend paradoxes.} Resolutions of extended Wigner's friend arguments fall into three broad categories.

\hspace{0.5cm} \textit{a. Refuting Universality of Quantum Theory.} 
A first category refutes the existence of superobservers, i.e., rejects the Universality of Quantum Theory. 
This would require further explanation of why quantum theory cannot be applied equally to all systems, and why a measurement cannot be modeled unitarily. 
We have not probed this possibility further. See also the paragraph \textit{Towards actual implementations of extended Wigner’s friend paradoxes} below.

\hspace{0.5cm} \textit{b. From Absoluteness to the Relativity of Observed Events.} A second category refutes the Absoluteness of Observed Events, meaning outcomes of performed measurements cannot be absolute. 
This requires a replacement principle.
As discussed in \Cref{chap:partI_refinedFR}, our refined Frauchiger--Renner paradox suggests a natural resolution, the Relativity of Observed Events, paraphrased à la Peres's dictum as `Unperformed measurements have no results, and unknown results have no values'. 
An appealing further direction is to explore whether quantum theory can be reformulated or reconstructed by taking the Relativity of Observed Events as a guiding principle, and connect with proposed relational reformulations~\cite{martin2023fact,rovelli1996relational,ormrod2024quantum,cavalcanti2021view}.
Our scenario combining Wigner's friend with the Pusey--Barrett--Rudolph theorem~\cite{pusey2012reality} of \Cref{chap:part1_EWF_PBR} moreover motivates to extend such relativity to the notion of free choices. 
As an extremely fundamental notion in physics, while at the same time deeply connected to foundational philosophical questions, rethinking the notion of free choice and `interventions' in this light may prove a valuable task (see, e.g., Ref.~\cite{di2021arrow} for work in this direction).

\hspace{0.5cm} \textit{c. Fine-tuning resolutions.} A third category for resolving extended Wigner's friend arguments is to hold on to the Absoluteness of Observed Events (AOE), at the cost of adopting a principle of fine-tuning, i.e., at the cost of a rather conspiratorial view of nature. 
In the case of the Local Friendliness no-go theorem, this involves refuting the assumption of Local Agency.
We have further elaborated on this fine-tuning underlying the Local Friendliness assumption of Local Agency in \Cref{sec:EWF_based_on_generalised_contextuality} by outlining a no-go theorem generalizing the Local Friendliness no-go theorem.
It would be interesting to connect in more detail to Ref.~\cite{yile2023}, who consider the Local Friendliness result in light of causal models and fine-tuning, as opposed to operational fine-tuning in our no-go result. 
On a technical level, the scenario employed in our result combined Wigner's friend with generalized noncontextuality, and still holds in background-independent theories, such as some theories of quantum gravity, as opposed to the Local Friendliness no-go theorem. 
Specifically, the Local Friendliness assumption of Local Agency can be seen as a generalization of no-signaling, and refuting it allows for some `unobserved' faster-than-light signaling. 
The fine-tuning then lies in the fact that, even though unobserved signaling would be allowed, this signaling can never be observed. 

In the Frauchiger--Renner paradox, and our refined version of it of \Cref{chap:partI_refinedFR}, the fine-tuning resolution lies in rejecting the possibility to formulate correct predictions using quantum theory for certain outcome values, even though those outcome values exist and are definite when assuming the Absoluteness of Observed Events.
To explain this somewhat more, consider the following setup.
Alice has performed a measurement, and Bob wants to formulate a prediction for her outcome value $a$. 
Assuming Absoluteness of Observed Events, Alice's outcome value exists and is absolute and single-valued, irrespective of anything else in the universe. 
Therefore, Alice's outcome $a$ exists, and is single-valued, for Bob as well. 
Suppose that Bob uses quantum theory, and finds that Alice should obtain $a=0$. 
The fine-tuned resolution then does not allow Bob to assert the prediction `I know that Alice obtained $a=0$'; in some cases, this prediction may be correct, but in other cases, such as when superobservers act as in the FR paradox, it may be wrong. 
How, then, should Bob formulate his prediction? One could argue that little room remains to do so, and one way would be to let Bob make the statement: `I know that I would find that Alice has $a=0$ if I ask for her outcome.' 
This, however, closely resembles the practical implementation of the Relativity of Observed Events, and one may question the practical benefit of assuming such hidden, fine-tuned additional layer of reality. 
Historically, simplicity is preferred, all else equal.

\vspace{0.5cm}

\textit{Nonclassicality and extended Wigner's friend arguments.} 
More technically, we have provided several arguments combining extended Wigner's friend paradoxes with notions of nonclassicality. 
In fact, also the Pusey--Barrett--Rudolph theorem can be seen as capturing a form of nonclassicality or quantum advantage in some tasks~\cite{ying2026contextualadvantageconclusiveexclusion}.
This motivates further exploration of such arguments, and to probe what is nonclassical about quantum theory assuming its universality, and about Wigner's friend paradoxes~\cite{jones_thinking_2024}. 
Moreover, it would be interesting to consider generalizations of the extended Wigner's friend scenarios based on fine-tunings that go beyond those linked to contextuality~\cite{catani2023mathematical,cavalcanti2018classical,Wood_2015,Pearl_2021}, such as violations of bounded ontological distinctness~\cite{Chaturvedi_2020,Khoshbin_2024}. 

\vspace{0.5cm}

\textit{On the Relativity of Observed Events and time symmetry.} Above, in category \textit{b. From Absoluteness to the Relativity of Observed Events}, we have argued that assuming the Universality of Quantum Theory and adhering to no fine-tuning, our refined FR paradox leads to the resolution of replacing the Absoluteness of Observed Events by the Relativity of Observed Events. 
A world governed by the Relativity of Observed Events rather than Absoluteness may be favored also on the basis of time symmetry for physical laws; under AOE, once an observer has performed a measurement, its outcome value is absolute and thus persists, even if its physical records are erased by a subsequent observer. The only way in which one could save time symmetry would then be by assuming it was pre-existing at all times, leading to even further requirements on this hidden layer of reality. 
A possibility of probing this argument in more detail would be to combine Wigner's friend with the `timelike Bell scenario' as was used to argue against no fine-tuned time-symmetric ontological models in Ref.~\cite{leifer_is_2017}; the scenario of \Cref{sec:EWF_based_on_generalised_contextuality} and Ref.~\cite{walleghem2025extendedwignersfriendnogo} may provide a starting point for this investigation. 
The view of the arrow of time as an emergent time asymmetry from time-symmetric physical laws~\cite{di2021arrow,reichenbach1991direction,zeh2007physical,hawking1985arrow,Chiribella2022IndefiniteTime,Parzygnat_2023} is in line with a practical world with abundance of decoherence and practical Absoluteness of Observed Events despite adhering to the Relativity of Observed Events on a fundamental level.

Another consideration regarding spacetime is the role of gravity and spacetime in the decoherence of quantum systems~\cite{gambini2007fundamental,Bassi_2017,gambini2019single,Anastopoulos_2013,Oppenheim_2023,Anastopoulos_2022,Petruzziello_2021,Aguiar_2025,Dukehart:2024bsn,bose2025massive,Pikovski_2015,Blencowe_2013,danielson2025horizonssoftquantuminformation,Danielson:2022tdw}---not necessarily invoking modifications to quantum theory or clashing with the equivalence principle as argued in Refs.~\cite{diosi1987universal,penrose1996gravity}, see, e.g., Ref.~\cite{bose2025massive,giacomini2022quantum}--- and, potentially, matter-gravity entanglement in the thermodynamic arrow of time~\cite{Kay_2023}. 
Although not necessarily a resolution to the measurement problem, these phenomena may contribute to the
practically observed decoherence of quantum systems and the emergence of classicality in our world, and the resulting apparent absence of superobservers~\cite{rivlin2025emergenceclassicalitywignersfriend}, see also the paragraph below.

\vspace{0.5cm}

\textit{On the reality of the wavefunction.} As discussed in \Cref{sec:nonclassicality}, studies of nonclassicality provide no-go results for a no fine-tuning interpretation of quantum theory within the ontological models framework of classical realism. 
Regarding the question whether the quantum wavefunction can be regarded as ontic or epistemic in this framework, the Pusey--Barrett--Rudolph theorem~\cite{pusey2012reality,leifer2014quantum} implies that it must be ontic, modulo some assumptions regarded as relatively weak. 
These nonclassicality results invite us to deepen our understanding of quantum theory beyond (classical) realism, with moreover extended Wigner's friend arguments suggesting a form of relationalism, even for free choices, as we have argued in \Cref{chap:part1_EWF_PBR} and category \textit{b.} above.
The Relativity of Observed Events is such a principle, and it would be interesting to probe what relational frameworks implementing this principle tell us about the nature of the wavefunction.

\vspace{0.5cm}

\textit{Towards actual implementations of extended Wigner's friend paradoxes and on the absence of superobservers.} 
Arguably, a simple qubit would not qualify as an observer, at least not in a naive way~\cite{brukner2021qubits}; a more complex system is required, while not losing its coherence to an uncontrollable environment, i.e., an environment the superobserver does not have (quantum) control over.  
Perhaps the closest to an observer playing the role of a friend in future experiments is a quantum artificial intelligence, i.e., an artificial intelligent program run on a quantum computer, passing some Turing tests if one insists.
We already have AIs; if we can build a sufficiently powerful quantum computer, implementing such a friend should not be a problem.
An experimental effort of realizing the Local Friendliness scenario with quantum systems serving as the friends, defining a `branch factor' to quantify the `observerness' of the system, can be found in Ref.~\cite{Zeng_2025}.

However, as discussed briefly in \Cref{sec:implications_WFBH}, we can ask ourselves, where are \textit{our} superobservers? Why can we trust our memories? If Wigner can manipulate the brain of his friend, Wigner could implant memories and experiences that would lead the friend to different physical laws, or even nonsensical results. Why are we not a Boltzmann brain~\cite{jones_thinking_2024}?
Does the concept of an observer require an emergent level of description that we cannot fully grasp yet?
Computational complexity, quantum error correction, chaos and random matrices have been associated with black holes, holography and the decoding of Hawking radiation~\cite{harlow_quantum_2013,aaronson_complexity_2016,kim_ghost_2020,Oppenheim_2014,susskind2013blackholecomplementarityharlowhayden,susskind2016computational,brown2019pythonslunchgeometricobstructions,harlow2016jerusalem,akers2022blackholeinteriornonisometric,Brown_2017,BAIGUERA20261,engelhardt2023algebraicereprcomplexitytransfer,Belin_2022,almheiri2018holographicquantumerrorcorrection,Verlinde_2013,Hayden_2019,Almheiri_2015,Pastawski_2015,polchinski2015chaosblackholesmatrix,Cotler_2017chaos}; it would be interesting to see whether these concepts play a role as well in the emergence of classicality and defining observers more rigorously.

The question of the practical absence of superobservers~\cite{rivlin2025emergenceclassicalitywignersfriend} is closely related to the emergence of kinematic and dynamical classicality, which can be approached through the decoherence approach~\cite{zurek1991decoherence,zurek2003decoherence,schlosshauer2007quantum,schlosshauer2005decoherence,Zurek:2009ymk,Brand_o_2015,Allahverdyan_2013} or, relatedly, the approach of coarse-grained observables~\cite{Kofler_2007,brukner2017quantum,bibak2025classicallimitquantummechanics,omnes1994interpretation,gell-mann1993classical,Lombardo_1996,Halliwell_1998}.
In fact, the concepts of (classical) chaos and random matrices have appeared in the decoherence and coarse-graining approaches to the emergence of classicality~\cite{zurek1994decoherencechaos,brukner2017quantum,Strasberg_2023}, and topics such as computational complexity, chaos, quantum error correction and scrambling in measurement-induced phase transitions~\cite{Choi_2020,Gullans_2020,Suzuki_2025,Skinner_2019,Li_2019}; it would be interesting to further probe connections between these areas.



\vspace{0.5cm}

\textit{Connecting to Gödel?} There are certainly analogies between Gödel's incompleteness theorem and the liar cycles that arise in extended Wigner's friend paradoxes (and quantum theory's Kochen--Specker contextuality they are based upon~\cite{abramsky2015contextuality}).
Prior work has explored this connection in some more detail~\cite{Dourdent_2021,Szangolies_2018}, but it remains to be seen whether this connection is more than a mere analogy~\cite{Franzen2005Godel}.   

\vspace{0.5cm}

\textit{What if quantum theory is not the final story?} Finally, it is, of course, possible that quantum theory is not the ultimate description of nature, a proposal being a `gravitized quantum theory', for example~\cite{Berglund_2022}; if this is the case and such theory has nontrivial implications for the definition of observers, perhaps contrary to current expectations, it might have nontrivial repercussions for Wigner's friend paradoxes.

\section{Part II: discussion and outlook}

In \Cref{part:universality_black_holes}, we have discussed some black hole paradoxes and considered a setup where quantum aspects of gravity are combined with quantum field theory.
In this part, we assumed that quantum theory also applies to gravity, another instance of assuming its universal validity. 
Our main contributions of \Cref{part:universality_black_holes} consist of combining Wigner's friend and black hole physics into a unified argument that strengthens existing black hole paradoxes, a critical analysis of standard assumptions therein, and a calculation of a detector's response near a superposed black hole.

\vspace{0.5cm}

In \Cref{chap:BH_prelims}, after briefly outlining some prominent black hole paradoxes, we have highlighted the many extra degrees of freedom in the form of soft hair that were first overlooked by Hawking when formulating his information loss argument for black holes---recognized by Hawking himself about fifty years later~\cite{hawking2016soft}.
Together with the assumption that quantum theory applies to gravity as well, a first argument by conservation laws suggests the black hole to become highly superposed, and provides a possibility for information recovery and information transfer to the radiation.

While providing a possibility towards resolving the black hole information puzzle, this does not yet resolve black hole paradoxes featuring infalling observers such as the cloning and firewall paradoxes~\cite{susskind1994gedanken,hayden2007black,almheiri2013black,braunstein2013better,braunstein2007quantum,mathur2009information,harlow2016jerusalem}, which was the topic of \Cref{chap:part2_WFBH}.
In this chapter, after formulating scenarios that strengthen these paradoxes by combining Wigner's friend and black hole physics into a unified argument, going beyond the analogy exposed by Hausmann and Renner~\cite{hausmann2025firewallparadoxwignersfriend}, we turned to a critical assessment of the underlying black hole assumptions.
These assumptions are inherited from the standard formulations of the black hole cloning and firewall paradoxes.
We have highlighted several important subtleties regarding these assumptions in \Cref{sec:implications_WFBH}, related to superposed spacetimes, gauge invariance and locality\footnote{By locality we mean here locality in the quantum theory itself, i.e., how to localize systems and events in quantum field theory and quantum gravity, and not the classical locality or classical local causality as in Bell's theorem.} concerning information recovery and observables for infalling observers. 
These nontrivialities are not problems of the formalism or setup used, but are crucial \textit{features} of any theory of quantum gravity.

Among the methods that can be employed for further probing these features and their implications for black hole paradoxes are quantum reference frames and detectors, or, more generally, a shift to focusing on operational questions, tied to the experience of specific observers.
We have presented a calculation combining quantum reference frames and a detector in \Cref{chap:part2_detector}, by analyzing the response of an Unruh--DeWitt detector near a position-superposed BTZ black hole, and identifying an `interference' term absent in classical gravity.
To obtain the interaction of a detector with the field on a superposed spacetime, we have used a quantum reference frame transformation to the case of a superposed detector near a black hole with definite position, which yields a non-superposed spacetime. 

The choice of a position-superposed 2+1d BTZ black hole, instead of other possible superposed spacetimes, was motivated by the study of a \textit{mass-superposed} black hole in Ref.~\cite{foo2022quantum}, where peaks in the detector response as a function of the ratio of the square root of the masses were found, argued to be reminiscent of black hole mass quantization.
The absence of peaks found in our response for a \textit{position-superposed} black hole further strengthens this claim. 
We have also provided an analytical expression for the spectrum probed by the detector, with peaks in the mass-superposed case corresponding to singularities that would give the Hadamard singularity structure in an otherwise classical two-point function.
Specifically, these singularities arise from an interplay of the black hole mass with a discrete image sum, which can be seen as arising from a `winding number' for functions of the angular coordinate on the 1-dimensional circle $S^1$ (as the BTZ black hole is 2+1-dimensional).
This connection arises in the BTZ case of course from the definition of the BTZ black hole as a quotient of AdS that defines its mass parameter. 
In the case of a Schwarzschild black hole, this points toward searching for singularities in superposed black hole two-point functions arising from an interplay between the black hole mass and spherical harmonic labels $l,m$ in terms leading otherwise to Hadamard singularities in non-superposed black hole two-point functions.

\vspace{0.5cm}

Whereas the calculation of \Cref{chap:part2_detector} exemplifies an operational setup of a detector and quantum reference frames, it marks only a first step in the unfolding story of this operational approach. 
Further development of these methods in physical setups concerning cosmology and black hole physics is required.
As argued above, and more generally in \Cref{part:universality_black_holes} of this thesis, its application may yield interesting insights into the physics of black holes and its paradoxes.

\vspace{0.5cm}














\ifSubfilesClassLoaded{ 
  \printbibliography[heading=bibintoc, title=References]
  }{}
\end{document}

\appendix

\ifSubfilesClassLoaded{ 
	\appendix \setcounter{chapter}{0} 
	\maketitle \tableofcontents \listoftodos
	}{}

\chapter[Extra black hole protocols of \Cref{chap:part2_WFBH}]{Extra black hole protocols of \Cref{chap:part2_WFBH}}\label{app:part2_WFBH}

\section{A Local Friendliness version of the first protocol} \label{sec:Appendix_LF}

The protocol 
is similar to our first protocol, but now Bob additionally makes a choice whether to reveal his outcome to Wigner or not before falling into the black hole, and Alice encodes her outcome $a$ in the computational $Z$-basis of a qubit $S_C$ which she sends to another observer Charlie before falling in the black hole. Charlie makes a choice $x=0,1$ whether to measure $S_C$ in the $Z$-basis (decoding Alice's outcome) for $x=0$ or in the $\ket{\pm}$ basis for $x=1$. 

\begin{theorem} \label{th:theorem_nosignalling}
   If a superobserver can decode the Hawking radiation and reconstruct the quantum state of an infalling observer, and still receive a short message from that observer sent inside the black hole and assumed to be unaltered by actions outside, an operational violation of no-signalling is found.
\end{theorem}

\begin{proof}[Proof of \Cref{th:theorem_nosignalling}.]
Consider the following protocol.
\begin{itemize}
    \item Alice and Bob, residing in sealed labs, share a system $S_A S_B$ in the entangled Hardy state of \cref{eq:hardy}, and each measure their qubit in the computational basis, modeled unitarily as $U_A,U_B$. 
    \item Alice encodes her outcome in the computational basis of a qubit $S_C$, which she sends to Charlie, and next Alice falls into the black hole.
    \item Charlie has a choice $x=0,1$ whether to measure the qubit in the computational $Z$-basis or in the $\ket{\pm}$ basis. 
    \item Somewhat later, Bob makes a choice $y=0,1$ to reveal his outcome to Wigner or not, and falls into the black hole. 
    \item Alice and Bob, once inside the black hole, try to send their outcomes to Ursula and Wigner (who will fall into the black hole somewhat later). 
    \item Before falling in, Ursula and Wigner retrieve the Hawking radiation containing the states of Alice's and Bob's labs, apply the `undoing operation' $U_A^\dagger,U_B^\dagger$, respectively, to the decoded lab states of Alice and Bob in the Hawking radiation and measure $S_A,S_B$, respectively, in the $\ket{\pm}$ basis. 
    \item Next, Ursula, Wigner and Charlie all fall into the black hole, aiming to still obtain the outcomes $a,b$ of Alice and Bob. 
    \end{itemize}

Regarding the spacetime arrangement (important as our argument will involve no-signalling), Alice falls in the black hole before Bob, such that Ursula can retrieve Alice's Hawking radiation outside of the future lightcone of Bob's choice $y$, and obtain $u,b$ outside of the future lightcone of $y$. Moreover, Charlie makes his measurement choice $x$ such that Wigner's measurement on Bob's Hawking radiation takes place outside of the future lightcone of $x$. The spacetime arrangement is sketched in \Cref{fig:LF_BH}.  

\begin{figure}[h]
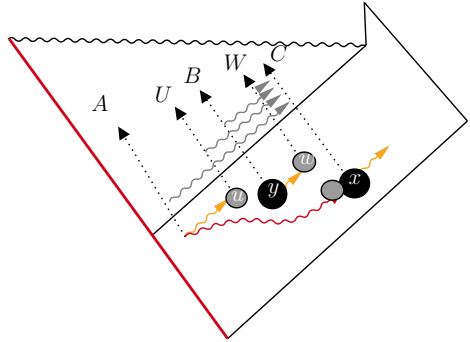

         \centering
\includestandalone[width=0.42\textwidth]{Figures/LF_BH}
         \caption[Combining Local Friendliness with black hole cloning]{The LF-black hole protocol in a Penrose diagram; $x,y$ denote choices by Charlie and Bob, respectively, and the smaller black dots denote operations by Alice, Ursula, Bob, Wigner and Charlie.}
         \label{fig:LF_BH}
\end{figure}

Alice and Bob measure $S_A,S_B$, which can be modeled as unitaries $U_A,U_B$, and, more specifically, as $CNOT$ gates if we assume their initial memory states are $\ket{0}_A,\ket{0}_B$, i.e. \begin{equation}
    U_A = CNOT_{S_A A} = \Pi^{S_A}_0 \otimes \mathbb{1}_A + \Pi^{S_A}_1 \otimes (|1\rangle\langle 0|+|0\rangle\langle 1|)_{A},  
\end{equation} with $\Pi_0 = \ket{0}\bra{0}$, and similarly for Bob's measurement and Alice's action of preparing her outcome in the qubit $S_C$ she sends to Charlie. Thus, the actions described thus far are \begin{equation}
    CNOT_{S_A S_C} CNOT_{S_A A} CNOT_{S_B B} \ket{\psi_0}_{S_A S_B} \ket{000}_{AB S_C}, 
\end{equation} where we are sloppy about the tensor product ordering in the initial state (we omit some tensor products $\otimes$ and swap operations to bring the initial state into the ordering of the operations that act on it, but this should be clear from the use of indices). 

Note that when Charlie obtains the outcome $c=+$ for $x=1$, we have that
\begin{equation} \label{eq:S_C+}
    \ket{+}\bra{+}_{S_C} CNOT_{S_A S_C} = \ket{+}\bra{+}_{S_C} \otimes \mathbb{1}_{S_A}.
\end{equation} Therefore, if Charlie obtains $c=+$, then Ursula's undoing of Alice's measurement will indeed reset her lab back to the initial state, the same that would be obtained if Alice did not prepare her outcome in the computational basis of the qubit $S_C$.

Assuming all involved outcomes $a,b,u,w,c$ can be brought together inside the black hole, the argument goes as follows.

First of all, as one can verify using quantum theory's predictions and \cref{eq:S_C+}, we have \begin{equation}
    p(u{=}-,w{=}-|x{=}1,y{=}1,c{=}+) > 0.
\end{equation} 

Next, quantum theory predicts that $$p(u{=}-,b{=}0|x{=}1,y{=}0,c{=}+)=0.$$ As all of $u,b,c$ are obtained spacelike separated from each other and not in the future lightcone of $y$, we can give a no-signalling argument that the joint outcomes $u,b,c$ should be independent of $y$. Namely, we could give Ursula and Bob extra choices $x_u,x_b \in \{0,1\}$ whether (for choice $1$) or not to encode their outcomes in the computational basis of extra qubits that they send to Charlie. Then by measuring these qubits in the computational basis, Charlie can obtain the correlations $u,b,c$ if Ursula and Bob choose $x_u=1=x_b$. However, these correlations should be independent of $x_u,x_b,y$ by no-signalling as these choices are all spacelike separated from where Charlie obtains these correlations $ubc$. Therefore, we obtain that $$ p(u{=}-,b{=}0|x{=}1,y{=}1,c{=}+)=0.$$

Using quantum theory and no-signalling we have that $$p(a{=}0,b{=}1|x{=}1,y{=}1)=p(a{=}0,b{=}1|x{=}0,y{=}0)=0,$$ and thus also $p(a{=}0,b{=}1|x{=}1,y{=}1,c{=}+)=0$.

Finally, $$p(a{=}1,w{=}-|x{=}1,y{=}1) = p(a{=}1,w{=}-|x{=}0,y{=}1)$$ by no-signalling, and quantum theory gives that \begin{equation*}
    \begin{split}
         p(a{=}1&,w{=}-|x{=}0,y{=}1) \\ &= p(a{=}1,c{=}a,w{=}-|x{=}0,y{=}1) = 0,  
    \end{split}
\end{equation*} meaning that  
$$p(a{=}1,w{=}-|x{=}1,y{=}1) = 0$$ and thus also $p(a{=}1,w{=}-|x{=}1,y{=}1,c{=}+) = 0$. 

Thus, we obtain that the distribution $p(a,b,u,w|x{=}1,y{=}1,c{=}+)$ should satisfy the conditions that \begin{equation}
    \begin{split}
            &p(u{=}-,w{=}-|x{=}1,y{=}1,c{=}+) > 0, \\
            &p(u{=}-,b{=}0|x{=}1,y{=}1,c{=}+)=0, \\
            &p(a{=}0,b{=}1|x{=}1,y{=}1,c{=}+)=0 \\
            &p(a{=}1,w{=}-|x{=}1,y{=}1,c{=}+) = 0.
    \end{split}
\end{equation} However, this cannot be true as then $u=-,w=-$ can occur, but $u=-$ implies that $b=1$ and $w=-$ implies that $a=0$, but $a=0,b=1$ cannot occur.   
\end{proof}

\textit{Comparison to the LF protocol.} 
Note that, compared to the LF protocol~\cite{bong2020strong}, which involves Alice, Bob, Ursula and Wigner, we have an additional party involving Charlie who gets a qubit from Alice. This is because we cannot simply let Ursula and Wigner perform their measurements spacelike separated from the choices $y,x$, respectively, and dive in to obtain $u,b$ jointly outside of the future lightcone of $y$ and $a,w$ jointly outside of the future lightcone of $x$. Namely, after Alice and Bob have performed their measurements, choose whether or not to reveal their measurements and fall in the black hole, Ursula and Wigner have to wait some time for the Hawking radiation containing the information about the lab states of Alice and Bob to come out of the black hole. Estimates of this timescale that Ursula and Wigner would have to wait for an old black hole are of the order of the scrambling time $r_s \log r_s$~\cite{hayden2007black} (with $r_s$ the Schwarzschild radius), and the furthest we could position Ursula and Wigner away from each other are at antipodal locations from the black hole. But then light rays sent at the locations where Alice and Bob make their choices $x$ and $y$ might reach Wigner or Ursula, respectively, before they can decode the Hawking radiation, so that signalling is possible. The trick with having this additional prepare-and-measure part is the same one as applied in \Cref{chap:part1_EWF_noncontextuality} to obtain EWF arguments based on contextuality rather than nonlocality.

\section{No-signalling firewall argument} \label{sec:strengthened_firewall}

Consider the following protocol, a slightly altered firewall paradox. 
Alice holds the early radiation, and distills a qubit (possibly precomputed using Oppenheim and Unruh's argument if its computational complexity would require so~\cite{Oppenheim_2014}), maximally entangled with the (late-radiation) mode $B$ that Bob holds near the horizon of an old black hole: $1/\sqrt{2}(\ket{00}+\ket{11})_{AB}$. The mode $B$ is also maximally entangled with Charlie's mode inside the black hole: $1/\sqrt{2}(\ket{01}-\ket{10})_{CB}$.
Bob chooses $y=0,1$ to measure his qubit $B$ in the $Z$ or $X$ basis. Alice always measures her qubit $A$ in the $Z$ basis. 
Charlie chooses $x=0,1$ to measure his qubit in a basis such that Charlie and Bob violate Bell nonlocality, for example by choosing $-(X+Z)/\sqrt{2}$- and $(-Z+X)/\sqrt{2}$-measurements, such that the empirical probabilities $p(c_x,b_y|x,y)$ violate Bell nonlocality~\cite{Bell_1976,brunner2014bell} and thus by Fine's theorem~\cite{abramsky2011sheaf,fine1982hidden,fine1982joint} a global distribution $p(c_0,c_1,b_0,b_1)$ reproducing the empirical correlations $p(c_x,b_y|x,y)$ cannot exist. 
Now, quantum theory predicts that Alice and Bob will always have the same outcome $a=b$ when $y=0$, as $\langle b, \lnot b |_{AB} (\ket{00}+\ket{11}) = 0$ for $b=0,1$. 
Therefore, we have that $p(a,b,c|x,y=0) = \delta_{a,b}p(b,c|x,y=0)$ and thus also \begin{equation} \label{eq:App_C_prob_cb_y0_is_prob_ca_y1}
\begin{split}
    p(c,b|x,y=0) &= p(c,b=a|x,y=0) \\ &= \delta_{a,b} p(c,a|x,y=0) \\ &= \delta_{a,b} p(c,a|x,y=1),     
\end{split}
\end{equation} where we used no-signalling to conclude that $p(c,a|x,y) = p(c,a|x)$; we can let Alice and Charlie perform their measurement before Bob's choice so that they obtain $a,c$ spacelike separated from Bob's choice $y$. 
This implies that the empirical correlations of $p(c,b|x,y=0)$ are those of $\delta_{a,b} p(c,a|x,y=1)$. Therefore, the joint correlations $p(c,b|x,y=0),p(c,b|x,y=1)$ are equal and 
should both be marginals from the global distribution $p(a,b,c|x,y=1)$, which is empirically accessible if Alice, Bob and Charlie can meet inside the black hole.
If such a distribution $p(a,b,c|x,y=1)$ exists, then by a gluing procedure~\cite{walleghem2024connecting,fine1982joint,haddara2024local} we can produce a global distribution reproducing all empirical correlations $p(c,a|x=1,y=1),p(c,a|x=0,y=1),p(c,b|x=1,y=1),p(c,b|x=0,y=1)$, which by \cref{eq:App_C_prob_cb_y0_is_prob_ca_y1} are the same empirical correlations as the empirical Bell scenario correlations $p(c,b|x,y)$ for $x,y \in \{0,1\}$ that violate Bell nonlocality with $c,a|x,y=1$ replaced by $c,b|x,y=0$.
However, as we explained a few lines ago, by Fine's theorem such global distribution cannot exist, and we find a contradiction. 

The gluing procedure unfolds as follows. Denoting $c|x$ by $c_x$, for example using $c_0$ for the outcome variable $c$ when $x=0$, the correlations $p(c_0,b,a|x=0,y=1)$ can be written as \begin{equation}
   \begin{split}
       p(c_0,&b,a|x{=}0,y{=}1) = p(a,b|x{=}0,y{=}1)p(c_0|x{=}0,y{=}1,b,a) \\ &= p(a,b|x{=}1,y{=}1)p(c_0|x{=}0,y{=}1,b,a),
   \end{split} 
\end{equation} using no-signalling for $p(a,b|x=0,y=1)=p(a,b|x=1,y=1)$, and noting that $p(c_0,b,a|x=0,y=1)$ is also empirically accessible if an observer as Charlie can obtain all of $c_0,b,a$ inside the black hole. 
We also have \begin{equation}
   \begin{split}
       p(c_0,b,a|x{=}0,y{=}1) = p(a,b|x{=}1,y{=}1)p(c_1|x{=}1,y{=}1,b,a),
   \end{split} 
\end{equation}
Therefore, the empirical correlations $p(c_0,a|x=0,y=1),p(c_0,b|x=0,y=1),p(c_1,a|x=1,y=1),p(c_1,b|x=1,y=1)$ can all be obtained from the ``glued'' global distribution \begin{equation}
    p(c_0,c_1,a,b) := p(c_1,a,b|x{=}1,y{=}1) p(c_0|a,b,x{=}0,y{=}1).
\end{equation}

We thus obtain the following result.

\begin{theorem}[No-signalling firewall argument]
    Assuming (Fi)-(Fiii) of \Cref{th:firewall}, an operational violation of no-signalling is found if Alice, Bob and Charlie can meet inside the black hole.
\end{theorem}

The meaning of no-signalling is determined by the semiclassical causal structure.
By an `operational' violation of no-signalling it is meant that this violation is experimental, witnessed by a single observer, i.e. that all relevant variables are jointly accessible (by Alice, Bob and Charlie if they can meet up inside the black hole). 
If Alice, Bob and Charlie cannot meet up inside the black hole, one could argue the violation is one of Local Agency~\cite{cavalcanti2021implications} rather than no-signalling.

\chapter{Appendix for \Cref{chap:part2_detector}}

\section{The BTZ black hole as a quotient of $AdS_3$} \label{sec:app_BTZ_AdS}
As we will later in this Appendix refer to the relation between $AdS_3$ coordinates and the BTZ black hole, we shortly review how the BTZ black hole is obtained as a quotient of $AdS_3$. We follow Ref.~\cite{foo2022quantum}, nearly verbatim.

Three dimensional anti-de Sitter $AdS_3$ can be obtained as the submanifold defined by \begin{equation}
    -l^2=-T_1^2-T_2^2+X_1^2+X_2^2
\end{equation} in the 2+2-dimensional flat space $\mathbb{R}^{2,2}$ with  metric \begin{equation}
\mathrm{d} s^2=-\mathrm{d} T_1^2-\mathrm{d} T_2^2+\mathrm{d} X_1^2+\mathrm{d} X_2^2.
\end{equation}
Taking the coordinates $\overline{t},\overline{r},\overline{\phi}$ on this submanifold, defined by \begin{equation}
    \begin{split}
T_1 & =l \sqrt{\frac{\overline{r}^2}{l^2}} \cosh \overline{\phi} \\
X_1 & =l \sqrt{\frac{\overline{r}^2}{l^2}} \sinh \overline{\phi} \\
T_2 & =l \sqrt{\frac{\overline{r}^2}{l^2}-1} \sinh \frac{\overline{t}}{l} \\
X_2 & =l \sqrt{\frac{\overline{r}^2}{l^2}-1} \cosh \frac{\overline{t}}{l}
\end{split}
\end{equation} with $l> 0$ and ranges $-\infty < \overline{t} < +\infty, l <\overline{r}< +\infty, -\infty < \overline{\phi} < + \infty$, one obtains the $AdS$--Rindler metric \begin{equation}
    \mathrm{d} s^2=-\left(\frac{\overline{r}^2}{l^2}-1\right) \mathrm{d} \overline{t}^2+\left(\frac{\overline{r}^2}{l^2}-1\right)^{-1} \mathrm{d} \overline{r}^2+\overline{r}^2 \mathrm{~d} \overline{\phi}^2.
\end{equation} Note that $\overline{\phi}$ is not an angular coordinate in the AdS--Rindler space. 
Defining new coordinates $r,t,\phi$ by  $r=\overline{r} \sqrt{M}, t=\overline{t}/\sqrt{M}, \phi = \overline{\phi}/\sqrt{M}$ for some $M > 0$, and periodically identifying $\phi$ by taking the quotient under $\Gamma: \phi \rightarrow \phi + 2\pi$, one obtains the BTZ metric \begin{equation}
    \mathrm{d} s^2=-\left(\frac{r^2}{l^2}-M \right) \mathrm{d} t^2+\left(\frac{r^2}{l^2}-M\right)^{-1} \mathrm{d} r^2+r^2 \mathrm{~d} \phi^2,
\end{equation} with now $\phi$ being an angular coordinate.

\section{BTZ correlation functions} \label{sec:app_W_BTZ}
The correlation functions in the BTZ black hole can be obtained from the correlation functions of $AdS_3$ using the method of images~\cite{smith1990gravitational,Carlip_1995}, as the BTZ solution is a quotient space thereof. We use the results of Refs.~\cite{Carlip_1995,Hodgkinson_2012,foo2022quantum}.
The correlation functions in the scalar vacuum of $AdS_3$ are given by  \begin{equation}
    \begin{split}
        W_{\mathrm{AdS}}\left(x, x^{\prime}\right)=\frac{1}{4 \pi l \sqrt{2}}\left[\frac{1}{\sqrt{\rho\left(x, x^{\prime}\right)}}-\frac{\zeta}{\sqrt{\rho\left(x, x^{\prime}\right)+2}}\right],
    \end{split}
\end{equation} with $\rho$ the squared geodesic distance in the 2+2 embedding space of $AdS_3$:
\begin{equation}
\begin{split}
    \rho\left(x, x^{\prime}\right)&=\frac{1}{2 l^2}\left[\left(X_1-X_1^{\prime}\right)^2-\left(T_1-T_1^{\prime}\right)^2+\left(X_2-X_2^{\prime}\right)^2-\left(T_2-T_2^{\prime}\right)^2\right].  
\end{split}
\end{equation} Furthermore, $\zeta \in\{-1,0,1\}$ specifies the boundary conditions at asymptotic infinity: Neumann, transparent or Dirichlet~\cite{carlip2003quantum,Lifschytz_1994,Shiraishi_1994}. We take transparent boundary conditions $\zeta=0$ for now, calculations for other boundary conditions are analogous. The effect of transparent boundary conditions is to ``recirculate the energy, angular momentum, etc lost to timelike infinity, resulting in a well-defined (but rather unusual) conservation law''~\cite{avis1978quantum}.
The BTZ correlation function is then given by \begin{equation}
\begin{split} \label{eq:BTZ_corr_1}
    W_{\mathrm{BTZ}}\left(x, x^{\prime}\right)&=\frac{1}{\sum_n \Upsilon^{2 n}} \sum_n \sum_m \Upsilon^n \Upsilon^m W_{\mathrm{AdS}}\left(\Gamma_M^n x, \Gamma_M^m x^{\prime}\right) \\ &= \sum_m \Upsilon^m W_{\mathrm{AdS}}\left(x, \Gamma_M^m x^{\prime}\right) ,    
\end{split}
\end{equation} with $\Gamma_M$ the identification action $\phi \rightarrow \phi+2 \pi  \sqrt{M}$ and $\Upsilon=+1,-1$ for untwisted and twisted fields; we will take $\Upsilon=1$.
The Wightman functions are actually distributions, and are given by taking the limit $\epsilon \rightarrow 0+$ of \cref{eq:BTZ_corr_1} with $\rho$ replaced by~\cite{foo2022quantum,preciado2024more}  
\begin{equation} \label{eq:BTZ_corr_ieps}
    \begin{aligned}
\rho_\epsilon\left({x}, \Gamma^n {x}^{\prime}\right) & =\frac{r r^{\prime}}{r_h^2} \cosh \left[\frac{r_h}{l}(\Delta \varphi-2 \pi n)\right]-1 
 -\frac{\sqrt{\left(r^2-r_h^2\right)\left(r^{\prime 2}-r_h^2\right)}}{r_h^2} \cosh \left(\frac{r_h}{l^2} \Delta t-\mathrm{i} \epsilon\right)
\end{aligned}
\end{equation} with $r_h=l\sqrt{M}$, with $\Delta \varphi, \Delta t$ the differences in angular and time coordinates of $x,x'$ whose radial components are denoted by $r,r'$ (with coordinates giving the metric in \cref{eq:BTZ_metric}). 

Using this expression for $\rho$ in the BTZ correlation function \cref{eq:BTZ_corr_1}, we obtain \begin{equation} \label{eq:BTZ_corr_app}
\begin{split}
    W_{BTZ}(x_1,x_2) &=  \frac{1}{4\pi l  \sqrt{2}}\sum_m \frac{1}{\scriptstyle \sqrt{ \frac{R_1 R_2}{r_h^2} \cosh \left[(\theta - 2 \pi m)\frac{r_h}{ l }\right]-1 -\sqrt{\left(\frac{R_1^2}{r_h^2}-1\right)\left(\frac{R_2^2}{r_h^2}-1\right)} \cosh \left(\frac{r_h}{ l^2} \Delta t-\mathrm{i} \epsilon\right) }} \\
    &=   \frac{1}{4\pi l \sqrt{2}} \frac{1}{\sqrt[\leftroot{-2}\uproot{2}4]{\left(\frac{R_1^2}{r_h^2}-1\right)\left(\frac{R_2^2}{r_h^2}-1\right)}} \sum_m \frac{1}{\sqrt{ \alpha_m^{(12)} - \cosh \left(\frac{r_h}{ l^2} \Delta t-\mathrm{i} \epsilon\right) }}
    \\ & =: W^{12}_{BTZ}(t-t'),
\end{split}
\end{equation} with \begin{equation} \label{eq:alpha_m}
    \alpha_m^{(12)} = \frac{ \frac{R_1 R_2}{r_h^2} \cosh \left[(\theta - 2 \pi m)\frac{r_h}{ l}\right]-1}{\sqrt{\left(\frac{R_1^2}{r_h^2}-1\right)\left(\frac{R_2^2}{r_h^2}-1\right)}}.
\end{equation} Recall that, unless stated otherwise, we are working under transparent boundary conditions, see \cref{eq:BTZ_correlation}; for Neumann and Dirichlet boundary conditions we will have an additional term, giving   \begin{equation} \label{eq:W_BTZ_app_NeuDiri}
\begin{split}
    W_{BTZ}(x_1,x_2) = &  \frac{1}{4\pi l \sqrt{2}} \frac{1}{\sqrt[\leftroot{-2}\uproot{2}4]{\left(\frac{R_1^2}{r_h^2}-1\right)\left(\frac{R_2^2}{r_h^2}-1\right)}} \sum_m \bigg( \frac{1}{\sqrt{ \alpha_m^{(12)} - \cosh \left(\frac{r_h}{ l^2} \Delta t-\mathrm{i} \epsilon\right) }} \\ &- \frac{\zeta}{\sqrt{ \alpha_m^{('12)} - \cosh \left(\frac{r_h}{ l^2} \Delta t-\mathrm{i} \epsilon\right) }}\bigg) 
\end{split}
\end{equation} with \begin{equation} \label{eq:alpha'_m}
    \alpha_m^{'(12)} = \frac{ \frac{R_1 R_2}{r_h^2} \cosh \left[(\theta - 2 \pi m)\frac{r_h}{ l}\right]+1}{\sqrt{\left(\frac{R_1^2}{r_h^2}-1\right)\left(\frac{R_2^2}{r_h^2}-1\right)}}.
\end{equation}

\section{Interaction picture calculation} \label{sec:app_interaction}
In this section we shortly sketch the calculation for the detector's transition amplitudes \cref{eq:single_detector_amplitude,eq:transition_amplitude_superp,eq:probabilities} in the interaction picture. The interaction picture (see for example \cite{birrell1984quantum,peskin2018introduction,tong2009university,Weinberg_1995}) is an intermediate picture between the Schrödinger representation, where states carry all time dependence but fields are time independent, and the Heisenberg representation, where operators carry all time dependence but states are time independent. 

More specifically, let the total Hamiltonian of the system be \begin{equation}
    \hat H = \hat H_{0,S}+ \hat H_\text{int}
\end{equation} where $\hat H_\text{int}$ is the interaction Hamiltonian between the detector and quantum fields in the Schrödinger picture; $\hat H_{0,S}$ is the free Hamiltonian in the Schrödinger picture.
In the interaction picture, operators evolve according to the free time evolution (without interaction, given by Hamiltonian $\hat H_{0,S}$) and states evolve according to the freely-time-evolved interaction Hamiltonian $\leftindex_I {\hat H}_\text{int}(t) = e^{i \hat H_{0,S} t} \hat H_\text{int} e^{-i\hat H_{0,S} t}$. Here the pre-subscript $I$ denotes the subscript for `interaction picture'. For simplicity of notation we will denote $\leftindex_I {\hat H}_\text{int}(t)$ simply by $\hat H_\text{int}(t)$ from now on as we will perform our calculations only in the interaction picture. 
States in the Schrödinger and interaction pictures are related by $\ket{\psi(t)}_I = e^{i\hat H_{0,S} t} \ket{\psi(t)}_S$.
Thus, in the interaction picture operators solve the equations of motion with respect to $\hat H_0$, and states evolve according to \begin{equation}
    i \frac{d}{dt} |\psi(t) \rangle_I =  {\hat H}_\text{int}(t) \ket{\psi(t)}_I
\end{equation} for time slicing $t$ defining the states $\ket{\psi(t)}_I$. Defining a unitary operator $U(t,t_0)$ such that \begin{equation} \label{eq:interaction_U}
    |\psi(t) \rangle_I = \hat U(t,t_0) \ket{\psi(t_0)}_I,
\end{equation} this unitary must satisfy 
\begin{equation}
     {\hat H}_\text{int}(t) \hat U(t,t_0) = i \frac{d}{dt} \hat U(t,t_0)
\end{equation} with initial condition $\hat U(t_0,t_0) = 1$. Simply exponentiating does not work because ${\hat H}_\text{int}$ is an operator and thus there is an ordering ambiguity. A formal solution of this equation can be written as the Dyson series \begin{equation}
    U(t,t_0) = T e^{-i  \int_{t_0}^t \hat H_\text{int}(t') dt'} 
\end{equation} with $T$ time ordering, which to second order in a perturbation expansion in the interaction is given by \begin{equation}
    \begin{split}
        {\hat U}(t_f,t_i)= \mathbb{1}-i  \int_{t_i}^{t_f} \mathrm{~d} t \hat H_{\text{int}}(t)- \int_{t_i}^{t_f} \mathrm{~d} t \int_{t_i}^t \mathrm{d} t^\prime \hat H_{\text{int}}(t) \hat H_{\text{int}}\left(t^{\prime}\right)+\mathcal{O}\left(\lambda^3\right)
    \end{split}
\end{equation} 
One can also think of the Dyson series as iteratively using the integral equation \begin{equation}
    \hat U\left(t_f, t_i\right)=\mathbb{1}-i \int_{t_i}^{t_f} d t' \hat H_\text{int}(t') \hat U\left(t', t_i\right).
\end{equation}

In our case, the free Hamiltonian governs the evolution of the (free) field $\phi$ and detector monopole moment $\hat \mu$, and the interaction Hamiltonian (with respect to proper time $\tau$) consists of a coupling between the detector's monopole moment and the field along the detector's path. We take the detector's path to be a classical variable here, and not a quantum operator. Below we will consider a detector in a background where a black hole is in a superposed location in a frame where the detector is at a single fixed location. If we were to move to the frame where the black hole is in a single location and the detector is in a superposed position, we would have to make the detector location an operator here, see also \Cref{sec:new_reference_fields}. Therefore, the interaction Hamiltonian is given by  \begin{equation}
    \hat H^\tau_{int} = \lambda \hat \mu \otimes \hat \phi(\mathbf{x})
\end{equation} in the Schrödinger picture, with $\mathbf{x}$ the position of the detector,
so in the interaction picture we have \begin{equation}
    \hat H^\tau_{int}(\tau) = \lambda \hat \mu(\tau) \otimes \hat \phi(x(\tau)).
\end{equation}
Let the total initial state be $\ket{\psi(\tau_i)}_S =|0\rangle_\phi |E_0\rangle_\mu$ of the scalar field vacuum and detector ground state (the detector's internal degree of freedom), and denote the final total state by $\ket{\psi(\tau_f)}_S$ in the Schrödinger picture. We take the detector to be held at a fixed spatial location rather than freely falling, as explained in \Cref{sec:setup_scenario,sec:new_reference_fields}. These states carry full time evolution dependence, being states in the Schrödinger picture. They are related to the states in the interaction picture by 
\begin{equation} \label{eq:interaction_pic_states1}
    \ket{\psi(\tau_i)}_I = e^{i \hat H^\tau_{0,S} \tau_i} \ket{\psi(\tau_i)}_S,
\end{equation} so that using \cref{eq:interaction_U} we have \begin{equation} \label{eq:interaction_pic_states2}
    \ket{\psi(\tau_f)}_S = e^{-i\hat H^\tau_{0,S} \tau_f} \hat U(\tau_f,\tau_i) e^{i\hat H^\tau_{0,S} \tau_i} \ket{\psi(\tau_i)}_S.
\end{equation} To first order in $\hat H_{\text{int}}$ we then have
\begin{equation} \label{eq:interaction_pic_states3}
    \begin{split}
    \left|\psi\left(\tau_f\right)\right\rangle_S=e^{-i \hat H^\tau_{0,S} \tau_f} e^{i \hat H^\tau_{0,S} \tau_i}\left|\psi\left(\tau_i\right)\right\rangle_S-e^{-i \hat H^\tau_{0, S} \tau_f} i \int_{\tau_i}^{\tau_f} \mathrm{~d} \tau \hat H_{\mathrm{int}}(\tau) e^{i \hat H^\tau_{0,S} \tau_i}\left|\psi\left(\tau_i\right)\right\rangle_S.
    \end{split}
\end{equation} 
Then, writing out $H_\text{int}$ and post-selecting on the detector making a transition to energy level $E$ we obtain at time $\tau_f$ 
\begin{equation} 
    \begin{split}
        i \lambda \langle E | \int_{\tau_i}^{\tau_f} \mathrm{d}  \tau \mu(\tau) \hat \phi[x(\tau)] \left|0, E_0\right\rangle,
    \end{split}
\end{equation} where coordinate times $t_i,t_f$ are related to the detector's proper time by $t_i = \tau_i/\gamma_J, t_f = \tau_f/\gamma_J$ with redshift factor $\gamma_J$ given in \cref{eq:tau_to_t}. As time evolution for $\hat \mu(\tau)$ is given by $$\hat \mu(\tau) = e^{i\hat H_{0,\mu} \tau} \hat \mu e^{-i\hat H_{0,\mu}\tau}$$ with $\hat H_{0,\mu}|E_i\rangle = E_i \ket{E_i}$ we obtain for the transition amplitude at time $t_f$ \begin{equation} \label{eq:single_UDW_amplitude}
    \begin{split}
        i \lambda   \langle E| \mu(0)  \left|E_0\right\rangle \int_{\tau_i}^{\tau_f} \mathrm{d} \tau \mathrm{e}^{\mathrm{i}\left(E-E_0\right) \tau}\langle\psi| \hat \phi(x(\tau))\left|0\right\rangle .
    \end{split}
\end{equation}
In our limited treatment, we do not model the dynamics of the black hole and control variable or apparatus keeping the detector on its trajectory. As a result, we do not acquire a branch-dependent phase associated with these degrees of freedom.
If we have a single path or geometry, an overall phase $\Phi$ is not important, but when we have a superposition of amplitudes a relative phase $\Delta \Phi$ might contribute.
The relative phase $\Delta \Phi$ in our case may have several potential origins or contributions in a more complete treatment. First, it may arise from the approximate evolution of the black hole as an approximate eigenstate of the gravitational Hamiltonian with its mass as eigenvalue with respect to coordinate time $t$ if we were to include the corresponding Hilbert space and state in our initial-state description; the gravitational Hamiltonian would then become a part of $\hat{H}^\tau_{0,S}$ in \cref{eq:interaction_pic_states1,eq:interaction_pic_states2,eq:interaction_pic_states3}, taking into account that $\hat{H}_{\text {grav }}^\tau=\frac{d t}{d \tau} \hat{H}_{\text {grav }}^t$. A relative phase difference when considering proper-time evolution might then appear as coordinate time relates differently to the detector's proper time in the two branches. 
The precise assumption in that case would thus be that for the purpose of the experiment the BTZ black hole spacetime remains approximately static and is an approximate eigenstate of the gravitational Hamiltonian, so that the phase coming from the black hole evolution is determined by its eigenvalue, presumably proportional to the black hole's ADM mass $M$ (in line with \cite{foo2022quantum})
i.e. $H^t_{0} \ket{M}_{BH} = M \ket{M}_{BH}$. 
Such phase might thus also be particularly relevant in the case of a mass-superposed black hole where a relative phase between the two branches of the superposition can also develop as the mass differs between the two branches. 
As we consider our black hole to have the same mass in both branches of the superposition, we do not assign an explicit Hilbert space for the mass to the black hole, only for its position. 
In a more complete treatment, the spacetime location and energy-momentum of the black hole may not be independent~\cite{louko_hamiltonian_1998,Eyheralde_2017}. 
Approximating the black hole as developing only a phase independent of spatial location might also be an issue because of noncovariance, which would require a dynamical analysis taking into account the gravitational constraints and backreaction, potential non-factorization, etc. We neglect these complications here.
Secondly, we also neglect any additional relative phase associated with the external control degree of freedom and the apparatus responsible for keeping the detector on the prescribed static trajectory.
If one were to take into account these phases, we could also simply absorb them in the definition of the states.
Whether an additional phase would be visible in the experimental data depends on the precise implementation of the interferometric black hole position measurement which we describe below.

\vspace{0.5cm}

Next we consider a detector in a superposition of classical geometries with black hole at positions $\mathbf{x}_1,\mathbf{x}_2$ with total initial state in the detector frame given by $\ket{\psi(\tau_i)}_S =  \ket{\mathbf{0}}_D \left(\left|\mathbf{x}_1\right\rangle_B +\left|\mathbf{x}_2\right\rangle_B  \right) \ket{0}_\phi |E_0\rangle_\mu / \sqrt{2}$ with $\ket{0}_\phi$ denoting the scalar field vacuum. 
In the case of a superposition of geometries (or detector location under a QRF transformation) one needs to be careful regarding coordinates, especially the choice of time parametrization used. 
In the detector QRF the proper time $\tau$ is single-valued and well-defined, but the black hole is in a superposition of locations so that the relation between proper time and coordinate times differs, whereas in the QRF with the black hole at the origin the coordinate time $t$ is well-defined but the proper time of the detector is indefinite. The relation between coordinate time $t$ and proper time $\tau$ is given by the time dilation of \cref{eq:tau_to_t}, $\tau = t \gamma(R)$ with $R$ the radial coordinate distance between the detector and black hole.

In the detector frame (i.e. putting the detector at the origin), the black hole is in a superposition of locations. Taking the interaction picture representation of the QRF-transformed UDW interaction \cref{eq:H_int_QRF_transformed} and post-selecting on the detector making a transition to energy level $E$ we obtain at time $\tau_f$ 
\begin{equation}  \label{eq:app_superp_detector_amplitude}
    \begin{split}
     \frac{1}{\sqrt{2}}\bigg( |\mathbf{x}_{1} \rangle_B  \Theta_1  + |\mathbf{x}_{2}\rangle_B  \Theta_2  \bigg),
    \end{split}
\end{equation} where \begin{equation}
    \Theta_J =   i \lambda \langle E |\hat \mu |E_0\rangle\int_{\tau_{i}}^{\tau_{f}} \mathrm{d}\tau  \eta(\tau) e^{i(E-E_0)\tau} \langle \vartheta| \hat \phi_J(x_J(\tau)) |0\rangle 
\end{equation} with $\phi_J$ the scalar field coupled to the metric $g_J$ with black hole at position $\mathbf{x}_J$ and $\ket{\vartheta}$ the final state of the scalar field which we will sum over. Note that if another clock were used to turn the detectors on and off, the proper time intervals for the UDW interaction could differ between the two branches.
We have omitted the detector's position state here as we are in the frame where this position is factorised as being the origin. 
As explained below \cref{eq:single_UDW_amplitude}, no additional branch-dependent relative phase is generated. Given the time-localized switching function, this also means that we could as well take $\tau_f, \tau_i \rightarrow \pm \infty$. 
If we now consider a joint measurement of the detector transitioning to the excited state $\ket{E}_\mu$ at time $\tau_f$ and an interferometric measurement on the black hole position, i.e. measuring in the basis $(\ket{\mathbf{x}_{1}}\pm \ket{\mathbf{x}_{2}})_B/\sqrt{2}$, we obtain the post-selected detector response given in \cref{eq:probabilities}.

\section{Calculating integrals for the measurement probabilities} \label{sec:app_integrals}

\subsection{Calculating the cross-term $\mathcal{F}_{12}(\Omega)$} 

We now insert the BTZ correlation function \cref{eq:BTZ_corr_app} (for transparent boundary conditions, calculations for other boundary conditions are analogous) into the cross-term for the probability \cref{eq:interference_term}, defining a response function $\mathcal{F}_{12} = P_{12}/\mathcal{N'}$ with $\mathcal{N'}=\lambda^2 |\langle E |\mu |E_0\rangle|^2$, obtaining\footnote{The calculations here resemble those of the Supplemental Material of Ref.~\cite{foo2022quantum} closely, but are slightly different as we consider a position-superposed BTZ black hole, as opposed to the mass-superposed case of Ref.~\cite{foo2022quantum}.} \begin{equation} \label{eq:cross_term_app_1}
    \begin{split}
         \mathcal{F}_{12}(\Omega) &= \int_{\tau_i}^{\tau_f} d\tau \int_{\tau_i}^{\tau_f} d\tau'  \eta(\tau) \eta(\tau') e^{-i\Omega(\tau-\tau')} W_{BTZ}(x_1(\tau),x_2(\tau')) \\
    \end{split}
\end{equation} where the paths $x_1(\tau)=(\tau/\gamma_1,R_1,0),x_2(\tau')=(\tau'/\gamma_2,R_2,\theta)$ with $\gamma_i=\gamma(R_i)$ and $\Omega = \Delta E= E-E_0$.
Now we move to coordinate time, obtaining
\begin{equation} \label{eq:cross_term_app_2}
    \begin{split}
         \mathcal{F}_{12}(\Omega) &= \gamma_1 \gamma_2 \int_{\tau_i/\gamma_1}^{\tau_f/\gamma_1} dt \int_{\tau_i/\gamma_2}^{\tau_f/\gamma_2} dt'  \eta(\gamma_1 t) \eta(\gamma_2 t') e^{-i\Omega(\gamma_1 t-\gamma_2 t')} W_{BTZ}(x_1(\tau),x_2(\tau')) \\
         &= \gamma_1 \gamma_2 \int_{-\tau_f/\gamma_1}^{\tau_f/\gamma_1} du \int_{-\tau_f/\gamma_2+u}^{\tau_f/\gamma_2+u} ds  \eta(\gamma_1 u) \eta(\gamma_2( u - s)) e^{-i\Omega \big(\gamma_1 u - \gamma_2(u- s)\big)} W_{BTZ}^{12}(s)
    \end{split}
\end{equation} where we have chosen $\tau_i=-\tau_f$ for simplicity, and changed variables $u=t$, $s = t-t'$ and $W^{12}_{BTZ}(t-t')$ is defined in \cref{eq:BTZ_corr_app}.
Next, we make the same approximation as in Appendix D (S31) of \cite{foo2022quantum}:
\begin{equation} \label{eq:approx_foo}
\begin{split}
        \int_{-t_{f1}}^{t_{f1}}& \mathrm{~d} u \int_{u-t_{f2}}^{u+t_{f2}} \mathrm{~d} s \eta\left(\gamma_D u\right) \eta\left(\gamma_D(u-s)\right) \ldots \\
        &\simeq \int_{-t_{f1}}^{t_{f1}} \mathrm{~d} u \int_{-t_{f2}}^{t_{f2}} \mathrm{~d} s \eta\left(\gamma_D u\right) \eta\left(\gamma_D(u-s)\right) \ldots,
\end{split}
\end{equation} with $t_{fJ}=\tau_f/\gamma_J$. 
This approximation would be exact if the integration limits were $-\infty$ to $\infty$; recall that we required that $\Delta \tau \gg \sigma$ such that the switching functions have small support outside $[-\tau_f/\gamma_J,\tau_f/\gamma_J]$.
Using \cref{eq:approx_foo} in \cref{eq:cross_term_app_2}, we find \begin{equation} \label{eq:cross_term_app_3}
    \begin{split}
        \mathcal{F}_{12}(\Omega) &\approx \gamma_1 \gamma_2 \int_{-\tau_f/\gamma_1}^{\tau_f/\gamma_1} du \int_{-\tau_f/\gamma_2}^{\tau_f/\gamma_2} ds  \eta(\gamma_1 u) \eta(\gamma_2( u - s)) e^{-i\Omega \big(\gamma_1 u - \gamma_2(u- s)\big)} W_{BTZ}^{12}(s) \\
        &= \gamma_1 \gamma_2  \int_{-\tau_f/\gamma_2}^{\tau_f/\gamma_2} ds I(s) e^{-\frac{\gamma_2^2 s^2}{2\sigma^2}}  e^{-i\Omega \gamma_2 s} W_{BTZ}^{12}(s)
    \end{split}
\end{equation} where $I(s)$ is an integral over $u$: \begin{equation} \label{eq:u_integration}
    \begin{split}
        I(s) &= \int_{-\tau_f/\gamma_1}^{\tau_f/\gamma_1} du  e^{-\frac{\gamma_1^2 u^2}{2\sigma^2}} e^{-\frac{\gamma_2^2 u^2 - 2 \gamma_2^2 u s}{2\sigma^2}} e^{-i\Omega(\gamma_1-\gamma_2)u}  \\
        &=  \int_{-\tau_f/\gamma_1}^{\tau_f/\gamma_1} du e^{-\frac{\big( \sqrt{\gamma_1^2 + \gamma_2^2} u - \frac{\gamma_2^2}{\sqrt{\gamma_1^2+\gamma_2^2}}s + i \Omega \frac{(\gamma_1-\gamma_2)}{\sqrt{\gamma_1^2+\gamma_2^2}} \sigma^2  \big)^2}{2\sigma^2}} e^{\frac{\gamma_2^4 s^2}{2\sigma^2(\gamma_1^2+\gamma_2^2)}} 
        e^{-\Omega^2 \sigma^2 \frac{(\gamma_1 - \gamma_2)^2}{2(\gamma_1^2+\gamma_2^2)}} e^{i\Omega \frac{\gamma_2(\gamma_1-\gamma_2)s}{\gamma_1^2+\gamma_2^2}}      \\
        &= \frac{\sqrt{2}\sigma}{\sqrt{\gamma_1^2+\gamma_2^2}} e^{\frac{\gamma_2^4 s^2}{2\sigma^2(\gamma_1^2+\gamma_2^2)}} e^{-\Omega^2 \sigma^2 \frac{(\gamma_1 - \gamma_2)^2}{2(\gamma_1^2+\gamma_2^2)}} e^{-i\Omega \frac{\gamma_2^2(\gamma_1-\gamma_2)s}{\gamma_1^2+\gamma_2^2}} \\ 
        & \int^{\frac{\sqrt{\gamma_1^2+\gamma_2^2}\tau_f}{\sqrt{2}\sigma \gamma_1}-\frac{\gamma_2^2 s}{\sqrt{\gamma_1^2+\gamma_2^2}\sqrt{2}\sigma}+i\Omega\sigma \frac{\gamma_1-\gamma_2}{\sqrt{2}\sqrt{\gamma_1^2+\gamma_2^2}}}_{-\frac{\sqrt{\gamma_1^2+\gamma_2^2}\tau_f}{\sqrt{2}\sigma \gamma_1}-\frac{\gamma_2^2 s}{\sqrt{\gamma_1^2+\gamma_2^2}\sqrt{2}\sigma}+i\Omega\sigma \frac{\gamma_1-\gamma_2}{\sqrt{2}\sqrt{\gamma_1^2+\gamma_2^2}}} dz e^{-z^2}           \\
        &=: \frac{\sqrt{2}\sigma}{\sqrt{\gamma_1^2+\gamma_2^2}} e^{\frac{\gamma_2^4 s^2}{2\sigma^2(\gamma_1^2+\gamma_2^2)}} e^{-\Omega\sigma^2 \frac{(\gamma_1 - \gamma_2)^2}{2(\gamma_1^2+\gamma_2^2)}} e^{-i\Omega \frac{\gamma_2^2(\gamma_1-\gamma_2)s}{\gamma_1^2+\gamma_2^2}}  \frac{\sqrt{\pi}}{2} H(s).
    \end{split}
\end{equation} Here we changed variables $z= u\frac{\sqrt{\gamma_1^2+\gamma_2^2}}{\sqrt{2}\sigma}+...$ and $H(s)$ is given by \begin{equation} \label{eq:app_H(s)}
    \begin{split}
        H(s) = &\text{erf}\bigg( \frac{(\gamma_1^2+\gamma_2^2)\tau_f - \gamma_1 \gamma_2^2 s + i\Omega \sigma^2 \gamma_1 (\gamma_1-\gamma_2) }{\sqrt{2}\sigma  \gamma_1 \sqrt{\gamma_1^2+\gamma_2^2}}\bigg)  \\ &+  \text{erf}\bigg( \frac{(\gamma_1^2+\gamma_2^2)\tau_f + \gamma_1 \gamma_2^2 s - i\Omega \sigma^2 \gamma_1 (\gamma_1-\gamma_2) }{\sqrt{2}\sigma  \gamma_1 \sqrt{\gamma_1^2+\gamma_2^2}}\bigg)
    \end{split}
\end{equation} with $\text{erf}$ the Gaussian error function.
Using \cref{eq:u_integration} in \cref{eq:cross_term_app_3} we find \begin{equation} \label{eq:cross_term_app_3.2}
    \begin{split}
         \mathcal{F}_{12}(\Omega) &\approx \frac{1}{4\pi l \sqrt{2}} \frac{\sqrt{M}}{\sqrt{\gamma_1 \gamma_2}}  \frac{\gamma_1 \gamma_2 \sigma}{\sqrt{\gamma_1^2+\gamma_2^2}} \sqrt{\frac{\pi}{2}} e^{-\Omega^2 \sigma^2\frac{(\gamma_1-\gamma_2)^2}{2(\gamma_1^2+\gamma_2^2)}} \\ &\phantom{ABCDDEF{}} \sum_n \int^{\tau_f/\gamma_2}_{-\tau_f/\gamma_2} ds \frac{ e^{-\frac{\gamma_1^2\gamma_2^2}{2\sigma^2(\gamma_1^2+\gamma_2^2)} s^2} e^{-iEs \frac{\gamma_1 \gamma_2 (\gamma_1+\gamma_2)}{\gamma_1^2+\gamma_2^2}}  H(s)}{\sqrt{\alpha_n^{(12)} - \cosh(\frac{r_h}{l^2} s - i\epsilon)}}
         \\ 
         &= \sigma \frac{ \sqrt{M}}{2l} Y_0 \sum_n \int^{\tau_f/\gamma_2}_{-\tau_f/\gamma_2} ds \frac{Z_0(s)H(s)}{\sqrt{\alpha_n^{(12)}-\cosh(\frac{r_h}{l^2} s - i\epsilon)}}
          \\ 
        &= \sigma {Y_0} \sum_n \text{ Re}  \int_{0}^{\frac{\tau_f \sqrt{M}}{ l\gamma_2}} dz \frac{Z_0( l z/\sqrt{M})H( l z/\sqrt{M})}{\sqrt{\alpha_n^{(12)}-\cosh( z - i\epsilon)}} 
    \end{split}
\end{equation} with $H(s)$ given in \cref{eq:app_H(s)}, $\alpha^{12}_n$ given in \cref{eq:alpha_m} and $Y_0$ and $Z_0(s)$ given by \begin{align} 
 Y_0 &=     \frac{\sqrt{\gamma_1 \gamma_2} \sqrt{\pi}}{4\pi \sqrt{\gamma_1^2+\gamma_2^2}} e^{-\Omega^2 \sigma^2\frac{(\gamma_1-\gamma_2)^2}{2(\gamma_1^2+\gamma_2^2)}}, \label{eq:Y_0_app} \\
 Z_0(s) &=  e^{-\frac{\gamma_1^2\gamma_2^2}{2\sigma^2(\gamma_1^2+\gamma_2^2)} s^2} e^{-i\Omega s \frac{\gamma_1 \gamma_2 (\gamma_1+\gamma_2)}{\gamma_1^2+\gamma_2^2}}. \label{eq:Z_0_app} 
\end{align} 
The cross-term will typically also contain singularities, but integrable ones.

For Neumann and Dirichlet boundary conditions, we have an extra term, namely \begin{equation} \label{eq:cross-term_NeuDiri}
    \begin{split}
          \mathcal{F}_{12}(\Omega) = \sigma Y_0 \sum_n & \text{ Re}  \int^{\frac{\tau_f \sqrt{M}}{ l\gamma_2}}_0 dz Z_0( l z/\sqrt{M})H( l z/\sqrt{M}) \bigg( \frac{1}{\sqrt{\alpha_n^{(12)}-\cosh( z - i\epsilon)}} \\ &\pm \frac{1}{\sqrt{\alpha'^{12}_n-\cosh( z - i\epsilon)}} \bigg)
    \end{split}
\end{equation} with $+,-$ for Neumann and Dirichlet boundary conditions, respectively.

\vspace{0.5cm}

\subsection{Calculating $\mathcal{F}_J(E)$} 

The integrals for the diagonal terms $\mathcal{F}_J(E)$ are exactly the ones of \cite{foo2022quantum}, we shortly repeat those here as slightly different normalizations were used; they are analogous to calculation of $\mathcal{F}_{12}(\Omega)$ just presented. We omit the label $J=1,2$ here as we are now in the case of a single non-superposed black hole and localized detector.

The response function is given by \begin{equation} \label{eq:appendix_response_single}
     \mathcal{F}(\Omega) = \int_{\tau_i}^{\tau_f} d\tau \int_{\tau_i}^{\tau_f} d\tau' \eta(\tau) \eta(\tau') e^{-i\Omega(\tau-\tau')}  W_{BTZ}(x(\tau), x(\tau'))  . 
\end{equation}
We first move to coordinate time, obtaining 
\begin{equation}
    \begin{split}
        \mathcal{F}(\Omega) =& \gamma^2 \int_{\tau_i/\gamma}^{\tau_f/\gamma} dt \int_{\tau_i/\gamma}^{\tau_f/\gamma}  dt' \eta(\gamma t) \eta(\gamma t') e^{-i\Omega\gamma(t-t')} W_{BTZ}(x(\gamma t), x(\gamma t'))  \\ &=  \gamma^2 \int_{\tau_i/\gamma}^{\tau_f/\gamma} du \int_{\tau_i/\gamma+u}^{\tau_f/\gamma+u}  ds \eta(\gamma u) \eta(\gamma (u-s)) e^{-i\Omega\gamma s} W_{BTZ}(s)
    \end{split}
\end{equation} with $\gamma = \gamma(R) = \sqrt{-g_{00}}$, we chose $\tau_i = -\tau_f$, and we have performed the change of variables $u=t, s= t-t'$.
Next, we require $\Delta \tau >> \sigma$ such that we can use the approximation of \Cref{eq:approx_foo}, taking $\tau_i=-\tau_f$ giving 
\begin{equation}
\begin{split}
        \mathcal{F}(\Omega) &\approx  \gamma^2 \int_{-\tau_f/\gamma}^{\tau_f/\gamma} du \int_{-\tau_f/\gamma}^{\tau_f/\gamma}  ds \eta(\gamma u) \eta(\gamma (u-s)) e^{-i\Omega\gamma s} W_{BTZ}(s) \\ &= \gamma^2 \int_{-\tau_f/\gamma}^{\tau_f/\gamma}  ds J(s) e^{-\gamma^2 \frac{s^2}{4\sigma^2}} e^{-i\Omega\gamma s} W_{BTZ}(s) 
\end{split}
\end{equation}  with  \begin{equation}
    \begin{split}
        J(s) &= \int_{-\tau_f/\gamma}^{\tau_f/\gamma} du e^{-\gamma^2\frac{(u-1/4 s)^2}{\sigma^2}} \\ &= \frac{\sigma \sqrt{\pi}}{ 2 \gamma } \bigg( \erf{\frac{\tau_f-\gamma  s /2 }{\sigma}} + \erf{\frac{ \tau_f+\gamma s /2}{\sigma}} \bigg) \\ &= \frac{\sigma \sqrt{\pi}}{ 2 \gamma } H_0(s). 
    \end{split}
\end{equation} Here we defined \begin{equation}
    \begin{split}
        H_0(s) &= \erf{\frac{\tau_f-\gamma s/2 }{\sigma}} + \erf{\frac{ \tau_f+\gamma s /2}{\sigma}}.
    \end{split}
\end{equation} If we take the integration limits $\tau_f$ to infinity we can approximate this integral $J(s)$ simply by $\sigma \sqrt{\pi}/\gamma$.

We thus find that 
\begin{equation}
    \begin{split} \label{eq:F_E_app_us}
        \mathcal{F}(\Omega) &\approx  \frac{\gamma \sigma \sqrt{\pi}}{2} \int_{-\tau_f/\gamma}^{\tau_f/\gamma}  ds H_0(s) e^{-\gamma^2 \frac{s^2}{4\sigma^2}} e^{-i\Omega\gamma s} W_{BTZ}(s) \\
        &=  \frac{\gamma \sigma \sqrt{\pi}}{2} \frac{\sqrt{M}}{4\pi \gamma l \sqrt{2}} \sum_m \int_{-\tau_f/\gamma}^{\tau_f/\gamma}  ds \frac{ H_0(s) e^{-\gamma^2 \frac{s^2}{4\sigma^2}} e^{-i\Omega\gamma s}}{\sqrt{\alpha_m-\cosh(\frac{r_h}{l^2}s - i\epsilon) }}
        \\ &= \frac{\sigma \sqrt{M}}{8 l \sqrt{2\pi}} \sum_m \int_{-\tau_f/\gamma}^{\tau_f/\gamma}  ds \frac{ H_0(s) e^{-\gamma^2 \frac{s^2}{4\sigma^2}} e^{-i\Omega\gamma s}}{\sqrt{\alpha_m-\cosh(\frac{r_h}{l^2}s - i\epsilon) }} 
    \end{split}
\end{equation} where we have inserted the BTZ correlation function of \Cref{eq:BTZ_corr_app} and $\alpha_m$ is given by 
\begin{equation}
    \alpha_m =  \frac{R^2}{\gamma^2 l^2} \cosh(2\pi m \sqrt{M}) -\frac{M}{\gamma^2} .
\end{equation}
Note that $Y_0,Z_0( l z)$ are the same as for the cross term for the superposed black hole mass case of Ref.~\cite{foo2022quantum}, but $H( l z)$ and $\alpha^{12}_n$ are different.

\subsection{Relating to the coordinates of Ref.~\cite{foo2022quantum} and final expressions} \label{sec:foo_time_rescaling} 
The calculations in Ref.~\cite{foo2022quantum} were done in a rescaled AdS time coordinate.
Namely, Ref.~\cite{foo2022quantum} works with the time coordinates $\overline{t}$ of $AdS_3$, of which the BTZ black hole is obtained by a periodic identification as explained in \Cref{sec:app_BTZ_AdS}.
This means that our coordinates $t,R$ are related to the coordinate $\overline{t},\overline{R}$ of 3d AdS--Rindler space of \Cref{sec:app_BTZ_AdS} by 
\begin{equation}
    \begin{split}
        \sqrt{M} t = \overline{t}, \\
        R/\sqrt{M} = \overline{R}.
    \end{split}
\end{equation} 
Our coordinates $t,R$ are referred to  as $\tilde{s},\tilde{R}$ in the Supplemental Materials of Ref.~\cite{foo2022quantum}, but the tildes are omitted there in the Supplemental Sections D and E.
Ref.~\cite{foo2022quantum} also defines a rescaled redshift factor which we refer to as $\widetilde{\gamma}$, defined by \begin{equation}
    \widetilde{\gamma}(R) = \sqrt{\frac{R^2}{ M l^2}-1} = \sqrt{\frac{\bar{R}^2}{  l^2}-1}, 
\end{equation} and thus related to our BTZ redshift factor $\gamma$ of \cref{eq:tau_to_t} by
\begin{equation}
     \gamma(R) = \widetilde{\gamma}(R) \sqrt{M}.
\end{equation}
Note that Ref.~\cite{foo2022quantum}  still uses the BTZ radial coordinate. Furthermore, because Ref.~\cite{foo2022quantum} rescaled the times by $\sqrt{M}$, the time span in $\tilde{t_f}$ is equal to the time span in coordinate $t$ times $\sqrt{M}$, so that, if one were to take a gravitational phase into account,  our definition of the phase being $M \cdot t$ is dimensionally consistent with their definition of the phase being $\sqrt{M} \cdot \overline{t}$. 

We can put our results for $\mathcal{F}_{12}(\Omega)$ and $\mathcal{F}(\Omega)$ in terms of the coordinates and conventions of Ref.~\cite{foo2022quantum} as follows. In the remaining part of this section, $t,\gamma$ should be understood as $\overline{t},\widetilde{\gamma}$, finding 
\begin{equation} \label{eq:F12_us_app_rescaled}
    \begin{split}
        \frac{\mathcal{F}_{12}(\Omega)}{\sigma} \approx Y_0 \sum_n \text{ Re}  \int^{\frac{\tau_f}{ l\widetilde{\gamma}_2}}_{0} dz \frac{Z_0( l z )H( l z)}{\sqrt{\alpha_n-\cosh( z - i\epsilon)}}
    \end{split}
\end{equation} with 
\begin{align} 
\alpha^{12}_n &= \frac{R_1 R_2}{\widetilde{\gamma}_1 \widetilde{\gamma}_2 \sqrt{M_1 M_2} l^2} \cosh (2 \pi n \sqrt{M})-\frac{1}{\widetilde{\gamma}_1 \widetilde{\gamma}_2}  \\
 Y_0 &=     \frac{\sqrt{\widetilde{\gamma}_1 \widetilde{\gamma}_2} \sqrt{\pi}}{4\pi \sqrt{\widetilde{\gamma}_1^2+\widetilde{\gamma}_2^2}} e^{-\Omega^2 \sigma^2\frac{(\widetilde{\gamma}_1-\widetilde{\gamma}_2)^2}{2(\widetilde{\gamma}_1^2+\widetilde{\gamma}_2^2)}}, \label{eq:Y_0_app2} \\
 Z_0(s) &=  e^{-\frac{\widetilde{\gamma}_1^2\widetilde{\gamma}_2^2}{2\sigma^2(\widetilde{\gamma}_1^2+\widetilde{\gamma}_2^2)} s^2} e^{-i \Omega s \frac{\widetilde{\gamma}_1 \widetilde{\gamma}_2 (\widetilde{\gamma}_1+\widetilde{\gamma}_2)}{\widetilde{\gamma}_1^2+\widetilde{\gamma}_2^2}}, \label{eq:Z_0_app2} \\
        H(s) &= \text{erf}\bigg( \frac{(\widetilde{\gamma}_1^2+\widetilde{\gamma}_2^2)\tau_f - \widetilde{\gamma}_1 \widetilde{\gamma}_2^2 s + i\Omega \sigma^2 \widetilde{\gamma}_1 (\widetilde{\gamma}_1-\widetilde{\gamma}_2) }{\sqrt{2}\sigma  \widetilde{\gamma}_1 \sqrt{\widetilde{\gamma}_1^2+\widetilde{\gamma}_2^2}}\bigg) \\ &\phantom{AB{}}+  \text{erf}\bigg( \frac{(\widetilde{\gamma}_1^2+\widetilde{\gamma}_2^2)\tau_f + \widetilde{\gamma}_1 \widetilde{\gamma}_2^2 s - i\Omega \sigma^2 \widetilde{\gamma}_1 (\widetilde{\gamma}_1-\widetilde{\gamma}_2) }{\sqrt{2}\sigma  \widetilde{\gamma}_1 \sqrt{\widetilde{\gamma}_1^2+\widetilde{\gamma}_2^2}}\bigg)
\end{align}  and $\gamma_i(R) = \sqrt{R_i^2/(M l^2) -1}$ for $i=1,2$.
If $\arccosh{\alpha_n} < \tau_f/(l \widetilde{\gamma}_2)$, we need to make sure to take the correct branch of the square root, i.e. 
\begin{equation} \label{eq:F12_us_app_rescaled_branch_squareroot}
    \begin{split}
        \frac{\mathcal{F}_{12}(\Omega)}{\sigma} \approx & Y_0 \sum_n \text{ Re}  \int^{\arccosh(\alpha_n)}_{0} dz \frac{Z_0( l z )H( l z)}{\sqrt{\alpha_n-\cosh( z - i\epsilon)}}  \\+ & Y_0 \sum_n \text{ Re}  \int_{\arccosh(\alpha_n)}^{\tau_f/(l \widetilde{\gamma}_2)} dz \frac{Z_0( l z )H( l z)}{i \sqrt{\cosh( z - i\epsilon)-\alpha_n}}
    \end{split}
\end{equation}

For $\mathcal{F}(\Omega)$ from \Cref{eq:F_E_app_us}, we take a single branch with detector at radial coordinate distance $R$ from the black hole (so $R=R_1,R_2$ for the two branches of our position-superposition), and $\widetilde{\gamma}$ is the related (rescaled) redshift, so $\widetilde{\gamma} = \widetilde{\gamma}_1,\widetilde{\gamma}_2$ for the two branches. Using the rescaled coordinates of Ref.~\cite{foo2022quantum} as explained above, we obtain
\begin{equation} \label{eq:F_E_Foo}
    \begin{split}
        \mathcal{F}(\Omega) \approx  \frac{\sigma}{8 l \sqrt{2\pi}} \sum_m \int_{-\tau_f /\widetilde{\gamma}}^{\tau_f  /\widetilde{\gamma}}  ds \frac{ H_0(s) e^{-\widetilde{\gamma}^2 \frac{s^2}{4\sigma^2}} e^{-i\Omega\widetilde{\gamma} s}}{\sqrt{\alpha_m-\cosh(s/l - i\epsilon) }},
    \end{split}
\end{equation} which, as shown in Ref.~\cite{foo2022quantum}, leads to the following 
\begin{equation} \label{eq:foo_FE_single}
\begin{split}
       \frac{\mathcal{F}(\Omega)}{\sigma} &\approx \frac{\sqrt{\pi} H_0(0)}{8}
    -\frac{i}{8 \sqrt{\pi}} \mathrm{PV} \int_{-\tau_f / (2 l \widetilde{\gamma})}^{\tau_f / (2 l \widetilde{\gamma})} \mathrm{~d} z \frac{X_0(2 l z) H_0(2 l z)}{\sinh (z)}
    \\ &+\frac{1}{4 \sqrt{2 \pi}} \sum_{n \neq 0} \operatorname{Re} \int_{0}^{\tau_f / (l\widetilde{\gamma})} \mathrm{~d} z \frac{X_0(l z) H_0(l z)}{\sqrt{\alpha_{n}-\cosh (z)}}.
\end{split}
\end{equation} with 
\begin{align}
    X_0(s)& = e^{-\frac{\widetilde{\gamma}^2 s^2}{4 \sigma^2}} e^{-i \Omega \widetilde{\gamma} s}, \\
    \alpha_n &= \frac{R^2}{M \widetilde{\gamma}^2 l^2} \cosh (2 \pi n \sqrt{M})-\frac{1}{\widetilde{\gamma}^2}, \label{eq:alpha_n_foo_corrected} \\
    H_0(s) &= \operatorname{erf}\left[\frac{\widetilde{\gamma} \left(s+2  \tau_f/\widetilde{\gamma} \right)}{2 \sigma}\right]-\operatorname{erf}\left[\frac{\widetilde{\gamma} \left(s-2 \tau_f / \widetilde{\gamma} \right)}{2 \sigma}\right].
\end{align} The first two terms of \Cref{eq:foo_FE_single} arise from a pole at $n=0$ using the Sokhotski formula 
\begin{equation}
    \frac{1}{\sinh (z-i \epsilon)}=i \pi \delta(z)+\mathrm{PV} \frac{1}{\sinh (z)}.
\end{equation}

Here we have corrected a small inconsistency in the Supplemental Material of Ref.~\cite{foo2022quantum}: in Eq. (S51) of Ref.~\cite{foo2022quantum}, the first term in $\beta_{nm}$ should be divided by $M$, i.e. the expression of $\beta_{nm}$ should be \begin{equation}
    \beta_{nm} = \frac{R_D^2}{\textcolor{blue}{M}\widetilde{\gamma}_D^2 l^2 } \cosh(2\pi(n-m)\sqrt{M}) - \frac{1}{\widetilde{\gamma}_D^2}
\end{equation} where we have highlighted the missing term $M$ in \textcolor{blue}{blue}. The expression $\beta_{n0}$ corresponds to our $\alpha_n$ in \cref{eq:alpha_n_foo_corrected}.

Similarly to \cref{eq:F12_us_app_rescaled_branch_squareroot}, if the term inside the square root in the denominator of the third term in \cref{eq:F_E_Foo} becomes negative, taking care of the square root leads to 
\begin{equation} \label{eq:third_term_F_E_Foo_squareroot}
    \begin{split}
        \frac{1}{4 \sqrt{2 \pi}} \bigg[ \sum_{n \neq 0} \operatorname{Re} \int_{0}^{\arccosh \alpha_n} \mathrm{~d} z \frac{X_0(l z) H_0(l z)}{\sqrt{\alpha_{n}-\cosh (z)}} + \sum_{n \neq 0} \operatorname{Re} \int_{\arccosh \alpha_n}^{\tau_f / (l\widetilde{\gamma})} \mathrm{~d} z \frac{X_0(l z) H_0(l z)}{i \sqrt{\cosh (z)-\alpha_n}} \bigg].
    \end{split}
\end{equation}

\section{Derivation of interaction Hamiltonian for superposed black hole} \label{app:H_int_superp_BH}
In this section we derive the form of the Unruh--DeWitt detector Hamiltonian in the laboratory frame, i.e. in the presence of a superposition of black holes. 
The quantum-controlled UDW interaction introduced in \cref{sec:UDW} in the black hole frame (but according to the detector's time) is, in the Schrödinger picture, given by  
\begin{equation}
    H_\text{int}^{(B)} = \lambda  \, \gamma(\hat{\mathbf{x}})  \, \mu \otimes \phi(\hat{\mathbf{x}}) = \lambda \, \mu \otimes \int d\nu(\mathbf{x})\,\ket{\mathbf{x}}  \bra{\mathbf{x}}_D \,  \gamma(\mathbf{x}) \otimes \phi(\mathbf{x})
\end{equation}
acting on the Hilbert space $\mathcal{H}_{D}\otimes \mathcal{H}_{\mu}\otimes \mathcal{H}_\phi$ corresponding to the external degrees of freedom of the detector (its position), the internal degrees of freedom of the detector (the $n$-level energy system) and the scalar field sector, respectively. 
Recall that $\gamma(\mathbf{x})$ is the redshift factor.
A QRF transformation maps the perspective of the black hole to the perspective of the detector (i.e. the laboratory), with the transformation 
\begin{equation}
    S^{(B\to D)} = \int d\nu(\mathbf{x}) \, \ket{-\mathbf{x}}_B\bra{\mathbf{x}}_{D} \otimes \mathbb{I}_\phi \otimes \mathbb{I}_{\mu}
\end{equation}
After the QRF transformation, the UDW interaction Hamiltonian in the detector frame becomes
\begin{equation}
\begin{split}
        H^{(D)}_{\text{int}} &= S^{(B\to D)}\gamma^{-1}(\hat{\mathbf{x}}_D) H^{(B)}_\text{int} S^{\dagger(B\to D)} = \lambda \int d\nu(\mathbf{x}) \, \ket{\mathbf{x}} \bra{\mathbf{x}}_B \,  \otimes \, \mu \otimes  \phi(-\mathbf{x})  \\&= \lambda \int d\nu(\mathbf{x}) \, \ket{\mathbf{x}} \bra{\mathbf{x}}_B \otimes H_{int,\mathbf{x}_B},
\end{split}
\end{equation} with $H_{int,\mathbf{x}_B}$ the Unruh--DeWitt interaction for a detector with a scalar field in a background with black hole at position $\mathbf{x}$ (and with detector now at the origin $\mathbf{0}$).

\section{Derivation of the analytical expressions of \Cref{sec:analytical}}
\label{sec:analytical_appendix}
\subsection{Summary of expressions} \label{sec:analytical_summary_app}
In this section we present the analytical expressions for the spectrum probed by the detector from \Cref{sec:analytical}. Before presenting detailed derivations, we provide a summary of the relevant expressions here. We consider a static detector near a (potentially superposed) BTZ black hole. 

\vspace{0.5cm}

\textit{Single-spacetime detector response.} In a non-superposed spacetime, the detector response from \cref{eq:transition_prob} can be expressed as \begin{equation} \label{eq:Fewster_eq_chi_hatW_app_expressions}
    \mathcal{F}(E)=\frac{1}{2 \pi} \int_{-\infty}^{\infty} d \omega|\widehat{\eta}({\scriptstyle\frac{\omega}{\gamma}})|^2 \widehat{W}_{\scriptstyle BTZ}(E \gamma +\omega),
\end{equation} 
The explicit form of the BTZ correlation function along the detector's path is given by taking $R_1=R_2=R$ to be the detector's radial coordinate and $\theta=0$ in \cref{eq:W_BTZ_app_NeuDiri}: 
\begin{equation}
     \begin{split}
       {W}_{\scriptstyle BTZ}(s) = \frac{1}{4 \pi l \sqrt{2} \widetilde{\gamma}} & \bigg[ \frac{1}{\sqrt{\alpha_m - \cosh(s \sqrt{M}/l - i \epsilon )}}  -  \frac{\zeta}{\sqrt{\alpha'_m - \cosh(s \sqrt{M}/l - i \epsilon)}} \bigg] 
    \end{split}
\end{equation} with limit $\epsilon \rightarrow 0^+$, $s= t-t', \widetilde{\gamma}=\gamma(R)/\sqrt{M}$ and
\begin{align}
     \alpha_m &= \frac{ \frac{R^2}{M l^2} \cosh [ 2 \pi m\sqrt{M}]-1}{\frac{R^2}{M l^2}-1 } \\
     \alpha'_m &= \frac{ \frac{R^2}{M l^2} \cosh [ 2 \pi m\sqrt{M}]+1}{\frac{R^2}{M l^2}-1 } .
\end{align}

Following Ref.~\cite{Lifschytz_1994}, an analytical expression for $\widehat{W}$ can be found, giving\footnote{Compared to Ref.~\cite{Lifschytz_1994} we have added the first term arising from the pole singularity using the Sokhotski formula.}
\begin{equation}
    \begin{split}
        \widehat{W}_{BTZ}(K) =  \frac{1}{4  {\gamma}} +  \frac{1}{2 {\gamma}} \frac{1}{e^{ K / T_H}+1} &\sum_n \bigg[ P_{\frac{i K }{2\pi T_H}-\frac{1}{2}}(\alpha_n)   -\zeta  P_{\frac{i K }{2 \pi T_H}-\frac{1}{2}}(\alpha'_n) \bigg]
    \end{split}
\end{equation} with Hawking temperature $T_H = \sqrt{M}/(2\pi l)$ and $P_\nu$ the Legendre function of the first kind.

\vspace{0.5cm}

\textit{Interference term in detector response.} The interference term $\mathcal{F}_{12}(E)$, for both mass- and position-superposed black holes now, can be written in a similar form as \cref{eq:Fewster_eq_chi_hatW_app_expressions}:
\begin{equation}
\begin{split}
    \mathcal{F}_{12}(E) = \frac{1}{2\pi} \int_{-\infty}^{+ \infty}  \mathrm{d}\omega  \hspace{0.05cm} \widehat{\eta}({\scriptstyle\frac{\omega}{\widetilde{\gamma_1}}})^* \hspace{0.05cm} \widehat{\eta}({\scriptstyle\frac{\omega}{\widetilde{\gamma_2}} + E [\frac{\widetilde{\gamma_1}}{\widetilde{\gamma_2}} -1] } )  \hspace{0.05cm} \widehat{W}^{(12)}_{BTZ}(\omega+\widetilde{\gamma_1} E). 
\end{split}
\end{equation}
To calculate the Fourier transform of $\widehat{W}^{(12)}$, we will first perform a change of coordinates in each branch of the superposition. 
Namely, we want to be able to treat also the mass-superposed case, in which case we need coordinates to compare spacetimes for different black hole masses. 
Following Ref.~\cite{foo2022quantum}, we do so by considering AdS time coordinates $\overline{t}$ related to the BTZ time coordinates $t$ by $\overline{t}=t \sqrt{M_i}$ in branch $i$. In the mass-superposed black hole, we now need the additional assumption that the Unruh--DeWitt interaction is a quantum-controlled version of the single-spacetime interaction, with control on the black hole properties, its position and mass, as in Ref.~\cite{foo2022quantum}. 
We have made this assumption more plausible by deriving this interaction form in the case of the black hole position in \cref{eq:H_int_QRF_transformed}.
We denote the position and mass by $R_1,M_1$ and $R_2,M_2$ in the two branches of the superposition. 
We also introduce $\widetilde{\gamma_i}(R_i) = \sqrt{R_i^2/(M_i l^2) -1}$.

The scalar field correlation function between two BTZ spacetimes is then given by~\cite{foo2022quantum}:  
\begin{equation} \label{eq:W_BTZ_mass_and_position_superp}
    \begin{split}
       \widehat{W}^{(12)}_{BTZ}(\overline{s}) =  \frac{1}{4 \pi  \sqrt{2 \widetilde{\gamma_1} \widetilde{\gamma_2}} } & \frac{1}{\sum_k \Upsilon^{2k}} \sum_{m,n} \bigg[ \frac{1}{\sqrt{\beta_{mn}^{(12)} - \cosh(\overline{s} /l)}}  -  \frac{\zeta }{\sqrt{\beta_{mn}^{'(12)} - \cosh(\overline{s} /l)}} \bigg], 
    \end{split}
\end{equation} with $\overline{s}=\overline{t}-\overline{t}' - i \epsilon$ and
\begin{align}
     \beta_{mn}^{(12)} &= \frac{ \frac{R_1 R_2}{\sqrt{M_1 M_2} l^2} \cosh [ 2 \pi (m\sqrt{M_1}- n \sqrt{M_2})]-1}{\sqrt{\frac{R_1^2}{M_1 l^2}-1}\sqrt{\frac{R_2^2}{M_2 l^2}-1} }, \\
     \beta_{mn}^{'(12)} &=  \frac{ \frac{R_1 R_2}{\sqrt{M_1 M_2} l^2} \cosh [ 2 \pi (m\sqrt{M_1}- n \sqrt{M_2})]+1}{\sqrt{\frac{R_1^2}{M_1 l^2}-1}\sqrt{\frac{R_2^2}{M_2 l^2}-1} }.
\end{align}

Using these coordinates $\overline{t}$, we can obtain an analytical expression of $\widehat{W}^{(12)}_{BTZ}$, with the additional subtlety that for $\sqrt{M_1/M_2} \in \mathbb{Q} \backslash \{1\}$ and $R_1/\sqrt{M_1} = R_2/\sqrt{M_2}$ an additional pole arises, giving a singular term $\textsf{sing.}$:
\begin{equation}
    \begin{split}
        \widehat{W}^{(12)}_{BTZ}(K) = \textsf{sing.} + \frac{1}{2 \sqrt{\widetilde{\gamma_1} \widetilde{\gamma_2}}} \frac{1}{\sum_k \Upsilon^{2k}} \hspace{0.07cm} \frac{1}{e^{K/T_H^{AdS}}+1} \\ \sum_{m,n} \bigg[ P_{\frac{i K}{2 \pi T_H^{AdS}}-\frac{1}{2}}\big(\beta_{mn}^{(12)} \big) - \zeta  P_{\frac{i K}{2 \pi T_H^{AdS}}-\frac{1}{2}}\big(\beta_{mn}^{'(12)}\big) \bigg],
    \end{split}
\end{equation} 
with $T_H^{AdS}$ the Hawking temperature of the AdS–Rindler horizon with respect to AdS–Rindler time, $T_H^{AdS} = 1/(2 \pi l)$.
If $\sqrt{M_1/M_2} \notin \mathbb{Q} \backslash \{ 1\}$, then we have $\textsf{sing.} = 0$,else it gives the following non-zero contribution if also $R_1 / \sqrt{M_1} = R_2 / \sqrt{M_2}$:
\begin{equation} \label{eq:sing.}
    \textsf{sing.} = \frac{1}{4  \sqrt{\widetilde{\gamma_1}\widetilde{\gamma_2}} pq}  \quad  \text{ if } \sqrt{M_1/M_2} = p/q \in \mathbb{Q} \backslash \{ 1\} \text{ and } R_1 / \sqrt{M_1} = R_2 / \sqrt{M_2}
\end{equation} with $\text{gcd}(p,q)=1$.

\subsection{Analytic expressions for single term}

Consider the detector response function 
\begin{equation}
\begin{split}
        \mathcal{F}(E) &= \int_{-\infty}^{\infty}   \int_{-\infty}^{\infty}  \mathrm{d}\tau \mathrm{d}\tau' \eta(\tau) \eta(\tau') e^{-iE (\tau-\tau')} W(\tau,\tau') 
    \\ 
    &=  \gamma^2 \int_{-\infty}^{\infty}   \int_{-\infty}^{\infty}  \mathrm{d}t \mathrm{d}t' \eta(\gamma t) \eta( \gamma t') e^{-iE \gamma (t-t')} W(t-t') 
    \end{split}
\end{equation} assuming $W(t,t') = W(t-t')$, with $\gamma = \gamma(R)$ with $R$ the fixed radial coordinate of the static detector. Here and everywhere below $W$ should be understood as $W_{BTZ}$.
By applying the inverse Fourier transform to $W$, and Fourier transforming the switching functions $\eta,\eta$, we obtain \begin{equation} \label{eq:Fewster_eq_chi_hatW_app}
    \mathcal{F}(E)=\frac{1}{2 \pi} \int_{-\infty}^{\infty} d \omega|\widehat{\eta}({\scriptstyle\frac{\omega}{\gamma}})|^2 \widehat{W}(E \gamma +\omega),
\end{equation} with the Fourier transform defined as \begin{equation}
    \widehat{f}(\omega)=\int_{-\infty}^{\infty} d s f(s) \mathrm{e}^{-\mathrm{i} \omega s} 
\end{equation} and inverse by \begin{equation}
    f(s) = \int_{-\infty}^{+\infty} \frac{d\omega }{2\pi} e^{i\omega s} \hat{f}(\omega).
\end{equation}  
See Ref.~\cite{Fewster_2016} for a more precise treatment.
For our Gaussian switching function, we have, for our detector at fixed location $r=R$,
\begin{equation}
    |\widehat{\eta}({\scriptstyle\frac{\omega}{\gamma}})|^2 = \frac{1}{N} 2 \sigma^2 e^{- \frac{\omega^2 \sigma^2}{ \gamma^2} } = \frac{1}{N}  2 \sigma^2 e^{- \frac{\omega^2 \sigma^2}{ \widetilde{\gamma}^2 M} }
\end{equation}  
with $N$ an appropriate normalization factor.
If we choose $\sigma$ to be very large, i.e. adiabatic switching, $\widehat{\eta}(\omega)$ becomes approximately a delta function,  and we are basically probing 
$\widehat{\mathcal{W}}(E \gamma)$ for frequencies $\omega/\gamma$. 
This $1/\gamma$ factor arising from the redshift is what leads to the local Tolman temperature $T(R)$ for a static detector at $R$ being related to the Hawking temperature $T_H$ by $T = T_H/\gamma(R)$ with $\gamma(R) = \sqrt{R^2/l^2 - M}$.

\vspace{0.5cm}

Now, to calculate $\widehat{\mathcal{W}}(K)$, we have, \begin{equation} \label{eq:w_K_1}
    \begin{split}
       \widehat{W}(K) &= \int_{-\infty}^{+\infty}  \mathrm{d}s e^{-i K s} W(s) \\ &=  \frac{1}{4 \pi l \sqrt{2} \widetilde{\gamma}} \sum_m \int_{-\infty}^{+\infty}  \mathrm{d}s \bigg[ \frac{e^{-iK s}}{\sqrt{\alpha_m - \cosh(s \sqrt{M}/l-i\epsilon)}}  \\ &\phantom{AAAAABBBBBBB{}} -  \frac{\zeta e^{-iK s}}{\sqrt{\alpha'_m - \cosh(s \sqrt{M}/l-i\epsilon)}} \bigg] 
       \\ 
       &=   \frac{1}{4 \pi  \sqrt{2} {\gamma}} \sum_m \int_{-\infty}^{+\infty}  \mathrm{d}z \bigg[ \frac{e^{-iK l z / \sqrt{M}}}{\sqrt{\alpha_m - \cosh(z-i\epsilon)}}   \\ &\phantom{AAAAABBBBBBB{}} -  \frac{\zeta e^{-iK l z / \sqrt{M} }}{\sqrt{\alpha'_m - \cosh(z-i\epsilon)}} \bigg].
    \end{split}
\end{equation} 
with $\widetilde{\gamma}= \gamma / \sqrt{M}$, 
\begin{equation}
     \alpha_m = \frac{ \frac{R^2}{M l^2} \cosh [ 2 \pi m\sqrt{M}]-1}{\frac{R^2}{M l^2}-1 }  = \frac{ \frac{R^2}{M l^2} \cosh [ 2 \pi m\sqrt{M}]-1}{\widetilde{\gamma}^2 }.
\end{equation} and \begin{equation}
     \alpha'_m = \frac{ \frac{R^2}{M l^2} \cosh [ 2 \pi m\sqrt{M}]+1}{\frac{R^2}{M l^2}-1 } = \frac{ \frac{R^2}{M l^2} \cosh [ 2 \pi m\sqrt{M}]+1}{\widetilde{\gamma}^2 } ,
\end{equation} the $z - i \epsilon$ prescription, and $\zeta = -1,0,1$ for Neumann, transparent or Dirichlet boundary conditions.

Note that we have $\alpha_{0} = 1$ so that the term with $\alpha_{0}$ has a pole at $z=0$. 
In this case, we first rewrite using $\cosh(z) - 1 = 2 \sinh^2(z/2)$, and using the Sokhotski formula \begin{equation}
    \frac{1}{\sinh (z-i \epsilon)}=i \pi \delta(z)+\mathrm{PV} \frac{1}{\sinh (z)},
\end{equation} 
We find that the singular part gives rise to the term $1/(4 l \widetilde{\gamma})$.

Following Ref.~\cite{Lifschytz_1994}, the nonsingular parts (including the principal value part of the singular term with $\alpha_0$ above) in the integrals in \cref{eq:w_K_1} can be performed, and give 
\begin{equation}
    \begin{split}
        \widehat{W}(K) &=  \frac{1}{4  {\gamma}} +  \frac{1}{2  {\gamma}} \frac{1}{e^{ K / T_H}+1} \sum_n \bigg[ P_{\frac{i K }{2\pi T_H}-\frac{1}{2}}(\alpha_n) - \zeta  P_{\frac{i K }{2 \pi T_H}-\frac{1}{2}}(\alpha'_n) \bigg]
    \end{split}
\end{equation} with $T_H = \sqrt{M}/(2\pi l)$ the Hawking temperature of the AdS–Rindler horizon with respect to AdS–Rindler time. 
Note that Ref.~\cite{Lifschytz_1994} left out the first term arising from the pole singularity.

\subsection{Analytic expressions for interference term}

For the interference term, we can obtain a similar expression.
We now consider the general situation that comprises both the position- and mass-superposed black holes, with position and mass $R_1,M_1$ and $R_2,M_2$ in the two branches of the superposition, respectively. 
Consider the interference term in the detector transition probability 
\begin{equation}
    \begin{split}
       \mathcal{F}_{12}(E) = \int_{\tau_i}^{\tau_f} d\tau d\tau'  \eta(\tau) \eta(\tau') e^{-i E (\tau -\tau')} W^{(12)}(x(\tau),x(\tau')),
    \end{split} 
\end{equation} with $W^{(12)}$ shorthand for $W^{(12)}_{BTZ}$. Using the AdS time coordinates $\overline{t}=t \sqrt{M_i} =\tau / \widetilde{\gamma_i}$ with $\widetilde{\gamma_i} = \sqrt{R_i^2/(M_i l^2)-1}$ in the branch with BTZ black hole mass $M_i$, we find \begin{equation}
    \begin{split}
       \mathcal{F}_{12}(E) = \widetilde{\gamma_1} \widetilde{\gamma_2} \int_{-\infty}^{+\infty} \int_{-\infty}^{+\infty}  \mathrm{d}\overline{t}  \mathrm{d}\overline{t}'  \eta(\widetilde{\gamma_1} \overline{t}) \eta(\widetilde{\gamma_2} \overline{t}') e^{-i (E \widetilde{\gamma_1} \overline{t} - E \widetilde{\gamma_2} \overline{t}')} W^{(12)}_{BTZ}(\overline{t}-\overline{t}').
    \end{split}
\end{equation} 
We will again try to write $\mathcal{F}_{12}$ in the form of \cref{eq:Fewster_eq_chi_hatW_app}. We start off by changing integration variables with $s= \overline{t}-\overline{t}'$ (we omit the integration limits here, all integrals, also under the changes of variables that occur, run from $-\infty$ to $+\infty$):
\begin{equation}
    \begin{split}
       \mathcal{F}_{12}(E) &=  \widetilde{\gamma_1} \widetilde{\gamma_2} \int \int  \mathrm{d}s  \mathrm{d}\overline{t}' \hspace{0.05cm} \eta(\widetilde{\gamma_1} [s+\overline{t}'] )  \hspace{0.05cm} \eta(\widetilde{\gamma_2} \overline{t}') e^{-i E \widetilde{\gamma_1} s}e^{-iE(\widetilde{\gamma_1}-\widetilde{\gamma_2})\overline{t}'} W^{(12)}_{BTZ}(s) 
       \\ 
       &=  \widetilde{\gamma_1} \widetilde{\gamma_2} \int \int  \mathrm{d}s  \mathrm{d}\overline{t}' \hspace{0.05cm} \eta(\widetilde{\gamma_1} [s+\overline{t}'] )  \hspace{0.05cm} \eta(\widetilde{\gamma_2} \overline{t}') e^{-i E \widetilde{\gamma_1} s}e^{-iE ( \widetilde{\gamma_1}-\widetilde{\gamma_2})\overline{t}'} \int  \frac{\mathrm{d}\omega }{2 \pi} \widehat{W}^{(12)}_{BTZ}(\omega) e^{i \omega s} 
       \\ 
       &= \frac{\widetilde{\gamma_1} \widetilde{\gamma_2}}{2\pi}  \int  \mathrm{d}\omega \int \int  \mathrm{d}s  \mathrm{d}\overline{t}' \hspace{0.05cm}   \eta(\widetilde{\gamma_1} [s+\overline{t}'] )  \hspace{0.05cm} \eta(\widetilde{\gamma_2} \overline{t}') e^{i \omega s}e^{-iE (\widetilde{\gamma_1}-\widetilde{\gamma_2})\overline{t}'} \widehat{W}^{(12)}_{BTZ}(\omega+\widetilde{\gamma_1} E)   
       \\ 
       &= \frac{\widetilde{\gamma_2}}{2\pi}  \int  \mathrm{d}\omega \int \int   \mathrm{d}\overline{t}' \mathrm{d}u \hspace{0.05cm} \eta( u )  e^{i \frac{\omega}{\widetilde{\gamma_1}} u} \eta(\widetilde{\gamma_2} \overline{t}') e^{-i [ \omega + E (\widetilde{\gamma_1}-\widetilde{\gamma_2}) ]\overline{t}'} \widehat{W}^{(12)}_{BTZ}(\omega+\widetilde{\gamma_1} E)  
       \\
       &= \frac{1}{2\pi} \int  \mathrm{d}\omega \int \mathrm{d}u \hspace{0.05cm} \eta( u )  e^{i \frac{\omega}{\widetilde{\gamma_1}} u} \int \mathrm{d} v \eta(v) e^{-i\big[ \frac{\omega}{\widetilde{\gamma_2}} + E \big(\frac{\widetilde{\gamma_1}}{\widetilde{\gamma_2}}-1\big) \big] v} \widehat{W}^{(12)}_{BTZ}(\omega+\widetilde{\gamma_1} E)  
       \\
       &=  \frac{1}{2\pi} \int_{-\infty}^{+ \infty}  \mathrm{d}\omega  \hspace{0.05cm} \widehat{\eta}({\scriptstyle\frac{\omega}{\widetilde{\gamma_1}}})^* \hspace{0.05cm} \widehat{\eta}({\scriptstyle\frac{\omega}{\widetilde{\gamma_2}} + E [\frac{\widetilde{\gamma_1}}{\widetilde{\gamma_2}} -1] } ) \hspace{0.05cm} \widehat{W}^{(12)}_{BTZ}(\omega+\widetilde{\gamma_1} E)
    \end{split}
\end{equation} where we changed variables to $u= \widetilde{\gamma_1} (s + \overline{t}'), v = \widetilde{\gamma_2} \overline{t}'$.

\vspace{0.5cm}

Next, the main calculation of this section, we  calculate $\widehat{W}^{(12)}(K)$: \begin{equation} \label{eq:w_K_1_app}
    \begin{split}
       \widehat{W}^{(12)}(K) &= \int_{-\infty}^{+\infty}  \mathrm{d}s e^{-i K s} W^{(12)}(s) 
       \\ 
       &=  \frac{1}{4 \pi l \sqrt{2 \widetilde{\gamma_1} \widetilde{\gamma_2}} } \frac{1}{\sum_k \Upsilon^{2k}} \sum_{m,n} \int_{-\infty}^{+\infty}  \mathrm{d}s \bigg[ \frac{e^{-iK s}}{\sqrt{\beta_{mn}^{(12)} - \cosh(s /l-i\epsilon)}}  \\ & \phantom{AABDDDEEEEEFFFFGG{}}-  \frac{\zeta e^{-iK s}}{\sqrt{\beta_{mn}^{'(12)} - \cosh(s /l-i\epsilon)}} \bigg] 
       \\ 
       &=   \frac{1}{4 \pi l \sqrt{2 \widetilde{\gamma_1} \widetilde{\gamma_2}} }  \frac{1}{\sum_k \Upsilon^{2k}}  \sum_{m,n} \int_{-\infty}^{+\infty}  \mathrm{d}z \bigg[ \frac{e^{-iK l z }}{\sqrt{\beta_{mn}^{(12)} - \cosh(z-i\epsilon)}}  \\ & \phantom{AABDDDEEEEEFFFFGG{}}-  \frac{\zeta e^{-iK l z }}{\sqrt{\beta_{mn}^{'(12)} - \cosh(z-i\epsilon)}} \bigg].
    \end{split}
\end{equation} 
with 
\begin{equation} \label{eq:alpha_12_m}
     \beta_{mn}^{(12)} = \frac{ \frac{R_1 R_2}{\sqrt{M_1 M_2} l^2} \cosh [ 2 \pi (m\sqrt{M_1}- n \sqrt{M_2})]-1}{\sqrt{\frac{R_1^2}{M_1 l^2}-1}\sqrt{\frac{R_2^2}{M_2 l^2}-1} }
\end{equation} and \begin{equation}
     \beta_{mn}^{'(12)} =  \frac{ \frac{R_1 R_2}{\sqrt{M_1 M_2} l^2} \cosh [ 2 \pi (m\sqrt{M_1}- n \sqrt{M_2})]+1}{\sqrt{\frac{R_1^2}{M_1 l^2}-1}\sqrt{\frac{R_2^2}{M_2 l^2}-1} },
\end{equation} and $\zeta = -1,0,1$ for Neumann, transparent or Dirichlet boundary conditions.
Note that if, for some $m,n$ we have that $\beta_{mn}^{(12)} = 1$, which occurs at rational values of $\sqrt{M_1/M_2}$ in the mass-superposed case (so $M_1 \neq M_2$) while $R_1 /\sqrt{M_1} = R_2 / \sqrt{M_2}$, then the term with $\beta_{mn}^{(12)}$ has a pole at $z=0$. 
In this case, we first rewrite using $\cosh(z) - 1 = 2 \sinh^2(z/2)$, and using the Sokhotski formula \begin{equation}
    \frac{1}{\sinh (z-i \epsilon)}=i \pi \delta(z)+\mathrm{PV} \frac{1}{\sinh (z)},
\end{equation} 
we find the singular part giving rise to the term $1/(4 l \sqrt{\widetilde{\gamma_1}\widetilde{\gamma_2}})$.
We have to divide this term by $1/\sum_k \Upsilon^{2k}$, and it occurs for all $m = \sqrt{M_1/M_2} n$ (and $R_1/\sqrt{M_1}=R_2/\sqrt{M_2}$), This yields a finite contribution which in general might be cut-off dependent, i.e. depending on how the infinite sums are dealt with. 
However, there is a natural way to deal with the sum, arising from the group structure of the groups generated by the identification $\Gamma_M$ that defines the BTZ spacetime as a quotient of AdS--Rindler (cf. Appendix~\ref{sec:app_W_BTZ} and Appendix of Ref.~\cite{foo2022quantum}). 
Namely, let $\Gamma_{M_i}$ denote the identification that defined the BTZ spacetime with mass $M_i$, $\Gamma_{M_i}^n: \varphi \mapsto \varphi + 2\pi n \sqrt{M_i}$. If $\sqrt{M_1/M_2}=p/q \in \mathbb{Q}$ with $\text{gcd}(p,q)=1$, then the groups $G_1,G_2$ generated by $\Gamma_{M_1},\Gamma_{M_2}$ share a subgroup $H$, with indices $[G_1:H]=q, [G_2:H]=p$, suggesting their volumes relate as $\text{Vol}(H)/\sqrt{\text{Vol}(G_1) \text{Vol}(G_2)} = 1/\sqrt{pq}$. 
The singularity is hit at every point of $H$ (assuming also $R_1/ \sqrt{M_1}= R_2/\sqrt{M_2}$), and a suggestive normalization of the BTZ two-point function is then \begin{equation}
    \begin{split} \label{eq:BTZ_corr_appF}
    W_{\mathrm{BTZ}}^{(12)}\left(x, x^{\prime}\right)&=\frac{1}{\sqrt{\text{Vol}(G_1) \text{Vol} (G_2)}}  \sum_{\Gamma_{M_1}^n \in G_1} \sum_{\Gamma_{M_2}^m \in G_2} W_{\mathrm{AdS}}\left(\Gamma_{M_1}^n x, \Gamma_{M_2}^m x^{\prime}\right)
\end{split}
\end{equation} for untwisted fields ($\Upsilon = 1$), arising from the identity for the automorphic field~\cite{foo2022quantum,langlois2005imprintsspacetimetopologyhawkingunruh}: \begin{equation}
    {\phi}_{\text{BTZ},M_i}({x}):=\frac{1}{\sqrt{\text{Vol}(G_i)}} \sum_n {\psi}_{\text{AdS--Rindler}}\left(\Gamma_{M_i}^n {x}\right)
\end{equation} for untwisted fields. As the singular term gives a contribution $1/(4\sqrt{\widetilde{\gamma_1}\widetilde{\gamma_2}})$ at each element of $H$, we get a contribution 
\begin{equation}
    \frac{1}{4\sqrt{\widetilde{\gamma_1}\widetilde{\gamma_2}}} \, \frac{ \text{Vol}(H)}{\sqrt{\text{Vol}(G_1) \text{Vol}(G_2)}} = \frac{1}{4\sqrt{\widetilde{\gamma_1}\widetilde{\gamma_2}}} \, \frac{1}{\sqrt{pq}} 
\end{equation} from the singularities. 
This thus gives the total contribution
\begin{equation} \label{eq:sing.app}
    \text{sing.} = \frac{1}{4  \sqrt{\widetilde{\gamma_1}\widetilde{\gamma_2} p q}}  \quad \text{for } \sqrt{M_1/M_2} = p/q \in \mathbb{Q}^+ \backslash \{1\} \text{ and } R_1/\sqrt{M_1}=R_2/\sqrt{M_2} 
\end{equation} with $\text{gcd}(p,q)=1$, to be inserted as the contribution of the (potential) singular part to $\widehat{W}^{(12)}(K)$. 
Had one taken a square-box regularisation of the sums instead (which is perhaps less well motivated than the above volume argument), i.e., taking $|m|, |n|, |k| \leq N$ such that the double sum becomes $1/(2N+1) \sum_{|m|,|n| \leq N}$ one would get a contribution $1/(4  \sqrt{\widetilde{\gamma_1}\widetilde{\gamma_2}} \max\{q,p\})$ for $\text{sing.}$ above instead.

Regarding all other terms without poles, including the principal value part of the term that was singular before, following Ref.~\cite{Lifschytz_1994}, the integrals in \cref{eq:w_K_1_app} can be performed, and give 
\begin{equation} \label{eq:widehat_interf_app}
    \begin{split}
        \widehat{W}^{(12)}(K) = \textsf{sing.} + \frac{1}{2  \sqrt{\widetilde{\gamma_1} \widetilde{\gamma_2}}} & \frac{1}{\sum_k \Upsilon^{2k}} \hspace{0.07cm} \frac{1}{e^{K/T_H^{AdS}}+1} \sum_{m,n} \bigg[ P_{\frac{i K}{2 \pi T_H^{AdS}}-\frac{1}{2}}\big(\beta_{mn}^{(12)} \big) \\ - \zeta  P_{\frac{i K}{2 \pi T_H^{AdS}}-\frac{1}{2}}\big(\beta_{mn}^{'(12)}\big) \bigg],
    \end{split}
\end{equation} with $T_H^{AdS}$ the Hawking temperature of the AdS–Rindler horizon with respect to AdS–Rindler time, $T_H^{AdS} = 1/(2 \pi l)$.
If $\sqrt{M_1/M_2} \notin \mathbb{Q} \backslash \{1\}$, we have $\textsf{sing.} = 0$, else it gives the contribution as in \cref{eq:sing.app}.

\vspace{0.5cm}

In fact, for the position-superposition, we can remove the double sum in \cref{eq:widehat_interf_app}. Namely, for the position-superposed case we have $M_1 = M_2$ and we can write \begin{equation}
     \begin{split}
        \widehat{W}^{(12)}(K) &= \frac{1}{2  \sqrt{\widetilde{\gamma_1} \widetilde{\gamma_2}}} \hspace{0.07cm} \frac{1}{e^{K/T_H^{AdS}}+1} \sum_{m} \bigg[ P_{\frac{i K}{2 \pi T_H^{AdS}}-\frac{1}{2}}\big(\alpha_{m}^{(12)} \big) - \zeta  P_{\frac{i K}{2 \pi T_H^{AdS}}-\frac{1}{2}}\big(\alpha_{m}^{'(12)}\big) \bigg],
    \end{split}
\end{equation} with \begin{equation}
    \alpha_{m}^{(12)} = \frac{ \frac{R_1 R_2}{M l^2} \cosh [ 2 \pi m \sqrt{M}]-1}{\sqrt{\frac{R_1^2}{M l^2}-1}\sqrt{\frac{R_2^2}{M l^2}-1} }.
\end{equation} 
To be precise, in case $R_1 = R_2$, an additional pole contribution $1/(4 \widetilde{\gamma})$ to $\widehat{W}^{(12)}$ arises, but at $R_1=R_2$ our initial `superposed' state of \cref{eq:initial_state_BH_superp} becomes an unnormalized non-superposed state, and similarly for the interferometric measurement effect, $\ket{+}_B = (\ket{\mathbf{x}_1}_B + \ket{\mathbf{x}_2}_B ) / \sqrt{2}$.

\vspace{0.5cm}

\textbf{A note about energy and time coordinates.} 
We shortly comment upon how to compare energy in different frames, choosing different time coordinates, important for example for establishing when the detector's energy gap is small compared to the black hole mass concerning the approximation of neglecting the detector's gravitational backreaction, or for comparing energies across different spacetimes. 
Let $H^{(D)}_i$ be the detector's free Hamiltonian in its rest frame, in branch $i$ of the superposition.
We take the detector in each branch of the superposition to be identical, i.e. $H^{(D)}_i=H^{(D)}$.
Then, acting on the detector energy eigenstate $\ket{E}_{D}$ with energy eigenvalue $E^{(D)}$ we have $ H \ket{E}_{D} = E^{(D)} \ket{E}_{D}$ in the detector frame. 
If we consider the detector in a superposition of locations, or in a superposed spacetime, with branches labelled by $i$, we need to be able to relate this detector frame energy to energy and Hamiltonians with respect to other time parametrizations.
The detector proper time and coordinate time are related by $\tau_i = \gamma_i t_i$ in the branch labelled by $i$. 
In such branch, the Hamiltonians are related by $H^{(D)} = H^{\tau_i} = dt/d\tau H^{t_i} = H^{t_i}/\gamma_i$. 
Therefore, the energies are related by \begin{equation}
    E^{(D)} = E^{\tau,i} = \frac{1}{\gamma_i(R_i)} E^{t_i}.
\end{equation} 
Moving to AdS--Rindler coordinates, we find \begin{equation}
     E^{(D)} = E^{\tau,i} = \frac{1}{\widetilde{\gamma_i}(R_i)} E^{\overline{t}_i}.
\end{equation}

\ifSubfilesClassLoaded{ 
  \printbibliography[heading=bibintoc, title=References]
  }{}
\end{document}

\backmatter


\printbibliography[heading=bibintoc, title=References]


\end{document}